%% file: tese.tex
\documentclass[11pt,twoside,a4paper]{book}

\usepackage{bm} 
\usepackage{amsmath}
\usepackage{amssymb}
\usepackage{epsfig}
\usepackage{xr}
\usepackage{colordvi}
\usepackage{float}
\usepackage{minitoc}
\usepackage{footnpag}
\usepackage{fancyheadings}
\usepackage{pifont}
\usepackage{rotating}
\usepackage{picinpar}
\def\spa#1{\phantom{\fbox{\rule[-#1cm]{0cm}{0cm}}}}

\addtocounter{secnumdepth}{1}

\newcommand{\clearemptydoublepage}{\newpage{\pagestyle{empty}\cleardoublepage}}

\begin{document}

\dominitoc

\renewcommand{\thepage}{\roman{page}}
\pagestyle{empty}

\begin{center}
\textsc{\bf \hskip 2cm \large UNIVERSIDADE T\'ECNICA DE LISBOA \\
\vskip 0.5cm \hskip 2cm INSTITUTO SUPERIOR T\'ECNICO}
\end{center}

\begin{figure}[h]
\epsfxsize=3cm \epsfysize=3cm
\begin{center}
\vskip -0.5cm
\end{center}
\end{figure}

\begin{center}
\vskip 3cm

\hskip 2cm \LARGE \bf Black Hole Solutions \\
\hskip 2cm and\\
\hskip 2cm Pair Creation of Black Holes \\
\hskip 2cm in\\
\hskip 2cm Three, Four and Higher Dimensional Spacetimes

\vspace{3cm}
{\hskip 2cm \LARGE \bf \'Oscar Jo\~ao Campos Dias} \\

\vskip 2cm {\hskip 2cm \large Disserta\c c\~ao para obten\c c\~ao
do Grau de Doutor em F\'{\i}sica}

\vskip 0.5cm \hskip 2cm (PhD Thesis)

 \vskip 1cm \hskip 2cm {\large December 2003}

\normalsize\rm


\vspace{1cm} \textsc{\bf Jury:}

 \vskip 0.5cm \vfill

{\bf
\begin{tabular}{ll}
Luis \textsc{Bento} &Examinator\\
Stanley  \textsc{Deser} &Examinator\\
Jorge \textsc{Dias de Deus} &President of the Jury\\
Alfredo Barbosa \textsc{Henriques} &Examinator\\
Jos\'e Pizarro de Sande e  \textsc{Lemos} &Supervisor \\
Jorge \textsc{Rom\~ao} &Examinator\\
\end{tabular}
}
 \vfill

\end{center}
\clearemptydoublepage

\thispagestyle{empty}

\begin{center}
\textsc{\bf \hskip 2cm \large UNIVERSIDADE T\'ECNICA DE LISBOA \\
\vskip 0.5cm \hskip 2cm INSTITUTO SUPERIOR T\'ECNICO}
\end{center}

\begin{figure}[h]
\epsfxsize=3cm \epsfysize=3cm
\begin{center}
\vskip -0.5cm
\end{center}
\end{figure}

\begin{center}
\vskip 3cm

\hskip 2cm \LARGE \bf Black Hole Solutions \\
\hskip 2cm and\\
\hskip 2cm Pair Creation of Black Holes \\
\hskip 2cm in\\
\hskip 2cm Three, Four and Higher Dimensional Spacetimes

\vspace{3cm}
{\hskip 2cm \LARGE \bf \'Oscar Jo\~ao Campos Dias} \\

\vskip 2cm {\hskip 2cm \large Disserta\c c\~ao para obten\c c\~ao
do Grau de Doutor em F\'{\i}sica}

\vskip 0.5cm \hskip 2cm (PhD Thesis)

 \vskip 1cm \hskip 2cm {\large December 2003}

\normalsize\rm


\vspace{1cm} \textsc{\bf Jury:}

 \vskip 0.5cm \vfill
{\bf
\begin{tabular}{ll}
Luis \textsc{Bento} &Examinator\\
Stanley  \textsc{Deser} &Examinator\\
Jorge \textsc{Dias de Deus} &President of the Jury\\
Alfredo Barbosa \textsc{Henriques} &Examinator\\
Jos\'e Pizarro de Sande e  \textsc{Lemos} &Supervisor \\
Jorge \textsc{Rom\~ao} &Examinator\\
\end{tabular}
}
 \vfill

\end{center}


\thispagestyle{empty}

\author{\textsc{\'Oscar Jo\~ao Campos Dias}}
 \title{ \huge{
{\bf
Black Hole Solutions \\
and\\
Pair Creation of Black Holes \\
in\\
Three, Four and Higher Dimensional Spacetimes
 }}}
\maketitle

\renewcommand{\thefootnote}{\alph{footnote}}

\pagestyle{fancyplain}
\addtolength{\headwidth}{\marginparsep}
\addtolength{\headwidth}{\marginparwidth}
\renewcommand{\chaptermark}[1]{\markboth{#1}{}}
\renewcommand{\sectionmark}[1]{\markright{\thesection\ #1}}
\lhead[]{\fancyplain{}{\bfseries\leftmark}}
\rhead[\fancyplain{}{\bfseries \rightmark}]{}
\lfoot[\fancyplain{}{\bfseries\thepage}]{}
\rfoot[]{\fancyplain{}{\bfseries\thepage}}
\cfoot{}

\setlength{\headrulewidth}{1pt}

\include{Resumo}

\include{Abstract}

\include{fct}

\include{Acknowledgements}


\clearemptydoublepage \lhead[]
      {\fancyplain{}{\bfseries Table of Contents}}
\rhead[\fancyplain{}{\bfseries Table of Contents}]
      {}
\tableofcontents \clearemptydoublepage



\normalsize

\nocite{*}


\clearemptydoublepage
\newpage
\pagestyle{fancyplain} \thispagestyle{empty}

\lhead[]
      {\fancyplain{}{\bfseries Preface}}
\rhead[\fancyplain{}{\bfseries Preface}]
      {}
\setlength{\headrulewidth}{1pt} \nocite{*} \thispagestyle{empty}
\addcontentsline{toc}{section}{\numberline{}{\bf \hskip -0.8cm
Preface}}

\begin{center}{\Large \bf Preface}
\end{center}
The research included in this thesis has been carried out at
Centro Multidisciplinar de Astrof\'{\i}sica (CENTRA) in the
Physics Department of Instituto Superior T\'ecnico.  I declare
that this thesis is not substantially the same as any that I have
submitted for a degree or diploma or other qualification at any
other University and that no part of it has already been or is
being concurrently submitted for any such degree or diploma or any
other qualification.

Chapter \ref{chap:BTZ family} was done in collaboration with
Professor Jos\'e Lemos and Dr. Carlos Herdeiro. Chapters
\ref{chap:3D Dilaton BH}-\ref{chap:Pair creation} are the outcome
of collaborations with Professor Jos\'e Lemos. Chapter
\ref{chap:Black holes in higher dimensions} was done in
collaboration with Professor Jos\'e Lemos, Vitor Cardoso and Nuno
Santos. Chapters \ref{chap:Pair creation in higher dimensions} and
\ref{chap:Grav Radiation} were done in collaboration with
Professor Jos\'e Lemos and Vitor Cardoso. All these chapters have
been submitted with minor modifications for publication.

A list of the works published included in this thesis are listed
below.

\vspace{0.1 cm} \noindent{\Large -}
 O. J. C. Dias, J. P. S.
Lemos, {\it Rotating magnetic solution in three dimensional
Einstein gravity}, JHEP {\bf 0201}: 006 (2002); {\tt
[hep-th/0201058]} (Chapter 2).

\vspace{0.1 cm} \noindent{\Large -} V. Cardoso, S. Yoshida, O. J.
C. Dias, J. P. S. Lemos, {\it Late-time tails of wave propagation
in higher dimensional spacetimes}, Phys. Rev. D {\bf 68}, 061503
(2003) [Rapid Communications]; {\tt [hep-th/0307122]} (mentioned
in Chapter 2).

\vspace{0.1 cm} \noindent{\Large -} O. J. C. Dias, J. P. S. Lemos,
{\it Static and rotating electrically charged black holes in
three-dimensional Brans-Dicke gravity theories}, Phys. Rev. D {\bf
64}, 064001 (2001); {\tt [hep-th/0105183]}  (Chapter 3).

\vspace{0.1 cm} \noindent{\Large -} O. J. C. Dias, J. P. S. Lemos,
{\it Magnetic point sources in three dimensional Brans-Dicke
gravity theories}, Phys. Rev. D {\bf 66}, 024034 (2002); {\tt
[hep-th/0206085]}  (Chapter 3).

\vspace{0.1 cm} \noindent{\Large -} O. J. C. Dias, J. P. S. Lemos,
{\it Magnetic strings in anti-de Sitter general relativity},
Class. Quantum Grav. {\bf 19}, 2265 (2002); {\tt [hep-th/0110202]}
(mentioned in Chapter 4).

\vspace{0.1 cm} \noindent{\Large -} O. J. C. Dias, J. P. S. Lemos,
{\it Pair of accelerated black holes in a anti-de Sitter
background: the AdS C-metric}, Phys. Rev. D {\bf 67}, 064001
(2003); {\tt [hep-th/0210065]}
 (Chapter 6).

\vspace{0.1 cm} \noindent{\Large -} O. J. C. Dias, J. P. S. Lemos,
{\it Pair of accelerated black holes in a de Sitter background:
the dS C-metric}, Phys. Rev. D {\bf 67}, 084018 (2003); {\tt
[hep-th/0301046]}
 (Chapter 6).

\vspace{0.1 cm} \noindent{\Large -} O. J. C. Dias, J. P. S. Lemos,
{\it The extremal limits of the C-metric: Nariai,
Bertotti-Robinson and anti-Nariai C-metrics}, Phys. Rev. D{\bf 68}
(2003) 104010; {\tt [hep-th/0306194]}
 (Chapter 7).

\vspace{0.1 cm} \noindent{\Large -} O. J. C. Dias, J. P. S. Lemos,
{\it False vacuum decay: Effective one-loop action for pair
creation of domain walls}, J. Math. Phys. {\bf 42}, 3292 (2001);
{\tt [hep-ph/0103193]}
  (Chapter 8).

\vspace{0.1 cm} \noindent{\Large -} O. J. C. Dias, {\it Pair
creation of particles and black holes in external fields}, {\tt
[gr-qc/0106081]}  (Chapter 8).

\vspace{0.1 cm} \noindent{\Large -} O. J. C. Dias, J. P. S. Lemos,
{\it Pair creation of de Sitter black holes on a cosmic string
background}, Phys. Rev. D{\bf 69} (2004) 084006; {\tt
[hep-ph/0310068]} (Chapter 9).

\vspace{0.1 cm} \noindent{\Large -}  O. J. C. Dias, {\it Pair
creation of anti-de Sitter black holes on a cosmic string
background}, Phys. Rev. D{\bf 70} (2004) 024007  (Chapter 9).

\vspace{0.1 cm} \noindent{\Large -} V. Cardoso, O. J. C. Dias, J.
P. S. Lemos, {\it Nariai, Bertotti-Robinson and anti-Nariai
solutions in higher dimensions}, Phys. Rev. D{\bf 70} (2004)
024002 (Chapter 10).

\vspace{0.1 cm} \noindent{\Large -} N. L. Santos, O. J. C. Dias, 
J. P. S. Lemos, {\it Global embedding Minkowskian spacetime
procedure in higher dimensional black holes: Matching between
Hawking temperature and Unruh temperature}, Phys. Rev. D, in press
(2004) (mentioned in Chapter 10).

\vspace{0.1 cm} \noindent{\Large -} O. J. C. Dias, J. P. S. Lemos,
{\it Pair creation of higher dimensional black holes on a de
Sitter background}, Phys. Rev. D, submitted (2004); {\tt
[hep-th/0410279]} (Chapter 11).

\vspace{0.1 cm} \noindent{\Large -} V. Cardoso, O. J. C. Dias, J.
P. S. Lemos, {\it Gravitational radiation in D-dimensional
spacetimes}, Phys. Rev. D {\bf 67}, 064026 (2003); {\tt
[hep-th/0212168]} (Chapter 12).



\renewcommand{\thepage}{\arabic{page}}

\setcounter{page}{1}

\include{Overview}
\part{\Large{Black holes and pair creation in 3 dimensions}} \label{part1}
 \include{Chapter1} 
 \include{Chapter2} 
 \include{Chapter3} 
\part{\Large{Black holes and pair creation in 4 dimensions}} \label{part2}
 \include{Chapter4} 
 \include{Chapter5} 
 \include{Chapter6} 
 \include{Chapter7} 
 \include{Chapter8} 
\part{\Large{Black holes and pair creation
in higher dimensions}} \label{part3}
 \include{Chapter9} 
 \include{Chapter10} 
 \include{Chapter11} 

\clearemptydoublepage
\setcounter{page}{261} \pagestyle{fancyplain}   
\thispagestyle{empty} \lhead[]
      {\fancyplain{}{\bfseries Bibliography}}
\rhead[\fancyplain{}{\bfseries Bibliography}]
      {}
\nocite{*} \thispagestyle{empty}
\addcontentsline{toc}{chapter}{\numberline{}Bibliography}
\bibliographystyle{acm}
\bibliography{BibThese}


\end{document}

%% file: Resumo.tex
\thispagestyle{empty}
 \chapter*{Resumo}


Buracos negros s\~ao importantes em astrof\'{\i}sica, uma vez que
resultam do colapso gravitacional de uma estrela massiva ou
aglomerados de estrelas, e em f\'{\i}sica porque revelam
propriedades da f\'{\i}sica fundamental, como propriedades
termodin\'amicas e qu\^anticas da gravita\c c\~ao.

Para melhor se entender a f\'{\i}sica dos buracos negros s\~ao
necess\'arias solu\c c\~oes exactas que des-crevem um ou v\'arios
buracos negros. Nesta tese estudamos solu\c c\~oes exactas em
tr\^es, quatro e dimens\~oes mais altas. O estudo em tr\^es
dimens\~oes justifica-se pela simplifica\c c\~ao do problema,
enquanto que a discuss\~ao em dimens\~oes superiores a quatro se
justifica porque diversas teorias indicam que existem dimens\~oes
extra no universo. Nesta tese, em qualquer das dimens\~oes acima
indicadas, estudamos solu\c c\~oes exactas com um s\'o buraco
negro e solu\c c\~oes exactas que representam um par de buracos
negros acelerados. Estas \'ultimas solu\c c\~oes s\~ao ent\~ao
usadas para estudar em detalhe o processo qu\^antico de cria\c
c\~ao de pares de buracos negros num campo externo. Tamb\'em
determinamos a radia\c c\~ao gravitacional emitida durante este
processo de cria\c c\~ao.

\vspace{4cm} {\normalfont\sffamily\bfseries%
\noindent PALAVRAS-CHAVE:} \\
\noindent Solu\c c\~oes exactas de buracos negros; Pares de
buracos negros acelerados, M\'etrica-C; Cria\c c\~ao qu\^antica de
pares de buracos negros; Radia\c c\~ao gravitacional; Espa\c
cos-tempo em $D$ dimens\~oes; Espa\c cos de fundo com constante
cosmol\'ogica.

%% file: Abstract.tex
\thispagestyle{empty}
 \chapter*{Abstract}

Black holes, first found as solutions of Einstein's General
Relativity, are important in astrophysics, since they result from
the gravitational collapse of a massive star or a cluster of
stars, and in physics since they reveal properties of the
fundamental physics, such as thermodynamic and quantum properties
of gravitation.

In order to better understand the black hole physics we need exact
solutions that describe one or more black holes. In this thesis we
study exact solutions in three, four and higher dimensional
spacetimes. The study in 3-dimensions is important due to the
simplification of the problem, while the discussion in higher
dimensions is essential due to the fact that many theories
indicate that extra dimensions exist in our universe. In this
thesis, in any of the dimensions  mentioned above, we study exact
solutions with a single black hole and exact solutions that
describe a pair of uniformly accelerated black holes, with the
acceleration source being well identified. This later solutions
are then used to study in detail the quantum process of black hole
pair creation in an external field. We also compute the
gravitational radiation released during this pair creation
process.


\vspace{4cm} {\normalfont\sffamily\bfseries%
\noindent KEY-WORDS:} \\
\noindent Exact black hole solutions; Pair of accelerated black
holes, C-metric; Pair creation of black holes; Gravitational
radiation; $D$-dimensional spacetimes; Cosmological constant
backgrounds.

%% file: fct.tex
\thispagestyle{empty}
 \chapter*{}

\vspace*{14cm}

 \noindent Este trabalho foi financiado pela
Funda\c{c}{\~a}o Para a Ci{\^e}ncia e Tecnologia (FCT),\\
sob o contrato Praxis XXI/BD/21282/99 (01/01/2000 - 31/12/2003).

\vspace*{2cm} \noindent This work was
supported by Funda\c{c}{\~a}o Para a Ci{\^e}ncia e Tecnologia (FCT),\\
under the grant Praxis XXI/BD/21282/99 (01/01/2000 - 31/12/2003).

%% file: Acknowledgements.tex
\thispagestyle{empty}
\chapter*{Acknowledgements}

In a PhD thesis an acknowledgement must be addressed to the
corresponding supervisor. My first acknowledgement goes indeed to
my supervisor, Jos\'e P. S. Lemos, not simply because I have to do
it but specially because I am really profoundly grateful to him.
This friend of mine taught me how research work is done, and when
difficulties appeared he was always present with his scientific
help, with his suggestions in the right direction, and with his
motivating word. I cannot avoid feeling that this acknowledgement
cannot express all my gratitude to him.

Almost fifteen years ago, a clever boy appeared in my high school
class. What begun as a stimulating loyal struggle for a good
school result, revealed to be some time later a very good
friendship. We have passed through a lot during this common path.
I have learned a lot with you my good friend Vitor Cardoso; I hope
I have given back something to you. Thank you also to you Ricardo
Marques, for the good moments during the high school.

My PhD research was done at Centro Multidisciplinar de
Astrof\'{\i}sica (CENTRA). I particularly acknowledge Ana
Mour\~ao, Alfredo Barbosa Henriques, Dulce, and Jorge Dias de Deus
for the excellent environment that you have proportionated to me,
and for being present in the moments I needed you.

Thank you to you Carlos Herdeiro for the nice discussions we had.
Although they occurred in a short period of time, I have learned a
lot with them. A special acknowledgement also to you Shijun
Yoshida for your friendship and for our clever comments and
discussions.

When I was in my fourth year of graduation, I started giving
practical lessons. I will never forget all the support received
from the professors to whom I first worked. Ana Branquinho, Pedro
Brogueira, Teresa Pe\~na, and later, M\'ario Pimenta, thank you
indeed for all the respect, for the care, and specially for the
advises you always gave to me. Unfortunately, a sincerely thanks
is probably all I have to give to you. I wish I could give more.

Thank you to you mother and father, specially for having supported
always my options, for your education and for your love. Thank you
also to you my brother, specially because you have taught me that
when we fight for something we really want, we can reach it.

Thank you my friends Alexandre Mestre, \'Alvaro Santos, Elisa Vaz,
Maria Rui, Miguel Meira, Nuno Cristino, Patr\'{\i}cia Maldito,
Pedro Rego, and Tiago Gomes. I have always counted on you. A warm
thank to you my good friend Miguel Quintas. I hope you know how
important you are to me.

Four years ago I decided not to go abroad because of you Ana Mei
Lin. Now, more than ever, I know I have taken the right decision.
Thank you for these beautiful years. Everything else, I tell you
personally.

%% file: Overview.tex
\thispagestyle{empty} \setcounter{minitocdepth}{2}
\chapter{\Large{Overview}} \label{Overview}
 \lhead[]{\fancyplain{}{\bfseries Chapter \thechapter. \leftmark}}
 \rhead[\fancyplain{}{\bfseries Overview \rightmark}]{}
  \minitoc \thispagestyle{empty}
\renewcommand{\thepage}{\arabic{page}}

\addtocontents{lof}{\textbf{Chapter Figures \thechapter}\\}


\section[Overview on black holes]{\large{Overview on black holes}}
 \label{sec:Historical Overview}

Black holes have begun to be entities of theoretical interest when
they have been found as exact solutions of the Einstein equations
with peculiar features. Although the actual universal name ``black
hole" has been introduced only in 1968 by Wheeler\footnote{In this
overview section when we refer to a work we will only specify the
author and the year of publication. Being reference works  they
can easily be found in the books.  For this purpose we suggest,
e.g., the book of Frolov and Novikov \cite{FrolovNovikov}.}, the
first black hole solution has been found by Schwarzschild in 1916,
just a few months after the publication of General Relativity by
Einstein. As proved by Birkhoff in 1923, the Schwarzschild
solution describes the external gravitational field generated by a
static spherical mass. It was much later found that one is in the
presence of a black hole when the mass $M$ is inside a critical
radius, $r_+=\frac{2GM}{c^2}\sim 3 \frac{M}{M_{\odot}}\,{\rm km}$
(where $G$ is Newton's constant and $c$ is the light velocity),
known as the event horizon of the black hole. This horizon acts
like an one-way membrane: particles and radiation can cross it
from outside, but they cannot escape from its interior. An
observer that is at rest at infinity sees an infalling observer
taking an infinite amount of time to approach the horizon but,
according to the infalling observer, his crossing through the
horizon is done in a finite amount of time and, once he has done
this fatal step, he is unavoidably pushed into the central
curvature singularity at $r=0$ where he suffers an infinite tidal
force. These properties of the Schwarzschild black hole are best
synthesized in its Carter-Penrose diagram, Fig. \ref{Fig
Schw_rays-introduction}. These kind of causal diagrams have been
progressively developed by works of Eddington, Finkelstein,
Kruskal, Carter and Penrose, and we will make a first contact with
them here since they will be very useful along this thesis.
\begin{figure}[H]
\centering
\includegraphics[height=4cm]{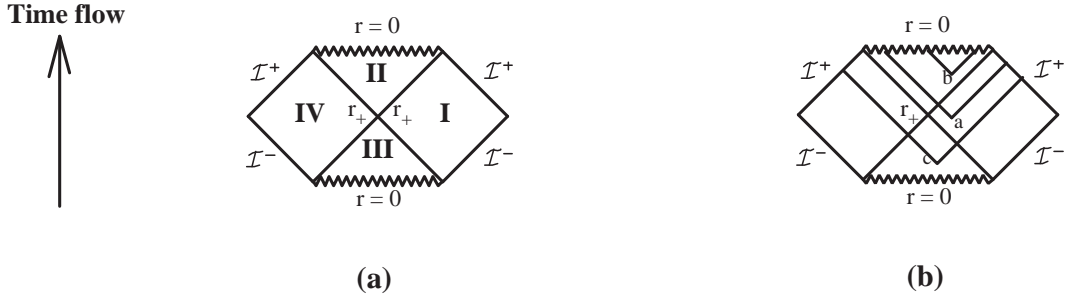}
   \caption{\label{Fig Schw_rays-introduction}
Carter-Penrose diagram of the Schwarzschild black hole. Time flows
into the top of the diagram, and light rays move necessarily along
$45^{\rm o}$ lines. The zigzag line represents the $r=0$ curvature
singularity, ${\cal I}$ represents the infinity $r=\infty$ (whose
points have been brought to a finite position by an appropriate
coordinate transformation), and the two mutually perpendicular
lines, $r_+$, represent the black hole event horizon. (a) Region
II represents the black hole interior, region III is the white
hole interior, and regions I and IV represent the exterior space,
$r_+<r<+\infty$. (b) The paths that a light ray can follow
determine the properties of the region from where they are
emitted. We observe that in fact these diagrams represent a
dynamic wormhole solution.
 }
\end{figure}
In the Carter-Penrose diagrams, a null particle (e.g., a light
ray) moves necessarily towards the top along $45^{\rm o}$ lines,
while timelike particles can move into the top of the diagrams
only along a curve whose tangent vector must do an angle less than
$45^{\rm o}$ with the vertical line. The curvature singularity,
$r=0$, of the Schwarzschild black hole is represented by a zigzag
line in Fig. \ref{Fig Schw_rays-introduction}. Region IV is
equivalent to region I, and both represent the region outside the
black hole horizon, $r_+<r<+\infty$. A null ray that is sent from
a point $a$ in these regions can move towards $r=0$, after
crossing $r_+$, but it is also free to move into the future
infinity ${\cal I}^+$ [see Fig. \ref{Fig
Schw_rays-introduction}.(b)]. Note that regions I and IV are
casually disconnected: no null or timelike particle can start in
region I and reach region IV. Region II is interpreted as the
interior of the black hole horizon, since a null ray emitted from
a point in its interior [e.g., point $b$ in Fig. \ref{Fig
Schw_rays-introduction}.(b)] necessarily hits the future curvature
singularity, and cannot cross the horizon towards region I or IV
(the discussion also applies to timelike particles). Region III is
interpreted as the interior of the white hole, since any particle,
e.g. the light ray that is emitted from point $c$ in Fig. \ref{Fig
Schw_rays-introduction}.(b), is necessarily expelled out from
region III, i.e., it necessarily crosses the horizon towards
region I or IV. We remark that regions III and IV were not present
in the original solution of Schwarzschild. They appear after the
realization that this original solution could be extended.

The inclusion of an electric charge $Q$ also yields an exact
static spherical solution which is known as the
Reissner-Nordstr\"{o}m black hole (1918) when $Q^2\leq G M^2$.
Finally, an angular momentum $J$ can be added to the system
yielding a stationary, axisymmetric solution which when $Q=0$ is
known as the Kerr solution (1963), and when $Q\neq 0$ is called
Kerr-Newman solution (1965). These solutions describe a black hole
when the condition $\frac{J^2 c^2}{G^2 M^2}+\frac{Q^2}{G}\leq M^2$
is satisfied. When the equality holds in the above relations we
have the extreme black hole solutions which have zero temperature.
When the above constraints are violated we have a naked
singularity, a visible singularity not surrounded by a horizon.
The uniqueness theorems (whose proofs have been given by Israel,
Robinson and Carter, among others in 1968 and in the following
decade) state that the only static or stationary solutions of the
Einstein-Maxwell equations that are asymptotically flat and have
regular horizons are the above solutions characterized only by the
parameters $M$, $Q$ and $J$. All the other parameters that
specified the initial state before the formation of the black hole
are radiated away during the creation process. This simple
description of a black hole is well summarized by the well-known
metaphoric statement of Wheeler: ``a black hole has no hair". The
construction of a formalism to compute these hairs $M$, $Q$ and
$J$ has been carried by Arnowitt, Deser and Misner (1963), by
Regge and Teiltelboim (1974), and by Brown and York (1993), among
others. In the seventies, Price, Wald, Chandrasekhar and Detweiler
have also proven that the above black holes are stable with
respect to small perturbations \cite{Chandrasekhar}.

The theory of black holes has been strongly connected to the
theory of gravitational collapse after the work of Oppenheimer and
Snyder (1939) and Penrose's theorem (1965): a realistic, slightly
non-spherical complete collapse leads unavoidably to the formation
of a black hole and a singularity (for a detailed description of
the theory of gravitational collapse see the work of Harrison,
Thorne, Wakano and Wheeler \cite{HarrisonThorneWakanoWheeler}).
Oppenheimer and Volkoff (1939) have shown that when a neutron star
(first predicted by Zwicky in 1934) forms, after the explosion of
the massive progenitor star, its mass must be smaller than $M_{\rm
OV} \sim 3 M_{\odot}$ (in fact Oppenheimer and Volkoff first found
the value $0.7 M_{\odot}$, but a more accurate account on the
nuclear processes yields the value $3 M_{\odot}$). This is the
maximum mass that the pressure of neutron degenerate matter can
support against further gravitational collapse. The existence of a
maximum mass for degenerate matter had been realized previously by
Chandrasekhar (1935), in the case of electron degenerate matter.
He has shown that white dwarfs must have a mass smaller than
$M_{\rm C}\sim 1.4 M_{\odot}$. Thus, black holes formed thought
the gravitational collapse of stellar matter have necessarily a
mass $M> 3M_{\odot}$. In the context of gravitational collapse
towards a black hole, two important conjectures have been
formulated, the cosmic censorship and the hoop conjectures. The
cosmic censorship conjecture (Penrose, 1969) forbids the existence
of naked singularities, while the hoop conjecture (Thorne, 1972)
states that black holes form when and only when a mass $M$ gets
compacted into a region whose circumference in every direction is
less than its Schwarzschild circumference $\frac{4\pi GM}{c^2}$.
During the last decade, a great effort has been initiated in order
to detect the energetic astrophysical processes predicted to be
powered by black holes. Mainly, the astrophysics are looking into
the X-ray emitting sources that can be found in the accretion of
matter into a binary stellar system and into an active galactic
nuclei. These efforts have by now already provided a set of
serious black hole candidates in a binary system with masses that
belong to the range $5 M_{\odot}-20 M_{\odot}$ (where $M_{\odot}$
is the solar mass), and in an active galactic nuclei with masses
that belong to the supermassive range $10^6 M_{\odot}-10^{10}
M_{\odot}$.

Black holes have entered in the domain of physics during the 1970s
when it has been shown that they are thermodynamic objects. Indeed
Bardeen, Carter and Hawking (1973), supported by previous work of
Christodoulou (1970) and Penrose (1971), have shown that black
holes obey the so called four laws of black hole mechanics. The
first one is known as the zero law and states that the surface
gravity is constant along the horizon of the black hole, even when
this is not spherically symmetric as occurs with the Kerr
solution. The surface gravity, $k_{\rm h}$, can be interpreted as
the force that must be exerted on a rope at infinite to hold a
unit mass at rest near the horizon of the black hole. The first
law (an energy conservation law) states that when one throws an
infinitesimal amount of matter into a stationary black hole
described by $M$, $J$ and $Q$, it will evolve into a new
stationary black hole in such a way that the change in the hairs
of the system satisfies
\begin{eqnarray}
c^2 dM= \frac{k_{\rm h} c^2}{8\pi G}\,d{\cal A} + \Omega_{\rm h}
dJ +\Phi_{\rm h} dQ \,,
 \label{first law}
\end{eqnarray}
where ${\cal A}$ is the surface area of the horizon, $\Omega_{\rm
h}$ is the angular velocity at the horizon, and $\Phi_{\rm h}$ is
the electric potential at the horizon. The second law states that
the area of the black hole horizon cannot decrease during any
physical process, $d {\cal A}\geq 0$ (this classical law was later
replaced by the generalized second law which states that the sum
of the black hole entropy plus the exterior matter entropy can
never decrease). Finally, the third law states that  it is
impossible to reduce the surface gravity to zero by a finite
number of processes. The surface gravity of the Kerr-Newman black
hole is given by
\begin{eqnarray}
k_{\rm h}= \frac{4\pi c^2}{{\cal
A}}\sqrt{\frac{G^2M^2}{c^4}-\frac{J^2}{M^2c^2}-\frac{GQ^2}{c^4}}
\,,
 \label{law 4}
\end{eqnarray}
and thus $k_{\rm h}=0$ implies $\frac{J^2 c^2}{G^2
M^2}+\frac{Q^2}{G}= M^2$, which as we saw above is the condition
for an extreme black hole. If the state $k_{\rm h}<0$ could be
reached we would have a naked singularity. Hence, the cosmic
censorship conjecture stated by Penrose plays the role of the
third law.

These laws are purely classical and resemble the usual four laws
of thermodynamics, if one admits that ${\cal A}$ is proportional
to the entropy, as suggested by Bekenstein (1973,1974), and that
$k_{\rm h}$ is proportional to the black hole temperature. At
first, this resemblance was only an analogy since if the black
hole had a temperature it would have to radiate, in a clear
contradiction with the known classical fact that nothing could
escape from the black hole horizon. However, in a revolutionary
work, Hawking (1975), using a semiclassical treatment in which the
gravitational field of the black hole is treated classically but
the matter is treated quantum mechanically, has shown that the
black holes do indeed radiate a spectrum characteristic of a
blackbody with a Hawking temperature given by
\begin{eqnarray}
T_{\rm h}=\frac{\hbar k_{\rm h}}{2\pi k_B} \,,
 \label{TempHawk}
\end{eqnarray}
where $\hbar=h/(2\pi)$ is the Planck constant, $k_B$ is the
Boltzmann constant, and the associated Bekenstein-Hawking entropy
is
\begin{eqnarray}
S_{\rm h}=\frac{{\cal A}_{\rm h}}{4}\frac{k_B c^3}{\hbar G} \,.
 \label{Entropy}
\end{eqnarray}
Actually, the spectrum is not a perfect blackbody one due to the
so called greybody factor: the emission rate of particles is
proportional to the cross-section for a particle to be absorbed by
the black hole and this cross-section is not constant, which leads
to deviations from the pure blackbody emission spectrum.
 For a Schwarzschild back hole we have $T_{\rm h}=\frac{\hbar
c^3}{8\pi G k_B}\frac{1}{M} \sim 10^{-7}\frac{M_{\odot}}{M}\: {\rm
K}$ and $S_{\rm h} = \frac{4\pi G k_B}{\hbar c}M^2\sim
10^{53}\left ( \frac{M}{M_{\odot}} \right )^2 \: {\rm Js^{-1}}$.
At this point some remarks are justified. First note that the four
fundamental constants of nature, $c$, $G$, $k_B$ and $\hbar$ (that
are the fingerprints of relativity, gravitation, statistical
mechanics and quantum mechanics) are, for the first time, unified
in a single formula. Second, in standard statistical mechanics,
the entropy of a system originates in the counting of quantum
states which are macroscopically indistinguishable. Now, the
Bekenstein-Hawking entropy of a black hole has not yet been fully
explained as a consequence of such state counting. Third, note
that the Hawking temperature of the Schwarzschild back hole is
inversely proportional to the mass. Thus its heat capacity is
negative and the black hole is quantum mechanically unstable since
as it loses mass, its temperature grows and the particle emission
rate grows (the black hole evaporates). The rate of mass loss per
unit time can be estimated using the Stefan-Boltzmann law,
$-\frac{dE}{dt} \sim \sigma {\cal A}_{\rm h} T_{\rm h}^4$ with
$E=Mc^2$ and $\sigma=\pi^2 k_B^4/(60\hbar^3 c^2)$,
\begin{eqnarray}
-\frac{dM}{dt} \sim \frac{\hbar c^6}{G^2} \frac{1}{M^2} \sim
10^{-44}\left ( \frac{M_{\odot}}{M} \right )^2\:{\rm kg\,
s^{-1}}\,,
 \label{mass loss}
\end{eqnarray}
to which corresponds an evaporation lifetime given by
\begin{eqnarray}
\tau\sim \frac{G^2}{\hbar c^4}M^3 \sim 10^{71}\left (
\frac{M}{M_{\odot}} \right )^3\:{\rm s} \,.
 \label{lifetime}
\end{eqnarray}
Third, note that although the Hawking evaporation would constitute
a definite proof of the black hole nature of a compact object, it
will be very difficult to be ever observed by astronomical means.
Indeed, an astrophysical black hole with a mass $M\sim M_{\odot}$
has a Hawking temperature of only $T_{\rm h}\sim 10^{-7}\,{\rm K}$
to which corresponds a total dissipating power of
$|d(Mc^2)/dt|\sim 10^{-27}\, {\rm W}$.

If one has some hope in observing the Hawking evaporation one must
then look for possible physical processes that lead to the
formation of mini-black holes with Planck sizes ($\ell_{\rm
Pl}\sim 1.616\times 10^{-33}\,{\rm cm}$, $m_{\rm Pl}\sim
2.177\times 10^{-5}\,{\rm g}$). For example, a black hole with
$M=10^3 m_{\rm Pl}$ would have a temperature $T_{\rm h}\sim
10^{28}\,{\rm K}$ and a total dissipating power of $\sim 10^{43}\,
{\rm W}$, and thus the Hawking evaporating process would be
extremely energetic and passible of being observed. A process that
might lead to the formation of such Planckian black holes has been
proposed by Zel'dovich and Novikov (1967), and by Hawking (1971).
These black holes are called primordial since they can have been
produced only in the very early universe when very large local
deviations from homogeneity were possible. Let $t$ be the time
elapsed since the Big Bang and $R$ the size of a fluctuation in
the metric, i.e., of an inhomogeneity in the metric that describes
the universe. Then, when $c t$ is of the order of $R$, the
gravitational forces can locally stop the cosmic expansion of an
agglomerate of matter and reverse it into complete collapse if the
self gravitational potential energy of the matter exceeds the
internal energy: $\frac{GM^2}{R}\geq p R^3$, where $p$ is the
pressure. Now, during the radiation era the relation between the
pressure and the density is $p\sim \rho c^2$ and, according to the
Einstein-de Sitter model of the early universe, the density and
the time are related by $G\rho \sim t^{-2}$. From these relations
one finds that the mass of the primordial black hole is related to
the cosmic time by $M\sim \frac{c^3}{G} t \sim 10^{35}\,t$. Thus,
at the Planck time, $t\sim 10^{-43} {\rm s}$, mini-black holes may
form with the Planck mass $M\sim 10^{-8} {\rm kg}$, at time $t\sim
10^{-4} {\rm s}$, black holes with $M_{\odot}\sim 10^{30} {\rm
kg}$ may be produced, and at the time of nucleosynthesis, $t\sim
100\,{\rm s}$, supermassive black holes with $10^7 M_{\odot}$ may
be created. Now, some primordial mini-black holes should by now be
in the last stage of the Hawking evaporation process, and thus we
should be detecting the highly energetic $\gamma$-ray bursts
released during this exploding process. These events have not been
observed, which indicates that the average density of the
primordial black holes is low or zero. On the other side the
hypothesis that the active galactic nuclei are feeded by the
accretion of matter into central supermassive black holes with
masses $10^6 M_{\odot}-10^{10} M_{\odot}$, as suggested by
Lynden-Bell (1969), may favor the rapid formation of supermassive
black holes in the early universe.

\begin{figure} [H]
\centering
\includegraphics[height=1.5in]{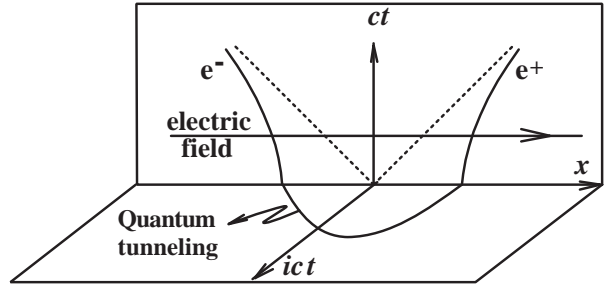}
\caption{\label{PairCreat_introduct-fig}
 During the imaginary time
$i t$ a quantum tunnelling process occurs, with the energy for the
materialization being provided by the external electric field.
This leads to the creation (at $t=0$) of a particle-antiparticle
pair ($e^-e^+$) at $x=\pm \frac{\ell}{2}$. This pair is then
separated apart by the external Lorentz force, describing the
hyperbolic uniform accelerated motion, $x^2-c^2t^2=(\ell /2)^2$.
The semi-circle trajectory that represents the tunnelling process
is obtained by euclideanizing the time ($t\rightarrow i t$) in the
hyperbolic equation of motion, yielding $x^2+c^2t^2=(\ell /2)^2$.
 }
\end{figure}

Another process that allows the formation of mini-black holes is
the gravitational analogue of the Schwinger (1951) quantum process
of pair creation of particles in an external electric field.
Recall that this Schwinger process is based on the fact that
virtual, short-lived, particles are constantly being created and
rapidly annihilated in the physical vacuum, so that it is stable.
However, in the presence of an external electric field, some of
these particle-antiparticle pairs may receive enough energy to
materialize and become real. This leads to the quantum creation of
the pair that is then accelerated away also by the external
Lorentz force. At the heuristic level one can estimate the
probability of this pair creation process as follows. Let ${\cal
E}$ be the external field strength, and $e$ and $m$ the charge and
mass of the particle, respectively. A pair separated by a distance
$\ell$ can be created if $\ell$ is of the order of the Compton
length of the particle, $\lambda=\frac{\hbar}{m c}$, and if the
work done by the field along this distance, $W=e {\cal E} \ell$,
is at least equal to the energy needed to materialize the pair,
$E_0=2m c^2$. An estimate for the probability of pair creation is
then given by the Boltzmann factor $\Gamma\sim
\exp{-\frac{E_0}{W}}\sim \exp{-\frac{2 m^2 c^3}{e{\cal E}
\hbar}}$. Unfortunately, in order to observe this effect one must
have a huge external critical field, ${\cal E}_{\rm cr}\sim
10^{18}\,{\rm V\,m^{-1}}$. So, at $t=0$ (say), a pair is
materialized at $x=\pm \frac{\ell}{2}\sim \frac{m c^2}{e {\cal
E}}$. Now, one can go further and analyze the subsequent evolution
of the pair. It is accelerated apart by the Lorentz force
describing the uniformly accelerated hyperbolic motion (the
Rindler motion), $x^2-c^2t^2=(\ell /2)^2$. Differentiating this
relation one obtains the velocity and thus, the energy of the
particle, $E=\frac{2m c^2}{\sqrt{1-v^2/c^2}}=2x e {\cal E}$.
Hence, after the creation, the work done by the field along the
distance between the pair, $2x$, is used to accelerate the pair.
The complete process is schematically represented in Fig.
\ref{PairCreat_introduct-fig}.

This Schwinger process also holds for other particles and
background fields. One example is the pair creation of solitons
and domain walls that accompany the decay of a false vacuum. We
will discuss this process in chapter \ref{chap:False vacuum}. The
gravitational analogue of the Schwinger process, that leads to the
pair creation of black holes in an external field, has proposed by
Gibbons (1986). In order to turn a pair of virtual black holes
into a real one, we also need a background field that provides the
energy needed to materialize the pair, and that furnishes the
force necessary to accelerate away the black holes once they are
created. This background field can be: (i) an external
electromagnetic field with its Lorentz force, (ii) the positive
cosmological constant $\Lambda$, or inflation, (iii) a cosmic
string with its tension, (iv) a combination of the above fields,
or (v) a domain wall with its gravitational repulsive energy. We
will make an overview of these processes in section \ref{sec:Pair
creation BHs introduction}, and we will analyze some of these
processes in great detail in chapter \ref{chap:Pair creation}. To
study these processes we must have exact solutions of the Einstein
equations that describe the pair of uniformly accelerated black
holes in the external field, after they are created. Fortunately,
these solutions exist and we will make a historical presentation
of them in section \ref{sec:Pair accelerated BHs introduction},
and we will discuss them in detail in chapter
\ref{chap:PairAccBH}. Finally, an important process that
accompanies the production of the black hole pair  is the emission
of electromagnetic and gravitational radiation. An estimate for
the amount of gravitational radiation released during the pair
creation period will be given in section \ref{sec:Energy in BH
pair creation introduction} and explicitly computed in chapter
\ref{chap:Grav Radiation}.

 \vspace{0.4 cm}

Nowadays, it is important to study black holes not only in an
asymptotically flat spacetime, but also in spacetimes with a
negative cosmological constant ($\Lambda<0$), i.e., asymptotically
anti-de Sitter (AdS) spacetimes, and in spacetimes with a positive
cosmological constant ($\Lambda>0$), i.e., asymptotically de
Sitter (dS) spacetimes. One of the motivations to study AdS
solutions, as we shall see in section \ref{sec:3D BHs
introduction}, comes from the fact that the only black holes that
3-dimensional gravity can provide live in an AdS background.
Another motivation is related to supergravities in $11$
dimensions, that provide a dynamic mechanism that spontaneously
compactifies 7 of the 11 dimensions and yields a vacuum state with
topology $AdS_4\times S^7$, with an energy density given by
$\Lambda=-3g^2/(4\pi G)$, where $g$ is the coupling constant of
the theory. Thus, its vacuum is described by an AdS spacetime. If
such theory is correct, it means that AdS should be considered as
a symmetric phase of the theory that must have been broken to
yield our actual background. String theory provides a further
motivation to analyze AdS backgrounds, through the AdS/CFT
correspondence conjecture of Maldacena \cite{AdS/CFT}. This
duality conjecture states a correspondence between supergravity on
an AdS space and a conformal field theory (CFT) on the boundary of
that space. The connection bridge between the string theory on AdS
and the gauge theory is supported by the large N limit of t'Hooft,
where N is the number of charged colors of the theory. This
relationship has been motivated by the studies on higher
dimensional black holes ($p$-branes), and by their full string
theory description as $D$-branes of Polchinski. In addition, AdS
spacetimes have other interesting features: (i) it is one of the
rare backgrounds yielding a consistent interaction with massless
higher spins, (ii) it allows a consistent string theory in any
dimension, not only at the critical ones, and (iii) it permits a
clear definition of mass, charge and angular momentum (a property
that is shared with the asymptotically flat case).

The properties of the dS solutions also deserve a detailed
investigation. The main motivation to study them comes from recent
astronomical observations that seem to indicate that our universe
is filled by a $\Lambda>0$ background. The dS background plays
also a crucial role in the inflationary era undergone by the early
universe, and supergravities theories in a dS space have also been
connected to a conformal field theory (CFT) in the boundary of the
space, through the dS/CFT duality conjecture of Strominger.
\vspace{0.2 cm}

Both the de Sitter and anti-de Sitter spacetimes allow the
existence of spherically symmetric black hole solutions that are
the direct counterparts of the asymptotically flat Schwarzschild,
Reissner-Nordstr\"{o}m, Kerr and Kerr-Newman black holes, and in
chapter \ref{chap:BH 4D} we will make a brief comparison between
these black holes in the $\Lambda<0$, $\Lambda=0$ and $\Lambda>0$
backgrounds. In this aspect, the AdS spacetime has a richer
structure than the other two. Indeed, the 4-dimensional AdS
background also allows black holes with a non-spherical topology.
This has been an unexpected result since a topological theorem of
Hawking \cite{HawkTopolCensor} states that if  the spacetime is
asymptotically flat and globally hyperbolic, and the dominant
energy condition holds, then the topology of the black holes must
be spherical. At the time of its formulation, this theorem was
supported by the spherical topology of the know black holes. The
reason why in the AdS background one can have black holes with
non-spherical horizons is obviously due to the presence of a
negative $\Lambda$. So, besides the spherical solutions, the AdS
case allows two other families of solutions, namely, solutions
with cylindrical, toroidal, or planar topology
\cite{Lemos}-\cite{Huang_L}, and solutions with hyperbolic
topology \cite{topological} (these are usually called
topological).

The first non-spherical AdS black hole, that we will discuss in
chapter \ref{chap:BH 4D}, has been found by Lemos \cite{Lemos}.
These solutions have a planar symmetry, i.e., they include black
holes with cylindrical, toroidal, and planar horizons, and can be
formed through gravitational collapse as shown by Lemos
\cite{Lemos3}. The interest on these spacetimes, and in AdS in
general, relies also in the fact that they can be used to test the
cosmic censorship and hoop conjectures. Indeed, as shown in
\cite{Lemos3}, (i) spherical collapse with $\Lambda<0$ may lead to
massless naked singularities (in contradiction with the cosmic
censorship conjecture), (ii) cylindrical, toroidal or planar
collapse with $\Lambda<0$ does not produce naked singularities (in
accordance with the cosmic censorship conjecture), and (iii)
cylindrical, toroidal or planar collapse with $\Lambda<0$ leads to
the formation of the black holes with cylindrical, toroidal, and
planar horizons found in \cite{Lemos}, which gives a clear
counter-example to the hoop conjecture. Static and rotating
electric charged black holes that are the electric counterparts of
the black holes found in \cite{Lemos1} have been obtained by Lemos
and Zanchin \cite{Zanchin_Lemos}, and magnetic solutions with
cylindrical or toroidal topology have been constructed by Dias,
and Lemos \cite{OscarLemos_string} (see chapter \ref{chap:BH 4D}).

\section[Black holes in 3-dimensional spacetimes]{\large{Black holes in 3-dimensional spacetimes}}
\label{sec:3D BHs introduction}

\subsection[Exact solutions]{\large{Exact solutions}} \label{sec:Exact solutions 3D introduction}

Einstein gravity in a 3-dimensional background has some unusual
features that clearly differentiate it from the 4-dimensional
Einstein gravity (see Gott and Alpert \cite{GA}, Giddings, Abbot
and Kucha\v{r} \cite{GAK}, Deser, Jackiw and t' Hooft
\cite{DJH_flat}, and Barrow, Burd and Lancaster \cite{BBL}). Any
vacuum solution with $\Lambda=0$ is flat, and any vacuum solution
with non-vanishing cosmological constant has constant curvature.
This follows from the fact that the Weyl tensor in 3 dimensions is
identically zero. The 3-dimensional spacetime has no local degrees
of freedom, and thus its dynamics is substantially different from
the one of the 4-dimensional case. In particular, there are no
gravitational waves, and no gravitons. Moreover, there is no
Newtonian limit, that is, there is no gravitational force between
masses. Note, however, that the absence of gravitational dynamics
in general relativity by no way means that 3-dimensional
spacetimes are trivial and out of interest. Indeed, as we shall
discuss, one can study solutions generated by point sources, one
can construct a time machine, and one can build a topological
black hole solution. Furthermore, quantization of the
gravitational field has been attempted with some successes in a
3-dimensional background.

 \vspace{0.2 cm} \noindent $\bullet$
 {\it Spacetimes generated by point sources in 3-dimensions} \vspace{0.2 cm}

The issue of spacetimes generated by point sources (no event
horizon) in 3-dimensional Einstein theory has been object of many
studies (for reviews see
\cite{Jackiw_book,Brown_book,Jackiw_Review}). These sources can be
viewed as normal sections of straight cosmic strings. In 1963,
Staruszkiewicz \cite{Star} has begun the analysis of 3-dimensional
Einstein gravity without cosmological constant ($\Lambda=0$)
coupled to a static massive point source. The corresponding space
has a conical geometry, i.e., it is everywhere flat except at the
location of the point source. The space can be obtained from the
Minkoswki space by suppressing a wedge and identifying its edges.
The wedge has an opening angle which turns to be proportional to
the source mass. Deser, Jackiw and t' Hooft \cite{DJH_flat} have
generalized the analysis of \cite{Star} in order to find the
spacetime solutions generated by an arbitrary number of  static
massive point sources. These static multi-source solutions are
possible because of the absence of gravitational interaction in
3-dimensional general relativity. Once more the geometry is
conical with a wedge angle suppressed at each source proportional
to its mass. They have also constructed the solution corresponding
to a massless spinning point source in 3-dimensional Einstein
gravity with $\Lambda=0$. The extension to include massive
spinning sources has been achieved by Cl\'ement \cite{Clem_spin}.
Their results indicate that (besides the conical geometry already
present in the spinless case) the spacetime can be seen as
characterized by a helical structure since a complete loop around
the source ends with a shift in time proportional to the angular
momentum. In the context of  Einstein-Maxwell theory but still
with $\Lambda=0$, Deser and Mazur \cite{Deser_Maz} and Gott, Simon
and Alpert \cite{GSA} have found the solutions produced by
electric charged point sources. These spacetimes once more have a
conical geometry with a helical structure. Due to the slow fall
off of the electric field in 3-dimensions, the system has infinite
total energy. That is, asymptotically the electric field goes as
$\frac{1}{r}$ (while in 4-dimensions it falls as $\frac{1}{r^2}$),
the energy density behaves as $\frac{1}{r^2}$ (while in
4-dimensions it goes as $\frac{1}{r^4}$), and therefore the
spatial integral of the electric field diverges not only at the
origin but also at infinity. Static charged multi-source solutions
are not possible because, due the absence of gravitational
interaction, the electric repulsion cannot be balanced by a
gravitational attraction. Barrow, Burd and Lancaster \cite{BBL}
found the horizonless spacetime generated by a magnetic charged
point source in a $\Lambda=0$ background. Melvin \cite{Melvin_3D}
also describes the exterior solution of electric and magnetic
stars in the above theory. 3-dimensional static spacetimes
generated by open or closed one-dimensional string sources with or
without tension (in a $\Lambda=0$ background) have been
constructed by Deser and Jackiw \cite{Deser_Jackiw_string}. The
extension to the rotating string source has been done by Grignani
and Lee \cite{Grig_Lee} and by Cl\'ement \cite{Clem_string}. Other
exact solutions in 3-dimensional Einstein gravity theory produced
by extended and stationary sources have been found by Menotti and
Seminara \cite{Men_Sem}.

The spacetimes generated by point sources in 3-dimensional
Einstein gravity with non-vanishing cosmological constant
($\Lambda \neq 0$) have been obtained by Deser and Jackiw
\cite{DJ_sitter} and by Brown and Henneaux \cite{Brown_Hen}. In
the de Sitter case ($\Lambda
>0$) there is no one-particle solution. The simplest solution
describes a pair of antipodal particles on a sphere with a wedge
removed between poles and with points on its great circle
boundaries identified. In the anti-de Sitter case ($\Lambda<0$),
the simplest solution describes a hyperboloid with a wedge removed
proportional to the source mass located at the vertex of the
wedge.

 \vspace{0.2 cm} \noindent $\bullet$
 {\it Time machines in 3-dimensional spacetimes} \vspace{0.2 cm}

In a clear example of the richness of 3-dimensional Einstein
gravity, Gott \cite{Gott}, in 1991, has shown that it is possible
to construct a time machine (see also Cutler \cite{Cutler}), i.e.,
a solution that allows the existence of closed timelike curves
(regions that violate causality and that allow an observer to
travel backwards in time). All that is needed is two particles in
a flat background passing through each other with a velocity
greater than a certain critical value, that is related with the
deficit angle (discussed above) associated with the particles.
However, Deser, Jackiw and t' Hooft \cite{DJH_flat}, and Carrol,
Fahri and Guth \cite{CarFahGuthGOTT} soon proved that Gott's
construction was tachyonic, i.e., the mass system that leads to
the appearance of closed timelike curves corresponds to an
effective mass travelling faster than light. Moreover, Carrol,
Fahri, Guth and Olum \cite{CarFahGuthOlumGOTT}, and Menotti and
Seminara \cite{MenotSeminGOTT} have shown that the total momentum
of Gott universe is spacelike. Holst \cite{HolstGOTT} has
constructed the AdS analogue of Gott universe, and concluded that
it still is tachyonic. Moreover, for closed universes, t'Hooft
\cite{HooftGOTT} has shown that if one starts with a spacetime
without closed timelike curves and tries to insert them through
Gott's construction, then the universe will always collapse before
they have an opportunity to appear. Thus, the analysis of
3-dimensional Einstein gravity has been important to show that
nature prevents the creation of time machines.

\vspace{1 cm}
 \vspace{0.2 cm} \noindent $\bullet$
 {\it Black holes in 3-dimensional Einstein gravity } \vspace{0.2 cm}

In what concerns black hole solutions in 3-dimensional Einstein
gravity, quite surprisingly (since the 3-dimensional spacetime is
quite poor at the dynamical level), Ba\~nados, Teitelboim and
Zanelli \cite{btz_PRL} have found a black hole solution (the BTZ
black hole), with mass and angular momentum, that is
asymptotically AdS. The existence of this black hole gets even
more remarkable when one realizes that the BTZ metric has constant
curvature and thus there can be no curvature singularity at the
origin. As discussed in detail by Ba\~nados, Henneaux, Teitelboim,
and Zanelli \cite{btz_PRD}, the BTZ black hole can be expressed as
a topological quotient of $AdS_3$ by a group of isometries. This
is, the BTZ black hole can be obtained through identifications
along an isometry of the $AdS_3$ spacetime, and in order to avoid
closed timelike curves, the origin must be a topological
singularity (a boundary of the spacetime). Qualitatively, the
reason why the black hole exists only in an AdS background can be
understood as follows. The radius of an event horizon ($r_+$) in a
$D$-dimensional spacetime is expected to be proportional to $G_D
M$, where $G_D$ is the $D$-dimensional Newton's constant. Now, in
mass units $G_D$ has dimension $M^{2-D}$. Thus, $G_D M$ (and
$r_+$) is dimensionless in $D=3$, and there is no length scale in
$D=3$. The cosmological constant provides this length scale for
the horizon, but only when $\Lambda<0$ (AdS case). One may argue
that this is due to the fact that the AdS background is
attractive, i.e., an analysis of the geodesic equations indicates
that particles in this background are subjected to a potential
well that attracts them (and so it is possible to concentrate
matter into a small region), while the dS background is repulsive
(in practice, if we try to construct a dS black hole through
identifications along an isometry of the $dS_3$ spacetime, we
verify that the possible horizon is inside the region that
contains closed timelike curves). In many ways, the BTZ black hole
has properties similar to the ones of the 4-dimensional black
holes. For example, it has an event horizon and, in the rotating
case, it has an inner horizon and the angular momentum has a
maximum bound. It can also be formed through the collapse of
matter as shown by Mann and Ross \cite{MannRoss_BTZ}. Matschull
\cite{Matschull}, Holst and Matschull \cite{HolstMatschull}, and
Birmingham and Sen \cite{BirmSen} have shown that the BTZ black
hole can be created when two point particles with sufficient
energy collide in the AdS background. We must however be careful
with some clear differences that exist between the BTZ black hole
and the 4-dimensional black holes, specially in what concerns
thermodynamic properties (see Carlip and Teitelboim
\cite{CarlipTeiltelboim}). Indeed, the temperature of the BTZ
black hole goes as $T_{\rm BTZ}\sim \sqrt{M}$ and tends to zero
when $M$ decreases. The BTZ black hole then has positive heat
capacity, and complete evaporation of it takes an infinite amount
of time. Stellar equilibrium in 3-dimensions has been discussed by
Cruz and Zanelli \cite{CruzZanelli}, and by Lubo, Rooman and
Spindel \cite{LuboRoomanSpindel}.

The extension to include a radial electric field in the BTZ black
hole has been done by Cl\'ement \cite{CL1} and Mart\'{\i}nez,
Teitelboim and Zanelli \cite{BTZ_Q} (this solution reduces to
those of \cite{Deser_Maz,GSA} when $\Lambda=0$). Again, due to the
slow fall off of the electric field in 3-dimensions, the system
has infinite total energy. The presence of the electric
energy-momentum tensor implies that the spacetime has no longer
constant curvature, and the electric BTZ black hole cannot be
expressed as a topological quotient of $AdS_3$ by a group of
isometries. The rotating charged black hole is generated from the
static charged solution (already found in \cite{btz_PRL}) through
the application of a rotation Lorentz boost. A BTZ solution with
an azimuthal electric field was found by Cataldo \cite{Cat}. This
solution is horizonless, and reduces to empty $AdS_3$ spacetime
when the charge vanishes. Pure magnetic solutions with
$\Lambda<0$, that reduce to the neutral BTZ black hole solution
when the magnetic source vanishes, also exist. Notice that, in
oppose to what occurs in 4-dimensions where the the Maxwell tensor
and its dual are 2-forms, in 3-dimensions the Maxwell tensor is
still a 2-form, but its dual is a 1-form (in practice, the Maxwell
tensor has only three independent components: two for the electric
vector field, and one for the scalar magnetic field). As a
consequence, the magnetic solutions are radically different from
the electric solutions in 3-dimensions. The static magnetic
solution has been found by Cl\'ement \cite{CL1}, Hirschmann and
Welch \cite{HW} and Cataldo and Salgado \cite{Cat_Sal}. This
spacetime generated by a static magnetic point source is
horizonless and has a conical singularity at the origin. The
extension to include rotation and a new interpretation for the
source of magnetic field has been made by Dias and Lemos
\cite{OscarLemos_BTZ}. Other black hole solutions of 3-dimensional
Einstein-Maxwell theory have also been found by Kamata and Koikawa
\cite{KK1,KK2}, Cataldo and Salgado \cite{CS} and  Chan
\cite{CHAN}, assuming self dual or anti-self dual conditions
between the electromagnetic fields. In chapter \ref{chap:BTZ
family} we will analyze in detail the BTZ family of solutions,
with a special focus on the topological construction leading to
the rotating massive BTZ black hole, and on the magnetic BTZ
solution.

 \vspace{0.2 cm} \noindent $\bullet$
 {\it Other 3-dimensional gravities and their black holes} \vspace{0.2 cm}

In order to turn the 3-dimensional dynamics more similar with the
realistic 4-dimensional one, we can manage a way by which we
introduce local degrees of freedom. This is done by coupling an
extra field to the Einstein general relativity. One way to do
this, as proposed by Deser, Jackiw and Templeton
\cite{Des_Jac_Temp} in 1982, is to add a Chern-Simmons term
yielding an alternative theory to Einstein gravity called
topological massive gravity. The new term appears as a counterterm
in the renormalization of quantum field theory in a 3-dimensional
gravitational background \cite{originTOPOLOG}. The attractive
interaction between masses is Yukawa-like, and is mediated by a
massive scalar graviton. This theory is perturbatively
renormalizable \cite{renormalTOPOLOG}. Spacetimes generated by
point sources in this theory have been obtained by Carlip
\cite{Carlip_source}, by Gerbert \cite{Gerbert_source}, and by
Deser and Steif \cite{DeserSteifLightSolution}. Cl\'ement
\cite{CL2}, Fernando and Mansouri \cite{FM} and Dereli and Obukhov
\cite{DO} have analyzed self-dual solutions for the
Einstein-Maxwell-Chern-Simons theory in 3-dimensions.

Another generalization can be obtained by coupling a scalar field
to Einstein gravity, yielding a so called dilaton gravity. The
most general form of this kind of gravity was proposed by Wagoner
\cite{scalarGravity} in 1970. The scalar field provides a local
dynamical degree of freedom to the theory, and models of this kind
(i.e., under certain choices of the parameters) appear naturally
in string theory. Some choices of the parameters of the theory
yield dilaton gravities that have black hole solutions. One of
these theories, that we will study in this thesis, is an
Einstein-dilaton gravity of the Brans-Dicke type in a $\Lambda <0$
background, first discussed by Lemos \cite{Lemos}. This theory is
specified by a Brans-Dicke parameter, $\omega$, and contains seven
different cases. Each $\omega$ can be viewed as yielding a
different dilaton gravity theory, with some of these being related
with other known special theories. For instance, for $\omega=-1$
one gets the simplest low-energy string action \cite{CFMP}, and
for $\omega=0$ one gets a theory related (through dimensional
reduction) to 4-dimensional general relativity with one Killing
vector \cite{Lemos,Zanchin_Lemos,OscarLemos_string}. This is, the
$\omega=0$ black holes are the direct counterparts of the
4-dimensional AdS black holes with toroidal or cylindrical
topology first discussed by Lemos, in the same way that a point
source in 3-dimensions is the direct cousin of a cosmic string in
4-dimensions. For $\omega=\pm \infty$ the theory reduces to the
pure 3-dimensional general relativity. This is, the $\omega=\pm
\infty$ black holes are the BTZ ones. Moreover, the case
$\omega>-1$ yields gravities whose black holes have a structure
and properties similar to the BTZ black hole, but with a feature
that might be useful: the $\omega>-1$ black holes have dynamical
degrees of freedom, which implies for example that the origin has
a curvature singularity and that gravitational waves can propagate
in the spacetime. The neutral black holes of this Brans-Dicke
theory were found and analyzed by S\'a, Kleber and Lemos
\cite{Sa_Lemos_Static,Sa_Lemos_Rotat}. The pure electric charged
black holes have been analysed by Dias and Lemos
\cite{OscarLemos}, and the pure magnetic solutions have been
discussed by Dias and Lemos \cite{OscarLemos-MagBD3D}. We will
analyze in detail these black holes in chapter
 \ref{chap:3D Dilaton BH}.
Other examples of dilaton black holes include the electric black
hole solutions of Chan and Mann \cite{CM}, the self dual solutions
of Fernando \cite{F}, the magnetic solutions of Kiem and Park
\cite{KiemP}, Park and Kim \cite{PKim} and Koikawa, Maki and
Nakamula \cite{KMK}, and the solutions found by Chen \cite{CHEN}.
For a detailed discussion on classical and quantum aspects of
3-dimensional gravity we refer the reader to Carlip's book
\cite{CarlipBOOK}, Jackiw's book \cite{Jackiw_book}, Brown's book
\cite{Brown_book}, and to the reviews \cite{CarlipREVIEW}.

\subsection[Pair creation of black holes in 3-dimensions]
 {\large{Pair creation of black holes in 3-dimensions}}
  \label{sec:Pair creation BHs 3D introduction}

In chapters \ref{chap:Pair creation} and \ref{chap:Pair creation
in higher dimensions} we will analyze in detail the pair creation
process of 4-dimensional black holes and of higher dimensional
black holes, respectively. The process of quantum pair creation of
black holes in an external field in a 3-dimensional background has
not been analyzed yet, as far as we know. In chapter
\ref{chap:Pair creation 3D}, we will try to understand the
difficulties associated with this issue, and we will also propose
a possible background in which the pair creation process in
3-dimensions might be analyzed.

\section[Black holes in 4-dimensional spacetimes]
{\large{Black holes in 4-dimensional spacetimes}} \label{sec:Part
4D}

\subsection[Exact single black hole solutions] {\large{Exact single black hole solutions}}
 \label{sec:Exact single BHs introduction}

In a 4-dimensional asymptotically flat background there four
well-known solutions that represent a single black hole, namely
the Schwarzschild black hole ($M\neq 0$, $Q=0$, $J=0$), the
Reissner-Nordstr\"{o}m black hole ($M\neq 0$, $Q\neq 0$, $J=0$),
the Kerr black hole ($M\neq 0$, $Q=0$, $J\neq 0$) and the
Kerr-Newman black hole ($M\neq 0$, $Q\neq 0$, $J\neq 0$). In an
asymptotically AdS or dS background the above solutions are also
present, and in the AdS case there are also black hole solutions
with non-spherical topology (see the end of section
\ref{sec:Historical Overview}).  In this thesis we will deal
essentially with the static solutions. For a review on some of the
properties of these black holes see, e.g., chapter
 \ref{chap:BH 4D} of this thesis.

\subsection[Pair of accelerated black holes: the C-metric and the
Ernst solution] {\large{Pair of accelerated black holes: the
C-metric and the Ernst solution}} \label{sec:Pair accelerated BHs
introduction}

\begin{figure}[H]
\centering
\includegraphics[height=3cm]{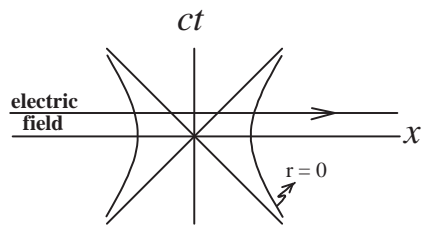}
   \caption{\label{Fig unif_accel_particles_introd}
Two oppositely charged particles without a horizon describe an
uniformly accelerated hyperbolic motion, approaching
asymptotically the velocity of light. This figure is simply an
extension to negative values of $t$ of the $ct-x$ plane of Fig.
\ref{PairCreat_introduct-fig}.
 }
\end{figure}

\begin{figure}[H]
\centering
\includegraphics[height=5cm]{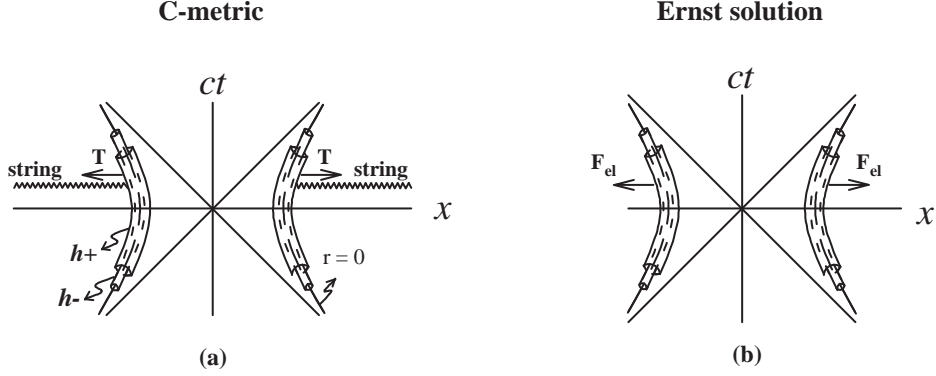}
   \caption{\label{Fig unif_accel_bh_introd}
(a) Schematic figure representing the C-metric, i.e., two black
holes with their horizons ($h_-$ and $h+$) that describe a
hyperbolic motion due to the strings tension, $T$. The string is
represented by the zigzag line. Instead of the strings, we could
alternatively have a single strut in between the black holes with
its outward tension providing the acceleration. (b) When the black
holes are charged, their acceleration can also be furnished by the
electromagnetic force $F_{\rm el}$, and this solution is exactly
described by the Ernst solution.
 }
\end{figure}

In 4-dimensional spacetime the Einstein equation (in vacuum or in
the presence of matter fields) allow a  wide variety of solutions
and, in particular, of black hole solutions, as displayed in
\cite{Kramer}. The main difficulty is then the appropriate
physical interpretation of these solutions, with many of them
being left apparently without known physical interest. One of the
best examples of this statement is the C-metric solution, found by
Levi-Civita \cite{LeviCivitaCmetric} and by Weyl
\cite{WeylCmetric} in 1918-1919, that has been interpreted only
fifty years after its discovery by Kinnersley and Walker
\cite{KW}, although some works during this gap period have dealt
with it. From this solution one can construct another exact
solution known as Ernst solution \cite{Ernst}. The C-metric and
the Ernst solution describe two uniformly accelerated black holes
in opposite directions, and the acceleration source is perfectly
identified. We can better understand these solutions going back to
the  pair of oppositely charged particles (without a horizon) that
are created in the Schwinger process, and then describe an
uniformly accelerated hyperbolic motion approaching asymptotically
the velocity of light. This accelerated 2-particle system is
represented in Fig. \ref{Fig unif_accel_particles_introd}, which
is simply an extension to negative values of $t$ of the $ct-x$
plane of Fig. \ref{PairCreat_introduct-fig}. During the negative
time the two particles approach each other until they come at rest
at $t=0$, and then they reverse their motion driving away from
each other. In Fig. \ref{Fig unif_accel_bh_introd}.(a), one
represents schematically the C-metric. To construct it, each
particle of Fig. \ref{PairCreat_introduct-fig} has been replaced
by a black hole with its horizons ($h_-$ and $h+$) that describes
a hyperbolic motion due to the string tension, $T$. When the black
holes are charged, their acceleration can also be furnished by an
external electromagnetic field, and this solution is exactly
described by the Ernst solution. The schematic figure representing
this last solution is sketched in Fig. \ref{Fig
unif_accel_bh_introd}.(b).

It is important to note that the C-metric and the Ernst solution
are one of the few exact solutions that contain already a
radiative term. This means, being accelerated, the black holes
necessarily release radiation and the information concerning the
radiative properties are already included in the metrics that
describe these solutions. Moreover, as we shall discuss in
subsection \ref{sec:Pair creation BHs introduction} these
solutions also describe appropriately the quantum process of black
hole pair creation in an external field and the consequent motion
of the created pair.


\begin{figure}[H]
\centering
\includegraphics[height=6.5cm]{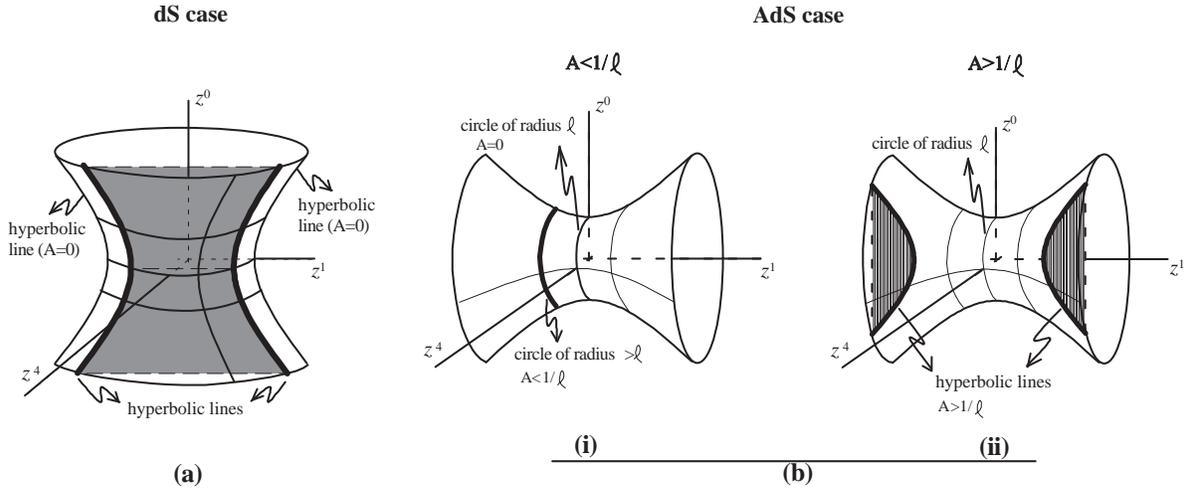}
   \caption{\label{Fig hyperb_introdud}
(a) The dS spacetime can be represented as a 4-hyperboloid,
$-(z^0)^2+(z^1)^2+(z^2)^2+(z^3)^2+(z^4)^2=\ell^2$
($\ell=\sqrt{3/\Lambda}$), embedded in a 5-dimensional Minkowski
spacetime with one timelike coordinate, $z^0$. The origin of the
dS C-metric describes the two hyperbolic trajectories. This
indicates that the solution represents two black holes being
accelerated apart. (b) The AdS spacetime can also be represented
as a 4-hyperboloid,
$-(z^0)^2+(z^1)^2+(z^2)^2+(z^3)^2-(z^4)^2=-\ell^2$
($\ell=\sqrt{3/|\Lambda|}$), embedded in a 5-dimensional Minkowski
spacetime, but this time with two timelike coordinates, $z^0$ and
$z^4$. (i) If $A<1/\ell$ the origin of the AdS C-metric moves in
the hyperboloid along a circle. This indicates that we have a
single black hole. (ii) If $A>1/\ell$ the origin of the AdS
C-metric moves along the two hyperbolic lines. These hyperbolic
lines indicate that the solution represents two black holes being
accelerated apart.
 }
\end{figure}

\begin{figure}[H]
\centering
\includegraphics[height=3.2cm]{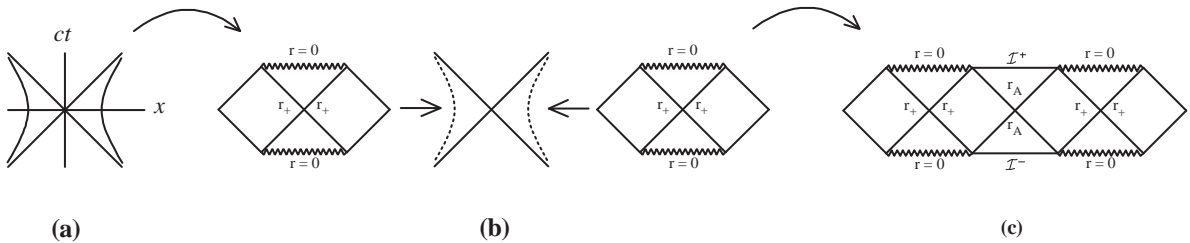}
   \caption{\label{Fig CPpair_bh_introd}
Schematic operation leading to the Carter-Penrose diagram of the
C-metric, where one identifies two accelerated black holes
approaching asymptotically the velocity of light, i.e., two black
holes separated by an acceleration horizon, $r_A$.
 }
\end{figure}

The extension of the Levi-Civita$-$Weyl solution
\cite{LeviCivitaCmetric,WeylCmetric} to the $\Lambda \neq 0$ case
has been done by Pleba\'nski and Demia\'nski \cite{PlebDem} in
1976. In this thesis (see chapter \ref{chap:PairAccBH}), we
analyze in detail the properties, and physical interpretation of
the C-metric in an AdS background \cite{OscLem_AdS-C}, and in a dS
background \cite{OscLem_dS-C}. We have thus extended for the
cosmological constant case the physical interpretation of the flat
case \cite{KW}. In particular we have concluded that while the
flat C-metric and dS C-metric describe a pair of accelerated black
holes with any acceleration, the AdS C-metric only describes a
pair of accelerated black holes if their acceleration satisfies
$A>\sqrt{|\Lambda|/3}$. We can interpret this as due to the fact
that the AdS background is attractive, i.e., an analysis of the
geodesic equations indicates that particles in this background are
subjected to a potential well that attracts them. Therefore, if we
want to have a pair of black holes accelerating apart, we will
have to furnish a sufficient force that overcomes this
cosmological constant background attraction. We then expect that a
pair of accelerated black holes is possible only if acceleration
$A$ is higher than a critical value. Our analysis in
\cite{OscLem_AdS-C,OscLem_dS-C} is essentially based on the
geometric description of the solution in the cosmological constant
4-hyperboloid embedded in a 5-dimensional Minkowski spacetime, and
on the study of the causal structure of the solution
(Carter-Penrose diagrams). The information provided by the
geometric description of the C-metric on the 4-hyperboloid is
displayed in Fig. \ref{Fig hyperb_introdud}. More precisely, in
Fig. \ref{Fig hyperb_introdud}.(a), the dS spacetime is
represented by the 4-hyperboloid
$-(z^0)^2+(z^1)^2+(z^2)^2+(z^3)^2+(z^4)^2=\ell^2$
($\ell=\sqrt{3/\Lambda}$) embedded in a 5-dimensional Minkowski
spacetime. The directions $z^2$ and $z^3$ are suppressed, $z_0$ is
a time coordinate, and $z^1$ and $z^2$ are space coordinates. In
the pure dS spacetime (acceleration $A=0$), its origin describes
the two hyperbolic trajectories that result from the intersection
of the hyperboloid surface with the $z^4=0$ plane. Basically, this
indicates that the dS solution already represents two black holes
being accelerated by the cosmological constant. In the dS C-metric
case ($A\neq 0$), its origin describes the two hyperbolic lines
lying on the dS hyperboloid that result from the intersection of
the hyperboloid surface with the $z^4$=constant$<\ell$ plane. This
indicates that the solution represents two black holes being
accelerated by the cosmological constant and, in addition, by the
string tension that provides the extra acceleration $A$. In Fig.
\ref{Fig hyperb_introdud}.(b), the AdS spacetime ($A=0$) is
represented by the 4-hyperboloid
$-(z^0)^2+(z^1)^2+(z^2)^2+(z^3)^2-(z^4)^2=-\ell^2$
($\ell=\sqrt{3/|\Lambda|}$) embedded in a 5-dimensional Minkowski
spacetime, but this time with two timelike coordinates, $z^0$ and
$z^4$. If $A<1/\ell$, see Fig. \ref{Fig hyperb_introdud}.(b.i),
the origin of the AdS C-metric moves in the hyperboloid along the
circle with $z^1$=constant$<0$. When $A=0$ this circle is at the
plane $z^1=0$ and has a radius $\ell$. In both cases this
indicates that we have a single black hole. If $A>1/\ell$, see
Fig. \ref{Fig hyperb_introdud}.(b.ii), the origin of the AdS
C-metric moves along the two hyperbolic lines, that result from
the intersection of the hyperboloid surface with the
$z^4$=constant$>\ell$ plane. These hyperbolic lines indicate that
in this case the solution represents two black holes being
accelerated.
An example of the information provided by the Carter-Penrose
diagrams of the C-metric, that leads to the physical
interpretation given to the C-metric, is schematically displayed
in Fig. \ref{Fig CPpair_bh_introd}. More precisely, in  Fig.
\ref{Fig CPpair_bh_introd}.(a), we show again two particles
describing an uniformly accelerated hyperbolic motion. In Fig.
\ref{Fig CPpair_bh_introd}.(b), the dashed lines represent the two
particles, and each one of these is now replaced by a black hole
represented here by its Carter-Penrose diagram, which was sketched
before in Fig. \ref{Fig Schw_rays-introduction}. Finally, in Fig.
\ref{Fig CPpair_bh_introd}.(c), the result of this operation is
shown. It yields the Carter-Penrose diagram of the C-metric, where
one identifies two accelerated black holes approaching
asymptotically the velocity of light, i.e., two black holes
separated by an acceleration horizon, $r_A$.

\begin{figure}[H]
\centering
\includegraphics[height=4.0cm]{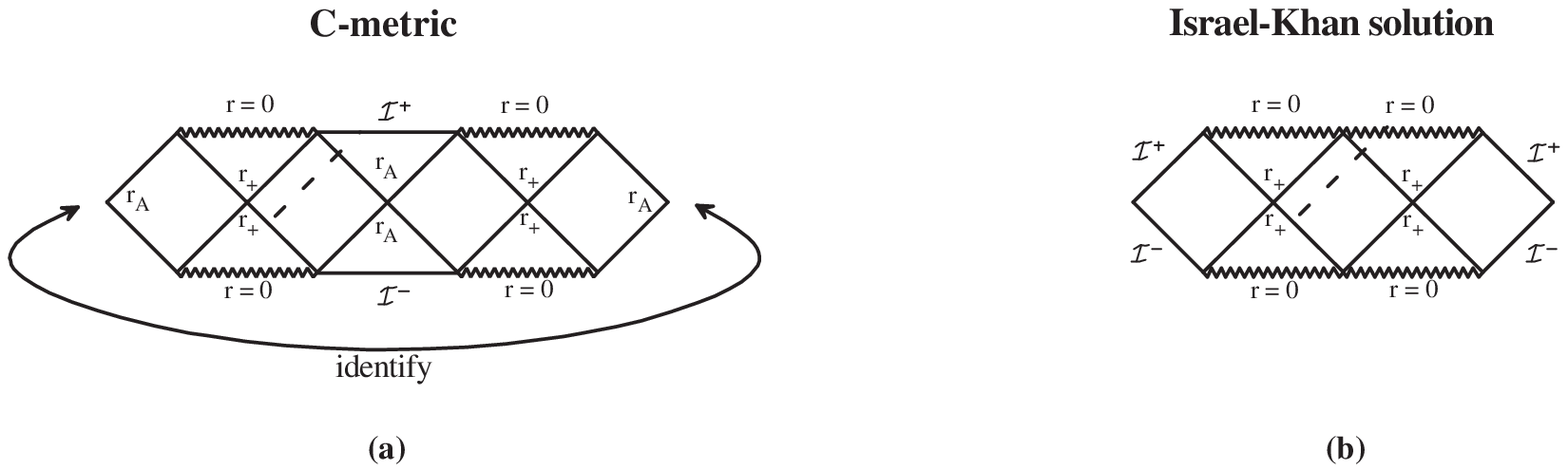}
   \caption{\label{Fig light flatC introduction}
Carter-Penrose diagram of the C-metric and of the Israel-Khan
solution along the direction that connects the two black holes.
The dashed line represents a null ray that is sent from the
vicinity of the event horizon of one of the black holes towards
the other black hole. (a) In the case of the C-metric this ray can
never reach the other black. Thus, there is no gravitational
interaction between the black holes.  (b) In the Israel-Khan
solution the null ray can reach the second black hole, and so they
attract each other gravitationally. In this solution, the two
black holes are connected by a strut that exerts an outward
pressure which cancels the inward gravitational attraction, and so
the distance between the two black holes remains fixed.
 }
\end{figure}

At this point, a remark is relevant. Israel and Khan \cite{bh_eq}
have found a $\Lambda=0$ solution that represents two (or more)
collinear Schwarzschild black holes interacting with each other in
such a way that allows dynamical equilibrium. In this solution,
the two black holes are connected by a strut that exerts an
outward pressure which cancels the inward gravitational
attraction, and so the distance between the two black holes
remains fixed, and they are held in equilibrium (see also Bach and
Weyl \cite{BachWeyl}, Aryal, Ford and Vilenkin
\cite{AryalFordVilenkin}, and Costa and Perry \cite{CostaPerry}).
Now, the flat C-metric solution reduces to a single
non-accelerated black hole free of struts or strings when the
acceleration parameter $A$ vanishes. Thus, when we take the limit
$A=0$, the flat C-metric does not reduce to the static solution of
Israel and Khan. The reason for this behavior can be found in the
causal diagrams of the two solutions (see Fig.
 \ref{Fig light flatC introduction}).
In particular, the black holes described by the C-metric do not
interact gravitationally. Their acceleration is provided only by
the tension of the strings or of the strut, without opposition
from gravitational attraction. A similar discussion applies in an
external background field. Indeed, Tomimatsu \cite{Tomimatsu} has
found a solution, that is the charged counterpart of the neutral
Israel-Khan solution \cite{bh_eq}, in which two
Reissner-Nordtr\"{o}m black holes are held in equilibrium. In this
case the gravitational attraction between the black holes is
cancelled by the electric repulsion, and this occurs only when the
black holes are extreme, $M_i=Q_i$. Tomimatsu has generalized for
the black hole case, the relativistic treatment used by Bonnor
\cite{Bonnor2ParticleEquil} and Ohta and Kimura \cite{OhtaKimura}
to study the equilibrium system of two charged particles. The
Ernst solution does not reduce to \cite{Tomimatsu} when $A=0$,
once again because there is no gravitational force between the two
Reissner-Nordtr\"{o}m black holes.

Now, an exact solution that exists in a dS background is the
Nariai solution, which can be connected with the near-extreme
Schwarzschild-dS solution by taking an appropriate extremal limit
introduced by Ginsparg and Perry \cite{GinsPerry}. Following the
procedure of \cite{GinsPerry}, we have further generated new
solutions \cite{OscLem_nariai} that are the C-metric counterparts
of the already known Nariai, Bertotti-Robinson and anti-Nariai
solutions. These solutions are conformal to the direct topological
product of two 2-dimensional manifolds of constant curvature. We
also give a physical interpretation to these solutions, e.g., in
the Nariai C-metric (with topology $dS_2\times \tilde{S}^2$) to
each point in the deformed 2-sphere $\tilde{S}^2$ corresponds a
$dS_2$ spacetime, except for one point which corresponds a $dS_2$
spacetime with an infinite straight strut or string. It is
unstable and decays into a slightly non-extreme black hole pair
accelerated by a strut or by strings.

In what follows we give a historical overview on the C-metric,
Ernst solution and Nariai, Bertotti-Robinson and anti-Nariai
solutions. But first we give a brief description of the properties
of strings, struts and domain walls.

\vspace{2 cm}
 \vspace{0.2 cm} \noindent $\bullet$
 {\it Strings, struts and domain walls} \vspace{0.2 cm}

We have already made reference (and we will do it again) to
gravitational objects other than black holes, namely
strings/struts and domain walls. Here we briefly comment on their
main properties. We work at the level of the Newtonian limit, but
the full general relativity analysis yields the same result.

Consider a static distribution of matter with an energy-momentum
tensor given by $T^{\mu}_{\:\:\nu}={\rm diag}(\rho,p_1,p_2,p_3)$,
where $\rho=T^0_{\:\:0}$ is the mass density of the system and
$p_i=T_i$ are the pressures along the three directions. The
Newtonian limit of Einstein equations for this distribution is
given by $\nabla^2\Phi=4\pi G(\rho-T^i_{\:\:i})$, where $\Phi$ is
the gravitational potential. For non-relativistic matter, $p_i\ll
\rho$, and we recover the classical Poisson equation,
$\nabla^2\Phi=4\pi G\rho$. For a straight string along the
$z$-axis one has $p_3=-\rho$ and $p_1=0$ and $p_2=0$, and thus one
gets $\nabla^2\Phi=0$. So, a straight string produces no
gravitational potential in its vicinity, and suffers a tension
along the $z$-axis that points inward. A general relativity
analysis carried by Vilenkin \cite{VilenkinString}, and by Ipser
and Sikivie \cite{IpserSikivie} shows that the geometry around the
straight string is conical, i.e., it can be obtained from the
Minkoswki space by suppressing a wedge (with a deficit angle
proportional to the line mass density, $\delta=8\pi G \rho$) and
identifying its edges. So its line element is equal to the
Minkowski one but the angle in the plane normal to the string
varies in the range $0\leq \phi <2\pi -\delta$ [see Fig. \ref{Fig
string-introduction}.(a)]. A straight strut has similar
properties, the only difference being the fact that its line mass
density is negative and its tension along the $z$-axis points
outward. Thus, instead of a deficit angle ($\delta<0$), it
produces an excess angle ($\delta>0$) [see Fig. \ref{Fig
string-introduction}.(b)].

In what concerns a domain wall, its properties are much different
from those of an usual massive wall. Indeed, a domain wall lying
in the $yz$-plane has a wall tension in the $y$ and $z$ directions
that is equal to the surface mass density of the domain wall,
$p_1=0$ and $p_2=p_3=\rho$. Its Newtonian limit is
$\nabla^2\Phi=-4\pi G\rho$ and thus it produces a repulsive
gravitational field. The general relativity analysis
\cite{VilenkinString,IpserSikivie} confirms these properties.
\begin{figure}[H]
\centering
\includegraphics[height=4cm]{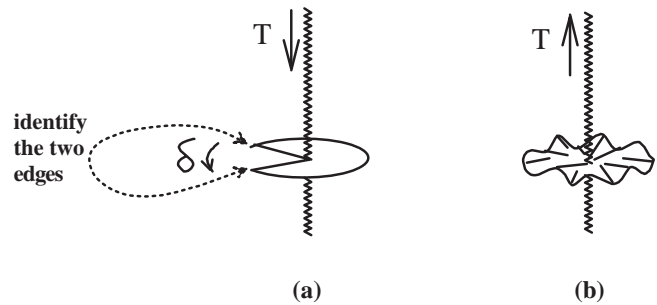}
   \caption{\label{Fig string-introduction}
(a) Schematic representation of a string (positive mass density,
$\rho>0$) with its tension that points inward, $T=-\rho$, and its
deficit angle $\delta>0$. The two edges are identified, so a
complete loop around a plane normal to it yields $\Delta \phi
=2\pi-\delta<2\pi$. (b) Schematic representation of a strut
(negative mass density, $\rho<0$) with its tension that points
outwards, $T=-\rho$, and its excess angle $\delta<0$. A complete
loop around a plane normal to it yields $\Delta \phi
=2\pi-\delta>2\pi$.
 }
\end{figure}

 \vspace{0.2 cm} \noindent $\bullet$
 {\it Historical overview on the C-metric and on the Ernst solution} \vspace{0.2 cm}

The original C-metric has been found Levi-Civita
\cite{LeviCivitaCmetric} and by Weyl \cite{WeylCmetric} in
1918-1919. During the following decades, many authors have
rediscovered it and studied its mathematical properties (see
\cite{Kramer} for references). In 1963 Ehlers and Kundt
\cite{EhlersKundt} have classified degenerated static vacuum
fields and put this Levi-Civita solution into the C slot of the
table they constructed. From then onwards this solution has been
called C-metric. This spacetime is stationary, axially symmetric,
Petrov type D, and is an exact solution which includes a radiative
term. Although the C-metric had been studied from a mathematical
point of view along the years, its physical interpretation
remained unknown until 1970 when Kinnersley and Walker \cite{KW},
in a pathbreaking work, have shown that the solution describes two
uniformly accelerated black holes in opposite directions. Indeed,
they noticed  that the original solution was geodesically
incomplete, and by defining new suitable coordinates they have
analytically extended it and studied its causal structure. The
solution has a conical singularity in one of its angular poles
that was interpreted by them as due to the presence of a strut in
between pushing the black holes away, or as two strings from
infinity pulling in each one of the black holes. The strut or the
strings lie along the symmetry axis and cause the acceleration of
the black hole pair. This work also included for the first time
the charged version of the C-metric. In an important development,
Ernst  in 1976 \cite{Ernst}, through the employment of an
appropriate transformation, has removed all the conical
singularities of the charged C-metric by appending an external
electromagnetic field (the magnetic solution is written in
\cite{Ernst}, while the explicit electric solution can be found in
Brown \cite{Brown}). Asymptotically, the Ernst solution reduces to
the Melvin universe \cite{Melvin}. In the Ernst solution the
acceleration of the pair of oppositely charged  black holes is
provided by the Lorentz force associated to the external field.
The geometrical properties of the C-metric were further
investigated by Farhoosh and Zimmerman \cite{FarhZimm1}, and the
asymptotic properties of the C-metric were analyzed by Ashtekar
and Dray \cite{AshtDray} who have shown that  null infinity admits
a conformal completion, has a spherical section, and moreover that
the causal diagrams drawn in \cite{KW} were not quite accurate.
The issue of physical interpretation of the C-metric has been
recovered by Bonnor \cite{Bonnor1}, but now following a different
approach. He transformed the C-metric into the Weyl form in which
the solution represents a finite line source (that corresponds to
the horizon of the black hole), a semi-infinite line mass
(corresponding to a horizon associated with uniform accelerated
motion) and a strut keeping the line sources apart. By applying a
further transformation that enlarges this solution, Bonnor
confirmed the physical interpretation given in \cite{KW}. Bonnor's
procedure has been simplified by Cornish and Uttley and extended
to include the massive charged solution \cite{CornUtt1}. More
recently, Yongcheng \cite{Yong}, starting from the metric of two
superposed Schwarzschild black holes, has derived the C-metric
under appropriate conditions. The black hole uniqueness theorem
for the C-metric has been proven by Wells \cite{Wells} and the
geodesic structure of the C-metric has been studied by Pravda and
Pravdova \cite{PravPrav2}. The limit when the acceleration goes to
infinity has been analyzed by Podolsk\' y and Griffiths
\cite{PodGrif1} who have shown that in this limit the solution is
analogous to the one which describes a spherical impulsive
gravitational wave generated by a snapping string. We note that
the C-metric is an important and explicit example of a general
class of asymptotically flat radiative spacetimes with
boost-rotation symmetry and with hypersurface orthogonal axial and
boost Killing vectors. The geometric properties of this general
class of spacetimes have been investigated by Bi\v c\' ak and
Schmidt \cite{BicSchm} and the radiative features were analyzed by
Bi\v c\' ak \cite{Bic} (see the recent review of Pravda and
Pravdova  \cite{PravPrav} on this class of spacetimes and the role
of the C-metric). 

Relevant generalizations to the C-metric were made by Pleba\'nski
and Demia\'nski in 1976 \cite{PlebDem} and by Dowker, Gauntlett,
Kastor and Traschen  in 1994 \cite{DGKT}. Pleba\'nski and
Demia\'nski, in addition to the mass ($m$) and electromagnetic
charge ($q$), have  included into the solution a NUT parameter, a
rotation and a cosmological constant term ($\Lambda$), and Dowker
et al have further included a dilaton field non-minimally coupled.
Thus, the most general C-metric has eight parameters, so far,
namely, acceleration, mass, electric and magnetic charges, NUT
parameter, rotation, cosmological constant and dilaton field. The
C-metric with mass and electromagnetic charges alone have been
extensively studied as shown above, and from now on we will refer
to it as the flat C-metric (i.e., C-metric with $\Lambda=0$). The
C-metric with a NUT parameter has not been studied, as far as we
know. The flat spinning C-metric has been studied by Farhoosh and
Zimmerman \cite{FarhZimm4}, Letelier and Oliveira \cite{LetOliv}
and by Bi\v c\' ak and Pravda \cite{BicPrav}. In particular, in
\cite{BicPrav} the flat spinning C-metric has been transformed
into the Weyl form and interpreted as two uniformly accelerated
spinning black holes connected by a strut. This solution
constitutes an example of a spacetime with an axial and boost
Killing vectors which are not hypersurface orthogonal. Dowker et
al generalized the flat C-metric and  flat Ernst solution to
include a dilaton field and applied these solutions for the first
time in the context of quantum pair creation of black holes, that
once created accelerate apart. Dowker and Thambyahpillai 
\cite{DowkerThamb} have found a C-metric family that describes 
multi-black holes being accelerated apart, and Hong and Teo 
\cite{HongTeo} have rewritten the C-metric in a new form that,
among other advantages, facilitates the computations that deal with 
the C-metric.

Our contribution in this field is related with the physical
interpretation of the cosmological constant C-metric introduced by
Pleba\'nski and Demia\'nski \cite{PlebDem}. The de Sitter (dS)
case ($\Lambda>0$) has been analyzed by Podolsk\' y and Griffiths
\cite{PodGrif2}, and discussed in detail by Dias and Lemos
\cite{OscLem_dS-C}. The anti-de Sitter (AdS) case ($\Lambda<0$)
has been studied, in special instances, by Emparan, Horowitz and
Myers \cite{EHM1} and by Podolsk\' y \cite{Pod}, and in its most
general case by Dias and Lemos \cite{OscLem_AdS-C}.  In general
the C-metric (either flat, dS or AdS) describes a pair of
accelerated black holes. Indeed, in the flat and dS backgrounds
this is always the case. However, in an AdS background the
situation is not so simple and depends on the relation between the
acceleration $A$ of the black holes and the cosmological length
$\ell=\sqrt{3/|\Lambda|}$.  One can divide the AdS study into
three cases, namely, $A<1/\ell$, $A=1/\ell$ and $A>1/\ell$. The
$A<1/\ell$ case was the one analyzed by Podolsk\' y \cite{Pod},
and the $A =1/\ell$ case has been investigated by Emparan,
Horowitz and Myers \cite{EHM1}, which has acquired an important
role since the authors have shown that, in the context of a lower
dimensional Randall-Sundrum model, it describes the final state of
gravitational collapse on the brane-world. The geodesic structure
of this solution has been studied by Chamblin \cite{Chamb}. Both
cases, $A < 1/\ell$ and $A= 1/\ell$, represent one  single
accelerated black hole.  The $A> 1/\ell$  AdS C-metric describes a
pair of accelerated black holes in an AdS background. The analysis
performed in \cite{OscLem_AdS-C,OscLem_dS-C} was supported on a
thorough analysis of the causal structure of the solution,
together with the description of the solution in the 4-hyperboloid
that describe the AdS and dS solutions, and with the study of the
physics of the strut or string.  When the acceleration is zero the
C-metric in any cosmological constant background reduces to a
single non-accelerated black hole with the usual properties.

We remark that in a cosmological constant background we cannot
remove the conical singularities through the application of the
Harrison transformation \cite{EmparanPrivCom} used by Ernst in the
flat case. Indeed, the Harrison transformation applied by Ernst
does not leave invariant the cosmological term in the action.
Therefore, applying the Harrison transformation to the
cosmological constant C-metric solutions does not yield a new
solution of the Einstein-Maxwell theory in a cosmological constant
background.

 \vspace{0.2 cm} \noindent $\bullet$
 {\it The extremal limits of the
C-metric: \newline
 Nariai, Bertotti-Robinson and anti-Nariai
C-metrics} \vspace{0.2 cm}

There are other very interesting solutions of general relativity
with generic cosmological constant, that are neither pure (like
the AdS, Minkowski and dS) nor contain a black hole (like the
Schwarzschild, Reissner-Nordstr\"om, Kerr, and Kerr-Newman), and
somehow can be considered intermediate type solutions. These are
the Nariai, Bertotti-Robinson, and anti-Nariai solutions.  The
Nariai solution \cite{Nariai,Nariai2} solves exactly the Einstein
equations with $\Lambda>0$, without or with a Maxwell field, and
has been discovered by Nariai in 1951 \cite{Nariai}. It is the
direct topological product of $dS_2 \times S^2$, i.e., of a
(1+1)-dimensional dS spacetime with a round 2-sphere of fixed
radius. The Bertotti-Robinson solution \cite{BertRob} is an exact
solution of the Einstein-Maxwell equations with any $\Lambda$, and
was found independently by Bertotti and Robinson in 1959. It is
the direct topological product of $AdS_2 \times S^2$, i.e., of a
(1+1)-dimensional AdS spacetime with a round 2-sphere of fixed
radius. The anti-Nariai solution, i.e., the AdS counterpart of the
Nariai solution, also exists \cite{CaldVanZer} and is an exact
solution of the Einstein equations with $\Lambda<0$, without or
with a Maxwell field. It is the direct topological product of
$AdS_2 \times H_2$, with $H_2$ being a 2-hyperboloid.

Three decades after Nariai's paper, Ginsparg and Perry
\cite{GinsPerry} connected the Nariai solution with the
Schwarzschild-dS solution. They showed that the Nariai solution
can be generated from a near-extreme dS black hole, through an
appropriate limiting procedure in which the black hole horizon
approaches the cosmological horizon. A similar extremal limit
generates the Bertotti-Robinson solution and the anti-Nariai
solution from an appropriate near extreme black hole (see, e.g.
\cite{CaldVanZer}). One of the aims of Ginsparg and Perry was to
study the quantum stability of the Nariai and the Schwarzschild-dS
solutions \cite{GinsPerry}.  It was shown that the Nariai solution
is in general unstable and, once created, decays through a quantum
tunnelling process into a slightly non-extreme black hole pair
(for a complete review and references on this subject see, e.g.,
Bousso \cite{Bousso60y}, and later discussions on chapter
\ref{chap:Extremal Limits} of this thesis).  The same kind of
process happens for the Bertotti-Robinson and anti-Nariai
solutions.

As we just saw, there is yet another class of related metrics, the
C-metric class, which represent not one, but two black holes,
being accelerate apart from each other. These black holes can also
inhabit a dS, flat or AdS background. It is therefore of great
interest to apply the Ginsparg-Perry procedure to these metrics in
order to find a new set of exact solutions with a clear physical
and geometrical interpretation. This has been done by Dias and
Lemos \cite{OscLem_nariai} (and will be recovered in chapter
\ref{chap:Extremal Limits}), where we addressed the issue of the
extremal limits of the C-metric with a generic $\Lambda$ following
\cite{GinsPerry}, in order to generate the C-metric counterparts
($A \neq 0$) of the Nariai, Bertotti-Robinson and anti-Nariai
solutions ($A=0$), among others.

\subsection[Quantum pair creation of  black holes in external
fields] {\large{Quantum pair creation of  black holes in external
fields}} \label{sec:Pair creation BHs introduction}

In nature there are few known processes that allow the production
of black holes. The most well known and better studied is the
gravitational collapse of a  massive star or cluster of stars. Due
to the fermionic degeneracy pressure these black holes cannot have
a mass below the Oppenheimer-Volkoff limiting mass. Another one,
as we already mentioned in subsection \ref{sec:Historical
Overview}, is the quantum Schwinger-like process \cite{Schwinger}
of black hole pair creation in an external field. These black
holes can have Plank dimensions and thus their evolution is ruled
by quantum effects. Moreover, gravitational pair creation involves
topology changing processes, and allows a study of the statistical
properties of black holes, namely: it favors the conjecture that
the number of internal microstates of a black hole is given by the
exponential of one-quarter of the area of the event horizon, and
it gives useful clues to the discussion of the black hole
information paradox.

In this thesis we study in detail the quantum process in which a
cosmic string breaks in an AdS background and in a dS background,
and a pair of black holes is created at the ends of the string
\cite{OscLem-PCdS,OscLem-PCAdS}. The energy to materialize and
accelerate the black holes comes from the strings tension, and we
have explicitly computed the pair creation rates. In Fig. \ref{Fig
string_cut_pc_introd}, we show a schematic description of the
process. We start, as indicated in Fig. \ref{Fig
string_cut_pc_introd}.(a), with a straight string that is going to
be broken. When the string breaks, as sketched in Fig. \ref{Fig
string_cut_pc_introd}.(b), two black holes are created at the ends
of the string, and the strings tension $T$ immediately accelerate
apart the black hole pair. The black holes described by the
C-metric have non-spherical horizons, due to the presence of a
conical singularity in at least one of their poles that signals
the presence of the string.
\begin{figure}[H]
\centering
\includegraphics[height=2.5cm]{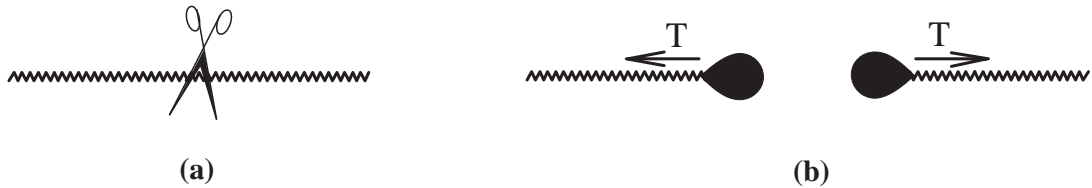}
   \caption{\label{Fig string_cut_pc_introd}
 Schematic description of the black hole pair
creation when a straight string breaks.
 }
\end{figure}

In the dS case the cosmological constant background acceleration
makes a positive contribution to the process, while in the AdS
case the cosmological constant background acceleration contributes
negatively. In particular, in the AdS case, pair creation of black
holes is possible only when the acceleration provided by the
string tension is higher than $\sqrt{|\Lambda|/3}$. We have thus
extended for the cosmological constant case a pair creation
process that in the flat case had been discussed in
\cite{HawkRoss-string}-\cite{GregHind}.
 We remark that in principle our explicit values for the pair creation rates
\cite{OscLem-PCdS,OscLem-PCAdS} also apply to the process of pair
creation in an external electromagnetic field, with the
acceleration being provided in this case by the Lorentz force
instead of being furnished by the string tension. There is no
Ernst solution in a cosmological constant background, and thus we
cannot discuss analytically the process. However, physically we
could in principle consider an external electromagnetic field that
supplies the same energy and acceleration as our strings and, from
the results of the $\Lambda=0$ case (where the pair creation rates
in the string and electromagnetic cases agree), we expect that the
results found in \cite{OscLem-PCdS,OscLem-PCAdS} do not depend on
whether the energy is being provided by an external
electromagnetic field or by strings.

As an example of our results and to understand the physical
interpretation that is associated with the process, we give the
pair creation rates $\Gamma$ of nonextreme black holes with $m=q$
when a string breaks in the dS, flat and AdS backgrounds,
respectively:
\begin{eqnarray}
& \Gamma_{\!\rm dS}^{\rm C-metric} \propto\exp \left [
\frac{3\pi}{64\Lambda} \frac{1}{mA}
 \left ( 1-\sqrt{\frac{1-4mA}{1+4mA}} \right )
  \left ( 1+\sqrt{1-(4mA)^2}-\frac{4m}{\sqrt{3}}\sqrt{\Lambda+3A^2}
   \right )-\frac{3\pi}{\Lambda}\frac{\sqrt{1-4mA}}{\sqrt{1+4mA}}\right ]\,,&
   \nonumber
\label{I-mag-luk-introduction}
 \end{eqnarray}
where the acceleration, $A>0$, and the mass parameter $m$ of the
black holes must satisfy $mA\leq \frac{\sqrt{3}}{4}
\frac{1}{\sqrt{\Lambda+3A^2}}$,
\begin{eqnarray}
& \Gamma_{\!\rm flat}^{\rm C-metric} \propto\exp \left [
-\frac{\pi}{4A^2}\frac{-1+4mA+\sqrt{1-(4mA)^2}}{\sqrt{1-(4mA)^2}}
\right ]\,, & \nonumber
 \label{area TOTAL-flat-introduction}
 \end{eqnarray}
where  the acceleration, $A>0$, and the mass parameter $m$ of the
black holes must satisfy $0<mA \leq\frac{1}{4}$,
\begin{eqnarray}
&\!\!\!\!\!\!\!\! \Gamma_{\!\rm AdS}^{\rm C-metric} \propto\exp
\left [-\frac{4\pi m^2}{\omega_+ (\omega_+^2-1)} {\biggl (}\!
 -\frac{\omega_+ -\omega_-}{(\omega_+ +\alpha)(\omega_- +\alpha)}
       +\frac{1}{\omega_+ -\alpha}
 + \frac{1}{ 1+\sqrt{1-\frac{8|\Lambda|m^2}{3}
\frac{3\omega_-^2-1}{\omega_-^2(1-\omega_-^2)^2} }}
 \frac{1-3\omega_-^2}{\omega_-(1-\omega_-^2)} {\biggr )} \right
 ],&
 \nonumber
 \label{area TOTAL-luk-introduction}
 \end{eqnarray}
where the acceleration $A$ and the mass $m$ of the black holes
must satisfy the conditions $A>\sqrt{\frac{|\Lambda|}{3}}$ and
$0<mA \leq\frac{1}{4}$, and we have $\omega_{\pm} = \sqrt{1\pm
4mA}$ and $\alpha
=\sqrt{1-\frac{4m}{\sqrt{3}}\sqrt{3A^2-|\Lambda|}}$. In the three
cases, $\Lambda<0$, $\Lambda=0$ and $\Lambda>0$, we conclude that
for a fixed $\Lambda$ and $A$ as the mass and charge of the black
holes increase, the probability they have to be pair created
decreases monotonically. This is the expected result since the
materialization of a higher mass implies the need of a higher
energy.

\noindent We can also discuss the behavior of the pair creation
rate as a function of the acceleration $A$ provided by the
strings, for a fixed $\Lambda$ and $m$. This evolution is
schematically represented in Fig. \ref{PC rates_AdS}, in the three
cosmological constant backgrounds. In general, for any $\Lambda$
background, the pair creation rate increases monotonically with
$A$. The physical interpretation of this result is clear: the
acceleration $A$ of the black hole pair is provided by the string.
When the energy of the string is higher (i.e., when its mass
density or the acceleration that it provides is higher), the
probability that it breaks and produces a black hole pair with a
given mass is also higher. This behavior is better understood if
we make a analogy with a thermodynamical system, with the mass
density of the string being the analogue of the temperature $T$.
Indeed, from the Boltzmann factor, $e^{-E_0/(k_{\rm B} T)}$ (where
$k_{\rm B}$ is the Boltzmann constant), one knows that a higher
background temperature $T$ turns the nucleation of a particle with
energy $E_0$ more probable. Similarly, a background string with a
higher mass density turns the creation of a black hole pair with
mass $2m$ more probable. In a flat background [see Fig. \ref{PC
rates_AdS-introd}.(a)], the pair creation rate is zero when $A=0$
\cite{HawkRoss-string}. In this case, the flat C-metric reduces to
a single black hole, and since we are studying the probability of
pair creation, the corresponding rate is naturally zero. This does
not mean that a single black hole cannot be materialized from the
quantum vacuum, it only means that this latter process is not
described by the C-metric. The creation probability of a single
black hole in a hot bath has been considered in
\cite{GrossPerryYaffe}. In a dS background [see Fig. \ref{PC
rates_AdS-introd}.(b)], the pair creation rate is not zero when
$A=0$ \cite{OscLem-PCdS}. This means that even in the absence of
the string, the positive cosmological constant is enough to
provide the energy to materialize the black hole pair
\cite{MannRoss}. If in addition one has an extra energy provided
by the string, the process becomes more favorable
\cite{OscLem-PCdS}. In the AdS case [see Fig. \ref{PC
rates_AdS-introd}.(c)], the negative cosmological constant makes a
negative contribution to the process, and black hole pair creation
is possible only when the acceleration provided by the strings
overcomes the AdS background attraction. The branch $0<A\leq
\sqrt{|\Lambda|/3}$ represents the creation probability of a
single black hole when the acceleration provided by the broken
string is not enough to overcome the AdS attraction, and was not
studied in this thesis.

Strictly speaking, the domain of validity of the pair creation
rates displayed above is $mA\ll 1$, for which the radius of the
black hole, $r_+ \sim m$, is much smaller than the typical
distance between the black holes at the creation moment, $\ell
\sim 1/A$ (this value follows from the Rindler motion
$x^2-t^2=1/A^2$ that describes the uniformly accelerated motion of
the black holes). So, for $mA\sim 1$ one has $r_+\sim \ell$ and
the black holes start interacting with each other.

\begin{figure} [h]
\centering
\includegraphics[height=2.3in]{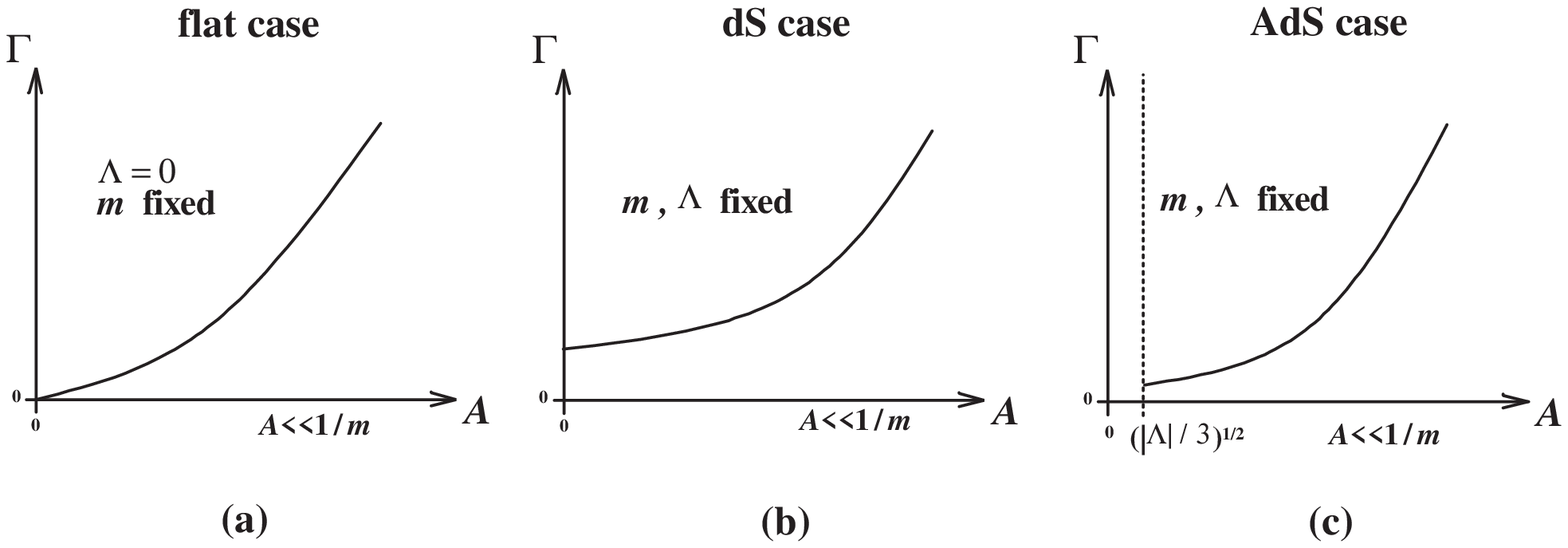}
\caption{\label{PC rates_AdS-introd}
 General behavior of the black hole pair creation rate $\Gamma$ as a function
of the acceleration $A$ provided by the strings, when a cosmic
string breaks: (a)  in a flat background ($\Lambda=0$)
\cite{HawkRoss-string}, (b)  in a dS background ($\Lambda>0$)
\cite{OscLem-PCdS}, and (c)  in an AdS background ($\Lambda<0$)
\cite{OscLem-PCAdS}.
 }
\end{figure}

 \vspace{0.2 cm} \noindent $\bullet$
 {\it Historical overview on quantum pair creation of particles in external fields} \vspace{0.2 cm}

The evaluation of the black hole pair creation rate has been done
at the semiclassical level using the instanton method (this method
is also called saddle point approximation). An instanton is an
Euclidean solution that interpolates between the initial and final
states of a classicaly forbidden transitation, and is a saddle
point for the Euclidean Feynmann path integral that describes the
pair creation rate. This instanton method has been introduced by
Langer in his work about decay of metastable termodynamical states
\cite{Langer}, and in the absence of gravity, it has been applied
to several different studies. For example, Coleman and Callan
\cite{Coleman,ColemanCallan}, Coleman and Luccia \cite{Luccia},
and Voloshin \cite{Voloshin2} have computed the bubble production
rate that accompanies the cosmological phase transitions in a
(3+1) dimensional scalar theory. Stone \cite{Stone}, Kiselev and
Selivanov \cite{Kiselev1,Kiselev2,Kiselev3}, and Voloshin
\cite{Voloshin1,Voloshin3} have discussed the soliton pair
production that accompanies the decay of false vacuum in a
(1+1)-dimensional scalar theory, and in Dias, Lemos
\cite{OscLem_FalVac} the study of this process has been recovered,
in the spirit of an effective one-loop action approach, and
extended to include domain wall pair creation (see Miller,
Cardenas, Garcia-Perez, More, Beckwith, and McCarten \cite{Miller}
for experimental evidence of this process). Affleck and Manton
\cite{Affleck2}, in the Yang-Mills-Higgs theory, have studied
monopole pair production in a weak external magnetic field, and
Affleck, Alvarez and Manton \cite{Affleck1}, have studied $e^+e^-$
boson pair production in a weak external electric field.

 \vspace{0.2 cm} \noindent $\bullet$
 {\it Historical overview on quantum pair creation of black holes
 in external fields} \vspace{0.2 cm}

The instanton method has been also introduced also as a framework
of quantum gravity, with successful results in the analysis of
gravitational thermodynamic issues and black hole pair production
processes, among others (see \cite{EQG-book}). The regular
instantons that describe the process we are interested in - the
pair creation of black holes in an external field - can be
obtained by analytically continuing (i) a solution found by Ernst
\cite{Ernst}, (ii) the de Sitter black hole solutions,  (iii) the
C-metric \cite{KW}, (iv) a combination of the above solutions, or
(v) the domain wall solution \cite{VilenkinString,IpserSikivie}.
To each one of these five families of instantons corresponds a
different way by which energy can be furnished in order to
materialize the pair of black holes and then to accelerate them
apart. In case (i) the energy is provided by the electromagnetic
Lorentz force, in case (ii) the strings tension furnishes this
energy, in case (iii) the energy is provided by the rapid
cosmological expansion associated to the positive cosmological
constant $\Lambda$, in case (iv) the energy is provided by a
combination of the above fields, and finally in case (v) the
energy is given by the repulsive gravitational field of the domain
wall.

It was believed that the only black hole pairs that could be
nucleated were those whose Euclidean sector was free of conical
singularities (instantons). This regularity condition restricted
the mass and charge of the black holes that could be produced, and
physically it meant that the only black holes that could be pair
produced were those that are in thermodynamic equilibrium.
However, Wu \cite{WuSubMax}, and Bousso and Hawking
\cite{BoussoHawkSubMax} have shown that Euclidean solutions with
conical singularities (sub-maximal instantons) may also be used as
saddle points for the pair creation process, as long as the
spacelike boundary of the manifold is chosen in order to contain
the conical singularity and the metric is specified there. In this
way, pair creation of black holes whose horizons are not in
thermodynamic equilibrium is also allowed.

We will now describe the studies that have been done on pair
creation of black holes in an external field.

(i) The suggestion that the pair creation of black holes could
occur in quantized Einstein-Maxwell theory has been given by
Gibbons \cite{Gibbons-book} in 1986, who has proposed that
extremal black holes could be produced in a background magnetic
field and that the appropriate instanton describing the process
could be obtained by euclideanizing the extremal Ernst solution.
This idea has been recovered by Giddings and Garfinkle
\cite{GarfGidd} who confirmed the expectation of
\cite{Gibbons-book} and, in addition, they have constructed an
Ernst instanton that describes pair creation of non-extremal black
holes. The explicit calculation of the rate for this last process
has lead Garfinkle, Giddings and Strominger
\cite{GarfGiddStrom_Sbh} to conclude that the pair creation rate
of non-extremal black holes is enhanced relative to the pair
creation of monopoles and extremal black holes by a factor
$e^{S_{\rm bh}}$, where $S_{\rm bh}={\cal A}_{\rm bh}/4$ is the
Hawking-Bekenstein entropy of the black hole and ${\cal A}_{\rm
bh}$ is the area of the black hole event horizon. This issue of
black hole pair creation in a background magnetic field and the
above relation between the pair creation rate and the entropy has
been further investigated by Dowker, Gaunlett, Kastor and Traschen
\cite{DGKT}, by Dowker, Gaunlett, Giddings and Horowitz
\cite{DGGH}, and by Ross \cite{RossU(1)}, but now in the context
of the low energy limit of string theory and in the context of
five-dimensional Kaluza-Klein theory. To achieve their aim they
have worked with an effective dilaton theory which, for particular
values of the dilaton parameter, reduces to the above theories,
and they have explicitly constructed the dilaton Ernst instantons
that describe the process. The one-loop contribution to the
magnetic black hole pair creation problem has been given by Yi
\cite{YiPConeLoop}. Brown \cite{Brown,Brown2} has analyzed the
pair creation of charged black holes in an electric external
field. Hawking, Horowitz and Ross \cite{HawHorRoss} (see also
Hawking and Horowitz \cite{HawkHor}) have related the action that
governs the rate of pair creation of extremal black holes with the
area of the acceleration horizon. In the non-extremal case, the
action has an additional contribution from the area of the black
hole horizon. From these relations emerges an explanation for the
fact, mentioned above, that the pair creation rate of non-extremal
black holes is enhanced relative to the pair creation of extremal
black holes by precisely the factor $e^{{\cal A}_{\rm bh}/4}$. For
a detailed discussion concerning the reason why this factor
involves only ${\cal A}_{\rm bh}$ and not two times this value see
also Emparan \cite{Emparan}. It has to do with the fact that the
internal microstates of two members of the black hole pair are
correlated.

(ii) The pair creation process of de Sitter (dS) black holes has
been also investigated. Notice that the dS black hole solution can
be interpreted as a pair of dS black holes that are being
accelerated apart by the positive cosmological constant. The
cosmological horizon can be seen as an acceleration horizon that
impedes the causal contact between the two black holes, and this
analogy is perfectly identified for example when we compare the
Carter-Penrose diagrams of the C-metric and of the dS black holes.
The study on pair creation of black holes in a dS background has
begun in 1989 by Mellor and Moss \cite{MelMos}, who have
identified the gravitational instantons that describe the process
(see also Romans \cite{Rom} for a detailed  construction of these
instantons). The explicit evaluation of the pair creation rates of
neutral and charged black holes accelerated by a cosmological
constant has been done by Mann and Ross \cite{MannRoss}. This
process has also been discussed in the context of the
 inflationary era undergone by the universe by Bousso and Hawking
 \cite{BoussoHawk}. Garattini \cite{GaratinniOneLoop},
 and Volkov and Wipf \cite{VolkovWipf} have
 computed the one-loop factor for this pair creation process,
 something that at the gravity quantum level is not an easy task.
Booth and Mann \cite{BooMann} have analyzed the cosmological pair
production of charged and rotating black holes. Pair creation of
dilaton black holes in a dS background has also been discussed by
Bousso \cite{BoussoDil}.

(iii) In 1995, Hawking and Ross \cite{HawkRoss-string} and
Eardley, Horowitz, Kastor and Traschen \cite{DougHorKastTras} have
discussed a process in which a cosmic string breaks and a pair of
black holes is produced at the ends of the string. The string
tension then pulls the black holes away, and the C-metric provides
the appropriate instantons to describe their creation. In order to
ensure that this process is physically consistent Ach\'ucarro,
Gregory and Kuijken \cite{AchGregKui}, and Gregory and Hindmarsh
\cite{GregHind} have shown that a conical singularity can be
replaced by a Nielson-Olesen vortex. This vortex can then pierce a
black hole \cite{AchGregKui}, or end at it \cite{GregHind}.
Moreover, it has been suggested that even topologically stable
strings can end at a black hole
\cite{HawkRoss-string}-\cite{PreskVil}.

(iv) We can also consider a pair creation process, analyzed by
Emparan \cite{Empar-string}, involving cosmic string breaking in a
background magnetic field. In this case the Lorentz force is in
excess or in deficit relative to the net force necessary to
furnish the right acceleration to the black holes, and this
discrepancy is eliminated by the string tension. The instantons
describing this process are a combination of the Ernst and
C-metric intantons. Another processes that involve black hole pair
creation in a combination of background fields are the ones,
already discussed above, in which a cosmic string breaks in a dS
background \cite{OscLem-PCdS}, and in an AdS background
\cite{OscLem-PCAdS}.

(v) The gravitational repulsive energy of a domain wall provides
another mechanism for black hole pair creation. This process has
been analyzed by Caldwell, Chamblin and Gibbons
\cite{CaldChamGibb}, and by Bousso and Chamblin \cite{BouCham} in
a flat background, while in an anti-de Sitter background the pair
creation of topological  black holes (with hyperbolic topology)
has been analyzed in by Mann \cite{MannAdS}.

Other studies concerning the process of pair creation in a
generalized background is done in \cite{Other,WuAdS}.

 \vspace{0.2 cm} \noindent $\bullet$
 {\it Pair creation of magnetic $vs$ electric black holes} \vspace{0.2 cm}

During a while it has been noticed that the pair creation of
electric black holes was apparently enhanced relative to the pair
creation of magnetic black holes. This was a consequence of the
fact that the Maxwell action has opposite signs in the two cases.
Now, this discrepancy between the two pair creation rates was not
consistent with the idea that electric and magnetic black holes
should have identical quantum properties. This issue has been
properly and definitively clarified by Hawking and Ross
\cite{HawkRoss} and by Brown \cite{Brown2}, who have shown that
the magnetic and electric solutions differ not only in their
actions, but also in the nature of the boundaries conditions that
can be imposed on them. More precisely, one can impose the
magnetic charge as a boundary condition at infinity but, in the
electric case, one instead imposes the chemical potential as a
boundary condition. As a consequence they proposed that the
electric action should contain an extra Maxwell boundary term.
This term cancels the opposite signs of the Maxwell action, and
the pair creation rate of magnetic and electric black holes is
equal.

 \vspace{0.2 cm} \noindent $\bullet$
 {\it Pair creation of black holes and the information loss problem} \vspace{0.2 cm}

The process of black hole pair creation gives also useful clues to
the discussion of the black hole information loss problem
\cite{InformationLossLOSS}. Due to the thermal Hawking radiation
the black holes evaporate. This process implies that one of the
following three scenarios occurs (see \cite
{InformationLossREVIEW} for nice reviews): (i) the information
previously swallowed to the black hole is destroyed, (ii) this
information is recovered to the exterior through the Hawking
radiation, or (iii) the endpoint of the evaporation is a Plank
scale remnant which stores the information. There are serious
difficulties associated to each one of this scenarios. Scenario
(i) implies non-unitarity and violation of energy conservation,
scenario (ii) implies violation of locality and causality, and the
main problem with scenario (iii) is that a huge energy is needed
in order to store all the information that has been swallowed by
the black hole, and a Planck scale remnant has very little energy.
Pair creation of black holes has been used to test these
scenarios. Indeed, it has been argued \cite{InformationLossREVIEW}
that if one demands preservation of unitarity and of locality then
a careful analysis of the one-loop contribution to the pair
creation process indicates that the Hawking process would leave
behind a catastrophic  infinite number of remnants. So the remnant
hypothesis seems to be discarded, although some escape solutions
can be launched \cite{InformationLossREVIEW}. On the other side,
Hawking, Horowitz and Ross \cite{HawHorRoss} have called attention
to the fact that the same instantons that describe pair creation
can, when reversed in time, describe their pair annihilation, as
long as the black holes have appropriate initial conditions such
that they come to rest at the right critical separation (this
annihilation process was also discussed by Emparan
\cite{Emparan}). One can then construct \cite{HawHorRoss} an
argument that  favors the information loss scenario: black holes
previously produced as a particle-antiparticle pair can accrete
information and annihilate, with their energy being given off as
electromagnetic and gravitational radiation. Therefore,  the
information loss scenario seems to occur at least in this
annihilation process.

\subsection[Energy released during black hole pair creation in
an external field] {\large{Energy released during black hole pair
creation in an external field}} \label{sec:Energy in BH pair
creation introduction}

If one tries to predict the spectrum of radiation coming from pair
creation, one expects of course a spectrum characteristic of
accelerated masses, but one also expects a previous signal
indicating pair creation. In other words, the process of pair
creation itself, which involves the sudden creation of particles
that suffer a violent acceleration that takes them to some final
velocity in a very short time, must imply emission of radiation.
The sudden creation of pairs can be viewed for our purposes as an
instantaneous creation of particles, i.e., the time reverse
process of instantaneous collisions. Now, the gravitational energy
released in this last process can be estimated using a technique
developed by Weinberg \cite{weinberg}, known as the hard collision
formalism. We argue that this formalism also applies to
instantaneous creation of particles (see \cite{VitOscLem}). The
total energy, i.e., the energy spectrum integrated over all
frequencies diverges, so one needs a physical cutoff frequency
that can be estimated by the uncertainty principle, $\omega_c \sim
\frac{E}{\hbar}$. We then find that the total gravitational energy
released during the 4-dimensional black hole pair creation process
is given by \cite{VitOscLem} (see chapter \ref{chap:Grav
Radiation})
\begin{equation}
\Delta E =\frac{4 G c}{\pi} \frac{\gamma^3M^3}{\hbar} \,,
\label{totenergpair4d-introduction}
\end{equation}
where $M$ is the mass of each one of the created black holes and
$\gamma=(1-v^2/c^2)^{-1/2}$ is the Lorentz factor. This value can
lead, under appropriate numbers of $M$ and $\gamma$ to huge
quantities, and is a very good candidate to emission of
gravitational radiation. For example, for black holes with 30
times the Planck mass and with $10\%$ of the velocity of light,
the gravitational energy released is $\Delta E\sim 10^{13}\,{\rm
J}$, which is about 100 times the rest energy of the pair.

This process is quite similar to another pure quantum-mechanical
process, the beta decay. The electromagnetic radiation emitted
during beta decay has been computed classically by Chang and
Falkoff \cite{chang} and is also presented in Jackson
\cite{jackson}.  The classical calculation is similar in all
aspects to the the instantaneous collision formalism, assuming the
sudden acceleration to energies $E$ of a charge initially at rest,
and requires also a cutoff in the frequency, which has been
assumed to be given by the uncertainty principle $\omega_c \sim
\frac{E}{\hbar}$. Assuming this cutoff one finds that the
agreement between the classical calculation and the quantum
calculation \cite{chang} is extremely good (specially in the low
frequency regime), and more important, it was verified
experimentally. Hence, we have a good degree of confidence to
believe that our estimate (\ref{totenergpair4d-introduction}) for
the black pair creation process is also good.

After the pair creation and after the emission of radiation
according to (\ref{totenergpair4d-introduction}), the created pair
of black holes accelerates away and, consequently, they continue
to release gravitational and electromagnetic energy. In a
$\Lambda=0$ background, the gravitational radiative properties of
the accelerated black holes described by the C-metric have been
analyzed by Bi\v c\'ak \cite{Bic}, and Pravda and Pravdova
\cite{PravPrav}. In a dS background, the gravitational radiation
emitted by uniformly accelerated sources without horizons has been
analyzed by Bi\v c\'ak and Krtou\v s \cite{BicKrt}, and the
radiative properties of accelerated black holes have been studied
by Krtou\v s and Podolsk\' y \cite{KrtPod}. In an AdS background
the energy released by a pair of accelerated black holes has been
discussed in detail by Podolsk\' y, Ortaggio and Krtou\v s
\cite{PodOrtKrtAdS}.

\section[Black holes in higher dimensional spacetimes]
{\large{Black holes in higher dimensional spacetimes}}
\label{sec:Part higher dimensional}

\subsection[Exact black hole solutions]
 {\large{Exact black hole solutions}} \label{sec:higher dimensional
BHs introduction}

By suggesting the existence of an extra dimension, besides the
usual (3+1)-dimensions that our daily experiments realize, Kaluza
and Klein in the 1920s unified for the first time gravity and
electromagnetism. Indeed, they realized that 5-dimensional vacuum
general relativity (i.e., satisfying $^5\!G_{AB}=0$ with $A$ and
$B$ taking the values $0,1,2,3,4$, and $G_{AB}$ being the usual
Einstein tensor), contained 4-dimensional general relativity in
the presence of an electromagnetic field (i.e., satisfying
$^4\!G_{\alpha \beta}=\frac{8\pi G}{c^4}\:^4\!T_{\alpha
\beta}^{\rm Max}$ with $\alpha$ and $\beta$ taking the values
$0,1,2,3$, and $T_{\alpha \beta}^{\rm Max}$ being the usual
Maxwell tensor) together with Maxwell's laws of electromagnetism
and an equation for a scalar field. The dominant view is then that
the extra dimension is not observed on experimentally accessible
energy scales because it is compactified. The Kaluza-Klein
mechanism has unified also matter and geometry since the photon
that is present in the 4-dimensional spacetime is a manifestation
of empty 5-dimensional spacetime. The key point to the
Kaluza-Klein unification is the realization that to a
4-dimensional gauge symmetry [e.g., the $U(1)$ gauge invariance of
Maxwell theory] corresponds a geometric symmetry (an invariance
with respect to coordinate transformations) in the extra
dimension. Thus, the main achievement of Kaluza-Klein proposal has
been the demonstration that different uncorrelated phenomena that
occur in 4-dimensional spacetime can be manifestations of the same
5-dimensional theory. From then onwards the theories of strong and
weak nuclear interactions have been developed, and several
attempts to unify the 4-dimensional theories in a single $D>5$
higher-dimensional theory have been tried with some successes (for
a nice review see Overduin and Wesson \cite{REVIEW-KaluzaKlein}).
Among others, two theories emerged, during the decade of 80, that
embodied the Kaluza-Klein compactification idea: supergravity and
superstring theory. Supergravity theories have the new ingredient
of supersymmetry, which states that each boson has a fermion
cousin and vice-versa. Mathematical consistency requires $D\leq
11$, and in these theories the case $D=11$ plays a very special
role due, among others reasons, to its uniqueness and to its
dynamical mechanism that spontaneously compactifies 7 of the 11
dimensions. However, supergravity theories also suffer from
serious problems, one of them being the fact that they are
non-renormalizable theories. The interest on supergravity theories
shifted to superstring theories, with the former being the low
energy limit of the later.  Superstring theories are
supersymmetric generalizations of string theories, in which the
notion that particles are point-like is abandoned and replaced by
the assumption that fundamental objects are excitations of
strings. Thus, the study of physics in higher dimensional
spacetimes is of great importance nowadays.

\vspace{0.4 cm}

Now, black holes play their role in this $D$-dimensional context.
We give two detailed examples. First, the idea that elementary
particles might behave as black holes and vice-versa is a very
attractive one. In principle one might expect such a scenario when
the Compton length, $\lambda=\frac{\hbar}{M c}$, associated to a
given mass is smaller than the corresponding Schwarzschild radius,
$r_+=\frac{2GM}{c^2}$, i.e., when the mass of the object is
smaller than the Planck mass, $M_{4}=\sqrt{\hbar c/G}$. Since this
is  a discussion at the quantum gravity level, superstring
theories provide  appropriate tools to analyze the problem. In
particular, some superstring states have been connected to extreme
$D$-dimensional black holes. This is an encouraging result since
if one wants to identify stable elementary particles with black
holes, these later must be extreme. As a second example, in which
we are particularly interested on, we point out to the recent
investigations, by Arkani-Hamed, Dimopoulos Dvali and Antoniadis
\cite{hamed}, that propose the existence of extra large dimensions
in our Universe in order to solve the hierarchy problem, i.e., the
huge difference between the electroweak scale, $M_{\rm EW}\sim
1{\rm TeV}$, and the 4-dimensional Planck scale, $M_4= \sqrt{\hbar
c/G}\sim 10^{16}\,{\rm TeV}$. The new main proposal of the so
called TeV-scale gravity (for reviews see, e.g.,
\cite{REVIEW-TeV}) is that the fundamental Planck scale, $M_{\rm
Pl}$, has the same size of the electroweak scale, $M_{\rm Pl}\sim
M_{\rm EW} \ll M_4$, but the 4-dimensional gravity is weak (i.e.,
the observed apparent Planck scale $M_4$ is large) due to dilution
of gravity in large or warped extra dimensions. At this point the
most accepted approach introduces a link to a brane-world
scenario. According to this scenario, matter and fields of
Standard Model  inhabit our 4-dimensional world, the brane,
whereas the gravitational degrees of freedom would propagate
throughout all dimensions, including the extra ones. By now,
experimental measurements involving submillimiter distances, and
studies on astrophysical and cosmological implications of large
extra dimensions, among others (see, e.g., references on
\cite{REVIEW-TeV}), require the existence of at least three extra
dimensions ($D>6$) in order that $M_{\rm Pl}\sim M_{\rm EW}$. More
precisely, the actual experimental bounds on the Planck
fundamental mass indicate that $M_{\rm Pl}\gtrsim 1.1\,{\rm
TeV}-0.8\,{\rm TeV}$ for $D=7$ to $D=10$.

One of the most spectacular consequences of this scenario would be
the production of black holes at the Large Hadron Collider (LHC)
at CERN, and the consequent experimental data relative to black
hole decay. This possibility has been pointed out by Argyres,
Dimopoulos and March-Russsel \cite{ArgDimMRussLHC}, by Banks and
Fischler \cite{BanksFiscLHC}, and by Emparan, Horowitz and Myers
\cite{EmpHorMyersLHC}, and has been quantified by Giddings and
Thomas \cite{GiddingsThomasLHC}, and by Dimopoulos and Landsberg
\cite{DimopoulosLandsbergLHC}. Now, the  LHC is designed to
operate at an center of mass energy of $14\,{\rm TeV}$, so if
$M_{\rm Pl}\sim 1\,{\rm TeV}$ we will indeed be able to produce
mini-black holes at LHC through the collisions of partons (quarks
and gluons). One can estimate the creation rate as follows. Due to
the fact that our description is only known at the semiclassical
level, one first sets that the black hole mass is above the Planck
mass, $M\sim 5\,{\rm TeV}-10\,{\rm TeV}$ (say). Then, in the
TeV-scale gravity/brane-world scenario one must work with two
approximations. First, one requires that the black holes are heavy
compared with $M_{\rm Pl}$ in order to ensure that the
gravitational field of the brane can be neglected. The second
approximation is to deal with black holes that are small compared
with the size of the extra dimensions. In the context of these
approximations the appropriate $D$-dimensional black holes are
described by the Tangherlini-Myers-Perry solution
\cite{tangherlini,myersperry}. To proceed, one needs to know the
parton density in a proton (which is known with some accuracy),
and the cross section $\sigma$ of black hole production in a
parton-parton collision. This cross section can be estimated from
pure geometric arguments following the Thorne's hoop conjecture: a
black hole is created when the impact parameter, $b$, of the
colliding partons is less than the Schwarzschild radius, $r_+$,
associated with the energy involved in the process. Hence
$\sigma\sim \pi r_+^2\sim E^{\frac{2}{D-3}}$. This estimate has
been further investigated by D'Eath, and Payne \cite{DEathPayne},
by Yoshino and Nambu \cite{yoshino}, and by Eardley and Giddings
\cite{eardley} who generalized a construction by Penrose to find
trapped surfaces on the union of two shock waves, describing
boosted Schwarzschild black holes. Their final conclusion is that
the above geometrical estimate for sigma is sufficiently accurate.
Putting everything together, the final result
\cite{GiddingsThomasLHC,DimopoulosLandsbergLHC} indicates that if
the TeV-scale gravity scenario is correct, then LHC will produce
black holes with masses larger than $5\,{\rm TeV}$ at the rate of
about 1 per second. One would really be in the presence of a black
hole factory. Moreover, these black holes will decay leaving
definite signatures.

\vspace{0.4 cm}
Since black holes in $D$-dimensions are important one should ask
which are the exact solutions that one can find.
 The natural higher dimensional generalization (which
contains the $D=4$ solution as a special case) of the
Schwarzschild black hole and of the Reissner-Nordstr\"{o}m black
hole has been obtained by Tangherlini \cite{tangherlini} (for a
AdS, flat and dS background). Myers and Perry \cite{myersperry}
have found the higher dimensional counterparts of the the Kerr and
Kerr-Newman black holes. The global structure of the non-rotating
solutions is similar to their 4-dimensional counterparts. In
particular, the black hole exists when $Q^2\leq G_D M^2$, where
$G_D$ is the $D$-dimensional Newton's constant. In what concerns
the rotating solutions, the properties of the higher dimensional
black holes are also similar to the ones of the 4-dimensional
case, with two important exceptions: for odd $D$ there are black
hole solutions with negative mass, and for $D\geq 6$ the angular
momentum is not bounded, i.e., it can have an arbitrary large
value. However, recently Emparan and Myers (2003) have shown that
there is a dynamical decay mechanism that turns these black holes
unstable at large enough rotation. Tangherlini \cite{tangherlini}
has also found the higher dimensional Schwarzschild and
Reissner-Nordstr\"{o}m black holes in an asymptotically AdS and dS
spacetime. In chapter \ref{chap:Black holes in higher dimensions},
we discuss in detail the extreme cases of the higher dimensional
dS black holes, i.e., we will give the explicit range of the mass
and charge for which one has an extreme black hole
\cite{VitorOscarLemos-BHdSDdim}. Moreover, we will find the higher
dimensional Nariai and Bertotti-Robinson solutions
\cite{VitorOscarLemos-BHdSDdim}. Besides the above black holes,
the higher dimensional spacetimes admit a richer variety of black
hole solutions. As examples, we mention the $D$-dimensional black
holes of Boulware, Deser \cite{BoulDeser}, of Callan, Myers and
Perry (1988), Gibbons and Maeda (1988), of Cvetic and Youm
\cite{CveticYoum}, and of Cvetic and Larsen \cite{CveticLarsen}
that are solutions of the string action. Supersymmetric rotating
black holes were found by Gibbons and Herdeiro
\cite{GibbonsHerdeiro}, and by Herdeiro \cite{Herdeiro}. With
non-spherical horizons, one has the 5-dimensional rotating black
ring solution of Emparan and Reall \cite{EmparanReall} 
(which has a horizon with topology $S^2\times S^1$), 
and the black $p$-branes whose
horizons have topology $S^{D-2}\times \mathbb{R}^p$.

\subsection[Quantum pair creation of higher dimensional black holes]
 {\large{Quantum pair creation of higher dimensional black holes}}
  \label{sec:higher dim pair creation BHs introduction}

We have already commented on the collision process at LHC that
might lead to the formation of higher dimensional  black holes.
Another process that can create black holes in a $D$-dimensional
spacetime is the Schwinger-like pair creation process in an
external field. As we saw in section \ref{sec:Part 4D}, in order
to study analytically this process one needs an exact solution
that describes a pair of accelerated black holes in the external
field. In a 4-dimensional background, the C-metric, the Ernst
solution and the dS black hole solution, among others, provided
such an exact solution. In spite of some efforts, the higher
dimensional generalization of the C-metric and of the Ernst
solution have not been found yet. However, the $D$-dimensional dS
black hole solution exists \cite{tangherlini}, and it describes a
pair of accelerated black holes. Its properties are discussed in
detail in \cite{OscarLemosNuno,VitorOscarLemos-BHdSDdim} (see
chapter \ref{chap:Black holes in higher dimensions}). We can then
analyze analytically the pair creation process of higher
dimensional dS black holes, which are accelerated apart by the
cosmological constant expansion \cite{VitorOscarLemosPChigherDim}
(see chapter \ref{chap:Pair creation in higher dimensions}).

\subsection[Gravitational radiation in higher dimensional spacetimes.
Energy released during black hole production]
{\large{Gravitational radiation in higher dimensional
spacetimes. \\
Energy released during black hole production}}
\label{sec:Gravitational radiation introduction}

Gravitational radiation is an essential feature related to black
hole physics. In 4-dimensional spacetimes there has been a
detailed analysis of the properties of gravitational radiation,
due to the possibility of detecting it from astrophysical sources.

In the context of black holes in $D$-dimensional spacetimes it is
important to extend the 4-dimensional results to higher
dimensional spacetimes. This has done  in \cite{VitOscLem} (see
also chapter \ref{chap:Grav Radiation} of this thesis). More
concretely, by solving the retarded gravitational potentials in
even D-dimensions, we were able to generalize many of the results
about gravitational waves (in a flat background) to higher
dimensions: we extended the quadrupole formula to higher
dimensions, we extended Weinberg's "zero time collision" formalism
to higher dimensions, and we have considered radiation processes
in the vicinities of higher dimensional  Tangherlini black holes
\cite{tangherlini}.

The analysis of gravitational radiation in higher dimensional
spacetimes is also important in the context of the TeV-scale
gravity and its prediction concerning the production of black
holes at the LHC. After the being produced, the black hole will
decay, i.e., it will shed its extra initial hair until it settles
into a Kerr or Schwarzschild black hole. Afterwards quantum
processes take place. In this thesis we are interested on the
first phase that accompanies the black hole production. One of the
experimental signatures of this phase will be a missing energy,
perhaps a large fraction of the center of mass energy
\cite{cardosolemos0}. This will happen because when the partons
collide to form a black hole, some of the initial energy will be
converted to gravitational waves, and due to the small amplitudes
involved, there is no gravitational wave detector capable of
detecting them, so they will appear as missing energy. Thus, the
collider will in fact serve indirectly as a gravitational wave
detector. This calls for the calculation of the energy given away
as gravitational waves when two high energy particles collide to
form a black hole, which lives in all the dimensions. In a
4-dimensional background this calculation has been done by D'Eath
and Payne \cite{DEathPayne}. In a higher dimensional background,
the calculation has been carried following distinct approaches by
 Yoshino and Nambu \cite{yoshino} (following a formalism developed
by \cite{eardley}), by Cardoso, Dias and Lemos \cite{VitOscLem}
(see chapter \ref{chap:Grav Radiation} of this thesis), and by
Berti, Cavagli\`{a} and Gualtieri \cite{BertiCavagliaGualteri}.
 We find that
the energy spectrum, i.e., the energy per unit of frequency and
per unit of solid angle is \cite{VitOscLem}
\begin{equation}
\frac{d^2E}{d\omega d\Omega}=\frac{2{\cal G}}{(2\pi)^{D-2}}
\frac{D-3}{D-2}\frac{\gamma_{1}^2m_{1}^2v_{1}^2(v_1+v_2)^2
\sin{\theta}^4}{(1-v_1\cos\theta)^2(1+v_2\cos\theta)^2}\times
\omega^{D-4}\,,
\label{energypersolidanglefreqinstcol-introduction}
\end{equation}
where $m_i$, $v_i$, and $\gamma_i=(1-v_i^2)^{-1/2}$ (with $i=1,2$)
are, respectively, the mass, the velocity, and the Lorentz factor
of the two colliding particles, and $\theta$ is the angle that the
line defined by the colliding particles makes with the direction
of the radiation. The formalism used in \cite{VitOscLem} assumes a
hard collision, i.e., a collision lasting zero seconds
\cite{weinberg}.

\section[Organization of the thesis]
{\large{Organization of the thesis}} \label{sec:Organization
thesis}

This thesis is divided in three parts. Briefly, part \ref{part1}
deals with black holes and pair creation in 3 dimensions. Part
\ref{part2} deals with the same subject but in 4-dimensional
spacetimes, and part \ref{part3} is devoted to the same issue but
in higher dimensional black holes.

The plan of the thesis is then as follows:

  \vspace{0.2 cm}
\noindent {\it $\bullet$ Part \ref{part1}. Black holes and pair
creation in 3 dimensions}
  \vspace{0.2 cm}

\noindent {\it Chapter \ref{chap:BTZ family}. The BTZ family of
black holes}. In this chapter, we review the main properties of
3-dimensional Einstein gravity, we show the explicit topological
construction that leads to the BTZ black hole, we briefly mention
the electric charged BTZ black hole, and we discuss in detail the
magnetic BTZ solution.

\noindent {\it  Chapter \ref{chap:3D Dilaton BH}. Three
dimensional dilaton black holes of the Brans-Dicke type}. In this
chapter, we analyze in detail the neutral, electric and magnetic
solutions of a 3-dimensional Einstein-dilaton gravity of the
Brans-Dicke type.

\noindent {\it  Chapter \ref{chap:Pair creation 3D}. Pair creation
of black holes in three dimensions}. The process of quantum pair
creation of black holes in an external field in a 3-dimensional
background has not been analyzed yet. In this chapter, we will try
to understand the difficulties associated with this issue, and we
will also propose a possible background in which the pair creation
process in 3-dimensions might be analyzed.

\vspace{0.2 cm}
 \noindent {\it $\bullet$ Part \ref{part2}. Black
holes and pair creation in 4 dimensions}
\vspace{0.2 cm}

\noindent {\it  Chapter \ref{chap:BH 4D}. Black holes in a
generalized  $\Lambda$ background}. In this chapter, we give an
overview of all the static 4-dimensional black holes that are
solutions of the Einstein equations in an AdS, flat, and dS
background. Special attention is dedicated to an AdS magnetic
solution with cylindrical/toroidal symmetry. This chapter
introduces also a description of some tools that will be used in
later chapters.

\noindent {\it Chapter \ref{chap:PairAccBH}. Pair of accelerated
black holes: the C-metric in a generalized $\Lambda$ background}.
In this chapter, we make a detailed discussion on the properties
and physical interpretation of the AdS C-metric, flat C-metric
(and Ernst solution), and dS C-metric.

\noindent {\it Chapter \ref{chap:Extremal Limits}. The extremal
limits of the C-metric: Nariai, Bertotti-Robinson and anti-Nariai
C-metrics}. In this chapter,  we apply the Ginsparg-Perry
procedure \cite{GinsPerry} to the C-metrics in order to find a new
set of exact solutions with a clear physical and geometrical
interpretation. In particular, we generate the C-metric
counterparts of the Nariai, Bertotti-Robinson and anti-Nariai
solutions, among others.

\noindent {\it Chapter \ref{chap:False vacuum}. False vacuum
decay: effective one-loop action for pair creation of domain
walls}. This chapter can be seen as an introductory toy model for
the black hole pair creation analysis. We propose an effective
one-loop action to describe the domain wall pair creation process
that accompanies the false vacuum decay of a scalar field (in the
absence of gravity). We compute the pair creation rate, including
the one-loop contribution, using the instanton method that is also
used to compute pair creation rates of black holes.

\noindent {\it Chapter \ref{chap:Pair creation}. Pair creation of
black holes on a cosmic string background}. In this chapter, we
discuss in detail the creation of a black hole pair when a cosmic
string breaks, in an AdS, flat and dS background. The instantons
that describe the process are constructed from the AdS, flat and
dS C-metrics. We explicitly compute the pair creation rates.

\vspace{0.2 cm}
 \noindent {\it $\bullet$ Part \ref{part3}. Black holes and pair creation in
higher dimensions}
 \vspace{0.2 cm}

\noindent {\it Chapter \ref{chap:Black holes in higher
dimensions}. Black holes in higher dimensional spacetimes}. In
this chapter, we discuss the higher dimensional Tangherlini black
holes in an AdS, flat, and dS background. In particular, we find
the explicit range of the mass and charge for which one has an
extreme black hole. Moreover, we will find the higher dimensional
Nariai and Bertotti-Robinson solutions.

\noindent {\it Chapter \ref{chap:Pair creation in higher
dimensions}. Pair creation of black holes in higher dimensional
spacetimes}. In this chapter, we discuss in detail the creation of
a higher dimensional Tangherlini black hole pair in a dS
background. The instantons that describe the process are
constructed from the dS Tangherlini solution. We explicitly
compute the pair creation rates.

\noindent {\it Chapter \ref{chap:Grav Radiation}. Gravitational
radiation in $D$-dimensional spacetimes and energy released during
black hole pair creation}. In this chapter, we construct the
formalism to study linearized gravitational waves in flat
$D$-dimensional spacetimes. This extension of the 4-dimensional
formalism is non-trivial, due to the behavior of the
$D$-dimensional Green's function. We find the $D$-dimensional
quadrupole formula, and we apply it to two cases: a particle in
circular motion in a generic background, and a particle falling
into a $D$-dimensional Schwarzschild black hole. We also consider
the hard collision between two particles, i.e., the collision
takes zero seconds, and introduce a cutoff frequency necessary to
have meaningful results. We then use it to compute the
gravitational energy released during the possible black hole
formation at the LHC, and to estimate the gravitational radiation
emitted during the process of black hole pair creation in an
external field.


%
%

%% file: Chapter1.tex
\thispagestyle{empty} \setcounter{minitocdepth}{1}
\chapter[The BTZ family of black holes]{\Large{The BTZ black hole}} \label{chap:BTZ family}
 \lhead[]{\fancyplain{}{\bfseries Chapter \thechapter. \leftmark}}
 \rhead[\fancyplain{}{\bfseries \rightmark}]{}
  \minitoc \thispagestyle{empty}
\renewcommand{\thepage}{\arabic{page}}
\addtocontents{lof}{\textbf{Chapter Figures \thechapter}\\}


In section \ref{sec:3Dtopological}, we will review some of the
unusual properties of 3-dimensional Einstein gravity.

Ba\~nados, Teitelboim and Zanelli \cite{btz_PRL} have found a
black hole solution (the BTZ black hole), with mass and angular
momentum, that is asymptotically AdS.  As discussed in detail by
Ba\~nados, Henneaux, Teitelboim, and Zanelli \cite{btz_PRD} (see
also Steif \cite{Steif}), the BTZ black hole can be expressed as a
topological quotient of $AdS_3$ by a group of isometries.  In
section \ref{sec:BTZ}, we will show in detail the topological
construction that leads to the rotating neutral BTZ black hole.

The extension to include a radial electric field in the BTZ black
hole has been done by Cl\'ement \cite{CL1} and Mart\'{\i}nez,
Teitelboim and Zanelli \cite{BTZ_Q} (this solution reduces to
those of \cite{Deser_Maz,GSA} when $\Lambda=0$).  A BTZ solution
with an azimuthal electric field was found by Cataldo \cite{Cat}.
In section \ref{sec:electric BTZ}, we will discuss briefly these
electric BTZ solutions.

Pure magnetic solutions with $\Lambda<0$, that reduce to the
neutral BTZ black hole solution when the magnetic source vanishes,
also exist.  The static magnetic solution has been found by
Cl\'ement \cite{CL1}, Hirschmann and Welch \cite{HW} and Cataldo
and Salgado \cite{Cat_Sal}.  The extension to include rotation and
a new interpretation for the source of magnetic field has been
made by Dias and Lemos \cite{OscarLemos_BTZ}. In section
\ref{sec:magnetic BTZ}, we will discuss in detail these solutions.

Other black hole solutions of 3-dimensional Einstein-Maxwell
theory have also been found by Kamata and Koikawa \cite{KK1,KK2},
Cataldo and Salgado \cite{CS} and  Chan \cite{CHAN}, assuming self
dual or anti-self dual conditions between the electromagnetic
fields.

\section{Properties of three dimensional general relativity}
\label{sec:3Dtopological}
 The Einstein equations $G_{\mu \nu}+\Lambda g_{\mu\nu}=8\pi G T_{\mu \nu}$
can also be written, in 3-dimensions, as
\begin{equation}
R_{\mu \nu}=2\Lambda g_{\mu\nu}+ 8\pi G \left (T_{\mu \nu}-g_{\mu
\nu}T \right ) \,.
 \label{ricci3D}
\end{equation}
The choice of the energy-momentum tensor $T_{\mu \nu}$ completely
determines the Ricci tensor $R_{\mu \nu}$, but in general it does
not determine the Riemann tensor $R_{\mu \nu \alpha \sigma}$.
However, and this is the fundamental difference between
3-dimensional and 4-dimensional gravities, in 3-dimensions the
Weyl tensor (the traceless part of $R_{\mu \nu \alpha \sigma}$)
vanishes, and so the Riemann tensor depends linearly on the Ricci
tensor,
\begin{equation}
R_{\mu \nu \alpha \sigma}=g_{\mu\alpha}R_{\nu \sigma}+ g_{\nu
\sigma}R_{\mu\alpha}-g_{\nu\alpha}R_{\mu \sigma}-g_{\mu
\sigma}R_{\nu\alpha} -\frac{1}{2}\left ( g_{\mu\alpha}g_{\nu
\sigma} - g_{\mu \sigma}g_{\nu\alpha} \right )R \,.
 \label{riemann3D}
\end{equation}
Inserting (\ref{ricci3D}) in this last relation one indeed
concludes that $R_{\mu \nu \alpha \sigma}$ is completely
determined by $T_{\mu \nu}$ and by the cosmological constant
$\Lambda$. In particular, regions of spacetime with $T_{\mu
\nu}=0$ are regions of constant curvature, with $R_{\mu \nu \alpha
\sigma}=\Lambda\left ( g_{\mu\alpha}g_{\nu \sigma} - g_{\mu
\sigma}g_{\nu\alpha} \right )$, and $R=6\Lambda$.

General relativity in a 3-dimensional spacetime has no Newtonian
limit in the sense that there is no gravitational force between
static point sources. In order to see this limit we start, has in
4-dimensions, from a slightly perturbed Minkowski background (for
simplicity, but without loss of generality, we deal this issue
only with the case $\Lambda=0$). To be more general let us work on
a $D$-dimensional spacetime described by a metric $g_{\mu\nu}$
that approaches asymptotically the $D$-dimensional Minkowski
metric $\eta_{\mu\nu}={\rm diag}(-1,+1,\cdots,+1)$, and write
\begin{equation}
g_{\mu\nu}=\eta_{\mu\nu}+h_{\mu\nu}\,
\hspace{1cm}\mu,\nu=0,1,\cdots,D-1 \,,
 \label{g-3D}
\end{equation}
where $h_{\mu\nu}$ is small, so that it represents small
corrections to the flat background. Then in later chapter
\ref{sec:inhomogeneous wave equation} we show that the first order
Einstein field equations yield
\begin{equation}
\square h_{\mu\nu}=-16\pi G_D \left (
T_{\mu\nu}-\frac{1}{D-2}\,\eta_{\mu\nu}\,T  \right ) \,,
 \label{ineq-3D}
\end{equation}
where $\square=\eta^{\mu\nu}\partial_{\mu}\partial_{\nu}$ is the
$D$-dimensional Laplacian, $T_{\mu\nu}$ is the energy-momentum
tensor, and $G_D$ is the $D$-dimensional Newton's constant. Now,
in the Newtonian limit we set as usual $T_{00}\sim \rho$ (where
$\rho$ is the mass density) and we neglect the other components of
$T_{\mu\nu}$. In these conditions, (\ref{ineq-3D}) yields
\begin{equation}
\square h_{00}=-16\pi G_D \frac{D-3}{D-2}\rho \,.
 \label{poisson-3D}
\end{equation}
In order to recover the Poisson equation, $\square \Phi=4\pi G_D
\rho$, where $\Phi$ is the gravitational potential, we set
$h_{00}=-4\frac{D-3}{D-2}\Phi$. To proceed, we apply the geodesic
equation, $\frac{d^2
x^{\alpha}}{d\tau^2}+\Gamma^{\alpha}_{\mu\nu}\frac{d
x^{\mu}}{d\tau}\frac{d x^{\nu}}{d\tau}=0$, to (\ref{g-3D})
yielding $\partial_t^2 x^i -\frac{1}{2}\partial_i h_{00}=0$.
Inserting (\ref{ineq-3D}) one finally has
\begin{equation}
\frac{d^2 x^i}{dt^2}+\frac{2(D-3)}{D-2}\frac{d \Phi}{dx^i}=0 \,.
 \label{Newtonian limit}
\end{equation}
For $D=4$ we get the usual Newtonian equation of motion, while for
$D=3$ we confirm that the acceleration is zero, that is static
point sources feel no gravitational force. The above discussion
can be extended to the $\Lambda \neq 0$ case, withe the Poisson
equation replaced by the Liouville equation, $\square
\Phi+2\Lambda e^{\Phi}=-16\pi G_D \rho$

Another unusual property of 3-dimensional gravity is the absence
of gravitational waves. In chapter \ref{sec:plane wave solutions}
we shall count the number of polarization states of a
gravitational wave in $D$ dimensions. We shall see that this
number is given by $D(D+1)/2-D-D=D(D-3)/2$. From this computation
we conclude that gravitational waves are indeed present only when
$D>3$.

Propagation of massless fields (other than the gravitational ones,
obviously) in a 3-dimensional spacetime has also a property
radically different from the propagation in a 4-dimensional
background. Indeed, in 3-dimensions, a pulse of electromagnetic or
scalar waves travels not only along the light cone but also
spreads out behind it, and slowly dies off in tails (this happens
even in the absence of any kind of black hole or matter and is due
only to the intrinsic dimension of the spacetime). That is, in
4-dimensions if one lights a lighter for five seconds and then
turn it off, any observer at rest relative to the candle will see
the light for exactly five seconds and then suddenly fade out
completely. However, in a 3-dimensional spacetime, after the light
source is turned off one will still see its shinning light, slowly
fading away but never completely  (see Soodak and Tiersten
\cite{SoodakTiersten}, and Cardoso, Yoshida, Dias and Lemos
\cite{CardosoYoshidaOscarLemos}). This is due to the fact that in
4-dimensions the retarded Green's function is
\begin{equation}
G^{\rm ret}(t,{\bf x})=\frac{1}{4\pi} \frac{\delta(t-r)}{r}\,,
\label{green-4D}
\end{equation}
where $\delta(t-r)$ is the delta function, while in 3-dimensions
it is given by
\begin{equation}
G^{\rm ret}(t,{\bf x})=\frac{\Theta(t)}{2\pi}
\frac{1}{\sqrt{t^2-r^2}}\,, \label{green-3D}
\end{equation}
where $\Theta(t)$ is the step Heaviside function defined as
\begin{equation}
\Theta(t)=\left\{ \begin{array}{ll}
             1   & \mbox{if $t>0$}\\
             0    & \mbox{if $t<0$}\,.
\end{array}\right.
\label{Heaviside-3D}
\end{equation}
Thus, while the 4-dimensional retarded Green's function is
supported only on the light cone, the 3-dimensional one this
support extends also to the interior of the light cone. So in
3-dimensions light propagates at the usual velocity of light, $c$,
but also with velocities smaller $c$ and this justifies the slow,
and never complete fading away.

 The gravitational field of a spinning point source (in a
$\Lambda=0$ background) is given by \cite{DJH_flat,Star,Clem_spin}
\begin{equation}
ds^2=-\left ( dt-4GJd\phi \right )^2 + dr^2 +(1-4GM)^2 r^2 d\phi^2
\,,
 \label{point source}
\end{equation}
where $M$ and $J$ are, respectively, the mass and the angular
momentum of the point source, and the identification of
coordinates, $(t,\phi)\sim (t,\phi+2\pi)$, holds. This line
element can be brought into the flat metric through the coordinate
transformation $\tilde{t}=t+4GJ\phi$, $\tilde{\phi}=(1-4GM)\phi$,
yielding $ds^2=-d\tilde{t}^2 + dr^2 +r^2 d\tilde{\phi}^2$.
However, due to the coordinate transformation, the usual
identification of coordinates no longer holds and has been
replaced by the new one,
$(\tilde{t},\tilde{\phi})\sim(\tilde{t}+8\pi G
J,\tilde{\phi}+2\pi[1-4GM])$. As a consequence, the spacetime
(\ref{point source}) has a conic and an helical structure. Indeed,
with the new coordinate identifications, one starts with the
Minkowski spacetime, one cuts out a wedge with an opening angle
$\delta=8\pi G M$, and one identifies the opposite edges to form a
cone with an extra time translation $8\pi G J$. This spacetime
also has the property that it admits closed timelike curves. This
occurs when $g_{\phi\phi}<0$, i.e., in a region in the vicinity of
the point source with radius $r<\frac{4G J}{1-4GM}$. Hence, the
spacetime (\ref{point source}) represents a physical solution for
distances greater than the above critical radius. For a spinless
point source this problem does not occur. Static multi-source
solutions also exist \cite{DJH_flat,Clem_spin} due to the absence
of gravitational interaction in 3-dimensions. Once more the
geometry has a conic and helical structure, with a wedge angle
suppressed at each source proportional to its mass.

In a $\Lambda \neq 0$, the spacetimes generated by point sources
in 3-dimensional Einstein gravity have also been found
\cite{DJ_sitter} and by Brown and Henneaux \cite{Brown_Hen}. In
the de Sitter case ($\Lambda>0$) there is no one-particle
solution. The simplest solution describes a pair of antipodal
particles on a sphere with a wedge removed between poles and with
points on its great circle boundaries identified \cite{DJ_sitter}.
In the anti-de Sitter case ($\Lambda<0$), the simplest solution
describes a hyperboloid with a wedge removed proportional to the
source mass located at the vertex of the wedge \cite{Brown_Hen}.

\section{The neutral BTZ black hole}\label{sec:BTZ}

\subsection{Identifications in $AdS_3$} \label{sec:IdentAdS3}

We will now consider \cite{HerdeiroLemos} what kind of
`topological spacetimes' can be obtained by considering
identifications in $AdS_3$. First, recall that $AdS_3$ is normally
defined by its embedding in ${\mathbb{M}}^{2,2}$, parameterised by
coordinates $x^i=(u,v,x,y)$ and with flat metric
\begin{equation} ds^2=-du^2-dv^2+dx^2+dy^2 \ . \label{flatm}
\end{equation} The $AdS_3$ `curve' is the hyperboloid
\begin{equation} -u^2-v^2+x^2+y^2=-\ell^2 \ , \label{hypers} \end{equation}
where $\ell=\sqrt{3/|\Lambda|}$ is the cosmological length. The
isometry group of ${\mathbb{M}}^{2,2}$ is $ISO(2,2)$, and the
Killing vector fields might be taken to be the timelike and
spacelike rotations \begin{equation}
J_{01}=v\partial_u-u\partial_v \ , \ \ \ \ \
J_{23}=y\partial_x-x\partial_y \ , \label{killingRotAdS3-4D}
 \end{equation} the four linearly independent
boosts \begin{equation} J_{02}=u\partial_x+x\partial_u \ , \ \
J_{03}=u\partial_y+y\partial_u \ , \ \
J_{12}=v\partial_x+x\partial_v \ , \ \
J_{13}=v\partial_y+y\partial_v \ , \label{killingBoostAdS3-4D}
 \end{equation}
and the four translations
\begin{equation}
P_u=\partial_u \ , \ \ P_v=\partial_v \ , \ \ P_x=\partial_x \ , \
\ P_y=\partial_y \ .
\end{equation}
Of these isometries, the translations do not leave the $AdS_3$
curve invariant. The isometry group of $AdS_3$ is only the
$SO(2,2)$ generated by the two rotations and the four boosts.

Let $\xi$ be one of the isometries of $AdS_3$. We wish to identify
points along the orbits of $\xi$. Such identifications define a
discrete subgroup of the isometry group, $G\subset SO(2,2)$, and
the quotient
\begin{equation}
{\mathcal{M}}^{iden}\equiv \frac{AdS_3\cong SL(2,{\mathbb{R}})}{G}
\ ,
\end{equation} defines
a new spacetime ${\mathcal{M}}^{iden}$. The universal covering
space of this spacetime, $\widetilde{{\mathcal{M}}^{iden}}$ is the
universal covering space of $AdS_3$.

In practice, the identification is implemented by defining an
angular coordinate $\phi$, such that
\begin{equation} \xi
=\partial_{\phi} \ . \label{identificacao}
\end{equation}
This angular direction will, of course, be an isometry of
${\mathcal{M}}^{iden}$. In order for the identification not to
create closed timelike curves, one might also require that $\xi$
should be spacelike. Thus one defines a `radial' coordinate, $r$,
such that
\begin{equation}
r^2=\xi \cdot \xi \ . \label{radial}
\end{equation}
The `origin' $r=0$ is when $\xi$ becomes null. It will be faced as
some kind of spacetime boundary, which may or may not be a conical
singularity depending on the the action of $\xi$ having fixed
points or being free.

\subsection{The construction leading to the BTZ black hole}
 \label{sec:non-extBTZ}

Take
\begin{equation}
\xi=\frac{r_+}{\ell}J_{02}-\frac{r_-}{\ell}J_{13}=
\frac{r_+}{\ell}\left(u\partial_x+x\partial_u\right)-\frac{r_-}{\ell}
\left(v\partial_y+y\partial_v\right) \ , \label{kvfne}
\end{equation}
where $r_{\pm}$ are constants, $r_+>r_-$. That is we will be
identifying points along a double boost: a boost in the
$(u,x)$-plane and a boost in the $(v,y)$-plane. For $r_{\pm}\neq
0$ the action of this vector field has a fixed point only at
$u=v=x=y=0$, which does not belong to the hyperboloid. Hence this
isometry acts freely on the hyperboloid. For $r_-=0$, the points
$u=x=0$ are fixed points of the isometry action. Hence the one
dimensional hyperboloid
\begin{equation} -v^2+y^2=-\ell^2 \ , \end{equation} is a surface of fixed
points. So we expect a conical singularity for $r_-=0$ but not for
$r_{\pm}\neq 0$.

From (\ref{identificacao}) the angular coordinate in
${\mathcal{M}}^{iden}$ is \begin{equation}
\partial_{\phi}=\frac{r_+}{\ell}\left(u\partial_x+x\partial_u\right)-\frac{r_-}{\ell}\left(v\partial_y+y\partial_v\right) \ . \end{equation}
Hence \begin{equation}
\left\{ \begin{array}{l} \displaystyle{\frac{\partial x}{\partial \phi}=u\frac{r_+}{\ell}} \spa{0.3}\\
\displaystyle{\frac{\partial u}{\partial \phi}=x\frac{r_+}{\ell}} \end{array} \right. \ \Rightarrow \ \left\{ \begin{array}{l} \displaystyle{\frac{\partial^2 x}{\partial \phi^2}=\left(\frac{r_+}{\ell}\right)^2 x} \spa{0.3}\\
\displaystyle{\frac{\partial^2 u}{\partial \phi^2}=\left(\frac{r_+}{\ell}\right)^2u} \end{array} \right. \ {\rm{and}} \ \ \ \left\{ \begin{array}{l} \displaystyle{\frac{\partial y}{\partial \phi}=-v\frac{r_-}{\ell}} \spa{0.3}\\
\displaystyle{\frac{\partial v}{\partial \phi}=-y\frac{r_-}{\ell}} \end{array} \right. \ \Rightarrow \ \left\{ \begin{array}{l} \displaystyle{\frac{\partial^2 y}{\partial \phi^2}=\left(\frac{r_-}{\ell}\right)^2 y} \spa{0.3}\\
\displaystyle{\frac{\partial^2 v}{\partial
\phi^2}=\left(\frac{r_-}{\ell}\right)^2v} \end{array} \right.\ .
\end{equation} Taking the parameters on the hyperboloid to be
$(t,r,\phi)$ the last equations imply that \begin{equation}
u,x=\alpha(t,r)\left\{\begin{array}{l}\displaystyle{\sinh\left(\frac{r_+}{\ell}\phi+\beta(r,t)\right)}
\spa{0.3}\\
\displaystyle{\cosh\left(\frac{r_+}{\ell}\phi+\beta(r,t)\right)}
\end{array} \right. , \ \
v,y=\gamma(t,r)\left\{\begin{array}{l}\displaystyle{\sinh\left(-\frac{r_-}{\ell}\phi+\delta(r,t)\right)}
\spa{0.3}\\
\displaystyle{\cosh\left(-\frac{r_-}{\ell}\phi+\delta(r,t)\right)}
\end{array} \ , \right. \label{embfun1} \end{equation} where, for each
of the variables $(u,v,x,y)$ either the `$\cosh$' or the `$\sinh$'
solution could be taken. Thus, using simply (\ref{identificacao})
for our choice of Killing vector field (\ref{kvfne}) we have
constrained the form of the embedding functions to be
(\ref{embfun1}).

Next, we use the second requirement, namely the definition of the
radial coordinate (\ref{radial}). This yields the condition
\begin{equation}
r^2=\left(\frac{r_+}{\ell}\right)^2(u^2-x^2)+\left(\frac{r_-}{\ell}\right)^2(v^2-y^2)
\ . \label{constraint2} \end{equation} It seems therefore natural
that if we take the solution $\cosh$ ($\sinh$) for $u$, then we
should take the solution $\sinh$ ($\cosh$) for $x$. The same
applies to the pair $(v,y)$. Thus, for the first pair we have the
two possible solutions\begin{equation} {\rm{Possibility \ 1 \ for
\ (u,x)}}\ , \ \
\left\{\begin{array}{l}\displaystyle{u=\alpha{(r,t)}\sinh\left(\frac{r_+}{\ell}\phi+\beta(r,t)\right)}
\spa{0.3}\\
\displaystyle{x=\alpha{(r,t)}\cosh\left(\frac{r_+}{\ell}\phi+\beta(r,t)\right)}
\end{array} \right. \Rightarrow \ \ \ \ u^2-x^2=-\alpha{(r,t)}^2 \ ,
\end{equation} \begin{equation} {\rm{Possibility \ 2 \ for \ (u,x)}}\ , \
\
\left\{\begin{array}{l}\displaystyle{u=\alpha{(r,t)}\cosh\left(\frac{r_+}{\ell}\phi+\beta(r,t)\right)}
\spa{0.3}\\
\displaystyle{x=\alpha{(r,t)}\sinh\left(\frac{r_+}{\ell}\phi+\beta(r,t)\right)}
\end{array} \right. \Rightarrow \ \ \ \ u^2-x^2=\alpha{(r,t)}^2 \ .
\end{equation} Similarly, for the second pair we have the two possibilities
\begin{equation} {\rm{Possibility \ 1 \ for \ (v,y)}}\ , \  \
\left\{\begin{array}{l}\displaystyle{v=\gamma{(r,t)}\sinh\left(-\frac{r_-}{\ell}\phi+\delta(r,t)\right)}
\spa{0.3}\\
\displaystyle{y=\gamma{(r,t)}\cosh\left(-\frac{r_-}{\ell}\phi+\delta(r,t)\right)}
\end{array} \right. \Rightarrow \ \ \ \ v^2-y^2=-\gamma{(r,t)}^2 \ ,
\end{equation} \begin{equation} {\rm{Possibility \ 2 \ for \ (v,y)}}\ , \
\
\left\{\begin{array}{l}\displaystyle{v=\gamma{(r,t)}\cosh\left(-\frac{r_-}{\ell}\phi+\delta(r,t)\right)}
\spa{0.3}\\
\displaystyle{y=\gamma{(r,t)}\sinh\left(-\frac{r_-}{\ell}\phi+\delta(r,t)\right)}
\end{array} \right. \Rightarrow \ \ \ \ v^2-y^2=\gamma{(r,t)}^2 \ .
\end{equation}

Denote by $i\otimes j$ choosing possibility $i$ for $(u,x)$ and
possibility $j$ for $(v,y)$. We will insert this several
possibilities in (\ref{constraint2}).  Since the left hand side of
(\ref{constraint2}) only depends on $r$ it is clear that the
simplest solutions for the functions $\alpha(r,t)$ and
$\gamma(r,t)$ will be functions of $r$ only. Thus we take
$\alpha(r)$ and $\gamma(r)$ from now on.

\begin{description}
\item[$\bullet$] $1\otimes 1$ yields for (\ref{constraint2}),
\begin{equation}
\ell^2r^2=-(r_+)^2\alpha(r)^2-(r_-)^2\gamma(r)^2 \ .
\end{equation}
Since the functions $\alpha(r,t)$ and $\gamma(r,t)$ are supposed
to be real functions this condition is impossible to solve.

\item[$\bullet$] $1\otimes 2$ yields for (\ref{constraint2}),
\begin{equation}
\ell^2r^2=-(r_+)^2\alpha(r)^2+(r_-)^2\gamma(r)^2 \ .
\end{equation}
The solution is
\begin{equation}
\alpha(r)=\sqrt{\frac{r_-^2-r^2}{r_+^2-r_-^2}}\, \ell \ , \ \ \ \
\gamma(r)=\sqrt{\frac{r_+^2-r^2}{r_+^2-r_-^2}}\, \ell \ ,
\end{equation}
which is only valid for $r<r_-<r_+$, which we call $region \ III$.

\item[$\bullet$] $2\otimes 1$ yields for (\ref{constraint2}),
\begin{equation}
\ell^2r^2=(r_+)^2\alpha(r)^2-(r_-)^2\gamma(r)^2 \ .
\end{equation}
The solution is
\begin{equation}
\alpha(r)=\sqrt{\frac{r^2-r_-^2}{r_+^2-r_-^2}}\, \ell \ , \ \ \ \
\gamma(r)=\sqrt{\frac{r^2-r_+^2}{r_+^2-r_-^2}}\, \ell \ ,
\end{equation}
which is only valid for $r_-<r_+<r$, which we call $region \ I$.

\item[$\bullet$] $2\otimes 2$ yields for (\ref{constraint2}),
\begin{equation}
\ell^2r^2=(r_+)^2\alpha(r)^2+(r_-)^2\gamma(r)^2 \ .
 \end{equation} The
solution is
\begin{equation}
\alpha(r)=\sqrt{\frac{r^2-r_-^2}{r_+^2-r_-^2}}\, \ell \ , \ \ \ \
\gamma(r)=\sqrt{\frac{r_+^2-r^2}{r_+^2-r_-^2}}\, \ell \ ,
\end{equation}
which is only valid for $r_-<r<r_+$, which we call $region \ II$.
\end{description}

Thus, the embedding functions of the BTZ black hole are, in Region
I
 \begin{equation}
\left\{\begin{array}{l}
\displaystyle{u=\alpha(r)=\sqrt{\frac{r^2-r_-^2}{r_+^2-r_-^2}}\,
\ell\cosh\left(\frac{r_+}{\ell}\phi+\beta(r,t)\right)}
\spa{0.3}\\
\displaystyle{x=\alpha(r)=\sqrt{\frac{r^2-r_-^2}{r_+^2-r_-^2}}\,
\ell\sinh\left(\frac{r_+}{\ell}\phi+\beta(r,t)\right)}
    \end{array} \right.
\left\{\begin{array}{l}
\displaystyle{v=\gamma(r)=\sqrt{\frac{r^2-r_+^2}{r_+^2-r_-^2}}\,
\ell\sinh\left(-\frac{r_-}{\ell}\phi+\delta(r,t)\right)}
\spa{0.3}\\
\displaystyle{y=\gamma(r)=\sqrt{\frac{r^2-r_+^2}{r_+^2-r_-^2}}\,
\ell\cosh\left(-\frac{r_-}{\ell}\phi+\delta(r,t)\right)}
\end{array} \right.
 \end{equation}
 in region II
 \begin{equation}
\left\{\begin{array}{l}
\displaystyle{u=\alpha(r)=\sqrt{\frac{r^2-r_-^2}{r_+^2-r_-^2}}\,
\ell\cosh\left(\frac{r_+}{\ell}\phi+\beta(r,t)\right)}
\spa{0.3}\\
\displaystyle{x=\alpha(r)=\sqrt{\frac{r^2-r_-^2}{r_+^2-r_-^2}}\,
\ell\sinh\left(\frac{r_+}{\ell}\phi+\beta(r,t)\right)}
\end{array} \right.
 \left\{\begin{array}{l}
 \displaystyle{v=\gamma(r)=\sqrt{\frac{r_+^2-r^2}{r_+^2-r_-^2}}\,
 \ell\cosh\left(-\frac{r_-}{\ell}\phi+\delta(r,t)\right)}
\spa{0.3}\\
\displaystyle{y=\gamma(r)=\sqrt{\frac{r_+^2-r^2}{r_+^2-r_-^2}}\,
\ell\sinh\left(-\frac{r_-}{\ell}\phi+\delta(r,t)\right)}
\end{array} \right.
 \end{equation}
  and in region III
  \begin{equation}
\left\{\begin{array}{l}
\displaystyle{u=\alpha(r)=\sqrt{\frac{r_-^2-r^2}{r_+^2-r_-^2}}\,
\ell\sinh\left(\frac{r_+}{\ell}\phi+\beta(r,t)\right)}
\spa{0.3}\\
\displaystyle{x=\alpha(r)=\sqrt{\frac{r_-^2-r^2}{r_+^2-r_-^2}}\,
\ell\cosh\left(\frac{r_+}{\ell}\phi+\beta(r,t)\right)}
\end{array} \right.
\left\{\begin{array}{l}
\displaystyle{v=\gamma(r)=\sqrt{\frac{r_+^2-r^2}{r_+^2-r_-^2}}\,
\ell\cosh\left(-\frac{r_-}{\ell}\phi+\delta(r,t)\right)}
\spa{0.3}\\
\displaystyle{y=\gamma(r)=\sqrt{\frac{r_+^2-r^2}{r_+^2-r_-^2}}\,
\ell\sinh\left(-\frac{r_-}{\ell}\phi+\delta(r,t)\right)}
\end{array} \right.
 \end{equation}

Choosing, for instance, the parameterization in region I, we can
now compute the metric induced on the hypersurface (\ref{hypers})
by (\ref{flatm}). Since our parameterization already obeys all the
constraints, still with $\beta(r,t)$ and $\gamma(r,t)$
unspecified, we take
\begin{equation}
\beta(r,t)=\beta t \ , \ \ \ \delta(r,t)=\delta t \ ,
\end{equation}
 with $\beta, \delta$ constants. The induced metric becomes
 \begin{equation}
\begin{array}{c}
\displaystyle{ds^2=\frac{r^2\ell^2dr^2}{(r^2-r^2_+)(r^2-r^2_-)}+r^2
d\phi^2+\left [ r^2(\beta r_+ +\delta r_-)-r_+r_-(\beta r_-+\delta
r_+)\right ]\frac{2\ell d\phi dt}{r_+^2-r_-^2}} \spa{0.4}\\
\displaystyle{~~~~~~~~~~~~~ +\frac{\ell^2dt^2}{r_+^2-r_-^2}\left [
r^2(\beta^2-\delta^2)+r_+^2\delta^2-r_-^2\beta^2\right ]}
\end{array} \ .
 \end{equation}
 Taking $\beta=-\chi r_-$ and $\delta=\chi
r_+$, for constant $\chi$, the cross terms becomes $r$
independent. Since $\beta$ and $\delta$ should have dimension
$(length)^{-1}$, take $\chi=1/\ell^2$. Then, the metric becomes
\begin{equation}
ds^2=-\frac{(r^2-r^2_+)(r^2-r^2_-)}{r^2\ell^2}dt^2+\frac{r^2\ell^2dr^2}
{(r^2-r^2_+)(r^2-r^2_-)}+r^2\left(d\phi-\frac{r_+r_-}{\ell
r^2}dt\right)^2 \ ,
\end{equation}
which is the standard BTZ metric. Alternatively, writing
\begin{equation}
m=\frac{r_+^2+r_-^2}{\ell^2} \ , \ \ \ \ J=\frac{2r_+r_-}{\ell} \
,
\end{equation} the metric becomes
\begin{equation}
ds^2=-\left(-m+\frac{r^2}{\ell^2}+\frac{J^2}{4r^2}\right)dt^2-
\frac{dr^2}{\left(-m+\frac{r^2}{\ell^2}+\frac{J^2}{4r^2}\right)}
+r^2\left(d\phi-\frac{J}{2r^2}\right)^2\ .
\end{equation}
m and J are interpreted as mass and angular momentum of the black
hole, respectively.

\section{The electric BTZ solutions}
 \label{sec:electric BTZ}

\subsection{The BTZ black hole with a radial electric field}
 \label{sec:radial electric BTZ}

\subsubsection{Static black hole with a radial electric field}
\begin{figure} [H]
\centering
\includegraphics[height=2.2in]{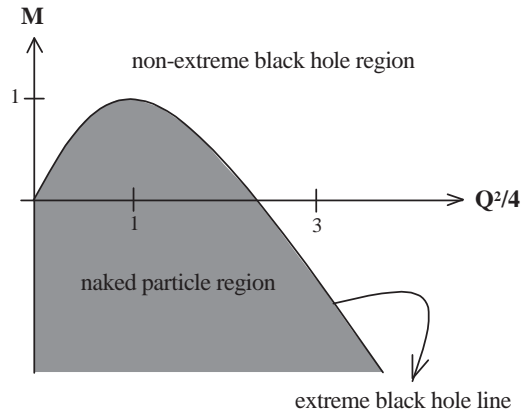}
\caption{\label{mq_Q BTZ}
 Values of the mass and of the charge of the static electric BTZ solution
for which one has a non-extreme black hole, an extreme black hole,
and a naked particle. The line represents the function
$m=(q/2)^2\left [ 1-\ln (q/2)^2\right ]$.}
\end{figure}

The gravitational field of the static electric BTZ black hole is
\cite{BTZ_Q}
\begin{equation}
ds^2=-f^2dt^2+ f^{-2} dr^2 +r^2 d\phi^2 \,, \qquad {\rm with}
\:\:\:\: f^2= r^2-m-\frac{q^2}{4} \ln\,r^2 \,,
 \label{static electric BTZ}
\end{equation}
where $m$, and $q$ are the mass and electric charge of the black
hole, and the electromagnetic vector potential 1-form is
\begin{equation}
A=-q \ln r \,dt \;.
                    \label{Max static electric BTZ}
\end{equation}
The function $f^2$ goes to $+\infty$ when $r\rightarrow 0$ and
when $r\rightarrow +\infty$. It has a minimum at $r_{\rm min}=\mid
q \mid/2$, and the value of $f^2$ at this minimum is
 $f^2_{\rm min}=-m+(q/2)^2\left [ 1-\ln (q/2)^2\right ]$. When
 $f^2_{\rm min}$ is negative, $f^2$ has two roots and the
 corresponding charged black hole has two horizons. When these two
 roots coincide one has an extreme charged black hole and,
 finally, when $f^2$ is positive one has a naked singularity.
The parameters $m$ and $q$ that represent these three regions are
represented in Fig. \ref{mq_Q BTZ}, withdrawn from \cite{BTZ_Q}.
When $0<m<1$, for small values of the charge there is a black hole
solution. Then, for a fixed mass in this interval, when the charge
increases slightly there is a finite range of $q$ where we find no
black hole solution and, surprisingly, when this charge grows
above a critical value we have again a black hole solution, no
matter how big the charge is. This is in clear contrast with the
usual charged solutions in (3+1)-dimensional spacetimes, for which
the black hole solutions exist only when the charge is below a
certain critical value. Moreover, and also in oppose to what
happens in (3+1)-dimensions, there exist black holes with negative
mass as long as the charge is big enough.

\subsubsection{Rotating black hole with a radial electric field}
The angular momentum is added to the solution (\ref{static
electric BTZ}) and (\ref{Max static electric BTZ}) through the
application of a Lorentz rotation boost in the $t$-$\varphi$ plane
\cite{BTZ_Q,HorWel}, $t \mapsto \frac{t-\varpi
\varphi}{\sqrt{1-\varpi^2}}$ and $\varphi \mapsto
\frac{\varphi-\varpi t}{\sqrt{1-\varpi^2}}$, where $\varpi^2 \leq
1$ is the rotation velocity parameter, and $\sqrt{1-\varpi^2}$ is
the usual Lorentz boost factor. The gravitational field of the
rotating electric BTZ black hole is then \cite{BTZ_Q}
\begin{eqnarray}
ds^2=-N^2 F^2 dt^2+F^{-2}dR^2+R^2(d\varphi+N^{\varphi}dt)^2 \,,
\label{rot electric BTZ}
\end{eqnarray}
where
\begin{eqnarray}
R^2=\frac{r^2-\varpi^2 f^2}{1-\varpi^2}\,, \quad \: F^2=\left (
\frac{d R}{dr}\right )^2 f^2\,, \quad \:
N=\frac{r}{R}\frac{dr}{dR}\,, \quad  {\rm and} \quad \:
N^{\varphi}=\frac{\varpi(f^2-r^2)} {R^2(1-\varpi^2)} \,,
\label{aux rot electric BTZ}
\end{eqnarray}
and $f^2$ is defined in (\ref{static electric BTZ}). Under the
above Lorentz boost the potential (\ref{Max static electric BTZ})
transforms into
\begin{equation}
A=-\frac{q \ln r}{\sqrt{1-\varpi^2}} (dt-\varpi d \varphi) \;.
                    \label{Max rot electric BTZ}
\end{equation}
The rotating electric BTZ black hole can have a maximum of two
horizons which are the direct counterparts of the two horizons of
the static electric BTZ solution. The ADM mass, angular momentum,
and electric charge of the solution are given \cite{BTZ_Q},
respectively, by $M=\left [ m(1+\varpi^2)-\varpi^2 q^2/2\right
]/(1-\varpi^2)$,
 $J=2\varpi (m-q^2/4)$, and $Q=q/\sqrt{1-\varpi^2}$.

\subsection{The BTZ solution with an azimuthal electric field}
 \label{sec:azimuthal electric BTZ}
To conclude this brief overview on the electric counterparts of
the BTZ solution, we mention that Cataldo \cite{Cat} has found a
horizonless BTZ solution with an azimuthal electric field, whose
gravitational field is given by
\begin{equation}
ds^2=-r^2dt^2+ g^{-2}dr^2 +r^2 d\phi^2 \,, \qquad {\rm with}
\:\:\:\: g^2= \frac{r^2}{\ell^2}+\frac{q^2}{r^2} \,,
 \label{azimuthal electric BTZ}
\end{equation}
where $q$ is the electric charge of the solution, and with
electromagnetic vector potential 1-form given by
\begin{equation}
A=-\frac{q}{r^2}\phi \,dt \;,
                    \label{Max azimuthal electric BTZ}
\end{equation}
to which corresponds the above mentioned azimuthal electric field.
When we set $q=0$ (\ref{azimuthal electric BTZ}) yields the empty
$AdS_3$ spacetime.
\section{The magnetic BTZ solution}
 \label{sec:magnetic BTZ}

In this section, we discuss in detail the static
\cite{CL1,HW,Cat_Sal} and rotating \cite{OscarLemos_BTZ} magnetic
solution, that reduces to the BTZ solution when the magnetic field
source vanishes. In particular, the conserved quantities (mass,
angular momentum and electric charge), as well as their upper
bounds, are defined, and we give a physical interpretation for the
magnetic field source.

The plan of this section is the following. In Section
\ref{sec:static magnet BTZ} we study the static magnetic solution
found in \cite{CL1,HW,Cat_Sal} and its properties. Section
\ref{sec:rotat magnet BTZ} is devoted to the rotating magnetic
solution. The angular momentum is added in section \ref{sec:rotat
magnet BTZ 1} through a rotational Lorentz boost. In section
\ref{sec:rotat magnet BTZ 2} we calculate the mass, angular
momentum, and electric charge of both the static and rotating
solutions and in section \ref{sec:rotat magnet BTZ 3} we set
relations between the conserved charges. The rotating magnetic
solution is written as a function of its hairs in section
\ref{sec:rotat magnet BTZ 4} and we show that it reduces to the
rotating BTZ solution when the magnetic source vanishes. In
section \ref{sec:Phys Interp magnetic BTZ} we give a physical
interpretation for the origin of the magnetic field source.

\subsection{Static solution. Analysis of its general structure}
\label{sec:static magnet BTZ}

\subsubsection{Static solution}

Einstein gravity with a negative cosmological constant and a
source of magnetic field (or Einstein--Maxwell--anti de Sitter
gravity) in three dimensions can be characterized by the action
\begin{equation}
S=\frac{1}{16\pi G} \int d^3x \sqrt{-g}
  {\bigl [} R -2 \Lambda - F^{\mu \nu}F_{\mu \nu}
  {\bigr ]}\,                                    \label{action BTZ}
\end{equation}
where $g$ is the determinant of the metric $g_{\mu\nu}$, $R$ the
Ricci scalar, $F_{\mu\nu}$ the electromagnetic tensor given in
terms of the vector potential $A_\mu$ by
$F_{\mu\nu}=\partial_{\mu} A_\nu - \partial_\nu A_\mu$, and
$\Lambda$ is the cosmological constant. In this subsection we
study the static point source solution of action
 (\ref{action BTZ}) found  by Hirschmann and Welch \cite{HW} and by Cataldo and
Salgado \cite{Cat_Sal}. The point source generates a gravitational
field and a magnetic field. Written in the gauge presented in
\cite{HW}, the static solution is given by the following metric
and vector potential 1-form
\begin{eqnarray}
ds^2 &=& - (r^2/\ell^2 -m)dt^2
        + [r^2+\chi^2_{\rm m}
           \ln{\mid r^2/\ell^2 - m \mid}] d\varphi^2
                            \nonumber \\
     & &   + r^2(r^2/\ell^2 -m)^{-1}
         [r^2+\chi^2_{\rm m}\ln{\mid r^2/\ell^2-m \mid}]^{-1}d r^2  \:,
                    \label{HW_0 BTZ}
\end{eqnarray}
\begin{equation}
A=\frac{1}{2}  \chi_{\rm m} \ln{\mid r^2/\ell^2-m \mid} d \varphi
\;,
                    \label{HW_Max BTZ}
\end{equation}
where $t$, $r$ and $\varphi$ are the time, the radial and the
angular coordinates, respectively, $\chi_{\rm m}$ is an
integration constant which measures the intensity of the magnetic
field source and $\ell\equiv -\frac{1}{\sqrt\Lambda}$ is the
cosmological length. This spacetime reduces to the three
dimensional BTZ black hole solution of Ba\~nados, Teitelboim and
Zanelli \cite{btz_PRL,btz_PRD} when the magnetic source vanishes.
The parameter $m$ is the mass of this uncharged solution.

The $g_{rr}$ function is negative for $r<r_+$ and positive for
$r>r_+$, where $r_+$ is such that
\begin{eqnarray}
r_+^2 + \chi^2_{\rm m} \ln{\mid r_+^2/\ell^2-m \mid}=0 \:,
                                  \label{r+ BTZ}
\end{eqnarray}
and the condition $r_+^2>m\ell^2$ is obeyed. One might then be
tempted to say that the solution has an horizon at $r=r_+$ and
consequently that one is in the presence of a magnetically charged
black hole. However, this is not the case. In fact, one first
notices that the metric components $g_{rr}$ and $g_{\varphi
\varphi}$ are related by $g_{\varphi \varphi}=
[g_{rr}(r^2-m\ell^2)/(\ell^{2} r^{2})]^{-1}$. Then, when $g_{rr}$
becomes negative (which occurs for $r<r_+$) so does $g_{\varphi
\varphi}$ and this leads to an apparent change of signature from
$+2$ to $-2$. This strongly indicates \cite{Hor_Hor} that an
incorrect extension  is being used and that one should choose a
different continuation to describe the region $r<r_+$. By
introducing a new radial coordinate, $\rho^2=r^2-r_+^2$, one
obtains a spacetime that is both null and timelike geodesically
complete for $r\geq r_+$ \cite{HW},
\begin{eqnarray}
ds^2 &=& -\frac{1}{\ell^2} (\rho^2+r_+^2 -m \ell^2)dt^2
        + \left [\rho^2+\chi^2_{\rm m}
           \ln{\left (1+
       \frac{\rho^2}{r_+^2 -m\ell^2}\right )}\right ]
            d\varphi^2                 \nonumber \\
     & &   + \ell^2 \rho^2(\rho^2+r_+^2-m\ell^2)^{-1}
      \left [ \rho^2+\chi^2_{\rm m}\ln{\left (1+
         \frac{\rho^2}{r_+^2 -m\ell^2}\right )}\right ]^{-1}
          d \rho^2  \:,
                    \label{HW_1 BTZ}
\end{eqnarray}
where $0\leq \rho < \infty$. This static spacetime has no
curvature singularity, but it presents a conical geometry and, in
particular, it has a conical singularity at $\rho=0$ which can be
removed if one identifies $\varphi$ with the period $T_{\varphi}=2
\pi \nu$ where \cite{HW}
 \begin{eqnarray}
\nu=\frac{\exp{(\beta/2)}}{[1+\chi^2_{\rm
m}\exp{(\beta)/\ell^2}]}\:,
                                \label{freq BTZ}
\end{eqnarray}
and $\beta=r_+^2/\chi^2_{\rm m}$. Near the origin, metric
(\ref{HW_1 BTZ}) describes a spacetime which is locally flat but
has a conical singularity at $\rho=0$ with an angle deficit
$\delta \varphi=2\pi(1-\nu)$.

\subsubsection{Geodesic structure}

We want to study the geodesic motion  and, in particular, to
confirm that the spacetime described by  (\ref{HW_1 BTZ}) is both
null and timelike geodesically complete, i.e., that every null or
timelike geodesic starting from an arbitrary point either can be
extended to infinite values of the affine parameter along the
geodesic or ends on a singularity. The equations governing the
geodesics can be derived from the
 Lagrangian
\begin{equation}
{\cal{L}}=\frac{1}{2}g_{\mu\nu}\frac{dx^{\mu}}{d \tau}
       \frac{dx^{\nu}}{d \tau}=-\frac{\delta}{2}\:,
                                 \label{LAG BTZ)}  \\
\end{equation}
where $\tau$ is an affine parameter along the geodesic which, for
a timelike geodesic, can be identified with the proper time of the
particle along the geodesic. For a null geodesic one has
$\delta=0$ and for a timelike geodesic $\delta=+1$. From the
Euler-Lagrange equations one gets that the generalized momenta
associated with the time coordinate and angular coordinate are
constants: $p_t=E$ and $p_{\varphi}=L$. The constant $E$ is
related to the timelike Killing vector $(\partial/\partial
t)^{\mu}$ which reflects the time translation invariance of the
metric, while the constant $L$ is associated to the spacelike
Killing vector $(\partial/\partial \varphi)^{\mu}$ which reflects
the invariance of the metric under rotation. Note that since the
spacetime is not asymptotically flat, the constants $E$ and $L$
cannot be interpreted as the  energy and angular momentum at
infinity.

 From the metric we can derive the radial geodesic,
\begin{eqnarray}
\dot{\rho}^2=-\frac{1}{g_{\rho\rho}}
      \frac{E^2 g_{\varphi \varphi}+L^2 g_{tt}}
              { g_{tt} g_{\varphi \varphi} }
       -\frac{\delta}{g_{\rho\rho}} \:.
                                        \label{GEOD_1 BTZ}
\end{eqnarray}
Using the two useful relations $g_{tt} g_{\varphi
\varphi}=-\rho^2/g_{\rho\rho}$ and $g_{\varphi
\varphi}=[g_{\rho\rho} (\rho^2+r_+^2-m\ell^2) /(\ell^2
\rho^2)]^{-1}$, we can write (\ref{GEOD_1 BTZ}) as
\begin{eqnarray}
\rho^2 \dot{\rho}^2= {\biggl [} \frac{\ell^2 E^2}{\rho^2 +
r_+^2-m\ell^2} -\delta {\biggr ]} \frac{\rho^2}{g_{\rho \rho}}
            +L^2 g_{tt} \:.
                                        \label{Geod_1 BTZ}
\end{eqnarray}
{\bf (i) Null geodesics} ($\delta=0$) $-$ Noticing that
$1/g_{\rho\rho}$ is always positive for $\rho>0$ and zero for
$\rho=0$, and that $g_{tt}$ is always negative we conclude the
following about the null geodesic motion. The first term in
(\ref{Geod_1 BTZ}) is positive (except at $\rho=0$ where it
vanishes),  while the second term is always negative. We can then
conclude that spiraling ($L \neq 0$) null particles coming in from
an arbitrary point are scattered at the turning point $\rho_{\rm
tp} > 0$ and spiral back to infinity. If the angular momentum L of
the null particle is zero it hits the origin (where there is a
conical singularity) with vanishing velocity.
 \newline{\bf (ii)
Timelike geodesics} ($\delta=+1$) $-$ Timelike geodesic motion is
possible only if the energy of the particle satisfies $E >
(r_+^2-ml^2)^{1/2}/\ell$. In this case, spiraling timelike
particles are bounded between two turning points that satisfy
$\rho_{\rm tp}^{\rm a} > 0$ and $\rho_{\rm tp}^{\rm b} <
\sqrt{\ell^2(E^2+m) - r_+^2}$, with $\rho_{\rm tp}^{\rm b} \geq
\rho_{\rm tp}^{\rm a}$. When the timelike particle has no angular
momentum ($L=0$) there is a turning point located at $\rho_{\rm
tp}^{\rm b}=\sqrt{\ell^2(E^2+m) - r_+^2}$ and it hits the conical
singularity at the origin $\rho=0$. Hence, we confirm that the
spacetime described by (\ref{HW_1 BTZ}) is both timelike and null
geodesically complete.


\subsection{Rotating magnetic solution}\label{sec:rotat magnet BTZ}

\subsubsection{Addition of angular momentum} \label{sec:rotat magnet BTZ 1}

Now, we want to endow the  spacetime solution (\ref{HW_1 BTZ})
with a global rotation, i.e.,  we want to add angular momentum to
the spacetime. In order to do so we perform the following rotation
boost in the $t$-$\varphi$ plane (see e.g.
\cite{BTZ_Q,OscarLemos,HorWel})
\begin{eqnarray}
 t &\mapsto& \gamma t-\ell \omega \varphi \:,
                                       \nonumber  \\
 \varphi &\mapsto& \gamma \varphi-\frac{\omega}{\ell} t \:,
                                       \label{TRANSF_J_HW BTZ}
\end{eqnarray}
where $\gamma$ and $\omega$ are constant parameters. Substituting
(\ref{TRANSF_J_HW BTZ}) into (\ref{HW_1 BTZ}) and
 (\ref{HW_Max BTZ}) we obtain the stationary spacetime generated by a  magnetic
source
\begin{eqnarray}
ds^2 &=& -\frac{1}{\ell^2} {\biggl [}(\gamma^2-\omega^2)\rho^2
         +\gamma^2(r_+^2 -m \ell^2) -\omega^2 \chi^2_{\rm m}
           \ln{{\biggl (}1+
         \frac{\rho^2}{r_+^2 -m\ell^2}{\biggr )}}{\biggr ]}dt^2
                                          \nonumber \\
     & &  -\frac{\gamma \omega}{\ell}
          {\biggl [} -(r_+^2 -m \ell^2) + \chi^2_{\rm m}
           \ln{{\biggl (}1+
         \frac{\rho^2}{r_+^2 -m\ell^2}{\biggr )}}{\biggr ]}
          2 dt d \varphi
                           \nonumber \\
     & &   + \ell^2 \rho^2(\rho^2+r_+^2-m\ell^2)^{-1}
      {\biggl [}\rho^2+\chi^2_{\rm m}\ln{{\biggl (}1+
         \frac{\rho^2}{r_+^2 -m\ell^2}{\biggr )}}{\biggr ]}^{-1}
          d \rho^2  \nonumber \\
& &   + {\biggl [}(\gamma^2-\omega^2)\rho^2
        -\omega^2(r_+^2 -m \ell^2)+\gamma^2
       \chi^2_{\rm m} \ln{{\biggl (}1+
         \frac{\rho^2}{r_+^2 -m\ell^2}{\biggr )}}{\biggr ]}
            d\varphi^2 \:,
                   \label{HW_Rot BTZ}  \\
A &=& -\frac{\omega}{\ell}A(\rho)dt + \gamma A(\rho) d \varphi \:,
\label{HW_Max_Rot BTZ}
\end{eqnarray}
with $A(\rho)=  \chi_{\rm m} \ln{[(\rho^2+r_+^2)/\ell^2-m]}/2$. We
set $\gamma^2-\omega^2=1$ because in this way when the angular
momentum vanishes ($\omega=0$) we have $\gamma=1$ and so we
recover the static solution.

Solution (\ref{HW_Rot BTZ}) represents a magnetically charged
stationary spacetime and also solves the three dimensional
Einstein-Maxwell-AdS gravity action (\ref{action BTZ}).
Transformations (\ref{TRANSF_J_HW BTZ}) generate a new metric
because they are not permitted global coordinate transformations
\cite{Stachel}. Transformations (\ref{TRANSF_J_HW BTZ}) can be
done locally, but not globally. Therefore, the metrics
 (\ref{HW_1 BTZ}) and (\ref{HW_Rot BTZ}) can be locally mapped into each other
but not globally, and as such they are distinct.

Chen \cite{CHEN} has applied $T$-duality to \cite{HW} in order to
write a rotating metric. However, the properties of the spacetime
were not studied.

\subsubsection{Mass, angular momentum and electric charge of the
solutions} \label{sec:rotat magnet BTZ 2}
 Both the static and rotating solutions are
asymptotically anti-de Sitter. This fact allows us to calculate
the mass, angular momentum and the electric charge of the static
and rotating solutions. To obtain these quantities  we apply the
formalism  of Regge and Teitelboim \cite{Regge} (see also
\cite{BTZ_Q,OscarLemos}). We first write  (\ref{HW_Rot BTZ}) in
the canonical form involving the lapse function $N^0(\rho)$ and
the shift function $N^{\varphi}(\rho)$
\begin{equation}
     ds^2 = - (N^0)^2 dt^2
            + \frac{d\rho^2}{f^2}
            + H^2(d\varphi+N^{\varphi}dt)^2 \:,
                               \label{MET_CANON BTZ}
\end{equation}
where $f^{-2}=g_{\rho\rho}$, $H^2=g_{\varphi \varphi}$,
 $H^2 N^{\varphi}=g_{t \varphi}$ and
$(N^0)^2-H^2(N^{\varphi})^2=g_{tt}$. Then, the action can be
written in the Hamiltonian form as a function of the energy
constraint ${\cal{H}}$, momentum constraint ${\cal{H}}_{\varphi}$
and Gauss constraint $G$
\begin{eqnarray}
S &=& -\int dt d^2x[N^0 {\cal{H}}+N^{\varphi} {\cal{H}_{\varphi}}
       + A_{t} G]+   {\cal{B}}          \nonumber \\
 &=&  -\Delta t \int d\rho N \nu
        {\biggl [} \frac{2 \pi^2}{H^3}
        +2f(fH_{,\rho})_{,\rho}
         +\frac{H}{\ell^2}
       +\frac{2H}{f}(E^2+B^2){\biggr ]}
                                        \nonumber \\
 & &     + \Delta t \int d\rho N^{\varphi}\nu{\biggl [}
      {\bigl (}2 \pi  {\bigr )}_{,\rho}
       +\frac{4H}{f}E^{\rho}B{\biggr ]}
       + \Delta t \int d \rho A_t \nu {\biggl [}-\frac{4H}{f}
       \partial_{\rho} E^\rho{\biggr ]} +{\cal{B}} \:,
                   \nonumber \\            \label{ACCAO_CANON BTZ}
\end{eqnarray}
where $N=\frac{N^0}{f}$, $\pi \equiv {\pi_{\varphi}}^{\rho}=
-\frac{fH^3 (N^{\varphi})_{,\rho}}{2N^0}$ (with $\pi^{\rho
\varphi}$ being the momentum conjugate to $g_{\rho \varphi}$),
$E^{\rho}$ and $B$ are the electric and magnetic fields and
${\cal{B}}$ is a boundary term. The factor $\nu$ comes from the
fact that, due to the angle deficit, the integration over
$\varphi$ is between $0$ and $2 \pi\nu$. Upon varying the action
with respect to $f(\rho)$, $H(\rho)$, $\pi(\rho)$
 and $E^{\rho}(\rho)$ one picks up additional surface terms.
Indeed,
\begin{eqnarray}
\delta S &=& - \Delta t N \nu{\biggl [}H_{,\rho}
         \delta f^2 -(f^2)_{,\rho}\delta H
                +2f^2 \delta (H_{,\rho}) {\biggr ]}
                                                        \nonumber \\
         & &    +\Delta t N^{\varphi} [2 \nu \delta \pi]
              + \Delta t A_t {\biggl [}
             - \nu\frac{4H}{f} \delta E^{\rho}{\biggr ]}
         + \delta {\cal{B}}         \nonumber \\
         & & +(\mbox{terms vanishing when the
                    equations of motion hold}).
                               \label{DELTA_ACCAO BTZ}
\end{eqnarray}
In order that the Hamilton's equations are satisfied, the boundary
term ${\cal{B}}$ has to be adjusted so that it cancels the above
additional surface terms. More specifically one has
\begin{equation}
  \delta {\cal{B}} = -\Delta t N \delta M  +\Delta t N^{\varphi}\delta J
                  + \Delta t A_t \delta Q_{\rm e} \:,
                              \label{DELTA_B BTZ}
\end{equation}
where one identifies $M$ as the mass, $J$ as the angular momentum
 and $Q_{\rm e}$ as the electric
charge since they are the terms conjugate to the asymptotic values
of $N$, $N^{\varphi}$ and $A_t$, respectively.

To determine the mass, the angular momentum  and the electric
charge of the solutions one must take the spacetime that we have
obtained and subtract the background reference spacetime
contribution, i.e., we choose the energy zero point in such a way
that the mass, angular momentum and charge vanish when the matter
is not present.

Now, note that (\ref{HW_Rot BTZ}) has an asymptotic metric given
by
\begin{equation}
-\frac{\gamma^2-\omega^2}{\ell^2}\rho^2 dt^2+
\frac{\ell^2}{\rho^2}d \rho^2+ (\gamma^2-\omega^2)
 \rho^2 d \varphi^2 \:,
                                          \label{ANTI_SITTER BTZ}
\end{equation}
where $\gamma^2-\omega^2 =1$ so, it is asymptotically an anti-de
Sitter spacetime. The anti-de Sitter spacetime is also the
background reference spacetime, since the metric (\ref{HW_Rot
BTZ}) reduces to (\ref{ANTI_SITTER BTZ}) if the matter is not
present ($m=0$ and $\chi_{\rm m}=0$).

Taking the subtraction of the background reference spacetime into
account we have that the mass, angular momentum and electric
charge are given by
\begin{eqnarray}
M &=& \nu {\bigl [}-H_{,\rho}(f^2-f^2_{\rm ref})
      +(f^2)_{,\rho}(H-H_{\rm ref})
  -2f^2 (H_{,\rho}-H_{,\rho}^{\rm ref}) {\bigr ]}\:,
                                               \nonumber \\
J &=&  -2\nu  (\pi-\pi_{\rm ref}) \:,
                                               \nonumber \\
Q_{\rm e} &=&  \frac{4H}{f} \nu
             (E^{\rho}-E^{\rho}_{\rm ref}) \:.
                                    \label{MQ_GERAL BTZ}
\end{eqnarray}
After taking the asymptotic limit, $\rho \rightarrow +\infty$, we
finally have that the mass and angular momentum are
\begin{eqnarray}
 M &=& \nu {\bigl [}(\gamma^2+\omega^2)(m - r_+^2/\ell^2)
       -2 \chi^2_{\rm m}/\ell^2 {\bigr ]}
      + {\rm Div_M}(\chi_{\rm m},\rho) \:,
                                        \label{M BTZ} \\
J &=& 2 \nu \gamma \omega(m\ell^2-r_+^2-\chi^2_{\rm m})/\ell
          + {\rm Div_J}(\chi_{\rm m},\rho)\:,
                                    \label{J BTZ}
\end{eqnarray}
where ${\rm Div_M}(\chi_{\rm m},\rho)$  and ${\rm Div_J}(\chi_{\rm
m},\rho)$ are logarithmic terms proportional to the magnetic
source $\chi_{\rm m}$ that diverge as $\rho \rightarrow +\infty$
(see also \cite{CHAN}). The presence of these kind of divergences
in the mass and angular momentum is a usual feature present in
charged solutions. They can be found for example in the
electrically charged point source solution \cite{Deser_Maz}, in
the electrically charged BTZ black hole \cite{BTZ_Q} and in the
electrically charged black holes of three dimensional Brans-Dicke
gravity \cite{OscarLemos}. Following
\cite{BTZ_Q,OscarLemos,Deser_Maz} the divergences on the mass can
be treated as follows. One considers a boundary of large radius
$\rho_0$ involving the system. Then, one sums and subtracts ${\rm
Div_M}(\chi_{\rm m},\rho_0)$ to (\ref{M BTZ}) so that the mass
(\ref{M BTZ}) is now written as
\begin{equation}
M = M(\rho_0)+ [{\rm Div_M}(\chi_{\rm m},\rho)-
     {\rm Div_M}(\chi_{\rm m},\rho_0)] \:,
           \label{M0_0 BTZ}
\end{equation}
where $M(\rho_0)=M_0+{\rm Div_M}(\chi_{\rm m},\rho_0)$, i.e.,
\begin{equation}
M_0=M(\rho_0)-{\rm Div_M}(\chi_{\rm m},\rho_0)\:.
                       \label{M0_0_v2 BTZ}
\end{equation}
The term between brackets in (\ref{M0_0 BTZ}) vanishes when $\rho
\rightarrow \rho_0$. Then $M(\rho_0)$ is the energy within the
radius $\rho_0$. The difference between $M(\rho_0)$ and $M_0$ is
$-{\rm Div_M}(\chi_{\rm m},\rho_0)$ which is interpreted as the
electromagnetic energy outside $\rho_0$ apart from an infinite
constant which is absorbed in $M(\rho_0)$. The sum
 (\ref{M0_0_v2 BTZ}) is then independent of $\rho_0$, finite and equal to the
total mass. In practice the treatment of the mass divergence
amounts to forgetting about $\rho_0$ and take as zero the
asymptotic limit: $\lim {\rm Div_M}(\chi_{\rm m},\rho)=0$.

To handle the angular momentum divergence, one first notices that
the asymptotic limit of the angular momentum per unit  mass
$(J/M)$ is either zero or one, so the angular momentum diverges at
a rate slower or equal to the rate of the mass divergence. The
divergence on the angular momentum can then be treated in a
similar way as the mass divergence. So, one can again consider a
boundary of large radius $\rho_0$ involving the system. Following
the procedure applied for the mass divergence one concludes that
the divergent term $-{\rm Div_J}(\chi_{\rm m},\rho_0)$ can be
interpreted as the electromagnetic angular momentum outside
$\rho_0$ up to an infinite constant that is absorbed in
$J(\rho_0)$.

Hence, in practice the treatment of both the mass and angular
divergences amounts to forgetting about $\rho_0$ and take as zero
the asymptotic limits: $\lim {\rm Div_M}(\chi_{\rm m},\rho)=0$ and
$\lim {\rm Div_J}(\chi_{\rm m},\rho)=0$ in (\ref{M BTZ}) and
(\ref{J BTZ}).

Now, we calculate the electric charge of the solutions. To
determine the electric field we must consider the projections of
the Maxwell field on spatial hypersurfaces. The normal to such
hypersurfaces is $n^{\nu}=(1/N^0,0,-N^{\varphi}/N^0)$ and the
electric field is given by $E^{\mu}=g^{\mu \sigma}F_{\sigma
\nu}n^{\nu}$. Then, from (\ref{MQ_GERAL BTZ}), the electric charge
is
\begin{equation}
 Q_{\rm e}=-\frac{4Hf}{N^0} \nu (\partial_{\rho}A_t-N^{\varphi}
    \partial_{\rho} A_{\varphi})=
        2 \nu \frac{\omega}{\ell}\chi_{\rm m} \:.
\label{CARGA BTZ}
\end{equation}
Note that the electric charge is proportional to $\omega \chi_{\rm
m}$. Since in three dimensions the magnetic field is a scalar
(rather than a vector) one cannot use Gauss's law to define a
conserved magnetic charge. In the next section we will propose a
physical interpretation for the origin of the magnetic field
source and discuss the result obtained in (\ref{CARGA BTZ}).

The mass, angular momentum and electric charge of the static
solutions can be obtained by putting $\gamma=1$ and $\omega=0$ on
the above expressions [see (\ref{TRANSF_J_HW BTZ})].

\subsubsection{Relations between the conserved charges} \label{sec:rotat magnet
BTZ 3}

Now, we want to cast the metric (\ref{HW_Rot BTZ}) in terms of
$M$, $J$, $Q_{\rm e}$ and $\chi_{\rm m}$. We can use
 (\ref{M BTZ}) and (\ref{J BTZ}) to solve a quadratic equation for $\gamma^2$
and $\omega^2$. It gives two distinct sets of solutions
\begin{equation}
\gamma^2=\frac{M\ell^2 + 2\chi_{\rm m}^2}{2(m\ell^2-r_+^2)}
\frac{(2-\Omega)}{\nu} \:,\:\:\:\:\:\:\: \omega^2= \frac{M\ell^2 +
2\chi_{\rm m}^2}{2(m\ell^2-r_+^2)} \frac{\Omega}{\nu} \:,
\label{DUAS_HW BTZ}
\end{equation}
\begin{equation}
\gamma^2=\frac{M\ell^2 + 2\chi_{\rm m}^2}{2(m\ell^2-r_+^2)}
\frac{\Omega}{\nu} \:,\:\:\:\:\:\:\: \omega^2= \frac{M\ell^2 +
2\chi_{\rm m}^2}{2(m\ell^2-r_+^2)} \frac{(2-\Omega)}{\nu} \:,
\label{DUAS_ERR_HW BTZ}
\end{equation}
where we have defined a rotating parameter $\Omega$, which ranges
between $0 \leq \Omega < 1$, as
\begin{equation}
\Omega \equiv 1- \sqrt{1-\frac{(m\ell^2-r_+^2)^2}
           {(M\ell^2 + 2\chi_{\rm m}^2)^2}
           \frac{\ell^2 J^2}{(m\ell^2-r_+^2-\chi^2_{\rm m})^2}} \:.
\label{OMEGA_HW BTZ}
\end{equation}
When we take $J=0$ (which implies $\Omega=0$), (\ref{DUAS_HW BTZ})
gives $\gamma \neq 0$ and $\omega= 0$ while
 (\ref{DUAS_ERR_HW BTZ}) gives the nonphysical solution $\gamma=0$ and $\omega \neq
0$ which does not reduce to the static original metric. Therefore
we will study the solutions found from (\ref{DUAS_HW BTZ}). The
condition that $\Omega$ remains real imposes a restriction on the
allowed values of the angular momentum
\begin{equation}
\ell^2 J^2 \leq
       \frac{(m\ell^2-r_+^2-\chi^2_{\rm m})^2}{(m\ell^2-r_+^2)^2}
       (M\ell^2 + 2\chi_{\rm m}^2)^2\:.
       \label{Rest_OMEGA_HW BTZ}
\end{equation}

The condition $\gamma^2-\omega^2=1$ allows us to write
$r_+^2-m\ell^2$ as a function of $M$, $\Omega$ and $\chi_{\rm m}$,
\begin{eqnarray}
r_+^2-m\ell^2= (M\ell^2 + 2\chi_{\rm m}^2) (\Omega-1)/\nu \:.
\label{b_HW BTZ}
\end{eqnarray}
This relation allows us  to achieve interesting conclusions about
the values that the parameters  $M$, $\chi_{\rm m}$ and $J$ can
have. Indeed, if we replace (\ref{b_HW BTZ}) into
 (\ref{DUAS_HW BTZ}) we get
\begin{equation}
\gamma^2=\frac{(2-\Omega)}{2(1-\Omega)} \:,\:\:\:\:\:\:\:
\omega^2= \frac{\Omega}{2(1-\Omega)} \:. \label{DUAS_HW_v2 BTZ}
\end{equation}
Since $\Omega$ ranges between $0 \leq \Omega < 1$, we have
$\gamma^2>0$ and $\omega^2>0$. Besides, one has that
$r_+^2>m\ell^2$ and $\nu>0$  so from (\ref{M0_0 BTZ}) we conclude
that both the static and rotating solutions have negative mass.
Therefore, from now one, whenever we refer to the mass of the
solution we will set
\begin{equation}
M=-|M|\:, \label{mod_m BTZ}
\end{equation}
unless otherwise stated.

Looking again to (\ref{b_HW BTZ}) we can also conclude that one
must have
\begin{equation}
 \chi_{\rm m}^2<\frac{|M|\ell^2}{2}\:,
\label{v_max_q BTZ}
\end{equation}
i.e., there is an upper bound for the intensity of the magnetic
field strength.

 From (\ref{J BTZ}) we also see that the angular momentum is always
negative indicating that the angular momentum and the angular
velocity, $\omega$, have opposite directions. This is the expected
result since $J$ is the inertial momentum times the angular
velocity and the inertial momentum is proportional to the mass
which is negative. Introducing  (\ref{b_HW BTZ}) into
(\ref{Rest_OMEGA_HW BTZ}) we find an upper bound for the angular
momentum
\begin{equation}
|J|\leq |M|\ell^2-2 \chi_{\rm m}^2+\nu\chi_{\rm m}^2/(1-\Omega)\:.
\label{J_max BTZ}
\end{equation}
Note that from (\ref{OMEGA_HW BTZ}) we can get the precise value
of $J$ as a function of $M$, $\Omega$ and $\chi_{\rm m}$.

Finally, we remark that the auxiliary equations (\ref{r+ BTZ}),
(\ref{freq BTZ}) and (\ref{DUAS_HW_v2 BTZ}) allow us to define the
auxiliary parameters $r_+$, $\nu$ and $m$ as a function of the
hairs $M$, $\Omega$ and $\chi_{\rm m}$.

\subsubsection{The rotating magnetic solution} \label{sec:rotat magnet BTZ 4}

We are now in position to write the stationary spacetime
(\ref{HW_Rot BTZ}) generated by a source of magnetic field in
three dimensional Einstein-Maxwell-anti de Sitter gravity as a
function of its hairs,
\begin{eqnarray}
ds^2 &=&  -\frac{1}{\ell^2} {\biggl [}\rho^2
        +\frac{1}{2\nu}(|M|\ell^2- 2\chi_{\rm m}^2)(2-\Omega)
        -\frac{Q_{\rm e}^2}{4 \nu}
           \ln{{\biggl (}1+
         \frac{\nu \rho^2}{(|M|\ell^2- 2\chi_{\rm m}^2)(1-\Omega)}
            {\biggr )}}{\biggr ]}dt^2
                                          \nonumber \\
     & &  +\frac{J}{\nu}\frac{
          (|M|\ell^2 - 2\chi_{\rm m}^2)(\Omega-1)
           + \nu \chi^2_{\rm m} \ln{{\biggl (}1+
         \frac{\nu\rho^2}{(|M|\ell^2 - 2\chi_{\rm m}^2)(1-\Omega)}
     {\biggr )}}}{(|M|\ell^2 - 2\chi_{\rm m}^2)(1-\Omega)
           +\nu \chi_{\rm m}^2}
               dt d \varphi
                           \nonumber \\
     & &   + \frac{\ell^2 \rho^2
        {\biggl [}\rho^2+\chi^2_{\rm m}\ln{{\biggl (}1+
         \frac{\nu \rho^2}{(|M|\ell^2 - 2\chi_{\rm m}^2)(1-\Omega)}
    {\biggr )}}{\biggr ]}^{-1}}{\rho^2+(|M|\ell^2 - 2\chi_{\rm m}^2)
           (1-\Omega)/\nu}
            d \rho^2  \nonumber \\
      & &   + {\biggl [}\rho^2
        -(|M|\ell^2 - 2\chi_{\rm m}^2)\frac{\Omega}{2 \nu}+
      \frac{2-\Omega}{1-\Omega}
       \frac{\chi^2_{\rm m}}{2} \ln{{\biggl (}1+
         \frac{\nu \rho^2}{(|M|\ell^2 - 2\chi_{\rm m}^2)(1-\Omega)}
            {\biggr )}}{\biggr ]}
            d\varphi^2 \:,      \label{HW_Rot_Fim BTZ}
\end{eqnarray}
as well as the vector potential 1-form (\ref{HW_Max_Rot BTZ})
\begin{eqnarray}
A=\frac{2}{\sqrt{1-\Omega}}{\biggl
[}-\frac{\sqrt{\Omega}}{\ell}A(\rho)dt +\sqrt{2-\Omega} A(\rho) d
\varphi{\biggr ]} \:, \label{HW_Max_Rot_Fim BTZ}
\end{eqnarray}
with $A(\rho)=(\chi_{\rm m}/2) \ln{[\rho^2/\ell^2+(|M|-2\chi_{\rm
m}^2/\ell^2) (1-\Omega)/\nu]}$.

If we set $\Omega=0$ (and thus $J=0$ and $Q_{\rm e}=0$) we recover
the static solution (\ref{HW_1 BTZ}) [see
 (\ref{TRANSF_J_HW BTZ})]. Finally if we set $\chi_{\rm m}=0$ (and so $\nu=1$) one
gets
\begin{eqnarray}
 ds^2 &=&  -\frac{1}{\ell^2} {\biggl [}\rho^2
        -M\ell^2\frac{2-\Omega}{2} {\biggr ]}dt^2
         -J dt d \varphi
         + \frac{\ell^2}
       {\rho^2-M\ell^2 (1-\Omega)} d \rho^2
                                      + {\biggl [}\rho^2
        +M\ell^2 \frac{\Omega}{2}{\biggr ]}
            d\varphi^2 \:,     \nonumber \\
     & &
                   \label{BTZ_0 BTZ}
\end{eqnarray}
where we have dropped the absolute value of $M$ since now the mass
can be positive. This is the  rotating uncharged  BTZ solution
written, however, in an unusual gauge. To write it in the usual
gauge we apply to (\ref{BTZ_0 BTZ}) the radial coordinate
transformation
\begin{eqnarray}
\rho^2=R^2- M\ell^2 \frac{\Omega}{2} \:\:\:\Rightarrow \:\:\:
d\rho^2=\frac{R^2}{R^2-M\ell^2 \Omega /2} dR^2
                   \label{transf_BTZ BTZ}
\end{eqnarray}
and use the relation $J^2=\Omega(2-\Omega)M\ell^2$ [see
(\ref{OMEGA_HW BTZ})] to obtain
\begin{eqnarray}
 ds^2 =  - {\biggl (}\frac{R^2}{\ell^2}
        -M{\biggr )}dt^2
         -J dt d \varphi
         + {\biggl (}\frac{R^2}{\ell^2}-M+\frac{J^2}{4R^2}
          {\biggr )}^{-1} d R^2
        + R^2 d\varphi^2 \:.
                   \label{BTZ BTZ}
\end{eqnarray}
So, as expected, (\ref{HW_Rot_Fim BTZ}) reduces to the rotating
uncharged  BTZ solution \cite{btz_PRL,btz_PRD} when the magnetic
field source vanishes.

\subsubsection{Geodesic structure}


The geodesic structure of the rotating spacetime is similar to the
static spacetime (see section II.2), although there are now direct
(corotating with $L>0$) and retrograde (counter-rotating with
$L<0$) orbits. The most important result that spacetime is
geodesically complete still holds for the stationary spacetime.

\vskip 0.5cm

\subsection{Physical interpretation of the magnetic source}
\label{sec:Phys Interp magnetic BTZ}

When we look back to the electric charge given in
 (\ref{CARGA BTZ}), we see that it is zero when $\omega=0$, i.e., when the
angular momentum $J$ of the spacetime vanishes. This is expected
since in the static solution we have imposed that the electric
field is zero ($F_{12}$ is the only non-null component of the
Maxwell tensor).

Still missing is a physical interpretation for the origin of the
magnetic field source. The magnetic field source is not a
Nielson-Oleson vortex solution since  we are working with the
Maxwell theory and not with an Abelian-Higgs model. We might then
think that the magnetic field is produced by a Dirac point-like
monopole. However, this is not also the case since a Dirac
monopole with strength $g_{\rm m}$ appears when one breaks the
Bianchi identity \cite{MeloNeto}, yielding $\partial_{\mu}
(\sqrt{-g} \tilde{F}^{\mu})= g_{\rm m} \delta^2 (\vec{x})$  (where
$\tilde{F}^{\mu}=\epsilon^{\mu \nu \gamma}F_{\nu \gamma}/2$ is the
dual of the Maxwell field strength), whereas in this work we have
that  $\partial_{\mu} (\sqrt{-g} \tilde{F}^{\mu})=0$. Indeed,  we
are clearly dealing with the Maxwell theory which satisfies
Maxwell equations and the  Bianchi identity
\begin{eqnarray}
& &  \frac{1}{\sqrt{-g}}\partial_{\nu}(\sqrt{-g}F^{\mu \nu})
        =\frac{\pi}{2}\frac{1}{\sqrt{-g}} j^{\mu} \:,
                                    \label{Max_j BTZ} \\
& &  \partial_{\mu}
     (\sqrt{-g} \tilde{F}^{\mu})=0 \:,
                                    \label{Max_bianchi BTZ}
\end{eqnarray}
respectively. In (\ref{Max_j BTZ}) we have made use of the fact
that the general relativistic current density is $1/\sqrt{-g}$
times the special relativistic current density $j^{\mu}=\sum q
\delta^2(\vec{x}-\vec{x}_0)\dot{x}^{\mu}$.

We then propose that the magnetic field source can be interpreted
as composed by a system of two symmetric and superposed electric
charges (each with strength $q$). One of the electric charges is
at rest with positive charge (say), and the other is spinning with
an angular velocity $\dot{\varphi}_0$ and negative electric
charge.  Clearly, this system produces no electric field since the
total electric charge is zero and the magnetic field is produced
by the angular electric current. To confirm our interpretation, we
go back to (\ref{Max_j BTZ}). In our solution, the only
non-vanishing component of the Maxwell field is $F^{\varphi \rho}$
which implies that only $j^{\varphi}$ is not zero. According to
our interpretation one has $j^{\varphi}=q
\delta^2(\vec{x}-\vec{x}_0)\dot{\varphi}$, which one inserts in
 (\ref{Max_j BTZ}). Finally, integrating over $\rho$ and
$\varphi$ we have
\begin{equation}
 \chi_{\rm m} \propto q \dot{\varphi}_0 \:.
\label{Q_mag BTZ}
\end{equation}
So, the magnetic source strength, $\chi_{\rm m}$, can be
interpreted as an electric charge $q$ times its spinning velocity.

Looking again to the electric charge given in (\ref{CARGA BTZ}),
one sees that after applying the rotation boost in the
$t$-$\varphi$ plane to endow the initial static spacetime with
angular momentum, there appears  a net electric charge. This
result was already expected since now, besides the scalar magnetic
field ($F_{\rho \varphi} \neq 0$), there is also an electric field
($F_{t \rho} \neq 0$) [see (\ref{HW_Max_Rot_Fim BTZ})]. A physical
interpretation for the appearance of the net electric charge is
now needed. To do so, we return to the static spacetime. In this
static spacetime there is a static positive charge and a spinning
negative charge of equal strength at the center. The net charge is
then zero. Therefore, an observer at rest ($S$) sees a density of
positive charges at rest which is equal to the density of negative
charges that are spinning. Now, we perform a local rotational
boost $t'= \gamma t-\ell \omega \varphi$ and $\varphi' = \gamma
\varphi-\frac{\omega}{\ell} t\:$ to an observer ($S'$) in the
static spacetime, so that $S'$ is moving relatively to $S$. This
means that $S'$ sees a different charge density since a density is
a charge over an area and this area suffers a Lorentz contraction
in the direction of the boost.  Hence, the two sets of charge
distributions that had symmetric charge densities in the frame $S$
will not have charge densities with equal magnitude in the frame
$S'$. Consequently, the charge densities will not cancel each
other in the frame $S'$ and a net electric charge appears.  This
was done locally. When we turn into the global rotational Lorentz
boost of (\ref{TRANSF_J_HW BTZ}) this interpretation still holds.
The local analysis above is similar to the one that occurs when
one has a copper wire with an electric current and we apply a
translation Lorentz boost to the wire: first, there is only a
magnetic field but, after the Lorentz boost, one also has an
electric field.  The difference is that in the present situation
the Lorentz boost is a rotational one and not a translational one.


\section{Summary and discussion}


The BTZ black hole \cite{btz_PRL}, with non-vanishing mass and
angular momentum, is asymptotically AdS. This solution has
constant curvature and thus there can be no curvature singularity
at the origin. As discussed in detail in \cite{btz_PRD}, the BTZ
black hole can be expressed as a topological quotient of $AdS_3$
by a group of isometries. This is, the BTZ black hole can be
obtained through identifications along an isometry of the $AdS_3$
spacetime, and in order to avoid closed timelike curves, the
origin must be a topological singularity (a boundary of the
spacetime). We have reviewed in detail this construction.
Qualitatively, the reason why the black hole exists only in an AdS
background can be understood as follows. The radius of an event
horizon ($r_+$) in a $D$-dimensional spacetime is expected to be
proportional to $G_D M$, where $G_D$ is the $D$-dimensional
Newton's constant. Now, in mass units $G_D$ has dimension
$M^{2-D}$. Thus, $G_D M$ (and $r_+$) is dimensionless in $D=3$,
and there is no length scale in $D=3$. The cosmological constant
provides this length scale for the horizon, but only when
$\Lambda<0$ (AdS case). One may argue that this is due to the fact
that the AdS background is attractive, i.e., an analysis of the
geodesic equations indicates that particles in this background are
subjected to a potential well that attracts them (and so it is
possible to concentrate matter into a small region), while the dS
background is repulsive (in practice, if we try to construct a dS
black hole through identifications along an isometry of the $dS_3$
spacetime, we verify that the possible horizon is inside the
region that contains closed timelike curves).

The extension to include a radial electric field in the BTZ black
hole has been done in \cite{CL1,BTZ_Q} (this solution reduces to
those of \cite{Deser_Maz,GSA} when $\Lambda=0$). Due to the slow
fall off of the electric field in 3-dimensions, the system has
infinite total energy. The presence of the electric
energy-momentum tensor implies that the spacetime has no longer
constant curvature, and the electric BTZ black hole cannot be
expressed as a topological quotient of $AdS_3$ by a group of
isometries. The rotating charged black hole is generated from the
static charged solution (already found in \cite{btz_PRL}) through
the application of a rotation Lorentz boost. A BTZ solution with
an azimuthal electric field was found in \cite{Cat}. This solution
is horizonless, and reduces to empty $AdS_3$ spacetime when the
charge vanishes.

Pure magnetic solutions with $\Lambda<0$, that reduce to the
neutral BTZ black hole solution when the magnetic source vanishes,
also exist and were reviewed in section \ref{sec:magnetic BTZ}.
Notice that, in oppose to what occurs in 4-dimensions where the
the Maxwell tensor and its dual are 2-forms, in 3-dimensions the
Maxwell tensor is still a 2-form, but its dual is a 1-form (in
practice, the Maxwell tensor has only three independent
components: two for the electric vector field, and one for the
scalar magnetic field). As a consequence, the magnetic solutions
are radically different from the electric solutions in
3-dimensions. The static magnetic solution has been found in
\cite{CL1,HW,Cat_Sal}. This spacetime generated by a static
magnetic point source is horizonless and has a conical singularity
at the origin. The extension to include rotation and a new
interpretation for the source of magnetic field has been made in
\cite{OscarLemos_BTZ}. In \cite{HW}, the static magnetic source
has been interpreted as a kind of magnetic monopole reminiscent of
a Nielson-Oleson vortex solution. In \cite{OscarLemos_BTZ}, we
prefer to interpret the static magnetic field source as being
composed by a system of two symmetric and superposed electric
charges. One of the electric charges is at rest and the other is
spinning. This system produces no electric field since the total
electric charge is zero and the scalar magnetic field is produced
by the angular electric current. When we apply a rotational
Lorentz boost to add angular momentum to the spacetime, there
appears an electric charge and an electric field.

%% file: Chapter2.tex
\thispagestyle{empty} \setcounter{minitocdepth}{1}
\chapter[Three dimensional dilaton black holes of the Brans-Dicke
type] {\Large{Three dimensional dilaton black holes of the
Brans-Dicke type}} \label{chap:3D Dilaton BH}
 \lhead[]{\fancyplain{}{\bfseries Chapter \thechapter. \leftmark}}
 \rhead[\fancyplain{}{\bfseries \rightmark}]{}
  \minitoc \thispagestyle{empty}
\renewcommand{\thepage}{\arabic{page}}

\addtocontents{lof}{\textbf{Chapter Figures \thechapter}\\}


In order to turn the 3-dimensional gravity dynamics more similar
with the realistic 4-dimensional one, we can manage a way by which
we introduce local degrees of freedom.  One way to do this, is  by
coupling a scalar field to Einstein gravity, yielding a so called
dilaton gravity. The most general form of this kind of gravity was
proposed by Wagoner \cite{scalarGravity} in 1970. The scalar field
provides a local dynamical degree of freedom to the theory, and
models of this kind (i.e., under certain choices of the
parameters) appear naturally in string theory. Some choices of the
parameters of the theory yield dilaton gravities that have black
hole solutions. One of these theories, which we will study in this
chapter, is an Einstein-dilaton gravity of the Brans-Dicke type in
a $\Lambda <0$ background, first discussed by Lemos \cite{Lemos}.
This theory is specified by a Brans-Dicke parameter, $\omega$, and
contains seven different cases. Each $\omega$ can be viewed as
yielding a different dilaton gravity theory, with some of these
being related with other known special theories. For instance, for
$\omega=-1$ one gets the simplest low-energy string action
\cite{CFMP}, and for $\omega=0$ one gets a theory related (through
dimensional reduction) to 4-dimensional general relativity with
one Killing vector \cite{Lemos,Zanchin_Lemos,OscarLemos_string}.
This is, the $\omega=0$ black holes are the direct counterparts of
the 4-dimensional AdS black holes with toroidal or cylindrical
topology first discussed by Lemos, in the same way that a point
source in 3-dimensions is the direct cousin of a cosmic string in
4-dimensions. For $\omega=\pm \infty$ the theory reduces to the
pure 3-dimensional general relativity. This is, the $\omega=\pm
\infty$ black holes are the BTZ ones. Moreover, the case
$\omega>-1$ yields gravities whose black holes have a structure
and properties similar to the BTZ black hole, but with a feature
that might be useful: the $\omega>-1$ black holes have dynamical
degrees of freedom, which implies for example that the origin has
a curvature singularity and that gravitational waves can propagate
in the spacetime. The neutral black holes of this Brans-Dicke
theory were found and analyzed by S\'a, Kleber and Lemos
\cite{Lemos,Sa_Lemos_Static,Sa_Lemos_Rotat}. The pure electric
charged black holes have been analysed by Dias and Lemos
\cite{OscarLemos}, and the pure magnetic solutions have been
discussed by Dias and Lemos \cite{OscarLemos-MagBD3D}. We will
analyze in detail these black holes in this chapter.

The plan of this chapter is as follows. In section
\ref{sec:Electric BH 3D} we briefly present the neutral black
holes of the Brans-Dicke theory
\cite{Lemos,Sa_Lemos_Static,Sa_Lemos_Rotat}. Then, in section
\ref{sec:Electric BH 3D}, we discuss in detail the electric
charged black holes (static and rotating) of the theory
\cite{OscarLemos} and, in section \ref{sec:Magnetic BH 3D}, we do
the same with the magnetic solutions \cite{OscarLemos-MagBD3D}.
Finally, in section \ref{sec:conc 3D}, concluding remarks are
given.

\section{Neutral Brans-Dicke dilaton black holes}
 \label{sec:BH 3D}

We work with an action of the Brans-Dicke type in three-dimensions
written in the string frame as
\begin{equation}
S=\frac{1}{2\pi} \int d^3x \sqrt{-g} e^{-2\phi}
  \left [ R - 4 \omega ( \partial \phi)^2
+ \Lambda \right ],
\label{ACCAO-neutral3D}
\end{equation}
where $g$ is the determinant of the 3-dimensional (3D) metric, $R$
is the curvature scalar, $\phi$ is a scalar field called dilaton,
$\Lambda<0$ is the cosmological constant,
 $\omega$ is the three-dimensional Brans-Dicke parameter.

From the field equations of (\ref{ACCAO-neutral3D}), one finds the
following solutions (with many of them being black holes),
\begin{eqnarray}
 ds^2 &=& -{\biggl [} (\alpha r)^2 - \frac{(\omega+1)}
          {2(\omega+2)} \frac{M(2-\Omega)}
          {(\alpha r)^{\frac{1}{\omega+1}}}{\biggr ]} d t^2
           -\frac{\omega+1}{2\omega+3}\,J\,
            \frac{1}{(\alpha r)^{\frac{1}{\omega+1}} }
               \,2dt d\varphi
                                      \nonumber \\
      & &
           + {\biggl [}(\alpha r)^2 -
           \frac{M[2(\omega+1)-
       (2\omega +3)\Omega]}{2(\omega +2)
            (\alpha r)^{\frac{1}{\omega+1}}}
        {\biggr ]}^{-1} dr^2
         + \frac{1}{\alpha^2}{\biggl [} (\alpha r)^2 +
           \frac{M \Omega}
           {2(\alpha r)^{\frac{1}{\omega+1}}}
            {\biggr ]} d\varphi^2,           \nonumber \\
      & &
      \hskip 9cm {\rm for} \:\:\:\: \omega \neq -2,-\frac32,-1\:,
                                        \label{MET_MJQ-neutral3D} \\
 ds^2 &=& -\left ( r^2 -\frac{J^2}{M}\: r \right ) dt^2
          -J\, r\, 2dt d\varphi + {\biggl [} r^2
          -M {\biggl (}\frac{J^2}{M^2}-1 {\biggl )}r
          {\biggr ]}^{-1} dr^2+ \left ( r^2+Mr \right ) d\varphi^2,\nonumber \\
      & &
      \hskip 9cm {\rm for} \:\:\:\: \omega = -2,
                                        \label{MET_MJQ_-2-neutral3D} \\
 ds^2 &=&  r^2 \Lambda \ln (br)\, dt^2
           -\frac{dr^2}
            {r^2 \Lambda \ln (br)}
           + r^2 d\varphi^2,           \hskip 3cm {\rm for} \:\:\:\:\omega=-\frac32,
                                             \label{MET-3/2-neutral3D}
\end{eqnarray}
where, for $\omega \neq -2,-\frac32,-1 \:$, $\alpha$ is defined as
$\alpha = \sqrt{ \left |
\frac{(\omega+1)^2\Lambda}{(\omega+2)(2\omega+3)}\right |}$. $M$
and $J$ are, respectively, the ADM mass and angular momentum of
the solutions, and we have defined the rotating parameter $\Omega
\equiv 1- \sqrt{1-\frac{4(\omega+1)(\omega+2)}
{(2\omega+3)^2}\frac{J^2 \alpha^2}{M^2}}$ in such a way that $J=0$
implies $\Omega=0$. The condition that $\Omega$ remains real
imposes, for $-2>\omega>-1$, a maximum bound for the angular
momentum: $|\alpha J|\leq \frac{|2\omega+3|M}{2\sqrt{(\omega+1)
(\omega+2)}}$. For $-\frac32$, one has $M=-\Lambda \ln b$ and
$J=0$, necessarily. For $\omega<-\frac32$ and $\omega>-1$, the
solutions have a curvature singularity located at $r=0$, and for
$-\frac32<\omega<-1$ the curvature singularity is at $r=+\infty$.
For $\omega = -\frac{3}{2}$, both $r=0$ and $r=+\infty$ are
singular. For $\omega=\pm \infty$ there is no curvature
singularity. Many of this theories (i.e., for different $\omega$)
have black hole solutions. A detailed analysis of the character of
the solutions can be found in
\cite{Sa_Lemos_Static,Sa_Lemos_Rotat}. Here we do not go further
in their study, since in next section we will study in detail
these solutions when a charge $Q$ is added to the system.

\section{Electric Brans-Dicke dilaton black holes}
 \label{sec:Electric BH 3D}

In this section we find and study in detail the static and
rotating electrically charged solutions of a
Einstein-Maxwell-Dilaton action of the Brans-Dicke type. So, these
are the electric counterparts of the solutions discussed in the
last section \cite{Sa_Lemos_Static,Sa_Lemos_Rotat}.

The electrically charged theory that we are going to study is
specified by the extra electromagnetic field $F^{\mu \nu}$. It
contains eight different cases. For $\omega=0$ one gets a theory
related (through dimensional reduction) to electrically charged
four dimensional General Relativity with one Killing vector
\cite{Zanchin_Lemos} and for $\omega=\pm \infty$ one obtains
electrically charged three dimensional General Relativity
\cite{CL1,BTZ_Q}.

Since magnetically charged solutions in (2+1) dimensions have
totally different properties from the electrically charged ones,
we leave their study for section \ref{sec:Magnetic BH 3D}.

The plan of this section is the following. In Section
\ref{sec:Field Eqs 3D} we set up the action and the field
equations. The static general solution of the field equations are
found in section \ref{sec:static solution 3D} and we write the
scalar $R_{\mu\nu}R^{\mu\nu} $ which, in 3-dimensions, signals the
presence of singularities. The angular momentum is added in
section \ref{sec:rotating solution 3D}. In section \ref{sec:MJQ
3D} we use an extension of the formalism of Regge and Teitelboim
to derive the mass, angular momentum, electric charge and dilaton
charge of the black holes. In section \ref{sec:Causal Geod 3D} we
study the properties of the different cases that appear naturally
from the solutions. We work out in detail the causal structure and
the geodesic motion of null and timelike particles for typical
values of $\omega$ that belong to the different ranges. The
Hawking temperature is computed in section \ref{sec:Haw Temp 3D}.

\subsection{Field equations of Brans-Dicke$-$Maxwell theory}
 \label{sec:Field Eqs 3D}

We are going to work with an action of the Maxwell-Brans-Dicke
type in three-dimensions written in the string frame as
\begin{equation}
S=\frac{1}{2\pi} \int d^3x \sqrt{-g} e^{-2\phi}
  {\bigl [} R - 4 \omega ( \partial \phi)^2
           + \Lambda - F^{\mu \nu}F_{\mu \nu}
  {\bigr ]},                                   \label{ACCAO}
\end{equation}
where $g$ is the determinant of the 3D metric, $R$ is the
curvature scalar, $\phi$ is a scalar field called dilaton,
$\Lambda$ is the cosmological constant,
 $\omega$ is the three-dimensional Brans-Dicke parameter and
$F_{\mu\nu} = \partial_\mu A_\nu-\partial_\nu A_\mu$ is the
Maxwell tensor, with $A_\mu$ being the vector potential. Varying
this action with respect to $g^{\mu \nu}$, $F^{\mu \nu}$ and
 $\phi$ one gets the Einstein, Maxwell and ditaton equations,
respectively
\begin{eqnarray}
   & \frac{1}{2} G_{\mu \nu}
       -2(\omega+1) \nabla_{\mu}\phi \nabla_{\nu}\phi
       +\nabla_{\mu} \nabla_{\nu}\phi
       -g_{\mu \nu} \nabla_{\gamma} \nabla^{\gamma}\phi
            +(\omega+2) g_{\mu \nu} \nabla_{\gamma}\phi
               \nabla^{\gamma}\phi-\frac{1}{4}g_{\mu \nu}\Lambda=
              \frac{\pi}{2} T_{\mu \nu},&
                                        \label{EQUACAO_MET} \\
   & \nabla_{\nu}(e^{-2 \phi}F^{\mu \nu})=0 \:,&
                                     \label{EQUACAO_MAX} \\
   & R -4\omega \nabla_{\gamma} \nabla^{\gamma}\phi
          +4\omega \nabla_{\gamma}\phi \nabla^{\gamma}\phi
          +\Lambda =-F^{\gamma \sigma}F_{\gamma \sigma},&
                                         \label{EQUACAO_DIL}
\end{eqnarray}
where $G_{\mu \nu}=R_{\mu \nu} -\frac{1}{2} g_{\mu \nu} R$ is the
 Einstein tensor, $\nabla$ represents the covariant derivative and
$T_{\mu \nu}=\frac{2}{\pi}(g^{\gamma \sigma}F_{\mu \gamma} F_{\nu
\sigma}-\frac{1}{4}g_{\mu \nu}F_{\gamma \sigma} F^{\gamma
\sigma})$ is the Maxwell energy-momentum tensor.

We want to consider now a spacetime which is both static and
rotationally symmetric, implying the existence of a timelike
Killing vector $\partial/\partial t$ and a spacelike Killing
vector $\partial/\partial\varphi$. The most general static metric
with a Killing vector $\partial/\partial\varphi$ with closed
orbits in three dimensions can be written as $ds^2 = - e^{2\nu(r)}
dt^2 + e^{2\mu(r)}dr^2 + r^2 d\varphi^2$, with $0\leq\varphi\leq
2\pi$. Each different $\omega$ has a very rich and non-trivial
structure of solutions which could be considered on its own. As in
\cite{Sa_Lemos_Static,Sa_Lemos_Rotat} we work in the Schwarzschild
gauge, $\mu(r)=-\nu(r)$, and compare different black hole
solutions in different theories. For this ansatz the metric is
written as
\begin{equation}
  ds^2 = - e^{2\nu(r)} dt^2 + e^{-2\nu(r)}dr^2 + r^2 d\varphi^2.
                               \label{MET_SYM}
\end{equation}
We also assume that the only non-vanishing components of the
vector potential are $A_t(r)$ and $A_{\varphi}(r)$, i.e. ,
\begin{equation}
A=A_tdt+A_{\varphi}d{\varphi}\:.  \label{Potential}
\end{equation}
This implies that the non-vanishing components of the symmetric
Maxwell tensor are $F_{tr}$ and $F_{r \varphi}$.

Inserting the metric (\ref{MET_SYM}) into equation
(\ref{EQUACAO_MET}) one obtains the following set of equations
\begin{eqnarray}
  & &   \phi_{,rr}
      + \phi_{,r} \nu_{,r}
      + \frac{\phi_{,r}}{r}
      - (\omega+2) (\phi_{,r})^2
      - \frac{\nu_{,r}}{2r}
      + \frac{1}{4} \Lambda e^{-2\nu}=
                \frac{\pi}{2}e^{-4\nu} T_{tt} \:,
                                    \label{MET_00}  \\
  & &  - \phi_{,r} \nu_{,r}
       - \frac{\phi_{,r}}{r}
       - \omega (\phi_{,r})^2
       + \frac{\nu_{,r}}{2r}
       - \frac{1}{4} \Lambda e^{-2\nu}=
                  \frac{\pi}{2} T_{rr} \:,
                                    \label{MET_11}  \\
  & &    \phi_{,rr}
       + 2 \phi_{,r} \nu_{,r}
       - (\omega+2) (\phi_{,r})^2
       - \frac{\nu_{,rr}}{2}
       - (\nu_{,r})^2+ \frac{1}{4} \Lambda e^{-2\nu}=
              -\frac{\pi}{2}\frac{e^{-2\nu}}{r^2}
                           T_{\varphi \varphi} \:,
                                           \label{MET_22}  \\
  & &   0=\frac{\pi}{2} T_{t \varphi}=
                           e^{-2\nu}F_{tr}F_{\varphi r} \:,
                                           \label{MET_02}
\end{eqnarray}
where ${}_{,r}$ denotes a derivative with respect to $r$. In
addition, inserting the metric (\ref{MET_SYM}) into equations
(\ref{EQUACAO_MAX}) and (\ref{EQUACAO_DIL}) yields
\begin{eqnarray}
 & &  \partial_r {\bigl [} e^{-2 \phi}r(F^{t r}+F^{\varphi r})
{\bigr ]}=0\:,
                                           \label{MAX_0} \\
& &  \omega \phi_{,rr}
      + 2 \omega \phi_{,r} \nu_{,r}
      + \omega \frac{\phi_{,r}}{r}
      - \omega (\phi_{,r})^2
      + \frac{\nu_{,r}}{r}
      + \frac{\nu_{,rr}}{2}
      + (\nu_{,r})^2
      -\frac{1}{4} \Lambda e^{-2\nu}=
      \frac{1}{4}e^{-2\nu}F^{\gamma \sigma}F_{\gamma \sigma} \:.
                                    \label{EQ_DIL}
\end{eqnarray}

\subsection{The general static solution}
 \label{sec:static solution 3D}
From the above equations valid for a static and rotationally
symmetric spacetime one sees that equation (\ref{MET_02}) implies
that the electric and magnetic fields cannot be simultaneously
non-zero, i.e., there is no static dyonic solution. In this work
we will consider the electrically charged case alone
($A_{\varphi}=0,\,A_t\neq 0$).

So, assuming vanishing magnetic field, one has from Maxwell
equation (\ref{MAX_0}) that
\begin{equation}
F_{tr}=-\frac{\chi}{4r}e^{2 \phi}, \label{MAX_1}
\end{equation}
where $\chi$ is an integration constant which, as we shall see in
(\ref{CARGA}), is the electric charge. One then has that
\begin{eqnarray}
 & & F^{\gamma \sigma}F_{\gamma \sigma}=
\frac{\chi^2}{8r^2}e^{4\phi}, \:\:\:\:\:
T_{tt}=\frac{\chi^2}{16 \pi r^2}e^{2\nu}e^{4\phi}, \nonumber \\
 & & T_{rr}=-\frac{\chi^2}{16 \pi r^2}e^{-2\nu}e^{4\phi}\:\:,
\:\:\:\:\:T_{\varphi \varphi}=\frac{\chi^2}{16 \pi} \: e^{4\phi}.
\label{MAX_2}
\end{eqnarray}
To proceed we shall first consider the case $\omega\neq -1$.
Adding equations (\ref{MET_00}) and (\ref{MET_11}) one obtains
$\phi_{,rr}=2(\omega+1)(\phi_{,r})^2$, yielding for the dilaton
field the following solution
\begin{equation}
\phi = -\frac{1}{2(\omega+1)}
        \ln [2(\omega+1)r+a_1] + a_2 \:, \quad\quad w \neq -1
                                             \label{DIL}
\end{equation}
where $a_1$ and $a_2$ are constants of integration. One can,
without loss of generality, choose $a_1=0$. Then, equation
(\ref{DIL}) can be written as
\begin{equation}
  e^{-2\phi}= a(\alpha r)^{\frac{1}{\omega+1}},
                       \quad\quad w \neq -1 \:,
                                    \label{DILATAO}
\end{equation}
where $\alpha$ is an appropriate constant that is proportional to
the cosmological constant [see equation (\ref{COSMOL})]. The
dimensionless constant $a$ can be viewed as a normalization to the
action (\ref{ACCAO}). Since it has no influence in our
calculations, apart a possible redefinition of the mass, we set
$a=1$. The vector potential $A=A_{\mu}(r)dx^{\mu}=A_t(r)dt$ with
$A_t(r)=\int F_{tr}dr$ is then
\begin{equation}
A=\frac{1}{4}\chi(\omega+1) (\alpha r)^{\frac{1}{\omega+1}} dt \:,
                           \quad\quad w \neq -1\:.
                                    \label{VEC_POTENT}
\end{equation}
Inserting the solutions (\ref{MAX_1})-(\ref{DILATAO}) in equations
(\ref{MET_00})-(\ref{EQ_DIL}), we obtain for the metric
\begin{eqnarray}
 ds^2 &=& -{\biggl [}
            (\alpha r)^2 -
           \frac{b}{(\alpha r)^{\frac{1}{\omega+1}}}
           +\frac{k\chi^2}
                      {(\alpha r)^{\frac{2}{\omega+1}}}
            {\biggr ]} dt^2
           + \frac{dr^2}{(\alpha r)^2 -
           \frac{b}{(\alpha r)^{\frac{1}{\omega+1}}}
           +\frac{k\chi^2}
                        {(\alpha r)^{\frac{2}{\omega+1}}} } + r^2 d\varphi^2,
                                        \nonumber \\
      & &  \hskip 8cm {\rm for} \:\:\:\: \omega \neq -2,-\frac32,-1,
                                 \label{MET_TODOS}  \\
 ds^2 &=& -{\biggl [} (1+\frac{\chi^2}{4}\ln{r})r^2
                 -br {\biggr ]} dt^2
           +\frac{dr^2}{(1+\frac{\chi^2}{4}
                \ln{r})r^2 -br}
           + r^2 d\varphi^2,                      \nonumber \\
      & &  \hskip 8cm {\rm for} \:\:\:\:\omega =-2,
                                         \label{MET-2}  \\
 ds^2 &=&  -r^2[-\Lambda \ln (br)+\chi^2r^2] dt^2
           + \frac{dr^2}
            {r^2[-\Lambda \ln (br)+\chi^2r^2]}
           + r^2 d\varphi^2,                      \nonumber \\
      & &  \hskip 8cm {\rm for} \:\:\:\:\omega=-\frac32,
                                             \label{MET-3/2}
\end{eqnarray}
where $b$ is a constant of integration related with the mass of
the solutions, as will be shown, and
$k=\frac{(\omega+1)^2}{8(\omega+2)}$. For $\omega \neq
-2,-\frac32,-1 \:$ $\alpha$ is defined as (we call your attention
to a typo in \cite{OscarLemos} in this definition)
\begin{equation}
 \alpha = \sqrt{\left | \frac{(\omega+1)^2\Lambda}
     {(\omega+2)(2\omega+3)} \right |}.
                                 \label{COSMOL}
\end{equation}
For $\omega=-2,-\frac32$ we set $\alpha=1$. For $\omega=-2$
equations (\ref{MET_00}) and (\ref{MET_11}) imply
$\Lambda=\chi^2/8$ so, in contrast with the uncharged case
\cite{Sa_Lemos_Static,Sa_Lemos_Rotat}, the cosmological constant
is not null.

Now, we consider the case $\omega=-1$. From equations
(\ref{MET_00})-(\ref{EQ_DIL})
 it follows that $\nu=C_1$,
$\phi=C_2$, where $C_1$ and $C_2$ are constants of integration,
and that the cosmological constant and electric charge are both
null, $\Lambda=\chi=0$. So, for $\omega=-1$ the metric gives
simply the three-dimensional Minkowski spacetime and the dilaton
is constant, as occurred in the uncharged case
\cite{Sa_Lemos_Static,Sa_Lemos_Rotat}.

In (2+1) dimensions, the presence of a curvature singularity is
revealed by the scalar $R_{\mu\nu}R^{\mu\nu}$
\begin{eqnarray}
 R_{\mu\nu}R^{\mu\nu}  &=&
       12\alpha^4+ \frac{4 \omega}{(\omega+1)^2}
       \frac{b \alpha^4}
       {(\alpha r)^{\frac{2\omega+3}{\omega+1}}}
     + \frac{(2\omega^2+4\omega+3)}{2(\omega+1)^4}
       \frac{b^2\alpha^4}
       {(\alpha r)^{\frac{2(2\omega+3)}{\omega+1}}}
                         \nonumber \\
         & & -\frac{(\omega-1)}{(\omega+1)^2}
       \frac{k \chi^2 \alpha^4}
       {(\alpha r)^{\frac{2(\omega+2)}{\omega+1}}}
              -\frac{(\omega^2+2\omega+2)}{(\omega+1)^4}
       \frac{k \chi^2 b \alpha^4}
       {(\alpha r)^{\frac{4\omega+7}{\omega+1}}}
                                           \nonumber \\
         & &    -\frac{(\omega^2+2\omega+3)}{(\omega+1)^4}
       \frac{k^2 \chi^4 \alpha^4}
       {(\alpha r)^{\frac{4(\omega+2)}{\omega+1}}} \:,  \nonumber \\
       & &
               \hskip 7cm {\rm for} \:\:\:\:\omega \neq -2,-\frac32,-1 \:,
                                \label{R-2-3/2-1}  \\
  R_{\mu\nu}R^{\mu\nu}  &=&
        8+\frac{32}{r}+\frac{6}{r^2}
        +\chi^2 {\biggl [}6 \ln r+ \frac{4 \ln r}{r}
        +\frac{3}{r}+5{\biggr ]}+\chi^4 {\biggl [}\frac{3}{4}\ln^2 r
         + \frac{5}{4} \ln r +9{\biggr ]} \:,
                                               \nonumber \\
       & &
              \hskip 7cm {\rm for} \:\:\:\:\omega =-2 \:,
                              \label{R-2}  \\
  R_{\mu\nu}R^{\mu\nu}  &=&
       \Lambda^2 [12\ln^2(br)+20 \ln(br)+9]
       + \Lambda \chi^2 r^2 {\biggr [} 5 \ln(br)+\frac{9}{2}
       {\biggr ]}+\frac{9}{16}\chi^4 r^4 \nonumber \\
       & &
        \hskip 7cm {\rm for} \:\:\:\:\omega=-\frac32 \:.
                                           \label{R-3/2}
\end{eqnarray}
An inspection of these scalars in (\ref{R-2-3/2-1})-(\ref{R-3/2})
reveals that for $\omega<-2$ and $\omega>-1$ the curvature
singularity is located at $r=0$ and for $-\frac32<\omega<-1$ the
curvature singularity is at $r=+\infty$. For $-2 \leq \omega \leq
-\frac{3}{2}$ both $r=0$ and $r=+\infty$ are singular. For
$\omega=\pm \infty$ spacetime has no curvature singularities. Note
that in the uncharged case \cite{Sa_Lemos_Static,Sa_Lemos_Rotat},
for $-2 \leq \omega < -\frac{3}{2}$ the curvature singularity is
located only at $r=0$.

%

\subsection{The general rotating solution}
 \label{sec:rotating solution 3D}

In order to add angular momentum to the spacetime we perform the
following coordinate transformations (see e.g.
\cite{Sa_Lemos_Rotat}-\cite{HorWel})
\begin{eqnarray}
 t &\mapsto& \gamma t-\frac{\theta}{\alpha^2} \varphi \:,
                                       \nonumber  \\
 \varphi &\mapsto& \gamma \varphi-\theta t \:,
                                       \label{TRANSF_J}
\end{eqnarray}
where $\gamma$ and $\theta$ are constant parameters. Substituting
(\ref{TRANSF_J}) into (\ref{MET_TODOS})-(\ref{MET-3/2}) we obtain
\begin{eqnarray}
 ds^2 &=& -{\biggl [} {\biggl (}\gamma^2-
           \frac{\theta^2}{\alpha^2} {\biggr )}
            (\alpha r)^2 -
     \frac{\gamma^2 b}{(\alpha r)^{\frac{1}{\omega+1}}}
           + \frac{\gamma^2 k\chi^2}
                      {(\alpha r)^{\frac{2}{\omega+1}}}
            {\biggr ]} dt^2       \nonumber \\
      & &
           -\frac{\gamma \theta}{\alpha^2}{\biggl [}
            \frac{b}
            {(\alpha r)^{\frac{1}{\omega+1}}}
           -\frac{k\chi^2}{(\alpha r)^
         {\frac{2}{\omega+1}}}  {\biggr ]} 2dt d\varphi
            + \frac{dr^2}{(\alpha r)^2 -
           \frac{b}{(\alpha r)^{\frac{1}{\omega+1}}}
           +\frac{k\chi^2}
                           {(\alpha r)^{\frac{2}{\omega+1}}} }
                                        \nonumber \\
      & &   + {\biggl [} {\biggl (}\gamma^2-
           \frac{\theta^2}{\alpha^2} {\biggr )}r^2 +
           \frac{\theta^2}{\alpha^4}\frac{b}
           {(\alpha r)^{\frac{1}{\omega+1}}}
           - \frac{\theta^2}{\alpha^4}\frac{k\chi^2}
                         {(\alpha r)^{\frac{2}{\omega+1}}}
            {\biggr ]} d\varphi^2,           \nonumber \\
      & &
     \hskip 7cm {\rm for} \:\:\:\:\omega \neq -2,-\frac32,-1,
                                 \label{MET_TODOS_J}  \\
 ds^2 &=&\!\! -{\biggl [} {\biggl (}(\gamma^2- \theta^2)+
                  \frac{\gamma^2 \chi^2}{4}\ln{r} {\biggr )}
            r^2    -\gamma^2 b r {\biggr ]} dt^2
           + \gamma \theta {\biggl [}
           \frac{\chi^2}{4}r^2\ln{r}
           - b r {\biggr ]} 2 dt d\varphi        \nonumber \\
      & &
           +\frac{dr^2}{(1+\frac{\chi^2}{4}
                \ln{r})r^2 -br}
           + {\biggl [} {\biggl (}(\gamma^2- \theta^2)-
              \frac{\theta^2 \chi^2}{4}\ln{r}
           {\biggr )}r^2 +\theta^2 b r {\biggr ]} d\varphi^2,
                                                 \nonumber \\
      & &  \hskip 7cm {\rm for} \:\:\:\: \omega =-2,
                                         \label{MET-2_J}  \\
 ds^2 &=&\!\!  -r^2[-\gamma^2 \Lambda \ln(br)-\theta^2+
                        \gamma^2 \chi^2r^2] dt^2
           -\gamma \theta r^2[\Lambda \ln(br)+1-
            \chi^2r^2] 2dt d\varphi      \nonumber \\
      & &
        +\frac{dr^2}{r^2[-\Lambda \ln (br)+ \chi^2r^2]}
           + r^2[\theta^2 \Lambda \ln(br)+\gamma^2
           -\theta^2 \chi^2r^2] d\varphi^2,
                                             \nonumber \\
      & &  \hskip 7cm {\rm for} \:\:\:\: \omega=-\frac32 \:.
                                             \label{MET-3/2_J}
\end{eqnarray}
Introducing transformations (\ref{TRANSF_J}) into
(\ref{VEC_POTENT}) we obtain that the vector potential
$A=A_{\mu}(r)dx^{\mu}$ is now given by
\begin{equation}
A=\gamma A(r)dt- \frac{\theta}{\alpha^2}A(r) d\varphi\:,
                           \quad\quad w \neq -1\:,
                                    \label{VEC_POTENT_J}
\end{equation}
where $A(r)=\frac{1}{4}\chi(\omega+1)(\alpha
r)^{\frac{1}{\omega+1}}$. Solutions
(\ref{MET_TODOS_J})-(\ref{VEC_POTENT_J}) represent electrically
charged stationary spacetimes and also solve (\ref{ACCAO}).
Analyzing the Einstein-Rosen bridge of the static solution one
concludes that spacetime is not simply connected which implies
that the first Betti number of the manifold is one, i.e., closed
curves encircling the horizon cannot be shrunk to a point. So,
transformations (\ref{TRANSF_J}) generate a new metric because
they are not permitted global coordinate transformations
 \cite{Stachel}. Transformations
(\ref{TRANSF_J}) can be done locally, but not globally. Therefore
metrics (\ref{MET_TODOS})-(\ref{MET-3/2}) and
(\ref{MET_TODOS_J})-(\ref{MET-3/2_J}) can be locally mapped into
each other but not globally, and such they are distinct.

\subsection{Mass, angular momentum and electric charge of the
solutions}
 \label{sec:MJQ 3D}

In this section we will calculate the mass, angular momentum,
electric charge and dilaton charge of the static and rotating
electrically charged black hole solutions. To obtain these
quantities we apply the formalism  of Regge and Teitelboim
\cite{Regge} (see also
\cite{BTZ_Q,Sa_Lemos_Static,Sa_Lemos_Rotat,Lemos}).

We first write the metrics (\ref{MET_TODOS_J})-(\ref{MET-3/2_J})
in the canonical form involving the lapse function $N^0(r)$ and
the shift function $N^{\varphi}(r)$
\begin{equation}
     ds^2 = - (N^0)^2 dt^2
            + \frac{dr^2}{f^2}
            + H^2(d\varphi+N^{\varphi}dt)^2 \:,
                               \label{MET_CANON}
\end{equation}
where $f^{-2}=g_{rr}$, $H^2=g_{\varphi \varphi}$,
 $H^2 N^{\varphi}=g_{t \varphi}$ and $(N^0)^2-H^2(N^{\varphi})^2=g_{tt}$.
Then, the action can be written in the hamiltonian form as a
function of the energy constraint ${\cal{H}}$, momentum constraint
${\cal{H}}_{\varphi}$ and Gauss constraint $G$
\begin{eqnarray}
S &=& -\int dt d^2x[N^0 {\cal{H}}+N^{\varphi} {\cal{H}_{\varphi}}+
A_{t} G]+
                     {\cal{B}}          \nonumber \\
 &=&  -\Delta t \int dr N
        {\biggl [} \frac{2 \pi^2}{H^3}e^{-2 \phi}-
        4f^2(H \phi_{,r}e^{-2 \phi})_{,r}-2H \phi_{,r}(f^2)_{,r}
        e^{-2 \phi}                                        \nonumber \\
 & &    +2f(fH_{,r})_{,r}e^{-2 \phi}+4 \omega H f^2
        (\phi_{,r})^2e^{-2 \phi}-\Lambda H e^{-2 \phi}+
\frac{2H}{f}e^{-2 \phi}(E^2+B^2){\biggr ]}
\nonumber \\
 & &
      + \Delta t \int dr N^{\varphi}{\biggl [}{\bigl (}2 \pi e^{-2 \phi}
       {\bigr )}_{,r}+\frac{4H}{f}e^{-2 \phi}E^rB{\biggr ]} \nonumber \\
 & &
       + \Delta t \int dr A_t {\biggl [}-\frac{4H}{f}
       e^{-2 \phi} \partial_r E^r{\biggr ]} +{\cal{B}} \:,
                               \label{ACCAO_CANON}
\end{eqnarray}
where $N=\frac{N^0}{f}$, $\pi \equiv {\pi_{\varphi}}^r=-\frac{fH^3
(N^{\varphi})_{,r}}{2N^0}$ (with $\pi^{r \varphi}$ being the
momentum conjugate to $g_{r \varphi}$),  $E^r$ and $B$ are the
electric and magnetic fields and ${\cal{B}}$ is a boundary term.
Upon varying the action with respect to $f(r)$, $H(r)$, $\pi(r)$,
$\phi(r)$ and $E^r(r)$ one picks up additional surface terms.
Indeed,
\begin{eqnarray}
\delta S &=& - \Delta t N {\biggl [}(H_{,r}-2H\phi_{,r})e^{-2\phi}
         \delta f^2 -(f^2)_{,r}e^{-2\phi}\delta H -4f^2 H
             e^{-2\phi} \delta(\phi_{,r})
                                               \nonumber \\
         & &
         +2H{\bigl [}(f^2)_{,r}+4(\omega+1)f^2 \phi_{,r}{\bigr ]}
         e^{-2\phi}\delta \phi
                +2f^2 e^{-2\phi}\delta (H_{,r}) {\biggr ]}
                                        \nonumber \\
         & & +\Delta t N^{\varphi}{\biggl [}2e^{-2\phi}\delta \pi
         -4 \pi e^{-2\phi}\delta \phi {\biggr ]}+ \Delta t A_t
{\biggl [}- \frac{4H}{f}e^{-2 \phi} \delta E^r{\biggr ]} + \delta
{\cal{B}}
          \nonumber \\
         & & +(\mbox{terms vanishing when the
                    equations of motion hold}).
                               \label{DELTA_ACCAO}
\end{eqnarray}
In order that the Hamilton's equations are satisfied, the boundary
term ${\cal{B}}$ has to be adjusted so that it cancels the above
additional surface terms. More specifically one has
\begin{equation}
  \delta {\cal{B}} = -\Delta t N \delta M +\Delta t N^{\varphi}\delta J+
             \Delta t A_t \delta Q \:,
                              \label{DELTA_B}
\end{equation}
where one identifies $M$ as the mass, $J$ as the angular momentum
and $Q$ as the electric charge since they are the terms conjugate
to the asymptotic values of $N$, $N^{\varphi}$ and $A_t$,
respectively.

To determine the $M$, $J$ and $Q$ of the black hole one must take
the black hole spacetime and subtract the background reference
spacetime contribution, i.e., we choose the energy zero point in
such a way that the mass, angular momentum and charge vanish when
the black hole is not present.

Now, note that for $\omega <-2$, $\omega>-3/2$ and $\omega \neq
-1$, spacetime (\ref{MET_TODOS_J}) has an asymptotic metric given
by
\begin{equation}
-{\biggl (}\gamma^2-\frac{\theta^2}{\alpha^2} {\biggr )}
 \alpha^2 r^2 dt^2+ \frac{d r^2}{ \alpha^2 r^2}+
{\biggl (}\gamma^2-\frac{\theta^2}{\alpha^2} {\biggr )}
 r^2 d \varphi^2 \:,
                                          \label{ANTI_SITTER}
\end{equation}
i.e., it is asymptotically an anti-de Sitter spacetime. In order
to have the usual form of the anti-de Sitter metric we choose
$\gamma^2-\theta^2 / \alpha^2=1$. For the cases $-2\leq \omega
\leq -3/2$ we shall also choose $\gamma^2-\theta^2 / \alpha^2=1$,
as has been done for the uncharged case
\cite{Sa_Lemos_Static,Sa_Lemos_Rotat}. For $\omega \neq -3/2,-1$
the anti-de Sitter spacetime is also the background reference
spacetime, since the metrics (\ref{MET_TODOS_J}) and
(\ref{MET-2_J})
 reduce to (\ref{ANTI_SITTER}) if the
black hole is not present ($b=0$ and $\varepsilon=0$). For
$\omega=-3/2$ the above described procedure of choosing the energy
zero point does not apply since for any value of $b$ and
$\varepsilon$ one still has a black hole solution. Thus, for
$\omega=-3/2$ the energy zero point is chosen arbitrarily to
correspond to the black hole solution with $b=1$ and
$\varepsilon=0$.

Taking the subtraction of the background reference spacetime into
account and noting that $\phi-\phi_{\rm ref}=0$ and that
$\phi_{,r}-\phi_{,r}^{\rm ref}=0$ we have that the mass, angular
momentum and electric charge are given by
\begin{eqnarray}
M &=& (2H\phi_{,r}-H_{,r})e^{-2\phi}(f^2-f^2_{\rm ref})
      +(f^2)_{,r}e^{-2\phi}(H-H_{\rm ref})
         -2f^2 e^{-2\phi}(H_{,r}-H_{,r}^{\rm ref})\:,
                                            \nonumber \\
J &=&  -2e^{-2\phi} (\pi-\pi_{\rm ref}) \:,
                                            \nonumber \\
Q &=&  \frac{4H}{f}e^{-2 \phi} (E^r-E^r_{\rm ref}) \:.
                                    \label{MJQ_GERAL}
\end{eqnarray}
Then, for $\omega>-3/2$  and $\omega \neq -1$, we finally have
that the mass and angular momentum are (after taking the
appropriate asymptotic limit: $r \rightarrow +\infty$ for
$\omega>-1$ and $r \rightarrow 0$ for $-3/2<\omega<-1$, see the
Penrose diagrams on section 6.3 to understand the reason for these
limits)
\begin{eqnarray}
M &=& b {\biggl [}\frac{\omega +2}{\omega +1}\gamma^2
  + \frac{\theta^2}{\alpha^2}{\biggl ]}=M_{{\rm Q}=0} \:,
    \nonumber \\
J &=&  \frac{\gamma \theta}{\alpha^2}b\frac{2\omega+3} {\omega
+1}=J_{{\rm Q}=0}\:,   \hskip 2cm  {\rm for} \:\:\:\: \omega>
-3/2, \: \omega \neq -1 \:,
                                    \label{MJ>-1}
\end{eqnarray}
where $M_{{\rm Q}=0}$ and $J_{{\rm Q}=0}$ are the mass and angular
momentum of the uncharged black hole. For $\omega<-3/2$, the mass
and angular momentum are (after taking the appropriate asymptotic
limit, $r \rightarrow +\infty$)
\begin{eqnarray}
M &=& b {\biggl [}\frac{\omega +2}{\omega +1}\gamma^2
  + \frac{\theta^2}{\alpha^2}{\biggl ]}
  + {\rm Div_M}(\chi,r)= M_{{\rm Q}=0}
  + {\rm Div_M}(\chi,r) \:,
            \label{M<-1}   \\
J &=&  \frac{\gamma \theta}{\alpha^2}b\frac{2\omega+3} {\omega +1}
+ {\rm Div_J}(\chi,r)= J_{{\rm Q}=0}+ {\rm Div_J}(\chi,r)
\:,   \nonumber \\
& & \hskip 6cm  {\rm for} \:\:\:\: \omega<-3/2 \:,
                                    \label{J<-1}
\end{eqnarray}
where ${\rm Div_M}(\chi,r)$ and ${\rm Div_J}(\chi,r)$ are terms
proportional to the charge $\chi$ that diverge at the asymptotic
limit. The presence of these kind of  divergences in the mass and
angular momentum is a usual feature present in charged solutions.
They can be found for example in the electrically charged point
source solution \cite{Deser_Maz}, in the electric counterpart of
the BTZ black hole \cite{BTZ_Q}, in the pure electric black holes
of 3D Brans-Dicke action \cite{OscarLemos} and in the magnetic
counterpart of the BTZ solution \cite{OscarLemos_BTZ}. Following
\cite{Deser_Maz,BTZ_Q}, the divergences on the mass can be treated
as follows. One considers a boundary of large radius $r_0$
involving the black hole. Then, one sums and subtracts ${\rm
Div_M}(\chi,r_0)$ to (\ref{M<-1}) so that the mass (\ref{M<-1}) is
now written as
\begin{equation}
M = M(r_0)+ [{\rm Div_M}(\chi,r)-
     {\rm Div_M}(\chi,r_0)] \:,
           \label{M0<-1}
\end{equation}
where $M(r_0)=M_{{\rm Q}=0}+{\rm Div_M}(\chi,r_0)$, i.e.,
\begin{equation}
M_{{\rm Q}=0}=M(r_0)- {\rm Div_M}(\chi,r_0)\:.
                       \label{M0_v2<-1}
\end{equation}
The term between brackets in (\ref{M0<-1}) vanishes when
$r\rightarrow r_0$. Then $M(r_0)$ is the energy within the radius
$r_0$. The difference between $M(r_0)$ and $M_{{\rm Q}=0}$ is
$-{\rm Div_M}(\chi,r_0)$ which is interpreted as the
electromagnetic energy outside $r_0$ apart from an infinite
constant which is absorbed in $M(r_0)$. The sum (\ref{M0_v2<-1})
is then independent of $r_0$, finite and equal to the total mass.

To handle the angular momentum divergence, one first notice that
the asymptotic limit of the angular momentum per unit  mass
$(J/M)$ is either zero or one, so the angular momentum diverges at
a rate slower or equal to the rate of the mass divergence. The
divergence on the angular momentum can then be treated in a
similar way as the mass divergence. So, the divergent term $-{\rm
Div_J}(\chi,r_0)$ can be interpreted as the electromagnetic
angular momentum outside $r_0$ up to an infinite constant that is
absorbed in $J(r_0)$.

In practice the treatment of the mass and angular divergences
amounts to forgetting about $r_0$ and take as zero the asymptotic
limits: $\lim {\rm Div_M}(\chi,r)=0$ and $\lim {\rm
Div_J}(\chi,r)=0$. So, for $\omega<-3/2$ the mass and angular
momentum are also given by (\ref{MJ>-1}).

Interesting enough, as has been noticed in \cite{BTZ_Q}, is the
fact that in four spacetime dimensions there occurs a similar
situation. For example, the $g_{tt}$ component of
Reissner-Nordstr\"{o}m solution can be written as
$1-\frac{2M(r_0)}{r}+Q^2(\frac{1}{r}-\frac{1}{r_0})$. The total
mass $M=M(r_0)+\frac{Q^2}{2r_0}$ is independent of $r_0$ and
$\frac{Q^2}{2r_0}$ is the electrostatic energy outside a sphere of
radius $r_0$. In this case, since $\frac{Q^2}{2r_0}$ vanishes when
$r_0 \rightarrow \infty$, one does not need to include an infinite
constant in $M(r_0)$. Thus, in this general Brans-Dicke theory in
3D we conclude that both situations can occur depending on the
value of $\omega$. The $\omega>-3/2, \:\omega \neq -1$ case is
analogous to the the Reissner-Nordstr\"{o}m black hole in the
sense that it is not necessary to include an infinite constant in
$M(r_0)$, while the case $\omega<-3/2$ is similar to the
electrically charged BTZ black hole \cite{BTZ_Q} since  an
infinite constant must be included in  $M(r_0)$.

For $\omega=-3/2$ the mass and angular momentum are ill defined
since the boundaries $r \rightarrow \infty$ and $r \rightarrow 0$
have logarithmic singularities in the mass term even in the
absence of the electric charge.

Now, we calculate the electric charge of the black holes. To
determine the electric field we must consider the projections of
the Maxwell field on spatial hypersurfaces. The normal to such
hypersurfaces is $n^{\nu}=(1/N^0,0,-N^{\varphi}/N^0)$ so the
electric field is $E^{\mu}=g^{\mu \sigma}F_{\sigma \nu}n^{\nu}$.
Then, from (\ref{MJQ_GERAL}), the electric charge is
\begin{equation}
 Q=-\frac{4Hf}{N^0}e^{-2 \phi}(\partial_rA_t-N^{\varphi}
    \partial_r A_{\varphi})=\gamma \chi\:,
\hskip 2cm  \omega \neq -1 \:. \label{CARGA}
\end{equation}

The mass, angular momentum and electric charge of the static black
holes can be obtained by putting $\gamma=1$ and $\theta=0$ on the
above expressions [see (\ref{TRANSF_J})].

Now, we want to cast the metric in terms of $M$, $J$ and $Q$. For
$\omega \neq -2,-3/2,-1$, we can use (\ref{MJ>-1}) to solve a
quadratic equation for $\gamma^2$ and $\theta^2 / \alpha^2$. It
gives two distinct sets of solutions
\begin{equation}
\gamma^2=\frac{\omega+1}{2(\omega+2)}\frac{M(2- \Omega)}{b}
\:,\:\:\: \:\:\:\: \frac{\theta^2}{\alpha^2}=\frac{M
\Omega}{2b}\:, \label{DUAS}
\end{equation}

\begin{equation}
\gamma^2=\frac{\omega+1}{2(\omega+2)}\frac{M \Omega}{b} \:,\:\:\:
\:\:\:\: \frac{\theta^2}{\alpha^2}=\frac{M(2- \Omega)}{2b}\:,
\label{DUAS_ERR}
\end{equation}
where we have defined a rotating parameter $\Omega$ as
\begin{equation}
\Omega \equiv 1- \sqrt{1-\frac{4(\omega+1)(\omega+2)}
{(2\omega+3)^2}\frac{J^2 \alpha^2}{M^2}} \:, \hskip 1cm  {\rm for}
\:\:\:\:  \omega \neq -2,-3/2,-1 \:.
                    \label{OMEGA}
\end{equation}
When we take $J=0$ (which implies $\Omega=0$), (\ref{DUAS}) gives
$\gamma \neq 0$ and $\theta= 0$ while (\ref{DUAS_ERR}) gives the
nonphysical solution $\gamma=0$ and $\theta \neq 0$ which does not
reduce to the static original metric. Therefore we will study the
solutions found from (\ref{DUAS}). For $\omega=-2$ we have
$\gamma^2=J^2/Mb$ and $\theta^2=M/b$.

The condition that $\Omega$ remains real imposes for
$-2>\omega>-1$ a restriction on the allowed values of the angular
momentum: $|\alpha J|\leq \frac{|2\omega+3|M}{2\sqrt{(\omega+1)
(\omega+2)}}$. For $-2>\omega>-1$ we have $0 \leq \Omega \leq 1$.
In the range $-2<\omega<-3/2$ and $-3/2<\omega<-1$ we have $\Omega
<0$. The condition $\gamma^2-\theta^2/\alpha^2=1$ fixes the value
of $b$ and from (\ref{CARGA}) we can write $k \chi^2$ as a
function of $b,M,\Omega,Q$. Thus,
\begin{eqnarray}
b &=&  \frac{M}{2(\omega +2)}{\biggl [}2(\omega+1)-
       (2\omega +3)\Omega  {\biggr ]} \:,
                                \label{b}  \\
k\chi^2&=&  \frac{b}{4}(\omega +1) \frac{Q^2}
                    {M(2-\Omega)}\:,
\hskip 2cm  {\rm for} \:\:\:\:  \omega \neq -2,-3/2,-1 \:,
                                    \label{c^2}
\end{eqnarray}
and
\begin{eqnarray}
b = \frac{J^2-M^2}{M}\:,\:\: \:\:\:\:\:\: \chi^2 =  \frac{Q^2 M
b}{J^2}\:, \hskip 2cm  {\rm for} \:\:\:\:  \omega=-2 \:.
\label{bc_2}
\end{eqnarray}
The metrics (\ref{MET_TODOS_J}) and (\ref{MET-2_J}) may now be
cast in the form
\begin{eqnarray}
 ds^2 &=& -{\biggl [} (\alpha r)^2 - \frac{(\omega+1)}
          {2(\omega+2)} \frac{M(2-\Omega)}
          {(\alpha r)^{\frac{1}{\omega+1}}}
           + \frac{(\omega+1)^2}{8(\omega+2)}
          \frac{Q^2} {(\alpha r)^{\frac{2}{\omega+1}}}
            {\biggr ]} dt^2       \nonumber \\
      & &
          -\frac{\omega+1}{2\omega+3}J{\biggl [}
            (\alpha r)^{-\frac{1}{\omega+1}}
           -\frac{(\omega+1)Q^2}{4M(2-\Omega)}
           (\alpha r)^{-\frac{2}{\omega+1}}  {\biggr ]}
               2dt d\varphi
                                      \nonumber \\
      & &
           + {\biggl [}(\alpha r)^2 -
           \frac{M[2(\omega+1)-
       (2\omega +3)\Omega]}{2(\omega +2)
            (\alpha r)^{\frac{1}{\omega+1}}}
                                       \nonumber \\
      & &  \:\:\:\:\:\:\:\:\:\:
       +\frac{(\omega +1)Q^2[2(\omega+1)-
       (2\omega +3)\Omega]}
           {8(\omega +2)(2-\Omega)(\alpha r)^{\frac{2}
           {\omega+1}}} {\biggr ]}^{-1} dr^2
                                        \nonumber \\
      & &   + \frac{1}{\alpha^2}{\biggl [} (\alpha r)^2 +
           \frac{M \Omega}
           {2(\alpha r)^{\frac{1}{\omega+1}}}
           - \frac{(\omega +1) \: \Omega \: Q^2} {8(2-\Omega)
            (\alpha r)^{\frac{2}{\omega+1}}}
            {\biggr ]} d\varphi^2,           \nonumber \\
      & &
      \hskip 7cm  {\rm for} \:\:\:\: \omega \neq -2,-\frac32,-1\:,
                                        \label{MET_MJQ} \\
 ds^2 &=& -{\biggl [}{\biggl (}1+\frac{Q^2}{4}\ln r{\biggl )}r^2
          -\frac{J^2}{M}\: r{\biggr ]} dt^2
          +J{\biggl [} \frac{Q^2 M}{4 J^2}r^2 \ln r-r
                {\biggr ]}  2dt d\varphi
                                       \nonumber \\
      & &
           + {\biggl [} {\biggl (}1+\frac{Q^2}{4}
          {\bigl (}1-\frac{M^2}{J^2}{\bigl )}\ln r{\biggl )}r^2
          -M {\biggl (}\frac{J^2}{M^2}-1 {\biggl )}r
          {\biggr ]}^{-1} dr^2
                                        \nonumber \\
      & & + {\biggl [}{\biggl (}1-\frac{Q^2 M^2}{4 J^2}\ln r
          {\biggl )}r^2+Mr
            {\biggr ]} d\varphi^2,
      \hskip 2cm  {\rm for} \:\:\:\: \omega = -2.
                                        \label{MET_MJQ_-2}
\end{eqnarray}
Analyzing the function $\Delta=g_{rr}^{-1}$ in
(\ref{MET_TODOS_J}), (\ref{MET-2_J}), and (\ref{b})-(\ref{bc_2})
we can set the conditions imposed on the mass and angular momentum
of the solutions obtained for the different values of $\omega \neq
-3/2,-1$, in order that the black holes might
 exist. These conditions are summarized on Table 1.

We can mention the principal differences between the charged and
uncharged theory. The charged theory has black holes which are not
present in the uncharged theory for the following range of
parameters: (i) $-\infty<\omega<-2$, $M<0$; (ii) $\omega=-2$,
$M>0$, $|J|<M$; (iii) $-3/2<\omega<-1$, $M<0$, $|\alpha J|>M$;
(iv)
 $-1<\omega<+\infty$, $M>0$, $M<|\alpha J|<
(2\omega+3)M/2\sqrt{(\omega+1)(\omega+2)}$ and (v)
$-1<\omega<+\infty$, $M<0$, $|\alpha J|<|M|$.

\begin{center}
 \vfill \vskip 5mm \noindent
\begin{tabular}{|l|l|l|}    \hline
     Range of $\omega$
        & Black holes with $M>0$
        & Black holes with $M<0$      \\ \hline\hline
     $-\infty<\omega<-2$
        &  $|\alpha J|\leq\frac{|2\omega+3|M}{2\sqrt{(\omega+1)
                                              (\omega+2)}}$
        &  $|\alpha J|\leq\frac{|(2\omega+3)M|}{2\sqrt{(\omega+1)
                                              (\omega+2)}}$
                                       \\ \hline
     $\omega=-2$
        &  might exist for any $J$
        &  $|J|<|M|$                    \\ \hline
     $-2<\omega<-\frac32$
        &  do not exist for any $J$
        &  might exist for any  $J$               \\ \hline
     $-\frac32<\omega<-1$
        &  $|\alpha J|>M$
        &  might exist for any $J$   \\ \hline
     $-1<\omega<+\infty$
        & $|\alpha J| \leq
                   \frac{(2\omega+3)M}{2\sqrt{(\omega+1)
                                         (\omega+2)}}$
        & $|\alpha J|  \leq \frac{(2\omega+3)|M|}
                        {2\sqrt{(\omega+1)(\omega+2)}}$
                                      \\ \hline
\end{tabular}
\end{center}
\vskip 1mm {\small \noindent {\bf Table 1.} \noindent Values of
the angular momentum for which  black holes with positive and
negative masses might exist. } \vskip 3mm

\subsection{Causal and geodesic structure of the charged
black holes}
 \label{sec:Causal Geod 3D}

\subsubsection{Analysis of the causal structure}
 \label{sec:Causal 3D}
In order to study the causal structure we follow the procedure of
Boyer and Lindquist \cite{Boyer} and Carter \cite{Carter} and
write the metrics (\ref{MET_TODOS_J})-(\ref{MET-3/2_J}) in the
form [see (\ref{TRANSF_J})]
\begin{equation}
ds^2=-\Delta {\biggl (}\gamma dt-\frac{\theta}{\alpha^2}
      d\varphi {\biggr )}^2+\frac{dr^2}{\Delta}
      +r^2{\biggr (}\gamma d\varphi-\theta dt{\biggr )}^2 \:,
                                    \label{MET_PENR}
\end{equation}
where
\begin{eqnarray}
 \Delta &=&
            (\alpha r)^2 -
           b(\alpha r)^{-\frac{1}{\omega+1}}
           +k\chi^2(\alpha r)^{-\frac{2}{\omega+1}} \:,
          \hskip 1cm  {\rm for} \:\:\:\:\omega \neq -2,-3/2,-1,
                                 \label{DELTA_TODOS}  \\
 \Delta &=&  (1+\frac{\chi^2}{4}\ln{r})r^2  -br \:,
                     \hskip 4cm  {\rm for} \:\:\:\: \omega=-2,
                                 \label{DELTA-2}  \\
\Delta &=&  r^2[-\Lambda \ln (br)+\chi^2r^2] \:,
               \hskip 4cm  {\rm for} \:\:\:\:\omega=-\frac32\:,
                                             \label{DELTA-3/2}
\end{eqnarray}
and in (\ref{DELTA_TODOS}) $b$ and $k\chi^2$ are given by
(\ref{b}) and (\ref{c^2}).

Bellow we describe the general procedure to draw the Penrose
diagrams. Following Boyer and Lindquist \cite{Boyer}, we choose a
new angular coordinate which straightens out the spiraling null
geodesics that pile up around the event horizon. A good choice is
\begin{equation}
\bar{\varphi}=\gamma \varphi-\theta t\:.
                                    \label{TRANSF_RECTA}
\end{equation}
Then (\ref{MET_PENR}) can be written as
\begin{equation}
ds^2=-\Delta {\biggl (} \frac{1}{\gamma} dt-\frac{\theta}
   {\alpha^2 \gamma} d \bar{\varphi} {\biggr )}^2
   +\frac{dr^2}{\Delta} +r^2 d \bar{\varphi}^2 \:.
                                    \label{MET_PENR_RECTA}
\end{equation}
Now the null radial geodesics are straight lines at $45^{\rm o}$.
The advanced and retarded null coordinates are defined by
\begin{equation}
u=\gamma t-r_{\ast}\:, \hskip 1cm v=\gamma t+r_{\ast}\:,
                                    \label{NULL_COORD}
\end{equation}
where $r_{\ast}=\int\Delta^{-1}dr$ is the tortoise coordinate. In
general, the integral defining the tortoise coordinate cannot be
solved explicitly for the solutions
(\ref{DELTA_TODOS})-(\ref{DELTA-3/2}). Moreover, the maximal
analytical extension depends critically on the values of $\omega$.
There are seven cases which have to be treated separately:
$\omega<-2$, $\omega=-2$, $-2<\omega<-3/2$, $\omega=-3/2$,
$-3/2<\omega<-1$, $\omega>-1$ and $\omega=\pm \infty$. As we shall
see, on some of the cases the $\Delta$ function has only one zero
and so the black hole has one event horizon, for other cases
$\Delta$ has two zeros and consequently two horizons are present.
If $\Delta$ has one zero, $r=r_+$, we proceed as follows. In the
region where $\Delta<0$ we introduce the Kruskal coordinates
$U=+e^{-k u}$ and $V=+e^{+k v}$ and so $UV=+e^{k(v-u)}$. In the
region where $\Delta>0$ we define the Kruskal coordinates as
$U=-e^{-k u}$ and $V=+e^{+k v}$ in order that $UV=-e^{k(v-u)}$.
The signal of the product $UV$ is chosen so that the factor
$\Delta/UV$, that appears in the metric coefficient $g_{UV}$, is
negative. The constant $k$ is introduced in order that the limit
of $\Delta/UV$ as $r\rightarrow r_+$ stays finite.

If $\Delta$ has two zeros, $r=r_-$ and $r=r_+$ (with $r_-<r_+$),
one has to introduce a Kruskal coordinate patch around each of the
zeros of $\Delta$. The first patch constructed around $r_-$ is
valid for $0<r<r_+$. For this patch, in the region where
$\Delta<0$ we introduce the Kruskal coordinates $U=+e^{+k_{-} u}$
and $V=+e^{-k_{-} v}$ and so  $UV=+e^{k_{-}(u-v)}$. In the region
where $\Delta>0$ we define the Kruskal coordinates as
$U=-e^{+k_{-} u}$ and $V=+e^{-k_{-} v}$ in order that
$UV=-e^{k_{-}(u-v)}$. The metric defined by this Kruskal
coordinates is regular in the patch $0<r<r_+$ and, in particular,
is regular at $r_-$. However, it is singular at $r_+$. To have a
metric non singular at $r_+$ one has to define new Kruskal
coordinates for the second patch which is constructed around $r_+$
and is valid for $r_-<r<\infty$. For this patch, in the region
where $\Delta<0$ we introduce the Kruskal coordinates
$U=+e^{-k_{+} u}$ and $V=+e^{+k_{+} v}$ and so
$UV=+e^{k_{+}(v-u)}$. In the region where $\Delta>0$ we define the
Kruskal coordinates as $U=-e^{-k_{+} u}$ and $V=+e^{+k_{+} v}$ in
order that $UV=-e^{k_{+}(v-u)}$. $k_{-}$ and $k_{+}$ are constants
obeying the same condition defined above for $k$ and the sign of
$UV$ is also chosen in order to have metric coefficient $g_{UV}
\propto \Delta/UV$ negative. Now, these two different patches have
to be joined together. Finally, to construct the Penrose diagram
one has to define the Penrose coordinates by the usual arctangent
functions of $U$ and $V$: ${\cal{U}}=\arctan U$ and
${\cal{V}}=\arctan V$.

The horizon is mapped into two mutual perpendicular straight null
lines at $45^{\rm o}$. In general, to find what kind of curve
describes the lines $r=0$ or $r=\infty$ one has to take the limit
of $UV$ as $r\rightarrow 0$, in the case of $r=0$, and the limit
of $UV$ as $r\rightarrow \infty$, in the case of $r=\infty$. If
this limit is $\infty$ the corresponding line is mapped into a
curved null line. If the limit is $-1$ the corresponding line is
mapped into a curved timelike line and finally, when the limit is
$+1$ the line is mapped into a curved spacelike line. The
asymptotic lines are drawn as straight lines although in the
coordinates ${\cal{U}}$ and ${\cal{V}}$ they should be curved
outwards, bulged. It is always possible to change coordinates so
that the asymptotic lines are indeed straight lines.

The lines of infinite redshift, $r=r_{\rm rs}$, are given by the
vanishing of the $g_{tt}$ metric component. There are closed
timelike curves, $r_{CTC}$, whenever $g_{\varphi \varphi}<0$.

The Penrose diagram for the static charged black hole is similar
to the one drawn for the corresponding rotating  charged
 black hole. The only difference is that the infinite redshift
lines coincide with the horizons and there are no closed timelike
surfaces. This similarity is due to the fact that the rotating
black hole is obtained from the static one by applying the
coordinate transformations (\ref{TRANSF_J}).

In practice, to find the curve that describes the asymptotic
limits $r=0$ and $r=\infty$, we can use a trick. We study the
behavior of $\Delta$ at  the asymptotic limits, i.e. we find which
term of $\Delta$ dominates as  $r\rightarrow 0$ and $r\rightarrow
\infty$. Then, in the asymptotic region, we take $\Delta \sim
\Delta_0$ in the vicinity of $r=0$ and $\Delta \sim
\Delta_{\infty}$ in the vicinity of $r=\infty$. The above
procedure of finding the Kruskal coordinates is then applied to
the asymptotic regions, e.g. in the vicinity of $r=\infty$ we can
take $r_{\ast} \sim
r_{\ast}^{\infty}=\int(\Delta_{\infty})^{-1}dr$ and find the
character of the $r=\infty$ curve from the limit of $UV$ as
$r\rightarrow \infty$.

\subsubsection{Analysis of the geodesic structure}
 \label{sec:Geod 3D}
Let us now consider the geodesic motion. The equations governing
the geodesics can be derived from the
 Lagrangian
\begin{equation}
{\cal{L}}=\frac{1}{2}g_{\mu\nu}\frac{dx^{\mu}}{d \tau}
       \frac{dx^{\nu}}{d \tau}=-\frac{\delta}{2}\:,
                                 \label{LAG)}  \\
\end{equation}
where $\tau$ is an affine parameter along the geodesic which, for
a timelike geodesic, can be identified with the proper time of the
particle along the geodesic. For a null geodesic one has
$\delta=0$ and for a timelike geodesic $\delta=+1$. From the
Euler-Lagrange equations one gets that the generalized momentums
associated with the time coordinate and angular coordinate are
constants: $p_t=E$ and $p_{\varphi}=L$. The constant $E$ is
related to the timelike Killing vector $(\partial/\partial
t)^{\mu}$ which reflects the time translation invariance of the
metric, while the constant $L$ is associated to the spacelike
Killing vector $(\partial/\partial \varphi)^{\mu}$ which reflects
the invariance of the metric under rotation. Note that since the
spacetime is not asymptotically flat, the constants $E$ and $L$
cannot be interpreted as the local energy and angular momentum at
infinity.

From the geodesics equations we can derive two equations  which
will be specially useful since they describe the behavior of
geodesic motion along the radial coordinate. For $\omega<-2$ and
$\omega>-1$ these are
\begin{eqnarray}
r^2 \dot{r}^2 &=& - {\biggl [}r^2 \delta
        +\frac{(\omega+2)c_0^2}{2(\omega+1)-(2\omega +3)\Omega}
       {\biggr ]}  \Delta +
        \frac{2(\omega+1)c_1^2}{2(\omega+1)-(2\omega +3)\Omega}
                 r^2
          \nonumber \\
                                 \label{GEOD_A}  \\
r^2 \dot{r}^2 &=& (E^2-\alpha^2L^2)r^2+\frac{M c_0^2}
    {2(\alpha r)^{\frac{1}{\omega+1}}} -
    \frac{\sqrt{2}(\omega+1)}
   {16 \sqrt{2-\Omega}}
   \frac{c_0^2 Q^2}{(\alpha r)^{\frac{2}{\omega+1}}}
   -r^2 \delta \Delta\:.
                                        \label{GEOD_A_2}
\end{eqnarray}
For $-2<\omega<-3/2$ and $-3/2<\omega<-1$ the two useful equations
are
\begin{eqnarray}
r^2 \dot{r}^2 &=&   - {\biggl [}r^2 \delta
   -\frac{(\omega+2)c_0^2}{2(\omega+1)-(2\omega +3)\Omega}
   {\biggr ]} \Delta +
        \frac{2(\omega+1)c_1^2}{2(\omega+1)-(2\omega +3)\Omega}
              r^2               \nonumber \\
                                 \label{GEOD_B}  \\
r^2 \dot{r}^2 &=& (E^2-\alpha^2L^2)r^2-\frac{M c_0^2} {2 (\alpha
r)^{\frac{1}{\omega+1}}} + \frac{\sqrt{2}(\omega+1)} {16
\sqrt{2-\Omega}}
 \frac{c_0^2 Q^2}{(\alpha r)^{\frac{2}{\omega+1}}}
 -r^2 \delta \Delta \:.
                                        \label{GEOD_B_2}
\end{eqnarray}
In the above equations $\Delta$ is the inverse of the metric
component $g_{rr}$  and we have introduced the definitions
\begin{eqnarray}
c_0^2  &=& {\biggl [}\frac{\sqrt{|\Omega|}}{\alpha}E \mp
    \sqrt{{\biggl |}\frac{\omega +1}{\omega +2}{\biggl |}}
    \sqrt{2-\Omega}\:L {\biggr ]}^2\:, \:\:\:  {\rm and}
                                        \label{c0}  \\
c_1^2 &=&  {\biggl [}\sqrt{\frac{2-\Omega}{2}}E-
    \sqrt{{\biggl |}\frac{\omega +2}{2(\omega +1)}{\biggl |}}
    \sqrt{|\Omega|} \:\alpha L {\biggr ]}^2 \:,
                                            \label{c1}
\end{eqnarray}
where in (\ref{c0}) the minus sign is valid for $\omega<-2$ and
$\omega>-1$ and the plus sign is applied when $-2<\omega<-3/2$ or
$-3/2<\omega<-1$. There are turning points, $r_{\rm tp}$, whenever
$\dot{r}=0$. If this  is the case, equations (\ref{GEOD_A}) and
(\ref{GEOD_B}) will allow us to make considerations about the
position of the turning point relatively to the position of the
horizon. For this purpose it will be important to note that for
$\omega<-2$ and $\omega>-1$ one has $0 \leq \Omega \leq 1$, and
for $-2<\omega<-3/2$ or $-3/2<\omega<-1$ we have that $\Omega <0$.
As a first and general example of the interest of equations
(\ref{GEOD_A}) and (\ref{GEOD_B}) note that the turning points
coincide with the horizons when the energy and the angular
momentum are such that $c_1=0$. From equations (\ref{GEOD_A_2}),
(\ref{GEOD_B_2}) and after some graphic computation we can reach
interesting conclusions.

 \subsubsection{Penrose diagrams and geodesics for each range of $\omega$}
 \label{sec:PD 3D}

We are now in position to draw the Penrose diagrams and study the
geodesic motion. Besides the cases $\omega=-2$ and $\omega=-3/2$,
for each different range of $\omega$ ($\omega<-2$,
$-2<\omega<-3/2$, $-3/2<\omega<-1$, $\omega>-1$ and $\omega=\pm
\infty$) we will consider a particular value of $\omega$. These
will be precisely the
 ones that have been analyzed in the uncharged study of
action (\ref{ACCAO}) \cite{Sa_Lemos_Static,Sa_Lemos_Rotat}.

We will study the solutions with positive mass, $M>0$, and
describe briefly the Penrose diagrams of the solutions with
negative mass.

\vskip 2mm {\bf A. Brans-Dicke theories with ${\bm{\omega<-2}}$}
\vskip 2mm

For this range of $\omega$ there is the possibility of having
black holes with positive mass whenever $|\alpha
J|\leq\frac{|2\omega+3|M}{2\sqrt{(\omega+1)(\omega+2)}}$. We are
going to analyze the typical case $\omega=-3$. The $\Delta$
function, (\ref{DELTA_TODOS}), is
\begin{equation}
 \Delta =(\alpha r)^2 -b\sqrt{\alpha r}+c(\alpha r)\:,
                                 \label{DEL_(-3)}  \\
\end{equation}
where from (\ref{b}) and (\ref{c^2}) one has that $b>0$ and
$c\equiv k\chi^2<0$ (if $M>0$). For $|\alpha
J|\leq\frac{3M}{2\sqrt{2}}$, $\Delta$ has always one and only one
zero given by
\begin{eqnarray}
& & r_+=\frac{1}{3\alpha}{\biggl [}-2c +\frac{c^2}{s}
    +s{\biggr ]}\:, \:\:\:\:\:{\rm where} \:\:\:\:
     s={\biggl [}\frac{1}{2}{\biggl (} 27b^2+2c^3+3\sqrt{3}b
    \sqrt{27b^2+4c^3}{\biggr )}{\biggr ]}^{\frac{1}{3}}\:,
                                 \label{Zero_(-3)}
\end{eqnarray}
$\Delta>0$ for $r>r_+$ and $\Delta<0$ for $r<r_+$. The curvature
singularity at $r=0$ is a spacelike line in the Penrose diagram
while $r=+\infty$ is a timelike line. The Penrose diagram is drawn
in figure 1. There is no extreme black hole for this case.

\begin{figure} [t]
\centering
\includegraphics[height=1.0in]{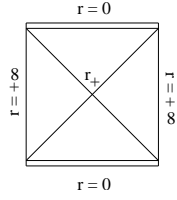}
\caption{\label{eBD-fig1}
 Penrose diagram for the $\omega=-3$,
$M>0$, $|\alpha J|\leq 3M/2\sqrt{2}$ black hole.
 }
\end{figure}
When we consider the solutions with negative mass  we conclude
that, for $|\alpha J|\leq\frac{3|M|}{2\sqrt{2}}$, one has black
holes with two horizons, with one (extreme case) or a spacetime
without black holes. For the black hole with two horizons the
Penrose diagram is shown in figure 2.(a) of and the extreme black
hole has a Penrose diagram which is  drawn in figure 2.(b).
\begin{figure} [t]
\centering
\includegraphics[height=1.7in]{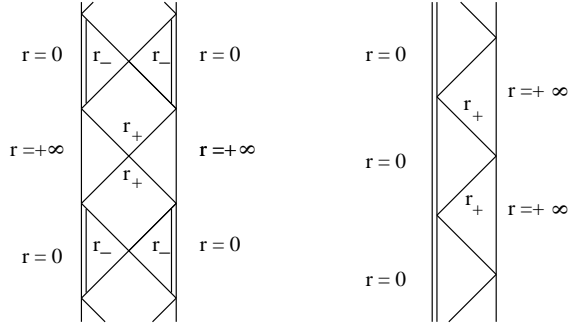}
\caption{\label{eBD-fig2}
 (a) Penrose diagram for the $\omega=-3$, $M<0$,
$|\alpha J|\leq 3|M|/2\sqrt{2}$ black hole with two horizons.
 (b) Penrose diagram for the $\omega=-3$, $M<0$, $|\alpha J|\leq
3|M|/2\sqrt{2}$ extreme black hole.
 }
\end{figure}

Let us now consider the geodesic motion. Analyzing (\ref{GEOD_A})
and noting that $0 \leq \Omega \leq 1$ we see that for the null
and timelike geodesics the coefficient of $\Delta$ is always
negative so, for $0<r<r_+$, the first term of (\ref{GEOD_A}) is
positive. Since the second term is positive or null we conclude
that whenever there are turning points, they are $r_{\rm tp}^1=0$
and $r_{\rm tp}^2\geq r_+$. From (\ref{GEOD_A_2}) we conclude the
following about the geodesic motion. (i) If $E^2-\alpha^2L^2 \geq
0$ null particles produced at $r=0$ escape to $r=+\infty$ and null
particles coming in from infinity are scattered at $r_{\rm
tp}^1=0$ and spiral back to infinity. (ii) Null geodesics with
$E^2-\alpha^2L^2< 0$ are bounded between the singularity $r_{\rm
tp}^1=0$ and a maximum ($r_{\rm tp}^2\geq r_+$) radial distance.
The turning point $r_{\rm tp}^2$ is exactly at the horizon $r_+$
if and only if the energy and the angular momentum are such that
$c_1=0$. (iii) Null geodesics with energy and angular momentum
such that $c_0=0$ can reach and ``stay'' at the curvature
singularity $r=0$. (iv) All the timelike geodesics present the
same features as the null geodesics with $E^2-\alpha^2L^2< 0$. So,
any timelike geodesic is bounded  within the region $r_{\rm
tp}^1\leq r \leq r_{\rm tp}^2$ (with $r_{\rm tp}^1=0$ and $r_{\rm
tp}^2\geq r_+$), and no timelike particle can either escape to
infinity or reach and ``stay'' at $r=0$. (v) Neither null or
timelike geodesics have stable or unstable circular orbits.

\vskip 2mm {\bf B. Brans-Dicke theory with $\bm{\omega=-2}$}
\vskip 2mm

For $\omega=-2$, $\Delta$ is given by (\ref{DELTA-2}) and, in the
case $M>0$, $\Delta$ has one zero given by
\begin{equation}
r_+=b {\biggl [} 4\chi^2{\rm ProdLog}{\biggl (}
\frac{be^{\frac{1}{4\chi^2}}}{4\chi^2}{\biggr )}{\biggr ]}^{-1},
                                 \label{ZERO_(-2)}  \\
\end{equation}
with ${\rm ProdLog}(x)=z$ being such that $x=ze^z$. The scalar
$R_{\mu\nu}R^{\mu\nu}$ (\ref{R-2}) diverge at $r=0$ and
$r=+\infty$. For $|J|>|M|$, $\Delta$ is positive for $r>r_+$  and
negative for $r<r_+$. At $r=+\infty$ the curvature singularity is
timelike while $r=0$ is a null curvature singularity.  The Penrose
diagram is represented in figure 3.(a). For $|J|<|M|$, unlike the
uncharged case, the theory also has a black hole with a horizon
located at $r_+$ given by (\ref{ZERO_(-2)}). $\Delta$ is negative
for $r>r_+$  and positive for $r<r_+$. The $r=+\infty$ curvature
singularity is spacelike and $r=0$ is a null curvature
singularity.  The Penrose diagram is drawn in figure 3.(b). For
$|J|=|M|$ one has $b= \chi=0$. Then, $r=0$ is a naked null
singularity and the boundary $r=\infty$ changes character and has
no singularity, being a timelike line in the Penrose diagram drawn
in figure 3.(c).

\begin{figure} [t]
\centering
\includegraphics[height=1.1in]{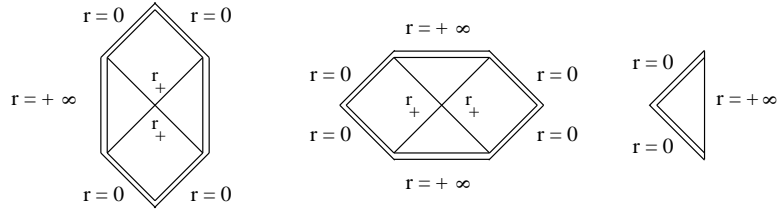}
\caption{\label{eBD-fig3}
 (a) Penrose diagram for the black hole
of: i) $\omega=-2$, $M>0$, $|J|>M$; ii) $\omega=-9/5$, $M<0$.
 (b) Penrose diagram for the  $\omega=-2$, $M>0$, $|J|<M$ naked
singularity. (c) Penrose diagram for the  $\omega=-2$, $M>0$,
$|J|=M$ naked singularity.
 }
\end{figure}

Now, we study the geodesic motion for $M>0$. The behavior of
geodesic motion along the radial coordinate can be obtained from
the following two equations
\begin{eqnarray}
r^2 \dot{r}^2 &=& - {\bigl [} r^2\delta +c_0^2 {\bigr ]} \Delta
+c_1^2r^2
                                \label{GEOD_(-2)}  \\
r^2\dot{r}^2 &=& (E^2-L^2)r^2-\frac{\chi^2c_0^2}{4}r^2 \ln r
 +\frac{J^2-M^2}{M} c_0^2 r
   -r^2 \delta \Delta\:,
                                        \label{GEOD_(-2)_2}
\end{eqnarray}
where
\begin{eqnarray}
c_0^2 &=& {\biggl [}{\biggl (}\frac{J^2}{M^2}-1 {\biggl )}^{-1/2}E
          - {\biggl (}1-\frac{M^2}{J^2} {\biggl )}^{-1/2} L
{\biggr ]}^2 \:,                           \label{GEOD_(-2)_c0}  \\
c_1^2 &=& {\biggl [}{\biggl (}1-\frac{M^2}{J^2} {\biggl )}^{-1/2}E
          - {\biggl (}\frac{J^2}{M^2}-1 {\biggl )}^{-1/2} L
{\biggr ]}^2 \:.
                                        \label{GEOD_(-2)_2_c1}
\end{eqnarray}
We first consider the case $|J|>M$. From equation
(\ref{GEOD_(-2)}) we conclude that whenever there are turning
points, they are given by $r_{\rm tp}^1=0$ and $r_{\rm tp}^2\geq
r_+$. The turning point $r_{\rm tp}^2$ is exactly at the horizon
$r_+$ if and only if the energy and the angular momentum are such
that $c_1=0$, i.e,  $E=M L/J$.
 From the graphic computation of (\ref{GEOD_(-2)_2}) we
conclude that: (i) the only particles that can escape to
$r=+\infty$ or $r=0$ are null particles that satisfy $c_0=0$ which
implies $E=J L/M$; (ii) all other null geodesics and all timelike
geodesics are bounded between $r_{\rm tp}^1=0$ and a maximum
($r_{\rm tp}^2\geq r_+$) radial distance.

 For $|J|<M$ one has that: (i) timelike and null spiraling
particles with  $E \neq J L/M$ start at $r_{\rm tp}\geq r_+$ and
reach infinity radial distances or timelike and null geodesics
start at $r=+\infty$ and spiral toward $r_{\rm tp}$ and then
return back to infinity; (ii)  null particles can escape to
$r=+\infty$ or $r=0$ if $c_0=0$, i.e., $E=J L/M$; (iii) timelike
geodesics with $E=J L/M$ can be bounded  between $r_{\rm tp}^1=0$
and a maximum ($r_{\rm tp}^2\geq r_+$) radial distance, or
 start at $r_{\rm tp}^3 \geq r_+$ and reach infinity radial
distances.

\vskip 2mm {\bf C. Brans-Dicke theories with
$\bm{-2<\omega<-3/2}$} \vskip 2mm

 In this range of $\omega$,
spacetime has no black holes since from (\ref{b}) and (\ref{c^2})
one has $b<0$ and $k \chi^2>0$ so the $\Delta$ function
(\ref{DELTA_TODOS}) is always positive. The curvature singularity
$r=0$ is a naked null singularity and the curvature singularity
$r=+\infty$ is a naked timelike singularity. The Penrose diagram
is drawn in figure 4.

When we consider the solutions with negative mass  we conclude
that one might have black holes with one horizon. If this is the
case, the Penrose diagram is exactly equal to the one shown in
figure 3.(a) (which represents the typical case $\omega=-9/5$).

\begin{figure} [t]
\centering
\includegraphics[height=0.7in]{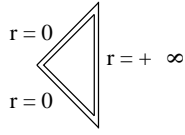}
\caption{\label{eBD-fig4}
 Penrose diagram for the spacetime of: i)
$\omega=-2$, $M>0$, $|J|=M$; ii) $-2<\omega<-3/2$, $M>0$; iii)
$\omega=-3/2$ (large $\Lambda/\chi^2$); iv) $\omega=-4/3$, $M<0$,
$|\alpha J|=M$ (the only difference is that $r=0$ is a topological
singularity rather than a curvature singularity).
 }
\end{figure}

Although there are no black holes with positive mass, it is
interesting to study the geodesics that we might expect for this
spacetime. From graphic computation of (\ref{GEOD_B_2}) we
conclude the following for null and timelike geodesics. (i) For
$E^2-\alpha^2L^2 \leq 0$ there is no possible motion. (ii) This
situation also occurs for small positive values of
$E^2-\alpha^2L^2$. (iii) However, when we increase the positive
value of $E^2-\alpha^2L^2$ there is a critical value for which a
stable circular orbit is allowed. (iii) And for positive values of
$E^2-\alpha^2L^2$ above the critical value, null and timelike
geodesics are bounded between a minimum ($r_{\rm tp}^1$) and a
maximum ($r_{\rm tp}^2$) radial distance. (iv) Null particles with
energy and angular momentum satisfying $c_0=0$ that are produced
at $r=0$ escape to $r=+\infty$ and null particles coming in from
infinity are scattered at $r_{\rm tp}=0$ and spiral back to
infinity. (vi) Timelike particles with energy and angular momentum
satisfying $c_0=0$ are bounded  within the region $0\leq r \leq
r_{\rm tp}$, with $r_{\rm tp}$ finite.

\vskip 2mm {\bf D. Brans-Dicke theory with $\bm{\omega=-3/2}$}
\vskip 2mm

In this case, $\Delta$ is given by (\ref{DELTA-3/2}). Depending on
the value of $\frac{\Lambda}{\chi^2}$ one has black holes with two
horizons (small $\frac{\Lambda}{\chi^2}$), with one (extreme case)
or a spacetime without black holes (large
$\frac{\Lambda}{\chi^2}$).

For the black hole with two horizons, we have $\Delta>0$ in
$r<r_-$ and $r>r_+$. For the patch $r_-<r<\infty$, the curvature
singularity $r=\infty$ is mapped into two symmetrical timelike
lines and the horizon $r=r_+$ is mapped into two mutual
perpendicular straight lines at $45^{\rm o}$. For the patch
$0<r<r_+$, the curvature singularity $r=0$ is mapped into a pair
of two null lines and the horizon $r=r_-$ is mapped into two
mutual perpendicular straight lines at $45^{\rm o}$. One has to
join these two different patches and then repeat them over in the
vertical. The resulting Penrose diagram is shown in figure 5.(a).
For the extreme black hole the two curvature singularities $r=0$
and $r=\infty$ are still null and timelike lines (respectively),
but the event and inner horizon join together in a single horizon
$r_+$. The Penrose diagram is like the one drawn in figure 5.(b).
The spacetime with no black hole present has a Penrose diagram
like the one represented in figure 4.
\begin{figure} [t]
\centering
\includegraphics[height=1.7in]{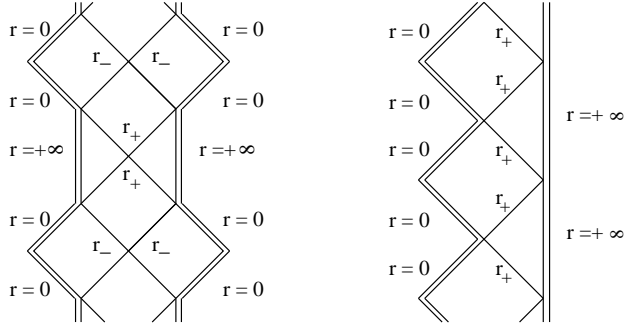}
\caption{\label{eBD-fig5}
 (a) Penrose diagram for the black hole
with two horizons of: i) the $\omega=-3/2$, (small
$\Lambda/\chi^2$); ii) $\omega=-4/3$, $M<0$, $|\alpha J|< |M|$
(the only difference is that $r=0$ is a topological singularity
rather than a curvature singularity).
 (b) Penrose diagram for the extreme black hole of: i)
$\omega=-3/2$; ii) $\omega=-4/3$, $M<0$, $|\alpha J|< |M|$ (the
only difference is that $r=0$ is a topological singularity rather
than a curvature singularity).
 }
\end{figure}
Now, we study the geodesic motion. The behavior of geodesic motion
along the radial coordinate can be obtained from the following two
equations
\begin{eqnarray}
r^2 \dot{r}^2 &=& - {\bigl [}r^2 \delta +c_0^2 {\bigr ]} \Delta
+c_0^2 r^2
                                                 \label{GEOD_(-3/2)}  \\
\dot{r}^2 &=& c_0^2 {\bigl [}\Lambda \ln(br)-\chi^2r^2+1{\bigr ]}
   - \delta \Delta\:,
                                        \label{GEOD_(-3/2)_2}
\end{eqnarray}
where $c_0=(\theta E-\gamma L)$. From equation (\ref{GEOD_(-3/2)})
we conclude that whenever there are turning points, they are given
by $r_{\rm tp}^1 \leq r_-$ and $r_{\rm tp}^2 \geq r_+$. The
turning points coincide with the horizons if and only if the
energy and the angular momentum are such that $c_0=0$. From the
graphic computation of (\ref{GEOD_(-3/2)_2}) we conclude that: (i)
Whenever $c_0=0$ null particles describe stable circular orbits
wherever they are located; (ii) Null geodesics with $c_0 \neq 0$
and all timelike geodesics describe a bound orbit between $r_{\rm
tp}^1$ and $r_{\rm tp}^2$.

For the extreme black hole the two horizons coincide so all null
geodesics  and all timelike geodesics describe a stable circular
orbit.

\vskip 2mm {\bf E. Brans-Dicke theories with
$\bm{-3/2<\omega<-1}$} \vskip 2mm

 For this range of $\omega$ there is the possibility of having
black holes with positive mass whenever $|\alpha J|>M$ (Table 1).
We are going to analyze the typical case $\omega=-4/3$. The
$\Delta$ function, (\ref{DELTA_TODOS}), is
\begin{equation}
\Delta =(\alpha r)^2 -b(\alpha r)^3+c(\alpha r)^6\:,
                                 \label{DEL_(-3/2)_(-1)}  \\
\end{equation}
where from (\ref{b}) and (\ref{c^2}) one has that $b>0$ and
$c\equiv k\chi^2<0$ if $M>0$ and $|\alpha J|>M$. $\Delta$ is
negative for $r>r_+$ and positive for $r<r_+$, where $r_+$ is the
only zero of $\Delta$ given by
\begin{eqnarray}
& & r_+=\frac{1}{2\alpha} {\biggl (}\frac{b}{c} {\biggl )}^{1/3}
{\biggl [}\sqrt{s} -\sqrt{\frac{2}{\sqrt{s}}-s}{\biggr ]}\:,
\:\:\:\:\:{\rm where}  \nonumber \\
& & s={\biggl [}\frac{1}{2}+\frac{1}{2}\sqrt{1-\frac{4^4 c} {3^3
b^4} }{\biggr ]}^{\frac{1}{3}}+{\biggl [}\frac{1}{2}-
\frac{1}{2}\sqrt{1-\frac{4^4 c}{3^3 b^4} }{\biggr
]}^{\frac{1}{3}}\:. \label{Zero_(-4/3)}
\end{eqnarray}

The physical curvature singularity is located inside the horizon
at $r=+\infty$ and is a spacelike line in the Penrose diagram. At
$r=0$ there is a null topological singularity.  The Penrose
diagram is sketched in figure 6.

\begin{figure} [t]
\centering
\includegraphics[height=0.9in]{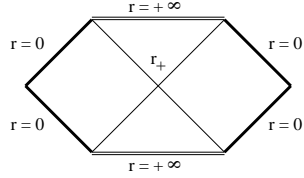}
\caption{\label{eBD-fig6}
 Penrose diagram for the black
hole of: i)  $\omega=-4/3$, $M>0$, $|\alpha J|> M$; ii)
$\omega=-4/3$, $M<0$, $|\alpha J|> |M|$.
 }
\end{figure}

When we consider the solutions with negative mass  we conclude
that, for $|\alpha J|>|M|$, one has black holes with a Penrose
diagram equal to the one drawn in figure 6. For $|\alpha J|<|M|$
one has black holes with two horizons, with one (extreme case) or
a spacetime without black holes. For the black hole with two
horizons the Penrose diagram is similar to the one shown in figure
5.(a)  and the extreme black hole has a Penrose diagram which is
similar to the one drawn in figure 5.(b). For $|\alpha J|=|M|$ the
Penrose diagram is similar to figure 4. The only difference is the
fact that $r=0$ is now a topological singularity rather than a
curvature singularity.

Let us now consider the geodesic motion for positive mass.
Analyzing (\ref{GEOD_B}) and noting that from (\ref{OMEGA}) the
condition $|\alpha J|>M$ implies $\Omega<-2$ we conclude that, for
null geodesics, whenever there are turning points, they are
$r_{\rm tp}^1=0$ and $r_{\rm tp}^2\geq r_+$. From (\ref{GEOD_B_2})
we conclude the following about the null geodesic motion. (i) For
$E^2-\alpha^2L^2 \leq 0$ there is no possible motion. (ii)  Null
geodesics with $E^2-\alpha^2L^2> 0$ are bounded between the
singularity $r_{\rm tp}^1=0$ and a maximum ($r_{\rm tp}^2\geq
r_+$) radial distance. The turning point $r_{\rm tp}^2$ is exactly
at the horizon $r_+$ if and only if the energy and the angular
momentum are such that $c_1=0$.

The timelike geodesic motion is radically different. (i) For
$E^2-\alpha^2L^2 \leq 0$ we have timelike spiraling particles that
start at $r_{\rm tp}$ and reach infinity radial distances or
timelike geodesics that start at $r=+\infty$ and spiral toward
$r_{\rm tp}$ and then return back to infinity. (ii) For small
positive values of $E^2-\alpha^2L^2$, timelike particles that are
produced at $r=0$ escape to $r=+\infty$ and timelike particles
coming in from infinity are scattered at $r_{\rm tp}=0$ and spiral
back to infinity. (iii) When we increase the positive value of
$E^2-\alpha^2L^2$ there is a critical value for which an unstable
circular orbit is allowed. (iv) And for positive values of
$E^2-\alpha^2L^2$ above the critical value,  timelike particles
are allowed to be bounded between $r_{\rm tp}^1=0$ and a maximum
($r_{\rm tp}^2$) radial distance or to start at $r_{\rm
tp}^3>r_{\rm tp}^2$ and escape to infinity.

\vskip 2mm {\bf F. Brans-Dicke theories with $\bm{\omega>-1}$}
\vskip 2mm

The range $\omega>-1$ is not discussed here since the properties
of the typical case $\omega=0$ have been presented in
\cite{Zanchin_Lemos}, where the three-dimensional gravity theory
of $\omega=0$ was obtained through dimensional reduction from
four-dimensional General Relativity with one Killing vector field.

\vskip 2mm {\bf G. Brans-Dicke theory with $\bm{\omega=\pm
\infty}$} \vskip 2mm

The case $\omega=\pm \infty$ is also not  discussed here since
this case reduces to the electrically charged BTZ black hole which
has been studied in detail in \cite{BTZ_Q,PY}.

\subsection{Hawking temperature of the charged black holes}
 \label{sec:Haw Temp 3D}

To compute the Hawking temperature of the rotating black holes,
one starts by writing the metric in the canonical form
(\ref{MET_CANON}). To proceed it is necessary to first perform the
coordinate transformation to coordinates $t$, $\tilde{\varphi}$
which corotate with the black hole. In other words, the angular
coordinate $\varphi$ must be changed to  $\tilde{\varphi}$ defined
by $\tilde{\varphi}=\varphi-\Omega_H t$, where
$\Omega_H=-\frac{g_{t \varphi}}{g_{\varphi \varphi}} {\bigl
|}_{r=r_+}=-N^{\varphi}(r_+)$ is the angular velocity of the black
hole. With this transformation the metric (\ref{MET_CANON})
becomes
\begin{equation}
     ds^2 = - (N^0)^2 dt^2
            + \frac{dr^2}{f^2}
            + H^2{\biggl [}d\tilde{\varphi}+{\biggl (}N^{\varphi}(r)
             -N^{\varphi}(r_+){\biggr )}dt{\biggr ]}^2 \:,
                               \label{MET_CANON_TEMP}
\end{equation}
Then, one applies the Wick rotation $t \rightarrow -i \tau$ in
order to obtain the euclidean counterpart of
(\ref{MET_CANON_TEMP}). Now, one studies the behavior of the
euclidean metric in the vicinity of the event horizon, $r_+$. In
this vicinity, one can write $N^{\varphi}(r)-N^{\varphi}(r_+) \sim
0$ and take the expansion $\Delta(r) \sim \frac{d\Delta(r_+)}{d
r}(r-r_+)+\cdots\,$. One proceeds applying the variable change
$\frac{1}{\Delta(r)}dr^2=d \rho^2$ so that  one has
$\rho=2[\sqrt{\frac{r-r_+}{d \Delta(r_+)/d r}}$
 and $\Delta(\rho) \sim [d \Delta(r_+)/dr]^2 \rho^2 /4$. With this
procedure the euclidean metric in the vicinity of the event
horizon can be cast in the form $ds^2 \sim (2 \pi / \beta_H)^2
\rho^2 d \tau^2 +d \rho^2+ H^2 d\tilde{\varphi}^2$. Applying a
final variable change, $\bar{\tau}=(2 \pi / \beta_H) \tau$, the
metric becomes $ds^2 \sim \rho^2 d \bar{\tau}^2+d \rho^2+H^2
d\tilde{\varphi}^2$. To avoid the canonical singularity at the
event horizon one must demand that the period of $\bar{\tau}$ is
$2 \pi$ which implies $0 \leq \tau \leq \beta_H$. Finally, the
Hawking temperature is defined as $T_H=(\beta_H)^{-1}$.

Applying the above procedure, one finds for the Hawking
temperature of the rotating black holes the following expressions
\begin{eqnarray}
T_H &=& \frac{1}{4 \pi}\sqrt{\frac{r_+^2 {\biggl [} 2 \alpha r_+
+\frac{b}{\omega+1}(\alpha r_+)^ {-\frac{\omega+2}{\omega+1}}
-\frac{2k \chi^2}{\omega+1}(\alpha r_+)^{-\frac{\omega+3}
{\omega+1}}{\biggr ]}^2}{r_+^2-\frac{\theta^2}{\alpha^4} {\biggl
[}-b (\alpha r_+)^{-\frac{1}{\omega+1}}+ k \chi^2(\alpha
r_+)^{-\frac{2}{\omega+1}} {\biggr ]}}}  \:,
  \nonumber \\
         & &     \hskip 5cm {\rm for} \:\:\:\:\omega \neq -2,-\frac32,-1 \:,
                                 \label{TEMP_TODOS}  \\
T_H &=& \frac{1}{4 \pi}\sqrt{\frac{r_+^2 {\biggl [}
\frac{r_+}{2}{\biggl (} 4+\frac{\chi^2}{2}+ \chi^2\ln r_ +{\biggl
)}-b{\biggr ]}^2}{{\biggl [}1-\frac{1}{4}\theta^2 \chi^2\ln
r_+{\biggr ]}r_+^2+\theta^2 b r_+}}  \:,
                           \:\:\:\:\:\:\:   {\rm for} \:\:\:\:\omega=-2   \:,
                                 \label{TEMP_(-2)} \\
T_H &=& \frac{1}{4 \pi}\sqrt{\frac{r_+^2 {\biggl [}  2\Lambda r_+
\ln(b r_+)+ \Lambda b r_+ -\frac{\chi^2r_+^3}{4}{\biggr ]}^2} {
\Lambda \theta^2 r_+^2 \ln(b r_+)+ \gamma^2 r_+^2
-\frac{\theta^2 \chi^2r_+^4}{16}}}  \:,   \nonumber \\
         & &     \hskip 5cm {\rm for} \:\:\:\:\omega=-3/2 \:,
                                 \label{TEMP_(-3/2)}
\end{eqnarray}
where in (\ref{TEMP_TODOS}), $b$, $k \chi^2$ and $\theta^2$ are
given by (\ref{b}), (\ref{c^2}) and (\ref{DUAS}), respectively.

The Hawking temperature of the static charged black holes can be
obtained from (\ref{TEMP_TODOS})-(\ref{TEMP_(-3/2)}) by taking
$\gamma=1$ and $\theta=0$ [see (\ref{TRANSF_J})] (see also
\cite{Lemos} for the $\omega=0$ uncharged black hole).

\section{Magnetic Brans-Dicke dilaton solutions}
 \label{sec:Magnetic BH 3D}

The issue of spacetimes generated by point sources in three
dimensional (3D) Einstein theory has been object of many studies,
as we reviewed in section \ref{sec:3D BHs introduction}.

The aim of this section is to find and study in detail the static
and rotating magnetic charged solutions generated by a magnetic
point source in the Einstein-Maxwell-Dilaton action of the
Brans-Dicke that we are studying in this chapter.  Here, we impose
that the only non-vanishing component of the vector potential is
$A_{\varphi}(r)$. For the $\omega=\pm \infty$ case, our solution
reduces to the spacetime generated by a magnetic point source in
3D Einstein-Maxwell theory with $\Lambda<0$, studied in
\cite{CL1}, \cite{HW}-\cite{OscarLemos_BTZ}. The $\omega=0$ case
is equivalent to 4D general relativity with one Killing vector
analysed by Dias and Lemos \cite{OscarLemos_string}.

The plan of this section is the following. In Section
\ref{sec:field eq Mag 3D} we set up the action and the field
equations. The static general solutions of the field equations are
found in section \ref{sec:static sol Mag 3D} and  we analyse in
detail the general structure of the solutions. The angular
momentum is added in section \ref{sec:rotat sol Mag 3D}  and we
calculate the mass, angular momentum, and electric charge of the
solutions. In section \ref{sec:phys interp Mag 3D} we give a
physical interpretation for the origin of the magnetic field
source.

\subsection{Field equations of Brans-Dicke$-$Maxwell theory}
\label{sec:field eq Mag 3D}

We are going to work with an action of the Brans-Dicke$-$Maxwell
type in 3D written in the string frame as
\begin{equation}
S=\frac{1}{2\pi} \int d^3x \sqrt{-g} e^{-2\phi}
  {\bigl [} R - 4 \omega ( \partial \phi)^2
           + \Lambda + F^{\mu \nu}F_{\mu \nu}
  {\bigr ]},                                   \label{ACCAO MagBD}
\end{equation}
where $g$ is the determinant of the 3D metric, $R$ is the
curvature scalar, $F_{\mu\nu} = \partial_\nu A_\mu - \partial_\mu
A_\nu$ is the Maxwell tensor, with $A_\mu$ being the vector
potential, $\phi$ is a scalar field called dilaton,
 $\omega$ is the 3D Brans-Dicke parameter and
$\Lambda$ is the cosmological constant. Varying this action with
respect to $g^{\mu \nu}$, $A_{\mu}$ and $\phi$ one gets the
Einstein, Maxwell and dilaton equations, respectively
\begin{eqnarray}
   & \frac{1}{2} G_{\mu \nu}
       -2(\omega+1) \nabla_{\mu}\phi \nabla_{\nu}\phi
       +\nabla_{\mu} \nabla_{\nu}\phi
       -g_{\mu \nu} \nabla_{\gamma} \nabla^{\gamma}\phi
       +(\omega+2) g_{\mu \nu} \nabla_{\gamma}\phi
               \nabla^{\gamma}\phi-\frac{1}{4}g_{\mu \nu}\Lambda=
              \frac{\pi}{2} T_{\mu \nu}, & \nonumber \\
      & &                           \label{EQUACAO_MET MagBD} \\
   & \nabla_{\nu}(e^{-2 \phi}F^{\mu \nu})=0 \:, &
                                     \label{EQUACAO_MAX MagBD} \\
   & R -4\omega \nabla_{\gamma} \nabla^{\gamma}\phi
          +4\omega \nabla_{\gamma}\phi \nabla^{\gamma}\phi
          +\Lambda =-F^{\gamma \sigma}F_{\gamma \sigma}, &
                                         \label{EQUACAO_DIL MagBD}
\end{eqnarray}
where $G_{\mu \nu}=R_{\mu \nu} -\frac{1}{2} g_{\mu \nu} R$ is the
 Einstein tensor, $\nabla$ represents the covariant derivative and
$T_{\mu \nu}=\frac{2}{\pi}(g^{\gamma \sigma}F_{\mu \gamma} F_{\nu
\sigma}-\frac{1}{4}g_{\mu \nu}F_{\gamma \sigma} F^{\gamma
\sigma})$ is the Maxwell energy-momentum tensor.

\subsection{General static solution. Analysis of its
structure} \label{sec:static sol Mag 3D}

\subsubsection{Field equations}

We want to consider now a spacetime which is both static and
rotationally symmetric, implying the existence of a timelike
Killing vector $\partial/\partial t$ and a spacelike Killing
vector $\partial/\partial\varphi$. We will start working with the
following ansatz for the metric
\begin{equation}
  ds^2 = -\alpha^2 r^2 dt^2 + e^{-2\mu(r)}dr^2
         +\frac{e^{2\mu(r)}}{\alpha^2} d\varphi^2 \:,
                               \label{MET_SYM MagBD}
\end{equation}
where the parameter $\alpha^2$ is, as we shall see, an appropriate
constant proportional to the cosmological constant $\Lambda$. It
is introduced in order to have metric components with
dimensionless units and an asymptotically anti-de Sitter
spacetime. The motivation for this choice for the metric gauge
[$g_{tt} \propto- r^2$ and $(g_{rr})^{-1} \propto g_{\varphi
\varphi}$] instead of the usual ``Schwarzschild'' gauge
[$(g_{rr})^{-1} =-g_{tt}$ and  $g_{\varphi \varphi}=r^2$] comes
from the fact that we are looking for magnetic solutions. Indeed,
let us first remember that the Schwarzschild gauge is usually an
appropriate choice when we are interested on electric solutions.
Now, we focus on the well known fact that the electric field is
associated with the time component, $A_t$, of the vector potential
while the magnetic field is associated with the angular component
$A_{\varphi}$. From the above facts, one can expect that a
magnetic solution can be written in a metric gauge in which the
components $g_{tt}$ and $g_{\varphi \varphi}$ interchange their
roles relatively to those present in the ``Schwarzschild''
 gauge used to describe electric solutions.
This choice will reveal to be a good one to find solutions since
the dilaton and graviton will decouple from each other on the
fields equations (\ref{EQUACAO_MET MagBD}) and
 (\ref{EQUACAO_DIL MagBD}).
 However, as we will see, for some values of the Brans-Dicke parameter
$\omega$ it is not the good coordinate system to interpret the
solutions.

We now assume that the only non-vanishing components of the vector
potential are $A_t(r)$ and $A_{\varphi}(r)$, i.e.,
\begin{equation}
A=A_tdt+A_{\varphi}d{\varphi}\:. \label{Potential MagBD}
\end{equation}
This implies that the non-vanishing components of the
anti-symmetric Maxwell tensor are $F_{tr}$ and $F_{r \varphi}$.
Use of metric (\ref{MET_SYM MagBD}) and equation
 (\ref{EQUACAO_MET MagBD}) yields the following set of equations
\begin{eqnarray}
 & & \hskip -0.4cm  \phi_{,rr}
       + 2 \phi_{,r} \mu_{,r}
       - (\omega+2) (\phi_{,r})^2
       - \frac{\mu_{,rr}}{2}
       - (\mu_{,r})^2+ \frac{1}{4} \Lambda e^{-2\mu}
       =\frac{\pi}{2 \alpha^2} \frac{e^{-2\mu}}{r^2} T_{tt} \:,
                                    \label{MET_00 MagBD}  \\
  & &  \hskip -0.4cm
       - \phi_{,r} \mu_{,r}
       - \frac{\phi_{,r}}{r}
       - \omega (\phi_{,r})^2
       + \frac{\mu_{,r}}{2r}
       - \frac{1}{4} \Lambda e^{-2\mu}=
                  \frac{\pi}{2} T_{rr} \:,
                                    \label{MET_11 MagBD}  \\
  & &   \hskip -0.4cm
       \phi_{,rr}
      + \phi_{,r} \mu_{,r}
      + \frac{\phi_{,r}}{r}
      - (\omega+2) (\phi_{,r})^2
      - \frac{\mu_{,r}}{2r}
      + \frac{1}{4} \Lambda e^{-2\mu}
     =-\frac{\pi}{2} \alpha^2 e^{-4\mu} T_{\varphi \varphi} \:,
                                           \label{MET_22 MagBD}  \\
  & &  \hskip -0.4cm
         0=\frac{\pi}{2} T_{t \varphi}=
                           e^{2\mu}F_{tr}F_{\varphi r} \:,
                                           \label{MET_02 MagBD}
\end{eqnarray}
where ${}_{,r}$ denotes a derivative with respect to $r$. In
addition, replacing the metric (\ref{MET_SYM MagBD}) into
equations (\ref{EQUACAO_MAX MagBD}) and (\ref{EQUACAO_DIL MagBD})
yields
\begin{eqnarray}
 & &  \partial_r {\bigl [} e^{-2 \phi}r(F^{t r}
      +F^{\varphi r}) {\bigr ]}=0\:,
                                 \label{MAX_0 MagBD} \\
& &  \omega \phi_{,rr}
      + 2 \omega \phi_{,r} \mu_{,r}
      + \omega \frac{\phi_{,r}}{r}
      - \omega (\phi_{,r})^2
      + \frac{\mu_{,r}}{r}
      + \frac{\mu_{,rr}}{2}
    + (\mu_{,r})^2
      -\frac{1}{4} \Lambda e^{-2\mu}=
       \frac{1}{4}e^{-2\mu}F^{\gamma \sigma}F_{\gamma \sigma} \:.
                                    \label{EQ_DIL MagBD}
\end{eqnarray}

\subsubsection{\label{sec:stat_sol}The general static solution.
Causal structure}

Equations (\ref{MET_00 MagBD})-(\ref{EQ_DIL MagBD}) are  valid for
a static and rotationally symmetric spacetime. One sees that
equation (\ref{MET_02 MagBD}) implies that the electric and
magnetic fields cannot be simultaneously  non-zero, i.e., there is
no static dyonic solution. In this work we will consider the
magnetically charged case alone ($A_t=0,\,A_{\varphi} \neq 0$).
For  purely electrically charged solutions of the theory see
\cite{OscarLemos}. So, assuming vanishing electric field, one has
from Maxwell equation (\ref{MAX_0 MagBD}) that
\begin{equation}
F_{\varphi r}=\frac{\chi_{\rm m}}{4 \alpha^2 r}e^{2 \phi},
\label{MAX_1 MagBD}
\end{equation}
where $\chi_{\rm m}$ is an integration constant which measures the
intensity of the magnetic field source. To proceed we shall first
consider the case $\omega\neq -1$. Adding equations
 (\ref{MET_11 MagBD}) and (\ref{MET_22 MagBD}) yields
$\phi_{,rr}=2(\omega+1)(\phi_{,r})^2$, and so the dilaton field is
given by
\begin{equation}
  e^{-2\phi}= (\alpha r)^{\frac{1}{\omega+1}},
                       \quad\quad \omega \neq -1 \:,
                                    \label{DILATAO MagBD}
\end{equation}
where $\alpha$ is a generic constant which will be appropriately
defined below in  equation (\ref{COSMOL MagBD}).
 The 1-form vector potential
$A=A_{\mu}(r)dx^{\mu}$  is then
\begin{equation}
A=-\frac{1}{4 \alpha^2}\chi_{\rm m}(\omega+1) (\alpha
r)^{-\frac{1}{\omega+1}} d \varphi \:,
                           \quad\quad \omega \neq -1\:.
                                    \label{VEC_POTENT MagBD}
\end{equation}
Replacing solutions (\ref{MAX_1 MagBD})-(\ref{VEC_POTENT MagBD})
into equations (\ref{MET_00 MagBD})-(\ref{EQ_DIL MagBD}) allows us
to find the $e^{2\mu(r)}$ function for $\omega\neq\{-2,-3/2,-1\}$;
$\omega =-2$ and $\omega =-3/2$, respectively
\begin{eqnarray}
e^{2\mu(r)} &=&
            (\alpha r)^2 +
           \frac{b}{(\alpha r)^{\frac{1}{\omega+1}}}
           -\frac{k\chi_{\rm m}^2}
            {(\alpha r)^{\frac{2}{\omega+1}}}\:,
                                 \label{MET_TODOS MagBD}  \\
 e^{2\mu(r)} &=&  {\biggl (}1-\frac{\chi_{\rm m}^2}{4}
           \ln{r}{\biggl )}r^2 -br \:,
                                         \label{MET-2 MagBD}  \\
 e^{2\mu(r)} &=&  r^2[-\Lambda \ln (br)-\chi_{\rm m}^2 r^2] \:,
                                         \label{MET-3/2 MagBD}
\end{eqnarray}
where $b$ is a constant of integration related with the mass of
the solutions, as will be shown, and
$k=\frac{(\omega+1)^2}{8\alpha^2(\omega+2)}$. For $\omega \neq
-2,-3/2,-1 \:$ $\alpha$ is defined as (we call your attention to a
typo in \cite{OscarLemos-MagBD3D} in this definition)
\begin{equation}
 \alpha =\sqrt{ {\biggl |} \frac{(\omega+1)^2\Lambda}
     {(\omega+2)(2\omega+3)} {\biggr |}}.
                                 \label{COSMOL MagBD}
\end{equation}
For $\omega=-2,-3/2$ we set $\alpha=1$. For $\omega=-2$ equations
(\ref{MET_00 MagBD}) and (\ref{MET_11 MagBD}) imply
$\Lambda=-\chi_{\rm m}^2/8$.

Now, we consider the case $\omega=-1$. From equations
 (\ref{MET_00 MagBD})-(\ref{EQ_DIL MagBD})
 it follows that $\mu=C_1$,
$\phi=C_2$, where $C_1$ and $C_2$ are constants of integration,
and that the cosmological constant and magnetic source are both
zero, $\Lambda=0=\chi_{\rm m}$. So, for $\omega=-1$ the metric
gives simply the 3D Minkowski spacetime and the dilaton is
constant, as occurred in the uncharged case
\cite{Sa_Lemos_Static,Sa_Lemos_Rotat}.

Now, we must analyze carefully the radial dependence of the
 $e^{2\mu(r)}$ function defined in equations
(\ref{MET_TODOS MagBD})-(\ref{MET-3/2 MagBD}) which, recall, is
related to the metric components through the relations
$g_{rr}=e^{-2\mu}$ and $g_{\varphi \varphi}=e^{2\mu}/\alpha^2$.
The shape of the $e^{2\mu(r)}$ function depends on the values of
the Brans-Dicke parameter $\omega$ and on the values of the
parameter $b$ (where $b$ is related to the mass of the solution as
we shall see in section \ref{sec:MJQ}). Nevertheless, we can group
the values of $\omega$ and $b$ into a small number of cases for
which the $e^{2\mu(r)}$ function
 has the same behavior. The general shape of the $e^{2\mu(r)}$ function
for these cases is drawn in the Appendix. Generally, the
$e^{2\mu(r)}$ function can take positive or negative values
depending on the value of the coordinate $r$. However, when
$e^{2\mu(r)}$ is negative, the metric components
$g_{rr}=e^{-2\mu}$ and $g_{\varphi \varphi}=e^{2\mu}/\alpha^2$
become simultaneously negative  and this leads to an apparent
change of signature from $+1$ to $-3$. This strongly indicates
that we are using an incorrect extension and that we should choose
a different continuation to describe the region where the change
of signature occurs \cite{HW,Hor_Hor}. Moreover, analysing the
null and timelike geodesic motion we conclude that null and
timelike particles can never pass through the
 zero of $e^{2\mu(r)}$, $r_+$ (say), from the region where $g_{rr}$ is
positive into the region where $g_{rr}$ is negative. This suggests
that one can introduce a new coordinate system in order to obtain
a  spacetime which is geodesically complete for the region where
both $g_{rr}$ and $g_{\varphi \varphi}$ are positive
\cite{HW,Hor_Hor}. That is our next step. Then, in section
\ref{sec:geod}, we will check the completeness of the spacetimes.
The Brans-Dicke theories can be classified into seven different
cases that we display and study below.

\vskip 2mm \noindent{{\bf (i) Brans-Dicke theories with
$\bm{-1<\omega<+\infty}$}} \vskip 2mm

The shape of the $e^{2\mu(r)}$ function is drawn in Fig. 1(a). We
see that for $0<r<r_+$ (where $r_+$ is the zero of $e^{2\mu(r)}$)
$g_{rr}$ and $g_{\varphi \varphi}$ become simultaneously negative
and this leads to an apparent change of signature. One can however
introduce a new radial coordinate $\rho$ so that the  spacetime is
geodesically complete for the region where both $g_{rr}$ and
$g_{\varphi \varphi}$ are positive,
\begin{eqnarray}
\rho^2=r^2-r_+^2 \:.
                                        \label{Transf_1 MagBD}
\end{eqnarray}
With this coordinate transformation, the spacetime generated by
the static magnetic point source is finally  given by
\begin{eqnarray}
 ds^2 = -\alpha^2 r^2(\rho) dt^2
     + \frac{\rho^2}{r^2(\rho)}\frac{1}{f(\rho)} d \rho^2
        + \frac{f(\rho)}{\alpha^2}
            d\varphi^2 \:,
                     \label{Met_1 MagBD}
\end{eqnarray}
where $0\leq\rho<\infty$, and function $f(\rho)$, which is always
positive except at $\rho=0$ where it is zero, is given by
\begin{eqnarray}
f(\rho)=  \alpha^2 r^2(\rho) +
      \frac{b}{[\alpha^2 r^2(\rho)]^{\frac{1}{2(\omega+1)}}}
           -\frac{k\chi_{\rm m}^2}
    {[\alpha^2 r^2(\rho)]^{\frac{1}{\omega+1}}} \:.
                            \label{f MagBD}
\end{eqnarray}
Along this section \ref{sec:stat_sol}, in cases {\bf(i)} and
{\bf(iii)} we will use the definition $r^2(\rho) \equiv \rho^2 +
r_+^2$ in order to shorten the formulas. This spacetime has no
horizons and so there are no magnetic black hole solutions, only
magnetic point sources. In three dimensions, the presence of a
curvature singularity is revealed by the scalar
$R_{\mu\nu}R^{\mu\nu}$
\begin{eqnarray}
 \hspace{-0.2cm} & & \hspace{-0.2cm}
        R_{\mu\nu}R^{\mu\nu}=
        - \frac{4 \omega}{(\omega+1)^2}
       \frac{b \alpha^4}
       {[\alpha^2 r^2(\rho)]^{\frac{2\omega+3}{2(\omega+1)}}}
    + \frac{(2\omega^2+4\omega+3)b^2\alpha^4}{2(\omega+1)^4
       [\alpha^2 r^2(\rho)]^{\frac{2\omega+3}{\omega+1}}}
                                                \nonumber \\
\hspace{-0.2cm} & &  \hspace{-0.2cm}
       +\frac{(\omega-1)}{(\omega+1)^2}
       \frac{8 k \chi_{\rm m}^2 \alpha^4}
       {[\alpha^2 r^2(\rho)]^{\frac{\omega+2}{\omega+1}}}
      -\frac{(\omega^2+2\omega+2)}{(\omega+1)^4}
       \frac{k \chi_{\rm m}^2 b \alpha^4}
       {[\alpha^2 r^2(\rho)]^{\frac{4\omega+7}{2(\omega+1)}}}
                                                  \nonumber \\
\hspace{-0.2cm} & &  \hspace{-0.2cm}
        -\frac{(\omega^2+2\omega+3)}{(\omega+1)^4}
             \frac{k^2 \chi_{\rm m}^4 \alpha^4}
      {[\alpha^2 r^2(\rho)]^{\frac{2(\omega+2)}{\omega+1}}}
       + 12\alpha^4   \:.
                             \label{R-2-3/2-1 MagBD}
\end{eqnarray}
This scalar does not diverge for any value of $\rho$ (if
$\omega>-1$). Therefore, spacetime (\ref{Met_1 MagBD}) has no
curvature singularities. However, it has a conic geometry with a
conical singularity at $\rho=0$. In fact, in the vicinity of
$\rho=0$, metric (\ref{Met_1 MagBD}) is written as
\begin{eqnarray}
 ds^2 \sim -\alpha^2  r_+^2 dt^2 + \frac{\nu}{\alpha r_+}
       d\rho^2+(\alpha r_+ \nu)^{-1} \rho^2  d\varphi^2 \:,
                                 \label{Met_1_0 MagBD}
\end{eqnarray}
with $\nu$ given by
\begin{eqnarray}
\nu=  {\biggl [}
        \alpha r_+ - \frac{b
         (\alpha r_+)^{-\frac{\omega+2}{\omega+1}}}
           {2(\omega+1)}
       +\frac{k\chi_{\rm m}^2}{\omega+1}
      (\alpha r_+)^{-\frac{\omega+3}{\omega+1}}
       {\biggr ]}^{-1}\:.
                                        \label{T_Con_1 MagBD}
\end{eqnarray}
So, there is indeed a conical singularity at $\rho=0$ since as the
radius $\rho$ tends to zero, the limit of the ratio
``circumference/radius'' is not $2\pi$. The period of  coordinate
$\varphi$ associated with this conical singularity is
\begin{eqnarray}
{\rm Period}_{\varphi}= 2 \pi {\biggl [}  \lim_{\rho \rightarrow
0} \frac{1}{\rho} \sqrt{\frac{g_{\varphi \varphi}} {g_{\rho
\rho}}} {\biggr ]}^{-1} =2 \pi \nu \:.
                                        \label{T_Con MagBD}
\end{eqnarray}
From (\ref{Met_1_0 MagBD})-(\ref{T_Con MagBD}) one concludes that
in the vicinity of the origin, metric (\ref{Met_1 MagBD})
describes a spacetime which is locally flat but has a conical
singularity with an angle deficit $\delta \varphi=2\pi(1-\nu)$.

Before closing this case, one should mention the particular
$\omega=0$ case of Brans-Dicke theory since this theory is related
(through dimensional reduction) to 4D General Relativity with one
Killing vector studied in \cite{OscarLemos_string}.

\vskip 2mm \noindent{{\bf (ii) Brans-Dicke theory with $\bm
{\omega=\pm \infty}$}} \vskip 2mm

The Brans-Dicke theory defined by $\omega=\pm \infty$ reduces to
the spacetime generated by a static magnetic point source in 3D
Einstein-Maxwell theory with $\Lambda<0$ studied in detail in
\cite{CL1}, \cite{HW}-\cite{OscarLemos_BTZ}. The behavior of this
spacetime is quite similar to those described in case {\bf (i)}.
It has a conical singularity at the origin and no horizons.

\vskip 2mm \noindent{{\bf (iii) Brans-Dicke theories with
$\bm{-\infty<\omega<-2}$}}\vskip 2mm

For this range of the Brans-Dicke parameter we have to consider
separately the case (1) $b>0$ and (2) $b<0$, where $b$ is the mass
parameter.

(1) If $b>0$ the shape of the $e^{2\mu(r)}$ function is drawn in
Fig. 1(b). Both $g_{rr}$ and $g_{\varphi \varphi}$ are always
positive (except at $r=0$) and there is no apparent change of
signature. Hence, for this range of parameters, the spacetime is
correctly described by equations (\ref{MET_SYM MagBD}) and
(\ref{MET_TODOS MagBD}). There are no horizons and so no magnetic
black holes,  but at $r=0$ the scalar $R_{\mu\nu}R^{\mu\nu}$
diverges (in equation (\ref{R-2-3/2-1 MagBD}) put $r_+=0$ and
replace $\rho$ by $r$). Therefore, at $r=0$ one has the presence
of a naked curvature singularity.

(2) If $b<0$ the shape of the $e^{2\mu(r)}$ function defined in
(\ref{MET_TODOS MagBD}) is sketched in Fig. 1(c). There occurs an
apparent change of signature for $0<r<r_+$. Proceeding exactly as
we did in case {\bf (i)} one can however introduce the new radial
coordinate $\rho$ defined in  (\ref{Transf_1 MagBD}) and obtain
the geodesically complete spacetime described by
 (\ref{Met_1 MagBD}) and (\ref{f MagBD}) (where now  $\omega<-2$). This
spacetime has no horizons and the scalar $R_{\mu\nu}R^{\mu\nu}$
given by (\ref{R-2-3/2-1 MagBD}) does not diverge for any value of
$\rho$ and so no curvature singularities are present. The
spacetime has a conical singularity at $\rho=0$ corresponding to
an angle deficit $\delta \varphi=2\pi(1-\nu)$, where $\nu$ is
defined in (\ref{T_Con_1 MagBD}).

\vskip 2mm \noindent{{\bf (iv) Brans-Dicke theory with
$\bm{\omega=-2}$}}\vskip 2mm

The shape of the $e^{2\mu(r)}$ function is drawn in Fig. 2(a).
There is an apparent change of signature for $r>r_+$, where $r_+$
is the zero of $e^{2\mu(r)}$. We can however introduce a new
coordinate system that will allow us to conclude that the
spacetime is complete for $0 \leq r \leq r_+$. First, we introduce
the radial coordinate $R=1/r$. With this new coordinate
$[g_{RR}(R)]^{-1}$ has a shape similar to the one shown in Fig.
1(a). Finally, we set a second coordinate transformation given by
$\rho^2=R^2-R_+^2$, where $R_+=1/r_+$ is the zero of
$(g_{RR})^{-1}$. Use of these coordinate transformations, together
with equations (\ref{MET_SYM MagBD}) and (\ref{MET-2 MagBD}),
allows us to write the spacetime generated by the static magnetic
point source as
\begin{eqnarray}
 ds^2  =
      -\frac{\alpha^2} {R^2(\rho)} dt^2
     + \frac{\rho^2}{[R^2(\rho)]^3} \frac{1}{h(\rho)} d\rho^2
          + \frac{h(\rho)}{\alpha^2}
            d\varphi^2 \:,
                                 \label{Met_2 MagBD}
\end{eqnarray}
where $0\leq\rho<\infty$ and the function $h(\rho)$ is given by
\begin{eqnarray}
h(\rho) = { {\biggl (} 1+\frac{\chi_{\rm m}^2}{8}
      \ln{[R^2(\rho)]} {\biggr )} R^{-2}(\rho)
            -b R^{-1}(\rho)}  \:.
                            \label{g_2 MagBD}
\end{eqnarray}
This function $h(\rho)$  is always positive except at $\rho=0$
where it is zero. Hence, the spacetime described by equation
(\ref{Met_2 MagBD}) and (\ref{g_2 MagBD}) has no horizon. Along
this section \ref{sec:stat_sol}, in cases {\bf(iv)}-{\bf(vii)} we
will use the definition $R^2(\rho) \equiv \rho^2 +R_+^2$ in order
to shorten the formulas.

The scalar $R_{\mu\nu}R^{\mu\nu}$ is given by
\begin{eqnarray}
R_{\mu\nu}R^{\mu\nu} \hspace{-0.1cm} &=&  \hspace{-0.1cm}
       \chi_{\rm m}^4 {\biggl [}\frac{3}{4}\ln^2[R(\rho)]
         + \frac{5}{4} \ln [R(\rho)] +9{\biggr ]}
         -\chi_{\rm m}^2 {\biggl [}6
\ln [R(\rho)]  +
   \frac{4 \ln [R(\rho)]}{R(\rho)} +\frac{3}{R(\rho)}+5{\biggr ]}+
                                               \nonumber \\
\hspace{-0.1cm} & &  \hspace{-0.1cm}
 + 8+\frac{32}{R(\rho)}+\frac{6}{R^2(\rho)}
   \:.
                                \label{R-2 MagBD}
\end{eqnarray}
This scalar diverges for $\rho=+\infty$ and so there is a
curvature singularity at $\rho=+\infty$. Besides, the  spacetime
described by (\ref{Met_2 MagBD}) and (\ref{g_2 MagBD}) has a
conical singularity at $\rho=0$ with coordinate $\varphi$ having a
period defined in equation (\ref{T_Con MagBD}),
\begin{eqnarray}
{\rm Period}_{\varphi}=2 \pi {\biggl [}r_+ {\biggl (}
           \frac{\chi_{\rm m}^2}{8}-1+\frac{b}{2r_+}
       -\frac{\chi_{\rm m}^2}{4} \ln{r_+}
       {\biggr )}  {\biggr ]}^{-1}\:.
                                        \label{T_Con_3 MagBD}
\end{eqnarray}
So, near the origin,  metric (\ref{Met_2 MagBD}) and
 (\ref{g_2 MagBD}) describe a spacetime which is locally flat but has a
conical singularity at $\rho=0$ with an angle deficit $\delta
\varphi=2\pi-{\rm Period}_{\varphi}$.

\vskip 2mm \noindent{{\bf (v)  Brans-Dicke theories with
$\bm{-2<\omega<-3/2}$}}\vskip 2mm

For this range of the Brans-Dicke parameter we have again  to
consider separately the case (1) $b>0$ and (2) $b<0$, where $b$ is
the mass parameter.

(1) If $b>0$ the shape of the $e^{2\mu(r)}$ function defined in
(\ref{MET_TODOS MagBD})  is similar to the one of case {\bf (iv)}
and sketched in Fig. 2(a). So, proceeding as in case  {\bf (iv)},
we find that the spacetime generated by the static magnetic point
source is given by (\ref{Met_2 MagBD}) with function $h(\rho)$
defined by
\begin{eqnarray}
h(\rho)  \hspace{-0.1cm}=  \hspace{-0.1cm}
     \frac{\alpha^2} {R^2(\rho)} +
                              \hspace{-0.1cm}
      b {\biggl (} \frac{\alpha^2} {R^2(\rho)}
      {\biggr )}^{-\frac{1}{2(\omega+1)}}
   \hspace{-0.4cm}
        -k\chi_{\rm m}^2  {\biggl (}
    \frac{\alpha^2} {R^2(\rho)}
     {\biggr )}^{-\frac{1}{\omega+1}}
      \hspace{-0.2cm} ,
                            \label{g MagBD}
\end{eqnarray}
which is always positive except at $\rho=0$ where it is zero.
Hence, the spacetime described by equations (\ref{Met_2 MagBD})
and (\ref{g MagBD}) has no horizons.

The scalar $R_{\mu\nu}R^{\mu\nu}$ is given by
 (\ref{R-2-3/2-1 MagBD}) as long as we replace function $r^2(\rho)$ by
$R^{-2}(\rho) \equiv (\rho^2 +R_+^2)^{-1}$. There is a curvature
singularity at $\rho=+\infty$.

Near the origin, equations (\ref{Met_2 MagBD}) and (\ref{g MagBD})
describe a spacetime which is locally flat but has a conical
singularity at $\rho=0$ with an angle deficit $\delta
\varphi=2\pi-{\rm Period}_{\varphi}$, with  ${\rm
Period}_{\varphi}$ defined in equation (\ref{T_Con MagBD}),
\begin{eqnarray}
 {\rm Period}_{\varphi} =
2 \pi   {\biggl [}
        \alpha r_+ - \frac{b
         (\alpha r_+)^{-\frac{\omega+2}{\omega+1}}}
           {2(\omega+1)}
       +\frac{k\chi_{\rm m}^2}{\omega+1}
      (\alpha r_+)^{-\frac{\omega+3}{\omega+1}}
       {\biggr ]}^{-1} \hspace{-0.3cm}.
                        \label{T_Con_2 MagBD}
\end{eqnarray}

(2) If $b<0$ the $e^{2\mu(r)}$ function can have a shape similar
to the one sketched in Fig. 2(b) or similar to Fig. 2(c),
depending on the values of the range. We will not proceed further
with the study of this case since it has a rather exotic spacetime
structure.

\vskip 2mm \noindent{{\bf (vi)  Brans-Dicke theory with
$\bm{\omega=-3/2}$}} \vskip 2mm

The shape of the $e^{2\mu(r)}$ function defined in
 (\ref{MET-3/2 MagBD}) is similar to the one of case  {\bf (iv)} and  sketched in
Fig. 2(a). So, proceeding as in case  {\bf (iv)}, we conclude that
the spacetime generated by the static magnetic point source is
given by (\ref{Met_2 MagBD}) with function $h(\rho)$ defined by
\begin{eqnarray}
\hspace{-0.2cm}  h(\rho) =  R^{-2}(\rho) {\biggl (}
      \frac{\Lambda}{2} \ln [b^{-2}R^2(\rho)]
      -\chi_{\rm m}^2 R^{-2}(\rho)
           {\biggr )} \:,
                             \label{g_3 MagBD}
\end{eqnarray}
which is always positive except at $\rho=0$ where it is zero.
Hence, the spacetime described by equations (\ref{Met_2 MagBD})
and (\ref{g_3 MagBD}) has no horizons.

The scalar $R_{\mu\nu}R^{\mu\nu}$ is
\begin{eqnarray}
 & R_{\mu\nu}R^{\mu\nu} =
       \Lambda^2 {\bigl [}12\ln^2[bR(\rho)]+20 \ln[bR(\rho)]+9{\bigr ]}+
        - \Lambda \chi_{\rm m}^2 R^2(\rho) {\biggr [}
     5 \ln[bR(\rho)]+\frac{9}{2}
   {\biggr ]}+\frac{9}{16}\chi_{\rm m}^4 R^4(\rho) \:.&
   \nonumber \\
   & &                             \label{R-3/2 MagBD}
\end{eqnarray}
There is a curvature singularity at $\rho=+\infty$.

Near the origin, equations (\ref{Met_2 MagBD}) and
 (\ref{g_3 MagBD}) describe a spacetime which is locally flat but has a
conical singularity at $\rho=0$ with an angle deficit $\delta
\varphi=2\pi-{\rm Period}_{\varphi}$, with ${\rm
Period}_{\varphi}$ defined in equation (\ref{T_Con MagBD}),
\begin{eqnarray}
{\rm Period}_{\varphi}=2 \pi {\biggl [}r_+ {\biggl (}
            \frac{\Lambda}{2} +
            \chi_{\rm m}^2 r_+^2 {\biggr )}  {\biggr ]}^{-1}  \:.
                                        \label{T_Con_4 MagBD}
\end{eqnarray}

\vskip 2mm \noindent{{\bf (vii) Brans-Dicke theories with
$\bm{-3/2<\omega<-1}$}} \vskip 2mm

The shape of the $e^{2\mu(r)}$ function is sketched in Fig. 2(a)
and is similar to the one of case  {\bf (iv)}. So, proceeding as
in case  {\bf (iv)}, we find that the spacetime generated by the
static magnetic point source is given by (\ref{Met_2 MagBD}) with
function $h(\rho)$ defined by (\ref{g MagBD}). There are no
horizons and there is no curvature singularity [the scalar
$R_{\mu\nu}R^{\mu\nu}$ is given by (\ref{R-2-3/2-1 MagBD}) if we
replace function $r^2(\rho)$ by $R^{-2}(\rho)$]. The spacetime has
a conical singularity at $\rho=0$ corresponding to an angle
deficit $\delta \varphi=2\pi-{\rm Period}_{\varphi}$,
 where ${\rm Period}_{\varphi}$ is defined in
(\ref{T_Con_2 MagBD}).

\subsubsection{\label{sec:geod}Geodesic structure}

We want to confirm that the spacetimes described by
 (\ref{MET_SYM MagBD}),
 (\ref{Met_1 MagBD}) and  (\ref{Met_2 MagBD})
are both null and timelike geodesically complete. The equations
governing the geodesics can be derived from the
 Lagrangian
\begin{equation}
{\cal{L}}=\frac{1}{2}g_{\mu\nu}\frac{dx^{\mu}}{d \tau}
       \frac{dx^{\nu}}{d \tau}=-\frac{\delta}{2}\:,
                                 \label{LAG MagBD)}  \\
\end{equation}
where $\tau$ is an affine parameter along the geodesic which, for
a timelike geodesic, can be identified with the proper time of the
particle along the geodesic. For a null geodesic one has
$\delta=0$ and for a timelike geodesic $\delta=+1$. From the
Euler-Lagrange equations one gets that the generalized momentums
associated with the time coordinate and angular coordinate are
constants: $p_t=E$ and $p_{\varphi}=L$. The constant $E$ is
related to the timelike Killing vector $(\partial/\partial
t)^{\mu}$ which reflects the time translation invariance of the
metric, while the constant $L$ is associated to the spacelike
Killing vector $(\partial/\partial \varphi)^{\mu}$ which reflects
the invariance of the metric under rotation. Note that since the
spacetime is not asymptotically flat, the constants $E$ and $L$
cannot be interpreted as the  energy and angular momentum at
infinity.

From the metric we can derive the radial geodesic,
\begin{eqnarray}
\dot{\rho}^2=-\frac{1}{g_{\rho\rho}}
      \frac{E^2 g_{\varphi \varphi}+L^2 g_{tt}}
              { g_{tt} g_{\varphi \varphi} }
       -\frac{\delta}{g_{\rho\rho}} \:.
                                        \label{GEOD_1 MagBD}
\end{eqnarray}
Next, we analyse this geodesic equation for each of the seven
cases defined in the last section. Cases {\bf (i)}, {\bf (ii)} and
{\bf (iii)} have identical geodesic structure, and cases {\bf
(iv)}-{\bf (vii)} also.

\vskip 2mm \noindent{{\bf (i) Brans-Dicke theories with
$\bm{-1<\omega<+\infty}$}}\vskip 2mm

Using the two useful relations $g_{tt} g_{\varphi
\varphi}=-\rho^2/g_{\rho\rho}$ and $g_{\varphi \varphi}=[\rho^2 /
(\rho^2+r_+^2)](\alpha^2 g_{\rho\rho})^{-1}$,
 we can write equation (\ref{GEOD_1 MagBD}) as
\begin{eqnarray}
\rho^2 \dot{\rho}^2= {\biggl [}
\frac{E^2}{\alpha^2}\frac{1}{\rho^2 + r_+^2} -\delta {\biggr ]}
\frac{\rho^2}{g_{\rho \rho}}
            +L^2 g_{tt} \:.
                                        \label{Geod_1 MagBD}
\end{eqnarray}
Noticing that $1/g_{\rho\rho}$ is always positive for $\rho>0$ and
zero for $\rho=0$, and that $g_{tt}<0$ we conclude the following
about the null geodesic motion ($\delta=0$). The first term in
(\ref{Geod_1 MagBD}) is positive (except at $\rho=0$ where it
vanishes), while the second term is always negative. We can then
conclude that spiraling ($L \neq 0$) null particles coming in from
an arbitrary point are scattered at the turning point $\rho_{\rm
tp}
> 0$ and spiral back to infinity. If the angular momentum L of the
null particle is zero it hits the origin (where there is a conical
singularity) with vanishing velocity.

Now we analyze the  timelike geodesics ($\delta=+1$). Timelike
geodesic motion is possible only if the energy of the particle
satisfies $E > \alpha r_+$. In this case, spiraling timelike
particles are bounded between two turning points that satisfy
$\rho_{\rm tp}^{\rm a} > 0$ and $\rho_{\rm tp}^{\rm b} <
\sqrt{E^2/\alpha^2 - r_+^2}$, with $\rho_{\rm tp}^{\rm b} \geq
\rho_{\rm tp}^{\rm a}$. When the timelike particle has no angular
momentum ($L=0$) there is a turning point located exactly at
$\rho_{\rm tp}^{\rm b}=\sqrt{E^2/\alpha^2 - r_+^2}$ and it hits
the origin $\rho=0$. Hence, we confirm that the spacetime
described by equation (\ref{Met_1 MagBD}) is both timelike and
null geodesically complete.

\vskip 2mm \noindent{{\bf (ii) Brans-Dicke theory with
$\bm{\omega=\pm \infty}$}} \vskip 2mm

The geodesic structure of the spacetime generated by a static
magnetic point source in 3D Einstein-Maxwell theory with
$\Lambda<0$ has been studied in detail in \cite{OscarLemos_BTZ}.
The behavior of this geodesic structure is quite similar to the
one described in case {\bf (i)}. In particular, the spacetime is
both timelike and null geodesically complete.

\vskip 2mm \noindent{{\bf (iii) Brans-Dicke theories with
$\bm{-\infty<\omega<-2}$}}

For this range of the Brans-Dicke parameter we have to consider
separately the case (1) $b>0$ and (2) $b<0$, where $b$ is the mass
parameter.

(1) If $b>0$ the motion of null and timelike geodesics is
correctly described by equation (\ref{Geod_1 MagBD}) if we replace
$\rho$ by $r$ and put $r_+ =0$. Hence, the null and timelike
geodesic  motion has the same feature as the one described in the
above case {\bf (i)}. In the above statements we just have to
replace $\rho$ by $r$, put $r_+ =0$ and remember that at the
origin there is now a naked curvature singularity rather than a
conical singularity.

 (2) If $b<0$ the motion of null and timelike geodesics is exactly
described by (\ref{Geod_1 MagBD}) and the statements presented in
case {\bf (i)} apply directly to this case.

\vskip 2mm \noindent{{\bf (iv), (v), (vi), (vii) Brans-Dicke
theories with $\bm{-2 \leq \omega < -1}$ }}\vskip 2mm

In order to study the geodesic motion of spacetime described by
equations (\ref{Met_2 MagBD}), one first notices that $g_{tt}
g_{\varphi \varphi}=-[g_{\rho\rho}(\rho^2+R_+^2)^2 / \rho^2]^{-1}$
and $g_{\varphi \varphi}=[\rho^2 / (\rho^2+R_+^2)^3] (\alpha^2
g_{\rho\rho})^{-1}$. Hence
 we can write equation (\ref{GEOD_1 MagBD}) as
\begin{eqnarray}
\rho^2 \dot{\rho}^2= {\biggl [}
\frac{E^2}{\alpha^2}\frac{1}{\rho^2 + R_+^2} -\delta {\biggr ]}
\frac{\rho^2}{g_{\rho \rho}}
            + (\rho^2 + R_+^2)^2 L^2 g_{tt} \:.
                                        \label{Geod_2 MagBD}
\end{eqnarray}
One concludes that the geodesic motion of null and timelike
particles has the same feature as the one described in case {\bf
(i)} after (\ref{Geod_1 MagBD}) if in the statements we replace
$r_+$ by $R_+$. The only difference is that on $\rho=+\infty$
there is now a curvature singularity for cases {\bf (iv)}
$\omega=-2$, {\bf (v)} $-2<\omega<-3/2$ and {\bf (vi)}
$\omega=-3/2$. In case {\bf (v)} $-2<\omega<-3/2$, if $b<0$ (as we
saw in last section) the spacetime has an exotic structure and so
we do not study it.

So, we confirm  that the spacetimes described by equations
(\ref{Met_2 MagBD}) are also  both timelike and null geodesically
complete.

\subsection{The general rotating solution}
\label{sec:rotat sol Mag 3D}

\subsubsection{The generating technique}

Now, we want to endow our spacetime solution with a global
rotation, i.e., we want to add angular momentum to the spacetime.
In order to do so we perform the following rotation boost in the
$t$-$\varphi$ plane (see e.g.
\cite{Sa_Lemos_Rotat,Zanchin_Lemos,HorWel})
\begin{eqnarray}
 t &\mapsto& \gamma t-\frac{\varpi}{\alpha^2} \varphi \:,
                                       \nonumber  \\
 \varphi &\mapsto& \gamma \varphi-\varpi t \:,
                                       \label{TRANSF_J MagBD}
\end{eqnarray}
where $\gamma$ and $\varpi$ are constant parameters.

\vskip 2mm \noindent{{\bf (i) Brans-Dicke theories with
$\bm{-1<\omega<+\infty}$}} \vskip 2mm

Use of equation (\ref{TRANSF_J MagBD}) and (\ref{Met_1 MagBD})
gives the gravitational field generated by the rotating magnetic
source
\begin{eqnarray}
 \hspace{-1cm}  & & \hspace{-1cm}
 ds^2 = -\alpha^2 (\rho^2 + r_+^2)
       (\gamma dt-\frac{\varpi}{\alpha^2} d\varphi)^2 +
                                            \nonumber \\
   \hspace{-0.4cm}  & & \hspace{-0.4cm}
     + \frac{\rho^2}{(\rho^2 + r_+^2)}\frac{1}{f(\rho)} d \rho^2
     + \frac{f(\rho)}{\alpha^2} (\gamma d\varphi-\varpi dt)^2,
                     \label{Met_1_J MagBD}
\end{eqnarray}
where $f(\rho)$ is defined in (\ref{f MagBD}).

The 1-form electromagnetic vector potential,
$A=A_{\mu}(\rho)dx^{\mu}$, of the rotating solution is
\begin{equation}
A=-\varpi A(\rho)dt +\gamma  A(\rho) d\varphi\:,
                                    \label{VEC_POTENT_J MagBD}
\end{equation}
where
\begin{equation}
A(\rho)=-\frac{1}{4\alpha^2}\chi_{\rm m}(\omega+1) [\alpha^2
(\rho^2+r_+^2)]^{-\frac{1}{2(\omega+1)}}\:.
                            \label{A_J MagBD}
\end{equation}

\vskip 10mm
\vskip 2mm \noindent{{\bf (ii) Brans-Dicke theory with $\bm
{\omega=\pm \infty}$}} \vskip 2mm

The spacetime generated by a rotating magnetic point source in 3D
Einstein-Maxwell theory with $\Lambda<0$ has been obtained and
studied in detail in \cite{OscarLemos_BTZ}.

\vskip 2mm \noindent{{\bf (iii) Brans-Dicke theories with
$\bm{-\infty<\omega<-2}$}}\vskip 2mm

Proceeding exactly as in case {\bf (i)}, we conclude that the
gravitational and electromagnetic fields generated by the rotating
magnetic source are also described by equations
 (\ref{Met_1_J MagBD})-(\ref{A_J MagBD}).

If $b>0$ we have to set $r_+ =0$ in equations
 (\ref{Met_1_J MagBD}) and (\ref{A_J MagBD}).

\vskip 2mm \noindent{{\bf (iv) Brans-Dicke theory with
$\bm{\omega=-2}$}}\vskip 2mm

Use of equations (\ref{TRANSF_J MagBD}) and (\ref{Met_2 MagBD})
yields the gravitational field generated by the rotating magnetic
source
\begin{eqnarray}
 \hspace{-1cm}  & & \hspace{-1cm}
      ds^2  =
      -\frac{\alpha^2} {(\rho^2 + R_+^2)}
     (\gamma dt-\frac{\varpi}{\alpha^2} d\varphi)^2+
                                          \nonumber \\
 \hspace{-0.4cm}  & & \hspace{-0.4cm}
     + \frac{\rho^2}{(\rho^2 + R_+^2)^3} \frac{1}{h(\rho)} d\rho^2
        + \frac{h(\rho)}{\alpha^2}
        (\gamma d\varphi-\varpi dt)^2,
                                 \label{Met_2_J MagBD}
\end{eqnarray}
where $h(\rho)$ is defined in (\ref{g_2 MagBD}).

The 1-form vector potential is also given by
 (\ref{VEC_POTENT_J MagBD}) but now  one has
\begin{equation}
A(\rho)=-\frac{1}{4 \alpha^2}\chi_{\rm m}(\omega+1) [\alpha^{-2}
(\rho^2 + R_+^2)]^{\frac{1}{2(\omega+1)}}\:.
                            \label{A_J_2 MagBD}
\end{equation}

%
\vskip 2mm \noindent{{\bf (v)  Brans-Dicke theories with
$\bm{-2<\omega<-3/2}$}}\vskip 2mm

If $b>0$, the gravitational and electromagnetic fields generated
by the rotating magnetic source are described by equations
(\ref{Met_2_J MagBD}) and (\ref{A_J_2 MagBD}), with $h(\rho)$
defined in (\ref{g MagBD}).

\vskip 2mm \noindent{{\bf (vi)  Brans-Dicke theory with
$\bm{\omega=-3/2}$}} \vskip 2mm

The gravitational and electromagnetic fields generated by the
rotating magnetic source are described by equations
 (\ref{Met_2_J MagBD}) and (\ref{A_J_2 MagBD}), with $h(\rho)$ defined in
(\ref{g_3 MagBD}).

\vskip 2mm \noindent{{\bf (vii) Brans-Dicke theories with
$\bm{-3/2<\omega<-1}$}} \vskip 2mm

The gravitational and electromagnetic fields generated by the
rotating magnetic source are described by equations
 (\ref{Met_2_J MagBD}) and (\ref{A_J_2 MagBD}), with $h(\rho)$
 defined in (\ref{g MagBD}).

\vskip 3mm

In equations (\ref{Met_1_J MagBD}), (\ref{VEC_POTENT_J MagBD}) and
(\ref{Met_2_J MagBD}) we choose $\gamma^2-\varpi^2 / \alpha^2=1$
because in this way when the angular momentum vanishes
($\varpi=0$) we have $\gamma=1$ and so we recover the static
solution.

Solutions (\ref{Met_1_J MagBD})-(\ref{A_J_2 MagBD}) represent
magnetically charged stationary spacetimes and also solve
(\ref{ACCAO MagBD}). Analyzing the Einstein-Rosen bridge of the
static solution one concludes that spacetime is not simply
connected which implies that the first Betti number of the
manifold is one, i.e., closed curves encircling the horizon cannot
be shrunk to a point. So, transformations (\ref{TRANSF_J MagBD})
generate a new metric because they are not permitted global
coordinate transformations \cite{Stachel}. Transformations
(\ref{TRANSF_J MagBD}) can be done locally, but not globally.
Therefore metrics (\ref{Met_1 MagBD}), (\ref{Met_2 MagBD}) and
(\ref{Met_1_J MagBD})-(\ref{A_J_2 MagBD}) can be locally mapped
into each other but not globally, and such they are distinct.

\subsubsection{\label{sec:MJQ} Mass, angular momentum and charge of
the solutions}

As we shall see the spacetime solutions describing the cases {\bf
(i)} $-1<\omega<+\infty$, {\bf (ii)}  $\omega=\pm \infty$ and {\bf
(iii)} $-\infty<\omega <-2$ are asymptotically anti-de Sitter.
This fact allows us to calculate the mass, angular momentum and
the electric charge of the static and rotating solutions. To
obtain these quantities  we apply the formalism  of Regge and
Teitelboim \cite{Regge}.

\vskip 10mm
\vskip 2mm \noindent{{\bf (i) Brans-Dicke theories with
$\bm{-1<\omega<+\infty}$}} \vskip 2mm

We first write  (\ref{Met_1_J MagBD}) in the canonical form
involving the lapse function $N^0(\rho)$ and the shift function
$N^{\varphi}(\rho)$
\begin{equation}
     ds^2 = - (N^0)^2 dt^2
            + \frac{d\rho^2}{f^2}
            + H^2(d\varphi+N^{\varphi}dt)^2 \:,
                               \label{MET_CANON MagBD}
\end{equation}
where $f^{-2}=g_{\rho\rho}$, $H^2=g_{\varphi \varphi}$,
 $H^2 N^{\varphi}=g_{t \varphi}$ and
$(N^0)^2-H^2(N^{\varphi})^2=g_{tt}$. Then, the action can be
written in the Hamiltonian form as a function of the energy
constraint ${\cal{H}}$, momentum constraint ${\cal{H}}_{\varphi}$
and Gauss constraint $G$
\begin{eqnarray}
S \hspace{-0.2cm}  &=& \hspace{-0.2cm}
     -\int dt d^2x[N^0 {\cal{H}}+N^{\varphi}
         {\cal{H}_{\varphi}}
       + A_{t} G]+   {\cal{B}}
                                     \nonumber \\
 \hspace{-0.2cm} &=& \hspace{-0.2cm}
         -\Delta t \int d\rho N \nu
        {\biggl [} \frac{2 \pi^2 e^{-2 \phi}}{H^3}
        -4f^2(H \phi_{,\rho}e^{-2 \phi})_{,\rho}
                                      \nonumber \\
\hspace{-0.5cm}  & & \hspace{-0.5cm}
        -2H \phi_{,\rho}(f^2)_{,\rho} e^{-2 \phi}
        +2f(fH_{,\rho})_{,\rho}e^{-2 \phi}
                                      \nonumber \\
 \hspace{-0.5cm}  & & \hspace{-0.5cm}
        +4 \omega H f^2 (\phi_{,\rho})^2e^{-2 \phi}
        -\Lambda H e^{-2 \phi}
       +\frac{2H}{f}e^{-2 \phi}(E^2+B^2){\biggr ]}
                                        \nonumber \\
\hspace{-0.5cm}  & & \hspace{-0.5cm}
  + \Delta t \int d\rho N^{\varphi}\nu{\biggl [}
      {\bigl (}2 \pi e^{-2 \phi} {\bigr )}_{,\rho}
       +\frac{4H}{f}e^{-2 \phi}E^{\rho}B{\biggr ]}
                                            \nonumber \\
\hspace{-0.5cm}  & & \hspace{-0.5cm}
       + \Delta t \int d \rho A_t \nu {\biggl [}-\frac{4H}{f}
       e^{-2 \phi} \partial_{\rho} E^\rho{\biggr ]} +{\cal{B}} \:,
                               \label{ACCAO_CANON MagBD}
\end{eqnarray}
where $N=\frac{N^0}{f}$, $\pi \equiv {\pi_{\varphi}}^{\rho}=
-\frac{fH^3 (N^{\varphi})_{,\rho}}{2N^0}$ (with $\pi^{\rho
\varphi}$ being the momentum conjugate to $g_{\rho \varphi}$),
$E^{\rho}$ and $B$ are the electric and magnetic fields and
${\cal{B}}$ is a boundary term. The factor $\nu$ [defined in
(\ref{T_Con_1 MagBD})] comes from the fact that, due to the angle
deficit, the integration over $\varphi$ is between $0$ and $2
\pi\nu$. Upon varying the action with respect to $f(\rho)$,
$H(\rho)$, $\pi(\rho)$, $\phi(\rho)$ and $E^{\rho}(\rho)$ one
picks up additional surface terms. Indeed,
\begin{eqnarray}
\delta S \hspace{-0.2cm}  &=& \hspace{-0.2cm}
       - \Delta t N \nu{\biggl [}(H_{,\rho}-2H\phi_{,\rho})e^{-2\phi}
         \delta f^2 -(f^2)_{,\rho}e^{-2\phi}\delta H
                                            \nonumber \\
 \hspace{-0.4cm}  & & \hspace{-0.4cm}
         -4f^2 H e^{-2\phi} \delta(\phi_{,\rho})
          +2f^2 e^{-2\phi}\delta (H_{,\rho})
                                               \nonumber \\
 \hspace{-0.4cm}  & & \hspace{-0.4cm}
     +2H{\bigl [}(f^2)_{,\rho}+4(\omega+1)f^2 \phi_{,\rho}{\bigr ]}
         e^{-2\phi}\delta \phi {\biggr ]}
                                                        \nonumber \\
  \hspace{-0.4cm}  & & \hspace{-0.4cm}
         +\Delta t N^{\varphi}\nu{\biggl [}2e^{-2\phi}\delta \pi
         -4 \pi e^{-2\phi}\delta \phi {\biggr ]}
                                                    \nonumber \\
  \hspace{-0.4cm}  & & \hspace{-0.4cm}
     + \Delta t A_t\nu {\biggl [}
             - \frac{4H}{f}e^{-2 \phi} \delta E^{\rho}{\biggr ]}
         + \delta {\cal{B}}
                                    \nonumber \\
  \hspace{-0.4cm}  & & \hspace{-0.4cm}
         +(\mbox{terms vanishing when
                    equations of motion hold}).
                                            \nonumber \\
                               \label{DELTA_ACCAO MagBD}
\end{eqnarray}
In order that the Hamilton's equations are satisfied, the boundary
term ${\cal{B}}$ has to be adjusted so that it cancels the above
additional surface terms. More specifically one has
\begin{equation}
  \delta {\cal{B}} = -\Delta t N \delta M  +\Delta t N^{\varphi}\delta J
                  + \Delta t A_t \delta Q_{\rm e} \:,
                              \label{DELTA_B MagBD}
\end{equation}
where one identifies $M$ as the mass, $J$ as the angular momentum
 and $Q_{\rm e}$ as the electric
charge since they are the terms conjugate to the asymptotic values
of $N$, $N^{\varphi}$ and $A_t$, respectively.

To determine the mass, the angular momentum  and the electric
charge of the solutions one must take the spacetime that we have
obtained and subtract the background reference spacetime
contribution, i.e., we choose the energy zero point in such a way
that the mass, angular momentum and charge vanish when the matter
is not present.

Now, note that for $\omega >-1$ (and $\omega <-2$), spacetime
(\ref{Met_1_J MagBD}) has an asymptotic metric given by
\begin{equation}
-{\biggl (}\gamma^2-\frac{\varpi^2}{\alpha^2} {\biggr )}
 \alpha^2 \rho^2 dt^2+ \frac{d \rho^2}{ \alpha^2 \rho^2}+
{\biggl (}\gamma^2-\frac{\varpi^2}{\alpha^2} {\biggr )}
 \rho^2 d \varphi^2 \:,
                                          \label{ANTI_SITTER MagBD}
\end{equation}
where $\gamma^2-\varpi^2 / \alpha^2=1$ so, it is asymptotically an
anti-de Sitter spacetime. For $\omega >-1$ (and $\omega <-2$) the
anti-de Sitter spacetime is also the background reference
spacetime, since the metric (\ref{Met_1_J MagBD}) reduces to
(\ref{ANTI_SITTER MagBD}) if the matter is not present ($b=0$ and
$\chi_{\rm m}=0$).

Taking the subtraction of the background reference spacetime into
account and noting that $\phi-\phi_{\rm ref}=0$ and that
$\phi_{,\rho}-\phi_{,\rho}^{\rm ref}=0$ we have that the mass,
angular momentum and electric charge are given by
\begin{eqnarray}
M &=& \nu {\bigl
[}(2H\phi_{,\rho}-H_{,\rho})e^{-2\phi}(f^2-f^2_{\rm ref})
     +(f^2)_{,\rho}e^{-2\phi}(H-H_{\rm ref})
      -2f^2 e^{-2\phi}(H_{,\rho}-H_{,\rho}^{\rm ref}) {\bigr ]}\:,
                                               \nonumber \\
J &=&  -2\nu e^{-2\phi} (\pi-\pi_{\rm ref}) \:,
                                               \nonumber \\
Q_{\rm e} &=&  \frac{4H}{f} \nu e^{-2 \phi}
             (E^{\rho}-E^{\rho}_{\rm ref}) \:.
                                    \label{MQ_GERAL MagBD}
\end{eqnarray}
Then, we finally have that the mass and angular momentum are
(after taking the asymptotic limit, $\rho \rightarrow +\infty$)
\begin{eqnarray}
 M &=& b \nu{\biggl [}\gamma^2
      + \frac{\omega +2}{\omega +1}
      \frac{\varpi^2}{\alpha^2}{\biggl ]}
      + {\rm Div_M}(\chi_{\rm m},\rho) \:,
                                        \label{M MagBD} \\
J &=&  \frac{\gamma \varpi}{\alpha^2} b \nu \frac{2\omega+3}
{\omega +1}+ {\rm Div_J}(\chi_{\rm m},\rho)\:,
                                    \label{J MagBD}
\end{eqnarray}
where ${\rm Div_M}(\chi_{\rm m},\rho)$  and ${\rm Div_J}(\chi_{\rm
m},\rho)$ are terms proportional to the magnetic source $\chi_{\rm
m}$ that diverge as $\rho \rightarrow +\infty$. The presence of
these kind of  divergences in the mass and angular momentum is a
usual feature present in charged solutions. They can be found for
example in the electrically charged point source solution
\cite{Deser_Maz}, in the electric counterpart of the BTZ black
hole \cite{BTZ_Q}, in the pure electric black holes of 3D
Brans-Dicke action \cite{OscarLemos} and in the magnetic
counterpart of the BTZ solution \cite{OscarLemos_BTZ}. Following
\cite{Deser_Maz,BTZ_Q}, the divergences on the mass can be treated
as follows. One considers a boundary of large radius $\rho_0$
involving the system. Then, one sums and subtracts ${\rm
Div_M}(\chi_{\rm m},\rho_0)$ to (\ref{M MagBD}) so that the mass
(\ref{M MagBD}) is now written as
\begin{equation}
M = M(\rho_0)+ [{\rm Div_M}(\chi_{\rm m},\rho)-
     {\rm Div_M}(\chi_{\rm m},\rho_0)] \:,
           \label{M0_0 MagBD}
\end{equation}
where $M(\rho_0)=M_0+{\rm Div_M}(\chi_{\rm m},\rho_0)$, i.e.,
\begin{equation}
M_0=M(\rho_0)-{\rm Div_M}(\chi_{\rm m},\rho_0)\:.
                       \label{M0_0_v2 MagBD}
\end{equation}
The term between brackets in (\ref{M0_0 MagBD}) vanishes when
$\rho \rightarrow \rho_0$. Then $M(\rho_0)$ is the energy within
the radius $\rho_0$. The difference between $M(\rho_0)$ and $M_0$
is $-{\rm Div_M}(\chi_{\rm m},\rho_0)$ which is interpreted as the
electromagnetic energy outside $\rho_0$ apart from an infinite
constant which is absorbed in $M(\rho_0)$. The sum
 (\ref{M0_0_v2 MagBD}) is then independent of $\rho_0$, finite and equal to the
total mass. In practice the treatment of the mass divergence
amounts to forgetting about $\rho_0$ and take as zero the
asymptotic limit: $\lim {\rm Div_M}(\chi_{\rm m},\rho)=0$.

To handle the angular momentum divergence, one first notices that
the asymptotic limit of the angular momentum per unit  mass
$(J/M)$ is either zero or one, so the angular momentum diverges at
a rate slower or equal to the rate of the mass divergence. The
divergence on the angular momentum can then be treated in a
similar way as the mass divergence. So, one can again consider a
boundary of large radius $\rho_0$ involving the system. Following
the procedure applied for the mass divergence one concludes that
the divergent term $-{\rm Div_J}(\chi_{\rm m},\rho_0)$ can be
interpreted as the electromagnetic angular momentum outside
$\rho_0$ up to an infinite constant that is absorbed in
$J(\rho_0)$.

Hence, in practice the treatment of both the mass and angular
divergences amounts to forgetting about $\rho_0$ and take as zero
the asymptotic limits : $\lim {\rm Div_M}(\chi_{\rm m},\rho)=0$
and $\lim {\rm Div_J}(\chi_{\rm m},\rho)=0$ in (\ref{M MagBD}) and
(\ref{J MagBD}).

Now, we calculate the electric charge of the solutions. To
determine the electric field we must consider the projections of
the Maxwell field on spatial hypersurfaces. The normal to such
hypersurfaces is $n^{\nu}=(1/N^0,0,-N^{\varphi}/N^0)$ and the
electric field is given by $E^{\mu}=g^{\mu \sigma}F_{\sigma
\nu}n^{\nu}$. Then, from (\ref{MQ_GERAL MagBD}), the electric
charge is
\begin{equation}
 Q_{\rm e}=-\frac{4Hf}{N^0} \nu e^{-2 \phi}(\partial_{\rho}A_t-N^{\varphi}
    \partial_{\rho} A_{\varphi})=\frac{\varpi}{\alpha^2} \nu
     \chi_{\rm m}\:.
\label{CARGA MagBD}
\end{equation}
Note that the electric charge is proportional to $\varpi \chi_{\rm
m}$. In section \ref{sec:Phys_Interp Mag3D} we will propose a
physical interpretation for the origin of the magnetic field
source and discuss the result obtained in (\ref{CARGA MagBD}).

The mass, angular momentum and electric charge of the static
solutions can be obtained by putting $\gamma=1$ and $\varpi=0$ on
the above expressions [see (\ref{TRANSF_J MagBD})].

\vskip 2mm \noindent{{\bf (ii) Brans-Dicke theory with $\bm
{\omega=\pm \infty}$}} \vskip 2mm

The mass, angular momentum  and electric charge of the spacetime
generated by a magnetic point source in 3D Einstein-Maxwell theory
with $\Lambda<0$ have been calculated in \cite{OscarLemos_BTZ}.
Both the static and rotating solutions have negative mass and
there is an upper bound for the intensity of the magnetic source
and for the value of the angular momentum.

\vskip 2mm \noindent{{\bf (iii) Brans-Dicke theories with
$\bm{-\infty<\omega<-2}$}}\vskip 2mm

The mass, angular momentum  and electric charge of the $\omega<-2$
solutions are also given by (\ref{M MagBD}), (\ref{J MagBD}) and
(\ref{CARGA MagBD}), respectively. If $b<0$ the factor $\nu$ is
defined in (\ref{T_Con_1 MagBD}) and if $b>0$ one has $\nu=1$.

\vskip 3mm

For $-2 \leq \omega \leq -1$ [cases {\bf (iv)}-{\bf (vii)}], the
asymptotic and background reference spacetimes have a very
peculiar form. In particular, they are not an anti-de Sitter
spacetime. Therefore, there are no conserved quantities for these
solutions.

\subsubsection{The rotating magnetic solution in final form}

For cases {\bf (i)} $-1<\omega<+\infty$, {\bf (ii)}  $\omega=\pm
\infty$ and {\bf (iii)} $-\infty<\omega<-2$  we may cast the
metric  in terms of $M$, $J$ and $Q_{\rm e}$.

\vskip 2mm \noindent{{\bf (i) Brans-Dicke theories with
$\bm{-1<\omega<+\infty}$}} \vskip 2mm

Use of (\ref{M MagBD}) and (\ref{J MagBD}) allows us to solve a
quadratic equation for $\gamma^2$ and $\varpi^2 / \alpha^2$. It
gives two distinct sets of solutions
\begin{equation}
\gamma^2=\frac{M(2- \Omega)}{2\nu b} \:,\:\:\: \:\:\:\:
\frac{\varpi^2}{\alpha^2}=
  \frac{\omega+1}{2(\omega+2)}\frac{M \Omega}{\nu b}\:,
\label{DUAS MagBD}
\end{equation}

\begin{equation}
\gamma^2=\frac{M \Omega}{2\nu  b} \:,\:\:\: \:\:\:\:
\frac{\varpi^2}{\alpha^2}=
      \frac{\omega+1}{2(\omega+2)}\frac{M(2- \Omega)}{\nu b}\:,
\label{DUAS_ERR MagBD}
\end{equation}
where we have defined a rotating parameter $\Omega$ as
\begin{equation}
\Omega \equiv 1- \sqrt{1-\frac{4(\omega+1)(\omega+2)}
{(2\omega+3)^2}\frac{J^2 \alpha^2}{M^2}} \:.
                    \label{OMEGA MagBD}
\end{equation}
When we take $J=0$ (which implies $\Omega=0$), (\ref{DUAS MagBD})
gives $\gamma \neq 0$ and $\varpi= 0$ while (\ref{DUAS_ERR MagBD})
gives the nonphysical solution $\gamma=0$ and $\varpi \neq 0$
which does not reduce to the static original metric. Therefore we
will study the solutions found from (\ref{DUAS MagBD}).

For $\omega>-1$ (and $\omega<-2$), the condition that $\Omega$
remains real imposes a restriction on the allowed values of the
angular momentum
\begin{equation}
|\alpha J|\leq \frac{|2\omega+3|M}{2\sqrt{(\omega+1) (\omega+2)}}
\:.
                    \label{Rest_OMEGA MagBD}
\end{equation}

The parameter $\Omega$ ranges between $0 \leq \Omega \leq 1$.
 The condition $\gamma^2-\varpi^2/\alpha^2=1$ fixes
the value of $b$ as a function of $M,\Omega,\chi_{\rm m}$,
\begin{eqnarray}
b &=& \frac{M}{\nu} {\biggl [} 1 - \frac{2\omega+3}{2(\omega +2)}
\Omega {\biggr ]}\:,
                                \label{b MagBD}
\end{eqnarray}
where
\begin{eqnarray}
\nu = \frac{2(\omega +1)(\alpha r_+)^{\frac{\omega +2}{\omega +1}}
+M{\biggl (}1- \frac{2\omega +3}{2(\omega +2)}\Omega {\biggr )} }
{2(\omega +1)(\alpha r_+)^{\frac{2\omega +3}{\omega +1}}+2k
\chi_{\rm m}^2 (\alpha r_+)^{-\frac{1}{\omega +1}}} \:.
                                  \label{b_2 MagBD}
\end{eqnarray}

The gravitational field (\ref{Met_1_J MagBD}) generated by the
rotating point source may now be cast in the form
\begin{eqnarray}
 ds^2 &=&
    -{\biggl [}\alpha^2 (\rho^2 + r_+^2)
           -\frac{\omega +1}{2(\omega +2)} \frac{M \Omega /\nu }
{[\alpha^2 (\rho^2 + r_+^2)]^{\frac{1}{2(\omega+1)}}}
    +\frac{(\omega +1)^2}{8(\omega +2)}\frac{Q^2_{\rm e}/\nu^2}
    {[\alpha^2 (\rho^2 + r_+^2)]^{\frac{1}{\omega+1}}}
          {\biggr ]} dt^2
                                             \nonumber \\
      & &
         -\frac{\omega +1}{2\omega +3}\frac{J}{\nu}{\biggl [}
         [\alpha^2 (\rho^2 + r_+^2)]^{-\frac{1}{2(\omega+1)}}
  -\frac{(\omega +1)Q^2_{\rm e}}{4M \Omega \nu}
    [\alpha^2 (\rho^2 + r_+^2)]^{\frac{1}{\omega+1}}
        {\biggr ]} 2dt d\varphi
                                                 \nonumber \\
      & &
       + \frac{\rho^2/(\rho^2 + r_+^2)}
       { {\biggl [}  \alpha^2 (\rho^2 + r_+^2) +\frac{M}{\nu}
      \frac{1-2(\omega+3)\Omega/[2(\omega+2)]}{[\alpha^2
      (\rho^2 + r_+^2)]^{\frac{1}{2(\omega+1)}}}
           -\frac{k\chi_{\rm m}^2}
    {[\alpha^2 (\rho^2 + r_+^2)]^{\frac{1}{\omega+1}}} {\biggr ]}}
          d\rho^2
                                              \nonumber \\
      & &
 + \frac{1}{\alpha^2}{\biggl [}
     \alpha^2 (\rho^2 + r_+^2)
   +\frac{M(2-\Omega)/(2\nu)}
   {[\alpha^2 (\rho^2 + r_+^2)]^{\frac{1}{2(\omega+1)}}}
    -\frac{(\omega+2)(2-\Omega)}{2(\omega+2)-(2\omega+3)\Omega}
     \frac{k\chi_{\rm m}^2}
    {[\alpha^2 (\rho^2 + r_+^2)]^{\frac{1}{\omega+1}}}
      {\biggr ]}   d\varphi^2 \:,  \nonumber \\
                                 \label{Met_MJQ MagBD}
\end{eqnarray}
with $0\leq\rho<\infty$ and the electromagnetic field generated by
the rotating point source can be written as
\begin{eqnarray}
A=\frac{A(\rho)}{\sqrt{2(\omega+2)-(2\omega+3)\Omega}}{\biggl [}
      -\alpha\sqrt{(\omega+1)\Omega} \: dt
+\sqrt{(\omega+2)(2-\Omega)} \: d \varphi{\biggr ]} \:,
\label{A_Fim MagBD}
\end{eqnarray}
with $A(\rho)$ defined in (\ref{A_J MagBD}).

The static solution can be obtained by putting $\Omega=0$ (and
thus $J=0$ and $Q_{\rm e}=0$) on the above expression [see
(\ref{TRANSF_J MagBD})].

\vskip 2mm \noindent{{\bf (ii) Brans-Dicke theory with $\bm
{\omega=\pm \infty}$}} \vskip 2mm

The spacetime generated by a rotating magnetic point source in 3D
Einstein-Maxwell theory with $\Lambda<0$ is written as a function
of its hairs in \cite{OscarLemos_BTZ}.

\vskip 2mm \noindent{{\bf (iii) Brans-Dicke theories with
$\bm{-\infty<\omega<-2}$}}\vskip 2mm

For this range of $\omega$, the gravitational and electromagnetic
fields  generated by a rotating magnetic point source are also
given by (\ref{Met_MJQ MagBD}) and (\ref{A_Fim MagBD}). If $b<0$
the factor $\nu$ is defined in (\ref{b_2 MagBD}) and if $b>0$ one
has $\nu=1$.

\subsubsection{Geodesic structure}

The geodesic structure of the rotating spacetime is similar to the
static spacetime (see section \ref{sec:geod}), although there are
now direct (corotating with $L>0$) and retrograde
(counter-rotating with $L<0$) orbits. The most important result
that spacetime is geodesically complete still holds for the
stationary spacetime.

\subsection{\label{sec:Phys_Interp Mag3D} Physical interpretation of the
magnetic source} \label{sec:phys interp Mag 3D}

When we look back to the electric charge given in
 (\ref{CARGA MagBD}), we see that it is zero when $\varpi=0$, i.e., when the
angular momentum $J$ of the spacetime vanishes. This is expected
since in the static solution we have imposed that the electric
field is zero ($F_{12}$ is the only non-null component of the
Maxwell tensor).

Still missing is a physical interpretation for the origin of the
magnetic field source. The magnetic field source is not a
Nielson-Oleson vortex solution since  we are working with the
Maxwell theory and not with an Abelian-Higgs model. We might then
think that the magnetic field is produced by a Dirac point-like
monopole. However, this is not also the case since a Dirac
monopole with strength $g_{\rm m}$ appears when one breaks the
Bianchi identity \cite{MeloNeto}, yielding $\partial_{\mu}
(\sqrt{-g} \tilde{F}^{\mu} e^{-2 \phi})= g_{\rm m} \delta^2
(\vec{x})$  (where $\tilde{F}^{\mu}=\epsilon^{\mu \nu
\gamma}F_{\nu \gamma}/2$ is the dual of the Maxwell field
strength), whereas in this work we have that  $\partial_{\mu}
(\sqrt{-g} \tilde{F}^{\mu} e^{-2 \phi})=0$. Indeed,  we are
clearly dealing with the Maxwell theory which satisfies Maxwell
equations and the  Bianchi identity
\begin{eqnarray}
& &  \frac{1}{\sqrt{-g}}\partial_{\nu}(\sqrt{-g}F^{\mu \nu} e^{-2
\phi})
        =\frac{\pi}{2}\frac{1}{\sqrt{-g}} j^{\mu} \:,
                                    \label{Max_j MagBD} \\
& &  \partial_{\mu}
     (\sqrt{-g} \tilde{F}^{\mu} e^{-2 \phi})=0 \:,
                                    \label{Max_bianchi MagBD}
\end{eqnarray}
respectively. In (\ref{Max_j MagBD}) we have made use of the fact
that the general relativistic current density is $1/\sqrt{-g}$
times the special relativistic current density $j^{\mu}=\sum q
\delta^2(\vec{x}-\vec{x}_0)\dot{x}^{\mu}$.

We then propose that the magnetic field source can be interpreted
as composed by a system of two symmetric and superposed electric
charges (each with strength $q$). One of the electric charges is
at rest with positive charge (say), and the other is spinning with
an angular velocity $\dot{\varphi}_0$ and negative electric
charge.  Clearly, this system produces no electric field since the
total electric charge is zero and the magnetic field is produced
by the angular electric current. To confirm our interpretation, we
go back to  (\ref{Max_j MagBD}). In our solution, the only
non-vanishing component of the Maxwell field is $F^{\varphi \rho}$
which implies that only $j^{\varphi}$ is not zero. According to
our interpretation one has $j^{\varphi}=q
\delta^2(\vec{x}-\vec{x}_0)\dot{\varphi}$, which one inserts in
 (\ref{Max_j MagBD}). Finally, integrating over $\rho$ and
$\varphi$ we have
\begin{equation}
 \chi_{\rm m} \propto q \dot{\varphi}_0 \:.
\label{Q_mag MagBD}
\end{equation}
So, the magnetic source strength, $\chi_{\rm m}$, can be
interpreted as an electric charge times its spinning velocity.

Looking again to the electric charge given in (\ref{CARGA MagBD}),
one sees that after applying the rotation boost in the
$t$-$\varphi$ plane to endow the initial static spacetime with
angular momentum, there appears  a net electric charge. This
result was already expected since now, besides the scalar magnetic
field ($F_{\rho \varphi} \neq 0$), there is also an electric field
($F_{t \rho} \neq 0$) [see (\ref{A_Fim MagBD})]. A physical
interpretation for the appearance of the net electric charge is
now needed. To do so, we return to the static spacetime. In this
static spacetime there is a static positive charge and a spinning
negative charge of equal strength at the center. The net charge is
then zero. Therefore, an observer at rest ($S$) sees a density of
positive charges at rest which is equal to the density of negative
charges that are spinning. Now, we perform a local rotational
boost $t'= \gamma t- (\varpi/\alpha^2) \varphi$ and $\varphi' =
\gamma \varphi-\varpi t\:$ to an observer ($S'$) in the static
spacetime, so that $S'$ is moving relatively to $S$. This means
that $S'$ sees a different charge density since a density is a
charge over an area and this area suffers a Lorentz contraction in
the direction of the boost. Hence, the two sets of charge
distributions that had symmetric charge densities in the frame $S$
will not have charge densities with equal magnitude in the frame
$S'$. Consequently, the charge densities will not cancel each
other in the frame $S'$ and a net electric charge appears.  This
was done locally. When we turn into the global rotational Lorentz
boost of  (\ref{TRANSF_J MagBD}) this interpretation still holds.
The local analysis above is similar to the one that occurs when
one has a copper wire with an electric current and we apply a
translation Lorentz boost to the wire: first, there is only a
magnetic field but, after the Lorentz boost, one also has an
electric field.  The difference is that in the present situation
the Lorentz boost is a rotational one and not a translational one.

\section{Summary and discussion}
\label{sec:conc 3D}

We have added the Maxwell action to the action of a generalized 3D
dilaton gravity specified by the Brans-Dicke parameter $\omega$
introduced in \cite{Sa_Lemos_Static,Sa_Lemos_Rotat} and discussed
in section \ref{sec:BH 3D}. We have concluded that for the static
spacetime the electric and magnetic fields cannot be
simultaneously  non-zero, i.e. there is no static dyonic solution.
Notice that, in oppose to what occurs in 4-dimensions where the
the Maxwell tensor and its dual are 2-forms, in 3-dimensions the
Maxwell tensor is still a 2-form, but its dual is a 1-form (in
practice, the Maxwell tensor has only three independent
components: two for the electric vector field, and one for the
scalar magnetic field). As a consequence, the magnetic solutions
are radically different from the electric solutions in
3-dimensions.

In section \ref{sec:Electric BH 3D} we have considered the
electrically charged case alone. We have found the static and
rotating black hole solutions of this theory. It contains eight
different cases that appear naturally from the solutions. For
$\omega=0$ one gets a theory related (through dimensional
reduction) to electrically charged four dimensional AdS general
relativity with one Killing vector \cite{Zanchin_Lemos}, and for
$\omega=\pm \infty$ one obtains electrically charged three
dimensional general relativity \cite{CL1,BTZ_Q}. For $\omega>-3/2$
the ADM mass and angular momentum of the solutions are finite,
well-behaved and equal to the ADM masses of the uncharged
solutions. However, for $\omega<-3/2$ the ADM mass and angular
momentum of the solutions have terms proportional to the charge
that diverge at the asymptotic limit, as frequently occurs in the
3-dimensional theories including a Maxwell field (see, e.g.
\cite{Deser_Maz,BTZ_Q,OscarLemos_BTZ,OscarLemos,KK2}). We have
shown how to treat this problem. For each range of $\omega$ we
have determined what conditions must be imposed on the ADM hairs
of the solutions in order to be possible the existence of  black
holes. Our results show that there is no upper bound on the
electric charge. The causal and geodesic structure of the charged
solutions has been analyzed in detail.

In section \ref{sec:Magnetic BH 3D} we have found geodesically
complete spacetimes generated by  static and rotating magnetic
point sources. These spacetimes are horizonless and many of them
have a conical singularity at the origin. These features are
common in spacetimes generated by point sources in 3D gravity
theories \cite{Jackiw_Review}-\cite{Brown_Hen}. The static
solution generates a scalar magnetic  field while the rotating
solution produces, in addition, a radial electric field. The
source for the magnetic field can be interpreted as composed by a
system of two symmetric and superposed electric charges. One of
the electric charges is at rest and the other is spinning. This
system produces no electric field since the total electric charge
is zero and the scalar magnetic field is produced by the angular
electric current. When we apply a rotational Lorentz boost to add
angular momentum to the spacetime, there appears a net electric
charge and a radial electric field. The $\omega=\pm \infty$
solution reduces to the magnetic counterpart of the BTZ solution
studied in chapter \ref{chap:BTZ family}. The solutions
corresponding to the theories described by a Brans-Dicke parameter
that belongs to the range $-1<\omega<+\infty$ or
$-\infty<\omega<-2$ have a
 behavior quite similar
to the  magnetic counterpart of the BTZ solution. For this range
of the Brans-Dicke parameter, the solutions are asymptotically
anti-de Sitter. This allowed us to calculate the mass, angular
momentum and charge of the solutions, and once again in we found
divergencies at spatial infinity.

The relation between spacetimes generated by sources in 3D and
cylindrically symmetric 4D solutions has been noticed by many
authors (see e.g.
\cite{Brown_book,DJH_flat,Cat,Lemos,OscarLemos_string}). The
$\omega=0$ solutions (electric and magnetic) considered in this
chapter are the 3D counterparts of the 4D cylindrical or toroidal
solutions studied in \cite{Lemos,Zanchin_Lemos,OscarLemos_string},
and that will be discussed in chapter \ref{chap:BH 4D}. Indeed,
the dimensional reduction of 4D general relativity with one
Killing vector yields the $\omega=0$ case of Brans-Dicke theory.

%% file: Chapter3.tex
\thispagestyle{empty} 
\chapter[Pair creation of black holes in three dimensions]
{\Large{Pair creation of black holes in three dimensions}}
\label{chap:Pair creation 3D}
 \lhead[]{\fancyplain{}{\bfseries Chapter \thechapter. \leftmark}}
 \rhead[\fancyplain{}{\bfseries \rightmark}]{}
\renewcommand{\thepage}{\arabic{page}}

\addtocontents{lof}{\textbf{Chapter Figures \thechapter}\\}


In chapters \ref{chap:Pair creation} and \ref{chap:Pair creation
in higher dimensions} we will analyze in detail the pair creation
process of 4-dimensional black holes and of higher dimensional
black holes, respectively. The process of quantum pair creation of
black holes in an external field in a 3-dimensional background has
not been analyzed yet, as far as we know. We will not do it here.
However, we will try to understand the difficulties associated
with this issue, and we will also propose a possible background in
which the pair creation process in 3-dimensions might be analyzed.

When we want to analyze analytically the gravitational pair
creation process, our first task is to ask if there is an exact
solution of Einstein gravity that represents a pair of accelerated
black holes. This solution is important for the analysis of the
process since it describes the subsequent motion of the pair,
after its creation. In 4-dimensions and higher dimensions these
solutions exist (see, respectively, chapters \ref{chap:PairAccBH}
and \ref{chap:Black holes in higher dimensions}), however in
3-dimensions these solutions are not known. In order to understand
this lack recall that in 3-dimensional Einstein gravity the only
black hole that exists is the BTZ one, and it lives in an AdS
background (see chapter \ref{chap:BTZ family}). So, black hole
pair creation in 3-dimensional Einstein gravity, if possible, must
occur in an AdS background. However, from the studies on
4-dimensional spacetimes (see chapter \ref{chap:Pair creation}),
we know that it is much more difficult in a sense to create a pair
in an AdS background than in a dS or flat background. Indeed, the
AdS background is attractive, i.e., it furnishes a cosmological
constant acceleration that pulls in the particles. Moreover, as we
saw in detail in chapter \ref{chap:BTZ family}, the BTZ black hole
has constant curvature and thus its construction is topological,
i.e., it is obtained from the pure $AdS_3$ solution through
identifications along an isometry of the $AdS_3$ spacetime.
Therefore, a possible exact solution describing an accelerating
pair of BTZ black holes would probably have to be also constructed
using a similar topological procedure.

Now, in what concerns the acceleration source of the black holes,
we could in principle manage a way to overcome the repulsive AdS
acceleration. Using again our knowledge from pair creation in
4-dimensions, we could try an external source such as (i) an
electromagnetic field,  (ii) a string with its tension, or (iii)
the 3-dimensional analogue of a domain wall with its repulsive
gravitational field. In what concerns the hypothesis (i), even in
4-dimensions, there is no AdS exact solution that describes a pair
of AdS black holes accelerated by an external Lorentz force. Thus,
we expect also serious difficulties when we try to find it in
3-dimensions. The hypothesis (ii) cannot be discarded since there
are string solutions in 3-dimensions. Hypothesis (iii) is
discarded since the analogue of a domain wall does not exist in
3-dimensions. The reason is simple: in 3-dimensional Einstein
gravity a matter distribution produces no gravitational field.

\begin{figure}[H]
\centering
\includegraphics[height=4cm]{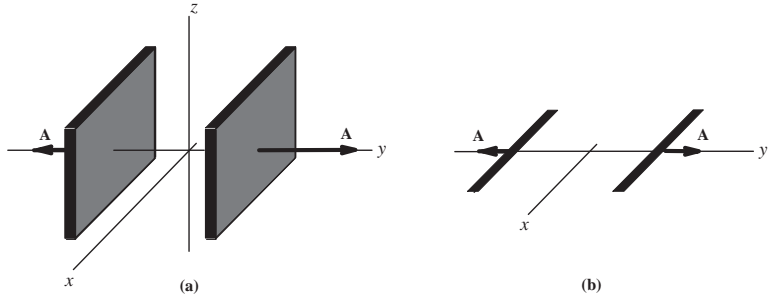}
  \caption{\label{Fig PairCreation toroidal 3D}
(a) Schematic representation of the planar AdS C-metric that
describes a pair of accelerated 4-dimensional black holes with
planar topology (black walls). (b) After dimensional reduction the
above solution might yield a 3-dimensional exact solution that
describes a pair of accelerated $\omega=0$ Brans-Dicke black
strings.
 }
\end{figure}

Another possibility, is to try to analyze the black hole pair
creation process in the context of an effective 3-dimensional
gravity theory, e.g., in the dilaton Brans-Dicke theory discussed
in chapter \ref{chap:3D Dilaton BH}. In particular, the case with
$\omega=0$ (where $\omega$ is the Brans-Dicke parameter) might be
interesting for our purposes. Indeed, recall that this $\omega=0$
case can be obtained through dimensional reduction from
4-dimensional AdS general relativity with one Killing vector
\cite{Lemos,Zanchin_Lemos,OscarLemos_string}. This is, the
cylindrical or toroidal 4-dimensional black hole found in
\cite{Lemos} yields, through dimensional reduction, the $\omega=0$
black hole. Now, in section \ref{sec:ExtLim AdS Toroidal} we study
the planar, cylindrical or toroidal AdS C-metric \cite{PlebDem},
which describes a pair of accelerated 4-dimensional black holes
with planar symmetry (i.e., with planar, cylindrical or toroidal
topology). The solution with planar black holes, or black walls,
is the most useful for our present discussion. Indeed, it is
possible that the dimensional reduction of this planar AdS
C-metric yields a 3-dimensional exact solution that describes a
pair of accelerated $\omega=0$ Brans-Dicke black strings (this is
sketched in Fig. \ref{Fig PairCreation toroidal 3D}). We will not
do this here. We leave it for future work. In chapter
\ref{chap:Pair creation} we will analyze in detail the pair
creation process of 4-dimensional black holes, and in chapter
\ref{chap:Pair creation in higher dimensions} we will study the
pair creation process of higher dimensional black holes.

We have been discussing pair creation of black holes in
3-dimensions, and gave reasons for the difficulty in studying the
problem. One could ask if there are works discussing particle pair
creation in 3-dimensions. As far as we know, the literarture is
also void in this subject. Some of the remarks gathered for the
difficulties in black hole pair creation also apply here. Although
point particle solutions in flat 3-dimensional general relativity
exist, perhaps there are still difficulties in finding exact
solutions with an accelerating field where the point particles
could be boxed in. We would like to mention that 2-dimensional
pair creation of particles has not been totally neglected. Indeed,
Brown and Teitelboim have discussed this problem
\cite{BrownTeitelboimPC2D}.

%% file: Chapter4.tex
\thispagestyle{empty} \setcounter{minitocdepth}{1}
\chapter[Black holes in a generalized $\bm \Lambda$ background]
{\Large{Black holes in a generalized $\bm \Lambda$ background}}
\label{chap:BH 4D}
 \lhead[]{\fancyplain{}{\bfseries Chapter \thechapter. \leftmark}}
 \rhead[\fancyplain{}{\bfseries \rightmark}]{}
  \minitoc \thispagestyle{empty}
\renewcommand{\thepage}{\arabic{page}}

\addtocontents{lof}{\textbf{Chapter Figures \thechapter}\\}


In this chapter we will briefly describe the main features of the
black hole solutions of the Einstein-Maxwell theory in a
background with a generalized cosmological constant $\Lambda$. We
stress that this description is by no way a complete one. We will
only focus on those properties that will be needed to better
understand the processes that will be discussed in later chapters.
Mainly, we shall discuss the range of parameters for which one has
black holes, we will analyze the causal structure of the
solutions, and we will briefly discuss the main difference between
the geodesic motion in an anti-de Sitter (AdS), in a flat, and in
a de Sitter (dS) background.

In sections \ref{sec:BH 4D AdS spherical}, \ref{sec:BH 4D flat}
and \ref{sec:BH 4D dS} we will describe black holes with spherical
topology in an AdS, flat and dS background, respectively. Their
properties will be useful for us in chapter \ref{chap:PairAccBH},
where we will analyze solutions that describe a pair of
accelerated black holes with spherical topology in an AdS, flat
and dS background. Then, in chapter \ref{chap:Pair creation} we
will discuss the pair creation process of these black holes in an
external field. Moreover, in chapter \ref{chap:Extremal Limits} we
will see that the solutions studied in sections \ref{sec:BH 4D
topological}, \ref{sec:BH 4D flat} and \ref{sec:BH 4D dS} of this
chapter allows us to generate the so called Nariai,
Bertotti-Robinson and anti-Nariai solutions. Subsection
\ref{sec:BH 4D AdS toroidal_magnetic} discusses work developed by
us \cite{OscarLemos_string}. We thus think that this introductory
chapter is well justified.

\vspace{0.4 cm}

\begin{figure}[H]
\centering
\includegraphics[height=3cm]{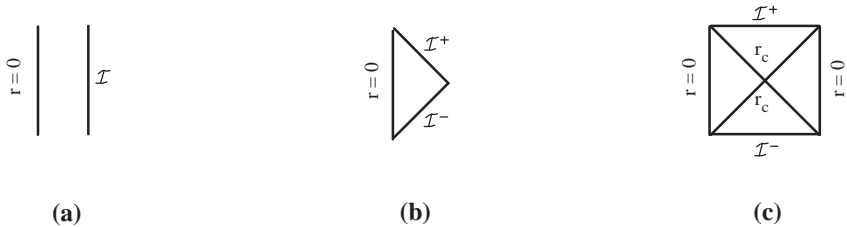}
   \caption{\label{Fig CPtodos}
Carter-Penrose diagram of the: (a) AdS solution, (b) Minkowski
solution, and (c) dS solution.  ${\cal I}$ represents the infinity
($r=\infty$). The infinity line is a timelike line in the AdS
case, a null line in the Minkowski case, and a spacelike line in
the dS case.
 }
\end{figure}

\begin{figure}[H]
\centering
\includegraphics[height=7cm]{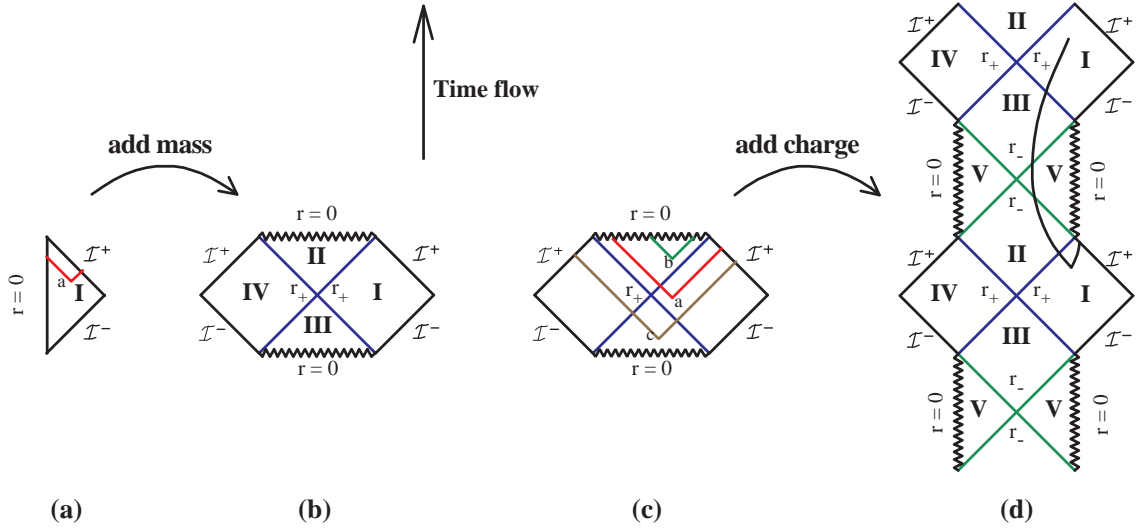}
   \caption{\label{Fig Schw_rays}
Carter-Penrose diagram of the: (a) Minkowski solution, $M=0$,
$Q=0$; (b) Schwarzschild ($Q=0$) black hole; and (c)
Reissner-Nordstr\"{o}m black hole. The zigzag line represents a
curvature singularity, ${\cal I}$ represents the infinity
($r=\infty$), $r_+$ represents a black hole event horizon, and
$r_-$ represents the Cauchy or inner black hole horizon.
 }
\end{figure}

In what concerns the causal structure, the main difference between
the Carter-Penrose diagrams of the $\Lambda<0$ (AdS), of the
$\Lambda=0$ (flat), and of the $\Lambda>0$ (dS) is at the level of
the line that represents infinity ($r=+\infty$). In the AdS case
the infinity is a timelike line (vertical line in the diagrams),
in the flat case it is represented by a null line (a $45^{\rm o}$
line), and in the dS case it is represented by a spacelike line
(an horizontal line in the diagram). These features are
represented in Fig. \ref{Fig CPtodos} for the pure AdS, Minkowski
and dS backgrounds (the character of the infinity line does not
change when a mass or a charge is added to the system). The
infinity line is then a signature of the cosmological background.
Note that in these Carter-Penrose diagrams, a null particle (e.g.,
a light ray) moves necessarily along $45^{\rm o}$ lines, while
timelike particles can move only into the top of the diagrams
along a curve whose tangent vector must do an angle less than
$45^{\rm o}$ with the vertical line. As a typical example of the
kind of information that can be withdrawn from the Carter-Penrose
diagrams, let us analyze the evolution of these diagrams in a
$\Lambda=0$ background when we start with the empty spacetime
(Minkowski solution with $M=0$, $Q=0$), and then progressively add
a mass $M$ (Schwarzschild solution) and then also a charge $Q$
(Reissner-Nordstr\"{o}m solution) to the background spacetime.
This evolution is sketched in Fig. \ref{Fig Schw_rays}. In Fig.
\ref{Fig Schw_rays}.(a), we have the diagram of the Minkowski
solution. The past infinity (${\cal I}^-$) and the future infinity
(${\cal I}^+$) are both represented by a null line, and the origin
of the radial coordinate, $r=0$, is represented by a timelike
line. A null ray that is emitted from a point $a$ is free to move
towards $r=0$ or into the future infinity ${\cal I}^+$. When we
add a mass $M$ to the system [see Fig. \ref{Fig Schw_rays}.(b)],
several changes occur. First of all, $r=0$ supports now a
curvature singularity (since any scalar polynomial of the
curvature, e.g. the square of the Riemann tensor, diverges there).
The presence of this curvature singularity is indicated by a
zigzag line. Moreover, when compared with Fig. \ref{Fig
Schw_rays}.(a), we see that $r=0$ suffers a $90^{\rm o}$ rotation
and is now represented by a spacelike line. This rotation is
accompanied by the appearance of two mutually perpendicular lines
at $45^{\rm o}$ that represent the black hole event horizon $r_+$.
The region IV is equivalent to region I [that was already present
in diagram (a)] and both represent the region outside the black
hole horizon, $r_+<r<+\infty$. A null ray that is sent from a
point $a$ in these regions can move towards $r=0$, after crossing
$r_+$, but it is also free to move into the future infinity ${\cal
I}^+$ [see Fig. \ref{Fig Schw_rays}.(c)]. Note that regions I and
IV are casually disconnected: no null or timelike particle can
start in region I and reach region IV. The adding of the mass also
leads to the appearance of two new regions, represented by regions
II and III in figure (b). Region II is interpreted as the interior
of the black hole horizon, since a null ray emitted from a point
in its interior [e.g., point $b$ in Fig. \ref{Fig Schw_rays}.(c)]
necessarily hits the future curvature singularity, and cannot
cross the horizon towards region I or IV (the discussion also
applies to timelike particles). Region III is interpreted as the
interior of the white hole, since any particle, e.g. the light ray
that is emitted from point $c$ in Fig. \ref{Fig Schw_rays}.(c), is
necessarily expelled out from region III, i.e., it necessarily
crosses the horizon towards region I or IV. In Fig. \ref{Fig
Schw_rays}.(d) we show the causal diagram of the solution when a
charge is added to the solution. Again, several changes occur.
When compared with Fig. \ref{Fig Schw_rays}.(b), we see that the
curvature singularity $r=0$ suffers again a $90^{\rm o}$ rotation
and is now represented by a timelike zigzag line. This rotation is
accompanied by the appearance of two new mutually perpendicular
lines at $45^{\rm o}$, that represent the black hole Cauchy
horizon $r_-$, together with two new equivalent regions that are
represented as region V in the figure. The properties of this
Cauchy horizon are quite interesting. As an example note that the
full history of the regions I, II and IV is in the causal past of
the Cauchy horizon, i.e., this horizon can have access to all the
information that is generated in those regions.
The causal diagrams provide a lot of information about the
solution, but there are also relevant features that cannot be
withdrawn from them. The study of the geodesic structure of the
solution provides some more information, that will be useful for
us in later chapters. The geodesic equation of motion, for a
general non-rotating black hole, can be written as
\begin{eqnarray}
\dot{r}^2 = E^2-\left (
\delta+\frac{L^2}{g_{\varphi\varphi}}\right
)\frac{1}{g_{rr}}=E^2-V(r)\, ,
 \label{geodesics-pedag}
\end{eqnarray}
where
\begin{eqnarray}
V(r)=\left ( \delta+\frac{L^2}{r^2}\right ) \left (
1-\frac{2M}{r}+\frac{Q^2}{r^2}-\frac{\Lambda}{3}r^2 \right )\, ,
 \label{geodesics-pedag-2}
\end{eqnarray}
and $\dot{r}=\frac{dr}{d\tau}$, with $\tau$ being an affine
parameter along the geodesic which, for a timelike geodesic, can
be identified with the proper time of the particle along the
geodesic. For a null geodesic one has $\delta=0$ and for a
timelike geodesic $\delta=+1$. $M$ and $Q$ are respectively the
mass and charge of the black holes, and $L$ is the angular
momentum of the particle subjected to the gravitational field of
the black hole.

\begin{figure}[H]
\centering
\includegraphics[height=5cm]{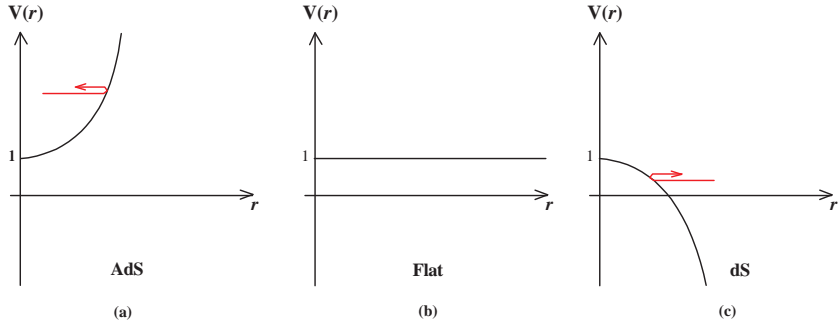}
   \caption{\label{Fig geodesic_M=0 Q=0}
Timelike geodesics for the black hole parameters $M=0$, and $Q=0$.
(a) Anti-de Sitter case ($\Lambda<0$), (b) flat case
($\Lambda=0$), and (c) de Sitter case ($\Lambda=0$).
 }
\end{figure}

In later chapters we will make a reference to the information
withdrawn from Fig. \ref{Fig geodesic_M=0 Q=0}, where we draw the
general form of $V(r)$ for timelike geodesics in a pure anti-de
Sitter, Minkowski and de Sitter spacetimes ($M=0$, $Q=0$,
$\delta=1$ and $L=0$). From it we conclude that the AdS spacetime
is attractive, in the sense that particles in this background are
subjected to a potential well that attracts them, i.e., if a
particle tries to escape to infinity, it will be reflected back,
no matter how big its energy. On the other side, the dS spacetime
is repulsive: if a low-energy particle tries to approach the
origin, it will be reflected back to infinity.
For massless particles, this information concerning the
attractive/repulsive character of the AdS/dS backgrounds could
also be withdrawn from the corresponding Carter-Penrose diagrams,
Fig. \ref{Fig CPtodos}. Indeed, we see that a null ray ($45^{\rm
o}$ line) that hits the timelike AdS infinity is necessarily
reflected back, and that a null ray that is sent towards the dS
spacelike infinity can never return back (unless we act on the ray
through an external process, a mirror, for example).

\begin{figure}[H]
\centering
\includegraphics[height=5cm]{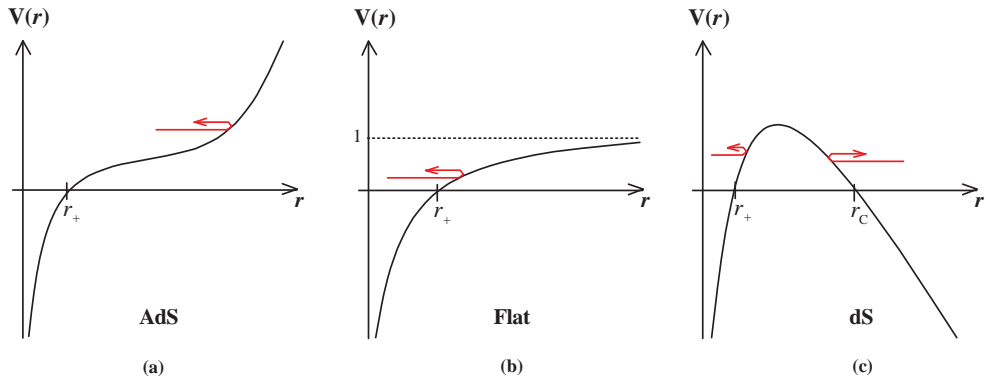}
   \caption{\label{Fig geodesic_M Q=0}
Timelike geodesics with vanishing angular momentum ($L=0$) for the
black hole parameters $M\neq 0$, and $Q=0$. (a) AdS case
($\Lambda<0$), (b) flat case ($\Lambda=0$), and (c) dS case
($\Lambda=0$).
 }
\end{figure}

As another example of the utility of the geodesic analysis, and of
the limitations associated with the causal diagrams, the general
form of $V(r)$ for $L=0$ timelike geodesics in a Schwarzschild
black hole and Reissner-Nordstr\"{o}m black hole is sketched,
respectively, in Figs. \ref{Fig geodesic_M Q=0} and \ref{Fig
geodesic_M Q} for the three cosmological backgrounds (AdS, flat,
dS). The important fact that we want to stress is that, in the
Schwarzschild black hole (AdS, flat or dS), when a massive
particle crosses inward the $r=r_+$ sphere it cannot come back, as
already expected. However, in the Reissner-Nordstr\"{o}m black
hole (AdS, flat or dS), a massive particle can never hit the
curvature singularity (note that this does not apply in the case
of massless particles). This information could not be withdrawn
from the analysis of the Carter-Penrose diagram.
\begin{figure}[H]
\centering
\includegraphics[height=5cm]{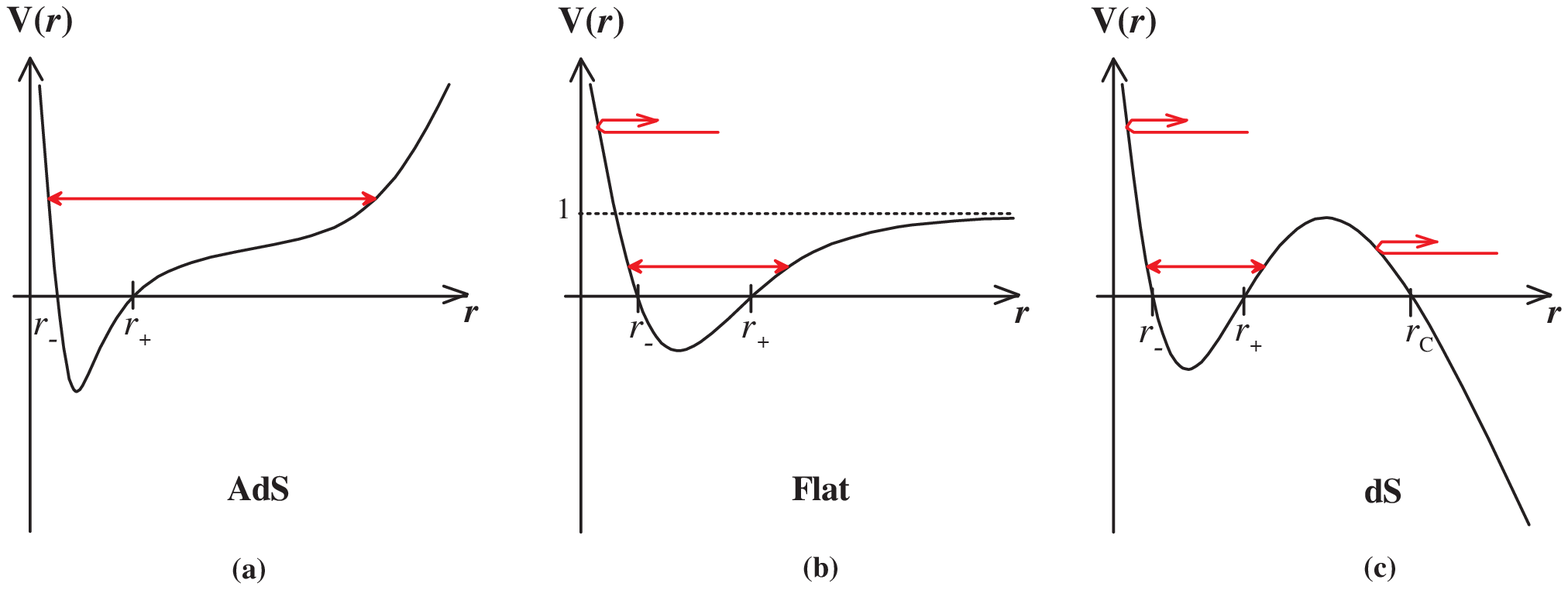}
   \caption{\label{Fig geodesic_M Q}
Timelike geodesics with vanishing angular momentum ($L= 0$) for
the black hole parameters $M\neq 0$, and $Q\neq 0$. (a) AdS case
($\Lambda<0$), (b) flat case ($\Lambda=0$), and (c) dS case
($\Lambda=0$).
 }
\end{figure}

\section{\label{sec:BH 4D AdS}Black holes in an anti-de Sitter background}

The Einstein-Maxwell equations in an AdS background ($\Lambda<0$)
admit a three-family of black hole solutions whose gravitational
field is described by
\begin{eqnarray}
d s^2 = - \left (
b-\frac{2M}{r}+\frac{Q^2}{r^2}-\frac{\Lambda}{3}r^2 \right )\,
dt^2
+\frac{dr^2}{b-\frac{2M}{r}+\frac{Q^2}{r^2}-\frac{\Lambda}{3}r^2}+r^2
d\Omega_b^2,
 \label{AdS bH}
\end{eqnarray}
where $b$ can take the values $1,0,-1$ and
\begin{eqnarray} \left\{ \begin{array}{l}
d\Omega_b^2=d \theta^2+\sin^2\theta\,d\phi^2
\:\:\:\:\:\:\:\:{\rm for}\:\: b=1\,,\\
d\Omega_b^2=d \theta^2+d\phi^2
\:\:\:\:\:\:\:\:\:\:\:\:\:\:\:\:\:\:\:\,{\rm for}\:\: b=0 \,,\\
d\Omega_b^2=d \theta^2+\sinh^2\theta\,d\phi^2 \:\:\:\:\:\: {\rm
for}\:\:  b=-1\,.
\end{array} \right.
\label{angular AdS bh}
\end{eqnarray}
These three solutions describe three different kind of AdS black
holes. The black holes with $b=1$ are the usual AdS black holes
with spherical topology. The black holes with $b=0$ have planar,
cylindrical  or toroidal (with genus $g\geq 1$) topology and were
introduced and analyzed in
\cite{Lemos,Zanchin_Lemos,OscarLemos_string}. The topology of the
$b=-1$ black holes is hyperbolic  or, upon compactification,
toroidal with genus $g\geq 2$, and they have been analyzed in
\cite{topological}. In the following subsections we shall describe
briefly which one of these families of black holes. The solutions
with non-spherical topology (cases $b=0$ and $b=-1$) do not have
counterparts in a $\Lambda=0$ or in a $\Lambda>0$ background.

\subsection{\label{sec:BH 4D AdS spherical}Black holes with spherical topology}
\begin{figure}[H]
\centering
\includegraphics[height=5cm]{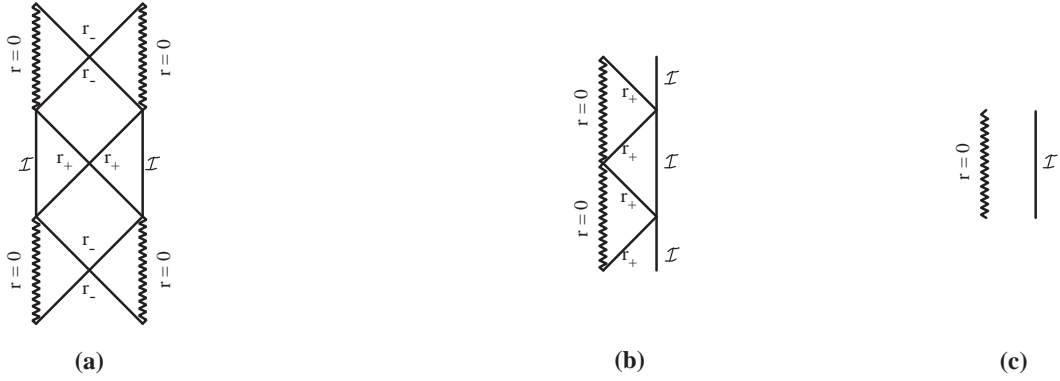}
   \caption{\label{Fig spherical AdS}
Carter-Penrose diagrams of the AdS$-$Reissner-Nordstr\"{o}m
($Q\neq 0$) black holes (with spherical topology) discussed in the
text of section \ref{sec:BH 4D AdS spherical}. The zigzag line
represents a curvature singularity, ${\cal I}$ represents the
infinity ($r=\infty$), $r_+$ represents a black hole event
horizon, and $r_-$ represents a Cauchy horizon.
 }
\end{figure}

\begin{figure}[H]
\centering
\includegraphics[height=2.5cm]{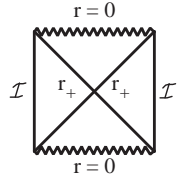}
   \caption{\label{Fig spherical Q=0 AdS}
Carter-Penrose diagrams of the AdS-Schwarzschild ($Q=0$) black
hole (with spherical topology) discussed in the text of section
\ref{sec:BH 4D AdS spherical}. The zigzag line represents a
curvature singularity, ${\cal I}$ represents the infinity
($r=\infty$), $r_+$ represents a black hole event horizon, and
$r_-$ represents a Cauchy horizon.
 }
\end{figure}

In this subsection we will briefly present the black hole
solutions with spherical symmetry of the Einstein-Maxwell
equations in a negative cosmological background ($\Lambda<0$).
 The gravitational field of the electrically charged black
hole solution is given by
\begin{eqnarray}
d s^2 = - \left (
1-\frac{\Lambda}{3}r^2-\frac{2M}{r}+\frac{Q^2}{r^2} \right )\,
dt^2 +\frac{dr^2}{
1-\frac{\Lambda}{3}r^2-\frac{2M}{r}+\frac{Q^2}{r^2} }+r^2 (d
\theta^2+\sin^2\theta\,d\phi^2)\:,
 \label{AdS spherical bH}
\end{eqnarray}
where  $M$ and $Q$ are, respectively, the ADM mass and charge of
the solution, while the electromagnetic field is
\begin{equation}
 A =-\frac{Q}{r} \,dt  \, , \qquad  A =Q \cos \theta \,d \phi\,,
                           \label{Max AdS bh4D}
\end{equation}
in the pure electric and in the pure magnetic cases, respectively.
The Carter-Penrose diagram of the nonextreme
AdS$-$Reissner-Nordstr\"{o}m black hole is sketched in Fig.
(\ref{Fig spherical AdS}).(a), of the extreme
AdS$-$Reissner-Nordstr\"{o}m black hole in Fig. (\ref{Fig
spherical AdS}).(b), and the naked particle in Fig. (\ref{Fig
spherical AdS}).(c). The Carter-Penrose diagram of the
AdS-Schwarzschild is drawn in Fig. (\ref{Fig spherical Q=0 AdS}).

\subsection{\label{sec:BH 4D AdS toroidal}Black holes with toroidal or cylindrical topology}

In this subsection we will consider solutions of the
Einstein-Maxwell equations with cylindrical symmetry in a negative
cosmological background ($\Lambda<0$).  The topology of the two
dimensional angular space can be (i) $R\times S^1$, the standard
cylindrically symmetric model, (ii) $S^1 \times S^1$ the flat
torus $T^2$ model , and (iii) $R^2$. We will focus upon (i) and
(ii). In the cylindrical model (i), we work with the cylindrical
coordinate system $({t}, r,{\varphi}, z)$ with
$-\infty<{t}<+\infty$, $0\leq r< +\infty$, $-\infty <z<+\infty$,
$0\leq{\varphi}< 2\pi$. In the toroidal model (ii), the range of
the coordinate $z$ is $0\leq \alpha z< 2\pi$, with $\alpha^2
\equiv -\frac{1}{3}\Lambda$.

\subsubsection{\label{sec:BH 4D AdS toroidal_electric}Neutral and electric charged black holes}

\begin{figure}[H]
\centering
\includegraphics[height=5cm]{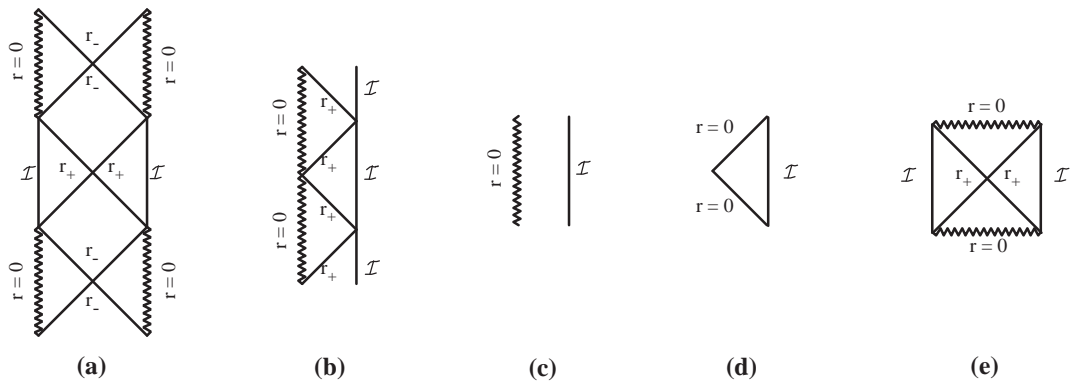}
   \caption{\label{Fig toroidal AdS}
Carter-Penrose diagrams of the  AdS black holes toroidal or
cylindrical topology discussed in the text of section
 \ref{sec:BH 4D AdS toroidal_electric}. The zigzag line represents a curvature
singularity, ${\cal I}$ represents the infinity ($r=\infty$),
$r_+$ represents a black hole event horizon, and $r_-$ represents
a Cauchy horizon.
 }
\end{figure}

The gravitational field of the
 rotating electrically charged black
hole solution can be written as \cite{Lemos,Zanchin_Lemos}
\begin{eqnarray}
& \!\!\!\!\!\!\!\!\!\!\!\!\!\!\!\! ds^{2} = -\left(\alpha^2 r^2
-\frac{4M(1-\frac{a^2\alpha^2}{2})}{\alpha r} +
\frac{4Q^2}{\alpha^2 r^2}\right) dt^2
-\frac{4aM\sqrt{1-\frac{a^2\alpha^2}{2}}}{\alpha r}\left(1-
\frac{Q^2}{M(1-\frac{a^2\alpha^2}{2})\alpha r}\right) 2dt d\varphi
+&
\nonumber \\
& \!\!\!\!\!\! + \left(\alpha^2 r^2
-\frac{4M(1-\frac{3}{2}a^2\alpha^2)}{\alpha r} +
\frac{4Q^2}{\alpha^2 r^2}
\frac{(1-\frac{3}{2}a^2\alpha^2)}{(1-\frac{a^2\alpha^2}{2})}
\right)^{-1}\!\!\!\! dr^2 + \left[r^2 + \frac{4Ma^2}{\alpha
r}\left(1- \frac{Q^2}{(1-\frac{a^2\alpha^2}{2})M\alpha
r}\right)\right] d\varphi^2 + \alpha^2 r^2 dz^2,&
\nonumber \\
& &
                             \label{eq:4004}
\end{eqnarray}
where $\alpha^2 \equiv -\frac{1}{3}\Lambda$, $M$ is the mass per
unit length, $Q$ is the electric charge per unit length, and
parameter $a$ (with units of angular momentum per unit mass, and
with range $0\leq a\alpha \leq 1$) is related to the angular
momentum per unit length of the black hole by
\begin{equation}
a=\frac{2}{3}\frac{J}{M}\left (1-\frac{a^2\alpha^2}{2}
 \right )^{-1/2}.
                          \label{eq:4003}
\end{equation}
The electromagnetic vector potential of the solution is given by
\begin{equation}
 A =\frac{2Q}{\alpha r} \left (-dt +\frac{a}{\sqrt{1-\frac{1}{2}\alpha^2 a^2}}
 d \varphi \right ) \, ,
                           \label{eq:26}
\end{equation}
At $r=0$ is located a curvature singularity. Depending on the
value of $Q$ and of $J$, there are five distinct cases to
consider, namely: (i) $0\leq a^2\alpha^2\leq \frac23 -
\frac{128}{81} \frac{Q^6}{M^4(1-\frac12a^2\alpha^2)^3}$, which
yields a black hole solution with event and Cauchy horizons; (ii)
$a^2\alpha^2 = \frac23 - \frac{128}{81}
\frac{Q^6}{M^4(1-\frac12a^2\alpha^2)^3}$, which corresponds to the
extreme case of the above black hole, where the two horizons
merge; (iii) $\frac23 - \frac{128}{81}
\frac{Q^6}{M^4(1-\frac12a^2\alpha^2)^3} < a^2\alpha^2<\frac23$,
which yields a naked singularity solution; (iv)
$a^2\alpha^2=\frac23$, which gives a null topological singularity;
and (v) $\frac23<a^2\alpha^2 <1$, which gives a pathological black
hole solution  with a single horizon. The most interesting
solutions are given by cases (i) and (ii), which present features
similar to the $\Lambda=0$ Kerr-Newman solution. For example, the
curvature singularity at $r=0$  has a ring structure, and closed
timelike curves are present but inside the inner horizon of the
solution. Solutions (iv) and (v) do not have partners in the
Kerr-Newman family. The black hole (v) is pathological since there
are closed timelike curves outside the horizon. The Carter-Penrose
diagrams of these solutions (i)-(v) are respectively sketched in
items (a)-(e) of Fig. \ref{Fig toroidal AdS}.

When we turn off the rotation, $J=0$ or $a=0$, only the
counterparts of the solutions (i)-(iii) survive. We then have (i)
a black hole solution with event and Cauchy horizons if
$Q^6<3^3M^4/4^3$, (ii) an extreme black hole if $Q^6=3^3M^4/4^3$,
and (iii) a naked singularity if $Q^6>3^3M^4/4^3$. All these three
solutions are free of closed timelike curves  and the curvature
singularity at $r=0$ looses the ring structure. The Carter-Penrose
diagrams of these solutions are identical to the ones sketched in
items (a)-(c) of Fig. \ref{Fig toroidal AdS}.

When we set $Q=0$ we have to consider three cases. If $0\leq
a^2\alpha^2< \frac23$, we have a black hole solution with a single
horizon and without closed timelike curves. The Carter-Penrose
diagrams of this solution is sketched in Fig. \ref{Fig toroidal
AdS}.(e), and it also describes the solution with $J=0$. If
$a^2\alpha^2= \frac23$ we have a null topological singularity
without an horizon [see Fig. \ref{Fig toroidal AdS}.(d)]. This
solution can be considered the extreme uncharged black hole of the
later black hole. Finally, when $a^2\alpha^2> \frac23$ we have a
naked timelike singularity [see Fig. \ref{Fig toroidal AdS}.(c)].

\subsubsection{\label{sec:BH 4D AdS toroidal_magnetic}Magnetic solutions}
The gravitational field of the rotating magnetic solution is given
by \cite{OscarLemos_string}
\begin{eqnarray}
& \!\!\!\!\!\!\!\!\!\!\!\!\!\!\!\!\!\!\!\!\!\!\!\!ds^2 =
   -{\biggl [}\alpha^2 (r^2 + \rho_0^2)-\frac{2M\Omega}{1-4\mu}
   [\alpha^2 (r^2 + \rho_0^2)]^{-1/2} +\frac{4Q^2_{\rm e}}{(1-4\mu)^2}
     [\alpha^2 (r^2 + \rho_0^2)]^{-1} {\biggr ]} dt^2&
                                             \nonumber \\
    & \!\!\!\!\!\!\!\!\!\!\!\!\!\!\!\!\!\!\!\!\!\!\!\!\!\!\!
         -\frac{8}{3}\frac{J}{1-4\mu}{\biggl [}[\alpha^2
        (r^2 + \rho_0^2)]^{-1/2}  -\frac{2Q^2_{\rm e}}
        {M \Omega (1-4\mu)}[\alpha^2 (r^2 + \rho_0^2)]^{-1}
        {\biggr ]} 2dt d\varphi  &
                                                 \nonumber \\
    &
    \!\!\!\!\!\!\!\!\!\!\!\!\!\!\!\!\!\!\!\!\!\!\!\!\!\!\!\!\!\!\!\!\!\!\!\!\!\!
       + \frac{r^2/(r^2 + \rho_0^2)}
    { {\biggl [}  \alpha^2 (r^2 + \rho_0^2) +\frac{2M (4-3\Omega)}
     {1-4\mu}  [\alpha^2 (r^2 + \rho_0^2)]^{-1/2}
     -4\chi_{\rm m}^2 [\alpha^4 (r^2 + \rho_0^2)]^{-1}
     {\biggr ]}}           dr^2  &
                                             \nonumber \\
    & \:\:\:\:\:\:\:\:\:\:\:\:\:\:\:
      + \frac{1}{\alpha^2}{\biggl [}\alpha^2 (r^2 + \rho_0^2)
       +\frac{4M(2-\Omega)/(1-4\mu)}
      {[\alpha^2 (r^2 + \rho_0^2)]^{1/2}}
    -\frac{8(2-\Omega)}{4-3\Omega}\frac{\chi_{\rm m}^2}
     {[\alpha^4 (r^2 + \rho_0^2)]}
         {\biggr ]}   d\varphi^2  + \alpha^2(r^2 + \rho_0^2) dz^2\:, &
                           \label{Met_MJQ}
\end{eqnarray}
and the vector potential is given by
\begin{equation}
A=-\frac{2 \chi_{\rm m}}{\alpha^3 \sqrt{r^2+\rho_0^2}}\frac{
1}{\sqrt{8-6\Omega}}{\biggl [}
   - \sqrt{2\Omega}dt +2\sqrt{2-\Omega}  d\varphi {\biggr ]}\:.
                                    \label{VEC_POTENT_J toroidal 4D}
\end{equation}
In these equations we have $\alpha^2 \equiv -\frac{1}{3}\Lambda$,
$\rho_0$ is the highest root of the equation $\alpha^2 \rho_0^2 +
\frac{8M/(1-4\mu)}{\alpha \rho_0} -\frac{4 \chi_{\rm m}^2}
{\alpha^4 \rho_0^2}=0$, $M$ is the mass per unit length, $Q_{\rm
e}$ is the electric charge per unit length, , $\chi_{\rm m}$ is a
constant that measures the intensity of the magnetic field source,
and we have defined a rotating parameter $\Omega$, which ranges
between $0 \leq \Omega < 1$, as
\begin{equation}
\Omega \equiv 1- \sqrt{1-\frac{8}{9}\frac{J^2 \alpha^2}{M^2}}\:.
                   \label{OMEGA Mag4D}
\end{equation}
The condition that $\Omega$ remains real imposes a restriction on
the allowed values of the angular momentum: $\alpha^2 J^2 \leq
\frac{8}{9}M^2$. Note also that (\ref{VEC_POTENT_J toroidal 4D})
indicates that the static solution ($\Omega=0$) produces only a
longitudinal magnetic field, while the rotating solution ($\Omega
\neq 0$) generates in addition a radial electric field.

The static solution can be obtained by setting $\Omega=0$ (and
thus $J=0$; this also implies that $Q_{\rm e}=0$, as we shall see
just below) on (\ref{Met_MJQ}) and
 (\ref{VEC_POTENT_J toroidal 4D}). This static solution has no horizons and no curvature
singularity. However, it has a conical singularity at $r=0$ with a
deficit angle given by
\begin{equation}
 \delta = 2\pi {\biggl (}1-\lim_{r \rightarrow 0}
\frac{1}{r} \sqrt{\frac{g_{\varphi \varphi}} {g_{r r}}} {\biggr )}
\:.
 \label{conical deficit Mag4D}
 \end{equation}
This conical deficit implies that the period of $\varphi$ is
$\Delta \varphi=2\pi-\delta$, and its presence is associated with
a string with mass density $\mu=\delta/(8\pi)$. The range of the
other coordinates is $-\infty<{t}<+\infty$, $0\leq r< +\infty$,
and $-\infty <z<+\infty$ (cylindrical model). In the toroidal
model, the range of the coordinate $z$ is $0\leq \alpha z< 2\pi$.

Now, the electric charge of the solution described by
(\ref{Met_MJQ}) and (\ref{VEC_POTENT_J toroidal 4D}) can be
written as a function of the other parameters of the solution as
\begin{eqnarray}
Q_{\rm e}=\frac{1-4\mu}{\alpha}\sqrt{\frac{\Omega}{4-3\Omega}}
\,\chi_{\rm m} \:.
                                           \label{Q Mag4D}
\end{eqnarray}
Thus we see that  $\Omega=0$ implies $Q_{\rm e}=0$. This feature
agrees with the conclusions withdrawn from
 (\ref{VEC_POTENT_J toroidal 4D}): the static solution produces no electric field
because its electric charge is zero.

In \cite{OscarLemos_string}, and following \cite{WittenMag4D}, we
have interpreted the magnetic field source as being composed by a
system of two symmetric and superposed electrically charged lines
along the $z$ direction. One of the electrically charged lines is
at rest with positive charge (say), and the other is spinning
around the $z$ direction with a negative electric charge. Clearly,
this system produces no electric field since the total electric
charge is zero and the magnetic field is produced by the angular
electric current. The value of the electric charge per unit length
(\ref{Q Mag4D}), says that after applying a rotation boost in the
$t$-$\varphi$ plane to endow our initial static spacetime with
angular momentum, there appears  a net electric charge. This
result was once again expected since now, besides the magnetic
field along the $z$ direction ($F_{r \varphi} \neq 0$), there is
also a radial electric field ($F_{t r} \neq 0$). In the same
spirit of the explanation presented in subsection
 \ref{sec:Phys Interp magnetic BTZ}, we can show that the rotational
boost induces an asymmetry in the charge densities of the two
above strings that is responsible for the appearance of the radial
electric field. This physical situation is similar to the one that
occurs when one has a copper wire with an electric current and we
apply a translation Lorentz boost to the wire: first, there is
only a magnetic field but, after the Lorentz boost, one also has
an electric field.  The difference is that in the present
situation the Lorentz boost is a rotational one and not a
translational one.

\subsection{\label{sec:BH 4D topological}Black holes with hyperbolic topology}

In this subsection we will briefly present the black hole
solutions of the $\Lambda<0$ Einstein-Maxwell equations with
hyperbolic topology or, upon compactification, with toroidal
topology with genus $g\geq 2$.
 The gravitational field of the electrically charged black
hole solution is given by \cite{topological}
\begin{eqnarray}
d s^2 = - \left (
-1-\frac{\Lambda}{3}r^2-\frac{2M}{r}+\frac{Q^2}{r^2} \right )\,
dt^2 +\frac{dr^2}{
-1-\frac{\Lambda}{3}r^2-\frac{2M}{r}+\frac{Q^2}{r^2} }+r^2 (d
\theta^2+\sinh^2\theta\,d\phi^2)\:,
 \label{AdS topolog bH}
\end{eqnarray}
while the electromagnetic field is still given by (\ref{Max AdS
bh4D}). $M$ and $Q$ are respectively the ADM mass and charge of
the solution.

\begin{figure}[H]
\centering
\includegraphics[height=5cm]{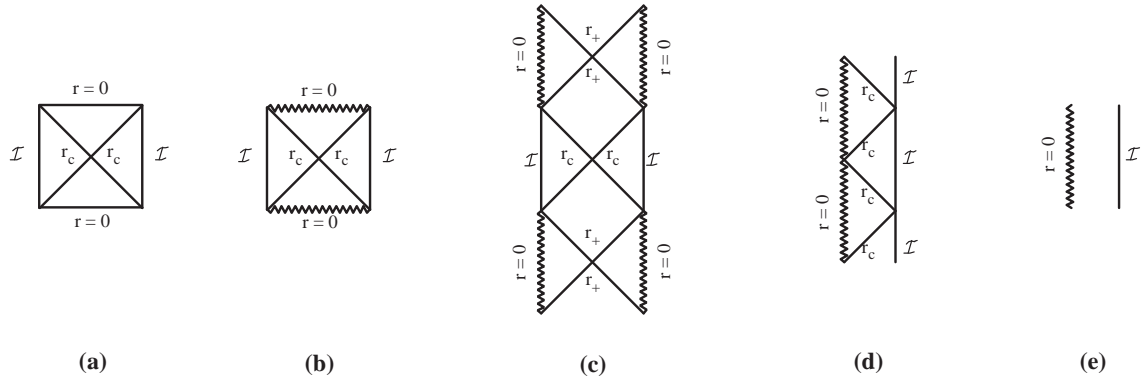}
   \caption{\label{Fig topolog AdS}
Carter-Penrose diagrams of the AdS black holes with hyperbolic
topology discussed in the text of section \ref{sec:BH 4D
topological}. The zigzag line represents a curvature singularity,
${\cal I}$ represents the infinity ($r=\infty$), $r_+$ represents
a black hole event horizon, and $r_{\rm c}$ represents a
cosmological horizon.
 }
\end{figure}

In order to describe the basic properties of these solutions let
us first define the quantity
\begin{eqnarray}
M_{\rm ext}=\frac{1}{3\sqrt{2}|\Lambda|}\left (
\sqrt{1+\frac{4}{3}|\Lambda|Q^2}-2 \right )\left (
\sqrt{1+\frac{4}{3}|\Lambda|Q^2}+1 \right )^{1/2} \:,
                                           \label{M crit topolog}
\end{eqnarray}
which is negative when $Q=0$. When $M=0$ and $Q=0$, the solution
has an horizon that we identify as a cosmological horizon ($r_{\rm
c}$) since it is present when the mass and charge vanish. In this
case $r=0$ is not a curvature singularity, but can be regarded as
a topological singularity (see Brill, Louko, and Peldan in
\cite{topological} for a detailed discussion). The Carter-Penrose
diagram of this solution is drawn in Fig.
 \ref{Fig topolog AdS}.(a). When $Q=0$ and $M>0$, the solution still has a single
horizon, the same cosmological horizon that is present in the
latter case. However, now a curvature singularity is present at
$r=0$.  The corresponding Carter-Penrose diagram of this solution
is represented in Fig. \ref{Fig topolog AdS}.(b). The most
interesting $Q=0$ solutions are present when the mass of the
solution is negative. We have three distinct cases, namely: (i)
$M_{\rm ext}<M<0$, which yields a black hole solution with an
event horizon and a cosmological horizon; (ii) $M=M_{\rm ext}$,
which corresponds to the extreme case of the above black hole,
where the two horizons merge; (iii) $M<M_{\rm ext}$, which yields
a naked timelike singularity solution. The Carter-Penrose diagrams
of these solutions (i)-(iii) are sketched in items (c)-(e) of Fig.
\ref{Fig topolog AdS}. When $Q\neq 0$, one has three cases,
namely: (i) $M>M_{\rm ext}$, which yields a black hole solution
with an event horizon and a cosmological horizon; (ii) $M=M_{\rm
ext}$, which corresponds to the extreme case of the above black
hole; (iii) $M<M_{\rm ext}$, which yields a naked timelike
singularity solution. Note that the presence of the charge does
not introduce an extra horizon, contrary to what usually  occurs
in other black hole solutions. The Carter-Penrose diagrams of
these charged solutions (i)-(iii) are also identical to the ones
sketched in items (c)-(e) of Fig. \ref{Fig topolog AdS}.

\section{\label{sec:BH 4D flat}Black holes in a flat background}

\begin{figure}[H]
\centering
\includegraphics[height=2.8cm]{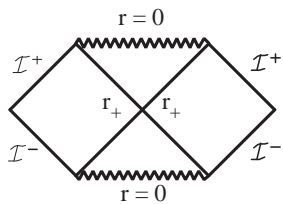}
   \caption{\label{Fig spheric Q=0 flat}
Carter-Penrose diagram of the Schwarzschild ($Q=0$) black hole
discussed in the text of section \ref{sec:BH 4D flat}. The zigzag
line represents a curvature singularity, ${\cal I}$ represents the
infinity ($r=\infty$), and $r_+$ represents a black hole event
horizon.
 }
\end{figure}

In this section we will briefly present the black hole solutions
 of the Einstein-Maxwell equations in a
flat background ($\Lambda=0$). These black holes have a spherical
topology and the  gravitational field of the electrically charged
black hole solution is given by (\ref{AdS spherical bH}), as long
as we set $\Lambda=0$. The electromagnetic field is given by
(\ref{Max AdS bh4D}). In these equations $M$ and $Q$ are still the
ADM mass and charge of the solution. The Reissner-Nordstr\"{o}m
solution ($Q\neq 0$) has three distinct cases, namely: (i) $Q<M$,
which yields a nonextreme black hole solution with event and
Cauchy horizons; (ii) $Q=M$, which corresponds to the extreme case
of the above black hole, where the two horizons merge; and (iii)
$Q>M$, which yields a naked singularity solution. The
Carter-Penrose diagrams of these solutions (i)-(iii) are
respectively sketched in items (a)-(c) of Fig. \ref{Fig spheric
flat}. The Schwarzschild solution ($Q=0$) has a Carter-Penrose
diagram sketched in Fig. \ref{Fig spheric Q=0 flat}.

\begin{figure}[H]
\centering
\includegraphics[height=5cm]{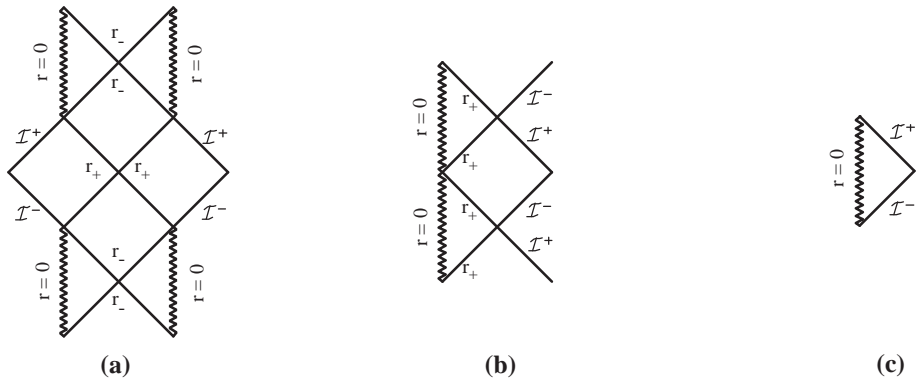}
   \caption{\label{Fig spheric flat}
Carter-Penrose diagrams of the Reissner-Nordstr\"{o}m ($Q\neq 0$)
black holes discussed in the text of section \ref{sec:BH 4D flat}.
The zigzag line represents a curvature singularity, ${\cal I}$
represents the infinity ($r=\infty$), $r_+$ represents a black
hole event horizon, and $r_-$ represents a Cauchy horizon.
 }
\end{figure}

\section{\label{sec:BH 4D dS}Black holes in a de Sitter background}

\begin{figure}[H]
\centering
\includegraphics[height=5cm]{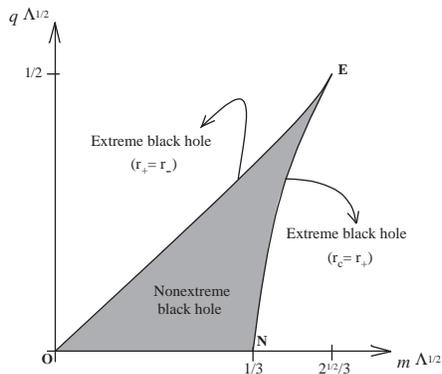}
   \caption{\label{range mq dS bh 4D}
Range of $M$ and $Q$ for which one has a nonextreme black hole, an
extreme black hole with $r_+ = r_{\rm c}$, and an extreme black
hole with $r_- =r_+$. The black holes represented by line $NE$ are
sometimes called as Nariai Reissner-Nordstr\"{o}m black holes. The
line $ON$ represents the nonextreme dS-Schwarzschild, and point
$N$ represents the extreme Nariai Schwarzschild black hole. Point
$E$ represents an extreme black hole with $r_-=r_+ = r_{\rm c}$.
 }
\end{figure}

\begin{figure}[H]
\centering
\includegraphics[height=6cm]{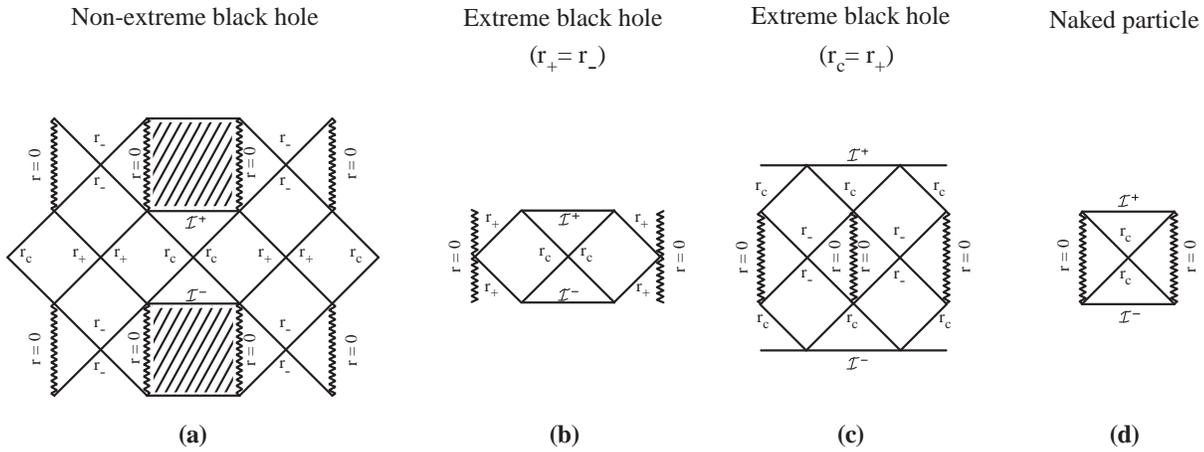}
   \caption{\label{Fig spheric dS}
Carter-Penrose diagrams of the dS$-$Reissner-Nordstr\"{o}m ($Q\neq
0$) black holes discussed in the text of section \ref{sec:BH 4D
dS}. The zigzag line represents a curvature singularity, ${\cal
I}$ represents the infinity ($r=\infty$), $r_{\rm c}$ represents a
cosmological horizon, $r_+$ represents a black hole event horizon,
and $r_-$ represents a Cauchy horizon.
 }
\end{figure}

\begin{figure}[H]
\centering
\includegraphics[height=4cm]{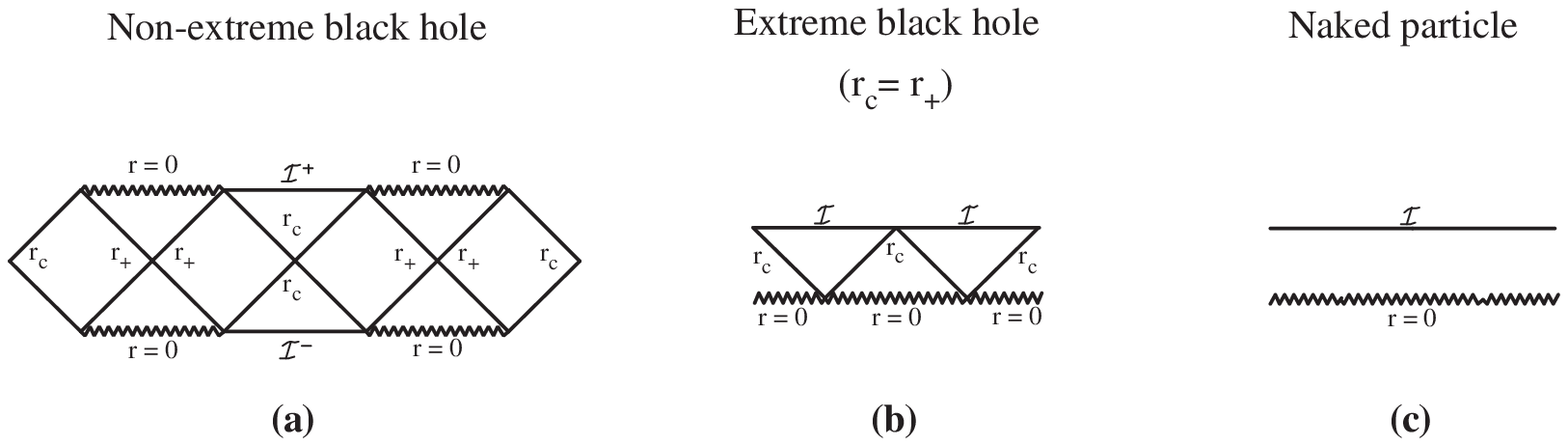}
   \caption{\label{Fig spheric Q=0 dS}
Carter-Penrose diagram of the dS$-$Schwarzschild ($Q=0$) black
hole discussed in the text of section \ref{sec:BH 4D dS}. The
zigzag line represents a curvature singularity, ${\cal I}$
represents the infinity ($r=\infty$), and $r_+$ represents a black
hole event horizon.
 }
\end{figure}

In this section we will briefly present the black hole solutions
 of the Einstein-Maxwell equations in a
de Sitter background ($\Lambda>0$). These black holes have a
spherical topology and the  gravitational field of the charged
black hole solution is given by (\ref{AdS spherical bH}), as long
as we set $\Lambda>0$. The electromagnetic field is given by
(\ref{Max AdS bh4D}). In these equations $M$ and $Q$ are still the
ADM mass and charge of the solution. The
dS$-$Reissner-Nordstr\"{o}m solution ($Q\neq 0$) has four distinct
cases, namely: (a) a nonextreme black hole solution with a
cosmological horizon ($r_{\rm c}$), and with an event ($r_+$) and
Cauchy horizons ($r_-$), where $r_-<r_+<r_{\rm c}$; (b) an extreme
case in which the cosmological horizon merges with the black hole
event horizon ($r_+ = r_{\rm c}$); (c) an extreme case in which
the Cauchy horizon merges with the black hole event horizon ($r_-
=r_+$); and (d) a naked singularity solution. The ranges of $M$
and $Q$ that represent which one of the above black holes is
sketched in Fig. \ref{range mq dS bh 4D}. The Carter-Penrose
diagrams of these solutions (a)-(d) are respectively sketched in
items (a)-(d) of Fig. \ref{Fig spheric dS}. The dS-Schwarzschild
solution ($Q=0$) has three distinct cases, namely: (a) a
nonextreme black hole solution with a cosmological horizon
($r_{\rm c}$), and with an event horizon ($r_+$); (b) an extreme
case in which the cosmological horizon merges with the black hole
event horizon ($r_+ = r_{\rm c}$); and (c) a  naked singularity
solution. The Carter-Penrose diagrams of these solutions (a)-(c)
are respectively sketched in items (a)-(c) of Fig. \ref{Fig
spheric Q=0 dS}. Case (b) of Figs. \ref{Fig spheric dS} and
\ref{Fig spheric Q=0 dS} is sometimes called as Nariai black hole,
although the this nomenclature is not the most appropriate (see
chapter \ref{chap:Extremal Limits}).

%% file: Chapter5.tex

\thispagestyle{empty} \setcounter{minitocdepth}{1}
\chapter[Pair of accelerated black
holes: the C-metric in a generalized $\bm{\Lambda}$
background]{\Large{Pair of accelerated black holes:\\ the C-metric
in a generalized $\bm{\Lambda}$ background}}
\label{chap:PairAccBH} \lhead[]{\fancyplain{}{\bfseries Chapter
\thechapter. \leftmark}} \rhead[\fancyplain{}{\bfseries
\rightmark}]{} \minitoc \thispagestyle{empty}
\renewcommand{\thepage}{\arabic{page}}

\addtocontents{lof}{\textbf{Figures du Chapitre \thechapter}\\}


Kinnersley and Walker \cite{KW}, in 1970, have interpreted the
C-metric, found by Levi-Civita \cite{LeviCivitaCmetric} and by
Weyl \cite{WeylCmetric} in 1918-1919, has describing a pair of
accelerated black holes, with the energy being provided by the
tension of the strings or strut that are present in the solution.
We will review their interpretation in section \ref{sec:flat
C-metric}. From the charged C-metric, Ernst \cite{Ernst} has
generated a knew exact solution in which the black hole pair is
accelerated by an external electromagnetic field. This solution
will also be reviewed in section \ref{sec:flat C-metric}. These
are one of the few exact solutions that contain already a
radiative term. I.e., being accelerated, the black holes
necessarily release radiation and the information concerning the
radiative properties are already included in the metric. The
cosmological C-metric has been found by  Pleba\'nski and
Demia\'nski in 1976 \cite{PlebDem}. In section \ref{sec:AdS
C-metric} we will study in detail the properties and physical
interpretation of the C-metric in an anti-de Sitter (AdS)
background. We follow our work \cite{OscLem_AdS-C} which
complements earlier work done by Emparan, Horowitz and Myers
\cite{EHM1} and by Podolsk\' y \cite{Pod}. Then, in section
\ref{sec:dS C-metric} we will discuss in detail the C-metric in a
de Sitter (dS) background, following our work \cite{OscLem_dS-C}
(which has complemented previous work of Podolsk\' y and Griffiths
\cite{PodGrif2}). The analysis of this chapter will be supported
by a thorough analysis of the causal structure of the solution,
together with the description of the solution in the AdS and dS
4-hyperboloid, and the study of the strings and strut's physics.
We follow the approach of Kinnersley and Walker \cite{KW} and
Ashtekar and Dray \cite{AshtDray}. Although the alternative
approach of Bonnor simplifies in a way the interpretation, we
cannot use the case of the AdS C-metric and dS C-metric, since the
cosmological constant prevents the coordinate transformation of
Bonnor into the Weyl form.
 In section \ref{sec:Conc C-metric} we will summarize the results,
 and in particular we will compare the C-metric in the three
 cosmological backgrounds, AdS, flat and dS.

For a more complete historical overview on the C-metric we ask the
reader to go back to subsection \ref{sec:Pair creation BHs
introduction}.  In chapter \ref{chap:Pair creation} we will use
the solutions analyzed in this chapter to construct the instantons
that describe the quantum process of black hole pair creation in
an external field.

\section[{\bf Pair of accelerated black holes in an anti-de Sitter
background: the AdS C-metric}]{Pair of accelerated black holes in
an anti-de Sitter
background: \\
the AdS C-metric} \label{sec:AdS C-metric}

The plan of this section is as follows. In subsection
\ref{sec:General properties AdS} we present the AdS C-metric and
analyze its curvature and conical singularities. In subsection
\ref{sec:PD AdS} we study the causal diagrams of the solution. In
subsection \ref{sec:Phys_Interp AdS} we give and justify a
physical interpretation to the solution. The description of the
solution in the AdS 4-hyperboloid and the physics of the strut are
analyzed. These two subsections, \ref{sec:PD AdS} and
\ref{sec:Phys_Interp AdS}, are highly related and so, in order to
fully understand each of them, the reading of the other is
required.

\subsection{\label{sec:General properties AdS}General properties of the
A\lowercase{d}S C-metric}

\subsubsection{The AdS C-metric}
The AdS C-metric, i.e., the C-metric with negative cosmological
constant $\Lambda$, has been obtained by Pleba\'nski and
Demia\'nski \cite{PlebDem}. For zero rotation and zero NUT
parameter it is given, according to \cite{PlebDem} (see also
\cite{MannAdS}), by
\begin{equation}
 d s^2 = 1/(\tilde{x}+\tilde{y})^2 (-{\cal F}d\tilde{t}^2+
 {\cal F}^{-1}d\tilde{y}^2+{\cal G}^{-1}d\tilde{x}^2+
 {\cal G}d\tilde{z}^2)\:,
 \label{C-PD AdS}
 \end{equation}
 where
 \begin{eqnarray}
 & &{\cal F}(\tilde{y}) = -\Lambda/6-\tilde{A}^2+ \tilde{y}^2
              -2m\tilde{y}^3+q^2\tilde{y}^4, \nonumber\\
 & &{\cal G}(\tilde{x}) = -\Lambda/6+\tilde{A}^2
 -\tilde{x}^2-2m\tilde{x}^3-q^2\tilde{x}^4.
 \label{FG-1 AdS}
 \end{eqnarray}
The meaning of parameters $\tilde{A}$, $m$, and $q$ will be
clarified soon. For the benefit of comparison with the flat
C-metric, we note that when $\Lambda$ vanishes we have ${\cal
F}(\tilde{y})=-{\cal G}(-\tilde{y})$. It is now convenient to
redefine the parameter $\tilde{A}$ as
$-\Lambda/6+\tilde{A}^2\equiv A^2$, together with the coordinate
transformations: $\tilde{t}=t/A, \tilde{y}=Ay, \tilde{x}=Ax$ and
$\tilde{z}=\phi/A$. With these redefinitions, the gravitational
field of the AdS C-metric is written as
\begin{equation}
 d s^2 = [A(x+y)]^{-2} (-{\cal F}dt^2+
 {\cal F}^{-1}dy^2+{\cal G}^{-1}dx^2+
 {\cal G}d\phi^2)\:,
 \label{C-metric AdS}
 \end{equation}
 where
 \begin{eqnarray}
 & &{\cal F}(y) = {\biggl (}\frac{1}{\ell^2A^2}-1{\biggl )}
                     +y^2-2mAy^3+q^2A^2y^4, \nonumber\\
 & &{\cal G}(x) = 1-x^2-2mAx^3-q^2 A^2 x^4\:,
 \label{FG AdS}
 \end{eqnarray}
 and the Maxwell field in the magnetic case is given by
\begin{eqnarray}
 F_{\rm mag}=-q\, dx\wedge d\phi \:,
\label{F-mag}
\end{eqnarray}
while in the electric case it is given by
 \begin{eqnarray}
 F_{\rm el}=-q\, dt\wedge dy \:.
\label{F-el-Lorentz}
\end{eqnarray}
This solution depends on four parameters namely, $A$ which
 is the acceleration of the black hole,  $m$ which is
 interpreted  as the ADM mass  of the
 non-accelerated black hole, $q$ which is
 interpreted as the ADM electromagnetic charge of the
 non-accelerated black hole and, in general, $q^2=e^2+g^2$ with $e$ and $g$
 being the electric and magnetic charges, respectively,  and finally the cosmological
 length $\ell^2\equiv3/|\Lambda|$. The meaning attributed to
 the parameter $A$ will be understood in section \ref{sec:Phys_Interp AdS}, while the
 physical interpretation given to the parameters $m$ and $q$ is
 justified in the Appendix. We will consider the case $A>0$.

 The coordinates used in (\ref{C-metric AdS})-(\ref{F-el-Lorentz})
 to describe the AdS C-metric are useful to understand the geometrical
properties of the spacetime, but they hide the physical
interpretation of the solution. In order to understand the
physical properties of the source and gravitational field we will
introduce progressively new coordinates more suitable to this
propose, following the approach of Kinnersley and Walker \cite{KW}
and Ashtekar and Dray \cite{AshtDray}. Although the alternative
 approach of Bonnor simplifies in a way the interpretation, we
cannot use it were since the cosmological constant prevents such a
coordinate transformation into the Weyl form.

\subsubsection{\label{sec:CurvSing AdS}Radial Coordinate. Curvature
Singularities}

We start by defining a coordinate $r$ as
\begin{equation}
 r = [A(x+y)]^{-1} \:.
 \label{r}
 \end{equation}
In order to interpret this coordinate as being a radial
coordinate, we calculate a curvature invariant of the metric,
namely the Kretschmann scalar,
\begin{eqnarray}
        R_{\mu\nu\alpha\beta}R^{\mu\nu\alpha\beta} &=&
        \frac{24}{\ell^2}
        +\frac{8}{r^8}{\biggl [}6m^2r^2+12m q^2(2Axr-1)r
                                 \nonumber \\
        & &
        +q^4(7-24Axr+24A^2x^2r^2){\biggr ]}
          \:.
                             \label{R2}
\end{eqnarray}
Clearly, this curvature invariant is equal to $24/\ell^2$ when the
mass $m$ and charge $q$ are both zero. When at least one of these
parameters is not zero, the curvature invariant diverges at $r=0$,
revealing the presence of a curvature singularity. Moreover, when
we take the limit $r\rightarrow \infty$, the curvature singularity
approaches the expected value for a spacetime which is
asymptotically AdS. Therefore it is justified that $r$ is
interpreted as a radial coordinate.

\subsubsection{\label{sec:ConSing AdS}Angular Surfaces. Conical Singularities}

To gain more insight into the physical nature of the AdS C-metric
we now turn our attention into the angular surfaces $t=$constant
and $r=$constant, onwards labelled by $\Sigma$. In this section we
follow \cite{KW}. In order to have the AdS C-metric with correct
signature, $(-+++)$, one must restrict the coordinate $x$ to a
domain on which the function ${\cal{G}}(x)$ is non-negative [see
 (\ref{C-metric AdS})]. The shape of this function depends
crucially on the three parameters $A$, $m$, and $q$. In this work
we will select only the ranges of these three parameters for which
${\cal{G}}(x)$ has at least two real roots, $x_\mathrm{s}$ and
$x_\mathrm{n}$ (say), and demand $x$ to belong to the range
$[x_\mathrm{s},x_\mathrm{n}]$ where ${\cal{G}}(x)\geq 0$. This
restriction has the important advantage of allowing us to endow
the angular surfaces $\Sigma$ with the topology of a compact
surface. In these surfaces we now define two new coordinates,
\begin{eqnarray}
 \theta &=& \int_{x}^{x_\mathrm{n}}{\cal{G}}^{-1/2}dx \:,
                                 \nonumber \\
 \tilde{\phi} &=& \phi/\kappa \:,
                             \label{ang}
\end{eqnarray}
where $\tilde{\phi}$ ranges between $[0,2\pi]$ and $\kappa$ is an
arbitrary positive constant which will be needed later when
regularity conditions at the poles are discussed. The coordinate
$\theta$ ranges between the north pole,
$\theta=\theta_\mathrm{n}=0$, and the south pole,
$\theta=\theta_\mathrm{s}$ (not necessarily at $\pi$). With these
transformations the metric restricted to the surfaces $\Sigma$, $d
\sigma^2 = r^2 ({\cal G}^{-1}dx^2+{\cal G}d\phi^2)$, takes the
form
\begin{equation}
 d \sigma^2
 = r^2{\bigl (} d\theta^2 + \kappa^2{\cal{G}}d\tilde{\phi}^2{\bigr )}\:.
 \label{ang-metric AdS}
 \end{equation}
When $A=0$ or when both $m=0$ and $q=0$, (\ref{ang}) gives
$x=\cos{\theta}$, ${\cal G}=1-x^2=\sin^2{\theta}$ and if we use
the freedom to put $\kappa \equiv 1$, the metric restricted to
$\Sigma$ is given by $d\sigma^2= r^2 ( d\theta^2 +
\sin^2{\theta}\,d\tilde{\phi}^2 )$. This implies that in this case
the angular surface is a sphere and justifies the label given to
the new angular coordinates defined in (\ref{ang}). In this case
the north pole is at $\theta_\mathrm{n}=0$ or $x_\mathrm{n}=+1$
and the south pole is at $\theta_\mathrm{s}=\pi$ or
$x_\mathrm{s}=-1$. In the other cases $x$ and $\sqrt{\cal G}$ can
always be expressed as elliptic functions of $\theta$. The
explicit form of these functions is of no need in this work. All
we need to know is that these functions have a period given by
$2\theta_\mathrm{s}$.

As we shall see, the regularity analysis of the metric in the
region $[0,\theta_\mathrm{s}]$ will play an essential role in the
physical interpretation of the AdS C-metric. The function ${\cal
G}$ is positive and bounded in $]0,\theta_\mathrm{s}[$ and thus,
the metric is regular in this region between the poles. We must be
more careful with the regularity analysis at the poles, i.e., at
the roots of ${\cal G}$. Indeed, if we draw a small circle around
the north pole, in general, as the radius goes to zero, the limit
circunference/radius is not $2\pi$. Therefore, in order to avoid a
conical singularity at the north pole one must require that
$\delta_\mathrm{n}=0$, where
\begin{equation}
 \delta_\mathrm{n} \equiv 2\pi {\biggl (}1-\lim_{\theta \rightarrow 0}
 \frac{1}{\theta}\sqrt{\frac{g_{\phi\phi}}{g_{\theta\theta}}}{\biggr )}
 =2\pi{\biggl (}1- \frac{\kappa}{2} {\biggl |}\frac{d {\cal G}}{dx}
 {\biggl |}_{x_\mathrm{n}}{\biggr )}\:.
 \label{def-N AdS}
 \end{equation}
 Repeating the procedure, this time for the south pole,
 $x_\mathrm{s}$, we conclude that the conical singularity at
 this pole can also be avoid if
\begin{equation}
 \delta_\mathrm{s} \equiv 2\pi{\biggl (}1- \frac{\kappa}{2}
 {\biggl |}\frac{d{\cal G}}{dx}
 {\biggl |}_{x_\mathrm{s}}{\biggr )}=0\:.
 \label{def-S AdS}
 \end{equation}
The so far arbitrary parameter $\kappa$ introduced in (\ref{ang})
plays its important role here. Indeed, if we choose
\begin{equation}
\kappa^{-1}=\frac{1}{2}{\biggl |}\frac{d{\cal G}}{dx}
 {\biggl |}_{x=x_\mathrm{s}}\:,
\label{k-s AdS}
 \end{equation}
 Eq. (\ref{def-S AdS}) is satisfied. However, since we only have
a single constant $\kappa$ at our disposal and this has been fixed
to remove the conical singularity at the south pole, we conclude
that the conical singularity will be present at the north pole.
There is another alternative.  We can choose instead
$2\kappa^{-1}=|d_x {\cal G}|_{x=x_\mathrm{n}}$ (where $d_x$ means
derivative in order to $x$) and by doing so we avoid the deficit
angle at the north pole and leave a conical singularity at the
south pole. In section \ref{sec:Phys_Interp AdS} we will see that
in the extended Kruskal solution the north pole points towards the
other black hole, while the south pole points towards  infinity.
The first choice of $\kappa$ corresponds to a strut between the
black holes while the alternative choice corresponds to two
strings from infinity into each black hole. When we choose
$\kappa$ such that $\delta_{s}=0$, the period of $\phi$ is given
by
\begin{equation}
\Delta \phi=\frac{4 \pi}{|{\cal G}'(x_\mathrm{s})|}\:,
 \label{Period phi-strut}
 \end{equation}
while the choice $\delta_{n}=0$ implies that the period of $\phi$
is given by
\begin{equation}
\Delta \phi=\frac{4 \pi}{|{\cal G}'(x_\mathrm{n})|}\:.
 \label{Period phi-string}
 \end{equation}
We leave a further discussion on the physical nature of the
conical singularities and on the two possible choices for the
value of $\kappa$ to section \ref{sec:PI.2-BH AdS}. There is a
small number of very special cases for which the very particular
condition,
 $|d_x{\cal G}|_{x_\mathrm{n}}= |d_x{\cal G}|_{x_\mathrm{s}}$
is verified. In these special cases, the solution is free of
conical singularities. They will be mentioned bellow.

Since we have managed to put ${\cal G}(x)$ in a form equal to
\cite{KW}, we can now, following \cite{KW} closely, describe the
behavior of ${\cal G}(x)$ for different values of the parameters
$A$, $m$, and $q$. We can divide this discussion in three cases.

{\it 1. Massless uncharged solution} ($m =0$, $q=0$): in this
case, we have $x=\cos \theta$, ${\cal G}=1-x^2=\sin^2 \theta$, and
$\kappa=1$. The angular surface $\Sigma$ is a sphere and this is a
particular case for which both the north and south poles are free
of conical singularities.

{\it 2. Massive uncharged solution} ($m>0$, $q=0$): the massive
uncharged case  must be divided into $mA<3^{-3/2}$, and $mA \geq
3^{-3/2}$. When $mA<3^{-3/2}$, ${\cal G}(x)$ has three roots and,
as justified above, we require $x$ to lie between the two roots
for which ${\cal G}(x)\geq 0$. In doing so we maintain the metric
with the correct signature and have an angular surface $\Sigma$
which is compact. Setting the value of $\kappa$ given in
(\ref{k-s AdS}) one avoids the conical singularity at the south
pole but leave one at the north pole. When $mA \geq 3^{-3/2}$,
$\Sigma$ is an open angular surface. For this reason, onwards we
will analyze only the case $mA<3^{-3/2}$.

{\it 3. Massive charged solution} ($m>0$, $q\neq0$): for a general
massive charged solution, depending on the values of the
parameters $A$, $m$ and $q$, ${\cal G}(x)$ can be positive in a
single compact interval, $]x_\mathrm{s},x_\mathrm{n}[$, or in two
distinct compact intervals, $]x'_\mathrm{s},x'_\mathrm{n}[$ and
$]x_\mathrm{s},x_\mathrm{n}[$, say. In this latter  case we will
work only with the interval $[x_\mathrm{s},x_\mathrm{n}]$ (say)
for which the charged solutions reduce to the uncharged solutions
when $q=0$. These solutions have a conical singularity at one of
the poles. The only massive charged solutions that are totally
free of conical singularities are those which satisfy the
particular conditions $m=|q|$ and $mA>1/4$. This indicates that in
this case the AdS C-metric is an AdS black hole written in an
accelerated coordinate frame. In the massless charged solution
($m= 0$ and $q\neq 0$), ${\cal G}(x)$ is an even function, has two
symmetric roots and is positive between them. The angular surface
$\Sigma$ is therefore compact and there are no conical
singularities at both poles. Once again, this suggests that the
solution is written in an accelerated coordinate frame.

\subsubsection{\label{sec:CoordRange AdS} Coordinate ranges}

In this section we analyze  the important issue of the coordinate
ranges. Rewritten in terms of the new coordinates introduced in
 (\ref{r}) and  (\ref{ang}), the AdS C-metric is given by
\begin{equation}
 d s^2 = r^2 [-{\cal F}(y)dt^2+
 {\cal F}^{-1}(y)dy^2+d\theta^2 + \kappa^2{\cal{G}}(x_{(\theta)})d\tilde{\phi}^2]\:,
 \label{AdS C-metric}
 \end{equation}
where ${\cal F}(y)$ and ${\cal{G}}(x_{(\theta)})$ are given by
(\ref{FG AdS}). The time coordinate $t$ can take any value from
the interval $]-\infty,+\infty[$ and $\tilde{\phi}$ ranges between
$[0,2\pi]$. As we saw in section \ref{sec:CurvSing AdS}, when $m$
or $q$ are not zero there is a curvature singularity at $r=0$.
Therefore, we restrict the radial coordinate to the range
$[0,+\infty[$. On the other hand, in section \ref{sec:ConSing AdS}
we have decided to consider only the values of $A$, $m$, and $q$
for which ${\cal G}(x)$ has at least two real roots,
$x_\mathrm{s}$ and $x_\mathrm{n}$ (say) and have demanded $x$ to
belong to the range $[x_\mathrm{s},x_\mathrm{n}]$ where ${\cal
G}(x)\geq 0$. By doing this we guarantee that the metric has the
correct signature $(-+++)$ and that the angular surfaces $\Sigma$
($t=$constant and $r=$constant) are compact. From $Ar=(x+y)^{-1}$
we then conclude that $y$ must belong to the range $-x\leq y <
+\infty$. Indeed, $y=-x$ corresponds to $r=+\infty$, and
$y=+\infty$ to $r=0$. Note however, that when both $m$ and $q$
vanish there are no restrictions on the ranges of $r$ and $y$
(i.e., $-\infty < r < +\infty$ and $-\infty < y < +\infty$) since
in this case there is no curvature singularity at the origin of
$r$ to justify the constraint on the coordinates.

\subsubsection{\label{sec:Phys_Interp m,e AdS} Mass and charge
parameters}

In this subsection, one gives the physical interpretation of
parameters $m$ and $q$ that appear in the AdS C-metric. Applying
the coordinate transformations to (\ref{C-metric AdS}) (see
\cite{Pod}),
\begin{eqnarray}
& & T=\sqrt{1-\ell^2A^2}A^{-1} t \:,  \;\;\;\;\;
    R=\sqrt{1-\ell^2A^2}(Ay)^{-1} \:, \nonumber \\
& & \theta = \int_{x}^{x_\mathrm{n}}{\cal{G}}^{-1/2}dx \:,
     \;\;\;\;\; \tilde{\phi} = \phi/\kappa \:,
  \label{mq AdS}
  \end{eqnarray}
and setting $A=0$ (and $\kappa=1$) one obtains
 \begin{equation}
 d s^2 = - F(R)\, d T^2 +F^{-1}(R)\, d R^2 +R^2 (d \theta^2
  +\sin^2\theta\,d\tilde{\phi}^2) \:,
\label{mq2 AdS}
\end{equation}
where $F(R)=1+R^2/\ell^2 -2m/R + q^2/R^2$. So, when the
acceleration parameter vanishes, the AdS C-metric, (\ref{C-metric
AdS}), reduces to the AdS-Schwarzschild and
AdS$-$Reissner-Nordstr\"{o}m black holes and the parameters $m$
and $q$ that are present in the AdS C-metric are precisely the ADM
mass and ADM electromagnetic charge of these non-accelerated black
holes. It should however be emphasized that the accelerated black
holes lose mass through radiative processes and so the
determination of the mass of the accelerated black holes would
require the calculation of the Bondi mass, which we do not here.

\subsection{\label{sec:PD AdS} Causal Structure of the A\lowercase{d}S
C-metric}

In this section we analyze the causal structure of the solution.
As occurs with the original flat C-metric \cite{KW,AshtDray}, the
original AdS C-metric, (\ref{AdS C-metric}), is not geodesically
complete. To obtain the maximal analytic spacetime, i.e., to draw
the Carter-Penrose diagrams we will introduce the usual null
Kruskal coordinates.

We now look carefully to the AdS C-metric, (\ref{AdS C-metric}),
with ${\cal F}(y)$ given by  (\ref{FG AdS}). We first notice that,
contrarily to what happens in the $\Lambda\geq 0$ background where
the causal structure and physical nature of the corresponding
C-metric is independent of the relation between the acceleration
$A$ and $\ell \equiv \sqrt{3/|\Lambda}|$, in the $\Lambda < 0$
case we must distinguish and analyze separately the cases
$A>1/\ell$, $A=1/\ell$ and $A<1/\ell$. Later, in section
\ref{sec:Phys_Interp AdS}, we will justify physically the reason
for this distinction. The mathematical reason for this difference
is clearly identified by setting $m=0$ and $q=0$ in (\ref{FG AdS})
giving ${\cal F}(y)=y^2-[1-1/(\ell^2A^2)]$. Since the horizons of
the solution are basically given by the real roots of ${\cal
F}(y)$, we
 conclude that we have to treat separately the cases
(A) $A>1/\ell$ for which ${\cal F}(y)$ can have two real roots,
(B) $A=1/\ell$ for which $y=0$ is double root and (C) $A<1/\ell$
for which ${\cal F}(y)$ has no real roots (see discussion in the
text of Fig. \ref{g1 AdS}). We will consider each of these three
cases separately in  subsections \ref{sec:PD A AdS} and
\ref{sec:PI.2-BH AdS} ($A>1/\ell$ case), \ref{sec:PD B AdS} and
\ref{sec:PI.1-BH AdS} ($A=1/\ell$ case), and \ref{sec:PD C AdS}
and \ref{sec:PI.1-BH.1 AdS} ($A<1/\ell$ case). The description of
the solution depends crucially on the values of $m$ and $q$. In
each subsection, we will consider the three most relevant
solutions, namely: {\it 1. Massless uncharged solution} ($m =0$,
$q=0$), {\it 2. Massive uncharged solution} ($m>0$, $q=0$), and
{\it 3. Massive charged solution} ($m>0$, $q\neq0$).

\subsubsection[Causal Structure of the $A>1/\ell$ solutions]
{\label{sec:PD A AdS} Causal Structure of the $\bm{A>1/\ell}$
solutions}

 \vspace{0.2 cm} \noindent $\bullet$ {\bf Massless uncharged
solution ($\bm{m=0, q=0}$)} \vspace{0.2 cm}

In this case we have
\begin{eqnarray}
 {\cal F}(y) = y^2-y_+^2 \;\;\;\;\;\;\mathrm{with}\;\;\;\;\;\;
 y_+=\sqrt{1-\frac{1}{\ell^2A^2}} \:,
 \label{F1 AdS}
 \end{eqnarray}
and $x \in [x_\mathrm{s}=-1,x_\mathrm{n}=+1]$, $x=\cos \theta$,
${\cal G}=1-x^2=\sin^2 \theta$, $\kappa=1$ and
$\tilde{\phi}=\phi$, with $0\leq \phi\leq 2\pi$. The shapes of
${\cal F}(y)$ and ${\cal G}(x)$ are represented in Fig. \ref{g1
AdS}.

\begin{figure}[H]
\centering
\includegraphics[height=2.2in]{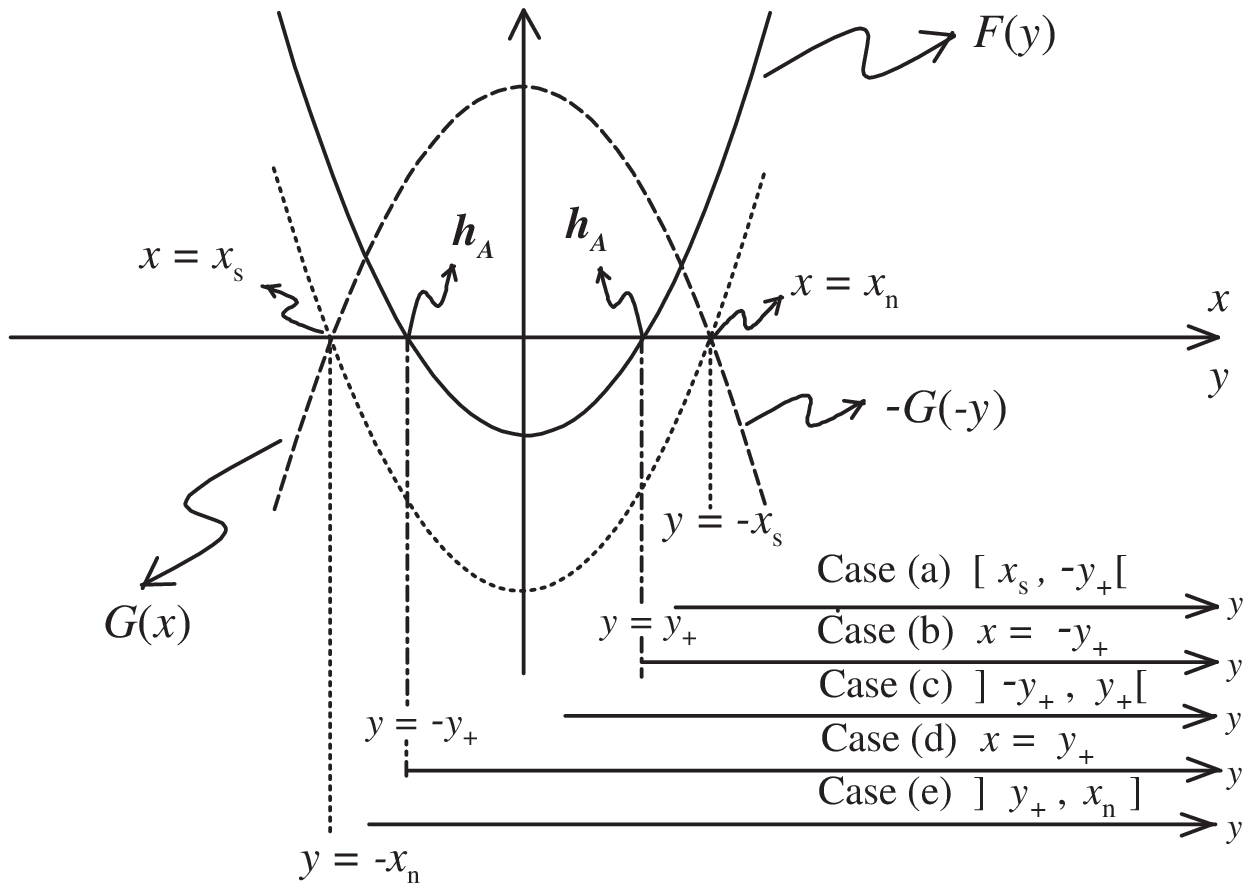}
\caption{\label{g1 AdS}
 Shape of ${\cal G}(x)$ and ${\cal F}(y)$ for the
 $A>1/\ell\,,\,m=0\,$ and $q=0$ C-metric studied in sections
\ref{sec:PD A AdS} and \ref{sec:PI.2-BH AdS}. The allowed range of
$x$ is between $x_\mathrm{s}=-1$ and $x_\mathrm{n}=+1$ where
${\cal G}(x)$ is positive and compact. The permitted range of $y$
depends on the angular direction $x$ ($-x\leq y < +\infty$) and is
sketched for the five cases (a)-(e) discussed in the text. The
presence of an accelerated horizon is indicated by $h_A$. [For
completeness we comment here on two other cases not represented in
the figure but analyzed in the text: for $A=1/\ell\,,\,m=0\,$ and
$q=0$ (this case is studied in sections \ref{sec:PD B AdS} and
\ref{sec:PI.1-BH AdS}), ${\cal F}(y)$ is zero at its minimum and
positive elsewhere.  For $A<1/\ell\,,\,m=0\,$ and $q=0$ (this case
is studied in sections \ref{sec:PD C AdS} and \ref{sec:PI.1-BH.1
AdS}), ${\cal F}(y)$ is always positive and only case (a)
survives.]
 }
\end{figure}

 The angular surfaces $\Sigma$ ($t=$constant and
$r=$constant) are spheres and both the north and south poles are
free of conical singularities. The origin of the radial
coordinate, $r=0$, has no curvature singularity and therefore both
$r$ and $y$ are in the range $]-\infty,+\infty[$. However, in the
general case, where $m$ or $q$ are non-zero, there is a curvature
singularity at $r=0$. Since the discussion of the present section
is only a preliminary to that of the massive general case,
following \cite{AshtDray}, the origin $r=0$ will treat as if it
had a curvature singularity and thus we admit that $r$ belongs to
the range $[0,+\infty[$ and $y$ lies in the region $-x\leq y <
+\infty$. We leave a discussion on the extension to negative
values of $r$ to section \ref{sec:PI.2-BH AdS}.

The general procedure to draw the Carter-Penrose diagrams is as
follows. First, we make use of the null condition
$g_{\mu\nu}k^{\mu}k^{\nu}=0$ (where $k^{\mu}$ is a geodesic
tangent) to introduce the advanced and retarded
Finkelstein-Eddington null coordinates,
\begin{eqnarray}
 u=t-y_* \:;    \;\;\;\;\;\;   v=t+y_* \:,
 \label{uv AdS}
 \end{eqnarray}
where the tortoise coordinate is
\begin{eqnarray}
y_*=\int {\cal F}^{-1}dy=\frac{1}{2y_+} \ln{{\biggl
|}\frac{y-y_+}{y+y_+}{\biggl |}}\:.
 \label{y* AdS}
 \end{eqnarray}
and both $u$ and $v$ belong to the range $]-\infty,+\infty[$. In
these coordinates the metric is given by
\begin{equation}
 d s^2 = r^2 [-{\cal F}dudv+
 d\theta^2 + \sin^2\!\theta \, d\phi^2]\:.
 \label{A.1.1 AdS}
 \end{equation}
 The metric still has coordinate
 singularities at the roots of ${\cal F}$. To overcome this
 unwanted feature we have to introduce Kruskal
 coordinates. Now, due to the lower restriction on the value of $y$
 ($-x\leq y$), the choice of the Kruskal coordinates (and therefore the
 Carter-Penrose diagrams) depends on the angular direction $x$ we
 are looking at. In fact, depending on the value of $x$, the
 region accessible to $y$ might contain two, one or zero roots of
 ${\cal F}$ (see Fig. \ref{g1 AdS}) and so the solution may have two,
 one or zero horizons, respectively. This angular dependence of the
 causal diagram is not new. The Schwarzschild and
  Reissner-Nordstr\"{o}m solutions being spherically symmetric do
 not present this feature but, in the Kerr solution, the
 Carter-Penrose diagram along the pole direction is different
 from the diagram along the equatorial direction. Such a dependence
 occurs also in the flat C-metric \cite{KW}. Back again to the
 AdS C-metric, we have to consider separately five distinct sets of
 angular directions, namely (a) $x_\mathrm{s}\leq x <-y_+$,  (b) $x =-y_+$,  (c)
 $ -y_+ < x <y_+$,  (d) $x=+y_+$ and (e)
 $y_+ < x \leq x_\mathrm{n}$, where $x_\mathrm{s}=-1$ and
 $x_\mathrm{n}=+1$ (see Fig. \ref{g1 AdS}).

\vspace{0.3 cm} (a) $x_\mathrm{s}\leq x <-y_+$: within this
interval of the angular direction, the restriction on the range of
$y$, $-x\leq y < +\infty$, implies that the function ${\cal F}(y)$
is always positive in the accessible region of $y$ (see Fig.
\ref{g1 AdS}), and thus the solution has no horizons. Introducing
the null coordinates defined in  (\ref{uv AdS}) followed by the
Kruskal coordinates $u'=-e^{-y_+ u}<0$ and $v'=+e^{+y_+ v}>0$
gives $u'v'=-e^{2y_+y_*}=-(y-y_+)/(y+y_+)<0$, and (\ref{A.1.1
AdS}) becomes
\begin{eqnarray}
 d s^2 =  r^2 {\biggl [}-\frac{(y+y_+)^2}{y_+^2}du'dv'+
 d\theta^2 + \sin^2\!\theta \,d\phi^2 {\biggr ]} \:,
 \label{A.1.2 AdS}
 \end{eqnarray}
where $y$ and $r=A^{-1}(x+y)^{-1}$ are regarded as functions of
$u'$ and $v'$,
\begin{eqnarray}
 y=y_+\frac{1-u'v'}{1+u'v'}\:, \;\;\;\;
 r=\frac{1}{A}\frac{1+u'v'}{(y_+ +x)-u'v'(y_+ -x)} \:.
 \label{y,r AdS}
 \end{eqnarray}
Now, let us find the values of the product $u'v'$ at $r=0$ and
$r=+\infty$,
\begin{eqnarray}
 \lim_{r \to 0} u'v'=-1\:, \;\;\;\;\;
 \lim_{r \to +\infty} u'v'=\frac{y_+ + x}{y_+ - x}<0 \;
 \mathrm{and \; finite} \:. \nonumber \\
 \label{lim u'v' AdS}
 \end{eqnarray}
So, for $x_\mathrm{s}\leq x <-y_+$, the original massless
uncharged AdS C-metric is described by  (\ref{A.1.2 AdS})
subjected to the following coordinates ranges,
\begin{eqnarray}
 \hspace{-0.5cm} & & \hspace{-0.5cm}
 0 \leq \phi < 2\pi\:, \;\;\; -1 \leq x \leq +1 \:,\;\;\; u'<0\:,
 \;\;\; v'>0 \:,\;\;\;          \\
 \hspace{-0.5cm} & & \hspace{-0.5cm}
  -1\leq u'v'<\frac{y_+ + x}{y_+ - x}  \:.
 \label{ranges u'v' AdS}
 \end{eqnarray}
 This spacetime is however geodesically incomplete. To obtain the
 maximal analytical extension one allows the Kruskal coordinates
 to take also the values $u'\geq 0$ and $v'\leq 0$ as long as
   (\ref{ranges u'v' AdS}) is satisfied.

\begin{figure} [H]
\centering
\includegraphics[height=4.0in]{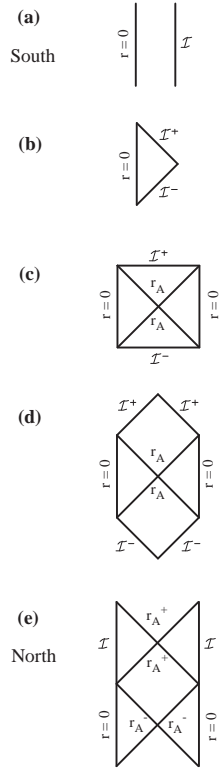}
   \caption{\label{Fig-a1 AdS}
Carter-Penrose diagrams of cases (a)-(e) discussed in the text of
section \ref{sec:PD A AdS} concerning the
$A>1/\ell\,,\,m=0\,,\,q=0$ C-metric. Case (a) describes the
solution seen from the vicinity of the south pole, case (c)
applies to the equatorial vicinity, and case (e) describes the
solution seen from the vicinity of the north pole. An accelerated
horizon is represented by $r_A$, and ${\cal I}^-$ and ${\cal I}^+$
represent respectively the past and future infinity ($r=+\infty$).
$r=0$ corresponds to $y=+\infty$ and $r=+\infty$ corresponds to
$y=-x$.
 }
\end{figure}

Finally, to construct the Carter-Penrose diagram one has to define
the Carter-Penrose coordinates by the usual arc-tangent functions
of $u'$ and $v'$: ${\cal{U}}=\arctan u'$ and ${\cal{V}}=\arctan
v'$, that bring the points at infinity into a finite position. In
general, to find what kind of curve describes the lines $r=0$ or
$r=+\infty$ one has to take the limit of $u'v'$ as $r\rightarrow
0$ (in the case of $r=0$) and the limit of $u'v'$ as $r\rightarrow
+\infty$ (in the case of $r=+\infty$). If this limit is $0$ or
$\infty$ the corresponding line is mapped into a curved null line.
If the limit is $-1$, or a negative and finite constant, the
corresponding line is mapped into a curved timelike line and
finally, when the limit is $+1$, or a positive and finite
constant, the line is mapped into a curved spacelike line. The
asymptotic lines are drawn as straight lines although in the
coordinates ${\cal{U}}$ and ${\cal{V}}$ they should be curved
outwards, bulged. It is always possible to change coordinates so
that the asymptotic lines are indeed straight lines. So, from
(\ref{lim u'v' AdS}) we draw the Carter-Penrose diagram sketched
in Fig. \ref{Fig-a1 AdS}.(a). There are no horizons and both $r=0$
and $r=+\infty$ (${\cal I}$) are timelike lines.

\vspace{0.3 cm}
 (b) $x =-y_+$: for this particular angular
direction, $y$ is restricted to be on $+y_+ \leq y < +\infty$ and
${\cal F}(y)$ is always positive except at $y=+y_+$ (which
corresponds to $r=+\infty$) where it is zero (see Fig. \ref{g1
AdS}). Therefore, the solution has no horizon and the Kruskal
construction is similar to the one described above in case (a).
The only difference is that now $\lim_{r \to +\infty} u'v'=0$ and
thus $r=+\infty$ (${\cal I}$) is represented by a null line in the
Carter-Penrose diagram which is shown in Fig. \ref{Fig-a1
AdS}.(b).

\vspace{0.3 cm}
 (c) $-y_+ < x <y_+$: the demand that $y$ must belong to the range
$[-x;+\infty[$ implies, for this range of the angular direction,
that we have a region I, $-x\leq y < +y_+$, where ${\cal F}(y)$ is
negative and a region II, $+y_+<y<+\infty$, where ${\cal F}(y)$ is
positive (see Fig. \ref{g1 AdS}). There is a single Rindler-like
acceleration horizon ($r_A$) at $y=+y_+$, so called because it is
is absent when $A=0$ and present even when $m=0$ and $q=0$. In
region I one sets the Kruskal coordinates $u'=+e^{-\alpha u}$ and
$v'=+e^{+\alpha v}$ so that
 $u'v'=+e^{2\alpha y_*}$.
In region II one defines $u'=-e^{-\alpha u}$ and $v'=+e^{+\alpha
v}$ in order that $u'v'=-e^{2\alpha y_*}$. We set $\alpha\equiv
y_+$. Thus, in both regions the product $u'v'$ is given by
\begin{eqnarray}
u'v'=-\frac{y-y_+}{y+y_+} \:,
 \label{u'v' AdS}
\end{eqnarray}
and (\ref{A.1.1 AdS}) expressed in terms of the Kruskal
coordinates is given by
\begin{eqnarray}
 d s^2 &=& r^2 {\biggl [}\frac{1}{y_+^2}\frac{{\cal F}}{u'v'}du'dv'+
 d\theta^2 + \sin^2\!\theta \,d\phi^2 {\biggr ]}    \label{F/u'v' AdS} \\
      &=&  r^2 {\biggl [}-\frac{(y+y_+)^2}{y_+^2}du'dv'+
 d\theta^2 + \sin^2\!\theta \,d\phi^2 {\biggr ]}. \nonumber \\
 \label{A.1.3 AdS}
 \end{eqnarray}
The Kruskal coordinates in both regions were chosen in order to
obtain a negative value for the factor ${\cal F}/(u'v')$, which
appears in the metric coefficient $g_{u'v'}$. The value of
constant $\alpha$ was selected in order that the limit of ${\cal
F}/(u'v')$ as $y \to y_+$ stays finite and different from zero. By
doing this, we have removed the coordinate singularity that was
present at the root $y_+$ of ${\cal F}$ [see  (\ref{A.1.1 AdS})].
So, the metric is now well-behaved in the whole range $-x\leq y
<+\infty$ or $0\leq r<+\infty$. The coordinates $y$ and $r$ are
expressed as functions of $u'$ and $v'$ by  (\ref{y,r AdS}) and at
the edges of the interval allowed to $r$, the product $u'v'$ takes
the values
\begin{eqnarray}
 \lim_{r \to 0} u'v'=-1\:, \;\;\;\;\;
 \lim_{r \to +\infty} u'v'=\frac{y_+ + x}{y_+ - x}>0 \;
 \mathrm{and \; finite} \:. \nonumber \\
 \label{lim u'v'.2 AdS}
 \end{eqnarray}
Once again, the maximal analytical extension is achieved by
allowing the Kruskal coordinates $u'$ and $v'$ to take all the
values on the range $]-\infty;+\infty[$, as soon as the condition
 $-1\leq u'v'<(y_+ + x)/(y_+ - x)$ is satisfied.
 The Carter-Penrose diagram for this range of the angular
 direction is drawn in Fig. \ref{Fig-a1 AdS}.(c). $r=0$ is represented by a timelike
line while $r=+\infty$ (${\cal I}$) is a spacelike line. The two
mutual perpendicular straight null lines at $45^{\rm o}$,
$u'v'=0$, represent the accelerated horizon at $y_A=+y_+$ or
$r_A=[A(x+y_+)]^{-1}$.

\vspace{0.3 cm}

 (d) $x = +y_+$: in this particular direction, the region accessible to
$y$ is $-y_+\leq y < +\infty$. ${\cal F}(y)$ is negative in region
I, $-y_+ < y < y_+$ and positive in region II, $y > y_+$. It is
zero at $y=+y_+$ where is located the only horizon ($r_A$) of the
solution and ${\cal F}(y)$ vanishes again at $y=-y_+$ which
corresponds to $r=+\infty$ (see Fig. \ref{g1 AdS}). The Kruskal
construction follows directly the procedure described in case (c).
The only difference is that now $\lim_{r \to +\infty}
u'v'=+\infty$ and thus the $r=+\infty$ line (${\cal I}$) is now
represented by a null line in the Carter-Penrose diagram which is
shown in Fig. \ref{Fig-a1 AdS}.(d).

\vspace{0.3 cm}
 (e) $y_+ < x \leq x_\mathrm{n}$: the region accessible to $y$ must
be separated into three regions (see Fig. \ref{g1 AdS}). In region
I, $-x < y < -y_+$, ${\cal F}(y)$ is positive; in region II, $-y_+
< y < +y_+$, ${\cal F}(y)$ is negative and finally in region III,
$y>+y_+$, ${\cal F}(y)$ is positive again. We have two
Rindler-like acceleration horizons, more specifically, an outer
horizon at $y=-y_+$ or $r_A^+ =[A(x-y_+)]^{-1}$ and an inner
horizon at $y=+y_+$ or $r_A^-=[A(x+y_+)]^{-1}$. Therefore one must
introduce a Kruskal coordinate patch around each of the horizons.
The first patch constructed around $-y_+$ is valid for $-x \leq y
< +y_+$ (thus, it includes regions I and II). In region I we
define $u'=-e^{+\alpha_- u}$ and $v'=+e^{-\alpha_- v}$ so that
$u'v'=-e^{-2\alpha_- y_*}$. In region II one defines
$u'=+e^{\alpha_- u}$ and $v'=+e^{-\alpha_- v}$ in order that
$u'v'=+e^{-2\alpha_- y_*}$. We set $\alpha_-\equiv y_+$. Thus, in
both regions, I and II, the product $u'v'$ is given by
\begin{eqnarray}
u'v'=-\frac{y+y_+}{y-y_+} \:,
 \label{u'v'.2 AdS}
\end{eqnarray}
and  (\ref{A.1.1 AdS}) expressed in terms of the Kruskal
coordinates is given by
\begin{eqnarray}
 d s^2 = r^2 {\biggl [}-\frac{(y-y_+)^2}{y_+^2}du'dv'+
 d\theta^2 + \sin^2\!\theta \,d\phi^2 {\biggr ]} \:,
 \label{A.1.4 AdS}
 \end{eqnarray}
which is regular in this patch $-x \leq y < +y_+$ and, in
particular, it is regular at the root $y=-y_+$ of ${\cal F}(y)$.
However, it is singular at the second root, $y=+y_+$, of
 ${\cal F}(y)$. To regularize the metric around $y=+y_+$, one has
 to introduce new Kruskal coordinates for the second patch which
is built around $y_+$ and is valid for $-y_+ < y < +\infty$ (thus,
it includes regions II and III). In region II we set
$u'=+e^{-\alpha_+ u}$ and $v'=+e^{+\alpha_+ v}$ so that
$u'v'=+e^{+2\alpha_+ y_*}$. In region III one defines
$u'=-e^{-\alpha_+ u}$ and $v'=+e^{+\alpha_+ v}$ in order that
$u'v'=-e^{+2\alpha_+ y_*}$. We set $\alpha_+\equiv y_+$. Thus, in
both regions, II and III, the product $u'v'$ is given by
\begin{eqnarray}
u'v'=-\frac{y-y_+}{y+y_+} \:,
 \label{u'v'.3 AdS}
\end{eqnarray}
and, in this second Kruskal patch,  (\ref{A.1.1 AdS}) is given by
\begin{eqnarray}
 d s^2 = r^2 {\biggl [}-\frac{(y+y_+)^2}{y_+^2}du'dv'+
 d\theta^2 + \sin^2\!\theta \,d\phi^2 {\biggr ]} \:,
 \label{A.1.5 AdS}
 \end{eqnarray}
which is regular in $ y > -y_+$ and, in particular, at the second
root $y=+y_+$ of ${\cal F}(y)$. Once again, in both patches, the
Kruskal coordinates were chosen in order to obtain a factor ${\cal
F}/(u'v')$ negative [see (\ref{F/u'v' AdS})]. The values of
constants $\alpha_-$ and $\alpha_+$ were selected in order that
the limit of ${\cal F}/(u'v')$ as $y \to \mp y_+$ stays finite and
different from zero. To end the construction of the Kruskal
diagram of this solution with two horizons, the two patches have
to be joined together in an appropriate way first defined by
Carter in the Reissner-Nordstr\"{o}m solution.

 From (\ref{u'v'.3 AdS}) and (\ref{u'v'.2 AdS}) we find the values of
product $u'v'$ at the edges $r=0$ and $r=+\infty$ of the radial
coordinate,
\begin{eqnarray}
 \lim_{r \to 0} u'v'=-1\:, \;\;\;\;\;
 \lim_{r \to +\infty} u'v'=\frac{y_+ - x}{y_+ + x}<0 \;
 \mathrm{and \; finite} \:. \nonumber \\
 \label{lim u'v'.4 AdS}
 \end{eqnarray}
and conclude that both $r=0$ and $r=+\infty$ (${\cal I}$) are
represented by timelike lines in the Carter-Penrose diagram
sketched in Fig. \ref{Fig-a1 AdS}.(e). The two accelerated
horizons of the solution are both represented as perpendicular
straight null lines at $45^{\rm o}$ ($u'v'=0$).

 \vspace{0.2 cm} \noindent $\bullet$ {\bf Massive uncharged
solution ($\bm{m >0}$, $\bm{q=0}$)} \vspace{0.2 cm}

\begin{figure} [H]
\centering
\includegraphics[height=2.2in]{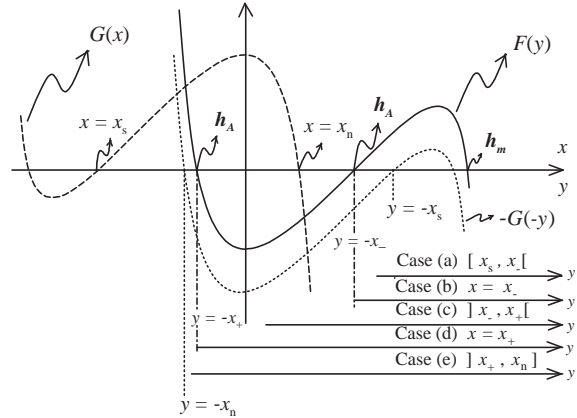}
\caption{\label{g2 AdS}
 Shape of ${\cal G}(x)$ and ${\cal F}(y)$ for the
 $A>1/\ell, mA<3^{-3/2}$, and $q=0$ C-metric studied in sections
\ref{sec:PD A AdS} and \ref{sec:PI.2-BH AdS}. The allowed range of
$x$ is between $x_\mathrm{s}$ and $x_\mathrm{n}$ where ${\cal
G}(x)$ is positive and compact. The permitted range of $y$ depends
on the angular direction $x$ ($-x\leq y < +\infty$) and is
sketched for the five cases (a)-(e) discussed in the text. The
presence of an accelerated horizon is indicated by $h_A$ and the
Schwarzschild-like horizon by $h_m$. [For completeness we comment
on two other cases not represented in the figure: for $A=1/\ell,
mA<3^{-3/2}$ and $q=0$ (this case is studied in sections
\ref{sec:PD B AdS} and \ref{sec:PI.1-BH AdS}), ${\cal F}(y)$ is
zero at its local minimum. For $A<1/\ell, mA<3^{-3/2}$ and $q=0$
(this case is studied in sections \ref{sec:PD C AdS} and
\ref{sec:PI.1-BH.1 AdS}), the local minimum of ${\cal F}(y)$ is
positive and only case (a) survives. For $mA=3^{-3/2}$, ${\cal
G}(x)$ is zero at its local minimum on the left and for
$mA>3^{-3/2}$ ${\cal G}(x)$ is positive between $-\infty$ and
$x_\mathrm{n}$. These two last cases are not studied in the text.]
 }
\end{figure}

Now that the causal structure of the AdS C-metric with $m=0$ and
$q=0$ has been studied, the construction of the Carter-Penrose
diagrams for the $m >0$ case follows up directly. As has justified
in detail in section \ref{sec:ConSing AdS}, we will consider only
the case with small mass or acceleration, i.e., we require
$mA<3^{-3/2}$, in order to have  compact angular surfaces (see
discussion on the text of Fig. \ref{g2 AdS}). We also demand $x$
to belong to the range $[x_\mathrm{s},x_\mathrm{n}]$ (see Fig.
\ref{g2 AdS}) where ${\cal G}(x)\geq 0$ and such that
$x_\mathrm{s} \to -1$ and $x_\mathrm{n} \to +1$ when $mA \to 0$.
By satisfying the two above conditions we endow the $t=$constant
and $r=$constant surfaces with the topology of a compact surface.

The technical procedure to obtain the Carter-Penrose diagrams is
similar to the one described along section \ref{sec:PD A AdS}. In
what concerns the physical conclusions, we will see that the
essential difference is the presence of an extra horizon, a
Schwarzschild-like horizon ($r_+$), due to the non-vanishing mass
parameter, in addition to the accelerated Rindler-like horizon
($r_A$) which has due to non-vanishing $A$. Another important
difference, as stated in section \ref{sec:CurvSing AdS}, is the
presence of a curvature singularity at $r=0$ and the existence of
a conical singularity at the north pole (see section
\ref{sec:ConSing AdS}).

 Once more the Carter-Penrose diagrams
depend on the angular direction we are looking at (see Fig.
\ref{g2 AdS}). We have to analyze separately five distinct cases,
namely (a)
 $x_\mathrm{s}\leq x < x_-$, (b) $x =x_-$,  (c)
 $ x_- < x < x_+$, (d) $x=x_+$ and (e)
 $x_+ < x \leq x_\mathrm{n}$, which are the massive counterparts of
 cases (a)-(e) that were considered in section
  \ref{sec:PD A AdS}.
 When $m \to 0$ we have $x_\mathrm{s} \to -1$, $x_\mathrm{n} \to +1$,
$x_- \to -y_+$ and $x_+ \to +y_+$.

\begin{figure}[H]
\centering
\includegraphics[height=13cm]{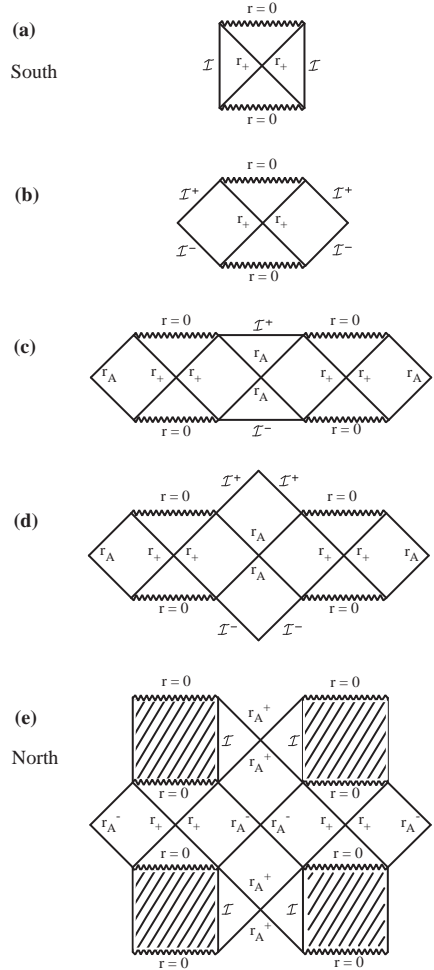}
   \caption{\label{Fig-a2 AdS}
Carter-Penrose diagrams of cases (a)-(e) discussed in the text of
section \ref{sec:PD A AdS} concerning the $A>1/\ell, mA<3^{-3/2}$,
and $q=0$ C-metric. Case (a) describes the solution seen from the
vicinity of the south pole, case (c) applies to the equatorial
vicinity, and case (e) describes the solution seen from the
vicinity of the north pole. The zigzag line represents a curvature
singularity, an accelerated horizon is represented by $r_A$, the
Schwarzschild-like horizon is sketched as $r_+$.  $r=0$
corresponds to $y=+\infty$ and $r=+\infty$ (${\cal I}$)
corresponds to $y=-x$. The hatched region does not belong to the
solution. In diagrams (c)-(e) we have to glue indefinitely copies
of the represented figure in the left and right sides of it. In
diagram (e) a similar gluing must be done in the top and bottom
regions.
 }
\end{figure}

\vspace{0.3 cm}
 (a) $x_\mathrm{s}\leq x <x_-$:
the Carter-Penrose diagram [Fig. \ref{Fig-a2 AdS}.(a)] for this
range of the angular direction has a spacelike curvature
singularity at $r=0$, a timelike line that represents $r=+\infty$
(${\cal I}$) and a Schwarzschild-like horizon ($r_+$) that was not
present in the $m=0$ corresponding diagram Fig. \ref{Fig-a1
AdS}.(a). The diagram is similar to the one of the
AdS-Schwarzschild solution, although now the curvature singularity
has an acceleration $A$, as will be seen in section
\ref{sec:Phys_Interp AdS}.

\vspace{0.3 cm}
 (b) $x =x_-$:
the curvature singularity $r=0$ is also a spacelike line in the
Carter-Penrose diagram [see Fig \ref{Fig-a2 AdS}.(b)] and there is
a Schwarzschild-like horizon ($r_+$). The infinity, $r=+\infty$
(${\cal I}$), is represented by a null line. The origin is being
accelerated (see section \ref{sec:Phys_Interp AdS}).

\vspace{0.3 cm}
  (c) $x_- < x <x_+$:
the Carter-Penrose diagram [Fig. \ref{Fig-a2 AdS}.(c)] has a more
complex structure that can be divided into left, middle and right
regions. The middle region contains the spacelike infinity (${\cal
I}$) and an accelerated Rindler-like horizon,
$r_A=[A(x-x_-)]^{-1}$, that is already present in the $m=0$
corresponding diagram [see Fig. \ref{Fig-a1 AdS}.(c)]. The left
and right regions both contain a spacelike curvature singularity
and a Schwarzschild-like horizon, $r_+$.

\vspace{0.3 cm}
 (d) $x = x_+$:
the Carter-Penrose diagram [Fig. \ref{Fig-a2 AdS}.(d)] for this
particular value of the angular direction is similar to the one of
above case (c). The only difference is that $r=+\infty$ (${\cal
I}$) is represented by a null line rather than a spacelike line.

\vspace{0.3 cm}
  (e) $x_+ < x \leq x_\mathrm{n}$:
the Carter-Penrose diagram [Fig. \ref{Fig-a2 AdS}.(e)] can again
be divided into left, middle and right regions. The middle region
consists of a timelike line representing $r=+\infty$ (${\cal I}$)
and two accelerated Rindler-like horizons, an inner one
($r_A^-=[A(x-x_-)]^{-1}$) and an outer one
($r_A^+=[A(x-x_+)]^{-1}$), that were already present in the $m=0$
corresponding diagram [Fig. \ref{Fig-a1 AdS}.(e)]. The left and
right regions both contain a spacelike curvature singularity and a
Schwarzschild-like horizon ($r_+$).

 \vspace{0.2 cm} \noindent $\bullet$ {\bf Massive charged solution
($\bm{m >0}$, $\bm{q\neq0}$)} \vspace{0.2 cm}

\begin{figure} [H]
\centering
\includegraphics[height=2.2in]{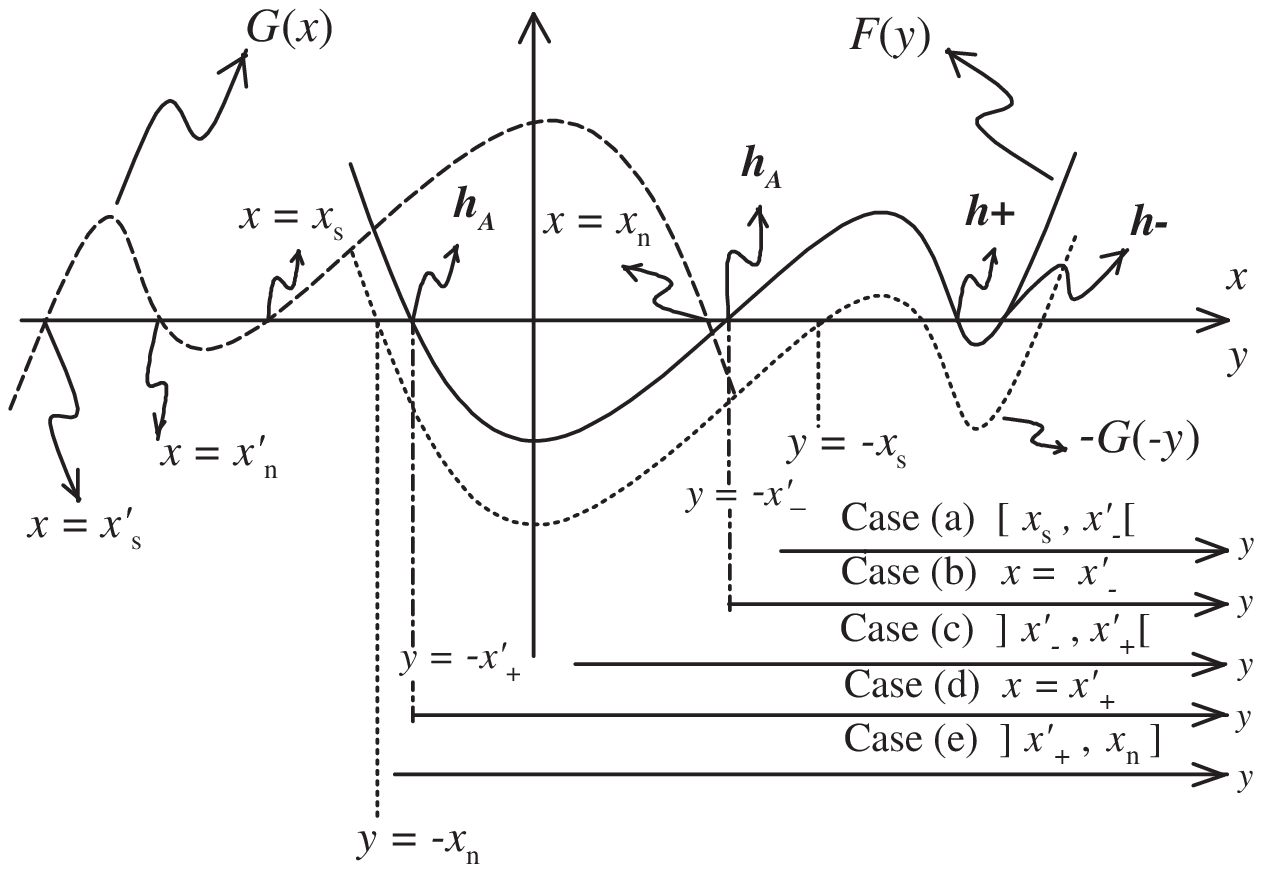}
\caption{\label{g3 AdS}
 Shape of ${\cal G}(x)$ and ${\cal F}(y)$ for the non-extremal
charged massive C-metric (with $A>1/\ell$) studied in sections
\ref{sec:PD A AdS} and \ref{sec:PI.2-BH AdS}. The allowed range of
$x$ is between $x_\mathrm{s}$ and $x_\mathrm{n}$ where ${\cal
G}(x)$ is positive and compact. The permitted range of $y$ depends
on the angular direction $x$ ($-x\leq y < +\infty$) and is
sketched for the five cases (a)-(e) discussed in the text. The
presence of an accelerated horizon is indicated by $h_A$ and the
inner and outer charged horizons by $h-$ and $h+$. In the extremal
case, $h-$ and $h+$ superpose each other and in the naked case
${\cal F}(y)>0$ in the local minimum on the right. [For
completeness we comment on two other cases not represented in the
figure: for $A=1/\ell$ (this case is studied in sections
\ref{sec:PD B AdS} and \ref{sec:PI.1-BH AdS}), ${\cal F}(y)$ is
zero at its local minimum on the left.  For $A<1/\ell$ (this case
is studied in sections \ref{sec:PD C AdS} and \ref{sec:PI.1-BH.1
AdS}), the local minimum on the left of ${\cal F}(y)$ is positive
and only case (a) survives.]
 }
\end{figure}

When both the mass and charge parameters are non-zero, depending
on the values of the parameters $A$, $m$ and $q$, ${\cal G}(x)$
can be positive in a single compact interval,
$]x_\mathrm{s},x_\mathrm{n}[$, or in two distinct compact
intervals, $]x'_\mathrm{s},x'_\mathrm{n}[$ and
$]x_\mathrm{s},x_\mathrm{n}[$, say (see Fig. \ref{g3 AdS}). In
this latter case we will work only with the interval
$[x_\mathrm{s},x_\mathrm{n}]$ (say) for which the charged
solutions are in the same sector of those we have analyzed in the
last two subsections when $q \to 0$.

Depending also on the values of $A$, $m$ and $q$, the function
${\cal F}(y)$ can have four roots, three roots (one of them
degenerated) or two roots (see the discussion on the text of Fig.
\ref{g3 AdS}). As will be seen, the first case describes a pair of
accelerated AdS$-$Reissner-Nordstr\"{o}m (AdS-RN) black holes, the
second case describes a pair of extremal AdS-RN black holes and
the third case describes a pair of naked AdS-RN singularities.

The essential differences between the Carter-Penrose diagrams of
the massive charged solutions and those of the massive uncharged
solutions are: (i) the curvature singularity is now represented by
a timelike line rather than a spacelike line, (ii) excluding the
extremal and naked cases, there are now (in addition to the
accelerated Rindler-like horizon, $r_A$) not one but two extra
horizons, the expected inner ($r_-$) and outer ($r_+$) horizons
associated to the charged character of the solution.

\begin{figure} [H]
\centering
\includegraphics[height=19cm]{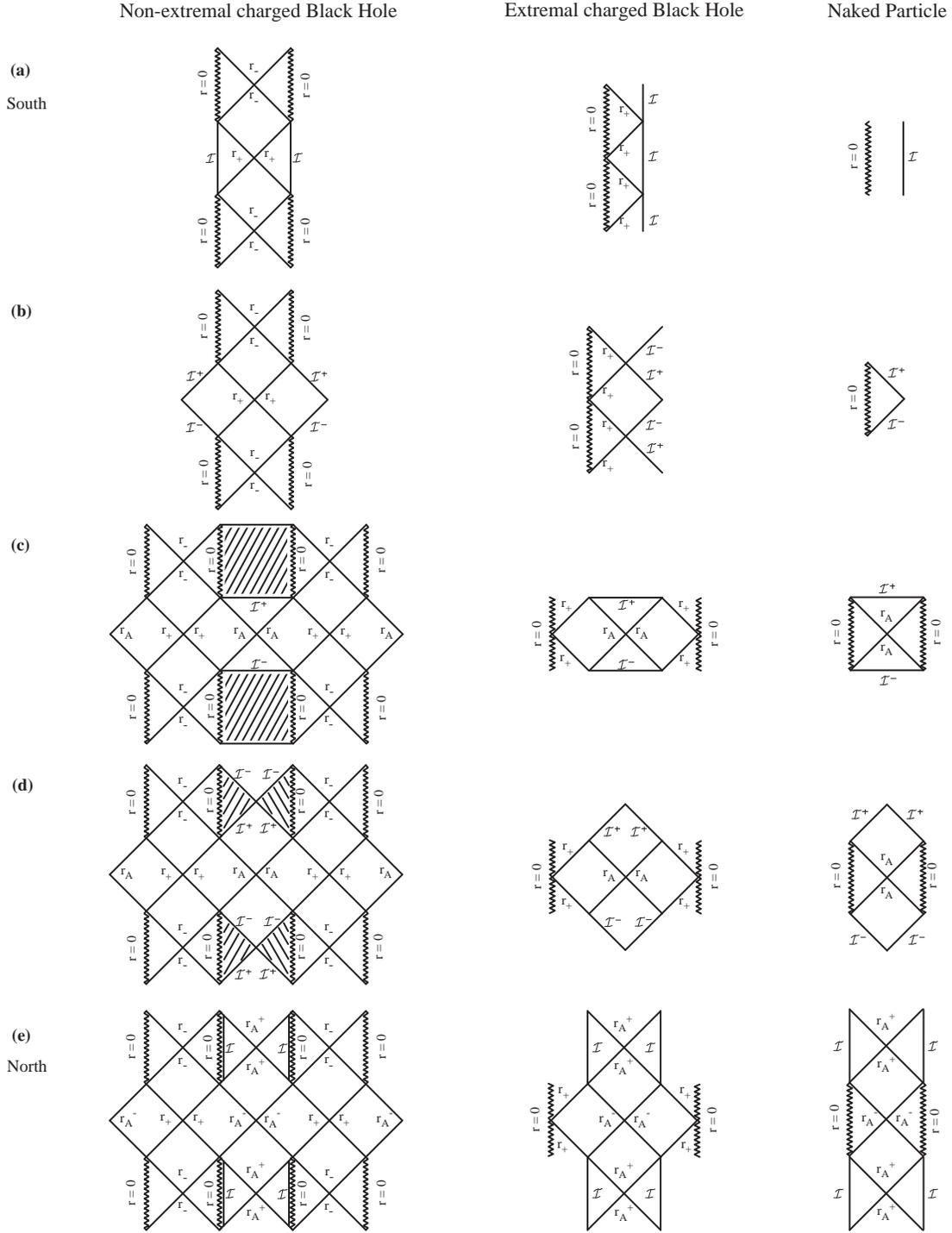}
   \caption{\label{Fig-a3 AdS}
Carter-Penrose diagrams of cases (a)-(e) discussed in the text of
section \ref{sec:PD A AdS} concerning the charged massive
C-metric. The left column represents the non-extremal case, the
middle column represents the extremal case and the right column
represents the naked charged case. The row (a) describes the
solution seen from the vicinity of the south pole, row (c) applies
to the equatorial vicinity, and row (e) describes the solution
seen from the vicinity of the north pole.
 }
\end{figure}

 Below, we study the causal structure of the electric or magnetic
 counterparts of cases (a)-(e) discussed in the two
 last sections (see Fig. \ref{g3 AdS}), namely (a)
 $x_\mathrm{s}\leq x < x'_-$, (b) $x =x'_-$,  (c)
 $ x'_- < x < x'_+$,  (d) $x=x'_+$ and  (e)
 $x'_+ < x \leq x_\mathrm{n}$.
 When $q \to 0$ we have $x'_- \to x_-$ and $x'_+ \to x_+$.
The Carter-Penrose diagrams are drawn in Fig. \ref{Fig-a3 AdS}. In
these diagrams, the left column represents the non-extremal case,
the middle column represents the extremal case and the right
column represents the naked charged case. The row (a) describes
the solution seen from the vicinity of the south pole, row (c)
applies to the equatorial vicinity, and row (e) describes the
solution seen from the vicinity of the north pole. The zigzag line
represents a curvature singularity, an accelerated horizon is
represented by $r_A$, the inner and outer charge associated
horizons are sketched as $r_-$ and $r_+$. ${\cal I}^-$ and ${\cal
I}^+$ represent respectively the past and future infinity
($r=+\infty$). $r=0$ corresponds to $y=+\infty$ and $r=+\infty$
corresponds to $y=-x$. The hatched region does not belong to the
solution. In diagrams (c)-(e) we have to glue indefinitely copies
of the represented figure in the left and right sides of it. In
some of the diagrams, a similar gluing must be done in the top and
bottom regions.

 \vspace{1 cm}

 \vspace{0.3 cm}
 (a) $x_\mathrm{s}\leq x <x'_-$:
both the curvature singularity, $r=0$, and $r=+\infty$ (${\cal
I}$) are represented by a spacelike line in the Carter-Penrose
diagram [Fig. \ref{Fig-a3 AdS}.(a)]. Besides, in the non-extremal
case, there is an inner horizon ($r_-$) and an outer horizon
($r_+$) associated to the charged character of the solution. In
the extremal case the two horizons $r_-$ and $r_+$ become
degenerate and so there is a single horizon, $r_+$ (say), and in
the naked case there is no horizon. The diagram is similar to the
one of the AdS$-$Reissner-Nordstr\"{o}m solution, although now the
curvature singularity has an acceleration $A$, as will be seen in
section \ref{sec:Phys_Interp AdS}.

\vspace{0.3 cm}
  (b) $x =x'_-$:
the curvature singularity $r=0$ is a spacelike line in the
Carter-Penrose diagram [see Fig. \ref{Fig-a3 AdS}.(b)] and
$r=+\infty$ (${\cal I}$) is represented by a null line. Again, in
the non-extremal case, there is an inner horizon ($r_-$) and an
outer horizon ($r_+$) associated to the charged character of the
solution. In the extremal case there is a single horizon, $r_+$,
and in the naked case there is no horizon. The origin is being
accelerated (see section \ref{sec:Phys_Interp AdS}).

\vspace{0.3 cm}
 (c) $x'_- < x <x'_+$:
the Carter-Penrose diagram [Fig. \ref{Fig-a3 AdS}.(c)] has a
complex structure. As before [see Fig \ref{Fig-a2 AdS}.(c)], it
can be divided into left, middle and right regions. The middle
region contains the spacelike infinity (${\cal I}$) and an
accelerated Rindler-like horizon, $r_A=[A(x-x'_-)]^{-1}$, that was
already present in the $m=0\,,\,q=0$ corresponding diagram [see
Fig. \ref{Fig-a1 AdS}.(c)]. The left and right regions both
contain a timelike curvature singularity ($r=0$). In addition,
these left and right regions contain, in the non-extremal case, an
inner horizon ($r_-$) and an outer horizon ($r_+$), in the
extremal case they contain is a single horizon ($r_+$), and in the
naked case they have no horizon.

\vspace{0.3 cm}
  (d) $x = x'_+$:
the Carter-Penrose diagram [Fig. \ref{Fig-a3 AdS}.(d)] for this
particular value of the angular direction is similar to the one of
above case (c). The only difference is that $r=+\infty$ (${\cal
I}$) is represented by a null line rather than a spacelike line.

\vspace{0.3 cm}
 (e) $x'_+ < x \leq x_\mathrm{n}$:
the Carter-Penrose diagram [Fig. \ref{Fig-a3 AdS}.(e)]. As before
[see Fig \ref{Fig-a2 AdS}.(e)], it can be divided into left,
middle and right regions. The middle region consists of a timelike
line representing $r=+\infty$ (${\cal I}$) and two accelerated
Rindler-like horizon, $r_A^-=[A(x-x'_-)]^{-1}$ and
$r_A^+=[A(x-x'_+)]^{-1}$, that were already present in the $m=0$
and $q=0$ corresponding diagram [see Fig. \ref{Fig-a1 AdS}.(e)].
The left and right regions both contain a timelike curvature
singularity ($r=0$). In addition, these left and right regions
contain, in the non-extremal case, an inner horizon ($r_-$) and an
outer horizon ($r_+$), in the extremal case they contain is a
single horizon ($r_+$), and in the naked case they have no horizon
(see however the physical interpretation of this case as a black
hole in the end of subsection \ref{sec:Phys_Interp AdS}).

\subsubsection[Causal Structure of the $A=1/\ell$ solutions]{\label{sec:PD B AdS}
 Causal Structure of the $\bm{A=1/\ell}$ solutions}

The $A=1/\ell$ case was studied in detail in \cite{EHM1}. In
particular,  the causal structure of the massive uncharged
solution was discussed. For completeness, we will also present the
causal diagrams of the massless uncharged solution and of the
massive charged solution.

Once more, due to the lower restriction on the value of $y$
($-x\leq y$), the causal diagrams depend on the angular direction
$x$ we are looking at. We have to consider
 separately three distinct sets of angular directions (see
discussion on the text of Figs. \ref{g1 AdS}, \ref{g2 AdS} and
\ref{g3 AdS}), namely (a) $x_\mathrm{s}\leq x <0$, (b) $x =0$ and
 (c) $0 < x \leq x_\mathrm{n}$, where $x_\mathrm{s}=-1$ and
 $x_\mathrm{n}=+1$ when $m=0$ and $q=0$.

 \vspace{0.2 cm} \noindent $\bullet$ {\bf Massless uncharged
solution ($\bm{m=0, q=0}$)} \vspace{0.2 cm}

\begin{figure} [H]
\centering
\includegraphics[height=4.8cm]{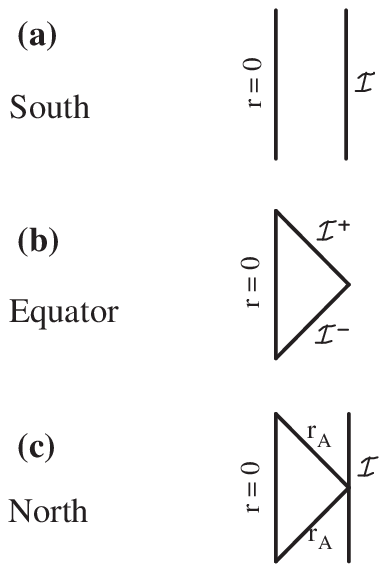}
   \caption{\label{Fig-b1 AdS}
Carter-Penrose diagrams of cases (a)-(c) discussed in the text of
section \ref{sec:PD B AdS} concerning the $A=1/\ell\,,\,m=0\,$,
and $q=0$ C-metric.  $r_A=(A x)^{-1}$.  In diagrams (a) and (c) we
have to glue indefinitely copies of the represented figure in the
top and bottom regions of it.
 }
\end{figure}

In this case we have $x \in [x_\mathrm{s}=-1,x_\mathrm{n}=+1]$,
$x=\cos \theta$, ${\cal G}=1-x^2=\sin^2 \theta$, $\kappa=1$ and
${\cal F}(y) = y^2$ (see discussion on the text of Fig. \ref{g1
AdS}). The angular surfaces $\Sigma$ ($t=$constant and
$r=$constant) are spheres free of conical singularities. The
origin of the radial coordinate $r$ has no curvature singularity
and therefore both $r$ and $y$ can lie in the range
$]-\infty,+\infty[$. However, in the general case, where $m$ or
$q$ are non-zero, there is a curvature singularity at $r=0$. Since
the discussion of the present section is only a preliminary to
that of the massive general case, following \cite{AshtDray}, we
will treat the origin $r=0$ as if it had a curvature singularity
and thus we admit that $r$ belongs to the range $[0,+\infty[$ and
$y$ lies in the region $-x\leq y < +\infty$. The Carter-Penrose
diagrams are drawn in Fig. \ref{Fig-b1 AdS}. In case (c) $0 < x
\leq x_\mathrm{n}$, and only in this case, there is an accelerated
horizon, $r_A=(Ax)^{-1}$.

 \vspace{0.2 cm} \noindent $\bullet$ {\bf Massive uncharged
solution ($\bm{m >0}$, $\bm{q=0}$)} \vspace{0.2 cm}

\begin{figure} [H]
\centering
\includegraphics[height=5.8cm]{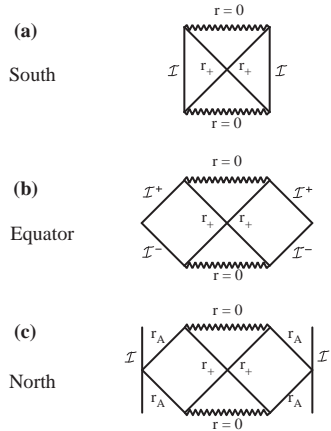}
   \caption{\label{Fig-b2 AdS}
 Carter-Penrose diagrams of cases (a)-(c) discussed in the text
of section \ref{sec:PD B AdS} concerning the $A=1/\ell,
mA<3^{-3/2}$, and $q=0$ C-metric.  $r_A=(A x)^{-1}$ is a
degenerated horizon (see \cite{EHM1}).  In diagram (c) we have to
glue indefinitely copies of the represented figure in the top and
bottom regions of it.
 }
\end{figure}

The causal diagrams of this solution were first presented in
\cite{EHM1} and are drawn in Fig. \ref{Fig-b2 AdS}. In the case
(c) $0 < x \leq x_\mathrm{n}$, and only in this case, there is an
accelerated horizon, $r_A=(Ax)^{-1}$ which is degenerated (see
\cite{EHM1}). The Schwarzschild-like horizon is at
$r_+=A^{-1}[x+1/(2mA)]^{-1}$.

 \vspace{0.2 cm} \noindent $\bullet$ {\bf Massive charged solution
($\bm{m >0}$, $\bm{q\neq0}$)} \vspace{0.2 cm}

\begin{figure}[H]
\centering
\includegraphics[height=11.5cm]{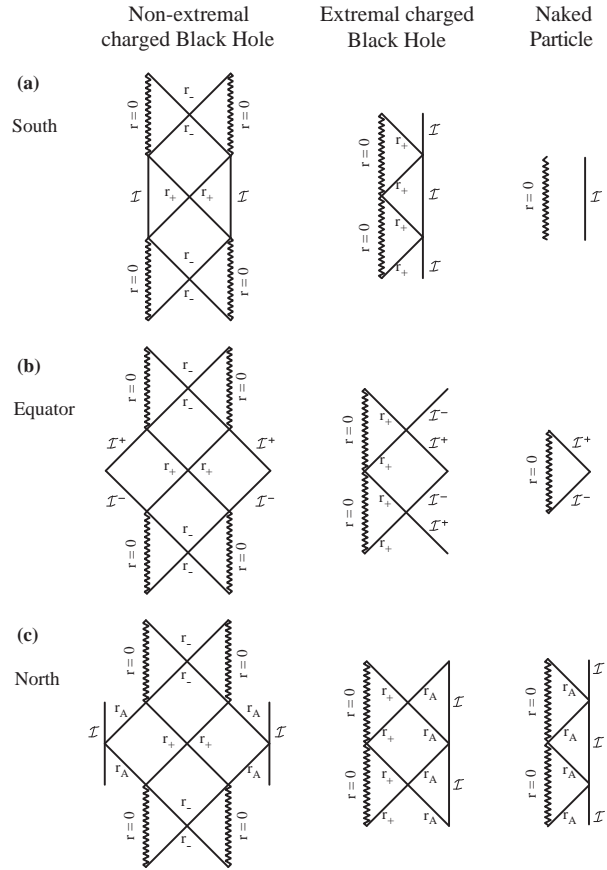}
   \caption{\label{Fig-b3 AdS}
Carter-Penrose diagrams of cases (a)-(c) discussed in the text of
section \ref{sec:PD B AdS} concerning the charged massive C-metric
with $A=1/\ell$. The left column represents the non-extremal black
hole, the middle column represents the extremal black hole and the
right column represents the naked charged particle.  $r_A=(A
x)^{-1}$ is an accelerated horizon and $r_-$ and $r_+$ are charged
associated horizons.  In these diagrams  we have to glue
indefinitely copies of the represented figure in the top and
bottom regions of it.
 }
\end{figure}

The Carter-Penrose diagrams of the solution for this range of
parameters is sketched in Fig. \ref{Fig-b3 AdS}. In these
diagrams, the left column represents the non-extremal black hole,
the middle column represents the extremal black hole and the right
column represents the naked charged particle. The row (a)
describes the solution seen from an angle that is between the
south pole (including) and the equator (excluding), row (b)
applies only to the equatorial direction, and row (c) describes
the solution seen from an angle between the equator (excluding)
and the north pole (including).

\subsubsection[Causal Structure of the $A<1/\ell$ solutions]{\label{sec:PD C AdS}
 Causal Structure of the $\bm{A<1/\ell}$ solutions}

The $A<1/\ell$ case was first analyzed in \cite{Pod}. We
complement it with the analysis of the causal structure.
Contrarily to the cases $A > 1/\ell$ and $A= 1/\ell$, the causal
diagrams of this spacetime do not depend on the angular direction
we are looking at. The reason for this feature is clearly
identified and explained in the discussion on the text of Figs.
\ref{g1 AdS}, \ref{g2 AdS} and \ref{g3 AdS}.

 \vspace{0.2 cm} \noindent $\bullet$ {\bf Massless uncharged
solution ($\bm{m=0, q=0}$)} \vspace{0.2 cm}

 The Carter-Penrose diagram is identical to the one of
the AdS solution ($A=0\,,\,m=0\,,\,q=0$) and is sketched in Fig.
\ref{Fig-b1 AdS}.(a). The origin has an acceleration $A$, as will
be seen in section \ref{sec:Phys_Interp AdS}.

\vspace{1 cm}
 \vspace{0.2 cm} \noindent $\bullet$ {\bf Massive uncharged
solution ($\bm{m >0}$, $\bm{q=0}$)} \vspace{0.2 cm}

The Carter-Penrose diagram is identical to the one of the
AdS-Schwarzschild solution ($A=0\,,\,m>0\,,\,q=0$) and is drawn in
Fig. \ref{Fig-b2 AdS}.(a). The origin has an acceleration $A$, as
will be seen in section \ref{sec:Phys_Interp AdS}.

 \vspace{0.2 cm} \noindent $\bullet$ {\bf Massive charged solution
($\bm{m >0}$, $\bm{q\neq0}$)} \vspace{0.2 cm}

The Carter-Penrose diagrams  are identical to those of the
AdS$-$Reissner-Nordstr\"{o}m solution ($A=0\,,\,m>0\,,\,q\neq0$)
and is represented in Fig. \ref{Fig-b3 AdS}.(a). In this figure,
the non-extremal black hole is represented in the left column, the
extremal black hole is represented in the middle column, and the
naked charged particle is represented in the right column. The
origin has an acceleration $A$, as will be seen in section
\ref{sec:Phys_Interp AdS}.

\subsection{\label{sec:Phys_Interp AdS} Physical interpretation of the
A\lowercase{d}S C-metric}

The parameter $A$ that is found in the AdS C-metric is interpreted
as being an acceleration and the AdS C-metric with $A>1/\ell$
describes a pair of black holes accelerating away from each other
in an AdS background, while the AdS C-metric with $A\leq 1/\ell$
describes a single accelerated black hole. In this section we will
justify this statement.

In subsection \ref{sec:Phys_Interp m,e AdS} we saw that, when
$A=0$, the general AdS C-metric,  (\ref{AdS C-metric}), reduces to
the AdS ($m=0\,,\,q=0$), to the AdS-Schwarzschild ($m>0\,,\,q=0$),
and to the AdS$-$Reissner-Nordstr\"{o}m solutions
($m=0\,,\,q\neq0$). Therefore, the parameters $m$ and $q$ are,
respectively, the ADM mass and ADM electromagnetic charge of the
non-accelerated black holes. Moreover, if we set the mass and
charge parameters equal to zero, even when $A\neq 0$, the
Kretschmann scalar
 [see (\ref{R2})] reduces to the value expected for the AdS spacetime.
This indicates that the massless uncharged AdS C-metric is an AdS
spacetime in disguise.

\subsubsection[$A>1/\ell$. Pair of accelerated black holes]
{\label{sec:PI.2-BH AdS} $\bm{A>1/\ell}$. Pair of accelerated
black holes}
In this section, we will first interpret case {\it 1. Massless
uncharged solution} ($m =0$, $q=0$), which is the simplest, and
then with the acquired knowledge we interpret cases {\it 2.
Massive uncharged solution} ($m>0$, $q=0$) and {\it 3. Massive
charged solution} ($m>0$, $q\neq0$). We will interpret the
solution following two complementary descriptions, the four
dimensional (4D) one and the five dimensional (5D).

 \vspace{0.2 cm} \noindent $\bullet$ {\bf The 4-Dimensional
description ($\bm{m =0}$, $\bm{q=0}$)} \vspace{0.2 cm}

\begin{figure}[H]
\centering
\includegraphics[height=1in]{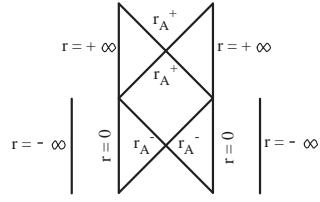}
   \caption{\label{Fig-e AdS}
Extending the Carter-Penrose diagrams of Fig. \ref{Fig-a1 AdS} to
negative values of $r$, we obtain the AdS spacetime with its
origin being accelerated. $r_A^+ =[A(x-y_+)]^{-1}>0$ and
$r_A^-=[A(x+y_+)]^{-1}>0$. We have to glue indefinitely copies of
the represented figure in the top and bottom regions.
 }
\end{figure}

As we said in \ref{sec:PD A AdS}, when $m=0$ and $q=0$ the origin
of the radial coordinate $r$ defined in  (\ref{r}) has no
curvature singularity and therefore $r$ has the range
$]-\infty,+\infty[$. However, in the realistic general case, where
$m$ or $q$ are non-zero, there is a curvature singularity at $r=0$
and since the discussion of the massless uncharged solution was
only a preliminary to that of the massive general case, following
\cite{AshtDray}, we have treated the origin $r=0$ as if it had a
curvature singularity and thus we admitted that $r$ belongs to the
range $[0,+\infty[$. In these conditions we obtained the causal
diagrams of Fig. \ref{Fig-a1 AdS}. Note however that one can make
a further extension to include the negative values of $r$,
enlarging in this way the range accessible to the Kruskal
coordinates $u'$ and $v'$. By doing this procedure we obtain the
causal diagram of the AdS spacetime. In Fig. \ref{Fig-e AdS} we
show the extension to negative values of coordinate $r$ (so
$-\infty <y<+\infty$) of the Carter-Penrose diagrams of Fig.
\ref{Fig-a1 AdS}. This diagram indicates that the origin of the
AdS spacetime, $r=0$, is accelerating. The situation is analogous
to the one that occurs in the usual Rindler spacetime,
$ds^2=-X^2dT^2+dX^2$. If one restricts the coordinate $X$ to be
positive one obtains an accelerated origin that approaches a
Rindler accelerated horizon. However, by making an extension to
negative values of $X$ one obtains the Minkowski spacetime.

Now, we want to clearly identify the parameter $A$ that appears in
the AdS C-metric with the acceleration of its origin. To achieve
this aim, we recover the massless uncharged AdS C-metric defined
by  (\ref{C-metric AdS}) and  (\ref{FG AdS}) (with $m=0$ and
$q=0$), and after performing the following coordinate
transformation
\begin{eqnarray}
& & \tau=\frac{\sqrt{\ell^2A^2-1}}{A} t \:,  \;\;\;\;\;
    \rho=\frac{\sqrt{\ell^2A^2-1}}{A} \frac{1}{y} \:, \nonumber \\
& & \theta = \arccos{x} \:,
     \;\;\;\;\; \phi = \tilde{\phi} \:,
  \label{transf-int AdS}
  \end{eqnarray}
we can rewrite the massless uncharged AdS C-metric as
\begin{eqnarray}
 d s^2 = \frac{1}{\gamma^2}
 {\biggl [}-(1-\rho^2/\ell^2)d\tau^2+
 \frac{d\rho^2}{1-\rho^2/\ell^2} +\rho^2 d\Omega^2 {\biggl ]},
\label{metric-int AdS}
 \end{eqnarray}
 with $d\Omega^2=d\theta^2+\sin^2\theta d \phi^2$ and
 \begin{eqnarray}
 \gamma=\sqrt{\ell^2A^2-1} + A\rho \cos\theta \:.
 \label{gamma AdS}
 \end{eqnarray}
The causal diagram of this spacetime is drawn in Fig. \ref{Fig-d
AdS}.
\begin{figure}[H]
\centering
\includegraphics[height=1.1in]{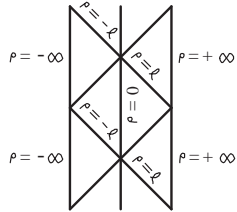}
   \caption{\label{Fig-d AdS}
Carter-Penrose diagram of metric (\ref{metric-int AdS}). We have
to glue indefinitely copies of the represented figure in the top
and bottom regions.
 }
\end{figure}
Notice that the origin of the radial coordinate $\rho$ corresponds
to $y=+\infty$ and therefore to $r=0$, where $r$ has been
introduced in  (\ref{r}). So, when we consider the massive AdS
C-metric there will be a curvature singularity at $\rho=0$ (see
section \ref{sec:CurvSing AdS}).

To discover the meaning of parameter $A$ we consider the 4D
timelike worldlines described by an observer with $\rho=$constant,
$\theta=0$ and $\phi=0$ (see \cite{Lemos2DTeiltJack}). These are
given by $x^{\mu}(\lambda)=(\gamma \ell
\lambda/\sqrt{\ell^2-\rho^2},\rho,0,0)$, were $\lambda$ is the
proper time of the observer since the 4-velocity
$u^{\mu}=dx^{\mu}/d\lambda$ satisfies $u_{\mu}u^{\mu}=-1$. The
4-acceleration of these observers,
$a^{\mu}=(\nabla_{\nu}u^{\mu})u^{\nu}$, has a magnitude given by
\begin{eqnarray}
 |a_4|=\sqrt{a_{\mu}a^{\mu}}=\frac{\rho\sqrt{\ell^2A^2-1}+\ell^2A}
 {\ell\sqrt{\ell^2-\rho^2}}\:.
 \label{a AdS}
 \end{eqnarray}
Since $a_{\mu}u^{\mu}=0$, the value $|a_4|$ is also the magnitude
of the 3-acceleration in the rest frame of the observer. From
(\ref{a AdS}) we achieve the important conclusion that the origin
of the AdS C-metric, $\rho=0$ (or $r=0$), is being accelerated
with a constant acceleration whose value is precisely given by the
parameter $A$ that appears in the AdS C-metric. Moreover, at
radius $\rho=\ell$ [or $y=y_+$ defined in equation (\ref{F1 AdS})]
the acceleration is infinite which corresponds to the trajectory
of a null ray. Thus, observers held at $\rho=$constant see this
null ray as an acceleration horizon and they will never see events
beyond this null ray.

 \vspace{0.2 cm} \noindent $\bullet$ {\bf The 5-Dimensional
description ($\bm{m =0}$, $\bm{q=0}$)} \vspace{0.2 cm}

In order to improve and clarify the physical aspects of the AdS
C-metric we turn now into the 5D representation of the solution.

The AdS spacetime can be represented as the 4-hyperboloid,
\begin{eqnarray}
-(z^0)^2+(z^1)^2+(z^2)^2+(z^3)^2-(z^4)^2=-\ell^2,
\label{hyperboloid AdS}
 \end{eqnarray}
in the 5D Minkowski (with two timelike coordinates) embedding
spacetime,
\begin{eqnarray}
 d s^2 = -(dz^0)^2+(dz^1)^2+(dz^2)^2+(dz^3)^2-(dz^4)^2.
 \label{AdS}
 \end{eqnarray}
Now, the massless uncharged AdS C-metric is an AdS spacetime in
disguise and therefore our next task is to understand how the AdS
C-metric can be described in this 5D picture. To do this we first
recover the massless uncharged AdS C-metric described by
(\ref{metric-int AdS}) and apply to it the coordinate
transformation
\begin{eqnarray}
  \hspace{-0.3cm} & & \hspace{-0.3cm}
  z^0=\gamma^{-1}\sqrt{\ell^2-\rho^2}\,\sinh(\tau/\ell)\:,
  \;\;\;\;\; z^2=\gamma^{-1} \rho \sin\theta \cos\phi \:,
  \nonumber \\
  \hspace{-0.3cm} & & \hspace{-0.3cm}
  z^1=\gamma^{-1}\sqrt{\ell^2-\rho^2}\,\cosh(\tau/\ell)\:,
  \;\;\;\;\;  z^3=\gamma^{-1} \rho \sin\theta \sin\phi \:,
  \nonumber \\
  \hspace{-0.3cm} & & \hspace{-0.3cm}
  z^4=\gamma^{-1}[\sqrt{\ell^2A^2-1} \,\rho \cos\theta
  +\ell^2A]\:,
\label{AdS to AdS-c}
  \end{eqnarray}
where $\gamma$ is defined in  (\ref{gamma AdS}). Transformations
(\ref{AdS to AdS-c}) define an embedding of the massless uncharged
AdS C-metric into the 5D description of the AdS spacetime since
they satisfy  (\ref{hyperboloid AdS}) and take directly
(\ref{metric-int AdS}) into  (\ref{AdS}).
\begin{figure}[H]
\centering
\includegraphics[height=2.6in]{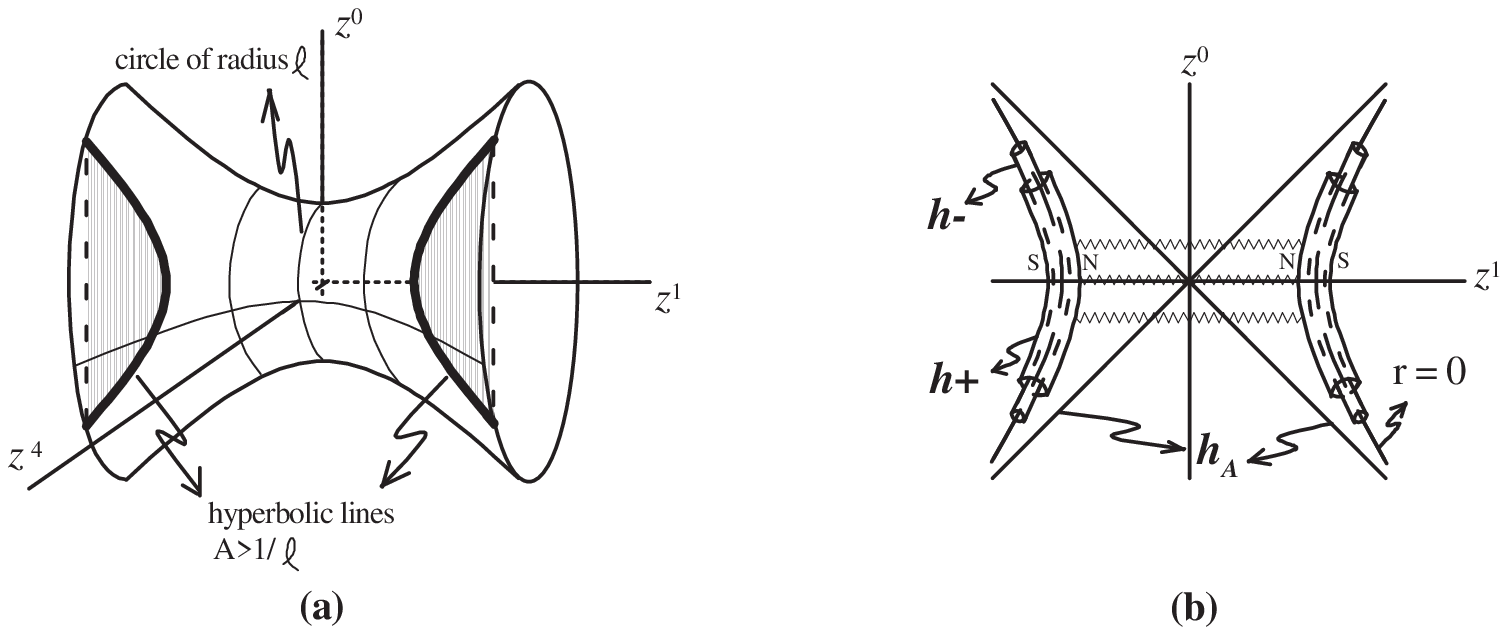}
\caption{\label{AdS-hyperb}
 (a) AdS 4-hyperboloid embedded in the 5D
Minkowski spacetime with two timelike coordinates, $z^0$ and
$z^4$. The directions $z^2$ and $z^3$ are suppressed. The two
hyperbolic lines lying on the AdS hyperboloid result from the
intersection of the hyperboloid surface with the
$z^4$=constant$>\ell$ plane. They describe the motion of the
origin of the AdS C-metric with $A>1/\ell$.
 (b) Schematic diagram representing the 5D
hyperbolic motion of two uniformly accelerating massive charged
black holes approaching asymptotically the Rindler-like
accelerated horizon ($h_A$). The inner and outer charged horizons
are represented by $h-$ and $h+$. The strut that connects the two
black holes is  represented by the zigzag lines. The north pole
direction is represented by $\rm{N}$ and the south pole direction
by $\rm{S}$.
 }
\end{figure}

 So, the massless uncharged AdS C-metric is an AdS spacetime, but
we can extract more information from this 5D analysis. Indeed, let
us analyze with some detail the properties of the origin of the
radial coordinate, $\rho=0$ (or $r=0$). This origin moves in the
5D Minkowski embedding spacetime according to [see (\ref{AdS to
AdS-c})]
\begin{eqnarray}
 & & z^2=0\;,\;\; z^3=0\;,\;\; z^4=\ell^2A /
 \sqrt{\ell^2A^2-1}\:>\ell
 \;\;\;\;\mathrm{and} \nonumber \\
 & & (z^1)^2-(z^0)^2=(A^2-1/\ell^2)^{-1}\equiv a_5^{-2} \:.
\label{rindler AdS}
  \end{eqnarray}
These equations define two hyperbolic lines lying on the AdS
hyperboloid which result from the intersection of this hyperboloid
surface defined by (\ref{hyperboloid AdS}) and the
$z^4$=constant$>\ell$ plane [see Fig. \ref{AdS-hyperb}.(a)]. They
tell us that the origin is subjected to a uniform 5D acceleration,
$a_5$, and consequently moves along a hyperbolic worldline in the
5D embedding space, describing a Rindler-like motion (see Fig.
\ref{AdS-hyperb}) that resembles the well-known hyperbolic
trajectory, $X^2-T^2=a^{-2}$, of an accelerated observer in
Minkowski space. But uniformly accelerated radial worldlines in
the 5D Minkowski embedding space are also uniformly accelerated
worldlines in the 4D AdS space \cite{DesLev}, with the 5D
acceleration $a_5$ being related to the associated 4D acceleration
$a_4$ by $a_5^2=a_4^2-1/\ell^2$. Comparing this last relation with
(\ref{rindler AdS}) we conclude that $a_4\equiv A$. Therefore, and
once again, we conclude that the origin of the AdS C-metric is
uniformly accelerating with a 4D acceleration whose value is
precisely given by the parameter $A$ that appears in the AdS
C-metric,  (\ref{C-metric AdS}), and this solution describes a AdS
space whose origin is not at rest as usual but is being
accelerated. Note that the origin of the usual AdS spacetime
describes the circle $(z^0)^2+(z^4)^2=\ell^2$ in the AdS
hyperboloid in contrast to the origin of the AdS C-metric with
$A>1/\ell$ whose motion is described by (\ref{rindler AdS}). This
discussion allowed us to find the physical interpretation of
parameter $A$ and to justify its label. Notice also that the
original AdS C-metric coordinates introduced in (\ref{C-metric
AdS}) cover only the half-space $z^1>-z^0$. The Kruskal
construction done in section \ref{sec:PD A AdS} extended this
solution to include also the $z^1<-z^0$ region and so, in the
extended solution, $r=0$ is associated to two hyperbolas that
represent two accelerated points (see Fig. \ref{AdS-hyperb}).
These two hyperbolas approach asymptotically the Rindler-like
acceleration horizon ($r_A$), so called because it is is absent
when $A=0$ and present even when $A\neq 0$, $m=0$ and $q=0$.

 \vspace{0.2 cm} \noindent $\bullet$ {\bf Pair of accelerated black
holes ($\bm{m >0}$, $\bm{q \neq 0}$)} \vspace{0.2 cm}

Now, we are in position to interpret the massive and charged
solutions that describe two black holes accelerating away from
each other. To see clearly this, let us look to the Carter-Penrose
diagrams near the equator, Fig. \ref{Fig-a1 AdS}.(c), Fig.
\ref{Fig-a2 AdS}.(c) and Fig. \ref{Fig-a3 AdS}.(c) (for the
discussion that follows we could, as well, look at the diagrams of
case (d) on these figures). Looking at these figures we can
compare the different features that belong to the massless
uncharge case [Fig. \ref{Fig-a1 AdS}.(c)], to the massive
uncharged case [Fig. \ref{Fig-a2 AdS}.(c)], and ending in the
massive charged case [Fig. \ref{Fig-a3 AdS}.(c)]. In Fig.
\ref{Fig-a1 AdS}.(c) we identify the two hyperbolas $r=0$
(represented by two timelike lines) approaching asymptotically the
Rindler-like acceleration horizon ($r_A$). When we add a mass to
the solution we conclude that each of these two simple hyperbolas
$r=0$ are replaced by the more complex structure that represents a
Schwarzschild black hole with its spacelike curvature singularity
and its horizon (these are represented by $r_+$ in the left and
right regions of Fig. \ref{Fig-a2 AdS}.(c)). So, the two
accelerating points $r=0$ have been replaced by two Schwarzschild
black holes that approach asymptotically the Rindler-like
acceleration horizon (represented by $r_A$ in the middle region of
Fig. \ref{Fig-a2 AdS}.(c)). The same interpretation can be
assigned to the massive charged solution. The two hyperbolas $r=0$
of Fig. \ref{Fig-a1 AdS}.(c) are replaced by two
Reissner-Nordstr\"{o}m black holes (with its timelike curvature
singularity and its inner $r_-$ and outer $r_+$ horizons; see the
left and right regions of Fig. \ref{Fig-a3 AdS}.(c)) that approach
asymptotically the Rindler-like acceleration horizon already
present in the $m=0$ and $q=0$ causal diagram.

\vspace{0.1 cm}

The Carter-Penrose diagrams of cases (a) and (b) of Fig.
\ref{Fig-a2 AdS} and Fig. \ref{Fig-a3 AdS} indicate that an
observer that is looking through an angular direction which is in
the vicinity of the south pole does not see the acceleration
horizon and notices the presence of a single black hole. This is
in agreement with Fig. \ref{AdS-hyperb}.(b). Indeed, in this
schematic figure, coordinates $z^0$ and $z^1$ can be seen as
Kruskal coordinates and we conclude that an observer, initially
located at infinity ($z^1=\infty$) and moving inwards into the
black hole along the south pole, passes through the black hole
horizons and hits eventually its curvature singularity. Therefore,
he never has the opportunity of getting in contact with the
acceleration horizon and with the second black hole. This is no
longer true for an observer that moves into the black hole along
an angular direction that is in the vicinity of the north pole. In
Fig. \ref{AdS-hyperb}.(b) this observer would be between the two
black holes, at one of the points of the $z^0<0$ semi-axis (say)
and moving into the black hole. Clearly, this observer passes
through the acceleration horizon before crossing the black hole
horizons and hitting its curvature singularity. This description
agrees with cases (c), (d) and (e) of Fig. \ref{Fig-a2 AdS} and
Fig. \ref{Fig-a3 AdS} which describe the solution along an angular
direction which includes the equatorial plane [case (c)] as well
as the north pole [case (e)].

\vspace{0.1 cm}

The diagrams of the third column of Fig. \ref{Fig-a3 AdS}
concerning the naked case of the $A>1/\ell$ massive charged
C-metric deserve a comment. First, we stress that the term naked
is employed in this situation because the values of parameters $m$
and $q$ are such that the solution has no charged associated
horizons, i.e., in the notation used along this paper, $r_-$ and
$r_+$ are not present in these diagrams. However, these diagrams
present an interesting new feature. Indeed, looking at at rows (a)
and (b) we have a single accelerated naked particle, in rows
(c)-(d) we find two naked singularities approaching asymptotically
the acceleration horizon $r_A$ but in row (e) we have no longer
two naked singularities. More precisely, we have a kind of a
single AdS$-$Reissner-Nordstr\"{o}m  black hole with the curvature
singularity being provided by the mass and charge but with the
horizons having their origin in the acceleration and cosmological
constant.

 \vspace{0.2 cm} \noindent $\bullet$ {\bf Source of acceleration.
The strut or the strings} \vspace{0.2 cm}

We can now ask what entity is causing the acceleration and where
it is localized. To achieve this aim, let us go back to the
massless uncharged AdS C-metric and consider radial worldlines
motions with $z^2=0$, $z^3=0$ and $z^4=$ constant or,
equivalently, with $\theta=0$, $\phi=$constant and
$\rho=$constant. These observers move along a Rindler-like
hyperbola described by [see  (\ref{AdS to AdS-c})]
\begin{eqnarray}
(z^1)^2-(z^0)^2 &=& \frac{\ell^2-\rho^2}{(\sqrt{\ell^2A^2-1} +
A\rho)^2}  \:.
 \label{rindler.2 AdS}
  \end{eqnarray}
Since the right hand side of  (\ref{rindler.2 AdS}) is smaller
than $a_5^{-2}$ defined in  (\ref{rindler AdS}), the north pole
$\theta_\mathrm{n}=0$ is localized between the hyperbolas
$(z^1)^2-(z^0)^2=a_5^{-2}$ in the $z^0,z^1$ diagram [see Fig.
\ref{AdS-hyperb}.(b)]. What does this means? When we put $m$ or
$q$ different from zero, each of the two hyperbolas assigned to
$r=0$ represent the accelerated motion of a black hole. Thus,
  (\ref{rindler.2 AdS}) tells us that the $\theta_\mathrm{n}=0$
axis points toward the other black hole, i.e., it is in the region
between the two black holes [see Fig. \ref{AdS-hyperb}.(b)]. The
south pole points along the symmetry axis towards spatial
infinity. Now, in section \ref{sec:ConSing AdS}, we saw that
parameter $\kappa$ has been chosen in order to avoid a conical
singularity at the south pole [see  (\ref{k-s AdS})] and, by doing
so, at the north pole is localized a conical singularity. This is
associated to a strut that joins the two black holes and provides
the acceleration of the black holes. To confirm this, recall that
either a straight string or a strut have a metric described by
\cite{VilenkinString}
\begin{eqnarray}
 d s^2 = -dt^2+ dZ^2+ d\varrho^2+\varrho^2 d\tilde{\varphi}^2,
\label{Vil-met AdS}
 \end{eqnarray}
where $\tilde{\varphi}=[1-\delta/(2\pi)]\varphi$ and
$0\leq\varphi<2\pi$. A string has $\delta>0$ and the geometry
around it is conic, i.e., it is a plane with a deficit angle
$\delta$, while a strut has $\delta<0$. Their mass per unit length
is $\mu=\delta/(8\pi)$ and their interior energy-momentum tensor
is
\begin{eqnarray}
T_{\alpha}^{\;\;\beta}=\mu\delta(X)\delta(Y)\mathrm{diag}(-1,0,0,-1),
\label{Vil-tens AdS}
 \end{eqnarray}
where $X=\varrho\cos\tilde{\varphi}$ and
$Y=\varrho\sin\tilde{\varphi}$ are the directions normal to the
strut, and $\delta(X)$ and $\delta(Y)$ are Dirac delta-functions.
The pressure on the string or in the strut satisfies $p=-\mu$. If
$\mu>0$ we have a string, if $\mu<0$ we have a strut. Now, turning
to our case, the AdS C-metric,  (\ref{AdS C-metric}), near the
north pole is given by
\begin{eqnarray}
d s^2 \sim -r^2{\cal F}dt^2 + r^2{\cal F}^{-1}dy^2
 + {\biggr (}r^2d\theta^2 +
 \frac{\kappa^2}{4}{\biggl |}\frac{d G}{dx}{\biggl |}_{x_\mathrm{n}}
 r^2\theta^2 d\tilde{\phi}^2{\biggr )}\:,
 \label{N-metric AdS}
 \end{eqnarray}
where $\kappa$ is defined in  (\ref{k-s AdS}) and the term between
the curved brackets is the induced metric in the plane normal to
the strut that connects the two black holes (along the $y$
direction) and will be labelled as $dX^2+dY^2$. The C-metric strut
has a mass per unit length given by
\begin{eqnarray}
\mu = \frac{1}{4}\frac{\delta_\mathrm{n}}{2\pi}=\frac{1}{4}
{\biggl (}1- {\biggl |}\frac{d {\cal G}}{dx}
 {\biggl |}^{-1}_{x_\mathrm{s}} {\biggl |}\frac{d {\cal G}}{dx}
 {\biggl |}_{x_\mathrm{n}}{\biggr )}  \:.
 \label{m.string AdS}
  \end{eqnarray}
We have $|d_x {\cal G}|_{x_\mathrm{s}} < |d_x {\cal
G}|_{x_\mathrm{n}}$ and so $\mu$ is negative. To obtain the
pressure of the C-metric strut, we write  (\ref{N-metric AdS}) in
a Minkowski frame,
 $ds^2=-\theta^{(0)2}+\theta^{(1)2}+\theta^{(2)2}+\theta^{(3)2}$,
 with $\theta^{(A)}=e^{(A)}_{\;\;\;\;\alpha} dx^{\alpha}$ and
 $e^{(0)}_{\;\;\;\;0}=r\sqrt{{\cal F}}$, $e^{(1)}_{\;\;\;\;1}=r$,
 $e^{(2)}_{\;\;\;\;2}=r\theta k|d_x {\cal G}|_{x_\mathrm{n}}/2$ and
 $e^{(3)}_{\;\;\;\;3}=r/\sqrt{{\cal F}}$. In this Minkowski frame the
energy-momentum tensor, $T_{(A)}^{\;\;\;(B)}$, of the C-metric
strut is given by (\ref{Vil-tens AdS}). In order to come back to
the coordinate basis frame and write the energy-momentum tensor of
the C-metric strut in this basis we use
 $T^{\alpha \beta}=e^{(A) \alpha}e_{(B)}^{\;\;\;\;\beta}T_{(A)}^{\;\;\;(B)}$
 and obtain
\begin{eqnarray}
 T^{\alpha \beta}=\mu (r^2{\cal F})^{-1}\delta(X)\delta(Y)
 \mathrm{diag}(1,0,0,-{\cal F}^2)\:.
\label{Vil-tens2 AdS}
 \end{eqnarray}
Defining the unit vector $\zeta=\partial/\partial y$ [so,
$\zeta^{\alpha}=(0,0,0,1)$], the pressure along the strut is
$T^{\alpha \beta}\zeta_{\alpha}\zeta_{\beta}$ and the pressure on
the C-metric strut is given by the integration over the $X$-$Y$
plane normal to the strut,
\begin{eqnarray}
p=\int dX dY \sqrt{^{(2)}g}\: T^{\alpha
\beta}\zeta_{\alpha}\zeta_{\beta}=-\mu\:.
 \label{Vil-pres AdS}
 \end{eqnarray}
So, the pressure and mass density of the C-metric strut satisfy
the relation $p=-\mu$. Since $\mu$ is negative, at both ends of
the strut, one has a positive pressure pushing away the two black
holes.

Alternatively, instead of  (\ref{k-s AdS}), we could have chosen
for $\kappa$ the value $\kappa^{-1}=(1/2)|d_x {\cal
G}|_{x_\mathrm{n}}$. By doing so we would avoid the deficit angle
at the north pole ($\delta_\mathrm{n}=0$) and leave a conical
singularity at the south pole ($\delta_\mathrm{s}>0$). This option
would lead to the presence of a semi-infinite string extending
from each of the black holes towards infinity along the south pole
direction, which would furnish the acceleration. The mass density
of both strings is
 $\mu =(1/4)(1-|d_x {\cal G}|^{-1}_{x_\mathrm{n}}
 |d_x {\cal G}|_{x_\mathrm{s}})>0$
and the pressure on the string, $p=-\mu$, is negative which means
that each string is pulling the corresponding black hole towards
infinity.

\vspace{0.2 cm}

At this point, we briefly remark that when we take the limit $A=0$
the AdS C-metric does not reduce to a static solution describing
two AdS black holes whose inward gravitational attraction is
cancelled by an outward pressure exerted by the strut or strings.
Indeed, when the acceleration parameter $A$ vanishes, the AdS
C-metric reduces into a single non-accelerated black hole free of
struts or strings (see subsection \ref{sec:Phys_Interp m,e AdS}
and section \ref{sec:PI.1-BH.1 AdS}). We shall return to this
issue in subsection \ref{sec:PI-mass flat}, where we will discuss
this note in detail and explain that the reason for this behavior
is due to the fact that the black holes do not interact
gravitationally. This conclusion is withdrawn from the
Carter-Penrose diagrams of the AdS C-metric.

\vspace{0.2 cm}

Ernst \cite{Ernst} has employed a Harrison-type transformation to
the $\Lambda=0$ charged C-metric in order to append a suitably
chosen external electromagnetic field (see discussion of this
solution in subsection \ref{sec:Ernst}). With this procedure the
so called Ernst solution is free of conical singularities at both
poles and the acceleration that drives away the two oppositely
charged Reissner-Nordstr\"{o}m black holes is totally provided by
the external electromagnetic field. In the AdS background we
cannot remove the conical singularities through the application of
the Harrison transformation \cite{EmparanPrivCom}. Indeed, the
Harrison transformation does not leave invariant the cosmological
term in the action. Therefore, applying the Harrison
transformation to
 (\ref{C-metric AdS})-(\ref{F-el-Lorentz}) does not yield a
new solution of the Einstein-Maxwell-AdS theory.

 \vspace{0.2 cm} \noindent $\bullet$ {\bf Radiative properties}
\vspace{0.2 cm}

The C-metric (either in the flat, de Sitter or anti-de Sitter
background) is an exact solution that is radiative. As noticed in
\cite{KW}, the gravitational radiation is present since the
complex scalar of the Newman-Penrose formalism,
$\Psi^4=-C_{\mu\nu\alpha\beta}
n^{\mu}\bar{m}^{\nu}n^{\alpha}\bar{m}^{\beta}$ (where
$C_{\mu\nu\alpha\beta}$ is the Weyl tensor and $\{l, n, m,
\bar{m}\}$ is the usual null tetrad of Newman-Penrose), contains a
term proportional to $r^{-1}$. Similarly, the charged version of
the C-metric includes, in addition, electromagnetic radiation. In
\cite{AshtDray}, it has been shown that the Bondi news functions
of the flat C-metric are indeed non-zero. These Bondi news
functions appear in the context of the Bondi method introduced to
study gravitational radiative systems. They are needed to
determine the evolution of the radiative gravitational field since
they carry the information concerning the changes of the system.
When at least one of them is not zero, the total Bondi mass of the
system decreases due to the emission of gravitational waves. The
Bondi news functions of the flat C-metric have been explicitly
calculated in \cite{Bic,PravPrav}. For a detailed review on the
radiative properties of the C-metric and other exact solutions see
\cite{PravPrav}. In an AdS background the energy released by a
pair of accelerated black holes has been discussed in detail by
Podolsk\' y, Ortaggio and Krtou\v s \cite{PodOrtKrtAdS}.

\subsubsection[$A= 1/\ell$. Single
accelerated black hole]{\label{sec:PI.1-BH AdS} $\bm{A= 1/\ell}$.
Single accelerated black hole}

When $A= 1/\ell$ the AdS C-metric describes a single accelerated
black hole. The absence of a second black hole is clearly
indicated by the Carter-Penrose diagrams of Figs. \ref{Fig-b2 AdS}
and \ref{Fig-b3 AdS}.

This case has been studied in detail in \cite{EHM1} where the
Randall-Sundrum model in a lower dimensional scenario has
analyzed. In this scenario, the brane-world is a 2-brane moving in
a 4D asymptotically AdS background. They have shown that the AdS
C-metric with $A=1/\ell$ describes a black hole bound to the
Minkowski 2-brane. The brane tension is fine tuned relative to the
cosmological background acceleration and thus, $A=1/\ell$ is
precisely the acceleration that the black hole has to have in
order to comove with the 2-brane. They concluded that the AdS
C-metric describes the final state of gravitational collapse on
the brane-world.
 The causal structure of the massive uncharged solution (Fig.
\ref{Fig-b2 AdS}) has been first discussed in \cite{EHM1}. For
completeness, we have also presented the causal diagrams of the
massless uncharged solution in Fig. \ref{Fig-b1 AdS} and of the
non-extremal, extremal, and naked massive charged solutions in
Fig. \ref{Fig-b3 AdS}.

In \cite{EHM1} the coordinate transformation that takes the
massless uncharged AdS C-metric with $A= 1/\ell$ into the known
description of the AdS spacetime in Poincar\'e coordinates is
given. From there one can easily go to the 5D description on the
AdS hyperboloid. This 5D description can be also understood
directly from the limits on the solutions $A>1/\ell$ and
$A<1/\ell$ when $A \to 1/\ell$. Indeed, if we take the limit $A
\to 1/\ell$ in section \ref{sec:PI.2-BH AdS} (where we have
studied the 5D description of case $A>1/\ell$), one sees that the
cut that generates the two hyperbolic lines degenerates into two
half circles which, on identifying the ends of the AdS hyperboloid
at both infinities, yields one full circle. This means that the
trajectory of the origin of the AdS C-metric in the $A= 1/\ell$
case is a circle (which when one unwraps the hyperboloid to its
universal cover yields a straight accelerated line). As we will
see in the next subsection, for $A<1/\ell$ the trajectory of the
origin is a circle which, on taking the limit $A \to 1/\ell$,
still yields a circle. The two limits give the same result as
expected.
\subsubsection[$A < 1/\ell$. Single
accelerated black hole]{\label{sec:PI.1-BH.1 AdS} $\bm{A <
1/\ell}$. Single accelerated black hole}

 The $A<1/\ell$ case was
first analyzed in \cite{Pod}. We have complemented this work with
the analysis of the causal structure. The causal diagrams of this
spacetime are identical to the ones of the AdS ($m=0\,,\,q=0$)
[see Fig. \ref{Fig-b1 AdS}.(a)], of the AdS-Schwarzschild ($m
>0\,,\,q=0$)  [see Fig. \ref{Fig-b2 AdS}.(a)], and of the
AdS$-$Reissner-Nordstr\"{o}m solutions ($m >0\,,\,q\neq 0$) [see
Fig. \ref{Fig-b3 AdS}.(a)]. However, the curvature singularity of
the single black hole of the solution is not at rest but is being
accelerated, with the acceleration $A$ provided by an open string
that extends from the pole into asymptotic infinity.

\begin{figure}[H]
\centering
\includegraphics[height=0.8in]{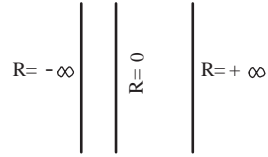}
   \caption{\label{Fig-f AdS}
Carter-Penrose diagram of metric (\ref{metric-int2 AdS}). }
\end{figure}

As was done with the $A>1/\ell$ case, it is useful to interpret
the solution following two complementary descriptions, the 4D one
and the 5D. One first recovers the massless uncharged AdS C-metric
defined by  (\ref{C-metric AdS}) and (\ref{FG AdS}) (with
$A<1/\ell$, $m=0$ and $q=0$), and after performing the following
coordinate transformation \cite{Pod}
\begin{eqnarray}
& & T=\frac{\sqrt{1-\ell^2A^2}}{A} t \:,  \;\;\;\;\;
    R=\frac{\sqrt{1-\ell^2A^2}}{A} \frac{1}{y} \:, \nonumber \\
& & \theta = \arccos{x} \:,
     \;\;\;\;\; \phi = \tilde{\phi} \:,
  \label{transf-int2 AdS}
  \end{eqnarray}
we can rewrite the massless uncharged AdS C-metric as
\begin{eqnarray}
 d s^2 = \frac{1}{\eta^2}
 {\biggl [}-(1+R^2/\ell^2)dT^2+
 \frac{dR^2}{1+R^2/\ell^2} +R^2 d\Omega^2 {\biggl ]},
 \nonumber \\
\label{metric-int2 AdS}
 \end{eqnarray}
with $\eta^{-1}=\sqrt{1-\ell^2A^2} + A R \cos\theta$ and
$d\Omega^2=d\theta^2+\sin^2\theta d \phi^2$. A procedure similar
to the one used to obtain (\ref{a AdS}) indicates that an observer
describing 4D timelike worldlines  with $R=$constant, $\theta=0$
and $\phi=0$ suffers a 4-acceleration with magnitude given by
\begin{eqnarray}
 |a_4|=\frac{\ell^2A-R\sqrt{1-\ell^2A^2}}
 {\ell\sqrt{\ell^2+R^2}}\:.
 \label{a2 AdS}
 \end{eqnarray}
Therefore, the origin of the AdS C-metric, $R=0$, is being
accelerated with a constant acceleration whose value is precisely
given by $A$. The causal diagram of this spacetime is drawn in
Fig. \ref{Fig-f AdS}. Notice that when we set $A=0$,
(\ref{metric-int2 AdS}) reduces to the usual AdS spacetime written
in static coordinates.
Now, to obtain the 5D description, one applies to
(\ref{metric-int2 AdS}) the coordinate
 transformation \cite{Pod},
\begin{eqnarray}
  \hspace{-0.3cm} & & \hspace{-0.3cm}
  z^0=\eta^{-1} \, \sqrt{\ell^2+R^2}\,\sin(T/\ell)\:,
  \;\;\;\;\;
  z^2=\eta^{-1} \, R \sin\theta \cos\phi \:,
  \nonumber \\
  \hspace{-0.3cm} & & \hspace{-0.3cm}
  z^4=\eta^{-1} \, \sqrt{\ell^2+R^2}\,\cos(T/\ell)\:,
  \;\;\;\;\;
  z^3=\eta^{-1} \, R \sin\theta \sin\phi \:,
  \nonumber \\
  \hspace{-0.3cm} & & \hspace{-0.3cm}
  z^1=\eta^{-1} \,[\sqrt{1-\ell^2A^2} \,R \cos\theta
  -\ell^2A]\:.
\label{AdS to AdS-c.2}
  \end{eqnarray}
Transformations (\ref{AdS to AdS-c.2}) define an embedding of the
massless uncharged AdS C-metric with $A<1/\ell$  into the 5D
description of the AdS spacetime since they satisfy
(\ref{hyperboloid AdS}) and take directly  (\ref{metric-int2 AdS})
into (\ref{AdS}).

\begin{figure}[H]
\centering
\includegraphics[height=2.3in]{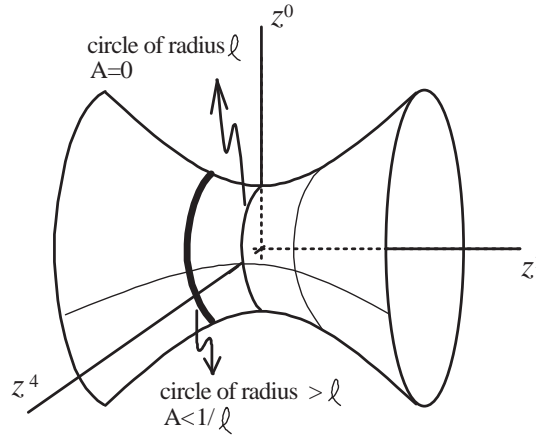}
\caption{\label{AdS-hyperb2} AdS 4-hyperboloid embedded in the 5D
Minkowski spacetime. The origin of the AdS C-metric with
$A<1/\ell$ moves in the hyperboloid along the circle with
$z^1$=constant$<0$. When $A=0$ this circle is at the plane $z^1=0$
and has a radius $\ell$.
 }
\end{figure}

The origin of the radial coordinate, $R=0$ moves in the 5D
Minkowski embedding spacetime according to [see  (\ref{AdS to
AdS-c.2})]
\begin{eqnarray}
 & & z^1=-\ell^2A/\sqrt{1-\ell^2A^2}\;,\;\; z^2=0\;,\;\; z^3=0
 \;\;\;\;\mathrm{and} \nonumber \\
 & & (z^0)^2+(z^4)^2=(1/\ell^2-A^2)^{-1}\equiv a_5^{-2} \:.
\label{circ AdS}
  \end{eqnarray}
So, contrarily to the case $A>1/\ell$ where the origin described a
Rindler-like hyperbolic trajectory [see  (\ref{rindler AdS})] that
suggests the presence of two black holes driving away from each
other in the extended  diagram, in the $A<1/\ell$ case the origin
describes a circle (a uniformly accelerated worldline) in the 5D
embedding space (see Fig. \ref{AdS-hyperb2}), indicating the
presence of a single trapped black hole in the AdS background.

To summarize and conclude, we present the global description on
the AdS hyperboloid of the AdS C-metric origin when the
acceleration $A$ varies from $+\infty$ to zero. When $A=+\infty$
the origin of the solution is represented in the hyperboloid by
two mutual perpendicular straight null lines at $45^{\rm o}$ that
result from the intersection of the hyperboloid surface defined by
(\ref{hyperboloid AdS}) and the $z^4=\ell$ plane (see Fig.
\ref{AdS-hyperb}). When $A$ belongs to $]1/\ell,+\infty[$, the
origin of the solution is represented by two hyperbolic lines [see
(\ref{rindler AdS})] lying on the AdS hyperboloid and result from
the intersection of (\ref{hyperboloid AdS}) and the
$z^4$=constant$>\ell$ plane [see Fig. \ref{AdS-hyperb}.(a)]. As
the acceleration approaches the value $A=1/\ell$ the separation
between the two hyperbolic lines increases. When $A=1/\ell$ the
separation between the two hyperbolic lines becomes infinite and
they collapse into two half circles which, on identifying the ends
of the AdS hyperboloid at both infinities, yields one full circle
in the $z^0-z^4$ plane at infinite $z^1$. At this point we see
again that the value $A=1/\ell$ sets a transition stage between
$A>1/\ell$ and $A<1/\ell$. When $A$ belongs to $]0,1/\ell[$ the
origin of the solution is described again by a circle [see
(\ref{circ AdS})] in the $z^0-z^4$ plane but now at a constant
$z^1<0$. As the acceleration approaches the value $A=0$, the
radius of this circle decreases and when $A=0$ the circle has a
radius with value $\ell$ and is at $z^1=0$ (see Fig.
\ref{AdS-hyperb2}) and we recover the usual AdS solution whose
origin is at rest.

\section[{\bf Pair of accelerated black holes in a flat
background: the flat C-metric and Ernst solution}]{Pair of
accelerated black holes in a flat background: \\ the flat C-metric
and Ernst solution}
 \label{sec:flat C-metric}

\subsection{\label{sec:General properties flat} General properties of the
flat C-metric and Ernst solution}

The original flat C-metric has been found Levi-Civita
\cite{LeviCivitaCmetric} and by Weyl \cite{WeylCmetric} in
1918-1919. In 1963 Ehlers and Kundt \cite{EhlersKundt} have given
the actual name to this solution.  The physical interpretation of
the flat C-metric remained unknown until 1970 when Kinnersley and
Walker \cite{KW}, in a pathbreaking work, have shown that the
solution describes two uniformly accelerated black holes in
opposite directions, with the acceleration being provided by
strings or by a strut. In this section we will review the work of
Kinnersley and Walker \cite{KW}. In this way, in the end of this
chapter we will be able to make a comparison between the C-metric
in an AdS, flat and dS background. From the charged flat C-metric,
Ernst \cite{Ernst} has generated a knew exact solution in which
the black hole pair is accelerated by an external electromagnetic
field. We will also review briefly this solution.

\subsubsection{\label{sec:flat C-metric sub} The flat C-metric}

The gravitational field of the flat C-metric can be written as
\cite{KW}
\begin{equation}
 d s^2 = [A(x+y)]^{-2} (-{\cal F}dt^2+
 {\cal F}^{-1}dy^2+{\cal G}^{-1}dx^2+
 {\cal G}d\phi^2)\:,
 \label{C-metric flat}
 \end{equation}
 where
 \begin{eqnarray}
 & &{\cal F}(y) = -1+y^2-2mAy^3+q^2A^2y^4, \nonumber\\
 & &{\cal G}(x) = 1-x^2-2mAx^3-q^2 A^2 x^4\:,
 \label{FG flat}
 \end{eqnarray}
and therefore we have now ${\cal F}(y)=-{\cal G}(-y)$. Note that
when $\Lambda=0$, the function ${\cal G}(x)$ behaves in the same
manner described in the AdS case. The Maxwell field in the
magnetic case is given by (\ref{F-mag}), while in the electric
case it is given by (\ref{F-el-Lorentz}). This solution depends on
three parameters namely, $A>0$ which is the acceleration of the
black holes (see section
 \ref{sec:Phys_Interp flat}), and $m$ and $q$ which are
interpreted  as the ADM mass and electromagnetic charge of the
non-accelerated black holes, respectively. To justify this
interpretation of $m$ and $q$, apply to (\ref{C-metric flat}) the
coordinate transformations \cite{DGKT}, $\tau=A^{-1} t$,
$\rho=(Ay)^{-1}$, together with (\ref{ang}) and setting $A=0$ (and
$\kappa=1$, $\tilde{\phi}=\phi$) one obtains
\begin{equation}
 d s^2 = - F(\rho)\, d \tau^2 +F^{-1}(\rho)\, d \rho^2
 +\rho^2 (d \theta^2+\sin^2\theta\,d\phi^2) \:,
 \label{mq2 flat}
\end{equation}
where $F(\rho)=1-2m/\rho + q^2/\rho^2$. So, when the acceleration
parameter vanishes, the flat C-metric, (\ref{C-metric flat}),
reduces to the Schwarzschild and Reissner-Nordstr\"{o}m black
holes and the parameters $m$ and $q$ that are present in the flat
C-metric are precisely the ADM mass and ADM electromagnetic charge
of these non-accelerated black holes. The physical properties and
interpretation of the flat C-metric have been analyzed in detail
by Kinnersley and Walker \cite{KW}.

Now, the general properties of the flat C-metric, in what concerns
the issues of radial coordinate and curvature singularities, the
analysis of the angular surfaces and conical singularities, and
the issue of coordinate ranges are straightforwardly similar to
the corresponding properties of the AdS C-metric analyzed in
subsections \ref{sec:CurvSing AdS}, \ref{sec:ConSing AdS}, and
\ref{sec:CoordRange AdS}. For this reason we will not analyze
again these properties in detail. We ask the reader to go back to
subsection \ref{sec:CurvSing AdS} or to see \cite{KW} for details.

\subsubsection{\label{sec:Ernst} The Ernst solution}

Ernst \cite{Ernst} has employed a Harrison-type transformation to
the charged flat C-metric in order to append a suitably chosen
external electromagnetic field. With this procedure the Ernst
solution is free of conical singularities at both poles and the
acceleration that drives away the two oppositely charged
Reissner-Nordstr\"{o}m black holes is provided by the external
electromagnetic field.  The gravitational field of the Ernst
solution is \cite{Ernst}
\begin{eqnarray}
 d s^2 = \frac{\Sigma^2 ( -{\cal F}dt^2+
 {\cal F}^{-1}dy^2+{\cal G}^{-1}dx^2
 +  \Sigma^{-4} {\cal G}\,d\phi^2 )} {[A(x+y)]^2} \:, \nonumber\\
 \label{Ernst}
 \end{eqnarray}
 where ${\cal F}(y)$ and ${\cal G}(x)$ are given by (\ref{FG flat}), and
 \begin{eqnarray}
 \Sigma(x,y)={\biggl (}1+\frac{1}{2}\, q \, {\cal E}_0\, x{\biggr )}^2
   + \frac{{\cal E}_0^2 \, {\cal G}(x)}{4A^2 \,(x+y)^2}\:.
 \label{factor}
 \end{eqnarray}
In the electric solution one has $q\equiv e$ and ${\cal E}_0\equiv
E_0$, i.e., $q$ and ${\cal E}_0$  are  respectively the electric
charge and the external electric field. In the magnetic solution
one has $q\equiv g$ and ${\cal E}_0\equiv B_0$, i.e., $q$ and
${\cal E}_0$ are respectively the magnetic charge and the external
magnetic field. The electromagnetic potential of the magnetic
Ernst solution is \cite{Ernst}
\begin{eqnarray}
 A_{\phi}=-\frac{2}{\Sigma \, B_0 }{\biggl (}1
+\frac{1}{2}\, g \,B_0\, x{\biggr )}\:.
 \label{pot-Ernst-B}
\end{eqnarray}
while for the electric Ernst solution it is given by \cite{Brown}
\begin{eqnarray}
A_t &=& qy -\frac{E_0}{2A^2} \frac{{\cal F}(y)}{(x+y)^2}
 \left ( 1+e \, E_0 \, x-\frac{1}{2} e \, E_0 \, y \right )
 \nonumber \\
 &-&\frac{E_0}{2A^2}(1+r_- A y)(1+r_+ A y) (1-e\,E_0\,y/2)
\:,  \nonumber \\
& & \label{pot-Ernst-E}
\end{eqnarray}
with $r_+ r_-=e^2$ and $r_+ +r_-=2m$. This exact solution
describes two oppositely charged Reissner-Nordstr\"{o}m black
holes accelerating away from each other in a magnetic Melvin
\cite{Melvin} or in an electric Melvin-like background,
respectively.

Technically, the Harrison-type transformation employed to generate
the Ernst solution introduces, in addition to the parameter
$\kappa$, a new parameter, the external field ${\cal E}_0$ that
when appropriately chosen allow us to eliminate the conical
singularities at both poles. Indeed, applying a procedure
analogous to the one employed in subsection \ref{sec:ConSing AdS},
but now focused in spacetime (\ref{Ernst}), one introduces the new
angular coordinates
\begin{eqnarray}
 \bar{\theta} = \int_{x_\mathrm{n}}^{x}\tilde{\cal{G}}^{-1/2}dx \:,
                        \:\:\:\:\:\: \:\:\:\:\:\:
\bar{\phi} = \phi/\bar{\kappa} \:,
                             \label{ang-Ernst}
\end{eqnarray}
where $\tilde{\cal{G}}(x,y)=\Sigma^{-2}(x,y){\cal{G}}(x)$. As
before, the conical singularity at the south pole is avoided by
choosing
\begin{equation}
\bar{\kappa}^{-1}=\frac{1}{2}\left | \tilde{\cal G}'(x_\mathrm{s})
\right | \,,
  \label{k-s-Ernst}
 \end{equation}
while the conical singularity at the north pole can now be also
eliminated by choosing the value of the external field
 ${\cal E}_0$ to satisfy
\begin{equation}
 \left | \tilde{\cal G}'(x_\mathrm{n}) \right | =
  \left | \tilde{\cal G}'(x_\mathrm{s}) \right | \:.
  \label{k=Ernst}
\end{equation}
An interesting support to the physical interpretation given to the
Ernst solution is the fact that in the particle limit, i.e.,  for
small values of $mA$, the condition (\ref{k=Ernst}) implies the
classical Newton's law \cite{Ernst}
\begin{equation}
q \, {\cal E}_0 \approx m \, A\:. \label{Lorentz}
\end{equation}
So, in this regime, the acceleration is indeed provided by the
Lorentz force.

Remark that in a cosmological background we cannot remove the
conical singularities through the application of the Harrison
transformation \cite{EmparanPrivCom}. Indeed, the Harrison
transformation applied by Ernst does not leave invariant the
cosmological term in the action. Therefore, applying the Harrison
transformation to the cosmological C-metric solutions does not
yield a new solution of the Einstein-Maxwell theory in a
cosmological background.

\subsection{\label{sec:PD flat} Causal Structure of the
flat C-metric}

For a similar reason to the one that occurs in the AdS C-metric,
due to the lower restriction on the value of $y$ ($-x\leq y$), the
choice of the Kruskal coordinates (and therefore the
Carter-Penrose diagrams) for the flat C-metric depends on the
angular direction $x$ we are looking at \cite{KW,AshtDray}. In
fact, depending on the value of $x$, the region accessible to $y$
might contain a different number roots of ${\cal F}$ (see Figs.
\ref{g1 flatC}, \ref{g2 flatC}, and \ref{g3 flatC}) and so the
solution may have a different number of horizons. We have to
consider separately three distinct sets of angular directions,
namely (a) $x=x_\mathrm{s}$, (c) $x_\mathrm{s} < x <x_\mathrm{n}$,
and (c) $x = x_\mathrm{n}$, where $x_\mathrm{s}=-1$ and
$x_\mathrm{n}=+1$ when $A\rightarrow 0$. These three cases are
perfectly identified in Figs. \ref{g1 flatC}, \ref{g2 flatC}, and
\ref{g3 flatC}. The technical procedures to obtain the
Carter-Penrose diagrams of these flat cases (a), (b) and (c) is
equal to the ones presented in the AdS cases (b), (c) and (d) of
subsection \ref{sec:PD A AdS}, respectively. Hence, we will not
present again the construction process that leads to the diagrams.
We will only draw the diagrams, and even the discussion of their
main features follows now directly from the analysis of subsection
\ref{sec:PD A AdS}. Once again, the description of the solution
depends crucially on the values of $m$ and $q$. We will consider
the three most relevant solutions, namely: {\it A. Massless
uncharged solution} ($m =0$, $q=0$), {\it B. Massive uncharged
solution} ($m>0$, $q=0$), and {\it C. Massive charged solution}
($m \geq0$, $q\neq0$). The causal diagrams of the Ernst solution
are equal to the ones that describe the flat C-metric.

\subsubsection[Massless uncharged solution ($m=0$,
$q=0$)] {\label{sec:PD 1 flat} Massless uncharged solution
($\bm{m=0}$, $\bm{q=0}$)}

\begin{figure}[H]
\centering
\includegraphics[height=2.2in]{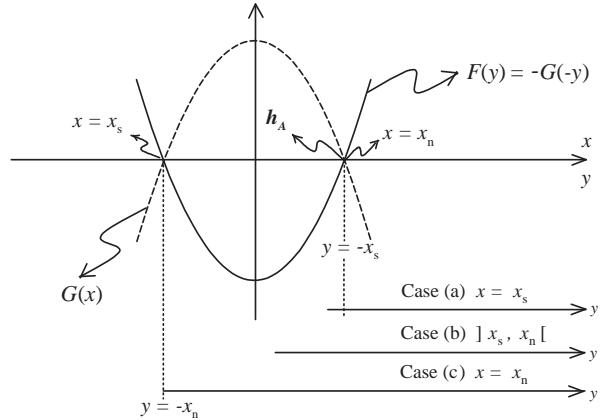}
\caption{\label{g1 flatC}
 Shape of ${\cal G}(x)$ and ${\cal F}(y)$ for the
 $m=0, q=0$ flat C-metric and Ernst solution. The allowed range of $x$ is between
$x_\mathrm{s}=-1$ and $x_\mathrm{n}=+1$ where ${\cal G}(x)$ is
positive and compact. The range of $y$ is restricted to $-x\leq y
< +\infty$. The presence of an accelerated horizon is indicated by
$h_A$.}
\end{figure}

\begin{figure}  [H]
\centering
\includegraphics[height=3in]{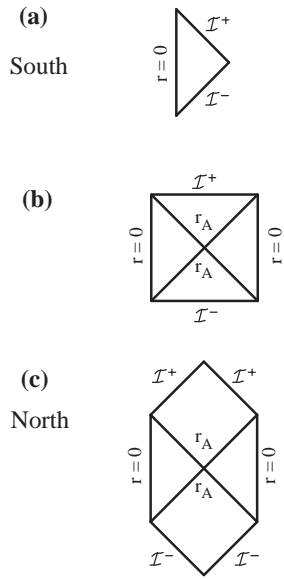}
   \caption{\label{Fig-1 flatC}
Carter-Penrose diagrams concerning the $m=0, q=0$ flat C-metric
and Ernst solution studied in section \ref{sec:PD 1 flat}. Case
(a) describes the solution seen from the south pole, case (c)
applies to the equatorial vicinity, and case (c) describes the
solution seen from the north pole. The accelerated horizon is
represented by $r_A$. ${\cal I}^-$ and ${\cal I}^+$ represent
respectively the past and future infinity ($r=+\infty$). $r=0$
corresponds to $y=+\infty$ and $r=+\infty$ corresponds to $y=-x$.
 }
\end{figure}

In this case we have $x \in [x_\mathrm{s}=-1,x_\mathrm{n}=+1]$,
$x=\cos \theta$, ${\cal G}=1-x^2=\sin^2 \theta$, $\kappa=1$,
$\tilde{\phi}=\phi$, and ${\cal F}(y) = y^2-1$. The general
behavior of these functions for this case is represented in Fig.
\ref{g1 flatC}.
The Carter-Penrose diagrams of the massless uncharged flat
C-metric and Ernst solution are sketched in Fig. \ref{Fig-1 flatC}
for the three angular directions, (a), (b) and (c), specified
above.

\subsubsection[Massive uncharged solution ($m >0$,
$q=0$)] {\label{sec:PD 2 flat} Massive uncharged solution ($\bm{m
>0}$, $\bm{q=0}$)}

\begin{figure} [H]
\centering
\includegraphics[height=2.2in]{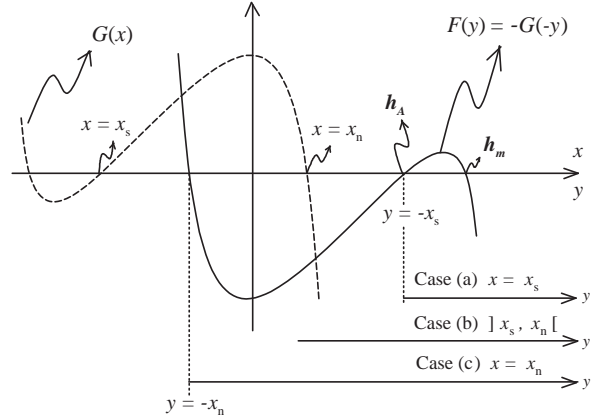}
\caption{\label{g2 flatC}
 Shape of ${\cal G}(x)$ and ${\cal F}(y)$ for the
 $27m^2A^2<1-9m^2\Lambda$, and $q=0$ flat C-metric and Ernst solution.
The allowed range of $x$ is between $x_\mathrm{s}$ and
$x_\mathrm{n}$ where ${\cal G}(x)$ is positive and compact. The
range of $y$ is restricted to $-x\leq y < +\infty$. The presence
of an accelerated horizon is indicated by $h_A$ and the
Schwarzschild-like horizon by $h_m$. For completeness we comment
on two other cases not studied in the text: for
$27m^2A^2=1-9m^2\Lambda$, ${\cal F}(y)$ is zero at its local
maximum, i.e., $h_A$ and $h_m$ coincide. For
$27m^2A^2>1-9m^2\Lambda$, ${\cal F}(y)$ is always negative in the
allowed range of $y$.
 }
\end{figure}

\begin{figure}[H]
\centering
\includegraphics[height=7.0cm]{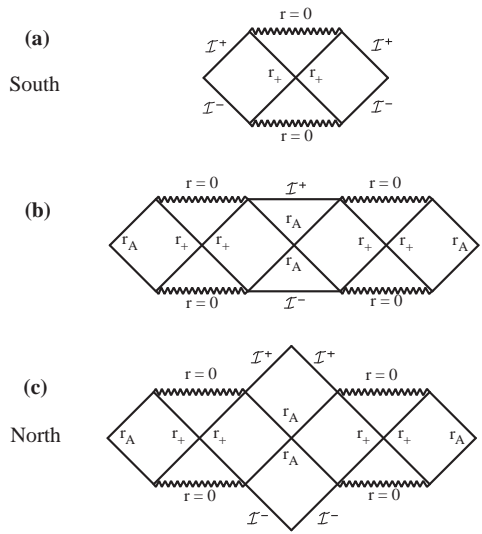}
   \caption{\label{Fig-2 flatC}
Carter-Penrose diagrams of the $27m^2A^2<1-9m^2\Lambda$, and $q=0$
flat C-metric discussed in  section \ref{sec:PD 2 flat}. Case (a)
describes the solution seen from the south pole, case (c) applies
to the equatorial vicinity, and case (c) describes the solution
seen from the north pole. The zigzag line represents a curvature
singularity, the accelerated horizon is represented by $r_A$. It
coincides with the cosmological horizon and has a non-spherical
shape. The Schwarzschild-like horizon is sketched as $r_+$. $r=0$
corresponds to $y=+\infty$ and $r=+\infty$ (${\cal I}^-$ and
${\cal I}^+$) corresponds to $y=-x$.
 }
\end{figure}

As occurs with the AdS case (see subsection \ref{sec:PD A AdS}),
we will consider only the case with small mass or acceleration,
i.e., we require $mA<3^{-3/2}$, in order to have compact angular
surfaces (see discussion on the text of Fig. \ref{g2 flatC}). We
also demand $x$ to belong to the range
$[x_\mathrm{s},x_\mathrm{n}]$ (see Fig. \ref{g2 flatC}) where
${\cal G}(x)\geq 0$ and such that $x_\mathrm{s} \to -1$ and
$x_\mathrm{n} \to +1$ when $mA \to 0$. By satisfying the two above
conditions we endow the $t=$constant and $r=$constant surfaces
with the topology of a compact surface.
The Carter-Penrose diagrams of the $27m^2A^2<1-9m^2\Lambda$, and
$q=0$ flat C-metric are sketched in Fig. \ref{Fig-2 flatC} for the
three angular directions, (a), (b) and (c), specified above.

\subsubsection[Massive charged
solution ($m >0$, $q\neq 0$)]{\label{sec:PD 3 flat} Massive
charged solution ($\bm{m >0}$, $\bm{q\neq0}$)}
\begin{figure} [H]
\centering
\includegraphics[height=2.2in]{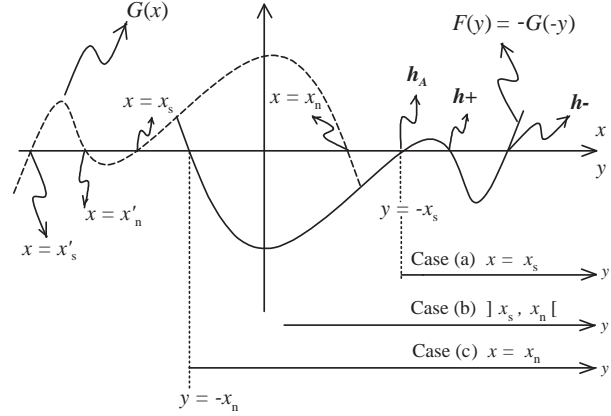}
\caption{\label{g3 flatC}
 Shape of ${\cal G}(x)$ and ${\cal F}(y)$
for the non-extreme charged massive flat C-metric and Ernst
solution. The allowed range of $x$ is between $x_\mathrm{s}$ and
$x_\mathrm{n}$ where ${\cal G}(x)$ is positive and compact.  The
presence of an accelerated horizon is indicated by $h_A$ and the
inner and outer charged horizons by $h-$ and $h+$. In the extreme
cases, $h-$ and $h+$ superpose each other. Note that when $y_+$
and $y_A$ coincide, the same occurs with $x_\mathrm{s}$ and $x_0$.
 }
\end{figure}

When both the mass and charge parameters are non-zero, depending
on the values of the parameters $A$, $m$ and $q$, ${\cal G}(x)$
can be positive in a single compact interval,
$]x_\mathrm{s},x_\mathrm{n}[$, or in two distinct compact
intervals, $]x'_\mathrm{s},x'_\mathrm{n}[$ and
$]x_\mathrm{s},x_\mathrm{n}[$, say (see Fig. \ref{g3 flatC}). In
this latter case we will work only with the interval
$[x_\mathrm{s},x_\mathrm{n}]$ (say) for which the charged
solutions are in the same sector of those we have analyzed in the
last two subsections when $q \to 0$.
The Carter-Penrose diagrams of the charged massive flat C-metric
 and Ernst solution are sketched in Fig. \ref{Fig-3 flatC} for the three angular
directions, (a), (b) and (c), specified above.

\subsection{\label{sec:Phys_Interp flat} Physical interpretation of
the  flat C-metric}

The parameter $A$ that is found in the flat C-metric is
interpreted as being an acceleration and the  flat C-metric
describes a pair of black holes accelerating away from each other.
In this section we will justify this statement.

\subsubsection[Description of the $m=0$, $q = 0$ solution]
{\label{sec:PI flat}Description of the $\bm{m=0}$, $\bm{q=0}$
solution}
Use of (\ref{r}), (\ref{ang}) with $\kappa=1$ and
$\tilde{\phi}=\phi$, together with $u=t+\int {\cal F}(y)^{-1}dy$
on (\ref{C-metric flat}) yields \cite{KW}
\begin{equation}
 d s^2 = \left (  1-2Ar\cos \theta -A^2 r^2 \sin^2 \theta \right )du^2+
 2 du dr -2A r^2 \sin \theta du d\theta -r^2  \left ( d\theta^2+\sin^2
 \theta d\phi^2\right ) \:.
 \label{metric-int flat}
 \end{equation}
As noticed by Kinnersly and Walker \cite{KW} this spacetime is
closely related to the ones discussed by Newman and Unti. There is
a coordinate transformation that allows to recast
 (\ref{metric-int flat}) into a Minkowski form, namely
 \begin{eqnarray}
 & &T = (A^{-1}-r \cos \theta)\sinh(Au)+r\cosh(Au)\,, \nonumber\\
 & &Z = (A^{-1}-r \cos \theta)\cosh(Au)+r\sinh(Au)\,, \nonumber\\
 & &X = r \sin \theta \cos\phi \,, \nonumber\\
 & &Y = r \sin \theta \sin\phi\:.
 \label{transf-int Mink}
 \end{eqnarray}
 \begin{figure} [H]
\centering
\includegraphics[height=12cm]{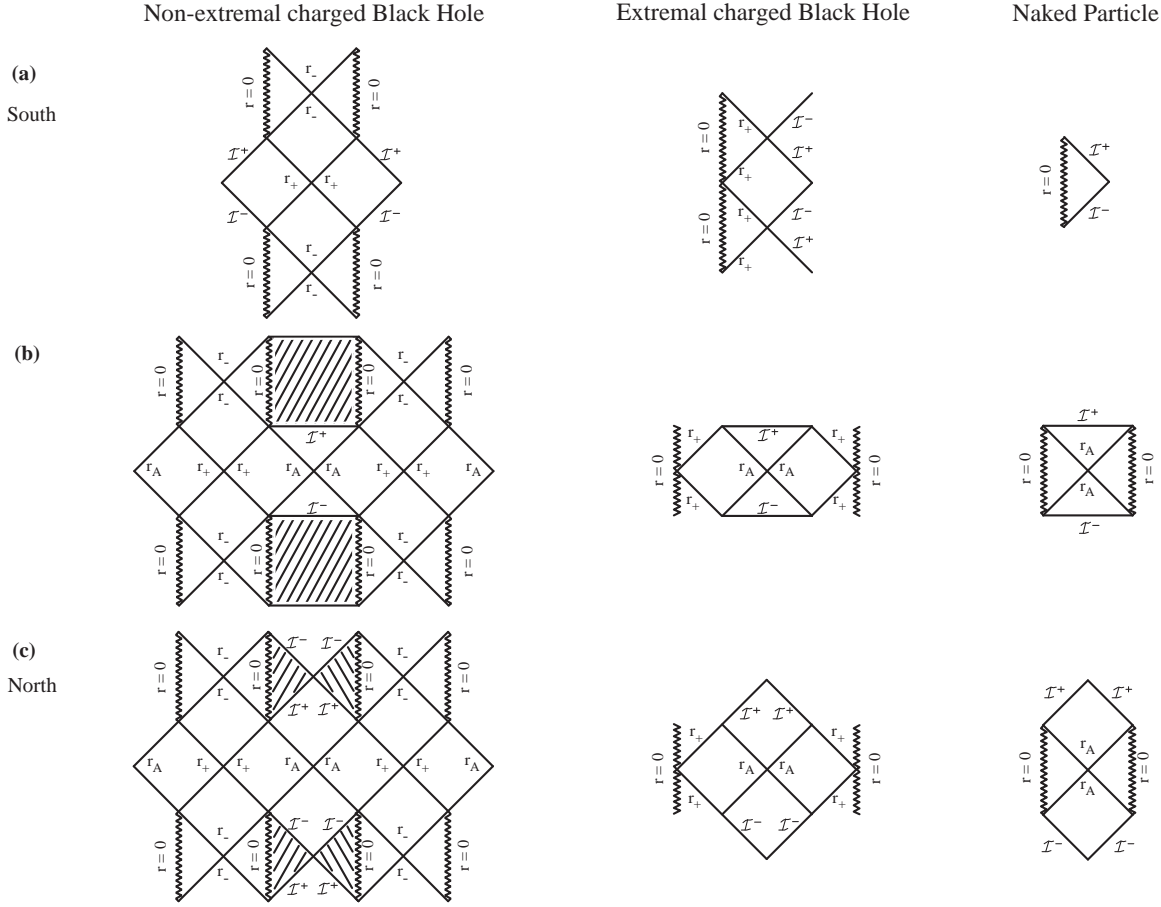}
   \caption{\label{Fig-3 flatC}
Carter-Penrose diagrams of the charged massive flat C-metric and
Ernst solution . Case (a) describes the solution seen from the
south pole, case (c) applies to the equatorial vicinity, and case
(c) describes the solution seen from the north pole.  The zigzag
line represents a curvature singularity, an accelerated horizon is
represented by $r_A$, the inner and outer charge associated
horizons are sketched as $r_-$ and $r_+$. ${\cal I}^-$ and ${\cal
I}^+$ represent respectively the past and future infinity
($r=+\infty$). $r=0$ corresponds to $y=+\infty$ and $r=+\infty$
corresponds to $y=-x$.
 }
\end{figure}
Indeed, under this transformation the massless neutral flat
C-metric transforms into
\begin{equation}
 d s^2 = -dT^2+dX^2+dY^2+dZ^2\:,
 \label{metric Mink}
 \end{equation}
and hence, the $m=0$ and $q=0$ flat C-metric is a Minkowski
spacetime. But we can extract more information from this analysis.
Indeed, let us analyze with some detail the properties of the
origin of the radial coordinate, $r=0$. This origin moves
according to [see (\ref{transf-int Mink})]
\begin{eqnarray}
 & & X=0\;,\;\; Y=0\;,\;\;
 \;\; \mathrm{and} \;\;\;\; Z^2-T^2=A^{-2}\:.
\label{rindler flat}
  \end{eqnarray}
These equations define two hyperbolic lines that tell us that the
origin is subjected to a uniform  acceleration, $A$, and
consequently moves along a hyperbolic worldline,  describing a
Rindler motion.

Let us now address the issue of the acceleration source and its
localization. In the massless uncharged dS C-metric, observers
that move along radial worldlines with $r=$const and
$\theta=$const describe the Rindler-like hyperbola [see (\ref{dS
to dS-c})]
\begin{eqnarray}
Z^2-T^2 = (A^{-1}-r \cos \theta)^2-r^2 \:.
 \label{rindler.2 flat}
  \end{eqnarray}
Thus, observers that move in the region $\theta=0$ describe the
the hyperbolic line $Z^2-T^2 < A^{-2}$ (which is in between the
two $r=0$ hyperbolas, $Z^2-T^2 =A^{-2}$), while observers in the
region $\theta=\pi$ follow the curve $Z^2-T^2
>A^{-2}$. When we put $m$ or $q$ different from zero, each of the
two hyperbolas assigned to $r=0$ represents the accelerated motion
of a black hole. Hence, from (\ref{rindler.2 flat}) we conclude
that the north pole axis is in the region between the two black
holes [see Fig. \ref{AdS-hyperb}.(b)]. When the conical
singularity is at the north pole we then have a strut in between
the black holes, while when we choose the parameters such that the
conical singularity is at the south pole we have two strings from
each of the black holes into infinity that accelerates them.
Notice also that the original flat C-metric coordinates introduced
in (\ref{C-metric flat}) cover only the half-space $Z>-T$. The
Kruskal construction extends this solution to include also the
$Z<-T$ region and so, in the extended solution, $r=0$ is
associated to two hyperbolas that represent two accelerated points
[see Fig. \ref{AdS-hyperb}.(b)] that approach asymptotically the
Rindler acceleration horizon ($r_A$).

\subsubsection[Pair of accelerated black holes ($m >0$, $q \neq 0$)]
{\label{sec:PI-mass flat}Pair of accelerated black holes
($\bm{m>0}$, $\bm{q \neq 0}$)}

The massive and charged solutions describe two black holes
accelerating away from each other. This interpretation follows
directly from the Carter-Penrose diagrams, Fig. \ref{Fig-1 flatC},
Fig. \ref{Fig-2 flatC} and Fig. \ref{Fig-3 flatC}, in a way
already explained in subsection \ref{sec:PI.2-BH AdS}. In Fig.
\ref{Fig-1 flatC} we identify the two hyperbolas $r=0$
(represented by two timelike lines) approaching asymptotically the
Rindler acceleration horizon ($r_A$). When we add a mass or a
charge to the solution we conclude that each of these two simple
hyperbolas $r=0$ are replaced by the more complex structure that
represents a Schwarzschild black hole (see Fig. \ref{Fig-2 flatC})
or a Reissner-Nordstr\"{o}m black hole (see Fig. \ref{Fig-3
flatC}).

\begin{figure}[H]
\centering
\includegraphics[height=4.0cm]{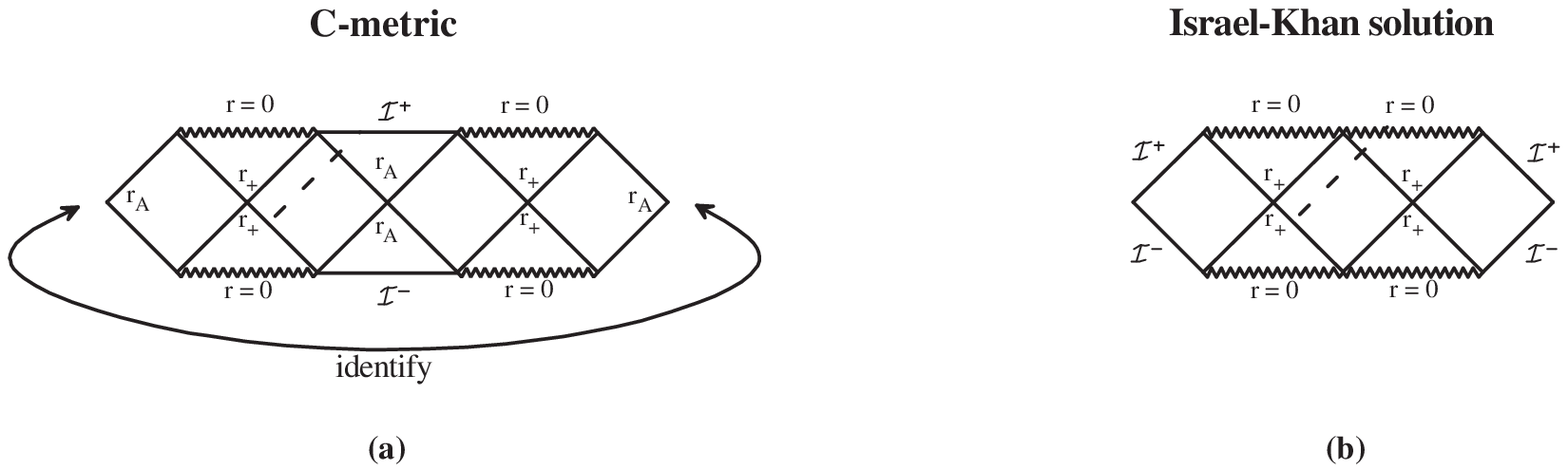}
   \caption{\label{Fig light flatC}
The dashed line represents a null ray that is sent from the
vicinity of the event horizon of one of the black holes towards
the other black hole. (a) In the case of the C-metric this ray can
never reach the other black. (b) In the Israel-Khan solution the
null ray can reach the second black hole.
 }
\end{figure}

At this point, a remark is relevant. Israel and Khan \cite{bh_eq}
(see also Bach and Weyl \cite{BachWeyl}, Aryal, Ford and Vilenkin
\cite{AryalFordVilenkin}, and Costa and Perry \cite{CostaPerry})
have found a $\Lambda=0$ solution that represents two (or more)
collinear Schwarzschild black holes interacting with each other in
such a way that allows dynamical equilibrium. In this solution,
the two black holes are connected by a strut that exerts an
outward pressure which cancels the inward gravitational attraction
and so the distance between the two black holes remains fixed
\cite{bh_eq}.  Now, the C-metric solution reduces to a single
non-accelerated black hole free of struts or strings when the
acceleration parameter $A$ vanishes (see subsection \ref{sec:flat
C-metric sub}). Thus, when we take the limit $A=0$, the C-metric
does not reduce to the static solution of Israel and Khan. The
reason for this behavior can be found in the Carter-Penrose
diagrams of the C-metric. For example, looking into Fig.
\ref{Fig-2 flatC}.(c) [redrawn again in Fig. \ref{Fig light
flatC}.(a) with the motion of a light ray that will be described
just below] which represents the massive uncharged flat C-metric
along the north pole, we conclude that a null ray sent from the
vicinity of one of the black holes can never cross the
acceleration horizon ($r_A$) into the other black hole. Indeed,
recall that in these diagrams the vertical axes represents the
time flow and that light rays move along $45^{\rm o}$ lines. Thus,
a null ray that is emitted from a region next to the horizon $r_+$
of the left black hole of Fig. \ref{Fig light flatC}.(a) can pass
through the acceleration horizon $r_A$ and, once it has done this,
we will necessarily proceed into infinity $\cal{I^+}$ [see Fig.
\ref{Fig light flatC}.(a)]. This null ray cannot enter the right
region of Fig. \ref{Fig light flatC}.(a) and, in particular, it
cannot hit the horizon $r_+$ of the right black hole. So, if the
two black holes cannot communicate through a null ray they cannot
interact gravitationally. The only interaction
 that is present in the system is between the strut and each one
 of the black holes, that suffer an acceleration
which is only furnished by the  strut's pressure. That the limit
$A=0$ does not yield the solution \cite{bh_eq} can also be
inferred from \cite{Yong}, where the flat C-metric is obtained
from the the two black hole solution of \cite{bh_eq} but through a
singular limit in which several quantities go appropriately to
infinity. The situation is completely different in the case of the
Israel-Khan solution, whose causal diagram is sketched in Fig.
\ref{Fig light flatC}.(b). In this case, a null ray that is
emitted from a region next to the horizon $r_+$ of the left black
hole can hit the horizon $r_+$ of the right black hole. Hence, in
the Israel-Khan solution the two black holes can communicate
through a null ray and so they interact gravitationally. Note also
that the acceleration horizon is absent in the causal diagram of
the Israel-Khan solution, as expected for a static solution.

The Israel-Khan solution \cite{bh_eq} is valid for $\Lambda=0$
but, although it has not been done, it can be extended in
principle for generic $\Lambda$. Hence, the above remark holds
also for generic $\Lambda$, as has already briefly mentioned in
the end of subsection \ref{sec:PI.2-BH AdS}.

A similar discussion also applies to the Ernst solution. In this
solution, there is no gravitational force between the two
Reissner-Nordtr\"{o}m black holes. They accelerate apart only
subjected to the Lorentz force provided for the external
electromagnetic field, and when $A=0$ the Ernst solution reduces
to a single non-accelerated Reissner-Nordtr\"{o}m black hole. On
the other side, Tomimatsu \cite{Tomimatsu} has found a solution,
that is the charged counterpart of the neutral Israel-Khan
solution \cite{bh_eq}, in which two Reissner-Nordtr\"{o}m black
holes are held in equilibrium. In this case the gravitational
attraction between the black holes is cancelled by the Coulomb
repulsion, and this occurs only when the black holes are extreme,
$M_i=Q_i$.  Tomimatsu has generalized for the black hole case, the
relativistic treatment used by Bonnor \cite{Bonnor2ParticleEquil}
and Ohta and Kimura \cite{OhtaKimura} to study the equilibrium
system of two charged particles.

\section[{\bf Pair of accelerated black holes in a de Sitter
background: the dS C-metric}]
 {Pair of accelerated black holes in a de Sitter background:
 \\ the dS C-metric} \label{sec:dS C-metric}

The plan of this section is as follows. In section
\ref{sec:General properties dS} we present the dS C-metric and
analyze its curvature and conical singularities. In section
\ref{sec:PD dS} we study the causal diagrams of the solution. In
section \ref{sec:Phys_Interp dS} we give and justify a physical
interpretation to the solution. The description of the solution in
the dS 4-hyperboloid and the physics of the strut are analyzed.

\subsection{\label{sec:General properties dS} General properties of the
\lowercase{d}S C-metric}
\subsubsection{\label{sec:dS C-metric sub} The \lowercase{d}S
C-metric}
The dS C-metric, i.e., the C-metric with positive cosmological
constant $\Lambda$, has been obtained by Pleba\'nski and
Demia\'nski \cite{PlebDem}. For zero rotation and zero NUT
parameter, the gravitational field of the dS C-metric can be
written as
\begin{equation}
 d s^2 = [A(x+y)]^{-2} (-{\cal F}dt^2+
 {\cal F}^{-1}dy^2+{\cal G}^{-1}dx^2+
 {\cal G}d\phi^2)\:,
 \label{C-metric}
 \end{equation}
 where
 \begin{eqnarray}
 & &{\cal F}(y) = -{\biggl (}\frac{1}{\ell^2A^2}+1{\biggl )}
                     +y^2-2mAy^3+q^2A^2y^4, \nonumber\\
 & &{\cal G}(x) = 1-x^2-2mAx^3-q^2 A^2 x^4\:.
 \label{FG}
 \end{eqnarray}
The Maxwell field in the magnetic case is given by (\ref{F-mag}),
while in the electric case it is given by (\ref{F-el-Lorentz}).
This solution depends on four parameters namely, the cosmological
length $\ell^2\equiv3/\Lambda$, $A>0$ which is the acceleration of
the black holes (see subsection
 \ref{sec:Phys_Interp dS}), and $m$ and $q$ which are
interpreted  as the ADM mass and electromagnetic charge of the
non-accelerated black holes, respectively (see subsection
\ref{sec:Phys_Interp m,e dS}). The physical
 properties and interpretation of this solution have been analyzed
 by Dias and Lemos \cite{OscLem_dS-C}, and
 by  Podolsk\'y and Griffiths \cite{PodGrif2}.

Now, the general properties of the dS C-metric, in what concerns
the issues of radial coordinate and curvature singularities, the
analysis of the angular surfaces and conical singularities, and
the issue of coordinate ranges are straightforwardly similar to
the corresponding properties of the AdS C-metric analyzed in
subsections \ref{sec:CurvSing AdS}, \ref{sec:ConSing AdS}, and
\ref{sec:CoordRange AdS}. For this reason we will not analyze
again these properties in detail. We will only briefly mention the
main features, and we ask the reader to go back to subsection
\ref{sec:CurvSing AdS} or to see \cite{OscLem_dS-C} for details.

As occurs with the AdS C-metric (see subsection \ref{sec:CurvSing
AdS}), we can define a radial coordinate given by (\ref{r}). The
Kretschmann scalar is also given by (\ref{R2}), and indicates that
the dS C-metric has a curvature singularity at $y=+\infty$, and
$y$ must belong to the range $-x\leq y <+\infty$. The point $y=-x$
corresponds to a point that is infinitely far away from the
curvature singularity, thus as $y$ increases we approach the
curvature singularity and $y+x$ is the inverse of a radial
coordinate. At most, ${\cal F}(y)$ can have four real zeros which
we label in ascending order by $y_{\rm neg}<0<y_A\leq y_+ \leq
y_-$. The roots $y_-$ and $y_+$ are respectively the inner and
outer charged black hole horizons, and $y_A$ is an acceleration
horizon which coincides with the cosmological horizon and has a
non-spherical shape, although the topology is spherical. The
negative root $y_{\rm neg}$ satisfies $y_{\rm neg}<-x$ and has no
physical significance, i.e., it does not belong to the range
accessible to $y$.

As occurs with the AdS C-metric (see subsection \ref{sec:ConSing
AdS}), we will demand that $x$ belongs to the interval
$x_\mathrm{s}\leq x\leq x_\mathrm{n}$, sketched onwards in the
appropriate figures, where ${\cal G}(x)\geq 0$. By doing this we
guarantee that the metric has the correct signature $(-+++)$ [see
 (\ref{C-metric})] and that the angular surfaces $t=$const and
$y=$const are compact.
 In these angular surfaces we can also define, through (\ref{ang}),
two new coordinates, $\theta$ and $\tilde{\phi}$, where
$\tilde{\phi}$ ranges between $[0,2\pi]$ and $\kappa$ is an
arbitrary positive constant which will be discussed later. The
coordinate $\theta$ ranges between the north pole,
$\theta=\theta_\mathrm{n}=0$, and the south pole,
$\theta=\theta_\mathrm{s}$ (not necessarily at $\pi$). Rewritten
as a function of these new coordinates, the angular part of the
metric becomes $d\theta^2+\kappa^2{\cal G}\,d\tilde{\phi}^2$. Note
that when we set $A=0$ we have $x_\mathrm{s}\leq x\leq
x_\mathrm{n}$, with $x_\mathrm{s}=-1$ and $x_\mathrm{n}=1$,
$x=\cos \theta$, ${\cal G}=1-x^2=\sin^2 \theta$, and $\kappa=1$.
Therefore, in this case the compact angular surface $\tilde{S}^2$
is a round $S^2$ sphere which justifies the label given to the new
angular coordinates. When we set $A \neq 0$ the compact angular
surface turns into a deformed 2-sphere that we represent onwards
by $\tilde{S}^2$.

Finally, note again that when $m$ or $q$ are not zero there is a
curvature singularity at $r=0$. Therefore, we restrict the radial
coordinate to the range $[0,+\infty[$. On the other hand, we  have
restricted $x$ to belong to the range
$[x_\mathrm{s},x_\mathrm{n}]$ where ${\cal G}(x)\geq 0$. From
$Ar=(x+y)^{-1}$ we then conclude that $y$ must belong to the range
$-x\leq y < +\infty$. Indeed, $y=-x$ corresponds to $r=+\infty$,
and $y=+\infty$ to $r=0$. However, when both $m$ and $q$ vanish
there are no restrictions on the ranges of $r$ and $y$ (i.e.,
$-\infty < r < +\infty$ and $-\infty < y < +\infty$) since in this
case there is no curvature singularity at the origin of $r$ to
justify the constraint on the coordinates.

\subsubsection{\label{sec:Phys_Interp m,e dS} Mass and charge
parameters}

In this subsection, one gives the physical interpretation of
parameters $m$ and $q$ that appear in the dS C-metric. Applying to
(\ref{C-metric}) the coordinate transformations \cite{PodGrif2},
$\tau=\sqrt{1+\ell^2A^2}A^{-1} t$,
$\rho=\sqrt{1+\ell^2A^2}(Ay)^{-1}$, together with (\ref{ang}) and
setting $A=0$ (and $\kappa=1$, $\tilde{\phi}=\phi$) one obtains
\begin{equation}
 d s^2 = - F(\rho)\, d \tau^2 +F^{-1}(\rho)\, d \rho^2
 +\rho^2 (d \theta^2+\sin^2\theta\,d\phi^2) \:,
 \label{mq2 dS}
\end{equation}
where $F(\rho)=1-\rho^2/\ell^2 -2m/\rho + q^2/\rho^2$. So, when
the acceleration parameter vanishes, the dS C-metric,
(\ref{C-metric}), reduces to the dS-Schwarzschild and
dS$-$Reissner-Nordstr\"{o}m black holes and the parameters $m$ and
$q$ that are present in the dS C-metric are precisely the ADM mass
and ADM electromagnetic charge of these non-accelerated black
holes.

\subsection{\label{sec:PD dS} Causal Structure of the
\lowercase{d}S C-metric}

In this section we analyze the causal structure of the solution.
The original dS C-metric, (\ref{C-metric}), is not geodesically
complete. To obtain the maximal analytic spacetime, i.e., to draw
the Carter-Penrose diagrams we have to introduce the usual null
Kruskal coordinates.  The technical procedure to obtain the
Carter-Penrose diagram is similar to the one presented in
subsection \ref{sec:PD AdS}. Hence, we will not present again the
construction process that leads to the diagrams. We will only
discuss the main features and the diagrams. The description of the
solution depends crucially on the values of $m$ and $q$. We will
consider the three most relevant solutions, namely: {\it A.
Massless uncharged solution} ($m =0$, $q=0$), {\it B. Massive
uncharged solution} ($m>0$, $q=0$), and {\it C. Massive charged
solution} ($m \geq0$, $q\neq0$).

\subsubsection[Massless uncharged solution ($m=0$, $q=0$)]{\label{sec:PD A.1 dS}
 Massless uncharged solution ($\bm{m=0}$, $\bf{q=0}$)}

\begin{figure}[H]
\centering
\includegraphics[height=1.6in]{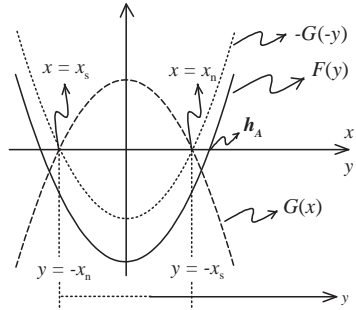}
\caption{\label{g1 dS}
 Shape of ${\cal G}(x)$ and ${\cal F}(y)$ for the
 $m=0, q=0$ dS C-metric. The allowed range of $x$ is between
$x_\mathrm{s}=-1$ and $x_\mathrm{n}=+1$ where ${\cal G}(x)$ is
positive and compact. The range of $y$ is restricted to $-x\leq y
< +\infty$. The presence of an accelerated horizon is indicated by
$h_A$. It coincides with the cosmological horizon of the solution
and has a non-spherical shape.}
\end{figure}

In this case we have $x \in [x_\mathrm{s}=-1,x_\mathrm{n}=+1]$,
$x=\cos \theta$, ${\cal G}=1-x^2=\sin^2 \theta$, $\kappa=1$ (and
so $\tilde{\phi}=\phi$)  and
\begin{eqnarray}
 {\cal F}(y) = y^2-y_+^2 \;\;\;\;\;\;\mathrm{with}\;\;\;\;\;\;
 y_+=\sqrt{1+\frac{1}{\ell^2A^2}} \:.
 \label{F1 dS}
 \end{eqnarray}
The general behavior of these functions for this case is
represented in Fig. \ref{g1 dS}.

 The angular surfaces $\Sigma$ ($t=$const and
$r=$const) are spheres and both the north and south poles are free
of conical singularities. The origin of the radial coordinate $r$
has no curvature singularity and therefore both $r$ and $y$ can
lie in the range $]-\infty,+\infty[$. However, in the realistic
general case, where $m$ or $q$ are non-zero, there is a curvature
singularity at $r=0$. Since the discussion of the present section
is only a preliminary to that of the massive general case we will
treat the origin $r=0$ as if it had a curvature singularity and
thus we admit that $r$ belongs to the range $[0,+\infty[$ and $y$
lies in the region $-x\leq y < +\infty$. We leave a discussion on
the extension to negative values of $r$ to section
 \ref{sec:PI dS}.

\begin{figure}  [H]
\centering
\includegraphics{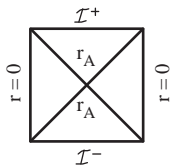}
   \caption{\label{Fig-1 dS}
Carter-Penrose diagram concerning the $m=0, q=0$ dS C-metric
studied in section \ref{sec:PD A.1 dS}. The accelerated horizon is
represented by $r_A$. It coincides with the cosmological horizon
and has a non-spherical shape. ${\cal I}^-$ and ${\cal I}^+$
represent respectively the past and future infinity ($r=+\infty$).
$r=0$ corresponds to $y=+\infty$ and $r=+\infty$ corresponds to
$y=-x$.
 }
\end{figure}

The construction of the Carter-Penrose diagram follows directly
the one given in case (c) of subsection \ref{sec:PD A AdS}. The
Carter-Penrose diagram of the massless uncharged dS C-metric is
sketched in Fig. \ref{Fig-1 dS}. $r=0$ is represented by a
timelike line while $r=+\infty$ is a spacelike line (with ${\cal
I}^-$ and ${\cal I}^+$ representing, respectively, the past and
future infinity). The two mutual perpendicular straight null lines
at $45^{\rm o}$, $u'v'=0$, represent a Rindler-like accelerated
horizon.

To end this subsection it is important to remark that, contrary to
what happens in the C-metric with  $\Lambda<0$ \cite{OscLem_AdS-C}
and with $\Lambda=0$ \cite{KW}, the presence of the acceleration
in the $\Lambda>0$ C-metric does not introduce an extra horizon
relatively to the $A=0$ solution. Indeed, in the dS C-metric the
acceleration horizon coincides with the cosmological horizon that
is already present in the $A=0$ solution. However, in the $A=0$
solution the cosmological horizon has the topology of a round
sphere, while in the dS C-metric ($A\neq 0$) the presence of the
acceleration induces a non-spherical shape in the acceleration
(cosmological) horizon. This conclusion is set from the expression
of the radius of the horizon, $r_A=A^{-1}(x+y_+)^{-1}$. It varies
with the angular direction $x=\cos \theta$ and depends on the
value of $A$ [see (\ref{F1 dS})]. Another important difference
between the causal structure of the dS C-metric and the causal
structure of the $\Lambda = 0$ and the $\Lambda<0$ cases is the
fact that the general features of the Carter-Penrose diagram of
the dS C-metric are independent of the angular coordinate $x=\cos
\theta$. Indeed, in the  $\Lambda<0$ case (see subsection
\ref{sec:PD A AdS} and \cite{OscLem_AdS-C}) and in the $\Lambda=0$
case (see subsection \ref{sec:PD flat} and \cite{KW}), the
Carter-Penrose diagram at the north pole direction is
substantially different from the one along the south pole
direction and different from the diagram along the equator
direction (see \cite{KW,OscLem_AdS-C}).

\subsubsection[Massive uncharged solution ($m >0$, $q=0$)]{\label{sec:PD A.2 dS}
 Massive uncharged solution ($\bm{m >0}$, $\bm{q=0}$)}
\begin{figure} [H]
\centering
\includegraphics[height=1.6in]{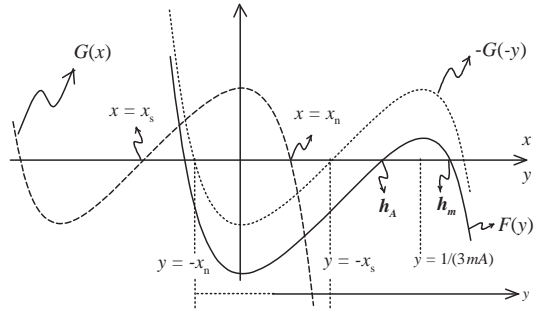}
\caption{\label{g2 dS}
 Shape of ${\cal G}(x)$ and ${\cal F}(y)$ for the
 $27m^2A^2<1-9m^2\Lambda$, and $q=0$ dS C-metric (case (i) in the text).
The allowed range of $x$ is between $x_\mathrm{s}$ and
$x_\mathrm{n}$ where ${\cal G}(x)$ is positive and compact. The
range of $y$ is restricted to $-x\leq y < +\infty$. The presence
of an accelerated horizon (which coincides with the cosmological
horizon and has a non-spherical shape) is indicated by $h_A$ and
the Schwarzschild-like horizon by $h_m$. For completeness we
comment on two other cases studied in the text: for
$27m^2A^2=1-9m^2\Lambda$ (case (ii) in the text), ${\cal F}(y)$ is
zero at its local maximum, i.e., $h_A$ and $h_m$ coincide. For
$27m^2A^2>1-9m^2\Lambda$ (case (iii) in the text), ${\cal F}(y)$
is always negative in the allowed range of $y$.
 }
\end{figure}
The construction of the Carter-Penrose diagram for the $m> 0$ dS
C-metric follows up directly from the last subsection. We will
consider the small mass or acceleration case, i.e., we require
$27m^2A^2<1$ and we also demand $x$ to belong to the range
$[x_\mathrm{n},x_\mathrm{s}]$ (represented in Fig. \ref{g2 dS} and
such that $x_\mathrm{s} \to -1$ and $x_\mathrm{n} \to +1$ when $mA
\to 0$) where ${\cal G}(x)\geq 0$. By satisfying the two above
conditions we endow the $t=$const and $r=$const surfaces $\Sigma$
with the topology of a compact surface. For $27m^2A^2 \geq 1$ this
surface is an open one and will not be discussed.

\begin{figure}[H]
\centering
\includegraphics[height=2cm]{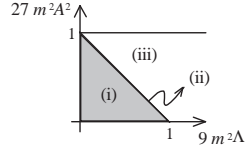}
   \caption{\label{Fig-RangeM dS}
Allowed ranges of the parameters $\Lambda, A$, and $m$ for the
cases (i), (ii), and (iii) of the uncharged massive dS C-metric
discussed in the text of section \ref{sec:PD A.2 dS}.
 }
\end{figure}

Now we turn our attention to the behavior of function ${\cal
F}(y)$. We have to consider three distinct cases (see Fig.
\ref{Fig-RangeM dS}), namely: (i) pair of non-extreme
dS-Schwarzschild black holes ($27m^2A^2<1-9m^2\Lambda$) for which
${\cal F}(y=1/3mA)>0$ (see Fig. \ref{g2 dS}), (ii) pair of extreme
dS-Schwarzschild black holes ($27m^2A^2=1-9m^2\Lambda$) for which
${\cal F}(y=1/3mA)=0$, and (iii) case $27m^2A^2>1-9m^2\Lambda$ for
which ${\cal F}(y)$ is always negative in the allowed range for
$y$. This last case represents a naked particle and will not be
discussed further. Notice that when we set $A=0$ in the above
relations we get the known results \cite{Lake} for the
non-accelerated dS spacetime, namely: for $9m^2\Lambda<1$ we have
the non-extreme dS-Schwarzschild solution and for $9m^2\Lambda=1$
we get the extreme dS-Schwarzschild solution. In what follows we
will draw the Carter-Penrose diagrams of cases (i) and (ii).

\begin{figure}[H]
\centering
\includegraphics[height=4.8cm]{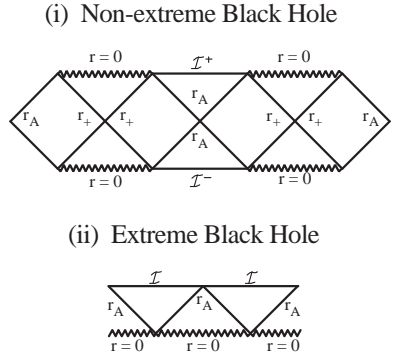}
   \caption{\label{Fig-2 dS}
(i) Carter-Penrose diagram of the $27m^2A^2<1-9m^2\Lambda$, and
$q=0$ dS C-metric discussed in case (i) of section \ref{sec:PD A.2
dS}. The zigzag line represents a curvature singularity, the
accelerated horizon is represented by $r_A$. It coincides with the
cosmological horizon and has a non-spherical shape. The
Schwarzschild-like horizon is sketched as $r_+$. $r=0$ corresponds
to $y=+\infty$ and $r=+\infty$ (${\cal I}^-$ and ${\cal I}^+$)
corresponds to $y=-x$. (ii) Carter-Penrose diagram of the
degenerate case (ii), $27m^2A^2=1-9m^2\Lambda$ and $q=0$,
discussed in the text of section \ref{sec:PD A.2 dS}. The
accelerated horizon $r_A$ coincides with the Schwarzschild-like
horizon $r_+$.
 }
\end{figure}

\vspace{0.1 cm} (i) {\it Pair of non-extreme dS-Schwarzschild
black holes} ($27m^2A^2<1-9m^2\Lambda$): the technical procedure
to obtain the Carter-Penrose diagram is similar to the one
described along section \ref{sec:PD A.1 dS}. In what concerns the
physical conclusions, we will see that the essential difference is
the presence of an extra horizon, a Schwarzschild-like horizon
($r_+$) due to the non-vanishing mass parameter, in addition to
the accelerated Rindler-like horizon ($r_A$). Another important
difference is the presence of a curvature singularity at the
origin of the radial coordinate and the existence of a conical
singularity at one of the poles. The Carter-Penrose diagram is
drawn in Fig. \ref{Fig-2 dS}.(i) and has a structure that can be
divided into left, middle and right regions. The middle region
contains the spacelike infinity (with ${\cal I}^-$ and ${\cal
I}^+$ representing, respectively, the past and future infinity)
and an accelerated Rindler-like horizon, $r_A=[A(x-x_-)]^{-1}$,
that was already present in the $m=0$ corresponding diagram [see
Fig. \ref{Fig-1 dS}]. The left and right regions both contain a
spacelike curvature singularity at $r=0$ and a Schwarzschild-like
horizon, $r_+$. This diagram is analogous to the one of the
non-accelerated ($A=0$) dS-Schwarzschild solution. However, in the
$A=0$ solution the cosmological and black hole horizons have the
topology of a round sphere, while in the dS C-metric ($A\neq 0$)
the presence of the acceleration induces a non-spherical shape in
the acceleration horizon (that coincides with the cosmological
horizon) and in the black hole horizon. Indeed, notice that once
we find the zero, $y_h$, of ${\cal F}(y)$ that corresponds to an
accelerated or black hole horizon, the position of these horizons
depends on the angular coordinate $x$ since $r_h=[A(x+y_h)]^{-1}$.
In section \ref{sec:PI-mass dS} we will justify that this solution
describes a pair of accelerated dS-Schwarzschild black holes.

\vspace{0.1 cm} (ii) {\it Pair of extreme dS-Schwarzschild black
holes} ($27m^2A^2=1-9m^2\Lambda$): for this range of values, we
have a degenerate case in which the size of the black hole horizon
approaches and equals the size of the acceleration horizon. In
this case, as we shall see in section \ref{sec:PI-mass dS}, the dS
C-metric describes a pair of accelerated extreme dS-Schwarzschild
black holes. The Carter-Penrose diagram of this solution is
sketched in Fig. \ref{Fig-2 dS}.(ii). It should be noted that for
this sector of the solution, and as occurs with the $A=0$ case,
there is an appropriate limiting procedure (see chapter
\ref{chap:Extremal Limits} and \cite{OscLem_nariai}) that takes
this solution into the Nariai C-metric, i.e., the accelerated
counterpart of the Nariai solution \cite{Nariai}.

\subsubsection[Massive charged
solution ($m >0$, $q\neq 0$)]{\label{sec:PD A.3 dS} Massive
charged solution ($\bm{m >0}$, $\bm{q\neq0}$)}
\begin{figure} [H]
\centering
\includegraphics[height=1.6in]{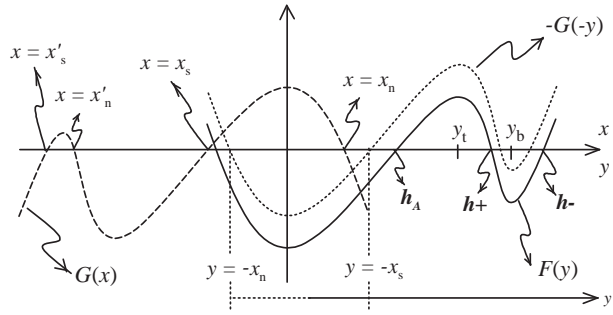}
\caption{\label{g3 dS}
 Shape of ${\cal G}(x)$ and ${\cal F}(y)$ for the non-extreme
charged massive dS C-metric (case (i) in the text of section
\ref{sec:PD A.3 dS}). The allowed range of $x$ is between
$x_\mathrm{s}$ and $x_\mathrm{n}$ where ${\cal G}(x)$ is positive
and compact.  The presence of an accelerated horizon is indicated
by $h_A$ and the inner and outer charged horizons by $h-$ and
$h+$. In the extreme cases, $h-$ and $h+$ [case (ii)] or $h-_A$
and $h+$ [case (iii)] superpose each other and in the naked case
[case (iv) and (v)] ${\cal F}(y)$ has only one zero in the allowed
range of $y$.
 }
\end{figure}

\begin{figure}[H]
\centering
\includegraphics[height=9cm]{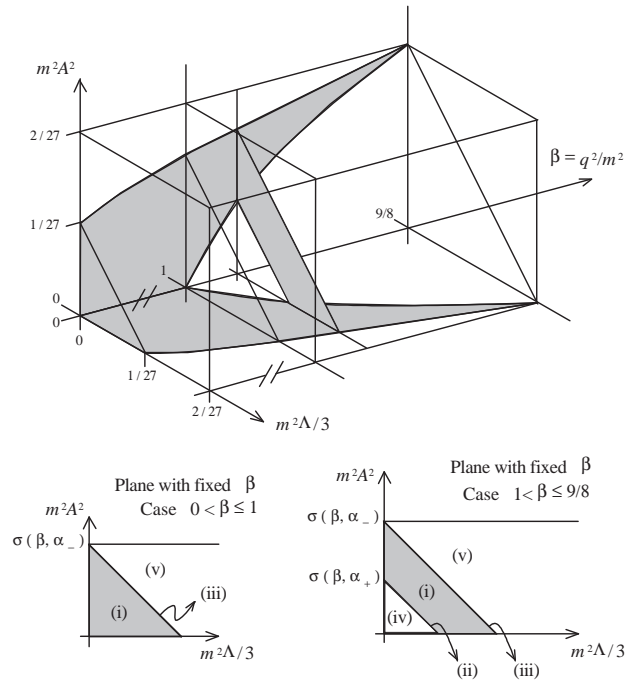}
   \caption{\label{Fig-RangeQ dS}
Allowed ranges of the parameters $\Lambda, A, m, \beta\equiv
q^2/m^2$ for the cases (i), (ii), (iii), (iv), and (v) of the
charged massive dS C-metric discussed in the text of section
\ref{sec:PD A.3 dS}.
 }
\end{figure}

When both the mass and charge parameters are non-zero, depending
on the values of the parameters, ${\cal G}(x)$ can be positive in
a single compact interval, $]x_\mathrm{s},x_\mathrm{n}[$, or in
two distinct compact intervals, $]x_\mathrm{s},x_\mathrm{n}[$ and
$]x'_\mathrm{s},x'_\mathrm{n}[$, say. We require that $x$ belongs
to the interval $[x_\mathrm{s},x_\mathrm{n}]$ (sketched in Fig.
\ref{g3 dS}) for which the charged solutions are in the same
sector of those we have analyzed in the last two subsections when
$q \to 0$. Defining
\begin{eqnarray}
& & \beta \equiv \frac{q^2}{m^2}\:,  \;\;\; 0<\beta\leq
\frac{9}{8} \:, \;\;\;\;\;\alpha_{\pm} \equiv
1 \pm \sqrt{1-\frac{8}{9}\beta} \:, \nonumber \\
 & &
\sigma(\beta,\alpha_{\pm})=
\frac{(4\beta)^2(3\alpha_{\pm})^2-8\beta(3\alpha_{\pm})^3+\beta(3\alpha_{\pm})^4}
{(4\beta)^4} \:,  \nonumber \\
 \label{beta dS}
\end{eqnarray}
the above requirement is fulfilled by the parameter range
$m^2A^2<\sigma(\beta,\alpha_-)$. Now we look into the behavior of
the function ${\cal F}(y)$. Depending on the sign of ${\cal F}(y)$
at $y_{\rm t}$ and $y_{\rm b}$  (with $y_{\rm
t}=\frac{3\alpha_-}{4\beta mA}$ and
 $y_{\rm b}=\frac{3\alpha_+}{4\beta mA}$ being the points represented in
Fig. \ref{g3 dS} where the derivative of ${\cal F}(y)$ vanishes)
we can group the solutions into five different relevant physical
classes, namely: (i) ${\cal F}(y_{\rm t})>0$ and ${\cal F}(y_{\rm
b})<0$, (ii) ${\cal F}(y_{\rm t})>0$ and ${\cal F}(y_{\rm b})=0$,
(iii) ${\cal F}(y_{\rm t})=0$ and ${\cal F}(y_{\rm b})<0$, (iv)
${\cal F}(y_{\rm t})>0$ and ${\cal F}(y_{\rm b})>0$, and (v)
${\cal F}(y_{\rm t})<0$ and ${\cal F}(y_{\rm b})<0$. The ranges of
parameters $\Lambda, A, m$, and $\beta$ that correspond to these
five cases are identified in Fig. \ref{Fig-RangeQ dS}.

Condition ${\cal F}(y_{\rm t})\geq 0$ requires
 $m^2A^2 \leq \sigma(\beta,\alpha_-)-m^2\Lambda/3$ and
 ${\cal F}(y_{\rm b})\leq 0$ is satisfied by
 $m^2A^2 \geq \sigma(\beta,\alpha_+)-m^2\Lambda/3$.
We have $\sigma(\beta,\alpha_-)>\sigma(\beta,\alpha_+)$ except at
$\beta=9/8$ where these two functions are equal;
$\sigma(\beta,\alpha_-)$ is always positive; and
$\sigma(\beta,\alpha_+)<0$ for $0<\beta<1$ and
$\sigma(\beta,\alpha_+)>0$ for $1<\beta\leq 9/8$. Case (i) has
three horizons, the acceleration horizon $h_A$ and the inner
($h_-$) and outer ($h_+$) charged horizons and is the one that is
exactly represented in Fig. \ref{g3 dS} ($h_A\neq h_+ \neq h_-$);
in case (ii) the inner horizon and outer horizon coincide ($h_+
\equiv h_-$) and are located at $y_{\rm b}$ ($h_A$ is also
present); in case (iii) the acceleration horizon and outer horizon
coincide ($h_A \equiv h_+$) and are located at $y_{\rm t}$ ($h_-$
is also present); finally in cases (iv) and (v) there is a single
horizon  $h_A \equiv h_+ \equiv h_-$.
  As will be seen, case (i) describes a pair of accelerated
dS$-$Reissner-Nordstr\"{o}m (dS-RN) black holes, case (ii)
describes a pair of extreme dS-RN black holes in which the inner
and outer charged horizons become degenerated, case (iii)
describes a pair of extreme dS-RN black holes in which
acceleration horizon and outer charged horizon  become
degenerated, and cases (iv) and (v) describe a pair of naked
charged particles. In chapter \ref{chap:Extremal Limits} we will
return to the study of the properties of the extreme cases of the
dS C-metric. We will give explicit expressions that give the mass
and charge of these extreme black holes as a function of $\Lambda$
and $A$. Moreover, we will see that for the sector (ii) of the
solution, and as occurs with the $A=0$ case,  there is an
appropriate limiting procedure \cite{OscLem_nariai} that takes
this solution into the Bertotti-Robinson C-metric, i.e., the
accelerated counterpart of the Bertotti-Robinson solution
\cite{BertRob}. We will also show that for the sector (iii) of the
solution, and as occurs with the $A=0$ case, there is also an
appropriate limiting procedure \cite{OscLem_nariai} that takes
this solution into the charged Nariai C-metric, i.e., the
accelerated counterpart of the charged Nariai solution
\cite{MannRoss,Nariai}.

The essential differences between the Carter-Penrose diagram of
the massive charged solutions and the diagram of the massive
uncharged solutions are: (1) the curvature singularity is now
represented by a timelike line rather than a spacelike line, (2)
excluding the extreme and naked cases, there are now (in addition
to the accelerated Rindler-like horizon, $r_A$) not one but two
extra horizons, the expected inner ($r_-$) and outer ($r_+$)
horizons associated to the charged character of the solution.

\begin{figure} [H]
\centering
\includegraphics[height=14cm]{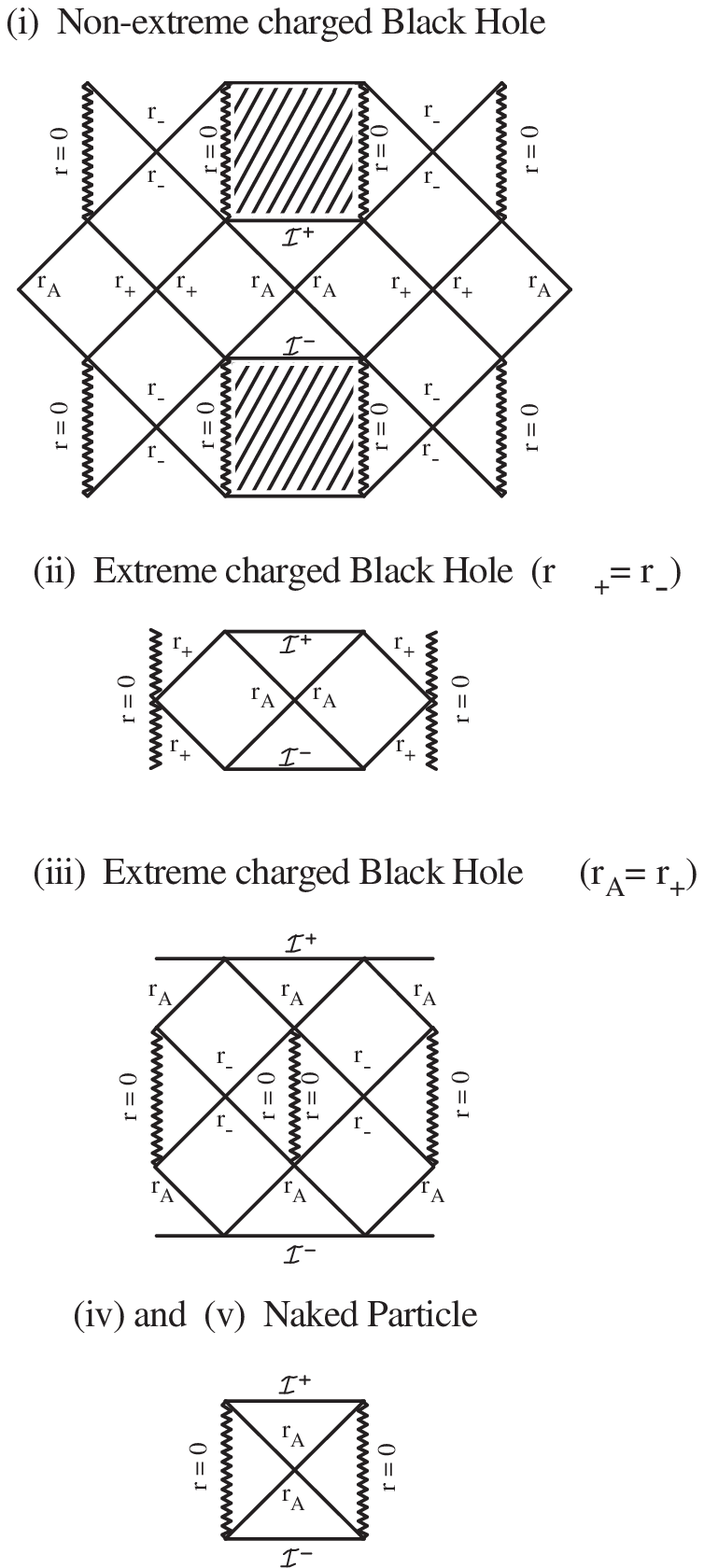}
   \caption{\label{Fig-3 dS}
Carter-Penrose diagrams of cases (i), (ii), (iii), and (iv) and
(v) of the charged massive dS C-metric. The zigzag line represents
a curvature singularity, an accelerated horizon is represented by
$r_A$, the inner and outer charge associated horizons are sketched
as $r_-$ and $r_+$. ${\cal I}^-$ and ${\cal I}^+$ represent
respectively the past and future infinity ($r=+\infty$). $r=0$
corresponds to $y=+\infty$ and $r=+\infty$ corresponds to $y=-x$.
 }
\end{figure}

The Carter-Penrose diagram of case (i) is drawn in Fig. \ref{Fig-3
dS}.(i) and has a structure that, as occurs in the massive
uncharged case, can be divided into left, middle and right
regions. The middle region contains the spacelike infinity and an
accelerated Rindler-like horizon, $r_A$, that was already present
in the $q=0=m$ corresponding diagram (see Fig. \ref{Fig-1 dS}).
The left and right regions both contain a timelike curvature
singularity ($r=0$), and an inner ($r_-$) and an outer ($r_+$)
horizons associated to the charged character of the solution. This
diagram is analogous the the one of the non-accelerated ($A=0$)
dS$-$Reissner-Nordstr\"{o}m solution. However, in the $A=0$
solution the cosmological and black hole horizons have the
topology of a round sphere, while in the dS C-metric ($A\neq 0$)
the presence of the acceleration induces a non-spherical shape in
the accelerated horizon (that coincides with the cosmological
horizon) and in the black hole horizons. Indeed, notice that once
we find the zero, $y_h$, of ${\cal F}(y)$ that corresponds to an
accelerated or black hole horizon, the position of these horizons
depends on the angular coordinate $x$ since $r_h=[A(x+y_h)]^{-1}$.
In Fig. \ref{Fig-3 dS} are also represented the other cases
(ii)-(v). Again the accelerated horizon is in between two
(extreme) black holes in cases (ii) and (iii) and in between two
naked particles in cases (iv) and (v).

\subsection{\label{sec:Phys_Interp dS} Physical interpretation of
the \lowercase{d}S C-metric}

The parameter $A$ that is found in the dS C-metric is interpreted
as being an acceleration and the dS C-metric describes a pair of
black holes accelerating away from each other in a dS background.
In this section we will justify this statement.

In subsection \ref{sec:Phys_Interp m,e dS} we saw that, when
$A=0$, the general dS C-metric, (\ref{C-metric}), reduces to the
dS ($m=0\,,\,q=0$), to the dS-Schwarzschild ($m>0\,,\,q=0$), and
to the dS$-$Reissner-Nordstr\"{o}m solutions ($m=0\,,\,q\neq0$).
Therefore, the parameters $m$ and $q$ are, respectively, the ADM
mass and ADM electromagnetic charge of the non-accelerated black
holes. Moreover, if we set the mass and charge parameters equal to
zero, even when $A\neq 0$, the Kretschmann scalar
 [see (\ref{R2})] reduces to the value expected for the dS spacetime.
This indicates that the massless uncharged dS C-metric is a dS
spacetime in disguise.

In this section, we will first interpret case {\it A. Massless
uncharged solution} ($m =0$, $q=0$), which is the simplest, and
then with the acquired knowledge we interpret cases {\it B.
Massive uncharged solution} ($m>0$, $q=0$) and {\it C. Massive
charged solution} ($m>0$, $q\neq0$). We will interpret the
solution following two complementary descriptions, the four
dimensional (4D) one and the five dimensional (5D).

\subsubsection[Description of the $m=0$, $q = 0$ solution]
{\label{sec:PI dS}Description of the $\bm{m=0}$, $\bm{q=0}$
solution}
\begin{figure} [H]
\centering
\includegraphics[height=2.3cm]{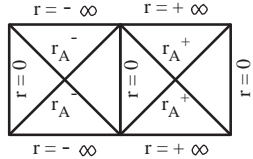}
   \caption{\label{Fig-dS}
Extending the Carter-Penrose diagram of Fig. \ref{Fig-1 dS} to
negative values of $r$, we obtain the dS spacetime with its origin
being accelerated. $r_A^- =[A(x-y_+)]^{-1}<0$ and
$r_A^+=[A(x+y_+)]^{-1}>0$.
 }
\end{figure}

\noindent {\bf The 4-Dimensional description}:

As we said in \ref{sec:PD A.1 dS}, when $m=0$ and $q=0$ the origin
of the radial coordinate $r$ defined in (\ref{r}) has no curvature
singularity and therefore $r$ has the range $]-\infty,+\infty[$.
However, in the realistic general case, where $m$ or $q$ are
non-zero, there is a curvature singularity at $r=0$ and since the
discussion of the massless uncharged solution was only a
preliminary to that of the massive general case, following
\cite{AshtDray}, we have treated the origin $r=0$ as if it had a
curvature singularity and thus we admitted that $r$ belongs to the
range $[0,+\infty[$. In these conditions we obtained the causal
diagram of Fig. \ref{Fig-1 dS}. Note however that one can make a
further extension to include the negative values of $r$, enlarging
in this way the range accessible to the Kruskal coordinates $u'$
and $v'$. By doing this procedure we obtain the causal diagram of
the dS spacetime, represented in Fig. \ref{Fig-dS}.

Now, we want to clearly identify the parameter $A$ that appears in
the dS C-metric with the acceleration of its origin. To achieve
this aim, we recover the massless uncharged dS C-metric defined by
(\ref{C-metric}) and (\ref{FG}) (with $m=0$ and $q=0$), and after
performing the following coordinate transformation \cite{PodGrif2}
\begin{eqnarray}
& & \tau=\frac{\sqrt{1+\ell^2A^2}}{A} t \:,  \;\;\;\;\;
    \rho=\frac{\sqrt{1+\ell^2A^2}}{A} \frac{1}{y} \:, \nonumber \\
& & \theta = \arccos{x} \:,
     \;\;\;\;\; \tilde{\phi} = \phi \:,
  \label{transf-int dS}
  \end{eqnarray}
we can rewrite the massless uncharged dS C-metric as
\begin{eqnarray}
 d s^2 = \frac{1}{\gamma^2}
 {\biggl [}-(1-\rho^2/\ell^2)d\tau^2+
 \frac{d\rho^2}{1-\rho^2/\ell^2} +\rho^2 d\Omega^2 {\biggl ]},
\label{metric-int dS}
 \end{eqnarray}
 with $d\Omega^2=d\theta^2+\sin^2\theta d \phi^2$ and
 \begin{eqnarray}
 \gamma=\sqrt{1+\ell^2A^2} + A\rho \cos\theta \:.
 \label{gamma dS}
 \end{eqnarray}
At this point some remarks are convenient. The origin of the
radial coordinate $\rho$ corresponds to $y=+\infty$ and therefore
to $r=0$, where $r$ has been introduced in (\ref{r}). So, when we
consider the massive dS C-metric there will be a curvature
singularity at $\rho=0$. Moreover, when we set $A=0$,
(\ref{metric-int dS}) reduces to the usual dS spacetime written in
static coordinates.

To discover the meaning of the parameter $A$ we consider the 4D
timelike worldlines described by an observer with $\rho=$const,
$\theta=0$ and $\phi=0$ (see \cite{PodGrif2}). These are given by
$x^{\mu}(\lambda)=(\gamma \ell
\lambda/\sqrt{\ell^2-\rho^2},\rho,0,0)$, were $\lambda$ is the
proper time of the observer and the 4-velocity
$u^{\mu}=dx^{\mu}/d\lambda$ satisfies $u_{\mu}u^{\mu}=-1$. The
4-acceleration of these observers,
$a^{\mu}=(\nabla_{\nu}u^{\mu})u^{\nu}$, has a magnitude given by
\begin{eqnarray}
 |a_4|=\sqrt{a_{\mu}a^{\mu}}=\frac{\rho\sqrt{1+\ell^2A^2}+\ell^2A}
 {\ell\sqrt{\ell^2-\rho^2}}\:.
 \label{a dS}
 \end{eqnarray}
Since $a_{\mu}u^{\mu}=0$, the value $|a_4|$ is also the magnitude
of the 3-acceleration in the rest frame of the observer. From
(\ref{a dS}) we achieve the important conclusion that the origin
of the dS C-metric, $\rho=0$ (or $r=0$), is being accelerated with
a constant acceleration $|a_4|$ whose value is precisely given by
the parameter $A$ that appears in the dS C-metric. Moreover, at
radius $\rho=\ell$ [or $y=y_+$ defined in equation (\ref{F1 dS})]
the acceleration is infinite which corresponds to the trajectory
of a null ray. Thus, observers held at $\rho=$const see this null
ray as an acceleration horizon and they will never see events
beyond this null ray. This acceleration horizon coincides with the
dS cosmological horizon and has a non-spherical shape. For the
benefit of comparison with the $A=0$ dS spacetime, we note that
when we set $A=0$,  (\ref{a dS}) says that the origin, $\rho=0$,
has zero acceleration and at radius $\rho=\ell$ the acceleration
is again infinite but now this is due only to the presence of  the
usual dS cosmological horizon which has a spherical shape.

\vspace{0.1 cm} \noindent {\bf The 5-Dimensional description}:

In order to improve and clarify the physical aspects of the dS
C-metric we turn now into the 5D representation of the solution.

The dS spacetime can be represented as the 4-hyperboloid,
\begin{eqnarray}
-(z^0)^2+(z^1)^2+(z^2)^2+(z^3)^2+(z^4)^2=\ell^2,
\label{hyperboloid dS}
 \end{eqnarray}
in the 5D Minkowski embedding spacetime,
\begin{eqnarray}
 d s^2 = -(dz^0)^2+(dz^1)^2+(dz^2)^2+(dz^3)^2+(dz^4)^2.
 \label{dS}
 \end{eqnarray}
Now, the massless uncharged dS C-metric is a dS spacetime in
disguise and therefore our next task is to understand how the dS
C-metric can be described in this 5D picture. To do this we first
recover the massless uncharged dS C-metric described by
(\ref{metric-int dS}) and apply to it the coordinate
transformation \cite{PodGrif2}
\begin{eqnarray}
  \hspace{-0.3cm} & & \hspace{-0.3cm}
  z^0=\gamma^{-1}\sqrt{\ell^2-\rho^2}\,\sinh(\tau/\ell)\:,
  \;\;\;\;\; z^2=\gamma^{-1} \rho \sin\theta \cos\phi \:,
  \nonumber \\
  \hspace{-0.3cm} & & \hspace{-0.3cm}
  z^1=\gamma^{-1}\sqrt{\ell^2-\rho^2}\,\cosh(\tau/\ell)\:,
  \;\;\;\;\;  z^3=\gamma^{-1} \rho \sin\theta \sin\phi \:,
  \nonumber \\
  \hspace{-0.3cm} & & \hspace{-0.3cm}
  z^4=\gamma^{-1}[\sqrt{1+\ell^2A^2} \,\rho \cos\theta
  +\ell^2A]\:,
\label{dS to dS-c}
  \end{eqnarray}
where $\gamma$ is defined in (\ref{gamma dS}). Transformations
(\ref{dS to dS-c}) define an embedding of the massless uncharged
dS C-metric into the 5D description of the dS spacetime since they
satisfy (\ref{hyperboloid dS}) and take directly (\ref{metric-int
dS}) into (\ref{dS}).

\begin{figure}[H]
\centering
\includegraphics[height=2.3in]{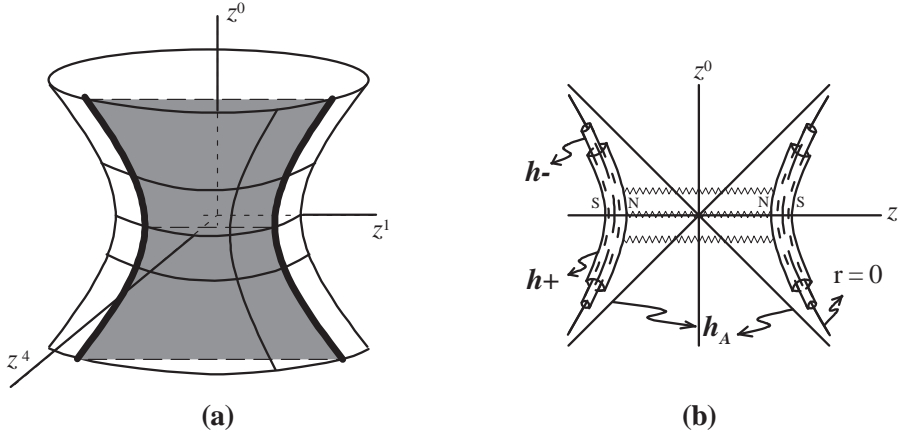}
\caption{\label{dS-hyperb} (a) The dS 4-hyperboloid embedded in
the 5D Minkowski spacetime. The directions $z^2$ and $z^3$ are
suppressed. The two hyperbolic lines lying on the dS hyperboloid
result from the intersection of the hyperboloid surface with the
$z^4$=constant$<\ell$ plane. They describe the motion of the
origin of the dS C-metric ($A \neq 0$). For $A=0$ the intersecting
plane is $z^4=0$.
 (b) Schematic diagram representing the 5D
hyperbolic motion of two uniformly accelerating massive charged
black holes approaching asymptotically the Rindler-like
accelerated horizon ($h_A$). This horizon coincides with the
cosmological horizon. The inner and outer charged horizons are
represented by $h-$ and $h+$. The strut that connects the two
black holes is represented by the zigzag lines. The north pole
direction is represented by $\rm{N}$ and the south pole direction
by $\rm{S}$.
 }
\end{figure}

 So, the massless uncharged dS C-metric is a dS spacetime, but
we can extract more information from this 5D analysis. Indeed, let
us analyze with some detail the properties of the origin of the
radial coordinate, $\rho=0$ (or $r=0$). This origin moves in the
5D Minkowski embedding spacetime according to [see  (\ref{dS to
dS-c})]
\begin{eqnarray}
 & & z^2=0\;,\;\; z^3=0\;,\;\; z^4=\ell^2A /
 \sqrt{1+\ell^2A^2}\:<\ell
 \;\;\;\;\mathrm{and} \nonumber \\
 & & (z^1)^2-(z^0)^2=(A^2+1/\ell^2)^{-1}\equiv a_5^{-2} \:.
\label{rindler dS}
  \end{eqnarray}
These equations define two hyperbolic lines lying on the dS
hyperboloid which result from the intersection of this hyperboloid
surface defined by  (\ref{hyperboloid dS}) and the
$z^4$=constant$<\ell$ plane (see Fig. \ref{dS-hyperb}). They tell
us that the origin is subjected to a uniform 5D acceleration,
$a_5$, and consequently moves along a hyperbolic worldline in the
5D embedding space, describing a Rindler-like motion [see Figs.
\ref{dS-hyperb}.(a) and \ref{dS-hyperb}.(b)] that resembles the
well-known hyperbolic trajectory, $X^2-T^2=a^{-2}$, of an
accelerated observer in Minkowski space.
But uniformly accelerated radial worldlines in the 5D Minkowski
embedding space are also uniformly accelerated worldlines in the
4D dS space \cite{DesLev}, with the 5D acceleration $a_5$ being
related to the associated 4D acceleration $a_4$ by
$a_5^2=a_4^2+1/\ell^2$. Comparing this last relation with
(\ref{rindler dS}) we conclude that $a_4\equiv A$. Therefore, and
once again, we conclude that the origin of the dS C-metric is
uniformly accelerating with a 4D acceleration whose value is
precisely given by the parameter $A$ that appears in the dS
C-metric, (\ref{C-metric}), and this solution describes a dS space
whose origin is not at rest as usual but is being accelerated. For
the benefit of comparison with the $A=0$ dS spacetime, note that
the origin of the $A=0$ spacetime describes the hyperbolic lines
$(z^1)^2-(z^0)^2=\ell^2$ which result from the intersection of the
$z^4=0$ plane with the dS hyperboloid. In this case we can say
that we have two antipodal points on the spatial 3-sphere of the
dS space accelerating away from each other due only to the
cosmological background acceleration. When $A \neq 0$ these two
points suffer an extra acceleration. This discussion allowed us to
find the physical interpretation of parameter $A$ and to justify
its label. Notice also that the original dS C-metric coordinates
introduced in (\ref{C-metric}) cover only the half-space
$z^1>-z^0$. The Kruskal construction done in section \ref{sec:PD
dS} extended this solution to include also the $z^1<-z^0$ region
and so, in the extended solution, $r=0$ is associated to two
hyperbolas that represent two accelerated points (see Fig.
\ref{dS-hyperb}.(b)). These two hyperbolas approach asymptotically
the Rindler-like acceleration horizon ($r_A$).

\subsubsection[Pair of accelerated black holes ($m >0$, $q \neq 0$)]
{\label{sec:PI-mass dS}Pair of accelerated black holes
($\bm{m>0}$, $\bm{q \neq 0}$)}
Now, we are in a position to interpret the massive and charged
solutions that describe two black holes accelerating away from
each other. To see clearly this, let us look to the Carter-Penrose
diagrams, Fig. \ref{Fig-1 dS}, Fig. \ref{Fig-2 dS} and Fig.
\ref{Fig-3 dS}. Looking at these diagrams we can compare the
different features that belong to the massless uncharge case (Fig.
\ref{Fig-1 dS}), to the non-extreme massive uncharged case [Fig.
\ref{Fig-2 dS}.(i)], and ending in the non-extreme massive charged
case [Fig. \ref{Fig-3 dS}).(i)]. In Fig. \ref{Fig-1 dS} we
identify the two hyperbolas $r=0$ (represented by two timelike
lines) approaching asymptotically the Rindler-like acceleration
horizon ($r_A$). When we add a mass to the solution we conclude
that each of these two simple hyperbolas $r=0$ are replaced by the
more complex structure that represents a Schwarzschild black hole
with its spacelike curvature singularity and its horizon [this is
represented by $r_+$ in the left and right regions of Fig.
\ref{Fig-2 dS}.(i)]. So, the two accelerating points $r=0$ have
been replaced by two Schwarzschild black holes that approach
asymptotically the Rindler-like acceleration horizon [represented
by $r_A$ in the middle region of Fig. \ref{Fig-2 dS}.(i)]. The
same interpretation can be assigned to the massive charged
solution. The two hyperbolas $r=0$ of Fig. \ref{Fig-1 dS} are
replaced by two Reissner-Nordstr\"{o}m black holes [with its
timelike curvature singularity and its inner $r_-$ and outer $r_+$
horizons; see the left and right regions of Fig. \ref{Fig-3
dS}.(i)] that approach asymptotically the Rindler-like
acceleration horizon already present in the $m=0$ and $q=0$ causal
diagram. An issue that is relevant here, is whether the Cauchy
horizons of the charged dS C-metric are stable. The Cauchy horizon
of the non-accelerated dS$-$Reissner-Nordstr\"{o}m black holes
 is stable to small perturbations, as shown in \cite{stability} (for a review on
Cauchy horizon instabilities see, e.g., Burko and Ori
\cite{BurkoOri}). Moreover, in \cite{HorShein} it has been shown
that nearly extremal accelerating black holes in the flat
background have stable Cauchy horizons, unlike the Cauchy horizon
of the non-accelerated flat Reissner-Nordstr\"{o}m black hole
which is unstable. Therefore, we expect that the Cauchy horizons
of the accelerated dS$-$Reissner-Nordstr\"{o}m black holes are
stable, although we do not have confirmed this result. The
discussion of this subsection also applies directly to the extreme
cases of the dS C-metric.

\subsubsection{\label{sec:PI-strut dS} Source of acceleration
and radiative properties}
In this subsection we address the issue of the acceleration source
and its localization. In the massless uncharged dS C-metric,
observers that move along radial worldlines with $\rho=$const and
$\theta=0$ describe the Rindler-like hyperbola [see (\ref{dS to
dS-c})]
\begin{eqnarray}
(z^1)^2-(z^0)^2 &=& \frac{\ell^2-\rho^2}{(\sqrt{1+\ell^2A^2} +
A\rho)^2}  \:.
 \label{rindler.2 dS}
  \end{eqnarray}
Moreover, when we put $m$ or $q$ different from zero, each of the
two hyperbolas assigned to $r=0$ represents the accelerated motion
of a black hole. Thus, from (\ref{rindler.2 dS}) we conclude
\cite{OscLem_AdS-C} that the north pole axis is in the region
between the two black holes (see Fig. \ref{dS-hyperb}.(b)). Now,
the value of the arbitrary parameter $\kappa$ introduced in
(\ref{ang}) can be chosen in order to avoid a conical singularity
at the south pole ($\delta_\mathrm{s}=0$), leaving a conical
singularity at the north pole ($\delta_\mathrm{n}<0$). This is
associated to a strut that joins the two black holes along their
north poles and provides their acceleration \cite{OscLem_AdS-C}.
This strut satisfies the relation $p=-\mu>0$, where $p$ and $\mu$
are respectively its pressure and its mass density
\cite{OscLem_AdS-C}. Alternatively, we can choose $\kappa$ such
that avoids the deficit angle at the north pole
($\delta_\mathrm{n}=0$) and leaves a conical singularity at the
south pole ($\delta_\mathrm{s}>0$). This option leads to the
presence of a string (with $p=-\mu<0$) that connects the two black
holes along their south poles, and furnishes the acceleration.

\vspace{0.2 cm}

The C-metric is an exact solution that emits gravitational and
electromagnetic radiation. In the flat background the Bondi news
functions have been explicitly calculated in
\cite{AshtDray,Bic,PravPrav}. In dS background these calculations
have  been carried in \cite{BicKrt,KrtPod}.

Finally, recall (see subsection \ref{Ernst}) that in a dS
background we cannot remove the conical singularities through
Ernst's trick.

\section{\label{sec:Conc C-metric}Summary and concluding remarks}

The C-metric in an AdS, flat, and dS backgrounds share many common
properties, but there are also many features that differentiate
them. In the flat ($\Lambda=0$) and dS ($\Lambda>0$) backgrounds,
the corresponding C-metric always describes a pair of accelerated
black holes. However, the AdS C-metric only describes a pair of
accelerated black holes if the acceleration parameter $A$
satisfies $A> 1/ \ell$, where $\ell=\sqrt{3/|\Lambda|}$ is the
cosmological length. We can interpret this as due to the fact that
the AdS background is attractive, i.e., an analysis of the
geodesic equations indicates that particles in this background are
subjected to a potential well that attracts them (see Fig.
\ref{Fig geodesic_M=0 Q=0} and the associated discussion in the
text). Therefore, if we want to have a pair of black holes
accelerating apart, we will have to furnish a sufficient force
that overcomes this cosmological background attraction. We then
expect that a pair of accelerated black holes in an asymptotically
AdS spacetime is possible only if acceleration $A$ is higher than
a critical value ($1/\ell$). For $A\leq 1/\ell$ the analysis of
the solution clearly indicates the absence of a second black hole
and so the solution describes a single accelerated black hole.

Among the properties that are shared by the three C-metrics, we
first point that the two black holes described by the solutions
cannot interact gravitationally. This feature is best displayed in
the Carter-Penrose diagram of the solutions, with one clarifying
example sketched in Fig. \ref{Fig light flatC}.(a). In this figure
one sees that a null ray sent from the vicinity of a black hole
can never reach the horizon of the second black hole. So, if the
two black holes cannot communicate through a null ray they cannot
interact gravitationally. The black holes accelerate away from
each other only due to the pressure of the strut/strings without
opposition from gravitational attraction. In particular this also
explains why the C-metric does not reduce to the Israel-Khan when
the acceleration vanishes. In the Israel-Khan solution [see Fig.
\ref{Fig light flatC}.(b)], the null ray sent from one of the
black holes can reach the second black hole, and so they attract
each other gravitationally. In this solution, the two black holes
are connected by a strut that exerts an outward pressure which
exactly cancels the inward gravitational attraction, and so the
distance between the two black holes remains fixed.

Another common feature between the three cases concerns the
physical source of the acceleration. This source is associated
with a conical singularity in the angular poles of the solutions.
The conical singularity can be located in between the two black
holes and along the symmetry axis (or alternatively from the black
holes out to infinity). When it is between the two black holes, it
is associated to a strut satisfying the relation $p=-\mu>0$, where
$p$ and $\mu$ are respectively the pressure on the strut and its
mass density. The pressure is positive, so it points outwards into
infinity and pulls the black holes apart, furnishing their
acceleration. When the conical singularity points from each of the
black holes into infinity, it is associated to a string from each
one of the black holes towards infinity with negative pressure
that pushes the black holes into infinity (to be more accurate, in
the dS case, there is only one string that connects the two black
holes along their south poles, but this difference is due to the
fact that the dS space is closed).

When we set the mass and charge to zero, the AdS C-metric and the
dS C-metric have a clarifying geometric interpretation in the
4-hyperboloids that represent the AdS and dS solutions in a
5-dimensional (5D) Minkowski spacetime. Indeed, the origin of the
AdS C-metric (with $A> 1/ \ell$) and the origin of the dS
C-metric, are subjected to a uniform acceleration, and describe a
hyperbolic Rindler-like worldline in the AdS/dS 4-hyperboloid
embedded in the 5D Minkowski space. To be more precise, the origin
is represented by two hyperbolic lines that approach
asymptotically the Rindler-like accelerated horizon, so called
because it is is absent when $A=0$ and present even when $A\neq
0$, $m=0$ and $q=0$. When we add a mass or a charge to the system
the causal diagrams indicate (see the clarifying explanation
sketched in Fig. \ref{Fig CPpair_bh_introd}) that now we have two
AdS/dS-Schwarzschild or two AdS/dS$-$Reissner-Nordstr\"{o}m black
holes approaching asymptotically the Rindler-like accelerated
horizon. In the case of the AdS C-metric with $A\leq 1/\ell$ the
above procedure indicates the absence of a second black hole and
so the solution describes a single black hole. In the AdS
4-hyperboloid, the origin of these solutions describes a circle in
the plane defined by the two timelike coordinates.

The general features of the Carter-Penrose diagram of the dS
C-metric are independent of the angular coordinate. This sets a
great difference between the causal diagrams of the dS C-metric
and the ones of the AdS case and of the flat case. Indeed, for the
$\Lambda\leq 0$ C-metric, the Carter-Penrose diagram at the north
pole direction is substantially different from the one along the
south pole direction and different from the diagram along the
equator direction. These angular differences between the causal
diagrams between the AdS, flat and dS cases are identified by
comparing the Figs. \ref{Fig-a2 AdS}, \ref{Fig-2 flatC} and
\ref{Fig-2 dS} (in the neutral black hole case), and by comparing
Figs. \ref{Fig-a3 AdS}, \ref{Fig-3 flatC} and \ref{Fig-3 dS} (in
the charged black hole case).

In the AdS and flat backgrounds, when $A\rightarrow 0$, the
acceleration horizon disappears, and the C-metric reduces to a
single black hole with the usual features. In the dS case, when
$A\rightarrow 0$, the solution still reduces to the usual
dS-Schwarzschild or dS$-$Reissner-Nordstr\"{o}m black holes. These
$A=0$ solutions, however, can also be interpreted as a pair of
black holes accelerated by the positive cosmological constant. The
cosmological horizon of the $A=0$ dS black hole plays the role of
a acceleration horizon.  In the $A=0$ dS solution the cosmological
horizon has the topology of a round sphere, while in the dS
C-metric ($A\neq 0$) the presence of the acceleration  induces a
non-spherical shape in the acceleration  horizon which coincides
with the cosmological horizon.

Ernst has removed the conical singularities of the flat charged
C-metric through the introduction of an appropriate external
electromagnetic field. Its new exact solution then describes a
pair of black holes accelerated by the external Lorentz force. In
a cosmological background we cannot remove the conical
singularities through the application of the Harrison
transformation used by Ernst in the flat case. Indeed, the
Harrison transformation applied by Ernst does not leave invariant
the cosmological term in the action. Therefore, applying the
Harrison transformation to the cosmological C-metric solutions
does not yield a new solution of the Einstein-Maxwell theory in a
cosmological background.

%% file: Chapter6.tex
\thispagestyle{empty} \setcounter{minitocdepth}{1}
\chapter[The extremal limits of the
C-metric:\\ Nariai, Bertotti-Robinson and anti-Nariai
C-metrics]{\Large{The extremal limits of the
C-metric:\\
Nariai, Bertotti-Robinson and anti-Nariai C-metrics}}
\label{chap:Extremal Limits}
 \lhead[]{\fancyplain{}{\bfseries Chapter \thechapter. \leftmark}}
 \rhead[\fancyplain{}{\bfseries \rightmark}]{}
  \minitoc \thispagestyle{empty}
\renewcommand{\thepage}{\arabic{page}}

\addtocontents{lof}{\textbf{Chapter Figures \thechapter}\\}

Three important exact solutions of general relativity are de
Sitter (dS) spacetime which is a spacetime with positive
cosmological constant ($\Lambda>0$), Minkowski (or flat) spacetime
with $\Lambda=0$, and anti-de Sitter (AdS) spacetime with negative
cosmological constant ($\Lambda<0$). These stainless spacetimes,
with trivial topology $\mathbb{R}^4$, have nonetheless a rich
internal structure best displayed through the Carter-Penrose
diagrams \cite{hawkingellis}.  They also serve as the background
to spacetimes containing black holes which are then asymptotically
dS, flat, or AdS. These black holes in background spacetimes with
a cosmological constant - Schwarzschild, Reissner-Nordstr\"om,
Kerr, and Kerr-Newman -  have a complex causal and topological
structure well described in \cite{carter}.

There are other very interesting solutions of general relativity
with generic cosmological constant, that are neither pure nor
contain a black hole, and somehow can be considered intermediate
type solutions. These are the Nariai, Bertotti-Robinson, and
anti-Nariai solutions \cite{Kramer}.  The Nariai solution
\cite{Nariai,Nariai2} solves exactly the Einstein equations with
$\Lambda>0$, without or with a Maxwell field, and has been
discovered by Nariai in 1951 \cite{Nariai}. It is the direct
topological product of $dS_2 \times S^2$, i.e., of a
(1+1)-dimensional dS spacetime with a round 2-sphere of fixed
radius. The Bertotti-Robinson solution \cite{BertRob} is an exact
solution of the Einstein-Maxwell equations with any $\Lambda$, and
was found independently by Bertotti and Robinson in 1959. It is
the direct topological product of $AdS_2 \times S^2$, i.e., of a
(1+1)-dimensional AdS spacetime with a round 2-sphere of fixed
radius. The anti-Nariai solution, i.e., the AdS counterpart of the
Nariai solution, also exists \cite{CaldVanZer} and is an exact
solution of the Einstein equations with $\Lambda<0$, without or
with a Maxwell field. It is the direct topological product of
$AdS_2 \times H_2$, with $H_2$ being a 2-hyperboloid.

Three decades after Nariai's paper, Ginsparg and Perry
\cite{GinsPerry} connected the Nariai solution with the
Schwarzschild-dS solution. They showed that the Nariai solution
can be generated from a near-extreme dS black hole, through an
appropriate limiting procedure in which the black hole horizon
approaches the cosmological horizon. A similar extremal limit
generates the Bertotti-Robinson solution and the anti-Nariai
solution from an appropriate near extreme black hole (see, e.g.
\cite{CaldVanZer}). One of the aims of Ginsparg and Perry was to
study the quantum stability of the Nariai and the Schwarzschild-dS
solutions \cite{GinsPerry}.  It was shown that the Nariai solution
is in general unstable and, once created, decays through a quantum
tunnelling process into a slightly non-extreme black hole pair
(for a complete review and references on this subject see, e.g.,
Bousso \cite{Bousso60y}, and later discussions on our paper).  The
same kind of process happens for the Bertotti-Robinson and
anti-Nariai solutions.

There is yet another class of related metrics, the C-metric class,
which represent not one, but two black holes, being accelerate
apart from each other. These black holes can also inhabit a dS,
flat or AdS background, as we saw in the last chapter. It is
therefore of great interest to apply the Ginsparg-Perry procedure
to these metrics in order to find a new set of exact solutions
with a clear physical and geometrical interpretation. In this
chapter, following our work \cite{OscLem_nariai}, we address this
issue of the extremal limits of the C-metric with a generic
$\Lambda$ following \cite{GinsPerry}, in order to generate the
C-metric counterparts ($A \neq 0$) of the Nariai,
Bertotti-Robinson and anti-Nariai solutions ($A=0$), among others.

These C-metric counterparts are conformal to the product of two
2-dimensional manifolds of constant curvature, the conformal
factor depending on the angular coordinate. In addition, the
C-metric extremal solutions have a conical singularity at least at
one of the poles of their angular surfaces.  We give a physical
interpretation to these solutions, e.g., in the Nariai C-metric
(with topology $dS_2\times \tilde{S}^2$) to each point in the
deformed 2-sphere $\tilde{S}^2$ corresponds a $dS_2$ spacetime,
except for one point which corresponds a $dS_2$ spacetime with an
infinite straight strut or string. There are other important new
features that appear. One expects that the solutions found in this
chapter are unstable and decay into a slightly non-extreme black
hole pair accelerated by a strut or by strings. Moreover, the
Euclidean version of some of these solutions mediate the quantum
process of black hole pair creation, that accompanies the decay of
the dS and AdS spaces, as we shall see in chapter \ref{chap:Pair
creation}.

The plan of this chapter is as follows. In section
\ref{sec:Nariai,BR}, we describe the main features of the Nariai,
Bertotti-Robinson and anti-Nariai solutions. In section
\ref{sec:ExtLim dS}, we analyze the extremal limits of the dS
C-metric. We generate the Nariai C-metric, the Bertotti-Robinson
dS C-metric, and the Nariai Bertotti-Robinson dS C-metric. We
study the topology and the causal structure of these solutions,
and we give a physical interpretation. In section \ref{sec:ExtLim
L=0} we present the extremal limits of the flat C-metric and Ernst
solution, which are obtained from the solutions of section
\ref{sec:ExtLim dS} by taking the direct $\Lambda=0$ limit. The
Euclidean version of one of these $\Lambda=0$ solutions has
already been used previously, but we discuss two new solutions
that have not been discussed previously. In section
\ref{sec:ExtLim AdS}, we discuss the extremal limits of the AdS
C-metric and, in particular, we generate the anti-Nariai C-metric.
Finally, in section \ref{sec:Conc} concluding remarks are
presented.

\section{\label{sec:Nariai,BR}Extremal limits of black hole solutions in
\lowercase{d}S, flat, and A\lowercase{d}S spacetimes: The Nariai,
Bertotti-Robinson and anti-Nariai solutions}
In this section we will describe the main features of the Nariai,
Bertotti-Robinson and anti-Nariai solutions. The extremal limits
of the C-metric that will be generated in later sections reduce to
these solutions when the acceleration parameter $A$ is set to
zero.

\subsection{\label{sec:Nariai}The Nariai solution}

The neutral Nariai solution has been first introduced by Nariai
\cite{Nariai,Nariai2}. It satisfies the Einstein field equations
in a positive cosmological constant $\Lambda$ background,
$G_{\mu\nu}+\Lambda g_{\mu\nu}=0$, and is given by
\begin{equation}
d s^2  = \Lambda^{-1}(- \sin^2\chi\, d\tau^2 +d\chi^2 +d
\theta^2+\sin^2\theta\,d\phi^2)\,,\label{mNariai}
\end{equation}
where $\chi$ and $\theta$ both run from $0$ to $\pi$, and $\phi$
has period $2\pi$.

The electromagnetic extension of the Nariai solution has been
introduced by Bertotti and Robinson \cite{BertRob}. Its
gravitational field is given by
\begin{equation}
d s^2 =  \frac{{\cal R}_0^{\:2}}{ {\cal K}_0}\left (- \sin^2\chi\,
d\tau^2 +d\chi^2 \right ) +{\cal R}_0^{\:2} \left ( d
\theta^2+\sin^2\theta\,d\phi^2 \right )\, \label{qNariai}
\end{equation}
where ${\cal R}_0$ is a positive constant and constant ${\cal
K}_0$ satisfies $0<{\cal K}_0\leq 1$, while the electromagnetic
field of the Nariai solution is
\begin{eqnarray}
F=q \sin \theta \, d\theta \wedge d\phi\,
 \label{MpotNariai}
\end{eqnarray}
in the purely magnetic case, and
\begin{eqnarray}
 F=\frac{q}{{\cal K}_0}\,\sin \chi \,d\tau \wedge d\chi \,
 \label{EpotNariai}
\end{eqnarray}
in the purely electric case, with $q$ being the electric or
magnetic charge, respectively. The cosmological constant and the
charge of the Nariai solution are related to ${\cal R}_0$ and
 ${\cal K}_0$ by
\begin{eqnarray}
\Lambda=\frac{1+{\cal K}_0}{2{\cal R}_0^{\:2}} \,, \qquad {\rm
and} \qquad q^2=\frac{1-{\cal K}_0}{2}\,{\cal R}_0^{\:2}\,.
\label{relations}
\end{eqnarray}
Note that $0<{\cal K}_0\leq 1$, otherwise the charge is a complex
number. The charged Nariai solution satisfies the field equations
of the Einstein-Maxwell action in a positive cosmological constant
background, $G_{\mu\nu}+\Lambda g_{\mu\nu}=8\pi T_{\mu\nu}$, with
$T_{\mu\nu}$ being the energy-momentum tensor of the Maxwell
field. The neutral Nariai solution (\ref{mNariai}) is obtained
from the charged solution (\ref{qNariai})-(\ref{MpotNariai}) when
one sets ${\cal K}_0=1$. The $\Lambda=0$ limit of the Nariai
solution, which is Minkowski spacetime, is taken in  subsection
\ref{sec:lim L=0 Nariai}. Through a redefinition of coordinates,
$\sin^2\chi=1-\frac{{\cal K}_0}{{\cal R}_0^{\:2}}\: R^2$ and
$\tau=\sqrt{\frac{{\cal K}_0}{{\cal R}_0^{\:2}}} \, T$, the
spacetime (\ref{qNariai}) can be rewritten in new static
coordinates as
\begin{eqnarray}
d s^2 = - N(R)\, dT^2 +\frac{dR^2}{N(R)} +{\cal R}_0^{\:2}\left (d
\theta^2+\sin^2\theta\,d\phi^2 \right ), \label{qNariai2}
\end{eqnarray}
with
\begin{eqnarray}
N(R)=1-\frac{{\cal K}_0}{{\cal R}_0^{\:2}} \:R^2 \, ,
 \label{N}
\end{eqnarray}
and the electromagnetic field changes also accordingly to the
coordinate transformation. Written in these coordinates, we
clearly see that the Nariai solution is the direct topological
product of $dS_2 \times S^2$, i.e., of a (1+1)-dimensional dS
spacetime with a round 2-sphere of fixed radius ${\cal R}_0$. This
spacetime is homogeneous with the same causal structure as
(1+1)-dimensional  dS spacetime, but it is not an asymptotically
4-dimensional dS spacetime since the radius of the 2-spheres is
constant (${\cal R}_0$), contrarily to what happens in the dS
solution where this radius increases as one approaches infinity.

Another way \cite{Nariai2,Ortag} to see clearly the topological
structure of the Nariai solution is achieved by defining it
through its embedding in the flat manifold ${\mathbb{M}}^{1,5}$,
with  metric
\begin{eqnarray}
d s^2 = -dz_0^{\:\:2}+dz_1^{\:\:2}+dz_2^{\:\:2}+
dz_3^{\:\:2}+dz_4^{\:\:2}+dz_5^{\:\:2} \,. \label{M1+5}
\end{eqnarray}
The Nariai 4-submanifold is determined by the two constraints
\begin{eqnarray}
& & -z_0^{\:\:2}+z_1^{\:\:2}+z_2^{\:\:2}= \ell^2 \:,
             \nonumber \\
& & z_3^{\:\:2}+z_4^{\:\:2}+z_5^{\:\:2}= {\cal R}_0^{\:2} \:,
\label{hypersurface}
\end{eqnarray}
where $\ell^2=  {\cal R}_0^{\:2}/{\cal K}_0$. The first of these
constraints defines the $dS_2$ hyperboloid and the second defines
the 2-sphere of radius ${\cal R}_0$. The parametrization of
${\mathbb{M}}^{1,5}$ given by $z_0=\sqrt{\ell^2-R^2}\sinh \left (
T/\ell \right )$, $z_1=\sqrt{\ell^2-R^2}\,\cosh \left ( T/\ell
\right)$, $z_2=R$, $z_3= {\cal R}_0 \sin\theta \cos\phi$, $z_4=
{\cal R}_0 \sin\theta \sin\phi$ and $z_5= {\cal R}_0 \cos\theta$,
induces the metric (\ref{qNariai2}) on the Nariai hypersurface
(\ref{hypersurface}).

Quite remarkably, Ginsparg and Perry \cite{GinsPerry} (see also
\cite{BoussoHawk}) have shown that the neutral Nariai solution
(\ref{mNariai}) can be obtained from the near-extreme
Schwarzschild-dS black hole through an appropriate limiting
procedure. By extreme we mean that the black hole horizon and the
cosmological horizon coincide. Hawking and Ross \cite{HawkRoss},
and Mann and Ross \cite{MannRoss} have concluded that a similar
limiting approach takes the near-extreme
dS$-$Reissner-Nordstr\"{o}m black hole into the charged Nariai
solution (\ref{qNariai}). In this case, by extreme we mean that
the cosmological and outer charged black hole horizons coincide.
We will make heavy use of this Ginsparg-Perry procedure later, so
we will not sketch it here. These relations between the
near-extreme dS black holes and the Nariai solutions are a priori
quite unexpected since (i) the dS black holes have a curvature
singularity while the Nariai solutions do not, (ii) the Nariai
spacetime is homogeneous unlike the dS black holes spacetimes,
(iii) the dS black holes approach asymptotically the 4-dimensional
dS spacetime while the Nariai solutions do not. The Carter-Penrose
diagram of the Nariai solution is equivalent to the diagram of the
(1+1)-dimensional dS solution, as will be discussed in subsection
\ref{sec:Gen Nariai}.

An important role played by the Nariai solution in physics is at
the quantum level (see Bousso \cite{Bousso60y} for a detailed
review of what follows). First, there is the issue of the
stability of the solution when perturbed quantically. Ginsparg and
Perry \cite{GinsPerry}, in the neutral case, and Bousso and
Hawking \cite{BoussoHawk} and Bousso \cite{BoussoDil}, in the
charged case, have shown that the Nariai solutions are quantum
mechanically unstable. Indeed, due to quantum fluctuations, the
radius ${\cal R}_0$ of the 2-spheres oscillates along the
non-compact spatial coordinate $\chi$ and the degenerate horizon
splits back into a black hole and a cosmological horizon. Those
2-spheres whose radius fluctuates into a radius smaller than
${\cal R}_0$ will collapse into the dS black hole interior, while
the 2-spheres that have a radius greater than ${\cal R}_0$ will
suffer an exponential expansion that generates an asymptotic dS
region. Therefore, the Nariai solutions are unstable and, once
created, they decay through the quantum tunnelling process into a
slightly non-extreme  dS$-$Reissner-Nordstr\"{o}m black hole pair.
This issue of the Nariai instability against perturbations and
associated evaporation process has been further analyzed by Bousso
and Hawking \cite{BoussoHawk-T}, by Nojiri and Odintsov
\cite{NojOd1}, and by Kofman, Sahni, and Starobinski
\cite{ShaKof}. Second, the Nariai Euclidean solution plays a
further role as an instanton, in the quantum decay of the dS
space. This decay of the dS space is accompanied by the creation
of a dS black hole pair, in a process that is the gravitational
analogue of the Schwinger pair production of charged particles in
an external electromagnetic field. Here, the energy necessary to
materialize the black hole pair and to accelerate the black holes
apart comes from the cosmological constant background. It is
important to note that not all of the dS black holes can be pair
produced through this quantum process of black hole pair creation.
Only those black holes that have regular Euclidean sections can be
pair created (the term regular is applied here in the context of
the analysis of the Hawking temperature of the horizons). The
Nariai instanton (regular Euclidean Nariai solution, that is
obtained from  (\ref{mNariai}) and (\ref{qNariai}) by setting
$\tau=i\bar{\tau}$), belongs to the very restrictive class of
solutions that are regular
\cite{MelMos,Rom,MannRoss,BoussoHawk,BooMann,VolkovWipf}, and can
therefore mediate the pair creation process in the dS background.
In the uncharged case the Nariai instanton is even the only
solution that can describe the pair creation of neutral black
holes. Another result at the quantum level by Medved \cite{Medv}
indicates that quantum back-reaction effects prevent a near
extreme dS black hole from ever reaching a Nariai state of precise
extremality.

Other extensions of the Nariai solution are the dilaton charged
Nariai solution found by Bousso \cite{BoussoDil}, the rotating
Nariai solution studied by Mellor and Moss \cite{MelMos,Rom}, and
by Booth and Mann \cite{BooMann}, and solutions that describe
non-expanding impulsive waves propagating in the Nariai universe
studied by Ortaggio \cite{Ortag}.

\subsection{\label{sec:BR}The Bertotti-Robinson solution}

The simplest Bertotti-Robinson solution \cite{BertRob} (see also
\cite{CaldVanZer,Lap}) is a solution of the $\Lambda=0$
Einstein-Maxwell equations. Its gravitational field is given by
\begin{eqnarray}
& & d s^2  = q^2(- \sinh^2\chi\, d\tau^2 +d\chi^2+ d
\theta^2+\sin^2\theta\,d\phi^2)\,, \nonumber \\
& &
 \label{br}
\end{eqnarray}
where $q$ is the charge of the solution, and $\chi$ is unbounded,
$\theta$ runs from $0$ to $\pi$ and $\phi$ has period $2\pi$. The
electromagnetic field of the Bertotti-Robinson solution is
\begin{eqnarray}
F=q \sin \theta \, d\theta \wedge d\phi \,, \label{EpotBRmagnetic}
\end{eqnarray}
and
\begin{eqnarray}
 F=-q \,\sinh \chi \,d\tau \wedge d\chi \,,
  \label{EpotBRelectric}
\end{eqnarray}
in the magnetic and electric cases, respectively. Through a
redefinition of coordinates, $\sinh^2 \chi=R^2/q^2-1$ and
$\tau=T/q$, the spacetime (\ref{br}) can be rewritten in new
static coordinates as
\begin{eqnarray}
d s^2 = - N(R)\, dT^2 +\frac{dR^2}{N(R)}
 +q^2(d \theta^2+\sin^2\theta\,d\phi^2),
 \label{br2}
\end{eqnarray}
with
\begin{eqnarray}
N(R)=R^2/q^2-1 \, ,
 \label{N-br2}
\end{eqnarray}
and the electromagnetic field changes also accordingly to the
coordinate transformation. Written in these coordinates, we
clearly see that the Bertotti-Robinson solution is the direct
topological product of $AdS_2 \times S^2$, i.e., of a
(1+1)-dimensional AdS spacetime with a round 2-sphere of fixed
radius $q$.  Another way to see clearly the topological structure
of the Bertotti-Robinson solution is achieved by defining it
through its embedding in the flat manifold ${\mathbb{M}}^{2,4}$,
with metric
\begin{eqnarray}
 d s^2 = -dz_0^{\:\:2}+dz_1^{\:\:2}-dz_2^{\:\:2}+dz_3^{\:\:2}
 +dz_4^{\:\:2}+dz_5^{\:\:2} \,.
 \label{M1+5-br}
 \end{eqnarray}
The Bertotti-Robinson 4-submanifold is determined by the two
constraints
\begin{eqnarray}
& & -z_0^{\:\:2}+z_1^{\:\:2}-z_2^{\:\:2}= -q^2 \:,
             \nonumber \\
& & z_3^{\:\:2}+z_4^{\:\:2}+z_5^{\:\:2}=q^2 \,.
\label{hypersurface-br}
\end{eqnarray}
The first of these constraints defines the $AdS_2$ hyperboloid and
the second defines the 2-sphere of radius $q$. The parametrization
of ${\mathbb{M}}^{2,4}$,
 $z_0=\sqrt{R^2-q^2}\,\sinh \left ( T/q \right)$,
 $ z_1=\sqrt{R^2-q^2}\,\cosh \left ( T/q \right )$,
 $ z_2= R$,
 $ z_3= q \sin\theta \cos\phi$,
 $ z_4= q\sin\theta \sin\phi$ and
$ z_5= q \cos\theta$, induces the metric (\ref{br2}) on the
Bertotti-Robinson hypersurface (\ref{hypersurface-br}). Note that
since the parametrization
 $z_0=\sqrt{q^2+\tilde{R}^2}\,\sin(\tilde{T}/q)$,
 $z_1=\tilde{R}$ and
 $z_2=\sqrt{q^2+\tilde{R}^2}\,\cos(\tilde{T}/q)$,
also obeys (\ref{M1+5-br}) and the first constraint of
(\ref{hypersurface-br}), the Bertotti-Robinson solution
(\ref{br2}) can also be written as
 $d s^2 = - N(\tilde{R})\, d\tilde{T}^2
+d\tilde{R}^2 / N(\tilde{R})
 +q^2(d \theta^2+\sin^2\theta\,d\phi^2)\,,$ with
$N(\tilde{R})=\tilde{R}^2/q^2+1\,.$

Following a similar procedure to the one applied in the Nariai
solution, the Bertotti-Robinson solution can be obtained from the
near-extreme Reissner-Nordstr\"{o}m black hole through an
appropriate limiting procedure (here,  by extreme we mean that the
inner black hole horizon and the outer black hole horizon
coincide). Later on, we will make heavy use of the Ginsparg-Perry
procedure, so we will not sketch it here. Generalizations of the
Bertotti-Robinson solution to include a cosmological constant
background also exist \cite{BertRob}, and are an extremal limit of
the near-extreme Reissner-Nordstr\"{o}m black holes with a
cosmological constant. The Carter-Penrose diagram of the
Bertotti-Robinson solution (with or without $\Lambda$) is
equivalent to the diagram of the (1+1)-dimensional AdS solution,
as will be discussed in subsection \ref{sec:Gen BR dS}.

The Hawking effect in the Bertotti-Robinson universe has been
studied by Lapedes \cite{Lap}, and its thermodynamic properties
have been analyzed by Zaslavsky \cite{Zasl}, and by Mann and
Solodukhin \cite{MannSolod}. In \cite{Navarro} the authors have
shown that quantum back-reaction effects prevent a near extreme
charged black hole from ever reaching a Berttoti-Robinson state of
precise extremality. Recently, Ortaggio and Podolsk\'y
\cite{OrtagPod} have found exact solutions that describe
non-expanding impulsive waves propagating in the Bertotti-Robinson
universe.

\subsection{\label{sec:anti-Nariai}The anti-Nariai solution}
The anti-Nariai solution has a gravitational field given by
\cite{CaldVanZer}
\begin{eqnarray}
ds^2  \!\!&=&\!\! \frac{{\cal R}_0^{\:2}}{{\cal K}_0} \left(
-\sinh^2\chi\, d\tau^2\!  + \! d\chi^2 \right )\!  + \! {\cal
R}_0^{\:2}\!  \left( d\theta^2\! +\! \sinh^2\theta\,d\phi^2
\right), \nonumber \\
& & \label{qNariai-a}
\end{eqnarray}
where $\chi$ and $\theta$ are unbounded, $\phi$ has period $2\pi$,
${\cal R}_0$ is a positive constant, and the constant ${\cal K}_0$
satisfies $1 \leq {\cal K}_0 <2$. The electromagnetic field of the
anti-Nariai solution is
\begin{eqnarray}
F=q \sinh \theta \, d\theta \wedge d\phi\,
 \label{MpotNariai-a}
\end{eqnarray}
in the purely magnetic case, and
\begin{eqnarray}
F=-\frac{q}{{\cal K}_0}\,\sinh \chi \, d\tau \wedge d\chi \,
\label{EpotNariai-a}
\end{eqnarray}
in the purely electric case, with $q$ being the magnetic or
electric charge, respectively. The cosmological constant,
$\Lambda<0$, and the charge of the anti-Nariai solution are
related to ${\cal R}_0$ and ${\cal K}_0$ by
\begin{eqnarray}
\Lambda=-\frac{1+{\cal K}_0}{2{\cal R}_0^{\:2}}<0 \,,  \qquad {\rm
and} \qquad q^2=\frac{{\cal K}_0-1}{2}\,{\cal R}_0^{\:2}\,.
\label{relations-a}
\end{eqnarray}
The neutral anti-Nariai solution ($q=0$) is obtained from the
charged solution (\ref{qNariai-a}) when one sets ${\cal K}_0=1$
which implies ${\cal R}_0^{\:2}=|\Lambda|^{-1}$.  The $\Lambda=0$
limit of the anti-Nariai solution, which is Minkowski spacetime,
is taken in subsection \ref{sec:lim L=0 Nariai}. The charged
anti-Nariai solution satisfies the field equations of the
Einstein-Maxwell action in a negative cosmological constant
background. Through a redefinition of coordinates, $\sinh^2\chi=1-
\frac{{\cal K}_0}{{\cal R}_0^{\:2}}\: R^2 $ and $\tau=
\sqrt{\frac{{\cal K}_0}{{\cal R}_0^{\:2}}} \: T$, the spacetime
(\ref{qNariai-a}) can be rewritten in static coordinates as
\begin{eqnarray}
d s^2 = - N(R)\, dT^2 +\frac{dR^2}{N(R)}
 +{\cal R}_0^{\:2}\left (d \theta^2+\sinh^2\theta\,d\phi^2 \right ),
  \nonumber \\
 \label{qNariai2-a}
\end{eqnarray}
with
\begin{eqnarray}
N(R)=-1+\frac{{\cal K}_0}{{\cal R}_0^{\:2}} \:R^2 \,,
 \label{N-a}
\end{eqnarray}
and the electromagnetic field changes also accordingly to the
coordinate transformation. Written in these coordinates, we
clearly see that the anti-Nariai solution is the direct
topological product of $AdS_2 \times H_2$, i.e., of a
(1+1)-dimensional AdS spacetime with a 2-hyperboloid of radius
${\cal R}_0$. It is a homogeneous spacetime with the same causal
structure as (1+1)-dimensional AdS spacetime, but it is not an
asymptotically 4-dimensional AdS spacetime since the size of the
2-hyperboloid is constant (${\cal R}_0$), contrarily to what
happens in the AdS solution where this radius increases as one
approaches infinity. Another way to see clearly the topological
structure of the anti-Nariai solution is achieved by defining it
through its embedding in the flat manifold ${\mathbb{M}}^{3,3}$
with metric
\begin{eqnarray}
d s^2 = -dz_0^{\:\:2}+dz_1^{\:\:2}-dz_2^{\:\:2}+dz_3^{\:\:2}
+dz_4^{\:\:2}-dz_5^{\:\:2} \,. \label{M1+5-a}
\end{eqnarray}
The anti-Nariai 4-submanifold is determined by the two constraints
\begin{eqnarray}
& & -z_0^{\:\:2}+z_1^{\:\:2}-z_2^{\:\:2}= -\ell^2 \, , \nonumber \\
& & z_3^{\:\:2}+z_4^{\:\:2}-z_5^{\:\:2}= -{\cal R}_0^{\:2} \,.
\label{hypersurface-a}
\end{eqnarray}
where $\ell^2= {\cal R}_0^{\:2}/{\cal K}_0$. The first of these
constraints defines the $AdS_2$ hyperboloid and the second defines
the 2-hyperboloid of radius ${\cal R}_0$. The parametrization of
${\mathbb{M}}^{3,3}$, $z_0=\sqrt{R^2-\ell^2}\,\sinh \left ( T/\ell
\right)$, $z_1=\sqrt{R^2-\ell^2}\,\cosh \left ( T/\ell \right)$,
$z_2=R$, $z_3= {\cal R}_0 \sinh\theta \cos\theta$, $z_4= {\cal
R}_0 \sinh\theta \sin\theta$ and $z_5= {\cal R}_0 \cosh\theta$
induces the metric (\ref{qNariai2-a}) on the anti-Nariai
hypersurface (\ref{hypersurface-a}).

Having in mind the example of the Nariai solution, we may ask if
the anti-Nariai solution can be obtained, through a similar
limiting Ginsparg-Perry procedure, from a near-extreme AdS black
hole. As we saw in subsection \ref{sec:BH 4D AdS}, in the AdS
background, there are black holes whose horizons have topologies
different from spherical, such as toroidal horizons
\cite{Lemos,Zanchin_Lemos,Huang_L,OscarLemos_string}, and
hyperbolical horizons \cite{topological}, also called topological
black holes. The AdS black hole that generates the anti-Nariai
solution is the hyperbolic one [as is clear from the angular part
of (\ref{qNariai2-a})], and has a cosmological horizon. Indeed,
the hyperbolic topology together with the presence of the
cosmological horizon turn the hyperbolic black holes (studied in
subsection \ref{sec:BH 4D topological}) into the appropriate
solutions that allow the generation of the anti-Nariai solution
(\ref{qNariai-a}) with the limiting Ginsparg-Perry procedure, when
the black hole horizon approaches the cosmological horizon.

A further study of the anti-Nariai solution was done in
\cite{OrtagPod} where  non-expanding impulsive waves propagating
in the anti-Nariai universe are described.

\subsection[$\Lambda=0$ limit
of the Nariai and anti-Nariai solutions]{\bm{$\Lambda=0$} limit of
the Nariai and anti-Nariai solutions}\label{sec:lim L=0 Nariai}
In this subsection we find the $\Lambda=0$ limit of the Nariai
solution, (\ref{qNariai}),  and of the anti-Nariai solution,
(\ref{qNariai-a}). In this limit the line element of both
solutions goes apparently to infinity since ${\cal
R}_0^{\:2}\rightarrow \infty$. To achieve the suitable limit of
the Nariai solution, we first make the coordinate rescales:
$\tilde{x}= ({\cal R}_0/ \sqrt{{\cal K}_0})\chi$, and
$\varrho={\cal R}_0 \,\theta$. Then, taking the $\Lambda=0$ limit,
the solution becomes
\begin{eqnarray}
d s^2 = (-\tilde{x}^2 d\tau^2 +d\tilde{x}^2)+ (d\varrho^2+
\varrho^2\,d\phi^2) \, .
 \label{Lim L=0 Nariai1}
\end{eqnarray}
The spacetime sector is a Rindler factor, and the angular sector
is a cylinder. Therefore, under the usual coordinate
transformation $\tilde{x}=\sqrt{\bar{x}^2-\bar{t}^2}$ and
$\tau={\rm arctanh(\bar{t}/\bar{x})}$, and unwinding the cylinder
($\bar{y}=\varrho\,\cos\phi $ and $\bar{z}=\varrho\,\sin\phi$), we
have
\begin{eqnarray}
d s^2 = (-d\bar{t}^{\,2} +d\bar{x}^2)+ (d\bar{y}^2+ d\bar{z}^2)\,
.
 \label{Lim L=0 Nariai2}
\end{eqnarray}
Therefore, while the Nariai solution is topologically $dS_2\times
S^2$, its $\Lambda=0$ limit  is topologically
${\mathbb{M}}^{1,1}\times {\mathbb{R}}^2$, i.e., ${\mathbb{R}}^4$.

A similar procedure shows that taking the $\Lambda=0$ limit of the
anti-Nariai solution (\ref{qNariai-a}) ($AdS_2\times H_2$) leads
to  (\ref{Lim L=0 Nariai2}).

\section{\label{sec:ExtLim dS}Extremal limits of the \lowercase{d}S C-metric}

In the last section we saw that there is an appropriate extremal
limiting procedure, introduced by Ginsparg and Perry
\cite{GinsPerry}, that generates from the near-extreme black hole
solutions the Nariai, Bertotti-Robinson and anti-Nariai solutions.
Analogously, we shall apply the procedure of \cite{GinsPerry} to
generate new exact solutions from the near-extreme cases of the dS
C-metric. Specifically the C-metric counterparts of the Nariai and
Bertotti-Robinson solutions are found using this method. When the
acceleration parameter $A$ is set to zero in these new solutions,
we will recover the Nariai and Bertotti-Robinson solutions. To
achieve our propose we will analyze with more detail the extreme
cases of the dS C-metric already discussed in subsections
\ref{sec:PD A.2 dS} and \ref{sec:PD A.3 dS}. Recall that the
gravitational field of the dS C-metric is given by
(\ref{C-metric}) and (\ref{FG}), and its electromagnetic field is
given by (\ref{F-mag}) and (\ref{F-el-Lorentz}).

\subsection{\label{sec:Gen Nariai}The Nariai C-metric}
We will generate the Nariai C-metric from a special extremal limit
of the dS C-metric. First we will describe this particular
near-extreme solution and then we will generate the Nariai
C-metric.

We are interested in a particular extreme dS C-metric, for which
the size of the acceleration horizon $y_A$ is equal to the size of
the outer charged horizon $y_+$. Let us label this degenerated
horizon by $\rho$, i.e.,  $y_A=y_+\equiv \rho$ with $\rho < y_-$.
 In this case, the function ${\cal F}(y)$ given by (\ref{FG}) can be written as
\begin{eqnarray}
{\cal F}(y) =\frac{\rho^2-3\gamma}{\rho^4}
 (y-y_{\rm neg})(y-y_-)(y-\rho)^2\:,
 \label{Fextreme}
 \end{eqnarray}
where
\begin{eqnarray}
\gamma =\frac{\Lambda+3A^2}{3A^2}\:,
 \label{gamma}
 \end{eqnarray}
and the roots $\rho$, $y_{\rm neg}$ and $y_-$ are given by
\begin{eqnarray}
& & \rho =\frac{3m}{4q^2A}
 \left ( 1- \sqrt{1-\frac{8}{9}\frac{q^2}{m^2}} \:\right )
 \:,  \label{zerosy1} \\
& & y_{\rm neg} =\frac{\gamma \rho}{\rho^2-3\gamma}
 \left ( 1- \sqrt{\frac{\rho^2-2\gamma}{\gamma}} \:\right )
 \:, \label{zerosy2} \\
& & y_- =\frac{\gamma \rho}{\rho^2-3\gamma}
 \left ( 1+ \sqrt{\frac{\rho^2-2\gamma}{\gamma}} \:\right )
 \:.
 \label{zerosy3}
 \end{eqnarray}
The mass and the charge of the solution are written as functions
of $\rho$ as
\begin{eqnarray}
& &m =\frac{1}{A\rho}
 \left ( 1- \frac{2\gamma}{\rho^2} \right )
 \:, \nonumber \\
& & q^2 =\frac{1}{A^2\rho^2}
 \left ( 1- \frac{3\gamma}{\rho^2} \right )
 \:.
 \label{mq}
 \end{eqnarray}
The conditions $\rho < y_-$ and $q^2 \geq 0$ require that the
allowed range of $\rho$ is
\begin{eqnarray}
 \sqrt{3\gamma} \leq \rho<\sqrt{6\gamma} \:.
 \label{range-gamma}
\end{eqnarray}
The value of $y_-$ decreases monotonically with $\rho$ and we have
$\sqrt{6\gamma}<y_-<+\infty$. The mass and the charge of the
Nariai solution, denoted now as $m_{\rm N}$ and $q_{\rm N}$
respectively, are monotonically increasing functions of $\rho$,
and as we go from $\rho=\sqrt{3\gamma}$ into $\rho=\sqrt{6\gamma}$
we have
\begin{eqnarray}
& &  \frac{1}{3}
  \frac{1}{\sqrt{\Lambda+3A^2}}\leq m_{\rm N}< \frac{\sqrt{2}}{3}
  \frac{1}{\sqrt{\Lambda+3A^2}}\:, \nonumber \\
& & 0 \leq q_{\rm N}< \frac{1}{2}
  \frac{1}{\sqrt{\Lambda+3A^2}}\:,
 \label{mq-cNariai}
 \end{eqnarray}
so the $A\neq 0$ extreme ($y_A=y_+$) solution has a lower maximum
mass and charge, and has a lower minimum mass than the
corresponding $A=0$ solution \cite{MelMos,Rom,MannRoss,BooMann},
and, for a fixed $\Lambda$, as the acceleration parameter $A$
grows these extreme values decrease monotonically. For a fixed
$\Lambda$ and for a fixed mass between $\sqrt{1/(9 \Lambda)}\leq
m<\sqrt{2/(9 \Lambda)}$, the allowed acceleration varies as
$\sqrt{1/(27m^2)-\Lambda/3} \leq A <\sqrt{2/(27m^2)-\Lambda/3}$.

We are now ready to obtain the Nariai C-metric.
 In order to generate it from the above
near-extreme dS C-metric we first set
\begin{eqnarray}
 y_A=\rho-\varepsilon, \:\:\:\:y_+=\rho+\varepsilon,
 \:\:\:\:\:\:{\rm with} \:\: \varepsilon<<1\:,
 \label{NariaiLimit}
\end{eqnarray}
in order that $\varepsilon$ measures the deviation from
degeneracy, and the limit $y_A\rightarrow y_+$ is obtained when
$\varepsilon \rightarrow 0$. Now, we introduce a new time
coordinate $\tau$ and a new radial coordinate $\chi$,
 \begin{eqnarray}
t= \frac{1}{\varepsilon {\cal K}}\,\tau \:, \:\:\:\:\:\:\:\:\:\:\:
y=\rho+\varepsilon \cos\chi \:,
 \label{NariaiCoord}
\end{eqnarray}
where
\begin{eqnarray}
{\cal K} = -\frac{\rho^2-3\gamma}{\rho^4} (\rho-y_{\rm
neg})(\rho-y_-) =\frac{2(\Lambda+3A^2)}{A^2\rho^2}-1\:, \nonumber
\\
& &
 \label{Kfactor}
\end{eqnarray}
and condition (\ref{range-gamma}) implies $0<{\cal K}\leq 1$ with
$q=0\Rightarrow {\cal K}=1$. In the limit $\varepsilon \rightarrow
0$, from (\ref{C-metric}) and (\ref{Fextreme}), we get the
gravitational field of the Nariai C-metric
\begin{eqnarray}
d s^2 &=& \frac{{\cal R}^2(x)}{{\cal K}} \left (-\sin^2\chi\,
d\tau^2 +d\chi^2\right )+ {\cal R}^2(x)\left [{\cal
G}^{-1}(x)dx^2+ {\cal G}(x)dz^2 \right ]
 \:,
 \label{Nariai-C-Metric}
\end{eqnarray}
where $\chi$ runs from $0$ to $\pi$ and
\begin{eqnarray}
& {\cal R}^2(x)=\left (Ax+\sqrt{\frac{2(\Lambda+3A^2)}{1+{\cal
K}}} \right )^{-2}\:,&
\nonumber \\
\!\!\!\!\!\!\!\!\!\!\!\!\!\!\!\!\!\!& \!\!\!\!\!\!\!\!\!
 {\cal G}(x) =1-x^2
 -\frac{2A}{3}\sqrt
{\frac{(1+{\cal K})(2-{\cal K})^2}{2(\Lambda+3A^2)}}\,x^3 -
\frac{A^2}{4}\frac{1-{\cal K}^2}{\Lambda+3A^2} \,x^4\,.&
 \nonumber \\
 & &
\label{Gfactor}
\end{eqnarray}
${\cal G}(x)$ has only two real roots, the south pole
$x_\mathrm{s}$ and the north pole $x_\mathrm{n}$. The angular
coordinate $x$ can range between these two poles, whose values are
calculated in section \ref{sec:angular}. Under the coordinate
transformation (\ref{NariaiCoord}), the Maxwell field for the
magnetic case is still given by  (\ref{F-mag}), while in the
electric case, (\ref{F-el-Lorentz}) becomes
 \begin{eqnarray}
 F=\frac{q}{{\cal K}}\,\sin \chi \,d\tau\wedge d\chi\:.
\label{F-el-Nariai}
\end{eqnarray}
 So, if we give the parameters $\Lambda$, $A$, and $q$ we can
construct the Nariai C-metric.
 The Nariai C-metric is
conformal to the topological product of two 2-dimensional
manifolds, $dS_2\times \tilde{S}^2$, with the conformal factor
${\cal R}^2(x)$ depending on the angular coordinate $x$, and
$\tilde{S}^2$ being a deformed 2-sphere.

 In order to obtain the $A=0$ limit, we first set
$\tilde{\rho}=A\rho$ [see  (\ref{zerosy1})], a parameter that has
a finite and well-defined value when $A\rightarrow 0$. Then when
$A\rightarrow 0$ we have ${\cal K} \rightarrow {\cal
K}_0=2\Lambda/\tilde{\rho}^2-1$ and ${\cal R}^2(x) \rightarrow
{\cal R}_0^{\:2}=\tilde{\rho}^{-2}$, with ${\cal R}_0^{\:2}$ and
${\cal K}_0$ satisfying relations (\ref{relations}). This,
together with transformations (\ref{ang}), show that the Nariai
C-metric transforms into the Nariai solution (\ref{qNariai}) when
$A=0$.

 The limiting procedure that has been applied in this
subsection has generated a new exact solution that satisfies the
Maxwell-Einstein equations in a positive cosmological constant
background.

In order to give a physical interpretation to this extremal limit
of the dS C-metric, we first recall the physical interpretation of
the dS C-metric. This solution describes a pair of uniformly
accelerated black holes in a dS background, with the acceleration
being provided by the cosmological constant and by a strut between
the black holes that pushes them away or, alternatively, by a
string that connects and pulls the black holes in. The presence of
the strut or of the string is associated to the conical
singularities that exist in the C-metric (see, e. g.,
\cite{OscLem_dS-C}). Indeed, in general, if we draw a small circle
around the north or south pole, as the radius goes to zero, the
limit circunference/radius is not $2\pi$. There is a deficit angle
at the north pole given by (see \cite{OscLem_AdS-C,OscLem_dS-C}),
$\delta_\mathrm{{n}} =
 2\pi\left (1- \frac{\kappa}{2} |{\cal G}'(x_\mathrm{n})|\right )$
(where the prime means derivative with respect to $x$) and,
analogously, a similar conical singularity ($\delta_\mathrm{{s}}$)
is present at the south pole. The so far arbitrary parameter
$\kappa$ introduced in  (\ref{ang}) plays its important role here.
Indeed, if we choose $\kappa^{-1}=\frac{1}{2}|{\cal
G}'(x_\mathrm{s})|$ we remove the conical singularity at the south
pole ($\delta_\mathrm{s}=0$). However, since we only have a single
constant $\kappa$ at our disposal and this has been fixed to avoid
the conical singularity at the south pole ($\delta_\mathrm{s}=0$),
we conclude that a conical singularity will be present at the
north pole with $\delta_\mathrm{n}<0$. This is associated to a
strut (since $\delta_\mathrm{n}<0$) that joins the two black holes
along their north poles and provides their acceleration. This
strut satisfies the relation $p=-\mu>0$, where $p$ and
$\mu=\delta_\mathrm{n}/(8\pi)$ are respectively its pressure and
its mass density (see \cite{OscLem_AdS-C}). There is another
alternative.  We can choose instead $\kappa^{-1}=\frac{1}{2}|{\cal
G}'(x_\mathrm{n})|$ and by doing so we avoid the deficit angle at
the north pole ($\delta_\mathrm{n}=0$), and leave a conical
singularity at the south pole ($\delta_\mathrm{s}>0$). This option
leads to the presence of a string (with $p=-\mu<0$) connecting the
black holes along their south poles that furnishes the
acceleration. Summarizing, when the conical singularity is at the
north pole, the pressure of the strut is positive, so it points
outwards and pushes the black holes apart, furnishing their
acceleration. When the conical singularity is at the south pole,
it is associated to a string between the two black holes with
negative pressure that pulls the black holes away from each other.

\begin{figure} [H]
\centering
\includegraphics[height=2.7in]{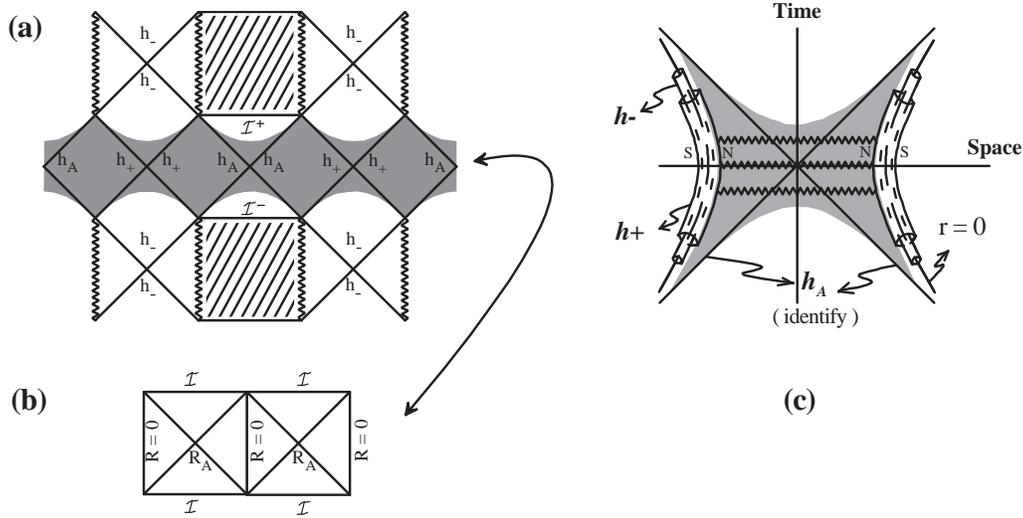}
\caption{\label{nariai-fig}
 (a) The whole figure is the Carter-Penrose
diagram of the dS C-metric.
 The shaded region represents the near-horizon area that, when the black hole
 horizon ($h_+$) approaches the acceleration horizon ($h_+$), gives the $dS_2$
 manifold of the Nariai C-metric solution. See discussion in the
 text.
 (b) Carter-Penrose diagram of the Nariai C-metric.
  (c) The whole figure represents schematically the two black holes of
 the dS C-metric approaching asymptotically the Rindler-like acceleration horizon.
 They are accelerated by a strut that joins them along their
 north poles.
 The shaded region represents the near-horizon area that, when the black hole
 horizon approaches the acceleration horizon, gives the Nariai C-metric
 solution [compare with Fig. \ref{nariai-fig}.(a)]. The strut survives
 to the limiting process.
 }
\end{figure}

The causal structure of the dS C-metric has been analyzed in
detail in \cite{OscLem_dS-C}. The Carter-Penrose diagram of the
non-extreme charged dS C-metric is sketched in Fig.
\ref{nariai-fig}.(a) (whole figure) and has a structure that,
loosely speaking, can be divided into left, middle and right
regions. The middle region contains the null infinity, the past
infinity, ${\cal I^-}$, and the future infinity, ${\cal I^+}$, and
an accelerated Rindler-like horizon, $h_A$ (that coincides with
the cosmological horizon). The left and right regions both contain
a timelike curvature singularity (the zig-zag line), and an inner
($h_-$) and an outer ($h_+$) horizons associated to the charged
character of the solution. This diagram represents two
dS$-$Reissner-Nordstr\"{o}m black holes that approach
asymptotically the Rindler-like acceleration horizon (for a more
detailed discussion see \cite{OscLem_dS-C}). This is also
schematically represented in Fig. \ref{nariai-fig}.(c) (whole
figure), where we explicitly show the strut that connects the two
black holes and provides their acceleration.

Now, as we have just seen, the Nariai C-metric can be
appropriately obtained from the vicinity of the black hole and
acceleration horizons in the limit in which the size of these two
horizons approach each other.  We now will see that the conical
singularity of the dS C-metric survives  the near-extremal
limiting procedure that generates the Nariai C-metric. Following
an elucidative illustration shown in \cite{MaldStrom} for the
Bertotti-Robinson solution (with $\Lambda<0$ and $A=0$), this
near-horizon region is sketched in Fig. \ref{nariai-fig}.(a) as a
shaded area, and from it we can identify some of the features of
the Nariai C-metric, e.g., the curvature singularity of the
original dS black hole is lost in the near-extremal limiting
procedure. But, more important, this shaded near-horizon region
also allows us to construct straightforwardly the Carter-Penrose
diagram of the Nariai C-metric drawn in Fig. \ref{nariai-fig}.(b).
The construction steps are as follows. First, as indicated by
(\ref{NariaiLimit}) and the second relation of
(\ref{NariaiCoord}), we let the cross lines that represent the
black hole horizon [$h_+$ in Fig. \ref{nariai-fig}.(a)] join
together with the cross lines that represent the acceleration
horizon [$h_A$ in Fig. \ref{nariai-fig}.(a)], and so on, i.e., we
do this joining ad infinitum with all the cross lines $h_+$ and
$h_A$. After this step all that is left from the original diagram
is a single cross line, i.e, all the spacetime that has originally
contained in the shaded area of Fig. \ref{nariai-fig}.(a) has
collapsed into two mutually perpendicular lines at $45^{\circ}$ at
$y=\rho$. Now, as indicated by the first relation of
(\ref{NariaiCoord}), when $\varepsilon \rightarrow 0$ the time
suffers an infinite blow up. To this blow up in the time
corresponds an infinite expansion in the Carter-Penrose diagram in
the vicinity of $y=\rho$. We then get again the shaded area of
Fig. \ref{nariai-fig}.(a), but now the cross lines of this shaded
area are all identified into a single horizon, and they no longer
have the original signature associated to $h_+$ and $h_A$ that
differentiated them. This is, in the shaded area of Fig.
\ref{nariai-fig}.(a) we must now erase the original labels $h_+$
and $h_A$. The Carter-Penrose diagram of the Nariai C-metric is
then given by Fig. \ref{nariai-fig}.(b), which is equivalent to
the diagram of the (1+1)-dimensional dS solution. Note that the
diagram of the $A=0$ dS$-$Reissner-Nordstr\"{o}m solution is
identical to the one of Fig. \ref{nariai-fig}.(a), as long as we
replace $h_A$ (acceleration horizon) by $h_c$ (cosmological
horizon). Applying the same construction process described just
above we find that the Carter-Penrose diagram of the Nariai
solution ($A=0$), described in subsection \ref{sec:Nariai}, is
also given by Fig. \ref{nariai-fig}.(b).

The Nariai near-horizon region is also sketched in Fig.
\ref{nariai-fig}.(c) as a shaded area. This schematic figure is
clarifying in the sense that it indicates that the strut that
connects the two original black holes along their north pole
directions survives to the near-extremal limiting process and will
be present in the final result of the process, i.e., in the Nariai
C-metric. Indeed, note that the endpoints of the strut are at the
north pole of the event horizons of the two black holes and
crosses the acceleration horizon. As we saw just above, this
region suffers first a collapse followed by an infinite blow up
and during the process the strut is not lost. Thus the Nariai
C-metric (\ref{Nariai-C-Metric})-(\ref{Gfactor}) describes a
spacetime that is conformal to the product $dS_2\times
\tilde{S}^2$. To each point in the deformed 2-sphere corresponds a
$dS_2$ spacetime, except for one point which corresponds a $dS_2$
spacetime with an infinite straight strut. This strut has negative
mass density given by
\begin{equation}
 \mu =\frac{1}{4}{\biggl (}1- {\biggl |}\frac{{\cal G}'(x_\mathrm{{n}})}
 {{\cal G}'(x_\mathrm{{s}})}
 {\biggl |}{\biggr )}\:,
 \label{mass-density}
 \end{equation}
where ${\cal G}'(x)$ is the derivative of  (\ref{Gfactor}), and
with a positive pressure $p=-\mu>0$. Alternatively, if we remove
the conical singularity at the north pole, the Nariai C-metric
describes a string with positive mass density
 $\mu =(1/4) \left ( 1- |{\cal G}'(x_\mathrm{{s}})/
 {\cal G}'(x_\mathrm{{n}})| \right )$
and negative pressure $p=-\mu<0$.

As we have said in subsection \ref{sec:Nariai}, the Nariai
solution ($A=0$) is unstable and, once created, it decays through
the quantum tunnelling process into a slightly non-extreme black
hole pair. We then expect that the Nariai C-metric is also
unstable and that it will decay into a slightly non-extreme pair
of black holes accelerated by a strut or by a string. The Nariai
C-metric instanton also plays an important role in the decay of
the dS space, since it can mediate the Schwinger-like quantum
process of pair creation of black holes in a dS background
\cite{OscLem-PCdS}. Indeed, as we said in subsection
\ref{sec:Nariai}, the Nariai instanton ($A=0$) has been used
\cite{MelMos,Rom,MannRoss,BooMann,VolkovWipf} to study the pair
creation of dS black holes materialized and accelerated by the
cosmological background field. Moreover, the Euclidean ``Nariai"
flat C-metric  and Ernst solution (see section \ref{sec:dS
C-metric}) have been used to analyze the process of pair
production of $\Lambda=0$ black holes, accelerated by a string or
by an electromagnetic external field, respectively. Therefore, it
is natural to expect that the Euclidean Nariai limit of the dS
C-metric mediates the process of pair creation of black holes in a
cosmological background, that are then accelerated by a string, in
addition to the cosmological background field. The picture would
be that of the nucleation, in a dS background, of a Nariai C
universe, whose string then breaks down and a pair of dS black
holes is created at the endpoints of the string. This expectation
is confirmed in \cite{OscLem-PCdS}.

\subsection{\label{sec:Gen BR dS}The Bertotti-Robinson dS C-metric}

Now, we are interested in another particular extreme dS C-metric
(usually called cold solution when $A=0$
\cite{MelMos,Rom,MannRoss,BooMann}) for which the size of the
outer charged black hole horizon $y_+$ is equal to the size of the
inner charged horizon $y_-$. Let us label this degenerated horizon
by $\rho$, such that, $y_+=y_-\equiv \rho$ and $\rho > y_A$. This
solution requires the presence of the electromagnetic charge. In
this case, the function ${\cal F}(y)$ given by (\ref{FG}) can be
written as
\begin{eqnarray}
{\cal F}(y) =\frac{\rho^2-3\gamma}{\rho^4}
 (y-y_{\rm neg})(y-y_A)(y-\rho)^2\:,
 \label{Fextreme-br}
 \end{eqnarray}
with $\gamma$ given by  (\ref{gamma}), the roots $\rho$ and
$y_{\rm neg}$ are defined by  (\ref{zerosy1}) and (\ref{zerosy2}),
respectively, and $y_A$ is given by
\begin{eqnarray}
 y_A =\frac{\gamma \rho}{\rho^2-3\gamma}
 \left ( 1+ \sqrt{\frac{\rho^2-2\gamma}{\gamma}} \:\right )
 \:.
 \label{zerosy3-br}
 \end{eqnarray}
Eq. (\ref{mq}) defines the mass and the charge of the solution as
a function of $\rho$, and, for a fixed $A$ and $\Lambda$, the
ratio $q/m$ is higher than $1$. The conditions $\rho> y_A$ and
$q^2 > 0$ require that the allowed range of $\rho$ is
\begin{eqnarray}
\rho>\sqrt{6\gamma} \:.
 \label{range-gamma-br}
\end{eqnarray}
The value of $y_A$ decreases monotonically with $\rho$ and we have
$\sqrt{\gamma}<y_A<\sqrt{6\gamma}$. Contrary to the Nariai case,
the mass and the charge of the Bertotti-Robinson solution, $m_{\rm
BR}$ and $q_{\rm BR}$, respectively, are monotonically decreasing
functions of $\rho$, and as we come from $\rho=+\infty$ into
$\rho=\sqrt{6\gamma}$ we have
\begin{eqnarray}
& &  0< m_{\rm BR}< \frac{\sqrt{2}}{3}
  \frac{1}{\sqrt{\Lambda+3A^2}}\:, \nonumber \\
& & 0< q_{\rm BR}< \frac{1}{2}
  \frac{1}{\sqrt{\Lambda+3A^2}}\:,
 \label{mq-cold}
 \end{eqnarray}
so the $A\neq 0$ extreme ($y_+=y_-$) solution has a lower maximum
mass and charge than the corresponding $A=0$ solution
\cite{MelMos,Rom,MannRoss,BooMann} and, for a fixed $\Lambda$, as
the acceleration parameter $A$ grows this maximum value decreases
monotonically. For a fixed $\Lambda$ and for a fixed mass below
$\sqrt{2/(9 \Lambda)}$, the maximum value of the acceleration is
$\sqrt{2/(27m^2)-\Lambda/3}$.

We are now ready to generate the Bertotti-Robinson dS C-metric
from the above cold dS C-metric. We first set
\begin{eqnarray}
 y_+=\rho-\varepsilon, \:\:\:\:y_-=\rho+\varepsilon,
 \:\:\:\:\:\:{\rm with} \:\: \varepsilon<<1\:,
 \label{NariaiLimit-br}
\end{eqnarray}
in order that $\varepsilon$ measures the deviation from
degeneracy, and the limit $y_+\rightarrow y_-$ is obtained when
$\varepsilon \rightarrow 0$. Now, we introduce a new time
coordinate $\tau$ and a new radial coordinate $\chi$,
 \begin{eqnarray}
t= \frac{1}{\varepsilon {\cal K}}\,\tau \:, \:\:\:\:\:\:\:\:\:\:\:
y=\rho+\varepsilon \cosh\chi \:,
 \label{NariaiCoord-br}
\end{eqnarray}
where
\begin{eqnarray}
{\cal K} =\frac{\rho^2-3\gamma}{\rho^4}
 (\rho-y_{\rm neg})(\rho-y_A)
= 1-\frac{2(\Lambda+3A^2)}{A^2\rho^2}\:, \nonumber \\
& &
 \label{Kfactor-br}
\end{eqnarray}
and condition (\ref{range-gamma-br}) implies $0<{\cal K}<1$. In
the limit $\varepsilon \rightarrow 0$, from (\ref{C-metric}) and
(\ref{Fextreme-br}), the metric becomes
\begin{eqnarray}
d s^2 = \frac{{\cal R}^2(x)}{{\cal K}} \left (-\sinh^2\chi\,
d\tau^2 +d\chi^2\right ) +
 {\cal R}^2(x)\left [{\cal G}^{-1}(x)dx^2+ {\cal G}(x)dz^2 \right ]
 \:,
 \label{br-ds-C-Metric}
\end{eqnarray}
where $\chi$ is unbounded and
\begin{eqnarray}
 &{\cal R}^2(x)=
\left(Ax+\sqrt{\frac{2(\Lambda+3A^2)}{1-{\cal K}}}\right )^{-2}
\:, & \nonumber \\
 & {\cal G}(x) = 1-x^2
 -\frac{2A}{3}
\sqrt{\frac{(1-{\cal K})(2+{\cal K})^2}{2(\Lambda+3A^2)}}\,x^3
  -\frac{A^2}{4}\frac{1-{\cal K}^2}{\Lambda+3A^2} \,x^4\:.& \nonumber \\
& &
 \label{Gfactor-br}
\end{eqnarray}
${\cal G}(x)$ has only two real roots, the south pole
$x_\mathrm{s}$ and the north pole $x_\mathrm{n}$. The angular
coordinate $x$ can range between these two poles whose value is
calculated in section \ref{sec:angular}. Under the coordinate
transformation (\ref{NariaiCoord-br}), the Maxwell field for the
magnetic case is still given by  (\ref{F-mag}), while in the
electric case, (\ref{F-el-Lorentz}) becomes
 \begin{eqnarray}
 F=-\frac{q}{{\cal K}}\,\sinh \chi \,d\tau\wedge d\chi\:.
 \label{F-el-br}
\end{eqnarray}
So, if we give the parameters $\Lambda$, $A$, and $q$ we can
construct the Bertotti-Robinson dS C-metric.  This solution is
conformal to the topological product of two 2-dimensional
manifolds, $AdS_2\times \tilde{S}^2$, with the conformal factor
${\cal R}^2(x)$ depending on the angular coordinate $x$.

In order to obtain the $A=0$ limit, we first set
$\tilde{\rho}=A\rho$ [see  (\ref{zerosy1})], a parameter that has
a finite and well-defined value when $A\rightarrow 0$. Then when
$A\rightarrow 0$ we have ${\cal K} \rightarrow {\cal
K}_0=1-2\Lambda/\tilde{\rho}^2$ and ${\cal R}^2(x) \rightarrow
{\cal R}_0^{\:2}=\tilde{\rho}^{-2}$. This, together with
transformations (\ref{ang}), show that the Bertotti-Robinson dS
C-metric transforms into the dS counterpart of the
Bertotti-Robinson solution discussed in subsection \ref{sec:BR},
when $A=0$. The limiting procedure that has been applied in this
subsection has generated a new exact solution that satisfies the
Maxwell-Einstein equations in a positive cosmological constant
background.
\begin{figure} [H]
\centering
\includegraphics[height=2.0in]{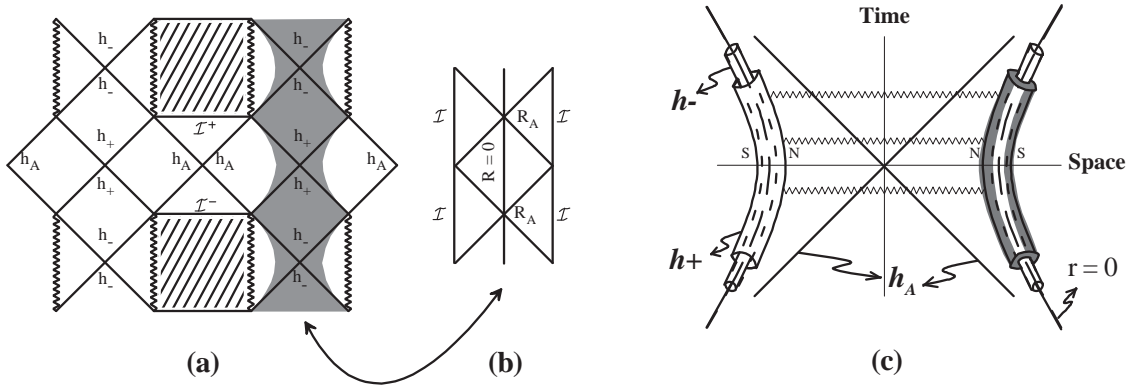}
\caption{\label{br-fig}
 (a) The whole figure is the Carter-Penrose diagram of the dS C-metric.
 The shaded region represents the near-horizon area that, when the
inner black hole
 horizon ($h_-$) approaches the outer black hole horizon ($h_+$),
gives the $AdS_2$
 manifold of the Bertotti-Robinson C-metric. See discussion in the
 text.
  (b) Carter-Penrose diagram of the Bertotti-Robinson C-metric.
  (c) The whole figure represents schematically the two accelerated black holes of
the dS C-metric. The shaded region represents the near-horizon
area that, when the inner black hole horizon approaches the outer
black hole horizon, gives the Bertotti-Robinson C-metric [compare
with Fig. \ref{br-fig}.(a)]. The strut does not survive to the
limiting process.
 }
\end{figure}

We have just seen that the Bertotti-Robinson dS C-metric can be
appropriately obtained from the vicinity of the inner and outer
black hole horizons in the limit in which the size of these two
horizons approach each other. This near-horizon region is sketched
in Fig. \ref{br-fig}.(a) as a shaded area, and from it we can
construct straightforwardly the Carter-Penrose diagram of the
Bertotti-Robinson dS C-metric drawn in Fig. \ref{br-fig}.(b). The
construction steps are as follows. First, as indicated by
(\ref{NariaiLimit-br}) and the second relation of
(\ref{NariaiCoord-br}), we let the cross lines that represent the
black hole Cauchy  horizon [$h_-$ in Fig. \ref{br-fig}.(a)] join
together with the cross lines that represent the  black hole event
horizon [$h_+$ in Fig. \ref{br-fig}.(a)], and so on, i.e., we do
this junction ad infinitum with all the cross lines $h_-$ and
$h_+$. After this step all that is left from the original diagram
is a single cross line, i.e, all the spacetime that has originally
contained in the shaded area of Fig. \ref{br-fig}.(a) has
collapsed into two mutually perpendicular lines at $45^{\circ}$ at
$y=\rho$. Now, as indicated by the first relation of
(\ref{NariaiCoord-br}), when $\varepsilon \rightarrow 0$ the time
suffers an infinite blow up. To this blow up in the time
corresponds an infinite expansion in the Carter-Penrose diagram in
the vicinity of $y=\rho$ that generates an $AdS_2$ region. We then
get again the shaded area of Fig. \ref{br-fig}.(a), but now the
cross lines of this shaded area are all identified into a single
line, and they no longer have the original labels associated to
$h_-$ and $h_+$ that differentiated them. This is, in the shaded
area of Fig. \ref{br-fig}.(a) we must now erase the original
labels $h_-$ and $h_+$. The Carter-Penrose diagram of the
Bertotti-Robinson dS C-metric is then given by Fig.
\ref{br-fig}.(b), which is equivalent to the diagram of the
(1+1)-dimensional AdS solution. Note that the diagram of the $A=0$
Reissner-Nordstr\"{o}m$-$dS solution is identical to the one of
Fig. \ref{br-fig}.(a). Therefore, applying the same construction
process described just above we find that the Carter-Penrose
diagram of the Bertotti-Robinson dS solution ($A=0$) is similar to
the one of Fig. \ref{br-fig}.(b). The diagram of the
Bertotti-Robinson solution with $\Lambda=0$ described by
(\ref{br}) is also given by  Fig. \ref{br-fig}.(b).

The Bertotti-Robinson near-horizon region is also sketched in Fig.
\ref{br-fig}.(c) as a shaded area. This schematic figure is
clarifying in the sense that it indicates that the strut that
connects the two original black holes along their north pole
directions does not survive to the near-extremal limiting process
and will not be present in the final result of the process, i.e.,
in the Bertotti-Robinson dS C-metric. However, a reminiscence of
this strut remains in the final solution. Indeed, the angular
factor of the Bertotti-Robinson dS C-metric [which, remind,
describes a deformed 2-sphere $\tilde{S}^2$ with a fixed size
${\cal R}^2(x)$ given by  (\ref{Gfactor-br})] has a conical
singularity at least at one of its poles. We can choose the
parameter $\kappa$, introduced in (\ref{ang}), in order to have a
conical singularity only at the north pole
($\delta_\mathrm{s}=0$), or only at the south pole
($\delta_\mathrm{n}=0$), however we cannot eliminate both. When
the parameter $A$ is set to zero, the conical singularities
disappear, the angular factor describes a round 2-sphere with
fixed radius, and the Bertotti-Robinson dS C-metric reduces into
the Bertotti-Robinson dS solution with topology $AdS_2 \times
S^2$.

\subsection{\label{sec:Gen Nariai-BR dS}The Nariai Bertotti-Robinson
dS C-metric}
\begin{figure} [H]
\centering
\includegraphics[height=2.1in]{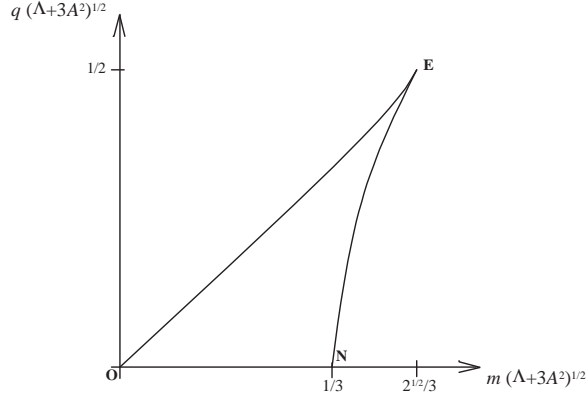}
\caption{\label{nbr-mq-fig}
 Relation $m\sqrt{\Lambda+3A^2}\times
q\sqrt{\Lambda+3A^2}$, for a fixed value of $A$ and $\Lambda$, for
the extremal limits of the dS C-metric. $NE$ represents the Nariai
C-metric (with point  $N$ representing the neutral case), $OE$
represents the Bertotti-Robinson dS C-metric, and point $E$
represents the Nariai Bertotti-Robinson dS C-metric.
 }
\end{figure}

As the previous sections and the label suggest, the Nariai
Bertotti-Robinson dS C-metric will be generated  from the extremal
limit of a very particular dS C-metric (usually called ultracold
solution  when $A=0$ \cite{MelMos,Rom,MannRoss,BooMann}) for which
the size of the three horizons ($y_A$, $y_+$ and $y_-$) are equal,
and let us label this degenerated horizon by $\rho$:
$y_A=y_+=y_-\equiv \rho$. In this case, the function ${\cal F}(y)$
is given by (\ref{Fextreme}) with $y_-=\rho$, and $\gamma$ defined
in (\ref{gamma}). The negative root $y_{\rm neg}$ is given by
(\ref{zerosy2}) and
\begin{eqnarray}
\rho=\sqrt{6\gamma} \:.
 \label{range-gamma-N-br}
\end{eqnarray}
The mass and the charge of the Nariai Bertotti-Robinson dS
C-metric solution , $m_{\rm NBR}$ and $q_{\rm NBR}$, respectively,
are given by
\begin{eqnarray}
& &m_{\rm NBR} =\frac{\sqrt{2}}{3}
 \sqrt{\frac{1}{\Lambda+3A^2}}
 \:, \nonumber \\
& & q_{\rm NBR} =\frac{1}{2}\sqrt{\frac{1}{\Lambda+3A^2}}
 \:,
 \label{mq-N-br}
\end{eqnarray}
and these values are the maximum values of the mass and charge of
both the Nariai C and Bertotti-Robinson C solutions,
(\ref{mq-cNariai}) and (\ref{mq-cold}), respectively. To clarify
the nature of these solutions, a diagram
$m\sqrt{\Lambda+3A^2}\times q\sqrt{\Lambda+3A^2}$ is plotted in
Fig. \ref{nbr-mq-fig}. For a fixed value of $A$ and $\Lambda$, the
allowed range of the mass and charge of the Nariai C-metric, of
the Bertotti-Robinson dS C-metric, and of the Nariai
Bertotti-Robinson dS C-metric is shown.

We are now ready to generate the Nariai Bertotti-Robinson dS
C-metric from the above ultracold dS C-metric. We first set
\begin{eqnarray}
 \rho=\sqrt{6\gamma}-\varepsilon, \:\:\:\:y_-=\sqrt{6\gamma}+\varepsilon \:,
 \:\:\:\:\:\:{\rm with} \:\: \varepsilon<<1\:.
 \label{NariaiLimit-N-br}
\end{eqnarray}
Now, we introduce a new time coordinate $\tau$ and a new radial
coordinate $\chi$,
 \begin{eqnarray}
t= \frac{1}{2 \varepsilon^2 {\cal K}}\,\tau \:,
\:\:\:\:\:\:\:\:\:\:\: y=\sqrt{6\gamma}+
 \varepsilon \cosh(\sqrt{2\varepsilon{\cal K}}\:\chi) \:,
 \label{NariaiCoord-N-br}
\end{eqnarray}
where
\begin{eqnarray}
{\cal K} &=& \frac{\rho^2-3\gamma}{\rho^4}
 (\rho-y_{\rm neg}) = \frac{1}{3}
 \sqrt{\frac{2A^2}{\Lambda+3A^2}}\:.
 \label{Kfactor-N-br}
\end{eqnarray}
 In the limit $\varepsilon \rightarrow
0$ the metric (\ref{C-metric}) becomes
\begin{eqnarray}
d s^2 \!\!&=&\!\! {\cal R}^2(x) \left [-\chi^2\, d\tau^2 +d\chi^2
 +{\cal G}^{-1}(x)dx^2+ {\cal G}(x)dz^2 \right ],
 \nonumber \\
 \label{Nariai-C-Metric-N-br}
\end{eqnarray}
with
\begin{eqnarray}
& {\cal R}^2(x) = \left (Ax+\sqrt{2(\Lambda+3A^2)}\right )^{-2}\:,
&
\nonumber \\
 &  {\cal G}(x) = 1-x^2
 -\frac{2A}{3}\sqrt{\frac{2}{\Lambda+3A^2}}\,x^3
 - \frac{A^2}{4(\Lambda+3A^2)} \,x^4\:.&
 \label{Gfactor-N-br}
\end{eqnarray}
${\cal G}(x)$ has only two real roots, the south pole
$x_\mathrm{s}$ and the north pole $x_\mathrm{n}$. The angular
coordinate $x$ can range between these two poles whose value is
calculated in section \ref{sec:angular}. Notice that the spacetime
factor $-\chi^2\, d\tau^2 +d\chi^2$ is just ${\mathbb{M}}^{1,1}$
in Rindler coordinates. Therefore, under the usual coordinate
transformation $\chi=\sqrt{\bar{x}^2-\bar{t}^{\,2}}$ and
$\tau={\rm arctanh(\bar{t}/\bar{x})}$, this factor transforms into
$-d\bar{t}^{\,2} +d\bar{x}^2$.
 Under the coordinate transformation (\ref{NariaiCoord-N-br}), the
Maxwell field for the magnetic case is still given by
(\ref{F-mag}), while in the electric case, (\ref{F-el-Lorentz})
becomes
 \begin{eqnarray}
 F=-q\,\chi \,d\tau\wedge d\chi\:.
 \label{F-el-N-br}
\end{eqnarray}
The Nariai Bertotti-Robinson dS C-metric is conformal to the
topological product of two 2-dimensional manifolds,
${\mathbb{M}}^{1,1}\times \tilde{S}^2$, with the conformal factor
${\cal R}^2(x)$ depending on the angular coordinate $x$.

In the $A=0$ limit, ${\cal R}^2(x) \rightarrow (2\Lambda)^{-1}$,
and one obtains the Nariai Bertotti-Robinson solution
$ds^2=(2\Lambda)^{-1}(-d\bar{t}^{\,2} +d\bar{x}^2+ d\theta^2+
\sin^2\theta\,d\phi^2)$, which has the topology
${\mathbb{M}}^{1,1}\times S^2$
\cite{MelMos,Rom,MannRoss,BooMann}).

\begin{figure} [H]
\centering
\includegraphics[height=1.7in]{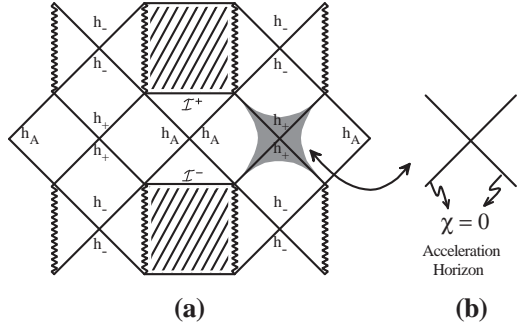}
\caption{\label{n-br-fig}
 (a) The whole figure is the Carter-Penrose diagram of the dS C-metric.
 The shaded region represents the near-horizon area that, when the
 inner black hole horizon ($h_-$) and the acceleration horizon ($h_A$)
 approach the outer black hole horizon ($h_+$), gives the Rindler manifold of
 the Nariai Bertotti-Robinson C-metric. This shaded region is the
 intersection of the shaded areas of Figs. \ref{nariai-fig}.(a) and
 \ref{br-fig}.(a). See discussion in the text.
 (b) Kruskal diagram of the Nariai Bertotti-Robinson dS C-metric.
 }
\end{figure}

We have just seen that the Nariai Bertotti-Robinson dS C-metric
can be appropriately obtained from the vicinity of the accelerated
and black hole horizons in the limit in which the size of these
three horizons approach each other. This near-horizon region is
sketched in Fig. \ref{n-br-fig}.(a) as a shaded area, and can be
viewed as the intersection of the shaded areas of Figs.
\ref{nariai-fig}.(a) and \ref{br-fig}.(a). From it we can
construct straightforwardly, following the construction steps
already sketched in subsections \ref{sec:Gen Nariai} and
\ref{sec:Gen BR dS}, the Kruskal diagram of the Nariai
Bertotti-Robinson dS C-metric drawn in Fig. \ref{n-br-fig}.(b).
This diagram is equivalent to the Kruskal diagram of the Rindler
solution. The strut that connects the two original black holes
along their north pole directions  survives to the near-extremal
limiting process. Thus the Nariai Bertotti-Robinson dS C-metric
describes a spacetime that is conformal to the product
${\mathbb{M}}^{1,1} \times \tilde{S}^2$. To each point in the
deformed 2-sphere corresponds a ${\mathbb{M}}^{1,1}$ spacetime,
except for one point which corresponds a ${\mathbb{M}}^{1,1}$
spacetime with an infinite straight strut or string, with a mass
density and pressure satisfying $p=-\mu$. In an analogous way to
the one that occurs with the Nariai C universe (see section
\ref{sec:Gen Nariai L=0}), we expect that the Nariai
Bertotti-Robinson dS C universe is unstable and, once created, it
decays through the quantum tunnelling process into a slightly
non-extreme black hole pair. The picture would be that of the
nucleation, in a dS background, of a Nariai Bertotti-Robinson dS C
universe, whose string then breaks down and a pair of dS black
holes is created at the endpoints of the string. This expectation
is confirmed in \cite{OscLem-PCdS}. When the parameter $A$ is set
to zero, the conical singularities disappear (and so the strut or
string are no longer present) and the angular factor describes a
round 2-sphere with fixed radius, and the topology of the Nariai
Bertotti-Robinson dS C-metric reduces into
${\mathbb{M}}^{1,1}\times S^2$.

\section{\label{sec:ExtLim L=0}Extremal limits of the flat C-metric
and of the Ernst solution}
The Euclidean version of the ``Nariai"  flat C-metric that will be
discussed in subsection \ref{sec:Gen Nariai L=0} has already been
used previously, \cite{GarfGidd}-\cite{Emparan} and
\cite{HawkRoss-string},  in the study of the quantum process of
pair creation of black holes. However, as far as we know, the
Bertotti-Robinson  flat C-metric (discussed in subsection
\ref{sec:Gen BR L=0}) and the Nariai Bertotti-Robinson flat
C-metric (discussed in subsection \ref{sec:Gen Nariai-BR L=0})
have not been written explicitly.

\subsection{\label{sec:Gen Nariai L=0}The ``Nariai"  flat C-metric}
The Nariai solution \cite{Nariai} is originally a solution in the
$\Lambda>0$ background which can be obtained from a near-extremal
limit of the dS black hole when the outer black horizon approaches
the cosmological horizon. Therefore, it would seem not appropriate
to use the name Nariai to label a $\Lambda=0$ solution. However,
in the flat C-metric the acceleration horizon plays the role of
the cosmological horizon. Moreover, the limit $A=0$ of the
solution discussed in this subsection is equal to the limit
$\Lambda=0$ of the Nariai solution (see subsection \ref{sec:lim
L=0 Nariai}). In this context, we find appropriate to label this
solution by ``Nariai" in between commas, in order to maintain the
nomenclature of the paper.

This solution can be obtained from a near-extremal limit of the
flat C-metric when the outer black horizon approaches the
acceleration horizon. This is the way it has been first
constructed in \cite{GarfGidd}-\cite{Emparan}. However, given the
Nariai C-metric ($\Lambda>0$) generated in subsection \ref{sec:Gen
Nariai}, we can construct the ``Nariai"  flat C-metric
($\Lambda=0$) by taking  directly the $\Lambda=0$ limit of
(\ref{Nariai-C-Metric})-(\ref{Gfactor}). The gravitational field
of the ``Nariai"  flat C-metric is then given by
(\ref{Nariai-C-Metric}) with
\begin{eqnarray}
& {\cal R}^2(x) = A^{-2}\left (x+\sqrt{\frac{6}{1+{\cal K}}}\right
)^{-2}\:,&
\nonumber \\
 & {\cal G}(x) = 1-x^2
 -\sqrt{\frac{2(1+{\cal K})(2-{\cal K})^2}{27}}\,x^3
- \frac{1-{\cal K}^2}{12} \,x^4\:, &
 \label{Gfactor L=0}
\end{eqnarray}
where $0<{\cal K}\leq 1$ and ${\cal K}=1$ when $q=0$. ${\cal
G}(x)$ has only two real roots, the south pole $x_\mathrm{s}$ and
the north pole $x_\mathrm{n}$. The angular coordinate $x$ can
range between these two poles whose value is calculated in section
\ref{sec:angular}.

There is a great difference between the Nariai C solution of the
$\Lambda>0$ case and the ``Nariai'' C solution of the $\Lambda=0$
case. In subsection \ref{sec:Gen Nariai} we saw that the Nariai
C-metric ($\Lambda>0$) has a conical singularity at one of the
poles of its deformed 2-sphere. This feature is no longer present
in the ``Nariai"  flat C-metric, i.e., it is free of conical
singularities. Indeed, in the flat C-metric we have ${\cal
F}(y)=-{\cal G}(-y)$. Therefore, if the outer black hole horizon
coincides with the acceleration horizon then the roots  $x_0$ and
$x_\mathrm{s}$ of ${\cal G}(x)$ also coincide (see Fig. \ref{g3
flatC}). This implies that the range of the angular coordinate $x$
becomes $x_\mathrm{s}<x\leq x_\mathrm{n}$ since the proper
distance between $x_\mathrm{s}$ and $x_\mathrm{n}$ goes to
infinity \cite{DGKT},\cite{GarfGidd}-\cite{Emparan}. The point
$x_\mathrm{s}$ disappears from the $x,z$ angular section which is
no longer compact but becomes topologically ${\mathbb{R}}\times
S^1$ or ${\mathbb{R}}^2$. So, we have a conical singularity only
at $x=x_\mathrm{n}$ which can be avoided by choosing
$2\kappa^{-1}=|{\cal G}'(x_\mathrm{n})|$. Therefore,  while the
Nariai C-metric ($\Lambda>0$) is topologically conformal to
$dS_2\times \tilde{S}^2$, the ``Nariai"  flat C-metric is
topologically conformal to $dS_2\times {\mathbb{R}}^2$. Its
Carter-Penrose diagram is given by Fig. \ref{nariai-fig}(b). We
could construct the ``Nariai" Ernst solution, however since its
main motivation is related to the removal of the conical
singularities present in the C-metric solution and in this case
our ``Nariai"  flat C-metric is free of conical singularities, we
will not do it.

At this point, it is appropriate to find the $A=0$ limit of the
``Nariai"  flat C-metric. This limit is not direct [see
(\ref{Gfactor L=0})]. To achieve the suitable limit we first make
the coordinate rescales: $\tilde{x}=\chi/A$, $\bar{y}=x/A$, and
$\bar{z}=z/A$. Then, setting $A=0$, the solution becomes
$ds^2=(1+{\cal K})/6\,[{\cal K}^{-1}(-\tilde{x}^2 d\tau^2
+d\tilde{x}^2)+ d\bar{y}^2+ d\bar{z}^2$]. This limit agrees with
the one taken from the $A=0$ Nariai solution (written in
subsection \ref{sec:Nariai}) in the limit $\Lambda=0$ (see
subsection \ref{sec:lim L=0 Nariai}). Therefore, while the limit
$A=0$ of the Nariai C-metric ($\Lambda>0$) is topologically
$dS_2\times S^2$, the $A=0$ limit of the ``Nariai" flat C-metric
is topologically ${\mathbb{M}}^{1,1}\times {\mathbb{R}}^2$. This
is a reminiscence of the fact that with $\Lambda=0$ when $A$ goes
to zero there is no acceleration horizon to play the role of
cosmological horizon that supports the extremal limit taken in
this subsection. As a final remark in this subsection, we note
that Horowitz and Sheinblatt \cite{HorowShein} have taken a
different extremal limit, which differs from the one discussed in
this subsection mainly because it preserves the asymptotic
behavior  and topology of the original flat C-metric solution.

\subsection{\label{sec:Gen BR L=0}The Bertotti-Robinson  flat C-metric}
Given the Bertotti-Robinson dS C-metric  generated in subsection
\ref{sec:Gen BR dS}, we can construct the Bertotti-Robinson  flat
C-metric by taking the direct $\Lambda=0$ limit of
(\ref{br-ds-C-Metric})-(\ref{Gfactor-br}). The gravitational field
of the Bertotti-Robinson flat C-metric is then given by
(\ref{br-ds-C-Metric}) with
\begin{eqnarray}
 &{\cal R}^2(x)=\left (Ax+\sqrt{\frac{6A^2}{1-{\cal K}}}\right
 )^{-2}\:,&
\nonumber \\
  &{\cal G}(x) = 1-x^2
 -\sqrt{\frac{2(1-{\cal K})(2+{\cal K})^2}{27}}\,x^3
 -\frac{1-{\cal K}^2}{12} \,x^4\:,&
 \label{Gfactor-br L=0}
\end{eqnarray}
and $0<{\cal K}< 1$. ${\cal G}(x)$ has only two real roots, the
south pole $x_\mathrm{s}$ and the north pole $x_\mathrm{n}$. The
angular coordinate $x$ can range between these two poles whose
value is calculated in section \ref{sec:angular}. As in the
Bertotti-Robinson dS C-metric, this solution has topology
conformal to  $AdS_2 \times \tilde{S}^2$, and a Carter-Penrose
diagram drawn in Fig. \ref{br-fig}.(b). The solution has a conical
singularity at one of the poles of its deformed 2-sphere
$\tilde{S}^2$. In the $A=0$ limit, we have ${\cal K} \rightarrow
1$ and ${\cal R}^2(x) \rightarrow q^2$ and we obtain the
Bertotti-Robinson solution (\ref{br}) discussed in subsection
\ref{sec:BR}.

From the above solution we can generate the Bertotti-Robinson
Ernst solution whose gravitational field is given by
\begin{eqnarray}
d s^2 = \Sigma^2(x)\frac{{\cal R}^2(x)}{{\cal K}} \left
(-\sinh^2\chi\, d\tau^2 +d\chi^2\right )+
 {\cal R}^2(x)\left [\frac{\Sigma^2(x)}{{\cal G}(x)}dx^2
 +\frac{{\cal G}(x)}{\Sigma^2(x)}\,dz^2 \right ]
 \:.
 \label{br-Ernst}
\end{eqnarray}
with ${\cal R}^2(x)$ and ${\cal G}(x)$ given by (\ref{Gfactor-br
L=0}), and with
 \begin{eqnarray}
 \Sigma(x)={\biggl (}1+\frac{1}{2}\, q \, {\cal E}_0\, x{\biggr )}^2
   + \frac{1}{4}{\cal E}_0^2 \, {\cal G}(x){\cal R}^2(x)\:.
 \label{factor-br}
 \end{eqnarray}
 Its electromagnetic field is given by   (\ref{pot-Ernst-B}), in
 the pure magnetic case, and by  (\ref{pot-Ernst-E}), in
 the pure electric case.
Choosing $\tilde{\kappa}$ satisfying  (\ref{k-s-Ernst}), with
$\tilde{\cal{G}}(x)=\Sigma^{-2}(x){\cal{G}}(x)$,  and ${\cal E}_0$
such that condition (\ref{k=Ernst}) is satisfied, the
Bertotti-Robinson Ernst solution is free of conical singularities.
As a final remark in this subsection, we note that Dowker,
Gauntlett, Kastor and Traschen \cite{DGKT} have taken a different
extremal limit, which differs from the one discussed in this
subsection mainly because it preserves the asymptotic behavior and
topology of the original flat C-metric solution.

\subsection{\label{sec:Gen Nariai-BR L=0}The ``Nariai" Bertotti-Robinson
flat C-metric}

From the Nariai Bertotti-Robinson dS C-metric  generated in
subsection \ref{sec:Gen Nariai-BR dS}, we can construct the
``Nariai" Bertotti-Robinson  flat C-metric by taking the direct
$\Lambda=0$ limit of
(\ref{Nariai-C-Metric-N-br})-(\ref{Gfactor-N-br}). We can also
construct the ``Nariai" Bertotti-Robinson Ernst solution. We do
not do these limits here because they are now straightforward.

\section{\label{sec:ExtLim AdS}Extremal limits of
the A\lowercase{d}S C-metric}
In the AdS background there are three different families of
C-metrics. When we set $A=0$, each one of these families reduces
to a different single AdS black hole. The main features of each
one of these three families of AdS black holes (the spherical, the
toroidal and the hyperbolic families) have been described in
section \ref{sec:BH 4D AdS}).

Now, to each one of these families corresponds a different AdS
C-metric which has been found by Pleba\'nski and Demia\'nski
\cite{PlebDem}. In what follows we will then describe each one of
these solutions and, in particular, we will generate a new
solution - the anti-Nariai C-metric.

\subsection{\label{sec:ExtLim AdS Sphere}Extremal limits of the
AdS C-metric with spherical horizons}
The spherical AdS C-metric has already been discussed in detail in
section \ref{sec:AdS C-metric}. Its gravitational field  is given
by (\ref{C-metric AdS}) with ${\cal F}(y)$ and ${\cal G}(x)$ given
by \ref{FG AdS}. Recall that this solution describes a pair of
accelerated black holes in the AdS background \cite{OscLem_AdS-C}
when the acceleration $A$ and the cosmological constant are
related by $A>|\Lambda|/3$. When we set $A=0$
\cite{EHM1,Pod,OscLem_AdS-C} we obtain a single non-accelerated
AdS black hole with spherical topology described subsection
\ref{sec:BH 4D AdS spherical}.

In a way analogous to the one described in last sections we can
generate new solutions from the extremal limits of the AdS
C-metric. Since this follows straightforwardly, we do not do it
here. The new relevant feature of this case is the fact that the
Nariai-like extremal solution only exists when $A>|\Lambda|/3$. In
this case, and only in this one, the acceleration horizon is
present in the AdS C-metric and we can have the outer black hole
horizon approaching it.

\subsection{\label{sec:ExtLim AdS Toroidal}Extremal limits of the
AdS C-metric with toroidal horizons }

The gravitational field of the massive charged toroidal AdS
C-metric (see \cite{PlebDem,MannAdS}) is given by (\ref{C-metric
AdS}) with
 \begin{eqnarray}
 & &{\cal F}(y) = \frac{|\Lambda|-3A^2}{3A^2}
                    -2mAy^3+q^2A^2y^4, \nonumber \\
 & &{\cal G}(x) = 1-2mAx^3-q^2 A^2 x^4\:,
 \label{FG-toroidal}
 \end{eqnarray}
and the electromagnetic field is given by (\ref{F-mag}) and
(\ref{F-el-Lorentz}). When we set $A=0$ we obtain the AdS black
hole with planar, cylindrical or toroidal topology
\cite{Lemos,Zanchin_Lemos,Huang_L,OscarLemos_string} described in
subsection \ref{sec:BH 4D AdS toroidal_electric}.

In a way analogous to the one described in section \ref{sec:ExtLim
dS} we can generate new solutions from the extremal limits of the
toroidal AdS C-metric, that are the toroidal AdS counterparts of
the Nariai and Bertotti-Robinson C-metrics. Since this follows
straightforwardly, we do not do it here.

\subsection{\label{sec:ExtLim AdS Topological}Extremal limits of the
AdS C-metric with hyperbolic horizons}

\begin{figure} [H]
\centering
\includegraphics[height=1.6in]{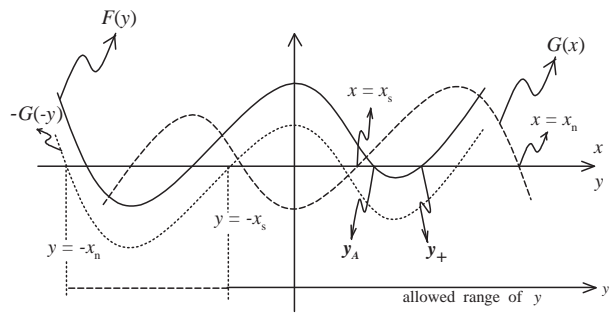}
\caption{\label{g3_topol}
 Shape of ${\cal G}(x)$ and ${\cal F}(y)$ for a general non-extremal
charged massive hyperbolic AdS C-metric studied in section
\ref{sec:ExtLim AdS Topological}. The allowed range of $x$ is
between $x_\mathrm{s}$ and $x_\mathrm{n}$ where ${\cal G}(x)$ is
positive and compact. The permitted range of $y$ is $-x\leq y <
+\infty$. The presence of an accelerated horizon is indicated by
$y_A$ and the black hole horizon by $y_+$. In the anti-Nariai case
considered in subsection \ref{sec:Gen antiNariai}, $y_A$ and $y_+$
coincide. [For completeness we comment on a case not represented
in the figure but discussed on the text: when $q=0$, the zero
$x_\mathrm{n}$ of ${\cal G}(x)$ disappears, and ${\cal G}(x)$
grows monotonically from $x=x_\mathrm{s}$ into $x=+\infty$.]
 }
\end{figure}

The gravitational field of the massive charged AdS C-metric with
hyperbolic horizons, the hyperbolic AdS C-metric, is given by
(\ref{C-metric AdS}) with  \cite{PlebDem,MannAdS}
\begin{eqnarray}
& &{\cal F}(y) = \frac{|\Lambda|+3A^2}{3A^2}
                     -y^2-2mAy^3+q^2A^2y^4, \nonumber \\
& &{\cal G}(x) = -1+x^2-2mAx^3-q^2 A^2 x^4\:
\label{FG-topological}
\end{eqnarray}
(represented in Fig. \ref{g3_topol}), and the electromagnetic
field is given by (\ref{F-mag}) and (\ref{F-el-Lorentz}).

This solution depends on four parameters namely, the cosmological
constant $\Lambda<0$, the acceleration parameter $A>0$, and $m$
and $q$ which are mass and electromagnetic charge parameters,
respectively. When $A=0$ this solution reduces to the hyperbolic
black holes \cite{topological} studied in subsection \ref{sec:BH
4D topological}.

\subsubsection{\label{sec:Gen antiNariai}The anti-Nariai C-metric}

We are interested in a particular extreme  hyperbolic AdS
C-metric, for which $y_A=y_+$ (see Fig. \ref{g3_topol}), and let
us label this degenerated horizon by $\rho$: $y_A=y_+\equiv \rho$.
In this case, the function ${\cal F}(y)$ can be written as
\begin{eqnarray}
{\cal F}(y) =-\frac{\rho^2-3\gamma}{\rho^4}
 (y-y_{\rm neg})(y-y'_{\rm neg})(y-\rho)^2\:,
 \label{Fextreme-a}
 \end{eqnarray}
where
\begin{eqnarray}
\gamma =\frac{|\Lambda|+3A^2}{3A^2}\:,
 \label{gamma-a}
 \end{eqnarray}
and the degenerate root $\rho$, and the negative roots $y_{\rm
neg}$ and $y'_{\rm neg}$ are given by
\begin{eqnarray}
& & \rho =\frac{3m}{4q^2A}
 \left ( 1+ \sqrt{1+\frac{8}{9}\frac{q^2}{m^2}} \:\right )
 \:,  \nonumber \\
& & y_{\rm neg} =-\frac{\gamma \rho}{3\gamma-\rho^2}
 \left ( 1+ \sqrt{\frac{\rho^2-2\gamma}{\gamma}} \:\right )
 \:, \nonumber \\
& & y'_{\rm neg} =-\frac{\gamma \rho}{3\gamma-\rho^2}
 \left ( 1- \sqrt{\frac{\rho^2-2\gamma}{\gamma}} \:\right )
 \:.
 \label{zerosy3-a}
 \end{eqnarray}
The mass parameter and the charge parameter of the solution are
written as a function of $\rho$ as
\begin{eqnarray}
& &m =-\frac{1}{A\rho}
 \left ( 1- \frac{2\gamma}{\rho^2} \right )
 \:,  \nonumber \\
& & q^2 =\frac{1}{A^2\rho^2}
 \left ( \frac{3\gamma}{\rho^2}-1 \right )
 \:.
 \label{mq-a}
 \end{eqnarray}
The requirement that $y_{\rm neg}$ and $y'_{\rm neg}$ are real
roots and the condition $q^2 \geq 0$ require that the allowed
range of $\rho$ is
\begin{eqnarray}
 2\gamma < \rho^2 \leq 3\gamma \:.
 \label{range-gamma-a}
\end{eqnarray}
The mass and the charge of the anti-Nariai type solution, $m_{\rm
aN}$ and $q_{\rm aN}$, respectively, are both monotonically
decreasing functions of $\rho$, and as one comes from
$\rho=\sqrt{3\gamma}$ into $\rho=\sqrt{2\gamma}$ one has,
\begin{eqnarray}
& &  -\frac{1}{3}
  \frac{1}{\sqrt{|\Lambda|+3A^2}}\leq m_{\rm aN}<0\:, \nonumber \\
& & 0\leq q_{\rm aN}< \frac{\sqrt{3}}{2}
  \frac{1}{\sqrt{|\Lambda|+3A^2}}\:.
 \label{mq-antiNariai}
 \end{eqnarray}

In order to generate the anti-Nariai C-metric from the
near-extreme topological AdS C-metric we first set
\begin{eqnarray}
 y_A=\rho-\varepsilon, \:\:\:\:y_+=\rho+\varepsilon,
 \:\:\:\:\:\:{\rm with} \:\: \varepsilon<<1\:,
 \label{NariaiLimit-a}
\end{eqnarray}
in order that $\varepsilon$ measures the deviation from
degeneracy, and the limit $y_A\rightarrow y_+$ is obtained when
$\varepsilon \rightarrow 0$. Now, we introduce a new time
coordinate $\tau$ and a new radial coordinate $\chi$,
 \begin{eqnarray}
t= \frac{1}{\varepsilon {\cal K}}\,\tau \:, \:\:\:\:\:\:\:\:\:\:\:
y=\rho+\varepsilon \cosh\chi \:,
 \label{NariaiCoord-a}
\end{eqnarray}
where
\begin{eqnarray}
& & \!\!\! {\cal K} = \frac{3\gamma-\rho^2}{\rho^4}
 (\rho-y_{\rm neg})(\rho-y'_{\rm neg})=
 \frac{2(|\Lambda|+3A^2)}{A^2\rho^2}-1\:,  \nonumber \\
& &
 \label{Kfactor-a}
\end{eqnarray}
and $2\gamma < \rho^2 \leq 3\gamma$ implies $1 \leq {\cal K}<2$
with $q=0\Rightarrow {\cal K}=1$. In the limit $\varepsilon
\rightarrow 0$, from (\ref{C-metric AdS}) and (\ref{Fextreme-a}),
the metric becomes
\begin{eqnarray}
d s^2 = \frac{{\cal R}^2(x)}{{\cal K}} \left (-\sinh^2\chi\,
d\tau^2 +d\chi^2\right ) +
 {\cal R}^2(x)\left [{\cal G}^{-1}(x)dx^2+ {\cal G}(x)dz^2 \right ]
 \:.
 \label{Nariai-C-Metric-a}
\end{eqnarray}
where
\begin{eqnarray}
 &{\cal R}^2(x)= \left (
Ax+\sqrt{\frac{2(|\Lambda|+3A^2)}{1+{\cal K}}}\right )^{-2}\:,& \\
\label{Rfactor-a}
  &  \!\!\!{\cal G}(x) = -1+x^2
+\frac{A}{3} \sqrt{\frac{2(1+{\cal K})(2-{\cal
K})^2}{(|\Lambda|+3A^2)}}\,x^3
  +\frac{A^2}{4}\frac{{\cal K}^2-1}{|\Lambda|+3A^2} \,x^4\:,&
  \nonumber \\
 \label{Gfactor-a}
\end{eqnarray}
Under the coordinate transformation (\ref{NariaiCoord-a}), the
Maxwell field for the magnetic case is still given by
(\ref{F-mag}), while in the electric case, (\ref{F-el-Lorentz})
becomes
 \begin{eqnarray}
 F=-\frac{q}{{\cal K}}\,\sinh \chi \,d\tau\wedge d\chi\:.
\label{F-el-antiNariai}
\end{eqnarray}
The Carter-Penrose diagram of the anti-Nariai C-metric is also
given by Fig. \ref{br-fig}.(b).

At this point, let us focus our attention on the angular surfaces
with $\tau=$constant and $\chi=$constant. When $q\neq 0$, we
choose $x$ such that it belongs to the range
$[x_\mathrm{s},x_\mathrm{n}]$ (sketched in Fig. \ref{g3_topol})
where ${\cal G}(x)\geq 0$. In this way, the metric has the correct
signature, the angular surfaces are compact, and the allowed range
of $y$ includes the acceleration ($y_A$) and black hole ($y_+$)
horizons (if we had chosen the other possible interval of $x$
where ${\cal G}(x)\geq 0$, sketched in Fig. \ref{g3_topol}, this
last condition would not be satisfied). We unavoidably have a
conical singularity at least at one of the poles of this compact
surface (that we label by $\tilde{S}^2$, say). Thus, the charged
anti-Nariai C-metric has a compact angular surface of fixed size
with a conical singularity at one of its poles. It is
topologically conformal to $AdS_2\times \tilde{S}^2$. When $q=0$,
the zero $x_\mathrm{n}$ of ${\cal G}(x)$ disappears, and ${\cal
G}(x)$ grows monotonically from $x=x_\mathrm{s}$ into $x=+\infty$.
Then, the angular surfaces are not compact, and we have a single
pole ($x=x_\mathrm{s}$) with a conical singularity, which can be
eliminated. Thus, the neutral anti-Nariai C-metric has a
non-compact angular surface of fixed size (a kind of a deformed
2-hyperboloid that we label by $\tilde{H}_2$, say) which is free
of conical singularities. It is topologically conformal to
$AdS_2\times \tilde{H}_2$.

In order to obtain the $A=0$ limit, we first set
$\hat{\rho}=A\rho$ [see first relation of (\ref{zerosy3-a})], a
parameter that has a finite and well-defined value when
$A\rightarrow 0$. Then when $A\rightarrow 0$ we have ${\cal K}
\rightarrow {\cal K}_0=2|\Lambda|/\hat{\rho}^2-1$ and ${\cal R}^2
\rightarrow {\cal R}_0^{\:2}=\hat{\rho}^{-2}$, with ${\cal
R}_0^{\:2}$ and ${\cal K}_0$ satisfying relations
(\ref{relations-a}). Moreover, when we set $A = 0$, $m\neq 0$ and
$q\neq 0$, the coordinate transformations $\theta =
\int_{x_\mathrm{s}}^{x}{\cal{G}}^{-1/2}dx$ and $\phi=z$ imply that
$x\in[x_\mathrm{s}=+1,+\infty[$, $x=\cosh \theta$, and ${\cal
G}=-1+x^2=\sinh^2 \theta$. The angular surface then reduces to a
2-hyperboloid, $H_2$, of fixed size with line element
$d\theta^2+\sinh^2{\theta}\,d\phi^2$. Therefore, when $A=0$ the
anti-Nariai C-metric reduces to the anti-Nariai solution
(\ref{qNariai-a}) described in subsection \ref{sec:anti-Nariai},
with topology $AdS_2 \times H_2$. The limiting procedure that has
been applied in this subsection has generated a new exact solution
that satisfies the Maxwell-Einstein equations in a negative
cosmological constant background.

\subsubsection{\label{sec:other limits} Other extremal limits}

We could also discuss other extremal limits of the charged
topological AdS C-metric (see Fig. \ref{g3_topol}), but these do
not seem to be so interesting.

\section{\label{sec:angular}Determination of the north and south poles}

In this section we discuss the zeros of the function ${\cal G}(x)$
that appears in the extremal limits of the dS C-metric, and in the
extremal limits of the flat C-metric. This function ${\cal G}(x)$
has only two real zeros in the cases discussed in this paper,
namely the Nariai, the Bertotti-Robinson, and the Nariai
Bertotti-Robinson (both for $\Lambda>0$ and $\Lambda=0$). These
two roots are the south pole and the north pole, and are
respectively given by
\begin{eqnarray}
x_\mathrm{s} & = & -p + \frac{h}{2} - \frac{a}{4\,b} < 0\, , \nonumber \\
x_\mathrm{n} & = & +p +
      \frac{h}{2} - \frac{a}{4\,b} > 0 \:,
\label{polos - cold}
    \end{eqnarray}
with
\begin{eqnarray}
p & = &\frac{1}{2}\left (-\frac{s}{3} + \frac{a^2}{2\,b^2} -
 \frac{1 - 12 \,b}{3\,s\,b^2} - \frac{4}{3\, b} + n\right)^{1/
          2} \:, \nonumber \\
n & = &\frac{-a^3 + 4\,a\, b}{4\,h \,b^3} \:, \nonumber \\
h & = &\sqrt{\frac{s}{3} + \frac{a^2}{4 \,b^2} +
   \frac{1 -12\, b}{3\,s\, b^2} - \frac{2}{3\,b}} \:, \nonumber \\
s & = &\frac{1}{2^{1/3}\, b}\left ( \lambda - \sqrt{\lambda^2 -
 4(1-12\, b)^3} \right)^{1/
          3} \:, \nonumber \\
        \lambda & = & 2 - 27 \,a^2 + 72\, b\:,
\label{acess - zeros - ang}
     \end{eqnarray}
where $a$ and $b$ are, respectively, the absolute values of the
coefficients of $x^3$ and $x^4$ in  (\ref{Gfactor}),
(\ref{Gfactor-br}), (\ref{Gfactor-N-br}), (\ref{Gfactor L=0}), and
(\ref{Gfactor-br L=0}). For the function ${\cal G}(x)$ written in
a different polynomial form that facilitates the determination of
its zeros see Hong and Teo \cite{HongTeo}.

\section{\label{sec:Conc}Summary and discussion}

Following the limiting approach first introduced by Ginsparg and
Perry \cite{GinsPerry}, we have analyzed the extremal limits of
the dS, flat and AdS C-metrics. Among other new exact solutions,
we have generated the Nariai C-metric, the Bertotti-Robinson
C-metric, the Nariai Bertotti-Robinson C-metric and the
anti-Nariai C-metric. These solutions are the C-metric
counterparts of the well know solutions found in the 1950's. They
are specified by an extra parameter: the acceleration parameter
$A$ of the C-metric from which they are generated. When we set
$A=0$ the solutions found in this paper reduce to the Nariai, the
Bertotti-Robinson, the Nariai Bertotti-Robinson and the
anti-Nariai solutions.

One of the features of these $A=0$ solutions is the fact that they
are topologically the direct product of two 2-dimensional
manifolds of constant curvature. Their C-metric counterparts are
conformal to this topology, with the conformal factor depending on
the angular coordinate. Moreover, the angular surfaces of these
new C-solutions have a fixed size, but they lose the symmetry of
the $A=0$ counterparts. For example, while the angular surfaces of
the Nariai and Bertotti-Robinson solutions are round 2-spheres,
the angular surfaces of their C-metric counterparts are deformed
2-spheres - they are compact but not round. Another important
difference between the $A=0$ and $A \neq 0$ solutions is the fact
that the $A \neq 0$ solutions have, in general, a conical
singularity at least at one of the poles of their angular
surfaces. This conical singularity is a reminiscence of the
conical singularity that is present in the C-metric from which
they were generated. In the C-metric these conical singularities
are associated to the presence of a strut or string that furnishes
the acceleration of the near-extremal black holes. In this
context, we find that the Nariai C-metric generated from a
extremal limit of the dS C-metric describes  a spacetime that is
conformal to the product $dS_2\times \tilde{S}^2$. To each point
in the deformed 2-sphere corresponds a $dS_2$ spacetime, except
for one point which corresponds a $dS_2$ spacetime with an
infinite straight strut or string, with a mass density and
pressure satisfying $p=-\mu$. Analogously, the Nariai
Bertotti-Robinson dS C-metric describes a spacetime that is
conformal to the product ${\mathbb{M}}^{1,1} \times \tilde{S}^2$.
To each point in the deformed 2-sphere corresponds a
${\mathbb{M}}^{1,1}$ spacetime, except for one point which
corresponds a ${\mathbb{M}}^{1,1}$ spacetime with an infinite
straight strut or string. In the case of the Bertotti-Robinson dS
C-metric (topologically conformal to $AdS_2\times \tilde{S}^2$),
the strut or string does not survive to the Ginsparg-Perry
limiting procedure, and thus in the end of the process we only
have a conical singularity.

In what concerns the causal structure, the Carter-Penrose diagrams
of the $A \neq 0$ solutions are equal to those of the $A=0$
solutions. For example, the diagram of the Nariai C-metric is
equal to the one that describes the (1+1)-dimensional dS solution,
the diagram of the Bertotti-Robinson C-metric and of the
anti-Nariai C-metric is equal to the one that describes the
(1+1)-dimensional AdS solution, and the diagram of the Nariai
Bertotti-Robinson C-metric is given by the Rindler diagram.

Some of these solutions, perhaps all, are certainly of physical
interest.  Indeed, it is known that the Nariai solution ($A=0$) is
unstable and, once created, it decays through the quantum
tunnelling process into a slightly non-extreme black hole pair
\cite{Bousso60y}. We then expect that the Nariai C-metric is also
unstable and that it will decay into a slightly non-extreme pair
of black holes accelerated by a strut or by a string. The
solutions found in this paper also play an important role in the
decay of the dS or AdS spaces, and therefore can mediate the
Schwinger-like quantum process of pair creation of black holes.
Indeed, the Nariai, and the dS Nariai Bertotti-Robinson instantons
($A=0$) are one of the few Euclidean solutions that are regular,
and have thus been used
\cite{MelMos,Rom,MannRoss,BooMann,VolkovWipf} to study the pair
creation of dS black holes materialized and accelerated by the
cosmological constant background field (an instanton is a solution
of the Euclidean field equations that smoothly connects the
spacelike sections of the initial state, the pure dS space in this
case, and the final state, the dS black hole pair in this case).
Moreover, the Euclidean ``Nariai" flat C-metric and Ernst solution
(see section \ref{sec:Pair Creation flat}) have been used to
analyze the process of pair production of $\Lambda=0$ black holes,
accelerated by a string or by an electromagnetic external field,
respectively. Therefore, its natural to expect that the Euclidean
extremal limits of the dS C-metric and AdS C-metric found in this
paper mediate the process of pair creation of black holes in a
cosmological background, that are then accelerated by a string, in
addition to the cosmological field acceleration. For the Nariai
case, e.g., the picture would be that of the nucleation, in a dS
background, of a Nariai C universe, whose string then breaks down
and a pair of dS black holes is created at the endpoints of the
string. This expectation is confirmed in \cite{OscLem-PCdS}.


%% file: Chapter7.tex
\thispagestyle{empty} \setcounter{minitocdepth}{1}
\chapter[False vacuum decay:
effective one-loop action for pair creation of domain walls]
{\Large{False vacuum decay:\\ effective one-loop action for pair
creation of domain walls}} \label{chap:False vacuum}
 \lhead[]{\fancyplain{}{\bfseries Chapter \thechapter. \leftmark}}
 \rhead[\fancyplain{}{\bfseries \rightmark}]{}
  \minitoc \thispagestyle{empty}
\renewcommand{\thepage}{\arabic{page}}

\addtocontents{lof}{\textbf{Chapter Figures \thechapter}\\}

This chapter can be seen as an introductory toy model for the
black hole pair creation analysis that will be done in chapter
\ref{chap:Pair creation}. We propose an effective one-loop action
to describe the domain wall pair creation process that accompanies
the false vacuum decay of a scalar field (in the absence of
gravity). We compute the pair creation rate of the process,
including the one-loop contribution, using the instanton method
that is also applied in the computation of black hole pair
creation rates.

\section{\label{sec:Introduction False Vacuum}The false vacuum decay process}

Stone \cite{Stone} has studied the problem of a scalar field
theory in (1+1)-dimensions with a metastable vacuum, i.e., with a
scalar potential that has a false vacuum, $\phi_+$, and a true
vacuum, $\phi_-$, separated by an energy density difference,
$\epsilon$. Stone has noticed that the decay process can be
interpreted as the false vacuum decaying into the true vacuum plus
a creation of a soliton-antisoliton pair: $\phi_+ \rightarrow
\phi_- +s +\bar{s}\:.$ The energy necessary for the
materialization of the pair comes from the energy density
difference between the two vacua. The soliton-antisoliton pair
production rate per unit time and length, $\Gamma/L$, can then be
identified with the decay rate of the false vacuum and is given by
($\hbar=c=1$) \cite{Stone}:
\begin{eqnarray}
\Gamma/L=A\:{\rm e}^{-S_0}=A\:{\rm e}^{-\frac{\pi
m^2}{\epsilon}}\:, \label{0.1}
\end{eqnarray}
where $m$ is the soliton mass and prefactor $A$ is a functional
determinant whose value was first calculated by Kiselev and
Selivanov \cite{Kiselev1,Kiselev2} and later by Voloshin
\cite{Voloshin1}. Extensions to this decay problem, such as
induced false vacuum decay, have been studied by several authors
(for a review and references see, e.g.,
\cite{Voloshin3,Kiselev3}).

The method used in \cite{Stone,Kiselev1,Kiselev2,Voloshin1} is
based on the instanton method introduced by Langer in his work
about decay of metastable termodynamical states \cite{Langer}.
This powerful method has been applied to several different
studies, namely: Coleman and Callan \cite{Coleman,ColemanCallan}
have computed the bubble production rate that accompanies the
cosmological phase transitions in a (3+1)-dimensional scalar
theory (this was indeed previously calculated by other methods by
Voloshin, Kobzarev and Okun \cite{Voloshin2}); Affleck and Manton
\cite{Affleck2} have studied monopole pair production in a weak
external magnetic field and Affleck, Alvarez and Manton
\cite{Affleck1}, have studied $e^+e^-$ boson pair production in a
weak external electric field. Recent developments studying pair
production of boson and spinorial particles in external Maxwell's
fields have been performed by several authors using different
methods \cite{gavri,solda,lin fv}. Similar results in the
Euler-Heisenberg theory, a modified Maxwell theory, have been also
obtained \cite{KS}.

In this chapter, following our work \cite{OscLem_FalVac}, we
propose an effective one-loop action built from the soliton field
itself to study the problem of Stone \cite{Stone}, Kiselev and
Selivanov \cite{Kiselev1,Kiselev2} and Voloshin \cite{Voloshin1}.
The action consists of the usual mass term and a kinetic term in
which the simple derivative of the soliton field is replaced by a
kind of covariant derivative.  In this effective action the
soliton charge is treated no longer as a topological charge but as
a Noether charge. This procedure of working with an effective
action for the soliton field itself has been introduced by Coleman
 \cite{Coleman1} where the equivalence between the
Sine-Gordon model and the Thirring model was shown, and by
Montonen and Olive \cite{Olive} who  have proposed an equivalent
dual field theory for the Prasad-Sommerfield monopole soliton.
More connected to our problem, Manton \cite{Manton} has proposed
an effective action built from the soliton field  itself which
reproduces the soliton physical properties of (1+1)-dimensional
nonlinear scalar field theories having symmetric potentials with
degenerate minima. In this paper we deal instead with a potential
with non-degenerate minima in a (1+1)-dimensional scalar field
theory. Thus, our effective action is new since Manton was not
dealing with the soliton pair production process.

Using the effective one-loop action and the method presented in
\cite{Affleck1}, we calculate the soliton-antisoliton pair
production rate, (\ref{0.1}). One recovers Stone's exponential
factor $S_0$ \cite{Stone} and the prefactor $A$ of Kiselev and
Selivanov \cite{Kiselev1,Kiselev2} and Voloshin \cite{Voloshin1}.

\section{\label{sec:Eff One Loop}Effective one-loop action}

In order to present some useful soliton properties let us consider
a scalar field theory in a (1+1)-dimensional  spacetime, whose
dynamics is governed by the action (see, e.g., \cite{Rajamaran}),
\begin{equation}
S[\phi(x,t)]=\int d^2x {\biggl [} \frac{1}{2}
\partial_{\mu}\phi \partial^{\mu}\phi -
U(\phi) {\biggl ]}, \label{1.1}
\end{equation}
where $U$ is a generic potential. A particular important case is
when $U$ is a symmetric potential, $U=U_{\rm s}(\phi)$, with two
or more degenerate minima. In the $\phi^4$ theory  the potential
is $U_{\rm s}(\phi)=\frac{1}{4}\lambda
{\bigl(}\phi^2-\mu^2/\lambda{\bigr)}^2$, with $\mu\geq0$ and
$\lambda \geq0\:$. Stationarizing the action one obtains the
solutions of the theory which have finite and localized energy.
The solutions are the soliton
\begin{equation}
\psi\equiv\phi_{\rm sol}=+
\frac{\mu}{\sqrt{\lambda}}\tanh{\biggl[} \frac{\mu}{\sqrt{2}}
\:\frac{(x-x_{0})-vt}{\sqrt{1-v^2}} {\biggr]}\:, \label{1.2}
\end{equation}
and the antisoliton $-\psi$. From the hamiltonian density, ${\cal
H}=\frac{1}{2}(\partial_{x}\phi)^2 + U_{\rm s}(\phi)$, one can
calculate the mass of the soliton and antisoliton
\begin{equation}
m=\int_{-\infty}^{+\infty} dx \,{\cal H}(x)=\frac{2\sqrt{2}}{3}
\frac{\mu^3}{\lambda}. \label{1.3}
\end{equation}
One can also define the topological charge, $Q=\frac{1}{2}
{\bigl[}\psi(x=+\infty)-\psi(x=-\infty){\bigr]}$, (conserved in
time) which has the positive value $Q_{\rm s}=+\mu/
\sqrt{\lambda}$ in the case of the soliton and the negative value
$Q_{\bar{\rm s}}=-\mu/ \sqrt{\lambda}$ in the case of the
antisoliton. To this charge one associates the topological current
$k^{\mu}=\frac{1}{2} \varepsilon^{\mu\nu} \partial_{\nu} \psi$
which is conserved, $\partial_{\mu}k^{\mu}=0$, and such that
$Q=\int_{-\infty}^{+\infty} dx k^0$.

Now, let us consider a non-degenerate potential $U$ in action
(\ref{1.1}) by adding to $U_{\rm s}$
 a small term that breaks its symmetry
\cite{Stone,ColemanCallan}: $U(\phi)=U_{\rm
s}(\phi)+\frac{\epsilon}{2\mu/ \sqrt{\lambda}}(\phi-\mu/
\sqrt{\lambda})$, where $\epsilon$ is the energy density (per unit
length) difference between the true ($\phi_-=-\mu/
\sqrt{\lambda}$) and false ($\phi_+=+\mu/ \sqrt{\lambda}$) vacua.
As noticed in \cite{Stone,Kiselev1,Kiselev2}, $\epsilon$ is
responsible for both the decay of false vacuum and
soliton-antisoliton pair creation.

We want to find an effective one-loop action built from the
soliton field itself and that describes the above pair creation
process. The soliton field should be a charged field since the
system admits  two charges, $Q_{\rm s}$ and $Q_{\bar{\rm s}}$.
Therefore, the action should contain the mass term $m^2\psi
\psi^*$, where $m$ is the soliton mass given in (\ref{1.3}), and
the kinetic term $(\partial_{\mu}\psi)(\partial^{\mu}\psi^*)$.
Thus, the free field effective action is $\int d^2x
[(\partial_{\mu}\psi)(\partial^{\mu}\psi^*)- m^2\psi \psi^*]$.
However, if one demands local gauge invariance one has to
introduce an ``electromagnetic'' 2-vector potential $A_{\mu}$
which transforms the common derivative $\partial_\mu \psi$ into a
covariant derivative $(\partial_\mu +i Q_{\rm s}A_{\mu})\psi$. As
is well known, the field $A_{\mu}$ itself should contribute to the
action. This contribution must be gauge invariant since the
covariant kinetic term plus the mass term are already gauge
invariant. This is achieved by defining the invariant 2-form
field, $F_{\mu\nu}=\partial_\mu A_{\nu}-\partial_\nu A_{\mu}$. In
two dimensions, an anti-symmetric field can only be of the form:
$F_{\mu\nu}=\sigma(t,x) \varepsilon_{\mu\nu}$, where
$\varepsilon_{\mu\nu}$ is the Levi-Civita tensor and $\sigma(t,x)$
a function. Therefore the effective one-loop action should be
$S^{\rm eff}=\int d^2x [(\partial_{\nu}\psi+i Q_{\rm
s}A_{\nu}\psi)(\partial^{\nu}\psi^* -i Q_{\rm s}A^{\nu}\psi^*)-
m^2\psi \psi^*-\frac{1}{4}F_{\mu\nu}F^{\mu\nu}]$.

Note now that the charged soliton acts also as a source, thus
modifying the surrounding field. As a first approximation we shall
neglect this effect and assume $A_\mu$ fixed by external
conditions. This allows us to drop the contribution of the term
$F_{\mu\nu}F^{\mu\nu}$ in the effective action. Moreover, the
external field responsible for the pair creation is essentially
represented by the energy density difference $\epsilon$ so we
postulate that $F^{\rm ext}_{\mu\nu}= \frac{\epsilon}{\mu/
\sqrt{\lambda}} \varepsilon_{\mu\nu}$. Therefore, $A^{\rm
ext}_{\mu}$ is given by $A^{\rm
ext}_{\mu}=\frac{1}{2}\frac{\epsilon}{\mu/ \sqrt{\lambda}}
\varepsilon_{\mu\nu}x^{\nu}$.

Finally, if the system is analytically continued  to Euclidean
spacetime $(t_{\rm Min} \rightarrow -it_{\rm Euc}$; $A_0
\rightarrow iA_2)$ one obtains the Euclidean effective one-loop
action for the soliton pair creation problem
\begin{eqnarray}
S^{\rm eff}_{\rm Euc}=\int d^2x {\biggl [} {\bigl
|}(\partial_{\mu}-\frac{1}{2} \epsilon
\:\varepsilon_{\mu\nu}x_\nu)\psi{\bigl |}^2+m^2|\psi|^2 {\biggr
]}\:. \label{1.4}
\end{eqnarray}

In the next section this Euclidean effective one-loop action is
going to be used to calculate the soliton-antisoliton pair
production rate (\ref{0.1}). Although the calculations are now
similar to those found in Affleck, Alvarez and Manton pair
creation problem \cite{Affleck1}, we present some important steps
and results since in two dimensions they are slightly different.

\section{\label{sec:Pair production rate}Pair production rate}

The soliton-antisoliton pair production rate per unit time is
equal to the false vacuum decay rate per unit time
\begin{eqnarray}
\Gamma=- 2\:{\rm Im}E_0\:, \label{2.1}
\end{eqnarray}
where the vacuum energy, $E_0$, is given by the Euclidean
functional integral
\begin{eqnarray}
{\rm e}^{-E_0 T} = \lim_{T \rightarrow \infty} \int [{\cal D}\psi]
[{\cal D}\psi^*] {\rm e}^{-S^{\rm eff}_{\rm Euc}(\psi;\psi^*) }\:.
\label{2.2}
\end{eqnarray}
As it will be verified, $E_0$ will receive a small imaginary
contribution from the negative-mode associated to the quantum
fluctuations about the instanton (which stationarizies the action)
and this fact is responsible for the decay. Combining (\ref{2.1})
and (\ref{2.2}) one has
\begin{eqnarray}
\Gamma = \lim_{T \rightarrow \infty} \frac{2}{T} {\rm Im}\ln \int
[{\cal D}\psi] [{\cal D}\psi^*] {\rm e}^{-S^{\rm eff}_{\rm
Euc}(\psi;\psi^*)}\:,
 \label{2.3}
\end{eqnarray}
where $S^{\rm eff}_{\rm Euc}$ is given by (\ref{1.4}). Integrating
out $\psi$ and $\psi^*$ in (\ref{2.3}) one obtains
\begin{eqnarray}
\Gamma=-\lim_{T \rightarrow \infty} \frac{2}{T}\:{\rm Im}\:{\rm
tr}\:\ln{\bigl [}(\partial_{\mu}-\frac{1}{2} \epsilon
\:\varepsilon_{\mu\nu}x_\nu)^2+ m^2 {\bigr ]}.  \label{2.5}
\end{eqnarray}
The logarithm in (\ref{2.5}) can be written as a ``Schwinger
proper time integral", $\ln u=-\int_{0}^{\infty}\frac{d
\cal{T}}{\cal{T}} \exp{\bigl (}-\frac{1}{2}u \cal{T}{\bigr )}$.
Taking $u={\bigl [}(\partial_{\mu}-\frac{1}{2} \epsilon
\:\varepsilon_{\mu\nu}x_\nu)^2+ m^2 {\bigr ]}$, yields
\begin{eqnarray}
\Gamma=\lim_{T \rightarrow \infty} \frac{2}{T}\:{\rm
Im}\int_{0}^{\infty}\frac{d \cal{T}}{\cal{T}} {\rm
e}^{-\frac{1}{2}m^2 \cal{T}} \:{\rm tr}\:\exp{\biggl
[}-\frac{1}{2}{\biggl (}P_\mu-\frac{1}{2} \epsilon
\:\varepsilon_{\mu\nu}x_\nu{\biggl )}^2 \cal{T}{\biggr ]}.
\label{2.6}
\end{eqnarray}
Notice that now the trace is of the form ${\rm tr}\:{\rm e}^{-H
\cal{T}}$, with $ H=\frac{1}{2}{\biggl [}P_\mu-\frac{1}{2}
\epsilon \:\varepsilon_{\mu\nu}x_\nu{\biggr ]}^2$ being the
Hamiltonian for
 a particle subjected to the interaction with the external
scalar field in a (2+1)-dimensional spacetime, and the proper time
playing the role of a time coordinate. One has started with a
scalar field theory in a Euclidean 2-dimensional spacetime and now
one has found an effective theory for particles in a 3D spacetime.
It is in this new context that the pair production rate is going
to be calculated. The gain in having the trace in the given form
is that it can be written as a path integral ${\rm tr}\:{\rm
e}^{-H \cal{T}}=\int [dx] \exp{\biggl [}-\int d {\cal{T}} \:L
{\biggr ]}$, where
$L=\frac{1}{2}\dot{x}_\mu\dot{x}_\mu+\frac{1}{2} \epsilon
\:\varepsilon_{\mu\nu}x_\nu\dot{x}_\mu$ is the Lagrangian
associated with our Hamiltonian. Thus,
\begin{equation}
\Gamma=\lim_{T \rightarrow \infty} \frac{2}{T}\:{\rm
Im}\int_{0}^{\infty}\frac{d \cal{T}}{\cal{T}} {\rm
e}^{-\frac{1}{2}m^2 \cal{T}} \:\int [dx] \exp{\biggl
[}\!\!-\!\!\int_{0}^{\cal{T}}\!\!d {\cal{T}}{\biggl (}
\frac{1}{2}\dot{x}_\mu\dot{x}_\mu+\frac{1}{2} \epsilon
\:\varepsilon_{\mu\nu}x_\nu\dot{x}_\mu{\biggr )} {\biggr ]}.
\label{2.7}
\end{equation}
Rescaling the proper time variable, $d {\cal{T}}
\rightarrow\frac{d\tau} {\cal{T}}$, and noticing that the path
integral is over all the paths, $x_\mu(\tau)$, such that
$x_\mu(1)=x_\mu(0)$, one has
\begin{equation}
\Gamma=\lim_{T \rightarrow \infty} \frac{2}{T}\:{\rm Im}\int
[dx]{\rm e}^{-\frac{1}{2} \epsilon \oint \varepsilon_{\mu\nu}x_\nu
dx_\mu}\! \int_{0}^{\infty}\frac{d {\cal{T}}}{\cal{T}} \exp{\biggl
[}\!\!-\!{\biggl (}\frac{1}{2}m^2 {\cal{T}}+\frac{1}{2
\cal{T}}\!\int_{0}^{1}\!\!d\tau \dot{x}_\mu\dot{x}_\mu{\biggr
)}{\biggr ]}\!. \label{2.8}
\end{equation}
The $\cal{T}$ integral can be calculated expanding the function
about the stationary point ${\cal{T}}_0^2=\frac{\int_{0}^{1}
d\tau\dot{x}^2}{m^2}$:
\begin{eqnarray}
\int\frac{d \cal{T}}{\cal{T}}{\rm e}^{-f(\cal{T})}\sim {\rm
e}^{-f({\cal{T}}_0)}\frac{1}{{\cal{T}}_0}\sqrt{\frac{\pi}
{\frac{1}{2}f''({\cal{T}}_0)}}\sim
 {\rm e}^{-m\sqrt{\int_{0}^{1}
d\tau\dot{x}^2}}\frac{1}{m}\sqrt{\frac{2\pi}{{\cal{T}}_0}}\:.
\label{2.9}
\end{eqnarray}
Then (\ref{2.8}) can be written as
\begin{eqnarray}
\Gamma = \lim_{T \rightarrow \infty}
\frac{1}{T}\:\frac{2}{m}\sqrt{\frac{2\pi}{{\cal{T}}_0}}{\rm
Im}\int [dx]{\rm e}^{-S_{\rm Euc}[x_\mu(\tau)]} \:, \label{2.10}
\end{eqnarray}
where $S_{\rm Euc}=m\sqrt{\int_{0}^{1}
d\tau\dot{x}_\mu\dot{x}_\mu}+\frac{1}{2} \epsilon \oint
\varepsilon_{\mu\nu}x_\nu dx_\mu$. This integral can be solved
using the instanton method. Stationarizing the action, one gets
the equation of motion in the (2+1)D spacetime
\begin{eqnarray}
\frac{m \ddot{x}_\mu(\tau')}{\sqrt{\int_{0}^{1}
d\tau\dot{x}^2}}=-\epsilon
\:\varepsilon_{\mu\nu}\dot{x}_\nu(\tau')\:;\hspace{.1in}{\rm
with}\:\:\mu=1,2\:\:{\rm and}\:\:\dot{x}_\mu=\frac{d
x_\mu}{d\tau}\:. \label{2.11}
\end{eqnarray}
The instanton, $x_\mu^{\rm cl}(\tau)$, i.e., the solution of the
Euclidean equation of motion that obeys the boundary conditions
$x_\mu(\tau=1)=x_\mu(\tau=0)$ is
\begin{eqnarray}
x_\mu^{\rm cl}(\tau)=R(\cos2 \pi \tau, \sin 2 \pi
\tau)\:;\:\:\:\:\:{\rm with}\:\:\:R=\frac{m}{\epsilon}\:.
\label{2.12}
\end{eqnarray}
The instanton represents a particle describing a loop of radius
$R$ in the plane defined by the time $x_2$ and by the direction
$x_1$. The loop is
 a thin wall that separates the true vacuum located inside the loop
 from the false vacuum outside.

The Euclidean action of the instanton is given by $S_0=S_{\rm
Euc}[x_\mu^{\rm cl}(\tau)]= m 2 \pi R-\epsilon\pi R^2$. The first
term is the rest energy of the particle times the orbital length
and the second term represents the interaction of the particle
with the external scalar field. The loop radius, $R=m/ \epsilon$,
stationarizies the instanton action. The action is then $S_0=\pi
m^2/ \epsilon$.

The second order variation operator is given by
\begin{eqnarray}
\hspace{-.2in}M_{\mu\nu} \!\!\!\!&\equiv&\!\!\!\! \frac{\delta^2
S}{\delta x_\nu(\tau') \delta x_\mu(\tau)} {\biggr |}_{x^{\rm
cl}}=\nonumber \\
&=&\!\!\!\! {\biggl [}-{\biggl
(}\frac{m\delta_{\mu\nu}}{\sqrt{\int_{0}^{1}
d\tau\dot{x}^2}}\frac{d^2}{d\tau^2} +\epsilon
\:\varepsilon_{\mu\nu}\frac{d }{d\tau}{\biggr
)}\delta(\tau-\tau')- \frac{m \ddot{x}_{\mu}(\tau)
\ddot{x}_{\nu}(\tau')}{{\bigl [}\int_{0}^{1} d\tau\dot{x}^2{\bigr
]}^{3/2}} {\biggr ]}_{x^{\rm cl}}
=\nonumber \\
&=&\!\!\!\! -{\biggl
[}\frac{\epsilon}{2\pi}\delta_{\mu\nu}\frac{d^2}{d\tau^2}
+\epsilon \:\varepsilon_{\mu\nu}\frac{d }{d\tau}{\biggr
]}\delta(\tau-\tau')- \frac{2\pi \epsilon x_{\mu}^{\rm cl}(\tau)
x_{\nu}^{\rm cl}(\tau')}{R^2}\:. \label{2.14}
\end{eqnarray}
The eigenvectors $\eta_\mu^n$, and the eigenvalues  $\lambda_n$,
associated with the operator $M_{\mu\nu}$ are such that
\begin{equation}
M_{\mu\nu}\:\eta_\nu^n(\tau')=\lambda_n
\:\eta_\mu^n(\tau')\:\delta(\tau-\tau')\:.
\end{equation}
From this one concludes that:
\newline
${\bf{(i)}}$ the positive eigenmodes are:
\newline
$(\cos 2n\pi \tau, \sin 2n\pi \tau)$ and $(\sin 2n\pi \tau, -\cos
2n\pi \tau)$ with $\lambda_n=2\pi \epsilon(n^2-n)$,  $n=2,3,4...$;
\newline
$(\sin 2n\pi \tau, \cos 2n\pi \tau)$ and $(\cos 2n\pi \tau, -\sin
2n\pi \tau)$ with $\lambda_n=2\pi \epsilon(n^2+n)$, $n=1,2,3...$;
\newline ${\bf{(ii)}}$ there are two zero-modes associated with the
translation of the loop along the $x_1$ and $x_2$ directions: $(1,
0)$ and $(0, 1)$ with $\lambda=0$;
\newline ${\bf{(iii)}}$ there is a zero-mode associated with the
translation along the proper time, $\tau$: $(\sin 2\pi \tau, -\cos
2\pi \tau)=-\frac{\dot{x}_\mu^{\rm cl}}{2\pi R}$ with $\lambda=0$;
\newline ${\bf{(iv)}}$ there is a single negative mode associated
to the change of the loop radius $R$: $(\cos 2\pi \tau, \sin 2\pi
\tau)=\frac{x_\mu^{\rm cl}}{R}$  with $\lambda_{-}=-2\pi
\epsilon$.

\begin{figure}[H]
\centering
\includegraphics[height=4cm]{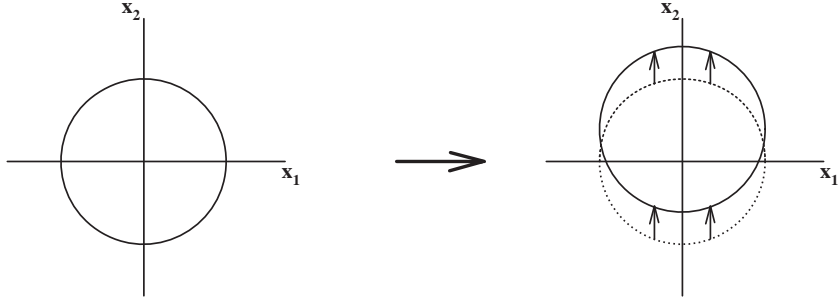}
   \caption{\label{Fig kink_translation}
The zero-mode $\eta_\mu^n(\tau)=(0,1)$ is associated with the
translation of the loop along the $x_2$ direction: $x_\mu^{\rm
cl}(\tau) \rightarrow x_\mu^{\rm cl}(\tau)+ \alpha \,
\eta_\mu^n(\tau)
 = R(\cos2 \pi \tau, \sin 2 \pi \tau)+\alpha (0,1)$
 }
\end{figure}

\begin{figure}[H]
\centering
\includegraphics[height=4cm]{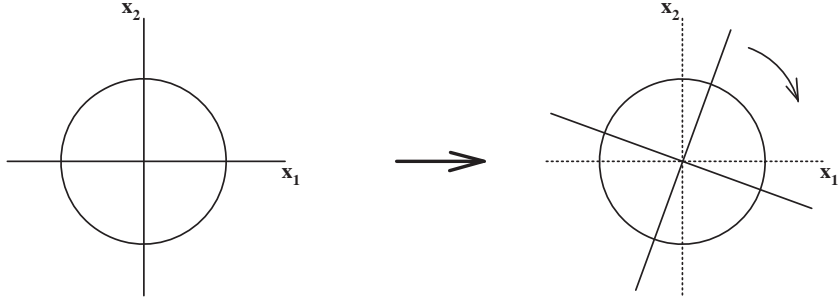}
   \caption{\label{Fig kink_proper time}
The zero-mode $\eta_\mu^n(\tau)=-\dot{x}_\mu^{\rm cl}/(2\pi R)$ is
associated with the translation of the loop along the proper time
$\tau$:
 $x_\mu^{\rm cl}(\tau) \rightarrow x_\mu^{\rm cl}(\tau)+
\alpha \, \eta_\mu^n(\tau) =
 x_\mu^{\rm cl}(\tau)+\alpha \dot{x}_\mu^{\rm cl}$
 }
\end{figure}

\begin{figure}[H]
\centering
\includegraphics[height=4cm]{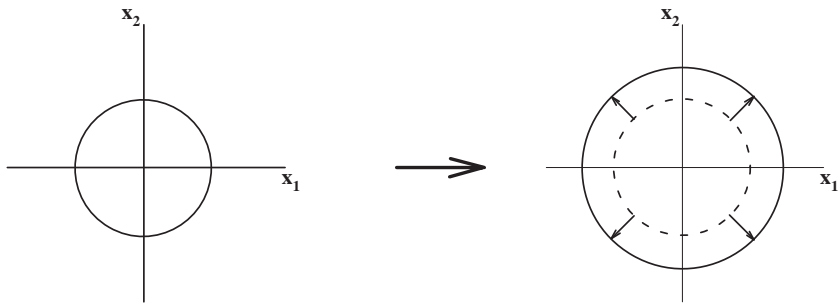}
   \caption{\label{Fig kink_negative mode}
The zero-mode $\eta_\mu^n(\tau)=x_\mu^{\rm cl}/R$ is associated
with the change of the loop radius $R$:
 $x_\mu^{\rm cl}(\tau) \rightarrow x_\mu^{\rm cl}(\tau)+
\alpha \, \eta_\mu^n(\tau) = (1+ \alpha)
 x_\mu^{\rm cl}(\tau)$
 }
\end{figure}

Now, we consider small fluctuations about the instanton, i.e., we
do $x_\mu(\tau)=x_\mu^{\rm cl}(\tau)+\eta_\mu(\tau)$. The
Euclidean action is expanded to second order so that the path
integral (\ref{2.10}) can be approximated by
\begin{equation}
\Gamma \simeq \lim_{T \rightarrow \infty}
\frac{1}{T}\frac{2}{m}\sqrt{\frac{2\pi}{{\cal{T}}_0}}\: {\rm
e}^{-S_0}\:{\rm Im} \int [d\eta(\tau)]\exp
{\biggl[}-\frac{1}{2}\int d\tau d\tau'\,\eta_\mu(\tau)\:
M_{\mu\nu} \:\eta_\nu(\tau'){\biggr]}. \label{2.20}
\end{equation}
The path integral in equation (\ref{2.20}) is the one-loop factor
and is given by ${\cal N}{\bigl (}{\rm Det}M{\bigr )}
^{-\frac{1}{2}} ={\cal N} \prod \,(\lambda_n)^{-\frac{1}{2}}$,
where $\lambda_n$ are the eigenvalues of $M_{\mu\nu}$ and ${\cal
N}$ is a normalization factor that will not be needed. To overcome
the problem that arises from having an infinite product of
eigenvalues, one compares our system with the free particle system
\begin{eqnarray}
\int [d\eta]\exp {\biggl
[}\!\!\!\!\!\!&-&\!\!\!\!\frac{1}{2}\int\! d\tau
d\tau'\,\eta_\mu(\tau)\: M_{\mu\nu} \:\eta_\nu(\tau'){\biggr ]}
= \nonumber \\
&=& \int [d\eta]\exp \!{\biggl [}\!\!-\frac{1}{2}\int\! d\tau
d\tau'\,\eta_\mu(\tau)\:M^0_{\mu\nu} \:\eta_\nu(\tau'){\biggr ]}
\frac{\prod \,(\lambda_n)^{-\frac{1}{2}}}{\prod
\,(\lambda'_n)^{-\frac{1}{2}}}\:, \label{2.21}
\end{eqnarray}
where $M^0_{\mu\nu} =-\frac{1}{{\cal{T}}_0}
\delta_{\mu\nu}\frac{d^2}{d\tau^2} \delta(\tau-\tau')$ is the
second variation operator of the free system with eigenvalues
$\lambda'_n =2 \pi \epsilon n^2\:,\:\:\:\:\:n=0,1,2,3...$ (each
with multiplicity 4). In equation (\ref{2.21}) the first factor is
the path integral of a free particle in a 3-dimensional Euclidean
spacetime
\begin{equation}
\int [d\eta]\exp \!{\biggl [}\!-\frac{1}{2}\int\! d\tau
d\tau'\,\eta_\mu\:M^0_{\mu\nu} \:\eta_\nu{\biggr ]}=\int
[d\eta]\exp \!{\biggl [}\!-\frac{1}{2{\cal{T}}_0}\int\! d\tau
\,\dot{\eta}_\mu\:\dot{\eta}_\mu{\biggr ]}= \frac{1}{2\pi
{\cal{T}}_0} \,. \label{2.22}
\end{equation}
In the productory, one omits the zero eigenvalues, but one has to
introduce the normalization factor $\frac{||dx^{\rm cl}_\mu/d
\tau||}{||\eta^0_\mu||}\sqrt{\frac{1}{ 2\pi}}=\sqrt{2 \pi}R$
 which is  associated with the proper time eigenvalue.
In addition, associated with the negative eigenvalue one has to
introduce a factor of $1/2$ which accounts for the loops that do
expand. The other half contracts (representing the annihilation of
recently created pairs) and so does not contribute to the creation
rate.  So, the one-loop factor becomes
\begin{equation}
\frac{1}{2\pi {\cal{T}}_0} \frac{\prod
\,(\lambda_n)^{-\frac{1}{2}}} {\prod
\,(\lambda'_n)^{-\frac{1}{2}}}= \frac{1}{2\pi {\cal{T}}_0}
\frac{1}{2}\frac{i}{\sqrt{2 \pi \epsilon}}\:\sqrt{2
\pi}R\:\frac{\prod_{\lambda> 0}
(\lambda_n)^{-\frac{1}{2}}}{\prod_{\lambda'> 0}
\,(\lambda'_n)^{-\frac{1}{2}}}= i\frac{1}{2\pi
{\cal{T}}_0}\frac{1}{2}\:\sqrt{2 \pi \epsilon}\: \sqrt{2 \pi}R\:.
\label{2.23}
\end{equation}
Written like this, the one loop factor accounts only for the
contribution of the instanton centered in $(x_1,x_2)=(0,0)$. The
translational invariance in the $x_1$ and $x_2$ directions
requires that one multiplies (\ref{2.23}) by the spacetime volume
factor$\int dx_2 \int dx_1=TL $, which represents the spacetime
region where the instanton might be localized. So, the correct
one-loop factor is given by
\begin{equation}
\int [d\eta(\tau)]\exp {\biggl[}\!-\frac{1}{2}\int\! d\tau
d\tau'\,\eta_\mu(\tau)\: M_{\mu\nu}
\:\eta_\nu(\tau'){\biggr]}=i\frac{L T}{2\pi
{\cal{T}}_0}\frac{1}{2}\:\sqrt{2 \pi \epsilon}\:\sqrt{2 \pi}R\:.
\label{2.24}
\end{equation}
Putting (\ref{2.24}) into (\ref{2.20}), using
${\cal{T}}_0^2=\frac{\int d\tau \dot{x}^2}{m^2}=\frac{(2 \pi
R)^2}{m^2}\:$, $R=\frac{m}{\epsilon}$ and $S_0=\pi m^2/\epsilon$,
one finally has that the soliton-antisoliton pair production rate
per unit time and length is given by
\begin{eqnarray}
 \Gamma/L = \frac{\epsilon}{2\pi}\:{\rm e}^{-\frac{\pi
m^2}{\epsilon}}\:. \label{2.25}
\end{eqnarray}
We have recovered Stone's exponential factor ${\rm e}^{-\frac{\pi
m^2}{\epsilon}}$ \cite{Stone} as well as the prefactor
$A=\epsilon/2\pi$ of Kiselev and Selivanov
\cite{Kiselev1,Kiselev2} and Voloshin \cite{Voloshin1}.

Note the difference to the 4D problem of Affleck {\it et al}
\cite{Affleck1} and Schwinger \cite{Schwinger}, who have found for
the factor $A$ the value $(eE)^2/(2 \pi)^3$ which is quadratic in
$eE$ and not linear, as in our case. This difference has to do
with the dimensionality of the problems.

It is well known that a one-particle system in 2D can be
transformed straightforwardly to a thin line in 3D and a thin wall
in 4D, where now the mass $m$ of the soliton should be interpreted
as a line density and surface density, respectively.  In fact, a
particle in (1+1)D, as well as an infinite line in (2+1)D, can be
considered as walls as seen from within the intrinsic space
dimension, justifying the use of the name wall for any dimension.
Our calculations apply directly to the domain wall pair creation
problem in any dimension.

\section{\label{sec:Conclusions FalseVacuum}Summary and discussion}

The equation for the loop of radius $R$ in 2-dimensional Euclidean
spacetime is given by $x^2+\tau^2=R^2$, where we have put $x=x_1$
and $\tau=x_2$. One can make an analytical continuation of the
Euclidean time ($\tau$) to the Minkowskian time ($\tau=it$) and
obtain the solution in 2-dimensional Minkowski spacetime
\begin{eqnarray}
x^2-t^2=R^2. \label{3.1}
\end{eqnarray}
At $\tau=t=0$ the system makes a quantum jump and a
soliton-antisoliton pair materializes at $x=\pm R=\pm m/\epsilon$.
After the materialization, the soliton and antisoliton are
accelerated, driving away from each other, as (\ref{3.1}) shows.
To check these statements note first that the energy necessary for
the materialization of the pair at rest is $E=2m$, where $m$ is
the soliton mass. This energy comes from the conversion of false
vacuum into true vacuum. Since $\epsilon$ is the energy difference
per unit length between the two vacua, we conclude that an energy
of value $E=2R \epsilon$ is released when this conversion occurs
in the region ($2R$) within the pair. So, the pair materialization
should occur only when $R$ is such that the energy released is
equal to the rest energy: $2R \epsilon=2m \Rightarrow R=m/
\epsilon$. This value agrees with the one that has been determined
in section \ref{sec:Pair production rate}.

After the materialization the pair is accelerated so that its
energy is now $E=2m/\sqrt{1-v^2}$. Differentiating (\ref{3.1}), we
get the velocity $v=\sqrt{1-R^2/x^2}$. The energy of the pair is
then given by $E=2\frac{m}{R}|x|=\epsilon\: 2|x|$. Notice now that
$\epsilon\: 2|x|$ is the energy released in the conversion of
false vacuum into true vacuum. So, after pair creation, all the
energy released in the conversion between the two vacua is used to
accelerate the soliton-antisoliton pair.

This discussion agrees with the interpretation of the process as
being the false vacuum decaying to the true vacuum plus a creation
of a soliton-antisoliton pair. It also justifies the presence of
the interaction term $\epsilon\:\varepsilon_{\mu\nu}x_\nu\psi$
present in the covariant derivative of the proposed effective
one-loop action, (\ref{1.4}), since $\epsilon \,x$ is the energy
released in the decay and responsible for the creation and
acceleration of the pair.

With the proposed effective one-loop action (\ref{1.4}) we have
recovered Stone's exponential factor $S_0$ \cite{Stone} of the
pair creation rate in (\ref{0.1}), and the prefactor $A$ of
Kiselev and Selivanov \cite{Kiselev1,Kiselev2} and Voloshin
\cite{Voloshin1}. In the proposed  effective one-loop action the
soliton charge is treated no longer as a topological charge but as
a Noether charge. Such an interchange between the topological and
the Noether charges was already present in \cite{Coleman1,Olive}.

The problem of false vacuum decay coupled to gravity has been
introduced in \cite{Luccia} and also discussed in
\cite{Garriga,Lavr}. With the proposed effective one-loop action
(\ref{1.4}) it would be interesting to further analyze this
problem.

%% file: Chapter8.tex
\thispagestyle{empty} \setcounter{minitocdepth}{1}
\chapter[Pair creation of black holes on a cosmic string background]
{\Large{Pair creation of black holes on a cosmic string
background}} \label{chap:Pair creation}
 \lhead[]{\fancyplain{}{\bfseries Chapter \thechapter. \leftmark}}
 \rhead[\fancyplain{}{\bfseries \rightmark}]{}
  \minitoc \thispagestyle{empty}
\renewcommand{\thepage}{\arabic{page}}

\addtocontents{lof}{\textbf{Chapter Figures \thechapter}\\}


A process that allows the formation of black holes (even with
Plank sizes) is the gravitational analogue of the Schwinger
quantum process of pair creation of particles in an external
electric field. This gravitational black hole pair creation
process has first proposed by Gibbons (1986). In order to turn the
pair of virtual black holes real one needs a background field that
provides the energy needed to materialize the pair, and that
provides the force necessary to accelerate away the black holes
once they are created. This background field can be: (i) an
external electromagnetic field with its Lorentz force
\cite{Gibbons-book}-\cite{HawkRoss}, (ii) the positive
cosmological constant $\Lambda$, or inflation
\cite{HawkRoss}-\cite{BoussoDil}, (iii) a cosmic string with its
tension \cite{HawkRoss-string}-\cite{Emparan}, (iv) a domain wall
with its gravitational repulsive energy
\cite{CaldChamGibb}-\cite{MannAdS}. One can also have a
combination of the above fields, for example, a process involving
cosmic string breaking in a background magnetic field
\cite{Empar-string}, or a scenario in which a cosmic string breaks
in a cosmological background \cite{OscLem-PCAdS,OscLem-PCdS}. We
have already made a historical overview of these processes in
section \ref{sec:Pair creation BHs introduction}, and we will now
analyze some of these processes in great detail. To study these
processes we must have exact solutions of the Einstein equations
that describe the pair of uniformly accelerated black holes in the
external field, after they are created. These solutions, the
C-metric and Ernst solution, exist and we have already studied
them in chapters \ref{chap:PairAccBH} and \ref{chap:Extremal
Limits}. Finally, an important process that accompanies the
production of the black hole pair  is the emission of radiation.
An estimate for the amount of gravitational radiation released
during the pair creation period will be explicitly computed in
chapter \ref{chap:Grav Radiation}.

The particular process we are going to study in this chapter is
the quantum process in which a cosmic string breaks, and a pair of
black holes is created at the ends of the string (see Fig.
\ref{Fig string_cut_pc_introd}). In a flat background this process
has been analyzed by Hawking and Ross \cite{HawkRoss-string}, with
the support provided by
\cite{DougHorKastTras,AchGregKui,GregHind}. We will discuss the
main results of \cite{HawkRoss-string} in section \ref{sec:Pair
Creation flat}, including the explicit values of the pair creation
rates which have not been written in \cite{HawkRoss-string}. In
section \ref{sec:Pair Creation AdS}, we will discuss in detail the
same process but in an AdS background, following your work
\cite{OscLem-PCAdS}. Finally, in section \ref{sec:Pair Creation
dS}, we will analyze the black hole pair creation probability when
a string breaks in an dS background, following your work
\cite{OscLem-PCdS}.

The energy to materialize and accelerate the black holes comes
from the strings' tension. In the dS case the cosmological
background acceleration also makes a positive contribution to the
process, while in the AdS case the cosmological background
acceleration contributes negatively. Thus, pair creation of black
holes in a dS background is possible even when there is no string,
and this process has been analyzed in
\cite{MelMos,Rom,MannRoss,BoussoHawk,VolkovWipf,BooMann}.  Due to
the negative cosmological background contribution, in the AdS
case, pair creation of black holes is possible only when the
acceleration provided by the strings tension is higher than
$\sqrt{|\Lambda|/3}$.

 We remark that in principle our explicit values for the pair creation rates
\cite{OscLem-PCAdS,OscLem-PCdS} also apply to the process of pair
creation in an external electromagnetic field, with the
acceleration being provided in this case by the Lorentz force
instead of being furnished by the string tension. There is no
Ernst solution in a cosmological constant background, and thus we
cannot discuss analytically the process. However, physically we
could in principle consider an external electromagnetic field that
supplies the same energy and acceleration as our strings and, from
the results of the $\Lambda=0$ case (where the pair creation rates
in the string and electromagnetic cases agree), we expect that the
results found in \cite{OscLem-PCdS,OscLem-PCAdS} do not depend on
whether the energy is being provided by an external
electromagnetic field or by strings.

\section{\label{sec:Pair Creation AdS}Pair creation of anti-de Sitter
black holes on a cosmic string background}

In this section we want to analyze the process in which a cosmic
string breaks and a pair of black holes is produced at the ends of
the string, in an anti-de Sitter (AdS) background ($\Lambda<0$).
Therefore, the energy to materialize and accelerate the pair comes
from the strings' tension. In an AdS background this will be the
only study done in the process of production of a pair of
correlated black holes with spherical topology.  The instantons
for this process can be constructed by analytically continuing the
AdS C-metric found in \cite{PlebDem} and analyzed in detail in
\cite{EHM1}-\cite{OscLem_AdS-C}. Contrary to the $\Lambda=0$
\cite{KW} and $\Lambda>0$ \cite{OscLem_dS-C} cases, the AdS
describes a pair of accelerated black holes only when the
acceleration supplied by the strings is greater than
$\sqrt{|\Lambda|/3}$ \cite{OscLem_AdS-C}. Hence we expect that
pair creation of black holes in an AdS background is possible only
when $A>\sqrt{|\Lambda|/3}$. We will confirm this expectation. The
quantum production of uncorrelated AdS black holes has been
studied in \cite{WuAdS}, and the pair creation process of
correlated topological AdS black holes (with hyperbolic topology)
has been analyzed in \cite{MannAdS} in a domain wall background.

The plan of this chapter is as follows. The AdS C-metric
represents two accelerating black holes in an AdS background (see
previous section \ref{sec:AdS C-metric}), and  in section
\ref{sec:AdS C-inst} we construct, from the AdS C-metric, the
regular instantons that describe the pair creation process. We
find an instanton that mediates pair creation of nonextreme black
holes and other that mediates the production of extreme black
holes. Then, in section \ref{sec:Calc-I AdS}, we explicitly
evaluate the pair creation rate for each one of the cases
discussed in section \ref{sec:AdS C-inst}.

\subsection{\label{sec:AdS C-inst} The A\lowercase{d}S C-metric
instantons}

The AdS C-metric has been discussed in detail in section
\ref{sec:AdS C-metric}. When $A>\sqrt{|\Lambda|/3}$, and only in
this case, the AdS C-metric describes a pair of uniformly
accelerated black holes in an anti-de Sitter background, with the
acceleration being provided by two strings, from each one of the
black holes towards infinity, that pulls them away. Since we are
interested in black hole pair creation, onwards we will deal only
with the $A>\sqrt{|\Lambda|/3}$ case. The presence of the string
is associated to the conical singularity that exists in the south
pole of the AdS C-metric (see subsections \ref{sec:ConSing AdS}
and \ref{sec:PI.2-BH AdS}).

Before we proceed, let us refresh some basic properties that will
be really needed later. The AdS C-metric (\ref{C-metric
AdS})-(\ref{F-el-Lorentz}) has a curvature singularity at
$y=+\infty$ where the matter source is and, in the Lorentzian
sector, $y$ must belong to the range $-x\leq y <+\infty$. The
point $y=-x$ corresponds to a point that is infinitely far away
from the curvature singularity, thus as $y$ increases we approach
the curvature singularity and $y+x$ is the inverse of a radial
coordinate. The south pole, $x=x_\mathrm{s}$, points towards the
infinity, while the north pole points towards the other black
hole. At most, ${\cal F}(y)$ can have four real zeros which we
label in ascending order by $y'_A<0<y_A\leq y_+ \leq y_-$. The
roots $y_-$ and $y_+$ are, respectively, the inner and outer
charged black hole horizons, and $y_A$ and $y'_A$ are acceleration
horizons. Later on it will be crucial to note that the number and
nature of the horizons crossed by an observer that travels into
the black hole singularity depends on the angular direction $x$
that he is following. This peculiar feature is due to the lower
restriction on the value of $y$ ($-x\leq y$). Thus, for
$A>\sqrt{|\Lambda|/3}$, we have to consider separately five
distinct sets of angular directions, namely (a) $x_\mathrm{s}\leq
x <-y_A$,  (b) $x =-y_A$, (c)
 $ -y_A < x <-y'_A$,  (d) $x=-y'_A$ and (e)
 $-y'_A < x \leq x_\mathrm{n}$.
For example, when the observer is travelling towards the the black
hole singularity following an angular direction in the vicinity of
the south pole [case (a)] we will cross only the outer ($y_+$) and
inner ($y_-$) black hole horizons. When he does this trip
following an angular direction in the vicinity of the equator
[case (c)], he crosses the acceleration horizon $y_A$ before
passing through the black hole horizons $y_+$ and $y_-$. If this
trip is done following an angular direction in the vicinity of the
north pole [case (e)] we will cross two accelerations horizons,
$y'_A$ and $y_A$ and then the black hole horizons $y_+$ and $y_-$.
The angular coordinate $x$ belongs to the range
$[x_\mathrm{s},x_\mathrm{n}]$ for which ${\cal G}(x)\geq 0$. By
doing this we guarantee that the Euclidean metric has the correct
signature $(++++)$, and that the angular surfaces are compact. In
order to avoid a conical singularity in the north pole, the period
of $\phi$ must be given by
\begin{equation}
\Delta \phi=\frac{4 \pi}{|{\cal G}'(x_\mathrm{n})|}\:,
 \label{Period phi PCAdS}
 \end{equation}
and this leaves a conical singularity in the south pole given by
\begin{equation}
 \delta=
 2\pi \left ( 1- \frac{{\cal G}'(x_{\mathrm s})}{ |{\cal G}'(x_\mathrm{n})|
 } \right ) \:,
 \label{conic-sing}
 \end{equation}
that signals the presence of a string with mass density, $\mu
=\frac{1}{4}\left (1- \left | {\cal G}'(x_\mathrm{{s}})/ {\cal
G}'(x_\mathrm{{n}}) \right | \right )$, and with pressure
$p=-\mu<0$.

Now, in order to evaluate the black hole pair creation rate we
need to find the instantons of the theory. I.e., we must look into
the euclidean section of the AdS C-metric and choose only those
euclidean solutions which are regular in a way that will be
explained soon. To obtain the euclidean section of the AdS
C-metric from the lorentzian AdS C-metric we simply introduce an
imaginary time coordinate $\tau=-it$ in
 (\ref{C-metric AdS}), (\ref{F-mag}), and (\ref{F-el-Lorentz}). Then
the gravitational field of the euclidean AdS C-metric is given by
\begin{equation}
 d s^2 = [A(x+y)]^{-2} ({\cal F}d\tau^2+
 {\cal F}^{-1}dy^2+{\cal G}^{-1}dx^2+
 {\cal G}d\phi^2)\:,
 \label{C-metric PCAdS}
 \end{equation}
 with
 \begin{eqnarray}
 & &{\cal F}(y) = -\frac{3A^2-|\Lambda|}{3A^2}
                     +y^2-2mAy^3+q^2A^2y^4, \nonumber \\
 & &{\cal G}(x) = 1-x^2-2mAx^3-q^2 A^2 x^4\:,
 \label{FG PCAdS}
 \end{eqnarray}
and the euclidean Maxwell field in the magnetic case is still
given by (\ref{F-mag}), while in the electric case it is now given
by $F_{\rm el}=-i\,q\, d\tau\wedge dy$.

To have a positive definite euclidean metric we must require that
$y$ belongs to $y_A \leq y \leq y_+$. In general, when $y_+ \neq
y_-$, one then has conical singularities at the horizons $y=y_A$
and $y=y_+$. In order to obtain a regular solution we have to
eliminate the conical singularities at both horizons, ensuring in
this way that the system is in thermal equilibrium. This is
achieved by imposing that the period of $\tau$ is the same for the
two horizons, and is equivalent to requiring that the Hawking
temperature of the two horizons be equal. To eliminate the conical
singularity at $y=y_A$ the period of $\tau$ must be $\beta=2 \pi/
k_A$ (where $k_A$ is the surface gravity of the acceleration
horizon),
\begin{equation}
\beta=\frac{4 \pi}{|{\cal F}'(y_A)|}\:.
 \label{Period tau-yA}
 \end{equation}
 This choice for the period of $\tau$ also eliminates
simultaneously the conical singularity at the outer black hole
horizon, $y_+$, if and only if the parameters of the solution are
such that  the surface gravities of the  black hole and
acceleration horizons are equal ($k_+=k_A$), i.e.
\begin{equation}
 {\cal F}'(y_+)=-{\cal F}'(y_A)\:.
 \label{k+=kA}
 \end{equation}
This condition is satisfied by a regular euclidean solution with
$y_A \neq y_+$ that will be referred to as nonextreme AdS
instanton. This solution requires the presence of an
electromagnetic charge.

We now turn our attention to the case $y_+ = y_-$ (and $y_A\neq
y_+$), which obviously requires the presence of charge. When this
happens the allowed range of $y$ in the euclidean sector is simply
$y_A \leq y < y_+$. This occurs because when $y_+ = y_-$ the
proper distance along spatial directions between $y_A$ and $y_+$
goes to infinity. The point $y_+$ disappears from the $\tau, y$
section which is no longer compact but becomes topologically $S^1
\times {\mathbb{R}}$. Thus, in this case we have a conical
singularity only at $y_A$, and so we obtain a regular euclidean
solution by simply requiring that the period of $\tau$ be equal to
(\ref{Period tau-yA}). We will label this solution by extreme AdS
instanton.

In a de Sitter background there is another $y_+ \neq y_-$
instanton  which satisfies $y_A=y_+$. It is called Nariai
instanton and exists with or without charge. Moreover, in the dS
background, there is also a special solution that satisfies
 $y_A=y_+=y_-$. It is called ultracold instanton. In the AdS
 C-metric case, the counterparts of these dS instantons are of no
 interest for the pair creation process because they are out of
 the allowed range of the angular direction $x$.

Below, we will describe in detail the nonextreme AdS instanton
with $m=q$ and the extreme AdS instanton. These instantons are the
natural AdS C-metric counterparts of the lukewarm dS C and cold dS
C instantons constructed in \cite{OscLem_dS-C}. Thus, these
instantons could also be labelled as lukewarm AdS C and cold AdS C
instantons. These two families of instantons will allow us to
calculate the pair creation rate of accelerated
AdS$-$Reissner-Nordstr\"{o}m black holes in section
\ref{sec:Calc-I AdS}.

As is clear from the above discussion, when the charge vanishes we
have no regular instanton available. Therefore, in the instanton
context, we cannot discuss the pair creation of accelerated
AdS-Schwarzchild black holes. In the dS background the instanton
that describes this process is the neutral Nariai instanton
\cite{GinsPerry,MannRoss,BoussoHawk,VolkovWipf}, whose AdS
counterpart is not well-behaved as we said.

\subsubsection{\label{sec:Lukewarm-inst AdS}The nonextreme AdS
instanton with $\bm{m=q}$}

As we said above, for the nonextreme AdS instanton the
gravitational field is given by (\ref{C-metric}) with the
requirement that ${\cal F}(y)$ satisfies ${\cal F}(y_+)=0={\cal
F}(y_A)$ and ${\cal F}'(y_+)=-{\cal F}'(y_A)$. In this case we can
then write
\begin{eqnarray}
{\cal F}(y)&=&-\left ( \frac{y_A \: y_+}{y_A+y_+} \right )^2 \left
( 1-\frac{y}{y_A} \right ) \left ( 1-\frac{y}{y_+} \right )
\nonumber\\
 & & \times
\left ( 1+\frac{y_A+y_+}{y_A \:y_+}\,y-\frac{y^2}{y_A \:y_+}
\right
 )\:,
 \label{F-luk PCAdS}
 \end{eqnarray}
with
\begin{eqnarray}
& & y_A =  \frac{1-\alpha}{2mA}\,, \:\:\:\:\:\:\:\: y_+ =
\frac{1+\alpha}{2mA}\,, \nonumber\\
 & & {\rm and} \:\:\:
 \alpha = \sqrt{1-\frac{4m}{\sqrt{3}}\sqrt{3A^2-|\Lambda|}} \:.
 \label{yA-luk PCAdS}
 \end{eqnarray}
 The parameters
$A$, $\Lambda$, $m$ and $q$, written as a function of $y_A$ and
$y_+$, are
\begin{eqnarray}
& & \frac{|\Lambda|}{3A^2} =  \left ( \frac{y_A \: y_+}{y_A+y_+}
\right )^2\,, \nonumber \\
& & mA=(y_A+y_+)^{-1}=qA \,.
 \label{zeros-luk PCAdS}
 \end{eqnarray}
Thus, the mass and the charge of the nonextreme AdS instanton are
necessarily equal, $m=q$, as occurs with its flat
\cite{DGKT,DGGH,HawkRoss} and dS counterparts
\cite{MelMos,MannRoss,OscLem_dS-C}. The demand that $\alpha$ is
real requires that
\begin{eqnarray}
  0< m \leq \frac{1}{4} \sqrt{\frac{3}{3A^2-|\Lambda|}}\:,
 \label{mq-luk-v0}
 \end{eqnarray}
and that
\begin{eqnarray}
 A>\sqrt{|\Lambda|/3}\:.
 \label{Amin}
 \end{eqnarray}
Therefore, as already anticipated, the nonextreme AdS instanton is
available only when (\ref{Amin}) is satisfied.

As we said, the allowed range of $y$ in the Euclidean sector is
$y_A \leq y \leq y_+$. Then, the period of $\tau$, (\ref{Period
tau-yA}), that avoids the conical singularity at both horizons is
\begin{equation}
\beta=\frac{8 \pi \,m A}{\alpha(1-\alpha^2)}\,,
 \label{beta-luk PCAdS}
 \end{equation}
and $T=1/\beta$ is the common temperature of the two horizons.

Using the fact that ${\cal G}(x)=|\Lambda|/(3A^2)-{\cal F}(-x)$
[see (\ref{FG PCAdS})] we can write
\begin{eqnarray}
{\cal G}(x) = 1-x^2 \left ( 1+m A\, x \right )^2 \:,
 \label{G-luk PCAdS}
 \end{eqnarray}
and the roots of ${\cal G}(x)$ we are interested in are the south
and north pole (represented, respectively, as $x_\mathrm{s}$ and
$x_\mathrm{n}$ in Fig. \ref{g3 AdS}),
\begin{eqnarray}
& & x_\mathrm{s} =  \frac{-1+\omega_-}{2mA}<0\,, \:\:\:\:\:\:\:\:
x_\mathrm{n}  = \frac{-1+\omega_+}{2mA}>0\,, \nonumber\\
 & & {\rm
with}\:\:\:\: \omega_{\pm} = \sqrt{1\pm 4mA} \:.
 \label{polos-luk PCAdS}
 \end{eqnarray}
When $m$ and $q$ go to zero we have $x_\mathrm{s}\rightarrow -1$
and $x_\mathrm{n}\rightarrow +1$. This is the reason why we
decided to work in between the roots $x_\mathrm{s}$ and
$x_\mathrm{n}$, instead of working in between the roots
$x'_\mathrm{s}$ and $x'_\mathrm{n}$ also represented in Fig.
\ref{g3 AdS}. Indeed, when $m\rightarrow 0$ and $q\rightarrow 0$
these two last roots disappear, and our instanton has no vacuum
counterpart. Now, the requirement that $\omega_-$ is real demands
that $mA<1/4$ [note that $1/(4A)<(1/4) \sqrt{3/(3A^2-|\Lambda|)}$,
see (\ref{mq-luk-v0})]. If this requirement is not fulfilled then
${\cal G}(x)$ has only two real roots that are represented as
$x'_\mathrm{s}$ and $x_\mathrm{n}$ in Fig. \ref{g3 AdS} and, as we
have just said, in this case the solution has no counterpart in
the $m=0$ and $q=0$ case. Therefore we discard the solutions that
satisfy $\frac{1}{4A}\leq m \leq \frac{1}{4}
\sqrt{\frac{3}{3A^2-|\Lambda|}}$, and hereafter when we refer to
the mass of the nonextreme AdS instanton we will be working in the
range
\begin{eqnarray}
  0< m \leq \frac{1}{4A}\:,
 \label{mq-luk PCAdS}
 \end{eqnarray}

The period of $\phi$, (\ref{Period phi}), that avoids the conical
singularity at the north pole (and leaves one at the south pole
responsible for the presence of the string) is
\begin{equation}
\Delta \phi=\frac{8 \pi\,m A}{\omega_+(\omega^2_+ -1)} <2 \pi\:.
 \label{Period phi-luk PCAdS}
 \end{equation}
When $m$ and $q$ go to zero we have $\Delta \phi \rightarrow 2
\pi$ and the conical singularity disappears.

The topology of the nonextreme AdS instanton is
 $S^2 \times S^2-\{ region \}$, where $S^2 \times S^2$ represents
 $0\leq \tau \leq \beta$,
 $y_A \leq y \leq y_+$, $x_\mathrm{s}\leq x \leq x_\mathrm{n}$,
and $0 \leq \phi \leq \Delta \phi$, but we have to remove the
region, $\{ region \}=\{ \{x, y \}: x_\mathrm{s} \leq x \leq -y_A
\:\: \wedge \:\: y+x=0 \}$. The Lorentzian sector of this
nonextreme instanton describes two charged AdS black holes being
accelerated by the strings, so this instanton describes pair
creation of nonextreme black holes with $m=q$.

\subsubsection{\label{sec:Cold-inst AdS}The extreme AdS instanton
 with $\bm{y_+=y_-}$}

The gravitational field of the extreme AdS instanton  is given by
(\ref{C-metric}) with the requirement that the size of the outer
charged black hole horizon $y_+$ is equal to the size of the inner
charged horizon $y_-$. Let us label this degenerated horizon by
$\rho$: $y_+=y_-\equiv \rho$ and $\rho > y_A$.  In this case, the
function ${\cal F}(y)$ can be written as
\begin{eqnarray}
{\cal F}(y)=\frac{\rho^2-3\gamma}{\rho^4}
 (y-y'_A)(y-y_A)(y-\rho)^2\:,
 \label{F-cold-PCads}
 \end{eqnarray}
with
\begin{eqnarray}
\gamma=\frac{3A^2-|\Lambda|}{3A^2}\:, \qquad {\rm and}\:\:\:
A>\sqrt{|\Lambda|/3}\:.
 \label{gamma-PCAdS}
 \end{eqnarray}
Note that, as occurred with the nonextreme AdS instanton, the
extreme AdS instanton is also available only when
$A>\sqrt{|\Lambda|/3}$. The roots $\rho$, $y'_A$ and $y_A$ are
given by
\begin{eqnarray}
& & \rho =\frac{3m}{4q^2A}
 \left ( 1+ \sqrt{1-\frac{8}{9}\frac{q^2}{m^2}} \:\right )
 \:,   \label{zerosy1-cold-PCads} \\
& & y'_A =\frac{\gamma \rho}{\rho^2-3\gamma}
 \left ( 1- \sqrt{\frac{\rho^2-2\gamma}{\gamma}} \:\right )
 \:,   \label{zerosy2-cold-PCads}\\
& & y_A =\frac{\gamma \rho}{\rho^2-3\gamma}
 \left ( 1+ \sqrt{\frac{\rho^2-2\gamma}{\gamma}} \:\right )
 \:.
 \label{zerosy3-cold-PCads}
 \end{eqnarray}
The mass and the charge of the solution are written as a function
of $\rho$ as
\begin{eqnarray}
& & m =\frac{1}{A\rho}
 \left ( 1- \frac{2\gamma}{\rho^2} \right )
 \:,  \nonumber \\
& & q^2 =\frac{1}{A^2\rho^2}
 \left ( 1- \frac{3\gamma}{\rho^2} \right )
 \:,
 \label{mq-PCads}
 \end{eqnarray}
and, for a fixed $A$ and $\Lambda$, the ratio $q/m$ is higher than
$1$. The conditions $\rho > y_A$ and $q^2 > 0$ require that, for
the extreme AdS instanton, the allowed range of $\rho$ is
\begin{eqnarray}
\rho>\sqrt{6\gamma} \:.
 \label{range-gamma-cold-PCads}
\end{eqnarray}
The value of $y_A$ decreases monotonically with $\rho$ and we have
$\sqrt{\gamma}<y_A<\sqrt{6\gamma}$.
 The mass and the charge of the extreme AdS instanton are also monotonically
decreasing functions of $\rho$, and as we come from $\rho=+\infty$
into $\rho=\sqrt{6\gamma}$ we have
\begin{eqnarray}
& &  0< m< \frac{\sqrt{2}}{3}
  \frac{1}{\sqrt{3A^2-|\Lambda|}}\:, \\
& & 0< q < \frac{1}{2}
  \frac{1}{\sqrt{3A^2-|\Lambda|}}\:,
 \label{mq-cold-PCads}
 \end{eqnarray}
so, for a fixed $\Lambda$, as the acceleration parameter $A$ grows
the maximum value of the mass and of the charge decreases
monotonically. For a fixed $\Lambda$ and for a fixed mass below
$\sqrt{2/(9 |\Lambda|)}$, the maximum value of the acceleration is
$\sqrt{2/(27m^2)+|\Lambda|/3}$.

As we have already said, the allowed range of $y$ in the Euclidean
sector is $y_A \leq y < y_+$ and does not include $y=y_+$. Then,
the period of $\tau$, (\ref{Period tau-yA}), that avoids the
conical singularity at the only  horizon of the extreme AdS
instanton is
\begin{equation}
\beta=\frac{2 \pi
\rho^3}{(y_A-\rho)^2\sqrt{\gamma(\rho^2-2\gamma)}}\:,
 \label{beta-cold-PCads}
 \end{equation}
and $T=1/\beta$ is the temperature of the acceleration horizon.

In what concerns the angular sector of the extreme AdS instanton,
${\cal G}(x)$ is given by (\ref{FG PCAdS}), and its only real
zeros are the south and north pole (represented, respectively, as
$x_\mathrm{s}$ and $x_\mathrm{n}$ in Fig. \ref{g3 AdS};
$x'_\mathrm{s}$ and $x'_\mathrm{n}$ also represented in this
figure become complex roots in the extreme case),
\begin{eqnarray}
x_\mathrm{s} &=&  -p + \frac{h}{2}-\frac{m}{2q^2A}<0\,, \nonumber \\
 x_\mathrm{n}  &=& +p +
\frac{h}{2}-\frac{m}{2q^2A}>0 \:,
 \label{polos-cold-PCads}
 \end{eqnarray}
with
\begin{eqnarray}
p &=& \frac{1}{2} \left (  -\frac{s}{3}+\frac{2m^2}{q^4A^2}
  -\frac{1-12 q^2A^2}{3s q^4A^4} -\frac{4}{3q^2A^2}
  + n\right )^{1/2} \:,  \nonumber \\
n &=&  \frac{-m^3+mq^2}{2hq^6 A^3} \:,  \nonumber \\
h &=& \sqrt{\frac{s}{3}+\frac{m^2}{q^4A^2}+\frac{1-12 q^2A^2}{3s
q^4A^4}
-\frac{2}{3q^2A^2}} \:, \nonumber \\
 s &=& \frac{1}{2^{1/3} q^2A^2}\left (
\lambda-\sqrt{\lambda^2-4(1-12 q^2A^2)^3} \right )^{1/3} \:,
\nonumber \\
\lambda &=& 2 - 108 m^2A^2 + 72 q^2A^2\:,
  \label{acess-zeros-ang}
 \end{eqnarray}
where $m$ and $q$ are fixed by (\ref{mq-PCads}), for a given $A$,
$\Lambda$ and $\rho$. When $m\rightarrow 0$ and $q\rightarrow 0$
we have $x_\mathrm{s}\rightarrow -1$ and $x_\mathrm{n}\rightarrow
+1$. The period of $\phi$ that avoids the conical singularity at
the north pole (and leaves one at the south pole responsible for
the presence of the strings) is given by (\ref{Period phi}) with
$x_\mathrm{n}$ defined in (\ref{polos-cold-PCads}).

The topology of the extreme AdS instanton is ${\mathbb{R}}^2
\times S^2-\{ region \}$, where ${\mathbb{R}}^2 \times S^2$
represents $0 \leq \tau \leq \beta$, $y_A \leq y < y_+$,
 $x_\mathrm{s}\leq x \leq x_\mathrm{n}$,
and $0 \leq \phi \leq \Delta \phi$, but we have to remove the
region, $\{ region \}=\{ \{x, y \}: x_\mathrm{s} \leq x \leq -y_A
\:\: \wedge \:\: y+x=0 \}$.  Since $y=y_+=\rho$ is at an infinite
proper distance, the surface $y=y_+=\rho$ is an internal infinity
boundary. The Lorentzian sector of this extreme case describes two
extreme ($y_+=y_-$) charged AdS black holes being accelerated by
the strings, and the extreme AdS instanton describes the pair
creation of these extreme black holes.
\subsubsection{\label{sec:submaximal-inst AdS}The nonextreme AdS
instanton with $\bm{m\neq q}$}

The AdS C-metric instantons studied in the two last subsections,
namely the nonextreme instantons with $m=q$ and the extreme
instantons with $y_+=y_-$,  are saddle point solutions free of
conical singularities both in the $y_+$ and $y_A$ horizons. The
corresponding black holes may then nucleate in the AdS background
when a cosmic string breaks, and we will compute their pair
creation rates in Sec. \ref{sec:Calc-I AdS}. However, these
particular black holes are not the only ones that can be pair
created. Indeed, it has been shown in
\cite{WuSubMax,BoussoHawkSubMax} that Euclidean solutions with
conical singularities may also be used as saddle points for the
pair creation process. These nonextreme instantons have $m\neq q$
and describe pair creation of nonextreme black holes with $m\neq
q$.

\begin{figure}[h]
\centering
\includegraphics[height=13cm]{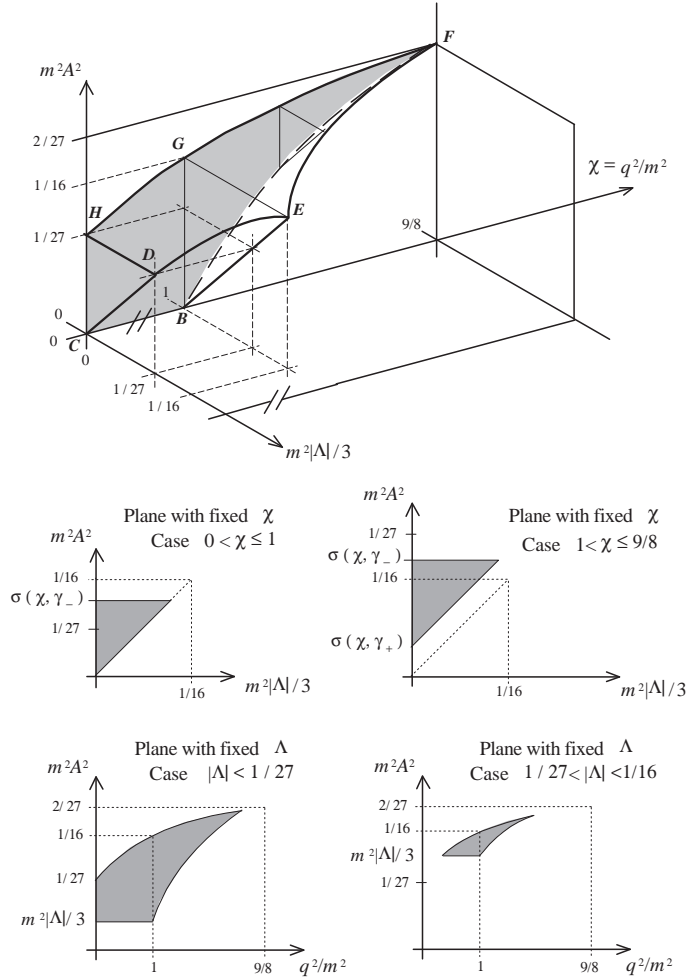}
   \caption{\label{Fig-range pc AdS}
Allowed ranges of the parameters $\Lambda, A, m, \chi\equiv
q^2/m^2$ for which one has a solution representing a pair of
accelerated black holes. The planar surface whose frontier is the
triangle $BEG$ represents the nonextreme AdS instanton with $m=q$.
The curved surface delimitated by the closed line $BEF$ represents
the extreme AdS instanton with $y_+=y_-$ and satisfies $m^2A^2 =
\sigma(\chi,\gamma_+)+m^2|\Lambda|/3$. The curved surface whose
frontier is $DEFGH$ satisfies $m^2A^2=\sigma(\chi,\gamma_-)$ [see
(\ref{rangeG<0})]. The plane surface with boundary given by $BCDE$
satisfies $A=\sqrt{|\Lambda|/3}$. Neutral AdS instantons ($q=0$)
are those that belong to the planar surface with the triangle
boundary $CDH$. The nonextreme AdS instantons with $m\neq q$ are
those whose parameters are in the volume with boundary defined by
$BCDE$, $BEF$, $CBFGH$, $CDH$ and $DEFGH$.
 }
\end{figure}

In what follows we will find the range of parameters for which one
has nonextreme black holes with conical singularities, i.e., with
$m\neq q$. First, when $m\neq 0$ and $q\neq 0$, we require that
$x$ belongs to the interval $[x_\mathrm{s},x_\mathrm{n}]$
(sketched in Fig. \ref{g3 AdS}) for which the charged solutions
are in the same sector of the $m=0$ and $q=0$ solutions. Defining
\begin{eqnarray}
& & \chi \equiv \frac{q^2}{m^2}\:,  \;\;\; 0<\chi\leq \frac{9}{8}
\:, \;\;\;\;\;\gamma_{\pm} \equiv
1 \pm \sqrt{1-\frac{8}{9}\chi} \:, \nonumber \\
 & &
\sigma(\chi,\gamma_{\pm})=
\frac{(4\chi)^2(3\gamma_{\pm})^2-8\chi(3\gamma_{\pm})^3+\chi(3\gamma_{\pm})^4}
{(4\chi)^4} \:,  \nonumber \\
 \label{beta dS-PCads}
\end{eqnarray}
the above requirement is fulfilled by the parameter range
\begin{equation}
m^2A^2<\sigma(\chi,\gamma_-)\:,
 \label{rangeG<0}
 \end{equation}
for which ${\cal G}(x=x_0)<0$, with
$x_0=-\frac{3\gamma_-}{4\chi}\frac{1}{mA}$ being the less negative
$x$ where the derivative of ${\cal G}(x)$ vanishes. Second, in
order to insure that one has a nonextreme solution we must require
that
\begin{equation}
m^2A^2 > \sigma(\chi,\gamma_+)+m^2|\Lambda|/3\:,
 \label{rangeF>0}
 \end{equation}
 for which ${\cal F}(y=y_0)<0$,
with $y_0=\frac{3\gamma_+}{4\chi}\frac{1}{mA}$ being the point in
between $y_+$ and $y_-$ where the derivative of ${\cal F}(y)$
vanishes. We have $\sigma(\chi,\gamma_-)>\sigma(\chi,\gamma_+)$
except at $\chi=9/8$ where these two functions are equal;
$\sigma(\chi,\gamma_-)$ is always positive; and
$\sigma(\chi,\gamma_+)<0$ for $0<\chi<1$ and
$\sigma(\chi,\gamma_+)>0$ for $1<\chi\leq 9/8$. The nonextreme
black holes with conical singularities are those that satisfy
(\ref{rangeG<0}), (\ref{rangeF>0}) and $A>\sqrt{|\Lambda|/3}$. The
range of parameters of these nonextreme black holes with $m\neq q$
are sketched in Fig. \ref{Fig-range pc AdS}.

In order to compute the pair creation rate of the nonextreme black
holes with $m\neq q$, we will need the relation between the
parameters $A$, $\Lambda$, $m$, $q$, and the horizons $y_A$, $y_+$
and $y_-$. In general, for a nonextreme solution with horizons
$y_A<y_+<y_-$, one has
\begin{eqnarray}
{\cal F}(y)= -\frac{1}{d}(y-y_A)(y-y_+)(y-y_-)(ay+b) \:,
 \label{F-nonext-sub-max pcAdS}
 \end{eqnarray}
with
\begin{eqnarray}
d\!\!&=&\!\! y_A y_+ y_- (y_A +y_+ +y_-) +(y_A y_+ +y_A y_- +y_+
y_-)^2 \nonumber \\
 a\!\!&=&\!\!
 \left ( y_A y_+ +y_A y_- +y_+y_- \right )  \nonumber \\
 b\!\!&=&\!\! y_A y_+y_-\:.
 \label{F-nonext-sub-max aux pcAdS}
 \end{eqnarray}
The parameters $A$, $\Lambda$, $m$ and $q$ can be expressed as a
function of $y_A$, $y_+$ and $y_-$ by
\begin{eqnarray}
\frac{|\Lambda|}{3A^2}\!\!&=&\!\! 1-d^{-2}(y_A y_+y_-)^2 \nonumber \\
 q^2A^2\!\!&=&\!\! d^{-1}(y_A y_+ +y_A y_-
+y_+ y_-) \nonumber \\
 mA\!\!&=&\!\! (2\,\sigma)^{-1}(y_A +y_+)(y_A +y_-)(y_+ +y_-) \nonumber \\
 \sigma\!\!&=&\!\! y_A^2 y_+y_- +y_A y_+^2 y_- +y_A y_+ y_-^2+ (y_A y_+)^2 \nonumber \\
  & & +(y_A y_-)^2 +(y_+ y_-)^2 \:.
 \label{relation parameters nonext-sub-max pcAdS}
 \end{eqnarray}
The allowed values of parameters $m$ and $q$ are those contained
in the interior region sketched in Fig. \ref{Fig-range pc AdS}.

The topology of these nonextreme AdS instantons with $m\neq q$ is
 $S^2 \times S^2-\{ region \}$, where $S^2 \times S^2$ represents
 $0\leq \tau \leq \beta$,
 $y_A \leq y \leq y_+$, $x_\mathrm{s}\leq x \leq x_\mathrm{n}$,
and $0 \leq \phi \leq \Delta \phi$, but we have to remove the
region, $\{ region \}=\{ \{x, y \}: x_\mathrm{s} \leq x \leq -y_A
\:\: \wedge \:\: y+x=0 \}$. These instantons describe the pair
creation of nonextreme black holes with $m \neq q$.

\subsubsection{\label{sec:background string AdS}Initial system: AdS
background with a string}

So far, we have described the solution that represents a pair of
black holes accelerated by the strings tension. This solution
describes the evolution of the black hole pair after its creation.
Now, we want to find a solution that represents a string in a AdS
background. This solution will describe the initial system, before
the breaking of the cosmic string that leads to the formation of
the black hole pair. In order to achieve our aim we note that at
spatial infinity the gravitational field of the Euclidean AdS
C-metric reduces to
\begin{eqnarray}
\!\!\!\!\!\!\!\!\! ds^2  = \frac{1}{[A_0(x+y)]^2}{\biggl [}
 -\left ( \frac{|\Lambda|}{3A_0^{\,2}}-1+y^2 \right )dt^2
  +  \frac{dy^2}{\frac{|\Lambda|}{3A_0^{\,2}}-1+y^2}
+\frac{dx^2}{1-x^2}+ (1-x^2)d\phi_0^{\,2} {\biggr ]},
 \label{backg metric}
\end{eqnarray}
and the Maxwell field goes to zero. $A_0$ is a constant that
represents a freedom in the choice of coordinates, and $-1 \leq x
\leq 1$. We want that this metric also describes the solution
before the creation of the black hole pair, i.e., we demand that
it describes a string with its conical deficit in an AdS
background. Now, if we want to maintain the intrinsic properties
of the string during the process we must impose that its mass
density and thus its conical deficit remains constant. After the
pair creation we already know that the conical deficit is given by
(\ref{conic-sing}). Hence, the requirement that the background
solution describes an AdS spacetime with a conical deficit angle
given exactly by (\ref{conic-sing}) leads us to impose that in
(\ref{backg metric}) one has
\begin{eqnarray}
\Delta \phi_0=2\pi-\delta=2\pi \frac{{\cal
G}'(x_\mathrm{s})}{|{\cal G}'(x_\mathrm{n})|} \,.
 \label{delta phi0}
 \end{eqnarray}
The arbitrary parameter $A_0$ can be fixed as a function of $A$ by
imposing a matching between (\ref{C-metric}) and
 (\ref{backg metric}) at large spatial distances. We will do this matching
in the following section.

\subsection{\label{sec:Calc-I AdS}Calculation of the black hole
pair creation rates}
The black hole pair creation rate is given by the path integral
 \begin{eqnarray}
\Gamma(g_{ij},A_{i})=\int d[g_{\mu\nu}]d[A_{\mu}]
 e^{-\left [ I(g_{\mu\nu},A_{\mu})-I_0(g^0_{\mu\nu},A^0_{\mu})\right ]}\:,
 \label{path integral}
 \end{eqnarray}
where $g_{ij}$ and $A_{i}$ are the induced metric and
electromagnetic potential on the boundary
 $\partial {\cal M}$ of a compact manifold ${\cal M}$,
 $d[g_{\mu\nu}]$ is a measure on the space of the metrics
$g_{\mu\nu}$ and $d[A_{\mu}]$ is a measure on the space of Maxwell
field $A_{\mu}$, and $I(g_{\mu\nu},A_{\mu})$ is the Euclidean
action of the instanton that mediates the process. In our case
this action is the Einstein-Maxwell action with a negative
cosmological constant $\Lambda$. The path integral is over all
compact metrics and potentials on manifolds ${\cal M}$ with
boundary $\partial {\cal M}$, which agree with the boundary data
on $\partial {\cal M}$. $I_0(g^0_{\mu\nu},A^0_{\mu})$ is the
action for the background reference spacetime, the AdS background
with the string, specified by $g^0_{\mu\nu}$ and $A^0_{\mu}$ (see
Sec. \ref{sec:background string AdS}). Its presence is required
because it describes the initial system. Moreover, the geometry of
the final system with the black hole pair is noncompact, that is
$I(g_{\mu\nu},A_{\mu})$ diverges. However, the physical action
$I(g_{\mu\nu},A_{\mu})-I_0(g^0_{\mu\nu},A^0_{\mu})$ is finite for
fields $g_{\mu\nu}$, $A_{\mu}$ that approach asymptotically
$g^0_{\mu\nu}$, $A^0_{\mu}$ in an appropriate way \cite{HawkHor}.
Specifically, one fixes a boundary near infinity,
$\Sigma^{\infty}$, and demands that $g_{\mu\nu}$, $A_{\mu}$ and
$g^0_{\mu\nu}$, $A^0_{\mu}$ induce fields on this boundary  that
agree to sufficient order, so that their difference does not
contribute to the action in the limit that $\Sigma^{\infty}$ goes
to infinity \cite{HawkHor}.

In the semiclassical instanton approximation, the dominant
contribution to the path integral (\ref{path integral}) comes from
metrics and Maxwell fields which are near the solutions
(instantons) that extremalize the Euclidean action and satisfy the
boundary conditions. In this approximation, the pair creation rate
of AdS black holes is then given by
\begin{eqnarray}
\Gamma \sim \eta \,e^{-I_{\rm inst}} \:,
 \label{PC-rate-PCads}
 \end{eqnarray}
where $I_{\rm inst}\equiv I-I_0$  includes already the
contribution from the background reference spacetime. $\eta$ is
the one-loop prefactor that accounts for the fluctuations in the
gravitational and matter fields, and its evaluation will not be
considered in this paper (see
\cite{GaratinniOneLoop,VolkovWipf,GrossPerryYaffe,YiPConeLoop}
\cite{GibbHawPer}-\cite{YoungdS} for a treatment of this factor in
some backgrounds).

Hawking and Horowitz \cite{HawkHor} (see also Brown and York
\cite{BrownYork}) have shown that the Euclidean action of the
instanton that mediates pair creation of nonextreme black holes
can be written in the form
 \begin{eqnarray}
I_{\rm inst}=\beta H-\frac{1}{4}
 \left(\Delta {\cal A}_{\rm ac}+{\cal A}_{\rm bh}\right )\:,
 \label{I-Non-Ext}
 \end{eqnarray}
where $\Delta {\cal A}_{\rm ac}$ is the difference in area of the
acceleration horizon between the AdS C-metric and the background,
and ${\cal A}_{\rm bh}$ is the area of the black hole horizon
present in the instanton. The Euclidean action of the instanton
that describes pair creation of extreme black holes can be written
as
\begin{eqnarray}
I_{\rm inst}=\beta H-\frac{1}{4}\Delta {\cal A}_{\rm ac}\:.
 \label{I-Ext}
 \end{eqnarray}
In these relations, $\beta$ is the period of the Euclidean time,
and $H$ represents the Hamiltonian of the system which, for static
solutions, is given by \cite{HawkHor,BrownYork}
\begin{eqnarray}
H\!\!&=&\!\!\int_{\Sigma_t}\!\!d^3x\sqrt{h}\,N{\cal
H}-\frac{1}{8\pi}\int_{\Sigma_t^{\infty}}\!\!\!d^2x\sqrt{\sigma}
\left ( N\, {}^2\!K-N_0\, {}^2\!K_0 \right ). \nonumber \\
 & &
 \label{hamiltonian}
 \end{eqnarray}
 The boundary $\partial M$ consists of an initial and final
spacelike surface, $\Sigma_t$, of constant $t$ with unit normal
$u^{\mu}$ ($u\cdot u=-1$) and with intrinsic metric
$h_{\mu\nu}=g_{\mu\nu}+u_{\mu}u_{\nu}$ plus a timelike surface
near infinity, $\Sigma^{\infty}$, with unit normal $n^{\mu}$
($n\cdot n=1$, and $u\cdot n=0$) and with intrinsic metric
$\sigma_{\mu\nu}=h_{\mu\nu}-n_{\mu}n_{\nu}$. This surface
$\Sigma^{\infty}$ needs not to be at infinity. It can also be at a
black hole horizon or at an internal infinity, but we will
generally label it by $\Sigma^{\infty}$. This surface
$\Sigma^{\infty}$ is foliated by a family of 2-surfaces
$\Sigma_t^{\infty}$ that result from the intersection between
$\Sigma^{\infty}$ and $\Sigma_t$. In (\ref{hamiltonian}), $N$ and
$N_0$ are the lapse functions of the system with the pair and of
the background, respectively, and $N=N_0$ in the boundary near
infinity, $\Sigma^{\infty}$.
 ${\cal H}$ is the hamiltonian constraint that contains
contributions from the gravitational and Maxwell fields, and
vanishes for solutions of the equations of motion. Then, the
Hamiltonian is simply given by the boundary surface term. In this
surface term, ${}^2\!K$ represents the trace of the extrinsic
curvature of the surface imbedded in the AdS C-metric, and
${}^2\!K_0$ is the extrinsic curvature of the surface imbedded in
the background spacetime.

We will now verify that the boundary surface term in the
Hamiltonian (\ref{hamiltonian}) is also zero, and thus the
Hamiltonian makes no contribution to (\ref{I-Non-Ext}) and
(\ref{I-Ext}). We follow the technical procedure applied by
Hawking, Horowitz and Ross \cite{HawHorRoss} in the Ernst solution
and by  Hawking and Ross \cite{HawkRoss-string} in the flat
C-metric. As we said above, one will require that the intrinsic
metric $ds^2_{(\Sigma_t^{\infty})}$ on the boundary
$\Sigma_t^{\infty}$ as embedded in the AdS C-metric agrees (to
sufficient order) with the intrinsic metric on the boundary
$\Sigma_t^{\infty}$ as embedded in the background spacetime, in
order to be sure that one is taking the same near infinity
boundary in the evaluation of the quantities in the two
spacetimes. In the AdS C-metric one takes this boundary to be at
$x+y=\varepsilon_c$, where $\varepsilon_c \ll 1$. The background
reference spacetime is described by (\ref{backg metric}),
subjected to $-1 \leq x \leq 1$, $y \geq -x$ and (\ref{delta
phi0}). In this background spacetime we take the boundary
$\Sigma_t^{\infty}$ to be at $x+y=\varepsilon_0$, where
$\varepsilon_0 \ll 1$.

We are now interested in writing the intrinsic metric in the
boundary $\Sigma_t^{\infty}$ (the 2-surface $t=$ const and
$x+y=\varepsilon$). In the AdS C-metric
 (\ref{C-metric}) one performs the coordinate transformation
\begin{eqnarray}
\phi=\frac{2}{|{\cal G}'(x_\mathrm{n})|}\,\tilde{\phi}\,, \qquad
t=\frac{2}{{\cal F}'(y_A)}\,\tilde{t}\,
 \label{redef phi t}
 \end{eqnarray}
in order that $\Delta \tilde{\phi}=2\pi$, and the analytic
continuation of $\tilde{t}$ has period $2\pi$, i.e., $\Delta
\tilde{\tau}=2\pi$. Furthermore, one takes
\begin{eqnarray}
x=x_\mathrm{s}+\varepsilon_c \chi \,, \qquad
y=-x_\mathrm{s}+\varepsilon_c (1-\chi)\,
 \label{def varepsilon chi C}
 \end{eqnarray}
where $0 \leq \chi \leq 1$. By making the evaluations up to second
order in $\varepsilon_c$ (since higher order terms will not
contribute to the Hamiltonian in the final limit $\varepsilon_c
\rightarrow 0$), the intrinsic metric on the boundary
$\Sigma_t^{\infty}$ is then
\begin{eqnarray}
ds^2_{(\Sigma_t^{\infty})} \sim \frac{2}{A^2\varepsilon_c {\cal
G}'(x_\mathrm{s}) }
 {\biggl [}
\frac{d \chi^2}{2\chi} + \left |
 \frac{{\cal G}'(x_\mathrm{s})}{{\cal G}'(x_\mathrm{n})}
 \right |^2 \!\!\!\left ( 2\chi+ \varepsilon_c
 \frac{{\cal G}''(x_\mathrm{s})}{{\cal G}'(x_\mathrm{s})}\chi^2 \right )
 d \tilde{\phi}^2  {\biggr ]}.
 \label{metric boundary C}
 \end{eqnarray}
Analogously, for the background spacetime (\ref{backg metric}) one
sets
\begin{eqnarray}
x=-1+\varepsilon_0 \chi\, , \qquad y=1+\varepsilon_0 (1-\chi)\,
 \label{def varepsilon chi bg}
 \end{eqnarray}
and the intrinsic metric on the boundary $\Sigma_t^{\infty}$
yields
\begin{eqnarray}
ds^2_{(\Sigma_t^{\infty})}\sim  \frac{1}{A_0^{\,2}\varepsilon_0}
 \left [ \frac{d \chi^2}{2\chi}+ \left ( 2\chi- \varepsilon_0
 \chi^2 \right )
 d\phi_0^{\,2}  \right ]
 \label{metric boundary bg}
 \end{eqnarray}
These two intrinsic metrics on the boundary will agree (up to
second order in $\varepsilon$) as long as we take the period of
$\phi_0$ to be given by (\ref{delta phi0}) and the following
matching conditions are satisfied,
\begin{eqnarray}
\varepsilon_0=- \frac{{\cal G}''(x_\mathrm{s})}{{\cal
G}'(x_\mathrm{s})}\,\varepsilon_c \,,
 \label{match condition epsilon}
 \end{eqnarray}
\begin{eqnarray}
A_0^{\,2}=- \frac{[{\cal G}'(x_\mathrm{s})]^2}{2{\cal
G}''(x_\mathrm{s})}\,A^2  \,.
 \label{match condition A}
 \end{eqnarray}
Note that the Maxwell fields of the two solutions agree trivially
at the near infinity boundary $\Sigma_t^{\infty}$.

In what concerns the lapse function of the AdS C-metric, we
evaluate it with respect to the time coordinate $\tilde{t}$
defined in (\ref{redef phi t}) and, using $[A(x+y)]^{-2}{\cal
F}dt^2=N^2 d\tilde{t}^{\,2}$, we find
\begin{eqnarray}
N\!\!&\sim&\!\!
\sqrt{\frac{|\Lambda|}{3}}\frac{2}{A^2\varepsilon_c {\cal F}'(y_A)
}
 \left ( 1+\frac{1}{2}\frac{3}{|\Lambda|}\left (1-\chi \right )
 A^2\varepsilon_c \,{\cal G}'(x_\mathrm{s})
  \right ) . \nonumber \\
 & &
 \label{lapse function C}
 \end{eqnarray}
Analogously, an evaluation with respect to the time coordinate $t$
defined in (\ref{backg metric}) yields
\begin{eqnarray}
N_0\sim \frac{{\cal G}'(x_\mathrm{s})}{{\cal F}'(y_A)}
\sqrt{\frac{|\Lambda|}{3}}\frac{1}{A_0^{\,2}\varepsilon_0}
 \left ( 1+\frac{3}{|\Lambda|}\left (1-\chi \right )
 A_0^{\,2}\varepsilon_0
  \right ) .
 \label{lapse function bg}
 \end{eqnarray}
Note that these two lapse functions are also matched by the
conditions (\ref{match condition epsilon}) and (\ref{match
condition A}).

The extrinsic curvature to $\Sigma_t^{\infty}$ as embedded in
$\Sigma_t$ is ${}^2\!K_{\mu\nu}=\sigma_{\mu}^{\:\:\:\alpha}
h_{\alpha}^{\:\:\:\beta}\nabla_{\beta}n_{\nu}$ (where
$\nabla_{\beta}$ represents the covariant derivative with respect
 to $g_{\mu\nu}$), and the trace of the extrinsic curvature is
${}^2\!K=g^{\mu\nu}\:{}^2\!K_{\mu\nu}=A\sqrt{{\cal F}(y)}$. The
extrinsic curvature of the boundary embedded in the AdS C-metric
is then
\begin{eqnarray}
{}^2\!K \sim \sqrt{\frac{|\Lambda|}{3}}
 \left ( 1+\frac{1}{2}\frac{3}{|\Lambda|}\left (1-\chi \right )
 A^2\varepsilon_c \,{\cal G}'(x_\mathrm{s})
  \right )\,,
 \label{extrinsic curvat C}
 \end{eqnarray}
while the extrinsic curvature of the boundary embedded in the
background reference spacetime is
\begin{eqnarray}
{}^2\!K_0\sim \sqrt{\frac{|\Lambda|}{3}}
 \left ( 1+\frac{3}{|\Lambda|}\left (1-\chi \right )
 A_0^{\,2}\varepsilon_0 \right ) .
 \label{extrinsic curvat bg}
 \end{eqnarray}

We are now in position to compute the contribution from the
surface boundary term in (\ref{hamiltonian}). The evaluation in
the AdS C-metric yields
\begin{eqnarray}
 \int_{\Sigma_t^{\infty}}\!\!\!d\chi d\tilde{\phi}
\sqrt{\sigma}\,
 N\,{}^2\!K  \sim \frac{8\pi}{|{\cal G}'(x_\mathrm{n})|{\cal F}'(y_A)}
 \frac{|\Lambda|}{3} \frac{1}{(A^2\varepsilon_c)^2}
 \left ( 1-\frac{1}{2}\frac{3}{|\Lambda|}
 A^2\varepsilon_c \,{\cal G}'(x_\mathrm{s})
  \right ) ,
 \label{int NK C}
 \end{eqnarray}
while for the background reference spacetime we have
\begin{eqnarray}
\int_{\Sigma_t^{\infty}}\!\!\!\sqrt{\sigma_0}d\chi d\phi_0
 N_0\,{}^2\!K_0 \sim
 \frac{8\pi \,[{\cal G}'(x_\mathrm{s})]^2}{|{\cal G}'(x_\mathrm{n})|{\cal F}'(y_A)}
 \frac{|\Lambda|}{3}  \frac{1}{(2A_0^{\,2}\varepsilon_0)^2}  \left ( 1-\frac{3}{|\Lambda|}
 A_0^{\,2}\varepsilon_0 \right ) .
 \label{int NK bg}
 \end{eqnarray}
From the matching conditions (\ref{match condition epsilon}) and
(\ref{match condition A}) we conclude that these two boundary
terms are equal. Hence, the surface term and the Hamiltonian
(\ref{hamiltonian}) vanish. The Euclidean action (\ref{I-Non-Ext})
of the nonextreme AdS instanton that mediates pair creation of
nonextreme black holes is then simply
\begin{eqnarray}
I_{\rm nonext}=-\frac{1}{4}
 \left(\Delta {\cal A}_{\rm ac}+{\cal A}_{\rm bh}\right )\:,
 \label{I-Non-Ext2}
 \end{eqnarray}
while the Euclidean action (\ref{I-Non-Ext}) of the extreme AdS
instanton that mediates pair creation of nonextreme black holes is
just given by
\begin{eqnarray}
I_{\rm ext}=-\frac{1}{4}\Delta {\cal A}_{\rm ac}\:.
 \label{I-Ext2}
 \end{eqnarray}
Thus as occurs in the Ernst case, in the flat C-metric case, in
the de Sitter case and in the dS C-metric case, the pair creation
of nonextreme black holes is enhanced relative to the pair
creation of extreme black holes by a factor of $e^{{\cal A}_{\rm
bh}}$.

In the next two subsections we will explicitly compute
(\ref{I-Non-Ext2}) using the results of subsection
\ref{sec:Lukewarm-inst AdS}, and (\ref{I-Ext2}) using the results
of subsection \ref{sec:Cold-inst AdS}.
 In subsection \ref{sec:submaximal-rate AdS} we analyze the pair creation
rate of black holes discussed in subsection
 \ref{sec:submaximal-inst AdS}.
 Remark that the only horizons that contribute with their
areas to (\ref{I-Non-Ext2}) and (\ref{I-Ext2})  are those that
belong to the instanton responsible for the pair creation, i.e,
only those horizons that are in the Euclidean sector of the
instanton will make a contribution.

The domain of validity of our results is the  particle limit,
$mA\ll 1$, for which the  radius of the black hole, $r_+ \sim m$,
is much smaller than the typical distance between the black holes
at the creation moment, $\ell \sim 1/A$ (this value follows from
the Rindler motion $x^2-t^2=1/A^2$ that describes the uniformly
accelerated motion of the black holes).

\subsubsection{\label{sec:Lukewarm-rate AdS}Pair creation rate  in
the nonextreme case with $\bm{m=q}$}

In the nonextreme case, the instanton has two horizons in its
Euclidean section, namely the acceleration horizon at $y=y_A$ and
the black hole horizon at $y=y_+$ (the horizons $y'_A$ and $y_-$
do not belong to the instanton, see Fig. \ref{g3 AdS}). The black
hole horizon covers the whole range of the angular coordinate $x$,
 $x_\mathrm{s} \leq x\leq x_\mathrm{n}$, and its area is
\begin{eqnarray}
\cal{A}_{\rm bh} \!&=&\!  \int_{y=y_+} \!\!\!\!\!\!
\sqrt{g_{xx}g_{\phi\phi}}\: dx \,d\phi  \nonumber \\
&=& \frac{1}{A^2}\int_{\Delta \phi}\!\! d\phi
  \int_{x_\mathrm{s}}^{x_\mathrm{n}} \!\! \frac{dx}{(x+y_+)^2}    \nonumber \\
&=&
 \frac{4\pi} {A^2 |{\cal G}'(x_\mathrm{n})|}\,
 \frac{ x_\mathrm{n}-x_\mathrm{s} }
   {(x_\mathrm{n}+y_+)(x_\mathrm{s}+y_+)}\,,
\label{area BH-luk PCAdS}
 \end{eqnarray}
where $y_+$ is given in (\ref{yA-luk PCAdS}), $x_\mathrm{s}$ and
$x_\mathrm{n}$ are defined by (\ref{polos-luk PCAdS}), and
 $\Delta \phi$ is given by (\ref{Period phi-luk PCAdS}).

  The acceleration horizon, $y_A$ defined in (\ref{yA-luk PCAdS}), of
the nonextreme AdS instanton covers the angular range $-y_A \leq
x\leq x_\mathrm{n}$ (i.e., it is not present in the vicinity of
the south pole), and is noncompact, i.e., its area is infinite. We
have to deal appropriately with this infinity. In order to do so,
we first introduce a boundary at $x=-y_A+\varepsilon_c$
($\varepsilon_c \ll 1$), and compute the area inside of this
boundary, which yields
\begin{eqnarray}
{\cal A}^{c}_{\rm ac} \!&=&\!  \int_{y=y_A} \!\!\!\!\!\!
\sqrt{g_{xx}g_{\phi\phi}}\: dx \,d\phi\nonumber \\
&=& \frac{1}{A^2}\int_{\Delta \phi}\!\! d\phi
  \int_{-y_A+\varepsilon_c}^{x_\mathrm{n}}  \frac{dx}{(x+y_A)^2}    \nonumber \\
&=& - \frac{4\pi} {A^2 |{\cal G}'(x_\mathrm{n})|}\,
 \left ( \frac{1}{ x_\mathrm{n}+y_A}
  -\frac{1}{\varepsilon_c} \right )\,,
\label{area Ac-luk PCAdS}
 \end{eqnarray}
When we let $\varepsilon_c \rightarrow 0$, the term
$1/\varepsilon_c$ diverges, and the acceleration horizon has an
infinite area. This area is renormalized with respect to the area
of the acceleration horizon of the background reference spacetime
(\ref{backg metric}). This background reference spacetime has  an
acceleration horizon at $y=\sqrt{1-|\Lambda|/(3A_0^{\,2})}$ which
is the direct counterpart of the above acceleration horizon of the
nonextreme AdS instanton, and it covers the angular range
$-\sqrt{1-|\Lambda|/(3A_0^{\,2})} \leq x\leq 1$. The area inside a
boundary at $x=-\sqrt{1-|\Lambda|/(3A_0^{\,2})}+\varepsilon_0$ is
\begin{eqnarray}
{\cal A}^{0}_{\rm ac} \!&=&\!
  \int_{y=\sqrt{1-\frac{|\Lambda|}{3A_0^{\,2}} }}
\sqrt{g_{xx}g_{\phi_0 \phi_0}}\: dx \,d\phi_0  \nonumber \\
\!&=&\! \frac{1}{A_0^{\,2} }\int_{\Delta \phi_0}\!\! d\phi_0
  \int_{-\sqrt{1-\frac{|\Lambda|}{3A_0^{\,2}} }+\varepsilon_0}^{1}
  \frac{dx}{\left ( x+\sqrt{1-\frac{|\Lambda|}{3A_0^{\,2}}}\right )^2}
   \nonumber \\
\!&=&\! - \frac{2\pi} {A_0^{\,2}} \frac{{\cal
G}'(x_\mathrm{s})}{|{\cal G}'(x_\mathrm{n})|}\,
 \left ( \frac{1}{1+\sqrt{1-\frac{|\Lambda|}{3A_0^{\,2}}} }
  -\frac{1}{\varepsilon_0} \right )   \,,
  \label{area Ac-luk-bg}
 \end{eqnarray}
where in the last step we have replaced $\Delta \phi_0$ by
(\ref{delta phi0}). When $\varepsilon_0 \rightarrow 0$, the term
$1/\varepsilon_0$ diverges, and thus the area of this background
acceleration horizon is also infinite. Now, we have found that the
intrinsic metrics at the near infinity boundary match together if
(\ref{match condition epsilon}) and (\ref{match condition A}) are
satisfied. Our next task is to verify that these matching
conditions between ($\varepsilon_c, A$) and ($\varepsilon_0, A_0$)
are such that the divergent terms in (\ref{area Ac-luk PCAdS}) and
(\ref{area Ac-luk-bg}) cancel each other. It is straightforward to
show that $\Delta {\cal A}_{\rm ac}={\cal A}^{c}_{\rm ac}-{\cal
A}^{0}_{\rm ac}$ yields a finite value if
\begin{eqnarray}
 2A_0^{\,2}\varepsilon_0= A^2\varepsilon_c {\cal
 G}'(x_\mathrm{s})\,.
  \label{match condition}
 \end{eqnarray}
Thus, the matching conditions (\ref{match condition epsilon}) and
(\ref{match condition A}) satisfy condition
 (\ref{match condition}), i.e., they indeed eliminate the divergencies in
$\Delta {\cal A}_{\rm ac}$. It is worthy to remark that with the
choices (\ref{match condition epsilon}) and (\ref{match condition
A}) the proper lengths of the boundaries $x=-y_A+\varepsilon_c$
and $x=-\sqrt{1-|\Lambda|/(3A_0^{\,2})}+\varepsilon_0$, (given,
respectively, by $l_c=\int \sqrt{g_{\phi\phi}}d\phi$ and $l_0=\int
\sqrt{g_{\phi_0\phi_0}}d\phi_0$) do not match. This is in contrast
with the flat case \cite{HawkRoss-string}, where the choice of the
matching parameters that avoids the infinities in $\Delta {\cal
A}_{\rm ac}$, also leads to $l_c=l_0$. In the AdS case, our main
goal was to remove the infinities in $\Delta {\cal A}_{\rm ac}$.
We have achieved this aim by comparing the appropriate
acceleration horizons in the instanton and in the reference
background. The fact that the matching relations then lead to $l_c
\neq l_0$ is not a problem at all\footnote{This question deserves
a physical interpretation. Both in the flat and AdS cases, the
proper lengths in the C-metric instanton and in the background are
equal, $l_c=l_0$, at the south pole boundaries, $x=x_{\rm s}
+\varepsilon_c$ and $x=-1+\varepsilon_0$, respectively. Now, a
string is present when the ratio between the perimeter of a circle
and its radius is not $2\pi$, i.e., one has a deficit angle. The
fact that for a same radius one has $l_c=l_0$ at the south pole
confirms that the string before and after the pair creation has
the same properties, as expected from the discussion of subsection
\ref{sec:background string AdS}. In the flat case, but not in the
AdS one, the south pole boundary coincides with the acceleration
boundary, i.e., $x=x_{\rm s} +\varepsilon_c\equiv
-y_A+\varepsilon_c$, and this is the reason why in the flat case
one has $l_c=l_0$ at the horizon boundary.}. Replacing (\ref{match
condition epsilon}) and (\ref{match condition A}) in (\ref{area
Ac-luk-bg}) yields for $\Delta {\cal A}_{\rm ac}={\cal A}^{c}_{\rm
ac}-{\cal A}^{0}_{\rm ac}$ the result
\begin{eqnarray}
 \Delta {\cal A}_{\rm ac} =
   -\frac{4\pi} {A^2 |{\cal G}'(x_\mathrm{n})|}\,
 {\biggl (} \frac{1}{ x_\mathrm{n}+y_A}
 + \frac{1}{ 1+\sqrt{1-\frac{|\Lambda|}{3A^2}
\frac{2|{\cal G}''(x_\mathrm{s})|}{[{\cal G}'(x_\mathrm{s})]^2} }}
 \frac{ {\cal G}''(x_\mathrm{s}) }{ {\cal G}'(x_\mathrm{s}) }
  {\biggr )},
  \label{diferenca area-luk PCAdS}
 \end{eqnarray}
where $2|{\cal G}''(x_\mathrm{s})| / [{\cal
G}'(x_\mathrm{s})]^2<1$.

\begin{figure} [H]
\centering
\includegraphics[height=2.3in]{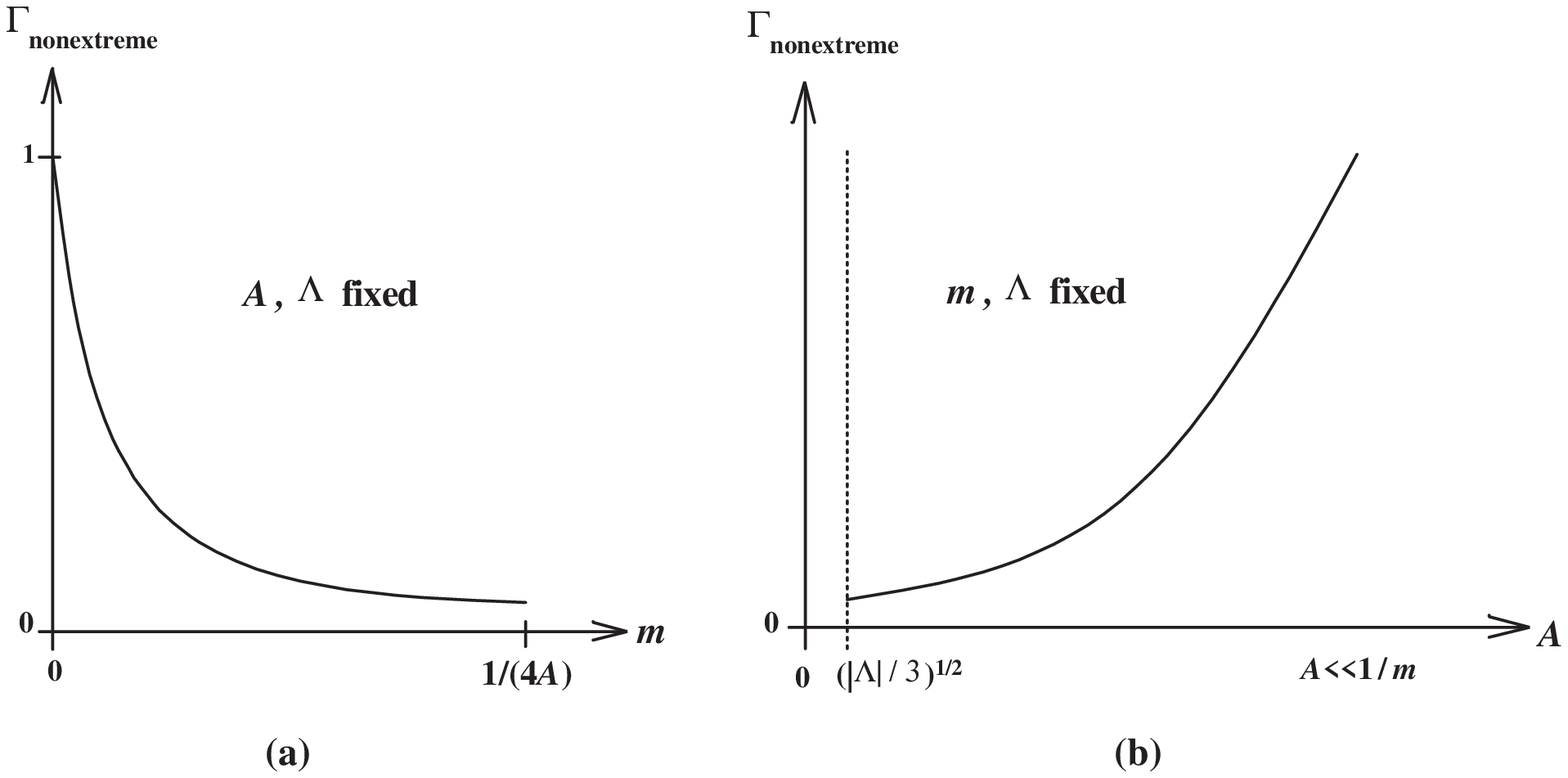}
\caption{\label{lukewarm AdS-fig}
 (a) Plot of the pair creation rate
of nonextreme black holes, $\Gamma_{\rm nonext}$, as a function of
$m$ for a fixed $A$ and $\Lambda$. The range of $m$ is
$0<m<\frac{1}{4m}$.
 (b) Plot of the pair creation rate
of nonextreme black holes, $\Gamma_{\rm nonext}$, as a function of
$A$ for a fixed $m$ and $\Lambda$. The range of $A$ is
$\sqrt{\frac{|\Lambda|}{3}}<A<\frac{1}{4m}$.
 See text of subsection \ref{sec:Lukewarm-rate AdS}.
 }
\end{figure}

Adding (\ref{area BH-luk PCAdS}) and
 (\ref{diferenca area-luk PCAdS}), and using the results of Sec.
\ref{sec:Lukewarm-inst AdS} yields finally the total area of the
nonextreme AdS instanton with $m=q$,
\begin{eqnarray}
& & \!\!\!\!\!\!\!\!\!\!{\cal A}_{\rm bh}^{\rm nonext}+\Delta
{\cal A}_{\rm ac}^{\rm nonext} = \nonumber \\
& & -\frac{16\pi m^2}{\omega_+ (\omega_+^2-1)} {\biggl (}\!
 -\frac{\omega_+ -\omega_-}{(\omega_+ +\alpha)(\omega_- +\alpha)}
       +\frac{1}{\omega_+ -\alpha}
 + \frac{1}{ 1+\sqrt{1-\frac{8|\Lambda|m^2}{3}
\frac{3\omega_-^2-1}{\omega_-^2(1-\omega_-^2)^2} }}
 \frac{1-3\omega_-^2}{\omega_-(1-\omega_-^2)} {\biggr )},
 \nonumber \\
 & &
 \label{area TOTAL-luk PCAdS}
 \end{eqnarray}
 where $\omega_+$ and $\omega_-$ are defined in (\ref{polos-luk PCAdS}),
 $\alpha$ is given by (\ref{yA-luk PCAdS}),
 and condition (\ref{mq-luk PCAdS}) must be satisfied. The pair
creation rate of nonextreme AdS black holes with $m=q$ is then
\begin{eqnarray}
\Gamma_{\rm nonext}\sim e^{\frac{1}{4}\left (
 {\cal A}_{\rm bh}^{\rm nonext}
 +\Delta {\cal A}_{\rm ac}^{\rm nonext} \right )}.
 \label{rate-luk PCAdS}
 \end{eqnarray}
Fixing $A$ and $\Lambda$ one concludes that the pair creation rate
decreases as the mass of the black holes increases (see Fig.
\ref{lukewarm AdS-fig}). Moreover, fixing $m$ and $\Lambda$, in
the domain of validity of our results, $mA\ll 1$ and
$A>\sqrt{|\Lambda|/3}$, as $A$ increases the pair creation rate
increases (see Fig. \ref{lukewarm AdS-fig}). Hence, the general
behavior of the pair creation rate of nonextreme black holes with
$m=q$ in the AdS case is analogous to the corresponding behavior
in the flat case \cite{HawkRoss-string} (see also section
\ref{sec:Pair Creation flat}).

\subsubsection{\label{sec:Cold-rate AdS}Pair creation rate in the
extreme case ($\bm{y_+=y_-}$)}

\begin{figure} [H]
\centering
\includegraphics[height=2.3in]{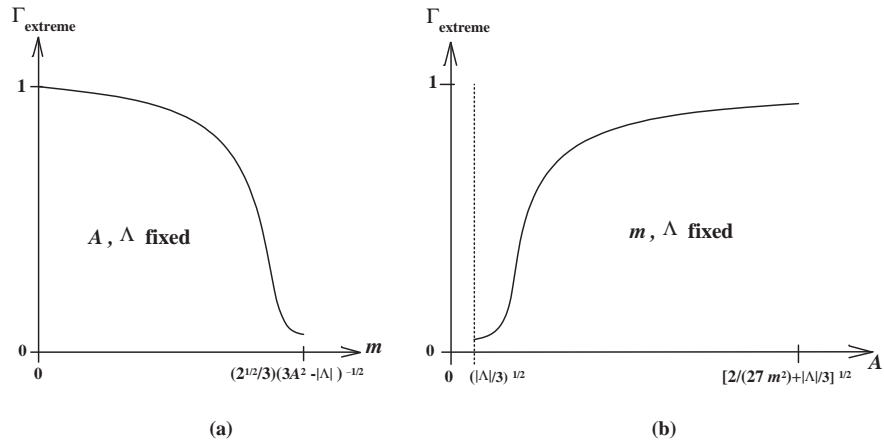}
\caption{\label{cold AdS-fig}
(a) Plot of the pair creation rate
of extreme black holes, $\Gamma_{\rm ext}$, as a function of $m$
for a fixed $A$ and $\Lambda$. The range of $m$ is
$0<m<\frac{\sqrt{2}}{3}
  \frac{1}{\sqrt{3A^2-|\Lambda|}}$.
(b) Plot of the pair creation rate of nonextreme black holes,
$\Gamma_{\rm ext}$, as a function of $A$ for a fixed $m$ and
$\Lambda$. The range of $A$ is
$\sqrt{\frac{|\Lambda|}{3}}<A<\sqrt{\frac{2}{27m^2}+\frac{|\Lambda|}{3}}$.
 See text of subsection \ref{sec:Cold-rate AdS}.
 }
\end{figure}

In the extreme AdS case, the instanton has a single horizon, the
acceleration horizon at $y=y_A$, in its Euclidean section, since
 $y=y_+$ is an internal infinity.
The pair creation rate of extreme AdS black holes with $y_+=y_-$
is then
\begin{eqnarray}
\Gamma_{\rm ext}\sim e^{\frac{1}{4}
 \Delta {\cal A}_{\rm ac}^{\rm ext}}\,,
 \label{rate-cold PCAdS}
 \end{eqnarray}
where $\Delta {\cal A}_{\rm ac}^{\rm ext}$ is given by
 (\ref{diferenca area-luk PCAdS}), with
 $y_A$ defined in (\ref{zerosy3-cold-PCads}), $x_\mathrm{s}$
and $x_\mathrm{n}$ given by (\ref{polos-cold-PCads}),
 and condition (\ref{mq-cold-PCads}) must be satisfied.
The pair creation rate decreases as the mass of the black holes
increases, and the pair creation rate increases when $A$ increases
(see Fig. \ref{cold AdS-fig}). The general behavior of the pair
creation rate of extreme black holes as a function of $m$ and $A$
in the AdS case is also analogous to the behavior of the flat
case, discussed in \cite{HawkRoss-string} (see also section
\ref{sec:Pair Creation flat}).

\subsubsection{\label{sec:submaximal-rate AdS} Pair creation rate
in the nonextreme case with $\bm{m\neq q}$}

The nonextreme instantons, that mediate the pair creation of
nonextreme black holes with $m\neq q$ (including the case $q=0$),
have two horizons in their Euclidean section, namely the
acceleration horizon at $y=y_A$ and the black hole horizon at
$y=y_+$. The pair creation rate of nonextreme AdS black holes with
$m\neq q$ is then
 $\Gamma \sim e^{( {\cal A}_{\rm bh}^{\rm nonext}
 +\Delta {\cal A}_{\rm ac}^{\rm nonext})/4}$, with
${\cal A}_{\rm bh}^{\rm nonext}$ given by
 (\ref{area BH-luk PCAdS}) and $\Delta {\cal A}_{\rm ac}^{\rm nonext}$
given by (\ref{diferenca area-luk PCAdS}), subjected to the
results found in Sec. \ref{sec:submaximal-inst AdS}. The pair
creation rate decreases as the mass of the black holes increases,
and the pair creation rate increases when $A$ increases.
\subsection{\label{sec:Heuristic AdS}Heuristic derivation
of the nucleation rates}
In order to clarify the physical interpretation of the results, in
this subsection we heuristically derive some results discussed in
the main body of section \ref{sec:Pair Creation AdS}. In
particular, we find heuristically the pair creation rates.

In an AdS background, pair creation of black holes is possible
only when the acceleration provided by the strings satisfies
$A>\sqrt{|\Lambda|/3}$. To understand this result one can argue as
follows. In general, the time-time component of the gravitational
field is given by $g_{00}=1-2\Phi$, where $\Phi$ is the Newtonian
potential. In the AdS spacetime, one has $\Phi=-|\Lambda| r^2/6$
and its derivative yields the force per unit mass or acceleration
of the AdS spacetime, $A_{\rm AdS}=-|\Lambda|r/3 \sim
-\sqrt{|\Lambda|/3}$, where we have replaced $r$ by the
characteristic AdS radius $\left (|\Lambda|/3 \right )^{-1/2}$.
The minus sign indicates that the AdS background is attractive and
thus, if one wants to have a pair of accelerated black holes
driving away from each other, the cosmic string will have to
provide a sufficient acceleration $A$ that overcomes the AdS
background attraction, i.e., $A>|A_{\rm AdS}|$.

An estimate for the black hole pair creation probability can be
given by the Boltzmann factor, $\Gamma \sim e^{-E_0/W_{\rm ext}}$,
where $E_0$ is the energy of the system that nucleates and $W_{\rm
ext}=F \ell$ is the work done by the external force $F$, that
provides the energy for the nucleation, through the typical
distance $\ell$ separating the created pair. First we ask what is
the probability that a black hole pair is created in a $\Lambda=0$
background when a string breaks. This process has been discussed
in \cite{HawkRoss-string} (see also section \ref{sec:Pair Creation
flat}) where it was found that the pair creation rate is $\Gamma
\sim e^{-m/A}$. In this case, $E_0 \sim 2m$, where $m$ is the rest
energy of the black hole, and $W_{\rm ext}\sim A$ is the work
provided by the strings. To derive $W_{\rm ext}\sim A$ one can
argue as follows. The  acceleration provided by the string is $A$,
the characteristic distance that separates the pair at the
creation moment is $1/A$ (this value follows from the Rindler
motion $x^2-t^2=1/A^2$ that describes the uniformly accelerated
motion of the black holes), and the characteristic mass of the
system is A by the Compton relation. Thus, the characteristic work
is $W_{\rm ext}={\rm mass}\times{\rm acceleration}\times{\rm
distance}\sim A A A^{-1}$. So, from the Boltzmann factor we indeed
expect that the creation rate of a black hole pair when a string
breaks in a $\Lambda=0$ background is given by $\Gamma \sim
e^{-m/A}$.

Now we ask what is the probability that a string breaks in an AdS
background and a pair of black holes is produced. As we saw just
above, the presence of the AdS background leads in practice to a
problem in which we have a net acceleration that satisfies
$A'\equiv \sqrt{A^2-|\Lambda|/3}$, this is, $\Lambda$ makes a
negative contribution to the process. Heuristically, we may then
apply the same arguments that have been used in the last
paragraph, with the replacement $A\rightarrow A'$. At the end, the
Boltzmann factor tells us that the creation rate for the process
is $\Gamma \sim e^{-m/\sqrt{A^2-|\Lambda|/3}}$. So, given  $m$ and
$\Lambda$, when the acceleration provided by the string grows the
pair creation rate increases, as the explicit calculations done in
the main body of section \ref{sec:Pair Creation flat}.

\subsection{\label{sec:Conc AdS}Summary and discussion}

We have studied in detail the quantum process in which a cosmic
string breaks in an anti-de Sitter (AdS) background and a pair of
black holes is created at the ends of the string. The energy to
materialize and accelerate the black holes comes from the strings'
tension. The analysis of this process in a flat background
($\Lambda=0$) has been carried in \cite{HawkRoss-string}, while in
a de Sitter background ($\Lambda>0$) it has been done in
\cite{OscLem-PCdS}. In an AdS background this is the only study
done in the process of production of a pair of correlated black
holes with spherical topology. Note that in a cosmological
background, the transformation used by Ernst to generate an exact
solution in an electromagnetic background does not work, since it
does not leave invariant the cosmological term in the action. Thus
Ernst's trick cannot be used, and we do not have an exact AdS
Ernst solution available to study analytically the process in
which a pair of black holes is produced and accelerated by the
electromagnetic force. However, in principle, we could find a
perturbative AdS Ernst solution that supplies the same energy and
acceleration as our strings and, from the results of the
$\Lambda=0$ case we expect that the results found in this paper
would not depend on whether the energy is being furnished by an
external electromagnetic field or by strings.

It is well known that the AdS background is attractive, i.e., an
analysis of the geodesic equations indicates that particles in
this background are subjected to a potential well that attracts
them. Therefore, if we have a virtual pair of black holes and we
want to turn them real, we will have to furnish a sufficient force
that overcomes this cosmological background attraction. We then
expect that pair creation is possible only if the strings' tension
and the associated acceleration $A$ is higher than a critical
value. We have confirmed that this is indeed the case: in the AdS
background, black holes with spherical topology can be produced
only with an acceleration higher than $\sqrt{|\Lambda|/3}$. This
result was also expected from the AdS C-metric properties, which
describes the evolution of the system after the creation process.
Indeed, in \cite{OscLem_AdS-C} we have shown that it only
describes a pair of black holes if $A>\sqrt{|\Lambda|/3}$,
otherwise it represents a single accelerated black hole.

We have constructed the instantons that mediate the pair creation
process through the analytic continuation of some special cases of
the AdS C-metric. The regularity condition imposed to these
instantons restricts the mass $m$ and the charge $q$ of the black
holes that are produced, and physically it means that the only
black holes that can be pair produced are those that are in
thermodynamic equilibrium. Concretely, we have found two charged
regular instantons. One mediates the pair creation of nonextreme
black holes with $m=q$, and the other mediates the pair creation
of extreme black holes. These instantons are the natural AdS
C-metric counterparts of the instantons found in previous works on
the subject. We note that the Nariai instanton, and the ultracold
instanton that are available in the de Sitter background, are not
present in the AdS case since they are out of the allowed range of
the angular direction $x$. The instantons constructed from the
$\Lambda<0$ and $\Lambda=0$ C-metric are noncompact (contrary to
what occurs with the $\Lambda>0$ instantons), in the sense that
they have an acceleration horizon with an infinite area. Thus when
dealing with them, we have to eliminate this infinity by
normalizing this area relative to the acceleration horizon area of
an appropriate background reference spacetime.

We have explicitly computed the pair creation rate for the
nonextreme and extreme black holes. In both cases, the AdS pair
creation rate reduces to the corresponding ones of the flat case
when we set $\Lambda=0$ \cite{HawkRoss-string} (see also section
\ref{sec:Pair Creation flat}). In the two cases, the pair creation
rate decreases monotonically, as the mass (and charge) of the
black holes increases  (see Figs. \ref{lukewarm AdS-fig}.(a) and
\ref{cold AdS-fig}.(a) for the nonextreme and extreme cases,
respectively). This is the physically expected result since an
higher mass demands an higher energy. In what concerns the
evolution of the pair creation rate with the acceleration $A$, we
found that the pair creation rate increases monotonically with $A$
(see Figs. \ref{lukewarm AdS-fig}.(b) and \ref{cold AdS-fig}.(b)
for the nonextreme and extreme cases, respectively). The physical
interpretation of this result is clear: the acceleration $A$ of
the black hole pair is provided by the string. When the energy of
the string is higher (i.e., when its mass density or the
acceleration that it provides is higher), the probability that it
breaks and produces a black hole pair with a given mass is also
higher. This behavior is better understood if we make a analogy
with a thermodynamical system, with the mass density of the string
being the analogue of the temperature $T$. Indeed, from the
Boltzmann factor, $e^{-E_0/(k_{\rm B} T)}$ (where $k_{\rm B}$ is
the Boltzmann constant), one knows that a higher background
temperature $T$ turns the nucleation of a particle with energy
$E_0$ more probable. Similarly, a background string with a higher
mass density turns the creation of a black hole pair with mass
$2m$ more probable.

We have also verified that (as occurs with pair creation in other
backgrounds) the pair production of nonextreme black holes is
enhanced relative to the pair creation of extreme black holes by a
factor of $e^{S_{\rm bh}}$, where $S_{\rm bh}={\cal A}_{\rm bh}/4$
is the gravitational entropy of the black hole.

\section{\label{sec:Pair Creation flat}Pair creation of flat
black holes on a cosmic string background}

In a flat background ($\Lambda=0$), the analysis of the process of
pair creation of black holes when a cosmic string breaks has been
analyzed by Hawking and Ross \cite{HawkRoss-string}. In this case,
there is a direct $\Lambda=0$ counterpart of the nonextreme AdS
instanton and of the extreme AdS instanton discussed in our AdS
case. These instantons describe pair creation of nonextreme (with
$m=q$) and extreme black holes with $y_+=y_-$, respectively
\cite{HawkRoss-string}. The total area of these $\Lambda=0$
instantons can be found in \cite{HawkRoss-string}, and can be
obtained by taking the direct $\Lambda=0$ limit of (\ref{area
BH-luk PCAdS}) and (\ref{diferenca area-luk PCAdS}), together with
the replacement $y_A \mapsto -x_{\mathrm s}$. This procedure
yields that (\ref{area BH-luk PCAdS}) also holds in the
$\Lambda=0$ case, while  $\Delta {\cal A}_{\rm ac}$ becomes
\begin{eqnarray}
\!\!\!\!\!\Delta {\cal A}_{\rm ac} =
   -\frac{4\pi} {A^2 |{\cal G}'(x_\mathrm{n})|}\,
 {\biggl (} \frac{1}{ x_\mathrm{n}-x_{\mathrm
s}}
 + \frac{1}{2}
 \frac{ {\cal G}''(x_\mathrm{s}) }{ {\cal G}'(x_\mathrm{s}) }
  {\biggr )} \,.
  \label{diferenca area-flat}
 \end{eqnarray}
In \cite{HawkRoss-string} the explicit numerical value of ${\cal
A}_{\rm bh}$ and $\Delta {\cal A}_{\rm ac}$ has not been computed.
We will do it here.

\begin{figure} [H]
\centering
\includegraphics[height=2.3in]{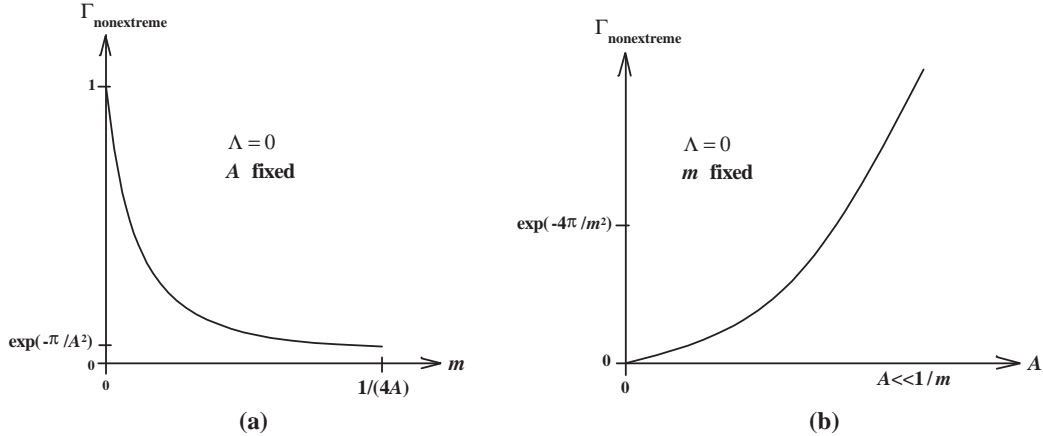}
\caption{\label{lukewarm flat-fig}
  (a) Plot of the pair creation
rate of nonextreme $\Lambda=0$ black holes, $\Gamma_{\rm nonext}$,
as a function of $m$ for a fixed $A$.
 (b) Plot of the pair
creation rate of nonextreme $\Lambda=0$ black holes, $\Gamma_{\rm
nonext}$, as a function of $A$ for a fixed $m$.
 }
\end{figure}

For the $\Lambda=0$ nonextreme case, the discussion of Sec.
\ref{sec:Lukewarm-inst AdS} applies generically, as long as we set
$\Lambda=0$ in the corresponding equations. With this data we find
the explicit value of the total area of the nonextreme flat
instanton
\begin{eqnarray}
 \!\!\!\!\!\!\!\!\!\!\!\!& &{\cal A}_{\rm bh}+\Delta
{\cal A}_{\rm ac} =
-\frac{\pi}{A^2}\frac{-1+4mA+\sqrt{1-(4mA)^2}}{\sqrt{1-(4mA)^2}}.\nonumber \\
& &
 \label{area TOTAL-flat}
 \end{eqnarray}
In the particle limit, $mA\ll 1$, the above relation reduces to
$-\pi m/(4A)$ and the mass density of the string is given by
$\mu\sim mA$. The pair creation rate is then $\Gamma\sim e^{-\pi
m^2/\mu}$. Thus, as occurs with the AdS case, the pair creation
rate decreases when $m$ increases and the rate increases when $A$
or $\mu$ increase [see Fig. \ref{lukewarm flat-fig}].

We remark that for $mA\sim 1$, and as occurs in the corresponding
AdS case, the pair creation rate associated to (\ref{area
TOTAL-flat}) starts decreasing when $A$ increases. This is a
physically unexpected result since a higher acceleration provided
by the string background should favor the nucleation of a fixed
black hole mass. The sector $mA\sim 1$ must then be discarded and
the reason is perfectly identified: the domain of validity of the
rates is $mA\ll 1$, for which the radius of the black hole, $r_+
\sim m$, is much smaller than the typical distance between the
black holes at the creation moment, $\ell \sim 1/A$ (this value
follows from the Rindler motion $x^2-t^2=1/A^2$ that describes the
uniformly accelerated motion of the black holes). So, for $mA\sim
1$ one has $r_+\sim \ell$ and the black holes start interacting
with each other.

In what concerns the pair creation rate of extreme and nonextreme
with $m=q$ $\Lambda=0$ black holes, an explicit computation shows
that its general behavior with $A$ and $m$ is also similar to the
one of the AdS case, i.e., the rate decreases when $m$ or $q$
increase, and the rate increases when $A$ increases.

\section{\label{sec:Pair Creation dS}Pair creation of de Sitter
black holes on a cosmic string background}

In this section we want to analyze the process in which a cosmic
string breaks and a pair of black holes is produced at the ends of
the string, in a de Sitter (dS) background. Therefore, the energy
to materialize and accelerate the pair comes from the positive
cosmological constant and, in addition, from the string tension.
This process is a combination of the processes considered in
\cite{MelMos}-\cite{VolkovWipf} and in
\cite{HawkRoss-string}-\cite{GregHind}. The instantons for this
process can be constructed by analytically continuing  the dS
C-metric found by Pleba\'nski and Demia\'nski \cite{PlebDem} and
analyzed by Podolsk\'y and Griffiths \cite{PodGrif2}, and in
detail  by Dias and Lemos \cite{OscLem_dS-C}.

The plan of this section is as follows. In section
\ref{sec:instanton-method}, we describe the semiclassical
instanton method used to evaluate the pair creation rate. In
section \ref{sec:dS C-inst} we construct, from the dS C-metric
studied in section \ref{sec:dS C-metric}, the regular instantons
that describe the pair creation process. Then, in section
\ref{sec:Calc-I}, we explicitly evaluate the pair creation rate
for each one of the cases discussed in section \ref{sec:dS
C-inst}. In section \ref{sec:Entropy} we verify that the usual
relation between pair creation rate, entropy and total area holds
also for the pair creation process discussed in this section.
Finally, in section \ref{sec:Conc dS} concluding remarks are
presented.

\subsection{\label{sec:instanton-method}Black hole pair creation
rate: the instanton method}
The pair creation of black holes in a de Sitter (dS) background is
described, according to the no-boundary proposal of Hartle and
Hawking \cite{HartleHawk}, by the propagation from nothing to a
3-surface boundary $\Sigma$. The amplitude for this process is
given by the wave function
 \begin{eqnarray}
\Psi(h_{ij},A_{i})=\int
d[g_{\mu\nu}]d[A_{\mu}]e^{-I(g_{\mu\nu},A_{\mu})}\:,
 \label{wave}
 \end{eqnarray}
where $h_{ij}$ and $A_{i}$ are the induced metric and
electromagnetic potential on the boundary
 $\Sigma=\partial {\cal M}$ of a compact manifold ${\cal M}$,
 $d[g_{\mu\nu}]$ is a measure on the space of the metrics $g_{\mu\nu}$
and $d[A_{\mu}]$ is a measure on the space of the Maxwell field
$A_{\mu}$, and $I(g_{\mu\nu},A_{\mu})$ is their Euclidean action.
The path integral is over all compact metrics and potentials on
manifolds ${\cal M}$ with boundary $\Sigma$, which agree with the
boundary data on $\Sigma$. For a nice detailed discussion of the
no-boundary proposal applied to the study of black hole pair
creation see Bousso and Chamblin \cite{BouCham}.

In the semiclassical instanton approximation, the dominant
contribution to the path integral comes from metrics and Maxwell
fields which are near the solutions (instantons) that extremalize
the Euclidean action and satisfy the boundary conditions. Thus,
considering small fluctuations around this solution, $g_{\mu\nu}
\rightarrow g_{\mu\nu}+\tilde{g}_{\mu\nu}$ and $A_{\mu}
\rightarrow A_{\mu}+\tilde{A}_{\mu}$, the action expands as
\begin{eqnarray}
I=I_{\rm inst}(g_{\mu\nu},A_{\mu})+
 \delta^2 I(\tilde{g}_{\mu\nu})+ \delta^2 I(\tilde{A}_{\mu})+\cdots\:,
 \label{Ifluct}
 \end{eqnarray}
where $\delta^2 I$ are quadratic in $\tilde{g}_{\mu\nu}$ and
$\tilde{A}_{\mu}$, and dots denote higher order terms. The wave
function, that describes the creation of a black hole pair from
nothing, is then given by $\Psi_{\rm inst}= B e^{-I_{\rm inst}}$,
where $I_{\rm inst}$ is the classical action of the gravitational
instanton that mediates the pair creation of black holes, and the
prefactor $B$ is the one loop contribution from the quantum
quadratic fluctuations in the fields, $\delta^2 I$. Similarly, the
wave function that describes the nucleation of a dS space with a
string from nothing is $\Psi_{\rm string} \propto e^{-I_{\rm
string}}$, and the wave function describing the nucleation of a dS
space from nothing is $\Psi_{\rm dS} \propto e^{-I_{\rm dS}}$.
 The nucleation probability of the dS
space from nothing, of the dS space with a string from nothing,
and of a space with a pair of black holes from nothing is then
given by $|\Psi_{\rm dS}|^2$, $|\Psi_{\rm string}|^2$ and
$|\Psi_{\rm inst}|^2$, respectively.

We may now ask four questions: what is the probability for (i)
pair creation of black holes in a dS spacetime, (ii) the
nucleation of a string in a dS background, (iii) the process in
which a string in a dS background breaks and a pair of black holes
is created, and (iv) the combined process (ii)+(iii). In the
process (i) the energy to materialize the pair comes only from the
positive cosmological constant background, $\Lambda$. The system
does not contain a string and the probability for this process has
been found in \cite{MannRoss}. The aim of the present paper is to
compute explicitly the probability for processes (ii)-(iv). It is
important to note that in the process (iii), one assumes that the
initial background contains a string, i.e. , the question that is
being asked is: given that the string is already present in our
initial system, what is the probability that it breaks and a pair
of black holes is produced and accelerated apart by $\Lambda$ and
by the string tension? On the other side, in (iv) one is asking:
starting from a pure dS background, what is the probability that a
string nucleates on it and then breaks forming a pair of black
holes? Naturally, the probability for process (iv) is  the product
of the probability for process (ii) and the probability for
process (iii).

According to the no-boundary proposal, the nucleation rate of a
string in a dS background is proportional to $|\Psi_{\rm
string}|^2 /|\Psi_{\rm dS}|^2$, i.e.,
\begin{eqnarray}
\Gamma_{\rm string/dS} \simeq \eta\bar{\eta} \, e^{- 2I_{\rm
string}+2I_{\rm dS} } \:.
 \label{PC-rate-string}
 \end{eqnarray}
The pair creation rate of black holes when a string breaks in a dS
background is given by
\begin{eqnarray}
\Gamma_{\rm BHs/string} \simeq \eta \, e^{- 2I_{\rm inst}+2I_{\rm
string} } \:,
 \label{PC-rate-breakstring}
 \end{eqnarray}
and the pair creation rate of black holes when process (iv) occurs
is given by the product of (\ref{PC-rate-string}) and
(\ref{PC-rate-breakstring}), i.e.,
\begin{eqnarray}
\Gamma_{\rm BHs/dS} \simeq \tilde{\eta} \, e^{- 2I_{\rm
inst}+2I_{\rm dS} } \:.
 \label{PC-rate}
 \end{eqnarray}
We will find $I_{\rm inst}$ and $I_{\rm string}$ in the next
subsections. In the three relations above, $\bar{\eta}$, $\eta$
and $\tilde{\eta}$ are one-loop prefactors which will not be
considered in this paper. The evaluation of this one-loop
prefactor has been done only in a small number of cases, namely
for the vacuum background by Gibbons, Hawking and Perry
\cite{GibbHawPer}, for the Schwarzschild instanton by Gross, Perry
and Yaffe \cite{GrossPerryYaffe}, for other asymptotically flat
instantons by Young \cite{Young}, for the dS background by Gibbons
and Perry \cite{GibbPer} and Christensen and Duff
\cite{ChristDuff}, for the dS-Schwarzschild instanton by Ginsparg
and Perry \cite{GinsPerry}, Young \cite{YoungdS}, Volkov and Wipf
\cite{VolkovWipf} and Garattinni \cite{GaratinniOneLoop}, and for
the Ernst instanton by Yi \cite{YiPConeLoop}.

At this point we must specify the Euclidean action needed to
compute the path integral (\ref{wave}). This issue was analyzed
and clarified in detail by Hawking and Ross \cite{HawkRoss} and by
Brown \cite{Brown2}. Now, due to its relevance for the present
paper, we briefly discuss the main results of
\cite{HawkRoss,Brown2}. One wants to use an action for which it is
natural to fix the boundary data on $\Sigma$ specified in
(\ref{wave}). That is, one wants to use an action whose variation
gives the Euclidean equations of motion when the variation fixes
these boundary data on $\Sigma$ \cite{BrownYork}. In the magnetic
case this Euclidean action is the Einstein-Maxwell action with a
positive cosmological constant $\Lambda$ given by
 \begin{eqnarray}
I=-\frac{1}{16\pi}\int_{\cal M} d^4x\sqrt{g} \left (
R-2\Lambda-F^{\mu\nu}F_{\mu\nu} \right
)-\frac{1}{8\pi}\int_{\Sigma=\partial {\cal M}} d^3x\sqrt{h}\, K
\:,
 \label{I}
 \end{eqnarray}
where $g$ is the determinant of the Euclidean metric, $h$ is the
determinant of the induced metric on the boundary $\Sigma$, $R$ is
the Ricci scalar, $K$ is the trace of
 the extrinsic curvature $K_{ij}$ of the boundary, and
 $F_{\mu\nu}=\partial_{\mu}A_{\nu}-\partial_{\nu}A_{\mu}$ is
the Maxwell field strength of the gauge field $A_{\nu}$. Variation
of (\ref{I}) yields $\delta I=(\cdots )+
 \frac{1}{4\pi}\int_{\Sigma}
d^3x\sqrt{h}\, F^{\mu\nu}n_{\mu}\delta A_{\nu}$, where $(\cdots)$
represents terms giving the equations of motion plus gravitational
boundary terms that are discussed in \cite{BrownYork}, and
$n_{\mu}$ is the unit outward normal to $\Sigma$. Thus, variation
of (\ref{I}) gives the equations of motion as long as it is at
fixed gauge potential $A_i$ on the boundary. Now, for magnetic
black hole solutions, fixing the potential fixes the charge on
each of the black holes, since the magnetic charge is just given
by the integral of $F_{ij}$ over a 2-sphere lying in the boundary.
However, in the electric case, fixing $A_i$ can be regarded as
fixing a chemical potential $\omega$ which is conjugate to the
charge \cite{HawkRoss}. Holding the electric charge fixed is
equivalent to fixing $n_{\mu}F^{\mu i}$ on $\Sigma$, as the
electric charge is given by the integral of the dual of $F$ over a
2-sphere lying in $\Sigma$. Therefore in the electric case the
appropriate Euclidean action is \cite{HawkRoss}
 \begin{eqnarray}
I_{\rm el}=I-\frac{1}{4\pi}\int_{\Sigma=\partial {\cal M}}
d^3x\sqrt{h}\, F^{\mu\nu}n_{\mu}A_{\nu}\:,
 \label{I-electric}
 \end{eqnarray}
 where $I$ is defined in (\ref{I}).
Variation of action (\ref{I-electric}) yields $\delta I_{\rm
el}=(\cdots)+
 \frac{1}{4\pi}\int_{\Sigma}
d^3x \delta (\sqrt{h}\, F^{\mu\nu}n_{\mu}) A_{\nu}$, and thus it
gives the equations of motion when $\sqrt{h}\,n_{\mu} F^{\mu i}$,
and so the electric charge, is held fixed. Since $\int_{\cal M}
d^4x\sqrt{g} F^{\mu\nu}F_{\mu\nu}$ has opposite signs for dual
magnetic and electric solutions, if we took (\ref{I}) to evaluate
both the magnetic and electric actions we would conclude that the
pair creation of electric black holes would be enhanced relative
to the pair creation of magnetic black holes. This physically
unexpected result does not occur when one considers the
appropriate boundary conditions and includes the extra Maxwell
boundary term in (\ref{I-electric}).

We have to be careful \cite{GibbHawk,HawkRoss} when computing the
extra Maxwell boundary term in the electric action
(\ref{I-electric}). Indeed, we have to find a vector potential,
$A_{\nu}$, that is regular everywhere in the instanton, including
at the horizons. Usually, as we shall see, this requirement leads
to unusual choices for $A_{\nu}$. The need of this requirement is
easily understood if we take the example of the electric
Reissner-Nordstr\"{o}m solution \cite{GibbHawk,HawkRoss}. In this
case, normally, the gauge potential in Schwarzschild coordinates
is taken to be $A=-\frac{q}{r} \,dt$. However, this potential is
not regular at the horizon $r=r_+$, since $dt$ diverges there. An
appropriate choice that yields a regular electromagnetic potential
everywhere, including at the horizon is $A=-q (\frac{1}{r}-
\frac{1}{r_+}) \,dt$ or, alternatively, $A=-\frac{q}{r^2}\,t\,dr$.
To all these potentials corresponds the field strength
$F=-\frac{q}{r^2} \,dt\wedge dr$.

Before we finish this section, note that since the dS C-metric
instantons that we will consider in this paper are all compact we
do not have to define their action relative an appropriate
background solution, contrarily to what happens with the
non-compact $\Lambda=0$ instantons
\cite{GarfGiddStrom_Sbh}-\cite{Emparan}.

\subsection{\label{sec:dS C-inst} The \lowercase{d}S C-metric instantons}

The dS C-metric has been studied in detail in section \ref{sec:dS
C-metric}. It describes a pair of uniformly accelerated black
holes in a dS background, with the acceleration being provided by
the cosmological constant and, in addition, by a string that
connects the two black holes along their south poles and pulls
them away. The presence of the string is associated to the conical
singularity that exists in the south pole of the dS C-metric.

Following section \ref{sec:instanton-method}, in order to evaluate
the black hole pair creation rate we need to find the instantons
of the theory, i.e., we must look into the Euclidean section of
the dS C-metric and choose only those Euclidean solutions which
are regular in a way that will be explained soon. To obtain the
Euclidean section of the dS C-metric from the Lorentzian dS
C-metric we simply introduce an imaginary time coordinate
$\tau=-it$. Then the gravitational field of the Euclidean dS
C-metric is given by (see, e.g., \cite{OscLem_dS-C})
\begin{equation}
 d s^2 = [A(x+y)]^{-2} ({\cal F}d\tau^2+
 {\cal F}^{-1}dy^2+{\cal G}^{-1}dx^2+
 {\cal G}d\phi^2)\:,
 \label{C-metric PCdS}
 \end{equation}
with ${\cal F}(y)$ and ${\cal G}(x)$ given by (\ref{FG}). The
Maxwell field in the magnetic case is still given by
(\ref{F-mag}), while in the electric case it is given by
 \begin{eqnarray}
F_{\rm el}=-i\,q\, d\tau\wedge dy \:.
 \label{F-el}
\end{eqnarray}
Recall some basic properties that will be needed. The solution has
a curvature singularity at $y=+\infty$ where the matter source is.
The point $y=-x$ corresponds to a point that is infinitely far
away from the curvature singularity, thus as $y$ increases we
approach the curvature singularity and $y+x$ is the inverse of a
radial coordinate. At most, ${\cal F}(y)$ can have four real zeros
which we label in ascending order by $y_{\rm neg}<0<y_A\leq y_+
\leq y_-$. The roots $y_-$ and $y_+$ are respectively the inner
and outer charged black hole horizons, and $y_A$ is an
acceleration horizon which coincides with the cosmological horizon
and has a non-spherical shape. The negative root $y_{\rm neg}$
satisfies $y_{\rm neg}<-x$ and thus has no physical significance.
The angular coordinate $x$ belongs to the range
$[x_\mathrm{s},x_\mathrm{n}]$ for which ${\cal G}(x)\geq 0$ (when
we set $A=0$ we have $x_\mathrm{s}=-1$ and $x_\mathrm{n}=+1]$). In
order to avoid a conical singularity in the north pole, the period
of $\phi$ must be given by
\begin{equation}
\Delta \phi=\frac{4 \pi}{|{\cal G}'(x_\mathrm{n})|}\:,
 \label{Period phi}
 \end{equation}
and this leaves a conical singularity in the south pole with
deficit angle
\begin{eqnarray}
\delta =2\pi \left ( 1-\frac{{\cal G}'(x_\mathrm{s})}{|{\cal
G}'(x_\mathrm{n})|} \right )\,.
 \label{conic-sing-dS}
 \end{eqnarray}
that signals the presence of a string with mass density $\mu
=\delta/(8\pi)$, and with pressure $p=-\mu<0$. When we set the
acceleration parameter $A$ equal to zero, the dS C-metric reduces
to the usual dS$-$Reissner-Nordstr\"{o}m or dS-Schwarzschild
solutions without conical singularities.

So far, we have described the solution that represents a pair of
black holes accelerated by the cosmological constant and by the
string tension. This solution describes the evolution of the black
hole pair after its creation. Now, we want to find a solution that
represents a string in a dS background. This solution will
describe the initial system, before the breaking of the cosmic
string that leads to the formation of the black hole pair. In
order to achieve our aim we note that at spatial infinity the
gravitational field of the Euclidean dS C-metric reduces to
\begin{eqnarray}
\!\!\!\!\!\!\!\!\! ds^2 \!\! &=& \!\!
\frac{1}{[A_0(x+y)]^2}{\biggl [}
 \left ( -\frac{\Lambda}{3A_0^{\,2}}-1+y^2 \right )dt^2
 \nonumber\\
 & & +  \frac{dy^2}{-\frac{\Lambda}{3A_0^{\,2}}-1+y^2}
+\frac{dx^2}{1-x^2}+ (1-x^2)d\phi_0^{\,2} {\biggr ]},
 \label{backg metric-dS}
\end{eqnarray}
and the Maxwell field goes to zero. $A_0$ is a constant that
represents a freedom in the choice of coordinates, and $-1 \leq x
\leq 1$. We want that this metric also describes the solution
before the creation of the black hole pair, i.e., we demand that
it describes a string with its conical deficit in a dS background.
Now, if we want to maintain the intrinsic properties of the string
during the process we must impose that its mass density and thus
its conical deficit remains constant. After the pair creation we
already know that the conical deficit is given by
(\ref{conic-sing-dS}). Hence, the requirement that the background
solution describes a dS spacetime with a conical deficit angle
given exactly by (\ref{conic-sing-dS}) leads us to impose that in
(\ref{backg metric-dS}) one has
\begin{eqnarray}
\Delta \phi_0=2\pi-\delta=2\pi \frac{{\cal
G}'(x_\mathrm{s})}{|{\cal G}'(x_\mathrm{n})|} \,.
 \label{delta phi0-dS}
 \end{eqnarray}
The arbitrary parameter $A_0$ can be fixed by imposing a matching
between (\ref{C-metric PCdS}) and (\ref{backg metric-dS}) at large
spatial distances \cite{HawHorRoss,HawkRoss-string}, yielding
$A_0^{\,2}=-A^2 [{\cal G}'(x_\mathrm{s})]^2/\left [ 2{\cal
G}''(x_\mathrm{s}) \right ]$.

Returning back to the euclidean dS C-metric (\ref{C-metric PCdS}),
in order to have a  positive definite Euclidean metric we must
require that $y$ belongs to $y_A \leq y \leq y_+$. In general,
when $y_+ \neq y_-$, one then has conical singularities at the
horizons $y=y_A$ and $y=y_+$. In order to obtain a regular
solution we have to eliminate the conical singularities at both
horizons. This is achieved by imposing that the period of $\tau$
is the same for the two horizons, and is equivalent to requiring
that the Hawking temperature of the two horizons be equal. To
eliminate the conical singularity at $y=y_A$ the period of $\tau$
must be $\beta=2 \pi/ k_A$ (where $k_A$ is the surface gravity of
the acceleration horizon),
\begin{equation}
\beta=\frac{4 \pi}{|{\cal F}'(y_A)|}\:.
 \label{Period tau-yA PCdS}
 \end{equation}
 This choice for the period of $\tau$ also eliminates
simultaneously the conical singularity at the outer black hole
horizon, $y_+$, if and only if the parameters of the solution are
such that  the surface gravities of the  black hole and
acceleration horizons are equal ($k_+=k_A$), i.e.
\begin{equation}
 {\cal F}'(y_+)=-{\cal F}'(y_A)\:.
 \label{k+=kA PCdS}
 \end{equation}
There are two ways to satisfy this condition. One is a regular
Euclidean solution with $y_A \neq y_+$, and will be called
lukewarm C instanton. This solution requires the presence of an
electromagnetic charge. The other way is to have $y_A=y_+$, and
will be called Nariai C instanton. This last solution exists with
or without charge. When we want to distinguish them, they will be
labelled by charged Nariai and neutral Nariai C instantons,
respectively.

We now turn our attention to the case $y_+ = y_-$ and $y_A\neq
y_+$, which obviously requires the presence of charge. When this
happens the allowed range of $y$ in the Euclidean sector is simply
$y_A \leq y < y_+$. This occurs because when $y_+ = y_-$ the
proper distance along spatial directions between $y_A$ and $y_+$
goes to infinity. The point $y_+$ disappears from the $\tau, y$
section which is no longer compact but becomes topologically $S^1
\times {\mathbb{R}}$. Thus, in this case we have a conical
singularity only at $y_A$, and so we obtain a regular Euclidean
solution by simply requiring that the period of $\tau$ be equal to
(\ref{Period tau-yA PCdS}). We will label this solution by cold C
instanton. Finally, we have a special solution that satisfies
 $y_A=y_+=y_-$ and that is regular when condition
(\ref{Period tau-yA PCdS}) is satisfied. This instanton will be
called ultracold C instanton and can be viewed as a limiting case
of both the charged Nariai C instanton and cold C instanton.

Below, we will describe in detail each one of these four C
instantons, following the order: (A) lukewarm C instanton, (B)
cold C instanton, (C) Nariai C instanton, and (D) ultracold C
instanton. These instantons are the C-metric counterparts ($A\neq
0$) of the $A=0$ instantons that have been constructed
 from the Euclidean section of
the dS$-$Reissner-Nordstr\"{o}m solution ($A=0$)
\cite{MelMos,Rom,MannRoss,BooMann}. The original name of the $A=0$
instantons is associated to the relation between their
temperatures: $T_{\rm lukewarm}>T_{\rm cold}>T_{\rm
ultracold}>T_{\rm Nariai}=0$. This relation is preserved by their
C-metric counterparts discussed in this paper, and we preserve the
$A=0$ nomenclature. The ultracold instanton could also, very
appropriately, be called Nariai Bertotti-Robinson instanton (see
\cite{OscLem_nariai}). These four families of instantons will
allow us to calculate the pair creation rate of accelerated
dS$-$Reissner-Nordstr\"{o}m black holes in section
\ref{sec:Calc-I}.

As is clear from the above discussion, when the charge vanishes
the only regular Euclidean solution that can be constructed is the
neutral Nariai C instanton. The same feature is present in the
$A=0$ case where only the neutral Nariai instanton is available
\cite{GinsPerry,MannRoss,BoussoHawk,VolkovWipf}.

\subsubsection{\label{sec:Lukewarm-inst}The lukewarm C instanton}

For the lukewarm C instanton the gravitational field is given by
(\ref{C-metric PCdS}) with the requirement that ${\cal F}(y)$
satisfies ${\cal F}(y_+)=0={\cal F}(y_A)$ and ${\cal
F}'(y_+)=-{\cal F}'(y_A)$. In this case we can then write (onwards
the subscript ``$\ell$" means lukewarm)
\begin{eqnarray}
{\cal F}_{\rm \ell}(y)=-\left ( \frac{y_A \: y_+}{y_A+y_+} \right
)^2 \left ( 1-\frac{y}{y_A} \right ) \left ( 1-\frac{y}{y_+}
\right )  \left ( 1+\frac{y_A+y_+}{y_A \:y_+}\,y-\frac{y^2}{y_A
\:y_+} \right
 )\:,
 \label{F-luk}
 \end{eqnarray}
with
\begin{eqnarray}
y_A =  \frac{1-\alpha}{2mA}\,, \:\:\:\:\:\:\:\: y_+ =
\frac{1+\alpha}{2mA}\,, \:\:\:\: {\rm and} \:\:\:\:
 \alpha = \sqrt{1-\frac{4m}{\sqrt{3}}\sqrt{\Lambda+3A^2}} \:.
 \label{yA-luk}
 \end{eqnarray}
 The parameters
$A$, $\Lambda$, $m$ and $q$, written as a function of $y_A$ and
$y_+$, are
\begin{eqnarray}
& & \frac{\Lambda}{3A^2} =  \left ( \frac{y_A \: y_+}{y_A+y_+}
\right )^2\,, \nonumber \\
& & mA=(y_A+y_+)^{-1}=qA \,.
 \label{zeros-luk}
 \end{eqnarray}
Thus, the mass and the charge of the lukewarm C instanton are
necessarily equal, $m=q$, as occurs with its $A=0$ counterpart,
the lukewarm instanton \cite{MelMos,Rom,MannRoss,BooMann}. The
demand that $\alpha$ is real requires that
\begin{eqnarray}
  0< m_{\rm \ell}\leq \frac{\sqrt{3}}{4} \frac{1}{\sqrt{\Lambda+3A^2}}\:,
 \label{mq-luk}
 \end{eqnarray}
so the lukewarm C instanton has a lower maximum mass and a lower
maximum charge than the $A=0$ lukewarm instanton
\cite{MelMos,Rom,MannRoss,BooMann} and, for a fixed $\Lambda$, as
the acceleration parameter $A$ grows this maximum value decreases
monotonically. For a fixed $\Lambda$ and for a fixed mass below
$\sqrt{3/(16 \Lambda)}$, the maximum value of the acceleration is
$\sqrt{1/(4m)^2-\Lambda/3}$.

As we said, the allowed range of $y$ in the Euclidean sector is
$y_A \leq y \leq y_+$. Then, the period of $\tau$, (\ref{Period
tau-yA PCdS}), that avoids the conical singularity at both
horizons is
\begin{equation}
\beta_{\rm \ell}=\frac{8 \pi \,m A}{\alpha(1-\alpha^2)}\,,
 \label{beta-luk}
 \end{equation}
and $T_{\rm \ell}=1/\beta_{\rm \ell}$ is the common temperature of
the two horizons.

Using the fact that ${\cal G}(x)=-\Lambda/(3A^2)-{\cal F}(-x)$
[see (\ref{FG})] we can write
\begin{eqnarray}
{\cal G}_{\rm \ell}(x) = 1-x^2 \left ( 1+m A\, x \right )^2 \:,
 \label{G-luk}
 \end{eqnarray}
 and the only real zeros of ${\cal G}_{\rm \ell}(x)$ are the south and
 north pole
\begin{eqnarray}
x_\mathrm{s} =  \frac{-1+\omega_-}{2mA}<0\,, \:\:\:\:\:\:\:\:
x_\mathrm{n}  = \frac{-1+\omega_+}{2mA}>0\,, \:\:\:\:{\rm
with}\:\:\:\: \omega_{\pm} \equiv \sqrt{1\pm 4mA} \:.
 \label{polos-luk}
 \end{eqnarray}
When $A$ goes to zero we have $x_\mathrm{s}\rightarrow -1$ and
$x_\mathrm{n}\rightarrow +1$.
 The period of $\phi$, (\ref{Period phi}), that avoids the
conical singularity at the north pole (and leaves one at the south
pole responsible for the presence of the string) is
\begin{equation}
\Delta \phi _{\rm \ell}=\frac{8 \pi\,m A}{\omega_+(\omega^2_+ -1)}
\leq 2 \pi\:.
 \label{Period phi-luk}
 \end{equation}
 When $A$ goes to zero we have $\Delta \phi _{\rm \ell} \rightarrow 2 \pi$
 and the conical singularity disappears.

The topology of the lukewarm C instanton is $S^2 \times S^2$
 ($0\leq \tau \leq \beta_{\rm \ell}$,
 $y_A \leq y \leq y_+$, $x_\mathrm{s}\leq x \leq x_\mathrm{n}$,
and $0 \leq \phi \leq \Delta \phi _{\rm \ell}$). The Lorentzian
sector describes two dS black holes being accelerated by the
cosmological background and by the string, so this instanton
describes pair creation of nonextreme black holes with $m=q$.

\subsubsection{\label{sec:Cold-inst}The cold C instanton}

The gravitational field of the cold C instanton  is given by
(\ref{C-metric PCdS}) with the requirement that the size of the
outer charged black hole horizon $y_+$ is equal to the size of the
inner charged horizon $y_-$. Let us label this degenerated horizon
by $\rho$: $y_+=y_-\equiv \rho$ and $\rho > y_A$.  In this case,
the function ${\cal F}(y)$ can be written as (onwards the
subscript ``${\rm c}$" means cold)
\begin{eqnarray}
{\cal F}_{\rm c}(y)=\frac{\rho^2-3\gamma}{\rho^4}
 (y-y_{\rm neg})(y-y_A)(y-\rho)^2\:,
 \label{F-cold}
 \end{eqnarray}
with
\begin{eqnarray}
\gamma=\frac{\Lambda+3A^2}{3A^2}\:,
 \label{gamma-PCdS}
 \end{eqnarray}
  and the roots $\rho$,
$y_{\rm neg}$ and $y_A$ are given by
\begin{eqnarray}
& & \rho =\frac{3m}{4q^2A}
 \left ( 1+ \sqrt{1-\frac{8}{9}\frac{q^2}{m^2}} \:\right )
 \:,   \label{zerosy1-cold} \\
& & y_{\rm neg} =\frac{\gamma \rho}{\rho^2-3\gamma}
 \left ( 1- \sqrt{\frac{\rho^2-2\gamma}{\gamma}} \:\right )
 \:,   \label{zerosy2-cold}\\
& & y_A =\frac{\gamma \rho}{\rho^2-3\gamma}
 \left ( 1+ \sqrt{\frac{\rho^2-2\gamma}{\gamma}} \:\right )
 \:.
 \label{zerosy3-cold}
 \end{eqnarray}
The mass and the charge parameters of the solution are written as
a function of $\rho$ as
\begin{eqnarray}
& & m =\frac{1}{A\rho}
 \left ( 1- \frac{2\gamma}{\rho^2} \right )
 \:,  \nonumber \\
& & q^2 =\frac{1}{A^2\rho^2}
 \left ( 1- \frac{3\gamma}{\rho^2} \right )
 \:,
 \label{mq PCdS}
 \end{eqnarray}
and, for a fixed $A$ and $\Lambda$, the ratio $q/m$ is higher than
$1$. The conditions $\rho > y_A$ and $q^2 > 0$ require that, for
the cold C instanton, the allowed range of $\rho$ is
\begin{eqnarray}
\rho>\sqrt{6\gamma} \:.
 \label{range-gamma-cold}
\end{eqnarray}
The value of $y_A$ decreases monotonically with $\rho$ and we have
$\sqrt{\gamma}<y_A<\sqrt{6\gamma}$.
 The mass and the charge of the cold C instanton are also monotonically
decreasing functions of $\rho$, and as we come from $\rho=+\infty$
into $\rho=\sqrt{6\gamma}$ we have
\begin{eqnarray}
& &  0< m_{\rm c}< \frac{\sqrt{2}}{3}
  \frac{1}{\sqrt{\Lambda+3A^2}}\:, \\
& & 0< q_{\rm c}< \frac{1}{2}
  \frac{1}{\sqrt{\Lambda+3A^2}}\:,
 \label{mq-cold PCdS}
 \end{eqnarray}
so the cold C instanton has a lower maximum mass and a lower
maximum charge than the $A=0$ cold instanton, and, for a fixed
$\Lambda$, as the acceleration parameter $A$ grows this maximum
value decreases monotonically. For a fixed $\Lambda$ and for a
fixed mass below $\sqrt{2/(9 \Lambda)}$, the maximum value of the
acceleration is $\sqrt{2/(27m^2)-\Lambda/3}$.

As we have already said, the allowed range of $y$ in the Euclidean
sector is $y_A \leq y < y_+$ and does not include $y=y_+$. Then,
the period of $\tau$, (\ref{Period tau-yA PCdS}), that avoids the
conical singularity at the only  horizon of the cold C instanton
is
\begin{equation}
\beta_{\rm c}=\frac{2 \pi
\rho^3}{(y_A-\rho)^2\sqrt{\gamma(\rho^2-2\gamma)}}\:,
 \label{beta-cold}
 \end{equation}
and $T_{\rm c}=1/\beta_{\rm c}$ is the temperature of the
acceleration horizon.

In what concerns the angular sector of the cold C instanton,
${\cal G}(x)$ is given by (\ref{FG}), and its only real zeros are
the south and north pole,
\begin{eqnarray}
x_\mathrm{s} &=&  -p + \frac{h}{2}-\frac{m}{2q^2A}<0\,, \nonumber \\
 x_\mathrm{n}  &=& p +
\frac{h}{2}-\frac{m}{2q^2A}>0 \:,
 \label{polos-cold}
 \end{eqnarray}
with
\begin{eqnarray}
p &=& \frac{1}{2} \left (  -\frac{s}{3}+\frac{2m^2}{q^4A^2}
  -\frac{1-12 q^2A^2}{3s q^4A^4} -\frac{4}{3q^2A^2}
  + n\right )^{1/2} \:,  \nonumber \\
n &=&  \frac{-m^3+mq^2}{2hq^6 A^3} \:,  \nonumber \\
h &=& \sqrt{\frac{s}{3}+\frac{m^2}{q^4A^2}+\frac{1-12 q^2A^2}{3s q^4A^4}-\frac{2}{3q^2A^2}} \:, \nonumber \\
 s &=& \frac{1}{2^{1/3} q^2A^2}\left (
\lambda-\sqrt{\lambda^2-4(1-12 q^2A^2)^3} \right )^{1/3} \:,
\nonumber \\
\lambda &=& 2 - 108 m^2A^2 + 72 q^2A^2\:,
  \label{acess-zeros-ang PCdS}
 \end{eqnarray}
where $m$ and $q$ are fixed by  (\ref{mq PCdS}), for a given $A$,
$\Lambda$ and $\rho$. When $A$ goes to zero we have
$x_\mathrm{s}\rightarrow -1$ and $x_\mathrm{n}\rightarrow +1$. The
period of $\phi$, $\Delta \phi _{\rm c}$, that avoids the conical
singularity at the north pole (and leaves one at the south pole
responsible for the presence of the string) is given by
(\ref{Period phi}) with $x_\mathrm{n}$ defined in
(\ref{polos-cold}).

The topology of the cold C instanton is ${\mathbb{R}}^2 \times
S^2$, since $y=y_+=\rho$ is at an infinite proper distance ($0
\leq \tau \leq \beta_{\rm c}$, $y_A \leq y < y_+$,
 $x_\mathrm{s}\leq x \leq x_\mathrm{n}$,
and $0 \leq \phi \leq \Delta \phi _{\rm c}$). The surface
$y=y_+=\rho$ is then an internal infinity boundary that will have
to be taken into account in the calculation of the action of the
cold C instanton (see section \ref{sec:Cold-rate}). The Lorentzian
sector of this cold case describes two extreme ($y_+=y_-$) dS
black holes being accelerated by the cosmological background and
by the string, and the cold C instanton describes pair creation of
these extreme black holes.

\subsubsection{\label{sec:Nariai-inst}The Nariai C instanton}
In the case of the Nariai C instanton, we  require that the size
of the acceleration horizon $y_A$ is equal to the size of the
outer charged horizon $y_+$. Let us label this degenerated horizon
by $\rho$: $y_A=y_+\equiv \rho$ and $\rho < y_-$.
 In this case, the function ${\cal F}(y)$ can be written as (onwards the subscript
 ``${\rm N}$" means Nariai)
\begin{eqnarray}
{\cal F}_{\rm N}(y) =\frac{\rho^2-3\gamma}{\rho^4}
 (y-y_{\rm neg})(y-y_-)(y-\rho)^2\:,
 \label{F-cNariai}
 \end{eqnarray}
where $\gamma$ is defined by (\ref{gamma-PCdS}), the roots $\rho$
and $y_{\rm neg}$ are given by (\ref{zerosy1-cold}) and
(\ref{zerosy2-cold}), and $y_-$ is given by $y_- =\frac{\gamma
\rho}{\rho^2-3\gamma}
 \left ( 1+ \sqrt{\frac{\rho^2-2\gamma}{\gamma}} \:\right )$.
The mass and the charge of the solution are defined as a function
of $\rho$ by (\ref{mq PCdS}). The conditions $\rho < y_-$ and $q^2
\geq 0$ require that for the Nariai C instanton, the allowed range
of $\rho$ is
\begin{eqnarray}
 \sqrt{3\gamma} \leq \rho<\sqrt{6\gamma}\:.
 \label{range-gamma-cNariai}
\end{eqnarray}
The value of $y_-$ decreases monotonically with $\rho$ and we have
$\sqrt{6\gamma}<y_-<+\infty$. Contrary to the cold C instanton,
the mass and the charge of the Nariai C instanton are
monotonically increasing functions of $\rho$, and as we go from
$\rho=\sqrt{3\gamma}$ to $\rho=\sqrt{6\gamma}$ we have
\begin{eqnarray}
& &  \frac{1}{3}
  \frac{1}{\sqrt{\Lambda+3A^2}}\leq m_{\rm N}< \frac{\sqrt{2}}{3}
  \frac{1}{\sqrt{\Lambda+3A^2}}\:, \nonumber \\
& & 0\leq q_{\rm N}< \frac{1}{2}
  \frac{1}{\sqrt{\Lambda+3A^2}}\:.
 \label{mq-cNariai PCdS}
 \end{eqnarray}
Note that $\rho= \sqrt{3\gamma}$ implies $q=0$.  For a fixed
$\Lambda$ and for a  mass fixed between $\sqrt{1/(9 \Lambda)}\leq
m<\sqrt{2/(9 \Lambda)}$, the acceleration varies as
$\sqrt{1/(27m^2)-\Lambda/3} \leq A <\sqrt{2/(27m^2)-\Lambda/3}$.

At this point, one has apparently a problem that is analogous to
the one that occurs with the $A=0$ neutral Nariai instanton
\cite{GinsPerry} and with the $A=0$ charged Nariai instanton
\cite{MannRoss,HawkRoss}. Indeed, as we said in the beginning of
this section, the allowed range of $y$ in the Euclidean sector is
$y_A \leq y \leq y_+$ in order to obtain a positive definite
metric. But in the Nariai case $y_A=y_+$, and so it seems that we
are left with no space to work with in the Euclidean sector.
However, as in \cite{GinsPerry,MannRoss,HawkRoss}, the proper
distance between $y_A$ and $y_+$ remains finite as $y_A
\rightarrow y_+$, as is shown in detail in \cite{OscLem_nariai}
where the Nariai C-metric is constructed and analyzed. In what
follows we briefly exhibit the construction. We first set
$y_A=\rho-\varepsilon$ and $y_+=\rho+\varepsilon$, in order that
$\varepsilon<<1$ measures the deviation from degeneracy, and the
limit $y_A\rightarrow y_+$ is obtained when $\varepsilon
\rightarrow 0$. Now, we introduce a new time coordinate
$\tilde{\tau}$, $\tau= \frac{1}{\varepsilon {\cal
K}}\,\tilde{\tau}$, and a new radial coordinate $\chi$,
$y=\rho+\varepsilon \cos\chi$, where $\chi=0$ and $\chi=\pi$
correspond, respectively, to the horizons $y_+$ and $y_A$, and
\begin{eqnarray}
{\cal K} = \frac{2(\Lambda+3A^2)}{A^2\rho^2}-1\:.
 \label{Kfactor PCdS}
\end{eqnarray}
Condition (\ref{range-gamma-cNariai})  implies $0<{\cal K}\leq 1$
with $q=0\Rightarrow {\cal K}=1$. Then, in the limit $\varepsilon
\rightarrow 0$, from  (\ref{C-metric PCdS}) and (\ref{F-cNariai}),
we obtain the gravitational field of the Nariai C instanton
\begin{eqnarray}
d s^2 = \frac{{\cal R}^2(x)}{{\cal K}} \left (\sin^2\chi\,
d\tilde{\tau}^2 +d\chi^2\right ) +
 {\cal R}^2(x)\left [{\cal G}^{-1}(x)dx^2+ {\cal G}(x)d\phi^2 \right ]
 \:.
 \label{Nariai-C-Metric PCdS}
\end{eqnarray}
where $\chi$ runs from $0$ to $\pi$, and
\begin{eqnarray}
 {\cal R}^2(x)&=& \left [ A(x+\rho) \right ]^{-2}\:.
\label{Rfactor PCdS}
\end{eqnarray}
In the new coordinates $\tilde{\tau}$ and $\chi$, the Maxwell
field for the magnetic case is still given by (\ref{F-mag}), while
in the electric case we have now
 \begin{eqnarray}
 F_{\rm el}=i\,\frac{q}{{\cal K}}\,\sin \chi \,d\tilde{\tau}\wedge d\chi\:.
\label{F-el-Nariai PCdS}
\end{eqnarray}
The period of $\tilde{\tau}$ of the Nariai C instanton is simply
$\beta_{\rm N}=2 \pi$.
The function ${\cal G}(x)$ is given by (\ref{FG}), with $m$ and
$q$ fixed by
 (\ref{mq PCdS}) for a given $A$, $\Lambda$ and $\sqrt{3\gamma} <
 \rho<\sqrt{6\gamma}$. Under these conditions the south and north
 pole (which are the only real roots) are also given by
 (\ref{polos-cold}) and (\ref{acess-zeros-ang PCdS}). The period of
$\phi$, $\Delta \phi _{\rm N}$, that avoids the conical
singularity at the north pole (and leaves one at the south pole
responsible for the presence of the string) is given by
(\ref{Period phi}) with $x_\mathrm{n}$ defined in
(\ref{polos-cold}).

The topology of the Nariai C instanton is $S^2 \times S^2$
 ($0 \leq \tilde{\tau} \leq \beta_{\rm N}$, $0\leq \chi \leq \pi$,
 $x_\mathrm{s}\leq x \leq x_\mathrm{n}$,
and $0 \leq \phi \leq \Delta \phi _{\rm N}$). The Nariai C
instanton transforms into the Nariai instanton when we take the
limit $A=0$ \cite{OscLem_nariai}.
 The Lorentzian sector is conformal to the direct topological product of
 $dS_2 \times S^2$, i.e., of a
(1+1)-dimensional de Sitter spacetime with a deformed 2-sphere of
fixed size. To each point in the sphere corresponds a $dS_2$
spacetime, except for one point - the south pole - which
corresponds a $dS_2$ spacetime with a string \cite{OscLem_nariai}.
When we set $A=0$ the $S^2$ is a round 2-sphere free of the
conical singularity and so, without the string. In this case it
has been shown \cite{GinsPerry,BoussoHawk} that the Nariai
solution decays through the quantum tunnelling process into a
slightly non-extreme dS black hole pair (for a complete review  on
this subject see, e.g., Bousso \cite{Bousso60y} and chapter
chapter \ref{chap:Extremal Limits} on this thesis). We then
naturally expect that an analogous quantum instability is present
in the Nariai C-metric. Therefore, the Nariai C instanton
describes the creation of a Nariai C universe that then decays
into a slightly non-extreme ($y_A \sim y_+$) pair of black holes
accelerated by the cosmological background and by the string.

We recall again that the neutral Nariai C instanton with
$m=\frac{1}{3} \frac{1}{\sqrt{\Lambda+3A^2}}$ is the only regular
Euclidean solution that can be constructed from the dS C-metric
when the charge vanishes. The same feature is present in the $A=0$
case where only the neutral Nariai instanton with $m=\frac{1}{3}
\frac{1}{\sqrt{\Lambda}}$ is available
\cite{GinsPerry,MannRoss,BoussoHawk,VolkovWipf}.
\subsubsection{\label{sec:Ultracold-inst}The ultracold C instanton}

In the case of the ultracold C instanton, we  require that  the
size of the three horizons ($y_A$, $y_+$ and $y_-$) is equal, and
let us label this degenerated horizon by $\rho$:
$y_A=y_+=y_-\equiv \rho$.
 In this case, the function ${\cal F}(y)$ can be written as (onwards
the subscript ``${\rm u}$" means ultracold)
\begin{eqnarray}
{\cal F}_{\rm u}(y) =\frac{\rho^2-3\gamma}{\rho^4}
 (y-y_{\rm neg})(y-\rho)^3\:,
 \label{F-ultra}
 \end{eqnarray}
where $\gamma$ is defined by (\ref{gamma-PCdS}) and the roots
$\rho$ and $y_{\rm neg}$ are given by (\ref{zerosy1-cold}) and
(\ref{zerosy2-cold}), respectively. Given the values of $A$ and
$\Lambda$, $\rho$ can take only the value
\begin{eqnarray}
\rho=\sqrt{6\gamma} \:.
 \label{range-gamma-ultra}
\end{eqnarray}
 The mass and the charge of the solution, defined by (\ref{mq-cold PCdS}), are then given by
\begin{eqnarray}
& &m_{\rm u} =\frac{\sqrt{2}}{3}
 \sqrt{\frac{1}{\Lambda+3A^2}}
 \:,  \nonumber \\
& & q_{\rm u} =\frac{1}{2}\sqrt{\frac{1}{\Lambda+3A^2}}
 \:,
 \label{mq-ultra}
 \end{eqnarray}
and these values are the maximum values of the mass and charge of
both the cold and charged Nariai instantons.

Being a limiting case of both the cold C instanton and of the
charged Nariai C instanton, the ultracold C instanton presents
similar features. The appropriate analysis of this solution (see
\cite{OscLem_nariai} for a detailed discussion) requires that we
first set $\rho=\sqrt{6\gamma}-\varepsilon$ and
$y_-=\sqrt{6\gamma}+\varepsilon$, with $\varepsilon<<1$. Then, we
introduce a new time coordinate $\tilde{\tau}$, $\tau= \frac{1}{2
\varepsilon^2 {\cal K}}\,\tilde{\tau}$, and a new radial
coordinate $\chi$, $y=\sqrt{6\gamma}+
 \varepsilon \cosh(\sqrt{2\varepsilon{\cal K}}\:\chi)$,
where ${\cal K} =\frac{1}{3}
 \sqrt{\frac{2A^2}{\Lambda+3A^2}}$.
Finally, in the limit $\varepsilon \rightarrow 0$, from
(\ref{C-metric PCdS}) and  (\ref{F-ultra}), we obtain the
gravitational field of the ultracold C instanton
\begin{eqnarray}
& & d s^2 = {\cal R}^2(x) \left [\chi^2\, d\tilde{\tau}^2 +d\chi^2
 +{\cal G}^{-1}(x)dx^2+ {\cal G}(x)d\phi^2 \right ],
 \nonumber \\
& & {\rm with} \qquad {\cal R}^2(x)=\left
(Ax+\sqrt{2(\Lambda+3A^2)}\right )^{-2}\:.
 \label{ultracold}
\end{eqnarray}
Notice that the spacetime factor $\chi^2\, d\tilde{\tau}^2
+d\chi^2$ is just Euclidean space in Rindler coordinates, and
therefore, under the usual coordinate transformation, it can be
putted in the form $dT^2+dX^2$. $\chi=0$ corresponds to the
Rindler horizon and $\chi=+\infty$ corresponds an internal
infinity boundary. In the new coordinates $\tilde{\tau}$ and
$\chi$,  the Maxwell field for the magnetic case is still given by
(\ref{F-mag}), while in the electric case we have now
 \begin{eqnarray}
 F_{\rm el}=-i\,q\,\chi \,d\tilde{\tau}\wedge d\chi\:.
\label{F-el-ultracold}
\end{eqnarray}
The period of $\tilde{\tau}$ of the ultracold C instanton is
simply $\beta_{\rm u}=2 \pi$.

 The function ${\cal G}(x)$ is given
by (\ref{FG}), with $m$ and $q$ fixed by
  (\ref{mq-ultra}) for a given $A$ and $\Lambda$. Under these
 conditions the south and north pole (which are the only real roots)
 are also given by  (\ref{polos-cold}) and (\ref{acess-zeros-ang PCdS}).
The period of $\phi$, $\Delta \phi _{\rm u}$, that avoids the
conical singularity at the north pole  is given by (\ref{Period
phi}) with $x_\mathrm{n}$ defined in (\ref{polos-cold}).

The topology of the ultracold C instanton is ${\mathbb{R}}^2
\times S^2$, since $\chi=+\infty$ is at an
 infinite proper distance
 ($0 \leq \tilde{\tau} \leq \beta_{\rm u}$, $0\leq \chi \leq \infty$,
 $x_\mathrm{s}\leq x \leq x_\mathrm{n}$,
 and $0 \leq \phi \leq \Delta \phi _{\rm u}$).
The surface $\chi=+\infty$ is then an internal infinity boundary
that will have to be taken into account in the calculation of the
action of the ultracold C instanton (see section
\ref{sec:Ultracold-rate}). The ultracold C instanton transforms
into the ultracold instanton when we take the limit $A=0$
\cite{OscLem_nariai}.
 The Lorentzian sector is conformal to the direct topological product of
 ${\mathbb{M}}^{1,1}\times S^2$, i.e., of a
(1+1)-dimensional Minkowski spacetime with a deformed 2-sphere of
fixed size. To each point in the sphere corresponds a
${\mathbb{M}}^{1,1}$ spacetime, except for one point - the south
pole - which corresponds a ${\mathbb{M}}^{1,1}$ spacetime with a
string \cite{OscLem_nariai}. We can, appropriately, label this
solution as Nariai Bertotti-Robinson C universe (see
\cite{OscLem_nariai}). When we set $A=0$ the $S^2$ is a round
2-sphere free of the conical singularity and so, without the
string. In an analogous way to the Nariai C universe, the Nariai
Bertotti-Robinson C universe is unstable and decays into a
slightly non-extreme ($y_A \sim y_+ \sim y_-$) pair of black holes
accelerated by the cosmological background and by the string. The
ultracold C instanton mediates this decay.

\begin{figure} [H]
\centering
\includegraphics[height=2.1in]{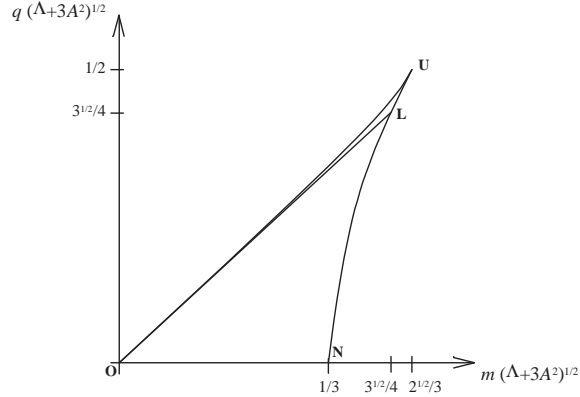}
\caption{\label{mq-fig}
 Relation $q\sqrt{\Lambda+3A^2}$ vs
$m\sqrt{\Lambda+3A^2}$ for a fixed value of $A$ and $\Lambda$ for
the four dS C instantons. $OL$ represents the lukewarm C instanton
($m=q$), $OU$ represents the cold C instanton, $NLU$ represents
the Nariai C instanton, and $U$ represents the ultracold C
instanton. Point $N$ represents the neutral Nariai C instanton,
the only instanton for the uncharged case. When we set $\Lambda=0$
we obtain the charge/mass relation of the flat C-metric and flat
Ernst instantons, and setting $A=0$ yields the charge/mass
relation for the dS instantons \cite{MannRoss}. This reveals the
close link that exists between the instantons that describe the
pair creation process in different backgrounds. $c_1=1/3$,
$c_2=\sqrt{3}/4$ and $c_3=\sqrt{2}/3$.
 }
\end{figure}

The allowed range of $m$ and $q$ for each one of the four C
instantons is sketched in Fig. \ref{mq-fig}.
\subsection{\label{sec:Calc-I}Calculation of the black hole pair creation rates}
In the last section we have found the regular dS C instantons that
can be interpreted as describing a circular motion in the
Euclidean sector of the solution with a period $\beta$. These
instantons are the Euclidean continuation of a Lorentzian solution
which describes a pair of black holes that start at rest at the
creation moment, and then run hyperbolically in opposite
directions by the strings that accelerate them. Now, in order to
compute the pair creation rate of the corresponding black holes,
we have to slice the instanton, the Euclidean trajectory, in half
along $\tau=0$ and $\tau=\beta/2$, where the velocities vanish.
The resulting geometry is precisely that of the moment of closest
approach of the black holes in the Lorentzian sector of the dS
C-metric, and in this way the Euclidean and Lorentzian solutions
smoothly match together. In particular, the extrinsic curvature
vanishes for both surfaces and therefore, they can be glued to
each other.

In this section, we will compute the black hole pair creation
rates given in (\ref{PC-rate}) for each one of the four cases
considered in section \ref{sec:dS C-inst}. Moreover, we will also
compute the pair creation rate of black holes whose nucleation
process is described by sub-maximal instantons. For the reason
explained in section \ref{sec:instanton-method}, the magnetic
action will be evaluated using (\ref{I}), while the electric
action will be computed using (\ref{I-electric}).  In general, in
(\ref{I}) and (\ref{I-electric}) we identify $\Sigma=\partial
{\cal{M}}$ with the surfaces of zero extrinsic curvature discussed
just above. It then follows that the boundary term $\int_{\Sigma}
d^3x\sqrt{h}\, K$ vanishes. However, as we already mentioned in
sections \ref{sec:Cold-inst} and \ref{sec:Ultracold-inst}, in the
cold and ultracold case we have in addition an internal infinity
boundary, $\Sigma^{\rm int}_{\infty}$, for which the boundary
action term does not necessarily vanish.

The domain of validity of our results is the  particle limit,
$mA\ll 1$, for which the  radius of the black hole, $r_+ \sim m$,
is much smaller than the typical distance between the black holes
at the creation moment, $\ell \sim 1/A$.

In order to compute the black hole pair creation rate given by
(\ref{PC-rate-breakstring}) or (\ref{PC-rate}), we need to find
$I_{\rm string}$ and $I_{\rm dS}$. The evaluation of the action of
the string instanton that mediates the nucleation of a string in
the dS spacetime is done using
 (\ref{backg metric-dS}) and (\ref{delta phi0-dS}), yielding
\begin{eqnarray}
I_{\rm string}&=&-\frac{1}{16\pi}\int_{\cal{M}} d^4x\sqrt{g}
\left ( R-2\Lambda \right )\nonumber \\
&=& -\frac{3\pi}{2\Lambda}\frac{{\cal G}'(x_{\rm s})}{|{\cal
G}'(x_{\rm n})|} \,,
 \label{I-string}
 \end{eqnarray}
where the integration over $\tau$ has been done in the interval
$[0,\beta_{\rm 0}/2]$ with $\beta_{\rm
0}=2\pi/\sqrt{1+\Lambda/(3A_0^{\,2})}$, the integration range of
$x$ was $[-1,1]$, the integration interval of $y$ was
$[\sqrt{1+\Lambda/(3A_0^{\,2})}\,,\infty]$, and $R=4\Lambda$.

The action of the $S^4$ gravitational instanton that mediates the
nucleation of the dS spacetime is \cite{BoussoHawk,VolkovWipf}
\begin{eqnarray}
I_{\rm dS}=-\frac{3\pi}{2\Lambda} \,.
 \label{I-dS}
 \end{eqnarray}

The nucleation rate of a string in a dS background is then given
by (\ref{PC-rate-string}). For the particle limit, $mA\ll 1$, the
mass density of the string, $\mu$, is given by $\mu \simeq mA$ and
\begin{eqnarray}
\Gamma_{\rm string/dS} \sim e^{- 12\pi \frac{\mu}{\Lambda}} \:.
 \label{PC-rate-stringend}
 \end{eqnarray}
Thus, the nucleation probability of a string in the dS background
decreases when its mass density increases.
\subsubsection{\label{sec:Lukewarm-rate}The lukewarm C pair creation rate}

We first consider the magnetic case, whose Euclidean action is
given by (\ref{I}), and then we consider the electric case, using
(\ref{I-electric}), and we verify that these two quantities give
the same numerical value. The boundary $\Sigma=\partial \cal{M}$
that appears in (\ref{I}) consists of an initial spatial surface
at $\tau=0$ plus a final spatial surface at $\tau=\beta_{\rm
\ell}/2$. We label these two 3-surfaces by $\Sigma_{\tau}$. Each
one of these two spatial 3-surfaces is delimitated by a 2-surface
at the acceleration horizon and by a 2-surface at the outer black
hole horizon. The two surfaces $\Sigma_{\tau}$ are connected by a
timelike 3-surface that intersects $\Sigma_{\tau}$ at the frontier
$y_A$ and by a timelike 3-surface that intersects $\Sigma_{\tau}$
at the frontier $y_+$. We label these two timelike 3-surfaces by
$\Sigma_{h}$. Thus $\Sigma=\Sigma_{\tau}+\Sigma_{h}$, and the
region $\cal{M}$ within it is compact. With the analysis of
section \ref{sec:Lukewarm-inst}, we can compute all the terms of
action (\ref{I}). We start with
\begin{eqnarray}
\!\!\!\!\!\!\!\!\!\!\!\!\!\!-\frac{1}{16\pi}\int_{\cal{M}}
d^4x\sqrt{g} \left ( R-2\Lambda \right )=
       -\frac{1}{16\pi}\int_{\Delta \phi _{\rm \ell}}
\!\!\!\!d\phi
 \int_0^{\beta_{\rm \ell}/2} \!\!\!\!d\tau
 \int_{x_\mathrm{s}}^{x_\mathrm{n}}\!\!\!\! dx
 \int_{y_A}^{y_+} \!\!\!\!dy \frac{2 \Lambda}{\left [A(x+y) \right ]^4} \:,
 \label{I1-luk}
 \end{eqnarray}
where we have used $R=4\Lambda$, and $y_A$ and $y_+$ are given by
(\ref{yA-luk}), $x_\mathrm{s}$ and $x_\mathrm{n}$ are defined by
(\ref{polos-luk}), and $\beta_{\rm \ell}$ and $\Delta \phi _{\rm
\ell}$ are respectively given by  (\ref{beta-luk}) and
(\ref{Period phi-luk}). The Maxwell term in the action yields
\begin{eqnarray}
 \!\!\!\!\!\!\!\!\!\!\!\!\!\! \frac{1}{16\pi}\int_{\cal{M}} d^4x\sqrt{g}
 \:F_{\rm mag}^2
 =\frac{q^2}{16\pi}\,\Delta \phi _{\rm \ell}\, \beta_{\rm
\ell}\,(x_\mathrm{n}-x_\mathrm{s})(y_+ -y_A) \:,
 \label{I2-luk}
 \end{eqnarray}
where we have used $F_{\rm mag}^2=2q^2A^4(x+y)^4$ [see
(\ref{F-mag})], and $\int_{\Sigma} d^3x\sqrt{h}\, K=0$. Adding all
these terms yields for the magnetic action (\ref{I})
\begin{eqnarray}
\!\!\!\!\!\!\!\!\!\! I_{\rm mag}^{\rm
\ell}=-\frac{3\pi}{16\Lambda} \frac{1}{8mA}
 \left ( 1-\sqrt{\frac{1-4mA}{1+4mA}} \right )
  \left ( 1+\sqrt{1-(4mA)^2}-\frac{4m}{\sqrt{3}}\sqrt{\Lambda+3A^2}
   \right )\,,
\label{I-mag-luk}
 \end{eqnarray}
and, given that the string is already present in the initial
system, the pair creation rate of nonextreme lukewarm black holes
when the cosmic string breaks is
\begin{eqnarray}
\Gamma_{\rm BHs/string}^{\rm \ell}=\eta\,e^{-2I_{\rm mag}^{\rm
\ell}+2I_{\rm string}}\,, \label{PC rate-luk}
 \end{eqnarray}
where (\ref{I-string}) yields $I_{\rm string}=
-\frac{3\pi}{2\Lambda}\frac{\sqrt{1-4mA}}{\sqrt{1+4mA}}$, and
$\eta$ is the one-loop contribution not computed here.
 $I_{\rm mag}^{\rm \ell}$ is a monotonically increasing function of
both $m$ and $A$ (for a fixed $\Lambda$). When we take the limit
$A=0$ on (\ref{I-mag-luk}) we get
\begin{eqnarray}
I_{\rm mag}^{\rm \ell}{\biggl |}_{A\rightarrow
0}=-\frac{3\pi}{2\Lambda}+\pi \,m \sqrt{\frac{3}{\Lambda}} \,,
\label{I-mag-luk-A=0}
 \end{eqnarray}
 recovering the action for the $A=0$ lukewarm
instanton \cite{MannRoss}, that describes the pair creation of two
non-extreme dS$-$Reissner-Nordstr\"{o}m black holes accelerated
only by the cosmological constant.

In the electric case, the Euclidean action is given by
(\ref{I-electric}) with  $F_{\rm el}^2=-2q^2A^4(x+y)^4$ [see
(\ref{F-el})]. Thus,
\begin{eqnarray}
\frac{1}{16\pi}\int_{\cal{M}} d^4x\sqrt{g}
 \:F_{\rm el}^2=- \frac{1}{16\pi}\int_{\cal{M}} d^4x\sqrt{g}
 \:F_{\rm mag}^2\:.
 \label{I2-luk-elect}
 \end{eqnarray}
In order to compute the extra Maxwell boundary term in
(\ref{I-electric}) we have to find a vector potential, $A_{\nu}$,
that is regular everywhere including at the horizons. An
appropriate choice in the lukewarm case is $A_y=- i\,q\,\tau$,
which obviously satisfies (\ref{F-el}). The integral over $\Sigma$
consists of an integration between $y_A$ and $y_+$ along the
$\tau=0$ surface and back along  $\tau=\beta_{\rm \ell}/2$, and of
an integration between $\tau=0$ and $\tau=\beta_{\rm \ell}/2$
along the $y=y_+$ surface and back along the $y=y_A$ surface. The
normal to $\Sigma_{\tau}$ is
$n_{\mu}=(\sqrt{{\cal{F}}}/[A(x+y)],0,0,0)$, and the normal to
$\Sigma_{h}$ is $n_{\mu}=(0,\sqrt{{\cal{F}}}/[A(x+y)],0,0)$. Thus
$F^{\mu\nu}n_{\mu}A_{\nu}=0$ on $\Sigma_{h}$, and the
non-vanishing contribution comes only from the integration along
the $\tau=\beta_{\rm \ell}/2$ surface. The Maxwell boundary term
in (\ref{I-electric}), $-\frac{1}{4\pi}\int_{\Sigma}
d^3x\sqrt{h}\, F^{\mu\nu}n_{\mu}A_{\nu}$, is then
 \begin{eqnarray}
-\frac{1}{4\pi}\int_{\Sigma_{\tau=\beta_{\rm \ell}/2} } \!\!\!\!
d^3x \sqrt{g_{yy}g_{xx}g_{\phi\phi}}\: F^{\tau y} n_{\tau} A_{y}=
 \frac{q^2}{8\pi}\,\Delta \phi _{\rm
\ell}\, \beta_{\rm \ell}\,(x_\mathrm{n}-x_\mathrm{s})(y_+ -y_A)\:.
 \label{I-electric-luk}
 \end{eqnarray}
Adding  (\ref{I1-luk}), (\ref{I2-luk-elect}) and
(\ref{I-electric-luk}) yields for the electric action
(\ref{I-electric})
\begin{eqnarray}
 I_{\rm el}^{\rm \ell}= I_{\rm mag}^{\rm \ell}\,,
\label{I-elect-luk}
 \end{eqnarray}
where $I_{\rm mag}^{\rm \ell}$ is given by  (\ref{I-mag-luk}).

\begin{figure} [H]
\centering
\includegraphics[height=2.1in]{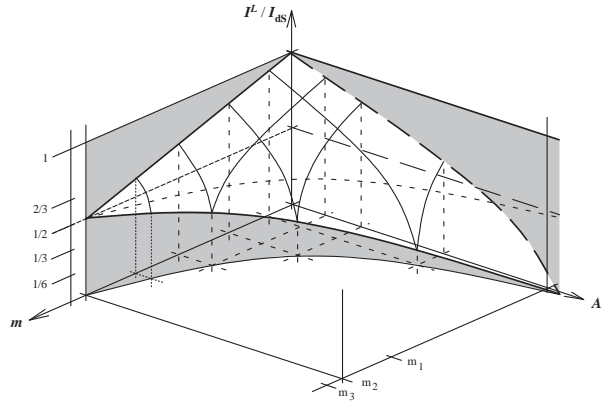}
\caption{\label{lukewarm-fig} Plot of $I^{\rm \ell}/I_{\rm dS}$ as
a function of $m$ and $A$ for a fixed $\Lambda$, where $I^{\rm
\ell}$ is the action of the lukewarm C instanton, given in
(\ref{I-mag-luk}), and $I_{\rm dS}=-\frac{3\pi}{2\Lambda}$ is the
action of de Sitter space.
 See text of section \ref{sec:Lukewarm-rate}.  $m_1=1/(3\sqrt{\Lambda})$,
$m_2=\sqrt{3}/(4\sqrt{\Lambda})$ and
$m_3=\sqrt{2}/(3\sqrt{\Lambda})$.
 }
\end{figure}

In Fig. \ref{lukewarm-fig} we show a plot of $I^{\rm \ell}/I_{\rm
dS}$ as a function of $m$ and $A$ for a fixed $\Lambda$, where
$I_{\rm dS}=-\frac{3\pi}{2\Lambda}$ is the action of de Sitter
space. Given the pair creation rate, $\Gamma_{\rm BHs/string}^{\rm
\ell}\propto e^{-2I_{\rm mag}^{\rm \ell}+2I_{\rm string}}$, we
conclude that, for a fixed $\Lambda$ and $A$, as the mass and
charge of the lukewarm black holes increase, the probability they
have to be pair created decreases monotonically. Moreover, for a
fixed mass and charge, this probability increases monotonically as
the acceleration provided by the string increases. Alternatively,
we can discuss the behavior of $\Gamma_{\rm BHs/dS}^{\rm
\ell}\propto e^{-2I_{\rm mag}^{\rm \ell}+2I_{\rm dS}}$. In this
case, for a fixed mass and charge, the probability decreases
monotonically as the acceleration of the black holes increases.

\subsubsection{\label{sec:Cold-rate}The cold C pair creation rate}

\begin{figure} [H]
\centering
\includegraphics[height=2.1in]{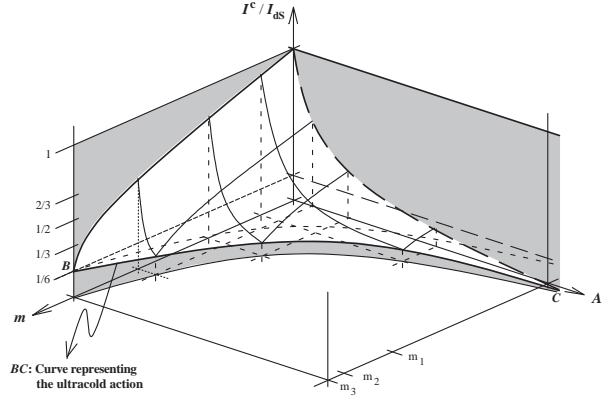}
\caption{\label{cold-fig}
 Plot of $I^{\rm c}/I_{\rm dS}$ as a
function of $m$ and $A$ for a fixed $\Lambda$, where $I^{\rm c}$
is the action of the cold C instanton, given in
(\ref{I-mag-cold}), and $I_{\rm dS}=-\frac{3\pi}{2\Lambda}$ is the
action of de Sitter space (see text of section
\ref{sec:Cold-rate}). The plot for the ultracold C action, given
in (\ref{I-mag-ultracold}), is also represented by the curve $BC$
(see text of section \ref{sec:Ultracold-rate}).
$m_1=1/(3\sqrt{\Lambda})$, $m_2=\sqrt{3}/(4\sqrt{\Lambda})$ and
$m_3=\sqrt{2}/(3\sqrt{\Lambda})$.
 }
\end{figure}

We first consider the magnetic case, whose Euclidean action is
given by (\ref{I}). The boundary that appears in (\ref{I}) is
given by $\Sigma=\Sigma_{\tau}+\Sigma_{h}+\Sigma^{\rm
int}_{\infty}$, where $\Sigma_{\tau}$ is a spatial surface at
$\tau=0$ and $\tau=\beta_{\rm c}/2$, $\Sigma_{h}$ is a timelike
3-surface at $y=y_A$, and the timelike 3-surface $\Sigma^{\rm
int}_{\infty}$ is an internal infinity boundary at $y=y_+=\rho$.
With the analysis of section \ref{sec:Cold-inst}, we can compute
all the terms of action (\ref{I}). We start with
\begin{eqnarray}
-\frac{1}{16\pi}\int_{\cal{M}} d^4x\sqrt{g} \left ( R-2\Lambda
\right )=
      -\frac{1}{16\pi}\int_{\Delta \phi _{\rm c}}
\!\!\!\!d\phi
 \int_0^{\beta_{\rm c}/2} \!\!\!\!d\tau
 \int_{x_\mathrm{s}}^{x_\mathrm{n}}\!\!\!\! dx
 \int_{y_A}^{\rho} \!\!\!\!dy \frac{2 \Lambda}{\left [A(x+y) \right ]^4} \:,
 \label{I1-cold}
 \end{eqnarray}
where we used $R=4\Lambda$, and $\rho$ and $y_A$ are respectively
given by (\ref{zerosy1-cold}) and (\ref{zerosy3-cold}),
$x_\mathrm{s}$ and $x_\mathrm{n}$ are defined by
(\ref{polos-cold}), and $\beta_{\rm c}$ and $\Delta \phi _{\rm c}$
are, respectively, given by (\ref{beta-cold}) and (\ref{Period
phi}). The Maxwell term in the action yields
\begin{eqnarray}
 \frac{1}{16\pi}\int_{\cal{M}} d^4x\sqrt{g}
 \:F_{\rm mag}^2
 =\frac{q^2}{16\pi}\,\Delta \phi _{\rm c}\, \beta_{\rm
c}\,(x_\mathrm{n}-x_\mathrm{s})(\rho -y_A) \:,
 \label{I2-cold}
 \end{eqnarray}
where we used $F_{\rm mag}^2=2q^2A^4(x+y)^4$ [see (\ref{F-mag})],
and $\int_{\Sigma} d^3x\sqrt{h}\, K=0$. Adding all these terms
yields for the magnetic action (\ref{I}) of the cold case
\begin{eqnarray}
I_{\rm mag}^{\rm c}=-\frac{\Delta \phi _{\rm c}} {8A^2}\,
 \frac{ x_\mathrm{n}-x_\mathrm{s} }
   {(x_\mathrm{n}+y_A)(x_\mathrm{s}+y_A)}\,,
    \label{I-mag-cold}
 \end{eqnarray}
where $y_A$ is given by (\ref{zerosy3-cold}), $x_\mathrm{s}$ and
$x_\mathrm{n}$ are defined by (\ref{polos-cold}), and
 $\Delta \phi _{\rm c}$ is
given by (\ref{Period phi}). Given that the string is already
present in the initial system, the pair creation rate of extreme
cold black holes when the string breaks is $\Gamma_{\rm
BHs/string}^{\rm c}=\eta\,e^{-2I_{\rm mag}^{\rm c}+2I_{\rm
string}}$, where $I_{\rm string}$ is given by (\ref{I-string}),
and $\eta$ is the one-loop contribution not computed here. In Fig.
\ref{cold-fig} we show a plot of $I_{\rm mag}^{\rm c}/I_{\rm dS}$
as a function of $m$ and $A$ for a fixed $\Lambda$. Given the pair
creation rate, $\Gamma_{\rm BHs/string}^{\rm c}$, we conclude that
for a fixed $\Lambda$ and $A$ as the mass and charge of the cold
black holes increases, the probability they have to be pair
created decreases monotonically. Moreover, for a fixed mass and
charge, this probability increases monotonically as the
acceleration of the black holes increases. Alternatively, we can
discuss the behavior of $\Gamma_{\rm BHs/dS}^{\rm c}\propto
e^{-2I_{\rm mag}^{\rm c}+2I_{\rm dS}}$. In this case, for a fixed
mass and charge, the probability decreases monotonically as the
acceleration of the black holes increases.
 When we take the limit $A=0$ we recover the action for the $A=0$
cold instanton \cite{MannRoss}, which lies in the range
$-\frac{3\pi}{2\Lambda} \leq I_{\rm mag}^{\rm c}{\bigl
|}_{A\rightarrow 0}\leq -\frac{\pi}{4\Lambda}$, and which
describes the pair creation of extreme dS$-$Reissner-Nordstr\"{o}m
black holes accelerated only by the cosmological constant.

In the electric case, the Euclidean action is given by
(\ref{I-electric}) with  $F_{\rm el}^2=-2q^2A^4(x+y)^4$ [see
(\ref{F-el})]. Thus,
\begin{eqnarray}
\frac{1}{16\pi}\int_{\cal{M}} d^4x\sqrt{g}
 \:F_{\rm el}^2=- \frac{1}{16\pi}\int_{\cal{M}} d^4x\sqrt{g}
 \:F_{\rm mag}^2\:.
 \label{I2-cold-elect}
 \end{eqnarray}
In order to compute the extra Maxwell boundary term in
(\ref{I-electric}) we have to find a vector potential, $A_{\nu}$,
that is regular everywhere including at the horizons. An
appropriate choice in the cold case is $A_y=- i\,q\,\tau$, which
obviously satisfies (\ref{F-el}). Analogously to the lukewarm
case, the non-vanishing contribution to the Maxwell boundary term
in (\ref{I-electric}) comes only from the integration along the
$\tau=\beta_{\rm c}/2$ surface, and is given by
 \begin{eqnarray}
-\frac{1}{4\pi}\int_{\Sigma_{\tau=\beta_{\rm c}/2}} \!\!\!\! d^3x
\sqrt{g_{yy}g_{xx}g_{\phi\phi}}\: F^{\tau y} n_{\tau} A_{y}=
\frac{q^2}{8\pi}\,\Delta \phi _{\rm c}\,\beta_{\rm
c}\,(x_\mathrm{n}-x_\mathrm{s})(\rho -y_A)\:.
 \label{I-electric-cold}
 \end{eqnarray}
Adding (\ref{I2-cold-elect}) and (\ref{I-electric-cold}) yields
(\ref{I2-cold}). Thus, the electric action (\ref{I-electric}) of
the cold instanton is equal to the magnetic action, $I_{\rm
el}^{\rm c}= I_{\rm mag}^{\rm c}$, and therefore electric and
magnetic cold black holes have the same probability of being pair
created.

\subsubsection{\label{sec:Nariai-rate}The Nariai C pair creation rate}
The Nariai C instanton is the only one that can have zero charge.
We will first consider the charged Nariai C instanton and then the
neutral Nariai C instanton.

\begin{figure} [H]
\centering
\includegraphics[height=2.1in]{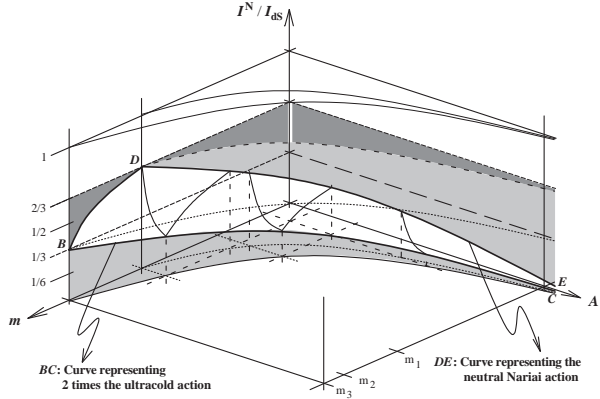}
\caption{\label{nariai-fig-PCAdS}
 Plot of $I^{\rm N}/I_{\rm dS}$ as a
function of $m$ and $A$ for a fixed $\Lambda$, where $I^{\rm N}$
is the action of the cold C instanton, given in
(\ref{I-mag-Nariai}), and $I_{\rm dS}=-\frac{3\pi}{2\Lambda}$ is
the action of de Sitter space. The neutral Nariai plot is sketched
by the curve $DE$ (see text of section \ref{sec:Nariai-rate}). The
curve $BC$ represents two times the ultracold C action, given in
(\ref{I-mag-ultracold}).  $m_1=1/(3\sqrt{\Lambda})$,
$m_2=\sqrt{3}/(4\sqrt{\Lambda})$ and
$m_3=\sqrt{2}/(3\sqrt{\Lambda})$.
 }
\end{figure}

We start with the magnetic case, whose Euclidean action is given
by (\ref{I}). The boundary that appears in (\ref{I}) is given by
$\Sigma=\Sigma_{\tilde{\tau}}+\Sigma_{h}$, where
$\Sigma_{\tilde{\tau}}$ is a spatial surface at $\tilde{\tau}=0$
and $\tilde{\tau}=\pi$, and $\Sigma_{h}$ is a timelike 3-surface
at $\chi=0$ and $\chi=\pi$. With the analysis of section
\ref{sec:Nariai-inst}, we can compute all the terms of action
(\ref{I}). We start with
\begin{eqnarray}
-\frac{1}{16\pi}\int_{\cal{M}} d^4x\sqrt{g} \left ( R-2\Lambda
\right )=
  -\frac{1}{16\pi}\int_{\Delta \phi _{\rm N}} \!\!\!\!d\phi
 \int_0^{\pi} \!\!\!\!d\tilde{\tau}
 \int_{x_\mathrm{s}}^{x_\mathrm{n}}\!\!\!\! dx
 \int_{0}^{\pi} \!\!\!\!d\chi \:
 \frac{2 \Lambda \sin \chi}{\left [A(x+\rho) \right ]^4 {\cal{K}} }\:,
 \label{I1-nariai}
 \end{eqnarray}
where we have used $R=4\Lambda$, $x_\mathrm{s}$ and $x_\mathrm{n}$
are defined by (\ref{polos-cold}), and $\Delta \phi _{\rm N}$ is
given by (\ref{Period phi}). The Maxwell term in the action yields
\begin{eqnarray}
 \frac{1}{16\pi}\int_{\cal{M}} d^4x\sqrt{g}
 \:F_{\rm mag}^2 =\frac{q^2}{4 \, \cal{K}}\,\Delta \phi _{\rm N}\,(x_\mathrm{n}-x_\mathrm{s}) \:,
 \label{I2-nariai}
 \end{eqnarray}
where we have used $F_{\rm mag}^2=2q^2A^4(x+\rho)^4$ [see
(\ref{F-mag})], and $\int_{\Sigma} d^3x\sqrt{h}\, K=0$. Adding
these three terms yields the magnetic action (\ref{I}) of the
Nariai case
\begin{eqnarray}
I_{\rm mag}^{\rm N}=
 -\,\frac{\Delta \phi _{\rm N}} {4A^2}\,
 \frac{ x_\mathrm{n}-x_\mathrm{s} }
   {(x_\mathrm{n}+\rho)(x_\mathrm{s}+\rho)}\,,
    \label{I-mag-Nariai}
 \end{eqnarray}
 where $\sqrt{3\gamma} \leq \rho<\sqrt{6\gamma}$, $x_\mathrm{s}$
and $x_\mathrm{n}$ are defined by (\ref{polos-cold}), and
 $\Delta \phi _{\rm N}$ is
given by (\ref{Period phi}), with $m$ and $q$ subjected to
(\ref{mq-cNariai PCdS}). Given that the string is already present
in the initial system, the pair creation rate of extreme Nariai
black holes when the string breaks is $\Gamma_{\rm
BHs/string}^{\rm N}=\eta\,e^{-2I_{\rm mag}^{\rm N}+2I_{\rm
string}}$, where $I_{\rm string}$ is given by (\ref{I-string}),
and $\eta$ is the one-loop contribution not computed here. In Fig.
\ref{nariai-fig-PCAdS} we show a plot of $I_{\rm mag}^{\rm
N}/I_{\rm dS}$ as a function of $m$ and $A$ for a fixed $\Lambda$.
Given the pair creation rate, $\Gamma_{\rm BHs/string}^{\rm N}$,
we conclude that for a fixed $\Lambda$ and $A$ as the mass and
charge of the Nariai black holes increases, the probability they
have to be pair created decreases monotonically. Moreover, for a
fixed mass and charge, this probability increases monotonically as
the acceleration of the black holes increases. Alternatively, we
can discuss the behavior of $\Gamma_{\rm BHs/dS}^{\rm N}\propto
e^{-2I_{\rm mag}^{\rm N}+2I_{\rm dS}}$. In this case, for a fixed
mass and charge, the probability decreases monotonically as the
acceleration of the black holes increases.
 When we take the limit $A=0$ we recover the action for the $A=0$
Nariai instanton \cite{MannRoss,HawkRoss}, which lies in the range
$-\frac{\pi}{\Lambda} \leq I_{\rm mag}^{\rm N}{\bigl
|}_{A\rightarrow 0}\leq -\frac{\pi}{2\Lambda}$, and that describes
the nucleation of a Nariai universe that is unstable
\cite{GinsPerry,BoussoHawk,Bousso60y} and decays through the pair
creation of extreme dS$-$Reissner-Nordstr\"{o}m black holes
accelerated only by the cosmological constant.

In the electric case, the Euclidean action is given by
(\ref{I-electric}) with  $F_{\rm el}^2=-2q^2A^4(x+\rho)^4$ [see
(\ref{F-el-Nariai PCdS})]. Thus,
\begin{eqnarray}
\frac{1}{16\pi}\int_{\cal{M}} d^4x\sqrt{g}
 \:F_{\rm el}^2=- \frac{1}{16\pi}\int_{\cal{M}} d^4x\sqrt{g}
 \:F_{\rm mag}^2\:.
 \label{I2-nariai-elect}
 \end{eqnarray}
In order to compute the extra Maxwell boundary term in
(\ref{I-electric}), the appropriate  vector potential, $A_{\nu}$,
that is regular everywhere including at the horizons is
$A_{\chi}=i\,\frac{q}{\cal{K}}\,\tilde{\tau}\, \sin \chi$, which
obviously satisfies (\ref{F-el-Nariai PCdS}). The integral over
$\Sigma$ consists of an integration between $\chi=0$ and
$\chi=\pi$ along the $\tilde{\tau}=0$ surface and back along
$\tilde{\tau}=\pi$, and of an integration between $\tilde{\tau}=0$
and $\tilde{\tau}=\pi$ along the $\chi=0$ surface, and back along
the $\chi=\pi$ surface.
 The unit normal to $\Sigma_{\tilde{\tau}}$ is
$n_{\mu}=(\frac{\sin \chi}{\sqrt{\cal{K}}A(x+\rho)},0,0,0)$, and
$F^{\mu\nu}n_{\mu}A_{\nu}=0$ on $\Sigma_{h}$. Thus, the
non-vanishing contribution to the Maxwell boundary term in
(\ref{I-electric}), $-\frac{1}{4\pi}\int_{\Sigma} d^3x\sqrt{h}\,
F^{\mu\nu}n_{\mu}A_{\nu}$, comes only from the integration along
the $\tilde{\tau}=\pi$ surface and is given by
 \begin{eqnarray}
-\frac{1}{4\pi}\int_{\Sigma_{\tilde{\tau}=\pi} } \!\!\!\!\!\! d^3x
\sqrt{h}\: F^{\tilde{\tau} \chi} n_{\tilde{\tau}} A_{\chi}=
\frac{q^2}{2 \, \cal{K}}\,\Delta \phi _{\rm
N}\,(x_\mathrm{n}-x_\mathrm{s})\:.
 \label{I-electric-nariai}
 \end{eqnarray}
Adding (\ref{I2-nariai-elect}) and (\ref{I-electric-nariai})
yields (\ref{I2-nariai}). Therefore, the electric action
(\ref{I-electric}) of the charged Nariai instanton is equal to the
magnetic action, $I_{\rm el}^{\rm N}= I_{\rm mag}^{\rm N}$, and
therefore electric and magnetic charged Nariai black holes have
the same probability of being pair created.

Now, we discuss the neutral Nariai C instanton. This instanton is
particulary important since it is the only regular Euclidean
solution available when we want to evaluate the pair creation of
neutral black holes. The same feature is present in the $A=0$ case
where only the neutral Nariai instanton is available
\cite{GinsPerry,MannRoss,BoussoHawk,VolkovWipf}. The action of the
neutral Nariai C instanton is simply given by (\ref{I1-nariai})
and, for a fixed $\Lambda$ and $A$, it is always smaller than the
action of the charged Nariai C instanton (see line $DE$ in Fig.
\ref{nariai-fig-PCAdS}): $I_{\rm charged}^{\rm N}> I_{\rm
neutral}^{\rm N}>I_{\rm dS}$. Thus the pair creation of charged
Nariai black holes is suppressed relative to the pair creation of
neutral Nariai black holes, and both are suppressed relative to
the dS space.

\subsubsection{\label{sec:Ultracold-rate}The ultracold C pair creation rate}
We first consider the magnetic case, whose Euclidean action is
given by (\ref{I}). The boundary that appears in (\ref{I}) is
given by $\Sigma=\Sigma_{\tilde{\tau}}+\Sigma_{h}+\Sigma^{\rm
int}_{\infty}$, where $\Sigma_{\tilde{\tau}}$ is a spatial surface
at $\tilde{\tau}=0$ and $\tilde{\tau}=\pi$, $\Sigma_{h}$ is a
timelike 3-surface at the Rindler horizon $\chi=0$, and the
timelike 3-surface $\Sigma^{\rm int}_{\infty}$ is an internal
infinity boundary at $\chi=\infty$. With the analysis of section
\ref{sec:Ultracold-inst}, we can compute all the terms of action
(\ref{I}). We start with $-\frac{1}{16\pi}\int_{\cal{M}}
d^4x\sqrt{g} \left ( R-2\Lambda \right )$ which yields  (using
$R=4\Lambda$)
\begin{eqnarray}
  -\frac{1}{16\pi}\int_{\Delta \phi _{\rm u}} \!\!\!\!d\phi
 \int_0^{\pi} \!\!\!\!d\tilde{\tau}
 \int_{x_\mathrm{s}}^{x_\mathrm{n}}\!\!\!\! dx
 \int_{0}^{\chi_0} \!\!\!\!d\chi \:
 \frac{2 \Lambda \chi}{(Ax+\rho)^4 }=
  -\frac{\Lambda}{16}\:\Delta \phi _{\rm u} \:\chi_0^2 \:
  \int_{x_\mathrm{s}}^{x_\mathrm{n}}\!\!\!\! dx
 \frac{1}{(Ax+A\rho)^4 }
  {\biggl |}_{\chi_0\rightarrow \infty}\:,
 \label{I1-ultracold}
 \end{eqnarray}
where  $x_\mathrm{s}$ and $x_\mathrm{n}$ are defined by
(\ref{polos-cold}) and (\ref{mq-ultra}), and $\Delta \phi _{\rm
u}$ is given by (\ref{Period phi}). The Maxwell term in the action
yields
\begin{eqnarray}
 \frac{1}{16\pi}\int_{\cal{M}} d^4x\sqrt{g}
 \:F_{\rm mag}^2 =\frac{q^2}{16}\,\Delta \phi _{\rm
u}\:\chi_0^2\,(x_\mathrm{n}-x_\mathrm{s})
  {\biggl |}_{\chi_0\rightarrow \infty}\!\!\!\!,
 \label{I2-ultracold}
 \end{eqnarray}
where we used $F_{\rm mag}^2=2q^2A^4(x+\rho)^4$ [see
(\ref{F-mag})] with $\rho=\sqrt{2(\Lambda+3A^2)}$. Due to the fact
that $\chi_0\rightarrow \infty$ it might seem that the
contribution from (\ref{I1-ultracold}) and (\ref{I2-ultracold})
diverges. Fortunately this is not the case since these two terms
cancel each other. Trying to verify this analytically is
cumbersome, but for our purposes we can simply fix any numerical
value for $\Lambda$ and $A$, and using (\ref{mq-ultra}) and
(\ref{polos-cold}) we indeed verify that (\ref{I1-ultracold}) and
(\ref{I2-ultracold}) cancel each other.

Now, contrary to the other instantons, the ultracold C instanton
has a non-vanishing extrinsic curvature boundary term,
$-\frac{1}{16\pi}\int_{\Sigma} d^3x\sqrt{h}\, K \neq 0$, due to
the internal infinity boundary ($\Sigma^{\rm int}_{\infty}$ at
$\chi=\infty$) contribution. The extrinsic curvature to
$\Sigma^{\rm int}_{\infty}$ is
$K_{\mu\nu}=h_{\mu}^{\:\:\:\alpha}\nabla_{\alpha}n_{\nu}$, where
 $n_{\nu}=(0,\frac{1}{A(x+\rho)},0,0)$ is the unit outward normal to
$\Sigma^{\rm int}_{\infty}$,
$h_{\mu}^{\:\:\:\alpha}=g_{\mu}^{\:\:\:\alpha}-n_{\mu}n^{\alpha}
 =(1,0,1,1)$ is the projection tensor onto $\Sigma^{\rm int}_{\infty}$,
 and $\nabla_{\alpha}$ represents the covariant derivative with respect
 to $g_{\mu\nu}$. Thus the trace of the extrinsic
curvature to $\Sigma^{\rm int}_{\infty}$ is
$K=g^{\mu\nu}K_{\mu\nu}=A(x+\rho)/\chi$, and
\begin{eqnarray}
 -\frac{1}{8\pi}\int_{\Sigma} d^3x\sqrt{h}\, K=
 -\frac{1}{8\pi}\,\int_{\Delta \phi _{\rm u}} \!\!\!\!d\phi
 \int_0^{\pi} \!\!\!\!d\tilde{\tau}
 \int_{x_\mathrm{s}}^{x_\mathrm{n}}\!\!\!\! dx \:
 \frac{1}{(Ax+A\rho)^2}.
 \label{I3-ultracold}
 \end{eqnarray}
The magnetic action (\ref{I}) of the ultracold C instanton is then
\begin{eqnarray}
 I_{\rm mag}^{\rm u}&=&
 -\frac{\pi}{4}
 \left [ x_\mathrm{n} \left ( 1+A\sqrt{\frac{2}{\Lambda+3A^2}}\,x_\mathrm{n}
 +\frac{A^2}{2(\Lambda+3A^2)}x_\mathrm{n}^2 \right ) \right ]^{-1}
  \nonumber \\
 & & \times
  \frac{x_\mathrm{n}-x_\mathrm{s}}
  {\left[ A x_\mathrm{n}+\sqrt{2(\Lambda+3A^2)}\right ]
   \left[ A x_\mathrm{s}+\sqrt{2(\Lambda+3A^2)}\right ]}\,,
\label{I-mag-ultracold}
 \end{eqnarray}
 where  $x_\mathrm{s}$ and $x_\mathrm{n}$ are defined by
(\ref{polos-cold}) and (\ref{mq-ultra}). When we take the limit
$A=0$ we get $x_\mathrm{s}=-1$ and $x_\mathrm{n}=1$, and
\begin{eqnarray}
I_{\rm mag}^{\rm u}{\biggl |}_{A\rightarrow 0}
=-\frac{\pi}{4\Lambda}\,,
 \label{I-mag-ultracold-A=0}
 \end{eqnarray}
and therefore we recover the action for the $A=0$ ultracold
instanton \cite{MannRoss}, that describes the pair creation of
ultracold black holes accelerated only by the cosmological
constant.

In Fig. \ref{lukewarm-fig} we show a plot of $I_{\rm mag}^{\rm
u}/I_{\rm dS}$ as a function of $m$ and $A$ for a fixed $\Lambda$.
When we fix $\Lambda$ and $A$ we also fix the mass and charge of
the ultracold black holes. For a fixed $\Lambda$, when $A$
increases the probability of pair creation of ultracold black
holes, $\Gamma_{\rm BHs/string}^{\rm u}$, increases monotonically
and they have a lower mass and charge. Alternatively, we can
discuss the behavior of $\Gamma_{\rm BHs/dS}^{\rm u}$. In this
case, the probability decreases monotonically as the acceleration
of the black holes increases.

In the electric case, the Euclidean action is given by
(\ref{I-electric}) with  $F_{\rm el}^2=-2q^2A^4(x+\rho)^4$ [see
 (\ref{F-el-ultracold})]. Thus,
\begin{eqnarray}
\frac{1}{16\pi}\int_{\cal{M}} d^4x\sqrt{g}
 \:F_{\rm el}^2=- \frac{1}{16\pi}\int_{\cal{M}} d^4x\sqrt{g}
 \:F_{\rm mag}^2\:.
 \label{I2-ultracold-elect}
 \end{eqnarray}
In the ultracold case the vector potential $A_{\nu}$, that is
regular everywhere including at the horizon, needed to compute the
extra Maxwell boundary term in (\ref{I-electric}) is
$A_{\tilde{\tau}}=i\,\frac{q}{2}\, \chi^2$, which obviously
satisfies (\ref{F-el-ultracold}). The integral over $\Sigma$
consists of an integration between $\chi=0$ and $\chi=\infty$
along the $\tilde{\tau}=0$ surface and back along
$\tilde{\tau}=\pi$, and of an integration between $\tilde{\tau}=0$
and $\tilde{\tau}=\pi$ along the $\chi=0$ surface, and back along
the internal infinity
 surface $\chi=\infty$. The non-vanishing contribution to the Maxwell
boundary term in (\ref{I-electric}) comes only from the
integration along the internal infinity boundary $\Sigma^{\rm
int}_{\infty}$, and is given by
 \begin{eqnarray}
-\frac{1}{4\pi}\int_{\Sigma^{\rm int}_{\infty}}  d^3x
\sqrt{g_{\tilde{\tau}\tilde{\tau}}g_{xx}g_{\phi\phi}}\:
F^{\chi\tilde{\tau}} n_{\chi} A_{\tilde{\tau}}=
\frac{q^2}{8}\,\Delta \phi _{\rm
u}\:\chi_0^2\,(x_\mathrm{n}-x_\mathrm{s})
  {\biggl |}_{\chi_0\rightarrow \infty}\:.
 \label{I-electric-ultracold}
 \end{eqnarray}
Adding (\ref{I2-ultracold-elect}) and (\ref{I-electric-ultracold})
yields (\ref{I2-ultracold}). Thus, the electric action
(\ref{I-electric}) of the ultracold C instanton is equal to the
magnetic action, $I_{\rm el}^{\rm u}= I_{\rm mag}^{\rm u}$, and
therefore electric and magnetic ultracold black holes have the
same probability of being pair created.

\subsubsection{\label{sec:Sub-Maximal-rate}Pair creation rate of nonextreme sub-maximal black holes}

The lukewarm, cold, Nariai and ultracold C-metric instantons are
saddle point solutions free of conical singularities both in the
$y_+$ and $y_A$ horizons. The corresponding black holes may then
nucleate in the dS background when a cosmic string breaks, and we
have computed their pair creation rates in the last four
subsections. However, these particular black holes are not the
only ones that can be pair created. Indeed, it has been shown in
\cite{WuSubMax,BoussoHawkSubMax} that Euclidean solutions with
conical singularities may also be used as saddle points for the
pair creation process. In this way, pair creation of nonextreme
sub-maximal black holes is allowed (by this nomenclature we mean
all the nonextreme black holes other than the lukewarm ones that
are in the region interior to the close line $NOUN$ in Fig.
\ref{mq-fig}), and their pair creation rate may be computed. In
order to calculate this rate, the action is given by (\ref{I}) and
(\ref{I-electric}) (in the magnetic and electric cases
respectively) and, in addition, it has now an extra contribution
from the conical singularity (c.s.) that is present in one of the
horizons ($y_+$, say) given by
\cite{ReggeGibbonsPerryAconSing,GinsPerry}
\begin{eqnarray}
\frac{1}{16\pi}\int_{\cal{M}} d^4x\sqrt{g}
 \:\left ( R-2\Lambda \right ){\biggl |}_{{\rm c.s.}\:{\rm at}\:y_+}
 \!\!\!= \frac{ {\cal A}_+\:\delta}{16\pi}\:,
 \label{I conical sing}
 \end{eqnarray}
where ${\cal A}_+=\int_{y=y_+} \sqrt{g_{xx}g_{\phi\phi}}\: dx
\,d\phi$ is the area of the 2-surface spanned by the conical
singularity, and
\begin{eqnarray}
\delta=2\pi \left ( 1-\frac{\beta_A}{\beta_+}\right )
 \label{delta concical sing PCdS}
 \end{eqnarray}
is the deficit angle associated to the conical singularity at the
horizon $y_+$, with $\beta_A=4 \pi / |{\cal F}'(y_A)|$ and
$\beta_+=4 \pi / |{\cal F}'(y_+)|$ being the periods of $\tau$
that avoid a conical singularity in the horizons $y_A$ and $y_+$,
respectively. The contribution from (\ref{I}) and
(\ref{I-electric}) follows straightforwardly in a similar way as
the one shown in subsection \ref{sec:Lukewarm-rate} with the
period of $\tau$, $\beta_A$,  chosen in order to avoid the conical
singularity at the acceleration horizon, $y=y_A$. The full
Euclidean action for general nonextreme sub-maximal black holes is
then
\begin{eqnarray}
& & \!\!\!\!\!\!\!\! I\!=\!
 \frac{\Delta \phi} {A^2}\,\left ( \!
 \frac{ x_\mathrm{n}-x_\mathrm{s} }
   {(x_\mathrm{n}+y_A)(x_\mathrm{s}+y_A)}
  \! + \!\frac{ x_\mathrm{n}-x_\mathrm{s} }
   {(x_\mathrm{n}+y_+)(x_\mathrm{s}+y_+)}\!
   \right ), \nonumber \\
& & \label{I-Sub-Maximal}
 \end{eqnarray}
where $\Delta \phi$ is given by (\ref{Period phi}), and the pair
creation rate of nonextreme sub-maximal  black holes is
$\Gamma=\eta\,e^{-2I+2I_{dS}}$, where $I_{\rm
dS}=-\frac{3\pi}{2\Lambda}$ is the action of de Sitter space, and
$\eta$ is the one-loop contribution not computed here. In order to
compute (\ref{I-Sub-Maximal}), we need the relation between the
parameters $A$, $\Lambda$, $m$, $q$, and the horizons $y_A$, $y_+$
and $y_-$. In general, for a nonextreme solution with horizons
$y_A<y_+<y_-$, one has
\begin{eqnarray}
{\cal F}(y)= -\frac{1}{\mu}(y-y_A)(y-y_+)(y-y_-)(ay+b) \:,
 \label{F-nonext-sub-max}
 \end{eqnarray}
with
\begin{eqnarray}
\mu\!\!&=&\!\! y_A y_+ y_- (y_A +y_+ +y_-) +(y_A y_+ +y_A y_- +y_+
y_-)^2 \nonumber \\
 a\!\!&=&\!\!
 \left ( y_A y_+ +y_A y_- +y_+y_- \right )  \nonumber \\
 b\!\!&=&\!\! y_A y_+y_-\:.
 \label{F-nonext-sub-max aux}
 \end{eqnarray}
The parameters $A$, $\Lambda$, $m$ and $q$ can be expressed as a
function of $y_A$, $y_+$ and $y_-$ by
\begin{eqnarray}
\frac{\Lambda}{3A^2}\!\!&=&\!\! \mu^{-2}(y_A y_+y_-)^2 -1 \nonumber \\
 q^2A^2\!\!&=&\!\! \mu^{-1}(y_A y_+ +y_A y_-
+y_+ y_-) \nonumber \\
 mA\!\!&=&\!\! (2\,\sigma)^{-1}(y_A +y_+)(y_A +y_-)(y_+ +y_-) \nonumber \\
 \sigma\!\!&=&\!\! y_A^2 y_+y_- +y_A y_+^2 y_- +y_A y_+ y_-^2+ (y_A y_+)^2  +(y_A y_-)^2 +(y_+ y_-)^2 \:.
 \label{relation parameters nonext-sub-max}
 \end{eqnarray}
The allowed values of parameters $m$ and $q$ are those contained
in the interior region defined by the close line $NOUN$ in Fig.
\ref{mq-fig}.

\subsection{\label{sec:Entropy}Entropy, area and pair creation rate}

In previous works on black hole pair creation in general
background fields it has been well established that the pair
creation rate is proportional to the exponential of the
gravitational entropy $S$ of the system, $\Gamma \propto e^S$,
with the entropy being given by one quarter of the the total area
$\cal{A}$ of all the horizons present in the instanton,
$S={\cal{A}}/4$. In what follows we will verify that these
relations also hold for the instantons of the dS C-metric.

\subsubsection{\label{sec:Lukewarm-S}The lukewarm C case. Entropy and area}
In the lukewarm case, the instanton has two horizons in its
Euclidean section, namely the acceleration horizon at $y=y_A$ and
the black hole horizon at $y=y_+$. So, the total area of the
lukewarm C instanton is
\begin{eqnarray}
\cal{A}^{\rm \ell} &=& \int_{y=y_A} \sqrt{g_{xx}g_{\phi\phi}}\: dx
\,d\phi + \int_{y=y_+} \sqrt{g_{xx}g_{\phi\phi}}\:
dx \,d\phi =          \nonumber \\
& &\!\!\!\!\!\!\!\!\!\!\!\!
 \frac{\Delta \phi _{\rm \ell}} {A^2}\,\left (
 \frac{ x_\mathrm{n}-x_\mathrm{s} }
   {(x_\mathrm{n}+y_A)(x_\mathrm{s}+y_A)}
   +\frac{ x_\mathrm{n}-x_\mathrm{s} }
   {(x_\mathrm{n}+y_+)(x_\mathrm{s}+y_+)}
   \right )\,, \nonumber \\
& &
 \label{area-luk}
 \end{eqnarray}
 where $y_A$ and $y_+$ are given by (\ref{yA-luk}), $x_\mathrm{s}$
and $x_\mathrm{n}$ are defined by (\ref{polos-luk}), and
 $\Delta \phi _{\rm \ell}$ is
given by (\ref{Period phi-luk}). It is straightforward to verify
that ${\cal{A}^{\rm \ell}}=-8I^{\rm \ell}$, where $I^{\rm \ell}$
is given by (\ref{I-mag-luk}), and thus $\Gamma^{\rm \ell} \propto
e^{S^{\rm \ell}}$, where $S^{\rm \ell}= {\cal{A}^{\rm \ell}}/4$.

\subsubsection{\label{sec:Cold-S}The cold C case. Entropy and area}

In the cold case, the instanton has a single horizon,  the
acceleration horizon at $y=y_A$, in its Euclidean section, since
 $y=y_+$ is an internal infinity. So, the total area of the
cold C instanton is
\begin{eqnarray}
{\cal{A}}^{\rm c} =
 \int_{y=y_A} \sqrt{g_{xx}g_{\phi\phi}}\: dx \,d\phi =
  \frac{\Delta \phi _{\rm c}} {A^2}\,
 \frac{ x_\mathrm{n}-x_\mathrm{s} }
   {(x_\mathrm{n}+y_A)(x_\mathrm{s}+y_A)}\,,
 \label{area-cold}
 \end{eqnarray}
 where $y_A$ is given by (\ref{zerosy3-cold}), $x_\mathrm{s}$
and $x_\mathrm{n}$ are defined by (\ref{polos-cold}), and
 $\Delta \phi _{\rm c}$ is
given by (\ref{Period phi}). Thus, ${\cal{A}^{\rm c}}=-8I^{\rm
c}$, where $I^{\rm c}$ is given by (\ref{I-mag-cold}), and thus
$\Gamma^{\rm c} \propto e^{S^{\rm c}}$, where $S^{\rm c}=
{\cal{A}^{\rm c}}/4$.

\subsubsection{\label{sec:Nariai-S}The Nariai C case. Entropy and area}

In the Nariai case, the instanton has two horizons in its
Euclidean section, namely the acceleration horizon $y_A$ and the
black hole horizon $y_+$, both at $y=\rho$, and thus they have the
same area. So, the total area of the Nariai C instanton is
\begin{eqnarray}
{\cal{A}}^{\rm N} = 2\,\int_{y=\rho} \sqrt{g_{xx}g_{\phi\phi}}\:
dx \,d\phi =
 2\,\frac{\Delta \phi _{\rm N}} {A^2}\,
 \frac{ x_\mathrm{n}-x_\mathrm{s} }
   {(x_\mathrm{n}+\rho)(x_\mathrm{s}+\rho)}\,,
 \label{area-nariai}
 \end{eqnarray}
 where $\sqrt{3\gamma} \leq \rho<\sqrt{6\gamma}$, $x_\mathrm{s}$
and $x_\mathrm{n}$ are defined by (\ref{polos-cold}), and
 $\Delta \phi _{\rm N}$ is
given by (\ref{Period phi}), with $m$ and $q$ subjected to
(\ref{mq-cNariai PCdS}). Thus, ${\cal{A}^{\rm N}}=-8I^{\rm N}$,
where $I^{\rm N}$ is given by (\ref{I-mag-Nariai}), and thus
$\Gamma^{\rm N} \propto e^{S^{\rm N}}$, where $S^{\rm N}=
{\cal{A}^{\rm N}}/4$.

\subsubsection{\label{sec:Ultracold-S}The ultracold C case. Entropy and area}

In the ultracold  case, the instanton has a single horizon, the
Rindler horizon at $\chi=0$, in its Euclidean section, since
 $\chi=\infty$ is an internal infinity. So, the total area of the
ultracold C instanton is
\begin{eqnarray}
{\cal{A}}^{\rm u} =
 \int_{\chi=0} \sqrt{g_{xx}g_{\phi\phi}}\: dx \,d\phi =
 \frac{\Delta \phi _{\rm u}} {A^2}\,
 \frac{ x_\mathrm{n}-x_\mathrm{s} }
   {(x_\mathrm{n}+\rho)(x_\mathrm{s}+\rho)}\,,
 \label{area-ultracold}
 \end{eqnarray}
with $\rho=\sqrt{2(\Lambda+3A^2)}$ [see
(\ref{range-gamma-ultra})], $x_\mathrm{s}$ and $x_\mathrm{n}$ are
defined by (\ref{polos-cold}), and
 $\Delta \phi _{\rm c}$ is
given by (\ref{Period phi}), with $m$ and $q$ subjected to
(\ref{mq-ultra}). It straightforward to verify that ${\cal{A}^{\rm
u}}=-8I^{\rm u}$, where $I^{\rm u}$ is given by
(\ref{I-mag-ultracold}), and thus $\Gamma^{\rm u} \propto
e^{S^{\rm u}}$, where $S^{\rm u}= {\cal{A}^{\rm u}}/4$.

As we have already said, the ultracold C instanton is a limiting
case of both the charged Nariai C instanton and the cold C
instanton (see, e.g., Fig. \ref{mq-fig}). Then, as expected, the
action of the cold C instanton  gives, in this limit, the action
of the ultracold C instanton (see Fig. \ref{cold-fig}). However,
the ultracold frontier of the Nariai C action is given by two
times the ultracold C action (see Fig. \ref{nariai-fig-PCAdS}).
From the results of this section we clearly understand the reason
for this behavior. Indeed, in the ultracold case and in the cold
case, the respective instantons have a single horizon (the other
possible horizon turns out to be an internal infinity). This
horizon gives the only contribution to the total area,
${\cal{A}}$, and therefore to the pair creation rate. In the
Nariai case, the instanton has two horizons with the same area,
and thus the ultracold limit of the Nariai action is doubled with
respect to the true ultracold action.

\subsubsection{\label{sec:Sub-Maximal-S}The nonextreme sub-maximal
case. Entropy and area}

In the lukewarm case, the instanton has two horizons in its
Euclidean section, namely the acceleration horizon at $y=y_A$ and
the black hole horizon at $y=y_+$. So, the total area of the
saddlepoint solution is
\begin{eqnarray}
\!\!\! {\cal A}=\int_{y=y_A}\!\!\! \sqrt{g_{xx}g_{\phi\phi}}\: dx
\,d\phi +\! \int_{y=y_+} \!\!\! \sqrt{g_{xx}g_{\phi\phi}}\: dx
\,d\phi \,,
 \label{area-Sub-Maximal}
 \end{eqnarray}
and once again one has ${\cal{A}}=-8I$, where $I$ is given by
(\ref{I-Sub-Maximal}), and thus $\Gamma \propto e^{S}$, where $S=
{\cal{A}}/4$.

\subsection{\label{sec:Heuristic dS}Heuristic derivation
of the nucleation rates}

The physical interpretation of our exact results of section
\ref{sec:Pair Creation dS} can be clarified with a heuristic
derivation of the nucleation rates. An estimate for the nucleation
probability is given by the Boltzmann factor,
 $\Gamma \sim e^{-E_0/W_{\rm ext}}$, where $E_0$ is the
energy of the system that nucleates and $W_{\rm ext}=F \ell$ is
the work done by the external force $F$,  that provides the energy
for the nucleation, through the typical distance $\ell$ separating
the created pair.

Forget for a moment the string, and ask what is the probability
that a black hole pair is created in a dS background. This process
has been discussed in \cite{MannRoss} where it was found that the
pair creation rate is $\Gamma \sim e^{-m/\sqrt{\Lambda}}$. In this
case, $E_0 \sim 2m$, where $m$ is the rest energy of the black
hole, and $W_{\rm ext}\sim \sqrt{\Lambda}$ is the work provided by
the cosmological background. To derive  $W_{\rm ext}\sim
\sqrt{\Lambda}$ one can argue as follows. In the dS case, the
Newtonian potential is $\Phi=\Lambda r^2/3$ and its derivative
yields the force per unit mass or acceleration, $\Lambda r$, where
$r$ is the characteristic dS radius, $\Lambda^{-1/2}$. The force
can then be written as $F= {\rm mass}\times{\rm acceleration}\sim
\sqrt{\Lambda}\sqrt{\Lambda}$, where the characteristic mass of
the system is $\sqrt{\Lambda}$. Thus, the characteristic work is
$W_{\rm ext}={\rm force}\times{\rm distance}\sim \Lambda
\Lambda^{-1/2}\sim \sqrt{\Lambda}$, where the characteristic
distance that separates the pair at the creation moment is
$\Lambda^{-1/2}$. So, from the Boltzmann factor we indeed expect
that the creation rate of a black hole pair in a dS background is
given by $\Gamma \sim e^{-m/\sqrt{\Lambda}}$ \cite{MannRoss}.

A question that has been answered in the present section
\ref{sec:Pair Creation dS} was: given that a string is already
present in our initial system, what is the probability that it
breaks and a pair of black holes is produced and accelerated apart
by $\Lambda$ and by the string tension? The presence of the string
leads in practice to a problem in which we have an effective
cosmological constant that satisfies $\Lambda'\equiv
\Lambda+3A^2$, that is, the acceleration $A$ provided by the
string makes a positive contribution to the process.
Heuristically, we may then apply the same arguments that have been
used in the last paragraph, with the replacement
$\Lambda\rightarrow \Lambda'$. At the end, the Boltzmann factor
tells us that the creation rate for the process is $\Gamma \sim
e^{-m/\sqrt{\Lambda+3A^2}}$. So, for a given black hole mass, $m$,
and for a given cosmological constant, $\Lambda$, the black hole
pair creation process is enhanced when a string is present, as the
explicit calculations done in section \ref{sec:Pair Creation dS}
show. For $\Lambda=0$ this heuristic derivation yields $\Gamma
\sim e^{-m/A}$ which is the pair creation rate found in
\cite{HawkRoss-string}.

Another question that we have dealt with in the present section
\ref{sec:Pair Creation dS} was: what is the probability for the
nucleation of a string in a dS background? Heuristically, the
energy of the string that nucleates is $E_0\sim \mu
\Lambda^{-1/2}$, i.e., its mass per unit length times the dS
radius, while the work provided by the cosmological background is
still given by $W_{\rm ext}\sim \sqrt{\Lambda}$. The Boltzmann
factor yields for nucleation rate the value $\Gamma \sim
e^{-\mu/\Lambda}$, in agreement with (\ref{PC-rate-stringend}).

\subsection{\label{sec:Conc dS}Summary and discussion}

We have studied in detail the quantum process in which a cosmic
string breaks in a de Sitter (dS) background and a pair of black
holes is created at the ends of the string. The energy to
materialize and accelerate the pair comes from the positive
cosmological constant and from the string tension. This process is
a combination of the processes considered in
\cite{MelMos}-\cite{VolkovWipf}, where the creation of a black
hole pair in a dS background has been analyzed, and in
\cite{HawkRoss-string}-\cite{GregHind}, where the breaking of a
cosmic string accompanied by the creation of a black hole pair in
a flat background has been studied.

We have constructed the saddle point solutions that mediate the
pair creation process through the analytic continuation of the dS
C-metric \cite{PlebDem,PodGrif2,OscLem_dS-C}, and we have
explicitly computed the nucleation rate of the process (see also a
heuristic derivation of the rate in the Appendix). Globally our
results state that the dS space is stable against the nucleation
of a string, or against the nucleation of a string followed by its
breaking and consequent creation of a black hole pair. In
particular, we have answered three questions. First, we have
concluded that the nucleation rate of a cosmic string in a dS
background $\Gamma_{\rm string/dS}$ decreases when the mass
density of the string increases. Second, given that the string is
already present in our initial system, the probability
$\Gamma_{\rm BHs/string}$ that it breaks and a pair of black holes
is produced and accelerated apart by $\Lambda$ and by the string
tension increases when the mass density of the string increases.
In other words, a string with a higher mass density makes the
process more probable, for a fixed black hole mass. Third, if we
start with a pure dS background, the probability $\Gamma_{\rm
BHs/dS}$ that a string nucleates on it and then breaks forming a
pair of black holes decreases when the mass density of the string
increases. These processes have a clear analogy with a
thermodynamical system, with the mass density of the string being
the analogue of the temperature $T$. Indeed, from the Boltzmann
factor, $e^{-E_0/(k_{\rm B} T)}$ (where $k_{\rm B}$ is the
Boltzmann constant), one knows that a higher background
temperature turns the nucleation of a particle with energy $E_0$
more probable. However, in order to have a higher temperature we
have first to furnish more energy to the background, and thus the
global process (increasing the temperature to the final value $T$
plus the nucleation of the particle) becomes energetically less
favorable as $T$ increases.

We have also verified that the relation between the rate, entropy
and area, which is satisfied for all the black hole pair creation
processes analyzed so far, also holds in the process studied in
this paper. Indeed, the pair creation rate is proportional to
$e^{S}$, where $S$ is the gravitational entropy of the system, and
is given by one quarter of the total area of all the horizons
present in the saddle point solution that mediates the pair
creation.

\section{\label{sec:Concluding Overview Pair Creation}Pair creation in AdS, flat and dS backgrounds: a comparing
discussion}

We have reviewed in detail the studies on the quantum process in
which a cosmic string breaks in an anti-de Sitter
\cite{OscLem-PCAdS}, in a flat
\cite{HawkRoss-string,DougHorKastTras,AchGregKui,GregHind} and in
a de Sitter background \cite{OscLem-PCdS}, and a pair of black
holes is created at the ends of the string (see Fig. \ref{Fig
string_cut_pc_introd}). The energy to materialize and accelerate
the black holes comes from the strings' tension, and we have
explicitly computed the pair creation rates. In the dS case the
cosmological background acceleration makes a positive contribution
to the process, while in the AdS case the cosmological background
acceleration contributes negatively. In particular, in the AdS
case, pair creation of black holes is possible only when the
acceleration provided by the string tension is higher than
$\sqrt{|\Lambda|/3}$: if we have a virtual pair of black holes and
we want to turn them real, we have to furnish a sufficient force
that overcomes the AdS background attraction.

We remark that in principle our explicit values for the pair
creation rates \cite{OscLem-PCAdS} in an AdS and dS
\cite{OscLem-PCdS} background also apply to the process of pair
creation in an external electromagnetic field, with the
acceleration being provided in this case by the Lorentz force
instead of being furnished by the string tension. Indeed, there is
no Ernst solution in a cosmological constant background, and thus
we cannot discuss analytically the process. However, physically we
could in principle consider an external electromagnetic field that
supplies the same energy and acceleration as our strings and, from
the results of the $\Lambda=0$ case (where the pair creation rates
in the string and electromagnetic cases agree), we expect that the
results found in \cite{OscLem-PCdS,OscLem-PCAdS} do not depend on
whether the energy is being provided by an external
electromagnetic field or by strings.

\begin{figure} [h]
\includegraphics*[height=2.3in]{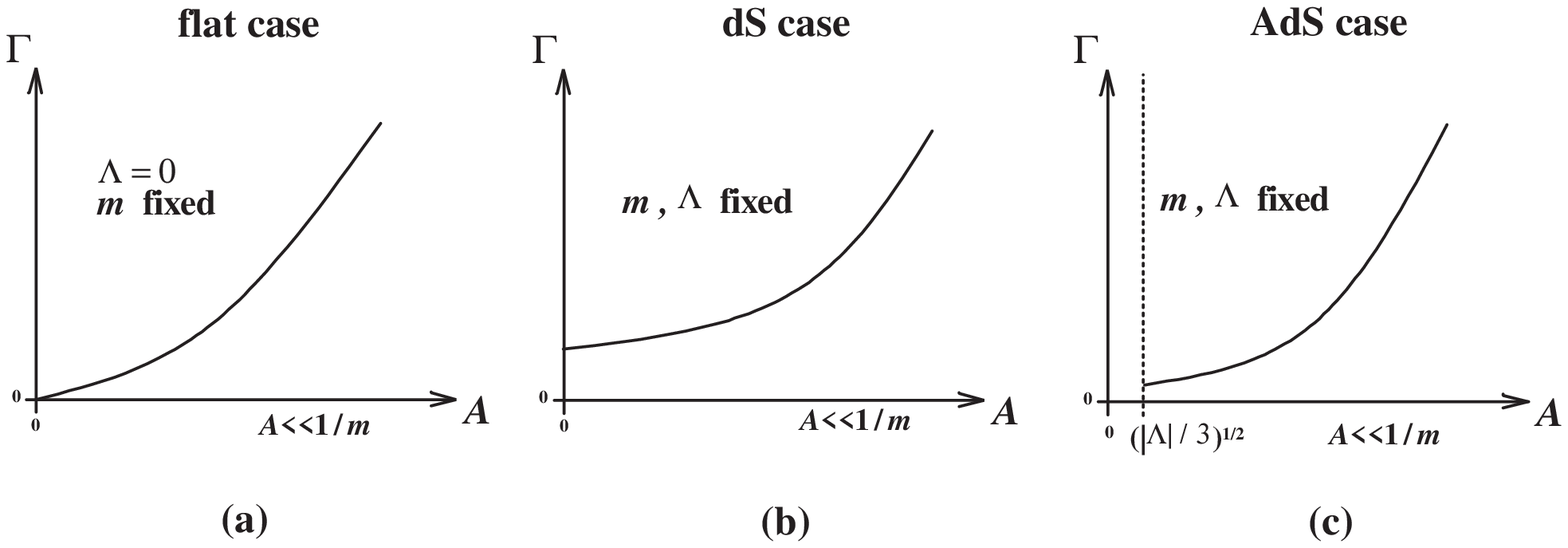}
\caption{\label{PC rates_AdS}
 General behavior of the black hole pair creation rate $\Gamma$ as a function
of the acceleration $A$ provided by the strings, when a cosmic
string breaks: (a)  in a flat background ($\Lambda=0$)
\cite{HawkRoss-string}, (b)  in a dS background ($\Lambda>0$)
\cite{OscLem-PCdS}, and (c)  in an AdS background ($\Lambda<0$)
\cite{OscLem-PCAdS}.
 }
\end{figure}

For the benefit of comparison, in Fig. \ref{PC rates_AdS} we
schematically represent the general behavior of the black hole
pair creation rate $\Gamma$ as a function of the acceleration $A$
provided by the strings, when a cosmic string breaks in the three
cosmological constant backgrounds. In a flat background [see Fig.
\ref{PC rates_AdS}.(a)], the pair creation rate is zero when $A=0$
\cite{HawkRoss-string}. In this case, the flat C-metric reduces to
a single black hole, and since we are studying the probability of
pair creation, the corresponding rate is naturally zero. This does
not mean that a single black hole cannot be materialized from the
quantum vacuum, it only means that this latter process is not
described by the C-metric. The creation probability of a single
black hole in a hot bath has been considered in
\cite{GrossPerryYaffe}. In a dS background [see Fig. \ref{PC
rates_AdS}.(b)], the pair creation rate is not zero when $A=0$
\cite{OscLem-PCdS}. This means that even in the absence of the
string, the positive cosmological constant is enough to provide
the energy to materialize the black hole pair \cite{MannRoss}. If
in addition one has an extra energy provided by the string, the
process becomes more favorable \cite{OscLem-PCdS}. In the AdS case
[see Fig. \ref{PC rates_AdS}.(c)], the negative cosmological
constant makes a negative contribution to the process, and black
hole pair creation is possible only when the acceleration provided
by the strings overcomes the AdS background attraction. The branch
$0<A\leq \sqrt{|\Lambda|/3}$ represents the creation probability
of a single black hole when the acceleration provided by the
broken string is not enough to overcome the AdS attraction, and
was not studied in this thesis.
\noindent We can also fix $\Lambda$ and $A$, and describe the
evolution of the pair creation rates when the mass and charge of
the black holes increase. In the three cases, $\Lambda<0$,
$\Lambda=0$ and $\Lambda>0$, we conclude that as the mass and
charge of the black holes increase, the probability they have to
be pair created decreases monotonically. This is the expected
result since the materialization of a higher mass implies the need
of a higher energy.

 In order to study the pair creation processes, we had to
construct the instantons that mediate the process. This has done
through the analytic continuation of some special cases of the
AdS, flat and dS C-metrics. The regularity condition imposed to
these instantons restricts the mass $m$ and the charge $q$ of the
black holes that are produced, and physically it means that the
only black holes that can be pair produced are those that are in
thermodynamic equilibrium. Concretely, in the three cases, AdS,
flat and dS, we have found two charged regular instantons. One
mediates the pair creation of nonextreme black holes with $m=q$,
and the other mediates the pair creation of extreme black holes
($y_+ = y_-$). The instantons constructed from the AdS and flat
C-metric are noncompact (contrary to what occurs with the
$\Lambda>0$ instantons), in the sense that they have an
acceleration horizon with an infinite area. Thus when dealing with
them, we had to eliminate this infinity by normalizing this area
relative to the acceleration horizon area of an appropriate
background reference spacetime. In the dS case, the acceleration
horizon has a finite area, and the above nonextreme instanton is
also called lukewarm and the extreme instanton is also called
cold. Moreover, in the dS case there are two other instantons
available: the Nariai instanton, and the ultracold instanton,
which are not present in the AdS and flat cases since they are out
of the allowed range of the angular direction $x$. The Nariai C
instanton describes the creation of a Nariai C universe (see
section \ref{sec:Gen Nariai}) that is unstable and, once created,
it decays through the quantum tunnelling process into a slightly
non-extreme ($y_A \sim y_+$) pair of black holes accelerated by
the cosmological background and by the string. Finally, the
ultracold C instanton describes the creation of a Nariai
Bertotti-Robinson C universe (see section \ref{sec:Gen Nariai-BR
dS}) that is unstable and, once created, it decays into a slightly
non-extreme ($y_A \sim y_+ \sim y_-$) pair of black holes
accelerated by the cosmological background and by the string.

In all the cases studied in this chapter, the pair creation rate
is proportional to $e^{S}$, where $S$ is the gravitational entropy
of the system, and is given by one quarter of the total area of
all the horizons present in the instanton that mediates the pair
creation.

To conclude let us recall that the C-metric allows two distinct
physical interpretations. In one of them one removes the conical
singularity at the north pole and leaves one at the south pole. In
this way the C-metric describes a pair of black holes accelerated
away by two strings with positive mass density from each one of
the black holes towards infinity. Alternatively, we can avoid the
conical singularity at the south pole and in this case the black
holes are pushed away by a strut (with negative mass density) in
between them, along their north poles. In this chapter, as in we
have adopted the first choice. Technically, the second choice
introduces minor changes. Essentially it changes the period of the
angular coordinate $\phi$: it would be given by $\Delta
\phi=\frac{4 \pi}{|{\cal G}'(x_\mathrm{s})|}$ instead of
(\ref{Period phi}). We have chosen the first choice essentially
for two reasons. First, the string has a positive mass density
and, in this sense, it is a more physical solution than the strut.
Second, in order to get the above string/pair configuration we
only have to cut the string in a point. The strings tension does
the rest of the work. However, if we desire the strut/pair system
described above we would have to cut the strut in two different
points. Then we would have to discard somehow the parts that join
each one of the black holes towards infinity along their south
poles.

%% file: Chapter9.tex
\thispagestyle{empty} \setcounter{minitocdepth}{1}
\chapter[Black holes in higher dimensional spacetimes]
{\Large{Black holes in higher dimensional spacetimes}}
 \label{chap:Black holes in higher dimensions}
 \lhead[]{\fancyplain{}{\bfseries Chapter \thechapter. \leftmark}}
 \rhead[\fancyplain{}{\bfseries \rightmark}]{}
  \minitoc \thispagestyle{empty}
\renewcommand{\thepage}{\arabic{page}}

\addtocontents{lof}{\textbf{Chapter Figures \thechapter}\\}

In this chapter we will describe the main features of the static
higher dimensional black hole solutions of the Einstein-Maxwell
theory in a background with a generalized cosmological constant
$\Lambda$. Some of these black holes were found by Tangherlini
\cite{tangherlini}, and are the higher dimensional cousins of the
Schwarzschild and of the Reissner-Nordstr\"{o}m black holes. We
work in the context of the Einstein-Maxwell action with a
cosmological constant $\Lambda$,
 \begin{eqnarray}
I=\frac{1}{16\pi}\int_{\cal M} d^Dx\sqrt{-g} \left ( R-2\lambda
\right )-\frac{1}{16\pi} \int_{\cal M} d^Dx\sqrt{-g}
F^{\mu\nu}F_{\mu\nu}  \:,
 \label{I D-dim}
 \end{eqnarray}
where $D$ is the dimension of the spacetime, $g$ is the
determinant of the metric, $R$ is the Ricci scalar,
$F_{\mu\nu}=\partial_{\mu}A_{\nu}-\partial_{\nu}A_{\mu}$ is the
Maxwell field strength of the gauge field $A_{\nu}$, and we have
defined
\begin{eqnarray}
\lambda=\frac{(D-1)(D-2)\,\Lambda}{6} \:.
 \label{D-dim:def lambda}
\end{eqnarray}
The coefficient of the $\Lambda$ term was chosen in order to
insure that, for any dimension $D$, the pure dS or AdS spacetime
is described by $g_{tt}=1-(\Lambda/3)r^2$, as occurs with $D=4$.
 We set the $D$-dimensional Newton's constant equal to
one, and $c=1$. The variation of (\ref{I D-dim}), $\delta I=0$,
yields the equations for the gravitational field and for the
Maxwell field, respectively,
\begin{eqnarray}
& & R_{\mu\nu}-\frac{1}{2}R\,g_{\mu\nu}+\lambda
g_{\mu\nu}=8\pi\,T_{\mu\nu} \:, \label{D-dim:eqs of motion Einstein} \\
& & \nabla_{\mu} F^{\mu\nu}=0 \:,
 \label{D-dim:eqs of motion Maxwell}
\end{eqnarray}
where $R_{\mu\nu}$ is the Ricci tensor and $T_{\mu\nu}$ is the
electromagnetic energy-momentum tensor,
\begin{eqnarray}
T_{\mu\nu}=\frac{1}{4\pi}\left ( g^{\alpha \beta} F_{\alpha \mu}
F_{\beta \nu}-\frac{1}{4}g_{\mu\nu}F_{\alpha\beta}F^{\alpha\beta}
\right ) \:.
 \label{D-dim:energy tensor}
\end{eqnarray}
The contraction of (\ref{D-dim:eqs of motion Einstein}) with
$g^{\mu\nu}$ yields for the Ricci tensor
\begin{eqnarray}
R=\frac{D(D-1)}{3}\,\Lambda-\frac{16\pi}{D-2}\,T \:,
 \label{D-dim:Ricci Tensor}
\end{eqnarray}
where $T$ is the trace of $T_{\mu\nu}$.
 We remark that in a general $D$-dimensional background the electromagnetic
energy-momentum tensor is not traceless. Indeed, the contraction
of (\ref{D-dim:energy tensor}) with $g^{\mu\nu}$ yields
\begin{eqnarray}
T=-\frac{D-4}{4\pi}\,F_{\mu\nu}F^{\mu\nu} \:,
 \label{D-dim:trace T}
\end{eqnarray}
which vanishes only for $D=4$.

\section{\label{sec:BH D-dim flat}Higher dimensional black holes in a flat background}
In a flat background, $\Lambda=0$, the most general static
solution with spherical symmetry is given by \cite{tangherlini}
(see also \cite{myersperry})
\begin{equation}
ds^{2}=-f(r)dt^{2}+f(r)^{-1}dr^{2}+r^{2}\,d\Omega_{D-2}^2
 \label{RN:metric D-dim}
\end{equation}
where $d\Omega_{D-2}^2$ is the line element on an unit
$(D-2)$-sphere,
\begin{equation}
d\Omega^2_{D-2}=d\theta_1^2+\sin^2\theta_1\,d\theta_2^2+
\sin^2\theta_1 \sin^2\theta_2\,d\theta_3^2+\cdots
+\prod_{i=1}^{D-3}\sin^2\theta_i\,d\theta_{D-2}^2\:,
 \label{metric sphere D-dim}
\end{equation}
and the function $f(r)$ is given by
\begin{equation}
f(r)=1-\frac{M}{r^{D-3}}+\frac{Q^2}{r^{2(D-3)}}\:.
 \label{RN:f D-dim}
\end{equation}
The mass parameter $M$ and the charge parameter $Q$ are related to
the ADM mass, $M_{\rm ADM}$, and ADM electric charge, $Q_{\rm
ADM}$, of the solution by \cite{myersperry}
\begin{eqnarray}
M_{\rm ADM}=\frac{(D-2)\Omega_{D-2}}{16\pi}\,M \:,\qquad {\rm
and}\:\:\:\: Q_{\rm ADM}=\sqrt{\frac{(D-3)(D-2)}{2}}\,Q \:,
 \label{ADM hairs D-dim}
\end{eqnarray}
where $\Omega_{D-2}$ is the area of an unit $(D-2)$-sphere,
\begin{equation}
\Omega_{D-2}=\frac{2\pi^{(D-1)/2}}{\Gamma[(D-1)/2]}\:.
\label{integratedsolidangle}
\end{equation}
Here, $\Gamma[z]$ is the Gamma function, whose definition and
properties are listed in \cite{stegun}. For our purposes we need
to know that
\begin{eqnarray}
& & \Gamma[z]=(z-1)! \:, \qquad
 {\rm when}\:\: z \:\: {\rm is}\:\:{\rm a}\:\: {\rm positive}\:\: {\rm integer,} \nonumber \\
& & \Gamma[1/2]=\sqrt{\pi}\:, \qquad {\rm and} \qquad
\Gamma[z+1]=z\Gamma[z]\:, \qquad {\rm when}\:\: z \:\:{\rm is}
\:\:{\rm a} \:\:{\rm multiple} \:\:{\rm of}\:\: 1/2 \:,
 \label{Gamma function D-dim}
\end{eqnarray}
The radial electromagnetic field produced by the electric charge
$Q_{\rm ADM}$ is given by
\begin{equation}
F=-\frac{Q_{\rm ADM}}{r^{D-2}}\,dt\wedge dr\:.
 \label{RN:maxwell D-dim}
\end{equation}

When $Q^2<\frac{M^2}{4}$ the solution (\ref{RN:metric D-dim})
(\ref{RN:maxwell D-dim}) represents a black hole solution with a
curvature singularity at the origin, and with an event horizon,
$r_+$, and a Cauchy horizon, $r_-$ which satisfy
\begin{equation}
 r_{\pm}^{D-3}=\frac{M}{2}\pm
\sqrt{\frac{M^2}{4}-Q^2}.
 \label{R:rh D-dim}
\end{equation}
such that $r_+^{D-3}+r_-^{D-3}=M$, and $r_+^{D-3} r_-^{D-3}=Q^2$.
When $Q^2=\frac{M^2}{4}$, one has an extreme black hole with
$r_+=r_-=(M/2)^{\frac{1}{D-3}}$, and when $Q^2>\frac{M^2}{4}$, one
has a naked singularity. When $Q=0$ we have a neutral black hole
with an event horizon at $r_+=M^{\frac{1}{D-3}}$.

Now, any $D$-dimensional geometry can be embedded into a higher
dimensional Minkowski spacetime \cite{goenner}, with one or more
timelike coordinates. This process is usually called as global
embedding Minkowskian spacetime procedure (GEMS). This procedure
allows, for example, to verify that the Hawking temperature and
the Unruh temperature can be matched (see, e.g.,
\cite{deser3,OscarLemosNuno}). The neutral Tangherlini black hole
($Q=0$) admits the following embedding in a $(D+1)$-dimensional
Minkowski spacetime \cite{OscarLemosNuno},
\begin{eqnarray}
z_{0} & = & k^{-1}_{+}\sqrt{f(r)} \sinh{(k_{+}t)} \nonumber\\
z_{1} & = & k^{-1}_{+}\sqrt{f(r)} \cosh{(k_{+}t)} \nonumber\\
z_{2} & = & r \prod^{D-2}_{i=1}\sin \theta_i \nonumber\\
&\cdots& \nonumber\\
z_{j} & = & r {\bigg (} \prod^{D-j}_{i=1}\sin \theta_i {\bigg )} \cos \theta_{D+1-j}\:,\:\: {\rm for} \:\: 3\leq j < D \nonumber\\
&\cdots& \nonumber\\
z_{D} & = & r\cos\theta_1 \nonumber\\
z_{D+1} & = & \int dr {\biggl (}\frac{\sum_{i=1}^{D-1}
M^{\frac{(D-3)+(i-1)}{D-3}}r^{(D-1)-i}}
   {\sum_{i=1}^{D-3} M^{\frac{i-1}{D-3}}r^{(D-3)-i}}\frac{1}{r^{D-1}}{\biggr
   )}^{\frac{1}{2}}\:,
\label{sch:gems}
\end{eqnarray}
where $k_{+}$ is the surface gravity of the event horizon $r_+$
given by $k_{+}=\frac{D-3}{2} M^{-\frac{1}{D-3}}$. Starting with
the $(D+1)$-dimensional Minkowski spacetime,
$ds^{2}=-dz_{0}^{2}+\sum_{i=1}^{D+1} dz_{i}^{2}$, the
transformations (\ref{sch:gems}) map it into the neutral
Tangherlini black hole (\ref{RN:metric D-dim}). When $D=4$,
(\ref{sch:gems}) yields the global embedding of the Schwarzschild
black hole found in \cite{deser3}.

Analogously the charged Tangherlini black hole ($Q\neq 0$) admits
the following embedding in a $(D+2)$-dimensional Minkowski
spacetime with two timelike coordinates
($ds^{2}=-dz_{0}^2+\sum_{i=1}^{D+1}dz_{i}^{2}-dz_{D+2}^{2}$)
\cite{OscarLemosNuno},
\begin{eqnarray}
 z^{D+1}  =  \int dr & \bigg(\frac{
 W(r_{+}^{D-3}+r_{-}^{D-3})+r r_{+}^{2(D-3)}+r_{+}^{2D-5}}
{r^{2}(r^{D-3}-r_{-}^{D-3}) W}\bigg)^{\frac{1}{2}};& \nonumber\\
 z^{D+2}  =  \int dr & \bigg(\frac{4r_{+}^{3D-7}r_{-}^{D-3}}
  {(r_{+}^{D-3}-r_{-}^{D-3})^{2}r^{2(D-2)}}\bigg)^{\frac{1}{2}};&
 \label{rn:gems}
\end{eqnarray}
with the coordinates $z^{0}$ to $z^{D}$ being the same as those
defined in (\ref{sch:gems}), and we have defined
$W=\sum_{i=1}^{D-3}r_{+}^{i-1}r^{D-3-i}$. When $D=4$,
(\ref{rn:gems}) yields the global embedding of the
Reissner-Nordstr\"{o}m black hole found in \cite{deser3}.

\section{\label{sec:Solutions D-dim cosmolog dS}Higher dimensional
exact solutions in a dS background}

\subsection{\label{sec:BH D-dim dS}Higher dimensional
black holes in an asymptotically dS background}

In an asymptotically de Sitter background, $\Lambda> 0$, the
static higher dimensional black hole solutions with spherical
topology were also found by Tangherlini \cite{tangherlini}. The
gravitational field and the electromagnetic field are still given
by (\ref{RN:metric D-dim}) and by (\ref{RN:maxwell D-dim}),
respectively, but now the function $f(r)$ is given by
\begin{equation}
f(r)=1-\frac{\Lambda}{3}r^2-\frac{M}{r^{D-3}}+\frac{Q^2}{r^{2(D-3)}}\:.
 \label{RN:f cosmolog D-dim}
\end{equation}
The mass parameter $M$ and the charge parameter $Q$ are related to
the ADM hairs, $M_{\rm ADM}$ and $Q_{\rm ADM}$, by (\ref{ADM hairs
D-dim}). The causal structure of the dS-Tangherlini black holes is
similar to the one of their 4-dimensional counterparts (see
chapter \ref{chap:BH 4D}). In particular, these solutions have a
curvature singularity at the origin, and the black holes can have
at most three horizons with one of them being the cosmological
horizon.

In section \ref{sec:BH D-dim extremal limits dS} and in chapter
\ref{chap:Pair creation in higher dimensions}, we will need to
have the range of parameters for which one has an extreme dS black
hole. We start with the five-dimensional case, $D=5$
\cite{VitorOscarLemos-BHdSDdim}. The non-extreme dS black hole has
three horizons, namely the Cauchy horizon, $r_-$, the event
horizon, $r_+$, and the cosmological horizon $r_{\rm c}$, with
$r_-\leq r_+ \leq r_{\rm c}$. We are interested in the extreme dS
black holes, for which two of the above horizons coincide. Let us
label this degenerated horizon by $\rho$. In this case (and for
$D=5$), the function $f(r)$ given by (\ref{RN:f cosmolog D-dim})
can be written as \cite{VitorOscarLemos-BHdSDdim}
\begin{eqnarray}
f(r)=-\frac{\Lambda}{3}\frac{1}{r^4}(r-\rho)^2(r+\rho)^2
 \left ( r-\sqrt{\frac{3}{\Lambda}-2\rho^2}\right )
 \left ( r+\sqrt{\frac{3}{\Lambda}-2\rho^2}\right )\:,
 \label{Fextreme 5D}
 \end{eqnarray}
and thus the other horizon of the extreme black hole is
\begin{eqnarray}
\sigma =\sqrt{\frac{3}{\Lambda}-2\rho^2} \:.
  \label{zero extra 5D}
 \end{eqnarray}
The mass parameter $M$ and the charge parameter $Q$ of the black
holes are written as functions of $\rho$ as
\begin{eqnarray}
M=\rho^2 (2-\Lambda \rho^2)\:, \qquad {\rm and} \:\:\:
 Q^2 =\rho^4 \left (1- \frac{2\Lambda}{3}\rho^2 \right )\:.
 \label{mq 5D}
 \end{eqnarray}
The condition $Q^2\geq 0$ implies that $\rho \leq
\sqrt{\frac{3}{2\Lambda}}$. At this point we note that $M$ and $Q$
first increase with $\rho$ (this stage corresponds to
$\sigma>\rho$), until $\rho$ reaches the critical value
$\rho=\sqrt{1/\Lambda}$
 (this stage corresponds to $\sigma=\rho$), and then $M$ and $Q$ start decreasing until $\rho$
reaches its maximum allowed value (this stage corresponds to
$\sigma<\rho$). These three stages are associated to three
distinct extreme dS black holes: the cold, the ultracold and the
Nariai\footnote{Please note that in this section we deal with the
Nariai black hole while in the next section we will generate the
Nariai solution which is not a black hole solution.} black holes,
respectively (here we follow the nomenclature used in the
analogous 4-dimensional black holes \cite{Rom}). More precisely,
for $0<\rho<\frac{1}{\sqrt{\Lambda}}$ one has the cold black hole
with $r_-=r_+ \equiv \rho$ and $r_{\rm c}\equiv \sigma$. The
ranges of the mass and charge parameters for the cold black hole
are $0<M<\frac{1}{\Lambda}$ and $0<Q<\frac{1}{\sqrt{3}\Lambda}$.
The case $\rho=\frac{1}{\sqrt{\Lambda}}$ gives the ultracold black
hole in which the three horizons coincide, $r_-=r_+ = r_{\rm c}$.
Its mass and charge parameters are $M=\frac{1}{\Lambda}$ and
$Q=\frac{1}{\sqrt{3}\Lambda}$. For
$\frac{1}{\sqrt{\Lambda}}<\rho\leq \sqrt{\frac{3}{2\Lambda}}$ one
has the Nariai black hole with $r_+=r_{\rm c} \equiv \rho$ and
$r_-\equiv \sigma$. The ranges of the mass and charge parameters
for the Nariai black hole are $\frac{3}{4\Lambda}\leq
M<\frac{1}{\Lambda}$ and $0\leq Q<\frac{1}{\sqrt{3}\Lambda}$.

Now, the above construction can be extended for $D$-dimensional
extreme dS black holes. In the extreme case the function $f(r)$
given by (\ref{RN:f cosmolog D-dim}) can be written as
\cite{VitorOscarLemos-BHdSDdim}
\begin{eqnarray}
f(r)= (r-\rho)^2 \frac{1}{r^2}\left [ 1- \frac{\Lambda}{3}
 \left [ r^2+h(r) \right ] \right ]\:,
 \label{Fextreme D-dim}
 \end{eqnarray}
where $r=\rho$ is the degenerated horizon of the black hole, and
\begin{eqnarray}
h(r)=a+br+\frac{c_1}{r}+\frac{c_2}{r^2}+\cdots+\frac{c_{2(D-4)}}{r^{2(D-4)}}
\:,
  \label{aux function h(r)}
 \end{eqnarray}
where $a, b, c_1,...,c_{2(D-4)}$ are constants that can be found
through the matching between (\ref{RN:f cosmolog D-dim}) and
(\ref{aux function h(r)}). This procedure yields the mass
parameter $M$ and the charge parameter $Q$ of the black holes as
functions of $\rho$,
\begin{eqnarray}
M=2\rho^{D-3} \left ( 1-\frac{D-2}{D-3}\,\frac{\Lambda}{3} \rho^2
\right)\:, \qquad {\rm and} \qquad
 Q^2 =\rho^{2(D-3)} \left ( 1-\frac{D-1}{D-3}\,\frac{\Lambda}{3} \rho^2
\right)\:.
 \label{mq D-dim}
 \end{eqnarray}
The condition $Q^2\geq 0$ implies that $\rho \leq \rho_{\rm max}$
with
\begin{eqnarray}
 \rho_{\rm max}=\sqrt{\frac{D-3}{D-1}\,\frac{3}{\Lambda}}\:.
 \label{def rho max}
\end{eqnarray}
For the $D$-dimensional cold black hole ($r_-=r_+$), $M$ and $Q$
increase with $\rho$, and one has
\begin{eqnarray}
 0<\rho<\rho_{\rm u}\,, \qquad
 0<M<\frac{4}{D-1}\,  \rho_{\rm u}^{\, D-3}\,, \qquad {\rm and} \qquad
 0<Q<\frac{1}{\sqrt{D-2}}\, \rho_{\rm u}^{\, D-3}\:,
 \label{ColdDdim:range}
\end{eqnarray}
where we have defined
\begin{eqnarray}
 \rho_{\rm u}=\sqrt{\frac{3}{\Lambda}}\,\frac{D-3}{\sqrt{(D-2)(D-1)}}\:.
 \label{def rho ultra}
\end{eqnarray}
For the $D$-dimensional ultracold black hole
 ($r_-=r_+=r_{\rm c}$), one has
 \begin{eqnarray}
 \rho=\rho_{\rm u}\,, \qquad M=\frac{4}{D-1} \,  \rho_{\rm u}^{\,D-3}\,,
 \qquad {\rm and} \qquad Q=\frac{1}{\sqrt{D-2}}\,  \rho_{\rm u}^{\,D-3}\:.
 \label{UltracoldDdim:range}
\end{eqnarray}
Finally, for the $D$-dimensional Nariai black hole
 ($r_+=r_{\rm c}$), $M$ and $Q$ decrease with $\rho$, and one has
\begin{eqnarray}
 \rho_{\rm u}< \rho \leq \rho_{\rm max}\,, \quad
 \frac{2}{D-1}\,\rho_{\rm max}^{D-3}\leq M<\frac{4}{D-1} \, \rho_{\rm u}^{\,D-3}\,,
 \quad {\rm and} \quad 0\leq Q<\frac{1}{\sqrt{D-2}} \, \rho_{\rm u}^{\,D-3}\:.&
 \label{NariaiDdim:range}
\end{eqnarray}

\begin{figure}[H]
\centering
\includegraphics[height=5cm]{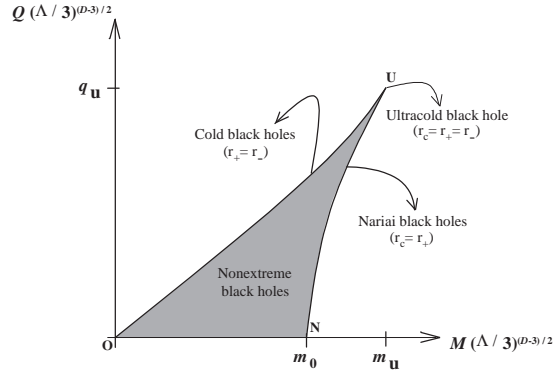}
   \caption{\label{range mq dS bh D-dim}
Range of $M$ and $Q$ for which one has a nonextreme black hole
(region interior to the closed line $ONUO$), an extreme Nariai
black hole with $r_+ = r_{\rm c}$ (line $NU$), an extreme cold
black hole with $r_- =r_+$ (line $OU$), and an extreme ultracold
black hole with $r_-=r_+ = r_{\rm c}$ (point $U$). The line $ON$
represents the nonextreme dS-Schwarzschild black hole, and point
$N$ represents the extreme Nariai Schwarzschild black hole.
$m_0=\frac{2}{D-1}\left ( \frac{D-3}{D-1}\right )^{(D-3)/2}$,
$m_{\rm u}=\frac{4}{D-1}\left ( \frac{(D-3)^2}{(D-2)(D-1)}\right
)^{(D-3)/2}$, and
 $q_{\rm u}=\frac{1}{\sqrt{D-2}}\left ( \frac{(D-3)^2}{(D-2)(D-1)}
  \right )^{(D-3)/2}$.
 }
\end{figure}

The ranges of $M$ and $Q$ that represent which one of the above
black holes are sketched in Fig. \ref{range mq dS bh D-dim}.

\subsection{\label{sec:BH D-dim extremal limits dS}Extremal limits of the higher dimensional dS black holes}

In chapter \ref{chap:Extremal Limits}, we saw that Ginsparg and
Perry \cite{GinsPerry} have connected the extreme Nariai black
hole with the Nariai solution \cite{Nariai} in a 4-dimensional
spacetime. This is, they have shown that the already known Nariai
solution (which is not a black hole solution) could be generated
from an extremal limit of a near-Nariai black hole. They realized
this connection while they were studying the quantum stability of
the Nariai and the Schwarzschild-dS solutions
\cite{GinsPerry,Bousso60y}. A similar procedure allows to generate
the Bertotti-Robinson solution and the Nariai$-$Bertotti-Robinson
solution from the near-cold black hole and near-ultracold black
hole, respectively. In this section, we will apply the near
extremal procedure of \cite{GinsPerry} to the extreme black holes
discussed in the last subsection, in order to find the higher
dimensional Nariai, dS$-$Bertotti-Robinson, Bertotti-Robinson and
Nariai$-$Bertotti-Robinson solutions
\cite{VitorOscarLemos-BHdSDdim}.

\subsubsection{\label{sec:BH D-dim Nariai}Higher dimensional Nariai
solution}

In order to generate the higher dimensional Nariai solution from
the near-Nariai black hole we first go back to (\ref{Fextreme
D-dim}) and rewrite it in the form $f(r)=-A(r)(r-\rho)^2$, where
$r=\rho$ is the degenerated horizon of the black hole, and $A(r)$
is a polynomial function of $r$. Then, we  set
$r_+=\rho-\varepsilon$ and $r_{\rm c}=\rho+\varepsilon$, in order
that $\varepsilon<<1$ measures the deviation from degeneracy, and
the limit $r_+\rightarrow r_{\rm c}$ is obtained when $\varepsilon
\rightarrow 0$. Now, we introduce a new time coordinate $T$, $t=
\frac{1}{\varepsilon A}\,T$, and a new radial coordinate $\chi$,
$r=\rho+\varepsilon \cos\chi$, where $\chi=0$ and $\chi=\pi$
correspond, respectively, to the horizons $r_{\rm c}$ and $r_+$,
and $A\equiv A(\rho)=\frac{1}{\rho^2}\left [ 1- \frac{\Lambda}{3}
 \left ( \rho^2+h(\rho) \right ) \right ]>0$, with
$h(\rho)$ is defined in (\ref{aux function h(r)}). Then, in the
limit $\varepsilon \rightarrow 0$, from  (\ref{RN:metric D-dim})
and (\ref{Fextreme D-dim}), we obtain the gravitational field of
the Nariai solution \cite{VitorOscarLemos-BHdSDdim}
\begin{eqnarray}
d s^2 = \frac{1}{A} \left (-\sin^2\chi\, dT^2 +d\chi^2 \right ) +
 \frac{1}{B}\, d\Omega_{D-2}^2 \:.
 \label{D-dim:Nariai solution}
\end{eqnarray}
where $\chi$ runs from $0$ to $\pi$, and $A$ and
$B=\frac{1}{\rho^2}$ are related to $\Lambda$ and $Q$ by
\begin{eqnarray}
 \Lambda = \frac{3}{(D-2)(D-1)}\left [ A+(D-3)^2 B \right ]\:,
 \quad {\rm and} \qquad Q^2=\frac{(D-3)B-A}{(D-3)(D-2)B^{D-2}}\:.
\label{D-dim:Nariai solution:range AB}
\end{eqnarray}
The Maxwell field (\ref{RN:maxwell D-dim}) of the higher
dimensional Nariai solution is
 \begin{eqnarray}
 F=Q_{\rm ADM}\,\frac{B^{(D-2)/2}}{A}\,\sin \chi \,dT \wedge d\chi\:.
\label{D-dim:Nariai Maxwell}
\end{eqnarray}
So, if we give the parameters $\Lambda$, and $Q$ we can construct
the higher dimensional Nariai solution. This solution is the
direct topological product of $dS_2 \times S^{D-2}$, i.e., of a
(1+1)-dimensional dS spacetime with a ($D-2$)-sphere of fixed
radius $B^{-1/2}$, and is an exact solution of Einstein-Maxwell
equations with $\Lambda>0$ in $D$-dimensions.

The neutral Nariai solution ($Q=0$) satisfies
$A=\frac{D-1}{3}\Lambda$ and $B=\frac{D-1}{3(D-3)}\Lambda$. The
$\Lambda=0$ limit of the Nariai solution is $D$-dimensional
Minkowski spacetime as occurs with the $D=4$ solution (see
subsection \ref{sec:lim L=0 Nariai}).

\subsubsection{\label{sec:BH D-dim dS Bertotti-Robinson}Higher
dimensional dS Bertotti-Robinson solution}

In order to generate the higher dimensional dS Bertotti-Robinson
solution from the near-cold black hole we first go back to
(\ref{Fextreme D-dim}) and rewrite it in the form
$f(r)=C(r)(r-\rho)^2$, where $r=\rho$ is the degenerated horizon
of the black hole, and $C(r)$ is a polynomial function of $r$.
Then, we set $r_-=\rho-\varepsilon$ and $r_+=\rho+\varepsilon$, in
order that $\varepsilon<<1$ measures the deviation from
degeneracy, and the limit $r_-\rightarrow r_+$ is obtained when
$\varepsilon \rightarrow 0$. Now, we introduce a new time
coordinate $T$, $t= \frac{1}{\varepsilon C}\,T$, and a new radial
coordinate $\chi$, $r=\rho+\varepsilon \cosh\chi$, where $C\equiv
 C(\rho)=\frac{1}{\rho^2}\left [ 1- \frac{\Lambda}{3}
 \left ( \rho^2+h(\rho) \right ) \right ]>0$, with
$h(\rho)$ is defined in (\ref{aux function h(r)}). Then, in the
limit $\varepsilon \rightarrow 0$, from  (\ref{RN:metric D-dim})
and (\ref{Fextreme D-dim}), we obtain the gravitational field of
the dS Bertotti-Robinson solution \cite{VitorOscarLemos-BHdSDdim}
\begin{eqnarray}
d s^2 = \frac{1}{C} \left (-\sinh^2\chi\, dT^2 +d\chi^2 \right ) +
 \frac{1}{B}\, d\Omega_{D-2}^2 \:.
 \label{D-dim:BertRob solution}
\end{eqnarray}
where $C$ and $B=\frac{1}{\rho^2}$ are related to $\Lambda$ and
$Q$ by
\begin{eqnarray}
 \Lambda = \frac{3}{(D-2)(D-1)}\left [-C+(D-3)^2 B \right ]\:,
 \quad {\rm and} \qquad Q^2=\frac{(D-3)B+C}{(D-3)(D-2)B^{D-2}}\:.
\label{D-dim:BertRob solution:range AB}
\end{eqnarray}
The Maxwell field (\ref{RN:maxwell D-dim}) of the higher
dimensional dS Bertotti-Robinson solution is
 \begin{eqnarray}
 F=-Q_{\rm ADM}\,\frac{B^{(D-2)/2}}{C}\,\sinh \chi \,dT \wedge d\chi\:.
\label{D-dim:BertRob Maxwell}
\end{eqnarray}
So, if we give the parameters $\Lambda$, and $Q$ we can construct
the higher dimensional dS Bertotti-Robinson  solution. This
solution is the direct topological product of $AdS_2 \times
S^{D-2}$, i.e., of a (1+1)-dimensional AdS spacetime with a
($D-2$)-sphere of fixed radius $B^{-1/2}$, and is an exact
solution of Einstein-Maxwell equations with $\Lambda>0$ in
$D$-dimensions. There is no neutral ($Q=0$) Bertotti-Robinson
solution.

\subsubsection{\label{sec:BH D-dim Bertotti-Robinson}Higher
dimensional flat Bertotti-Robinson solution}

From the $\Lambda=0$ limit of the dS Bertotti-Robinson, one can
generate the $\Lambda=0$ Bertotti-Robinson solution. It is
described by (\ref{D-dim:BertRob solution}) and
(\ref{D-dim:BertRob Maxwell}) with $C$ and $B$ being related to
$Q$ by \cite{VitorOscarLemos-BHdSDdim}
\begin{eqnarray}
 B = Q^{-2/(D-3)}\:,\quad {\rm and} \qquad C=(D-3)^2\,Q^{-2/(D-3)}\:.
\label{D-dim:BertRob L=0 solution:range AB}
\end{eqnarray}
Topologically this solution is also $AdS_2 \times S^{D-2}$, and is
an exact solution of Einstein-Maxwell equations with $\Lambda=0$
in $D$-dimensions.

\subsubsection{\label{sec:BH D-dim Nariai-Bertotti-Robinson}Higher
dimensional Nariai$-$Bertotti-Robinson solution}

In order to generate the higher dimensional
Nariai$-$Bertotti-Robinson solution from the near-ultracold black
hole we first go back to (\ref{Fextreme D-dim}) and rewrite it in
the form $f(r)=-P(r)(r-\rho)^2(r-\sigma)$, where $r=\rho$ is a
degenerated horizon of the black hole, $\sigma> \rho$ is the other
horizon, and $P(r)$ is a polynomial function of $r$. Then, we set
$\rho=\rho_{\rm u}-\varepsilon$ and $\sigma=\rho_{\rm u}+
\varepsilon$, with $\rho_{\rm u}$ defined in
 (\ref{def rho ultra}), in order that $\varepsilon<<1$ measures the deviation
from degeneracy, and the limit $\rho \rightarrow \sigma$ is
obtained when $\varepsilon \rightarrow 0$. Now, we introduce a new
time coordinate $T$, $t= \frac{1}{2\varepsilon^2 P}\,T$, and a new
radial coordinate $\chi$, $r=\rho_{\rm u}+\varepsilon \cos\left (
\sqrt{2\varepsilon P}\,\chi \right )$, where
 $P\equiv P(\rho_{\rm u})>0$.
Then, in the limit $\varepsilon \rightarrow 0$, from
(\ref{RN:metric D-dim}) we obtain the gravitational field of the
Nariai$-$Bertotti-Robinson solution
\cite{VitorOscarLemos-BHdSDdim}
\begin{eqnarray}
d s^2 =-\chi^2\, dT^2 +d\chi^2 +\rho_{\rm u}^{\,2}\,
d\Omega_{D-2}^2 \:.
 \label{D-dim:Nariai BertRob solution}
\end{eqnarray}
where $\chi$ runs from $0$ to $+\infty$, and $\rho_{\rm u}$
defined in (\ref{def rho ultra}). The Maxwell field
(\ref{RN:maxwell D-dim}) of the higher dimensional
Nariai$-$Bertotti-Robinson solution is
 \begin{eqnarray}
 F=\frac{Q_{\rm ADM}}{\rho_{\rm u}^{D-2}}\,\chi \,dT \wedge
 d\chi\:,
\label{D-dim:Nariai BertRob Maxwell}
\end{eqnarray}
where $Q_{\rm ADM}$ is given by (\ref{ADM hairs D-dim}) and
(\ref{UltracoldDdim:range}). So, if we give  $\Lambda$ we can
construct the higher dimensional Nariai$-$Bertotti-Robinson
solution. Notice that the spacetime factor $-\chi^2\, dT^2
+d\chi^2$ is just ${\mathbb{M}}^{1,1}$ (2-dimensional Minkowski
spacetime) in Rindler coordinates. Therefore, under the usual
coordinate transformation $\chi=\sqrt{x^2-t^2}$ and $T={\rm
arctanh(t/x)}$, this factor transforms into $-dt^2 +dx^2$. The
higher dimensional Nariai$-$Bertotti-Robinson solution is the
direct topological product of ${\mathbb{M}}^{1,1}\times S^{D-2}$,
and is an exact solution of Einstein-Maxwell equations with
$\Lambda>0$ in $D$-dimensions.

\section{\label{sec:Solutions D-dim cosmolog AdS}Higher dimensional
exact solutions in an AdS background}

\subsection{\label{sec:BH D-dim cosmolog AdS}Higher dimensional
black holes in an asymptotically AdS background}

In a higher dimensional asymptotically anti-de Sitter background,
$\Lambda< 0$, the Einstein-Maxwell equations allow a three-family
of static black hole solutions, parameterized by the constant $k$
which can take the values $1,0,-1$, and whose gravitational field
is described by
\begin{eqnarray}
d s^2 = - f(r)\, dt^2 +f(r)^{-1}\,dr^2+r^2 (d\,\Omega_{D-2}^k)^2,
 \label{AdS bH D-dim}
\end{eqnarray}
where
\begin{eqnarray}
f(r) =
k-\frac{\Lambda}{3}\,r^2-\frac{M}{r^{D-3}}+\frac{Q^2}{r^{2(D-3)}}\:,
 \label{RN:f AdS D-dim}
\end{eqnarray}
and
\begin{eqnarray}
(d\Omega_{D-2}^k)^2 \!\!\!&=& \!\!\!
d\theta_1^2+\sin^2\theta_1\,d\theta_2^2+ \sin^2\theta_1
\sin^2\theta_2\,d\theta_3^2+\cdots
+\prod_{i=1}^{D-3}\sin^2\theta_i\,d\theta_{D-2}^2\,,
\:\:\:\:\:\:\:\:{\rm for}\:\: k=1\,, \nonumber \\
(d\Omega_{D-2}^k)^2 \!\!\!&=& \!\!\! d\theta_1^2+d\theta_2^2+
d\theta_3^2+\cdots +d\theta_{D-2}^2\,, \qquad \qquad \qquad \qquad
\qquad \qquad
 \:\:\:\:\:\:\:\:\:\:\:\:\:\:\:\:\:\,{\rm for}\:\: k=0 \,,\nonumber \\
(d\Omega_{D-2}^k)^2 \!\!\!&=& \!\!\!
d\theta_1^2+\sinh^2\theta_1\,d\theta_2^2+ \sinh^2\theta_1
\sin^2\theta_2\,d\theta_3^2+\cdots +\sinh^2\theta_1
\prod_{i=2}^{D-3}\sin^2\theta_i\,d\theta_{D-2}^2\,, \:\:\:
{\rm for}\:\:  k=-1\,. \nonumber \\
& &
 \label{angular AdS bh D-dim}
\end{eqnarray}
Thus, the family with $k=1$ yields AdS black holes with spherical
topology found in \cite{tangherlini}. The family with $k=0$ yields
AdS black holes with planar, cylindrical  or toroidal (with genus
$g\geq 1$) topology that are the higher dimensional counterparts
(introduced in \cite{Birmingham-TopDdim} in the neutral case, and
in \cite{Awad-ToroidalDdim} in the charged case) of the
4-dimensional black holes found and analyzed in
\cite{Lemos,Zanchin_Lemos,OscarLemos_string}. Finally, the family
with $k=-1$ yields AdS black holes with hyperbolic, or toroidal
topology with genus $g\geq 2$ that are the higher dimensional
counterparts (introduced in \cite{Birmingham-TopDdim} in the
neutral case) of the 4-dimensional black holes analyzed in
\cite{topological}. The solutions with non-spherical topology
(i.e., with $k=0$ and $k=-1$) do not have counterparts in a
$\Lambda=0$ or in a $\Lambda>0$ background.

The mass parameter $M$ and the charge parameter $Q$ are related to
the ADM hairs, $M_{\rm ADM}$ and $Q_{\rm ADM}$, by (\ref{ADM hairs
D-dim}). The causal structure of these higher dimensional AdS
black holes is similar to the one of their 4-dimensional
counterparts (see chapter \ref{chap:BH 4D}). In particular, these
black holes can have at most two horizons. Following a similar
procedure as the one sketched in section \ref{sec:BH D-dim dS}, we
find the mass parameter $M$ and the charge parameter $Q$ of the
extreme black holes as functions of the degenerated horizon at
$r=\rho$ \cite{VitorOscarLemos-BHdSDdim}
\begin{eqnarray}
M=2\rho^{D-3} \left ( k-\frac{D-2}{D-3}\,\frac{\Lambda}{3} \rho^2
\right)\:, \qquad {\rm and} \qquad
 Q^2 =\rho^{2(D-3)} \left ( k-\frac{D-1}{D-3}\,\frac{\Lambda}{3} \rho^2
\right)\:.
 \label{mq AdS D-dim}
 \end{eqnarray}

\subsubsection{\label{sec:BH D-dim AdS spherical}Higher dimensional
AdS black holes with spherical topology}

\begin{figure}[H]
\centering
\includegraphics[height=5cm]{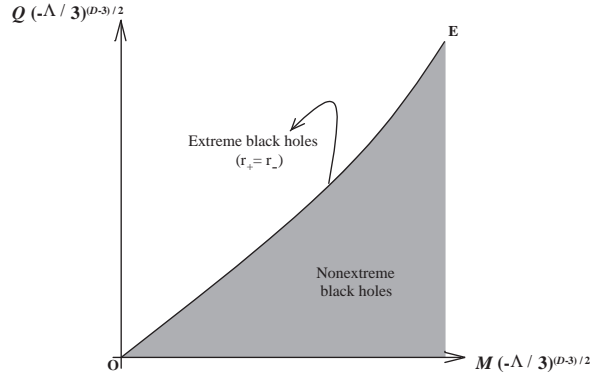}
   \caption{\label{range mq AdS spher+toroid bh D-dim}
Range of $M$ and $Q$ for which one has a nonextreme black hole
(region below the line $OE$), and an extreme black hole with $r_+
= r_-$ (line $OE$) in the AdS case with spherical topology ($k=1$)
or with planar, cylindrical or toroidal topology ($k=0$). The
region above the line $OE$ represents a naked singularity.
 }
\end{figure}

When $k=1$, one has $0<\rho<+\infty$ and $M$ and $Q$ in
 (\ref{mq AdS D-dim}) are always positive. The ranges of $M$ and $Q$
that represent extreme and nonextreme black holes are sketched in
Fig. \ref{range mq AdS spher+toroid bh D-dim}.

For $D=5$, the function $f(r)$ in the extreme case can be written
as
\begin{eqnarray}
f(r)=-\frac{\Lambda}{3}\frac{1}{r^4}(r-\rho)^2(r+\rho)^2
 \left ( r^2+2\rho^2-\frac{3}{\Lambda}\right )\:.
 \label{Fextreme 5D AdS spheric}
 \end{eqnarray}

\subsubsection{\label{sec:BH D-dim AdS toroidal}Higher dimensional
AdS black holes  with toroidal or cylindrical topology}

When $k=0$, one has $0<\rho<+\infty$ and $M$ and $Q$ in
 (\ref{mq AdS D-dim}) are also always positive. The ranges of $M$ and $Q$
that represent extreme and nonextreme black holes are sketched in
Fig. \ref{range mq AdS spher+toroid bh D-dim}.

For $D=5$, the function $f(r)$ in the extreme case can be written
as
\begin{eqnarray}
f(r)=-\frac{\Lambda}{3}\frac{1}{r^4}(r-\rho)^2(r+\rho)^2
  ( r^2+2\rho^2)\:.
 \label{Fextreme 5D AdS toroidal}
 \end{eqnarray}
\subsubsection{\label{sec:BH D-dim AdS topological}Higher dimensional
AdS black holes with hyperbolic topology}

\begin{figure}[H]
\centering
\includegraphics[height=5cm]{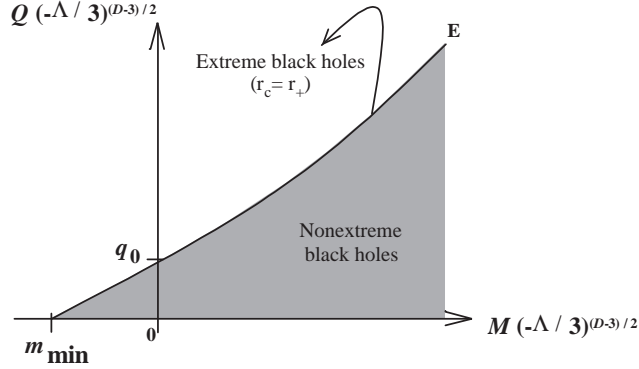}
   \caption{\label{range mq AdS hyp bh D-dim}
Range of $M$ and $Q$ for which one has a nonextreme black hole
(region below the line $OE$), and an extreme black hole with
$r_{\rm c} = r_+$ (line $OE$) in the AdS case with hyperbolic
topology ($k=-1$). The region above the line $OE$ represents a
naked singularity. $m_{\rm min}=-\frac{4}{D-1}\left (
-\frac{D-3}{D-1}\right )^{(D-3)/2}$, and
$q_0=\frac{1}{\sqrt{D-2}}\left ( -\frac{D-3}{D-2}
 \right )^{(D-3)/2}$.
 }
\end{figure}

When $k=-1$, the condition that $Q^2\geq 0$ demands that
 $-\frac{D-3}{D-1}\frac{3}{\Lambda}\leq \rho<+\infty$.
For  $\rho=-\frac{D-3}{D-1}\frac{3}{\Lambda}$, the extreme black
hole has no electric charge ($Q=0$) and its mass is negative,
$M=-\frac{4}{D-1}\left ( -\frac{D-3}{D-1}\right )^{(D-3)/2}$. For
$\rho=-\frac{D-3}{D-2}\frac{3}{\Lambda}$, the extreme black hole
has no mass ($M=0$) and its charge is given by
$Q=\frac{1}{\sqrt{D-2}}\left ( -\frac{D-3}{D-2}
 \right )^{(D-3)/2}$. The ranges of $M$ and $Q$
that represent extreme and nonextreme black holes are sketched in
Fig. \ref{range mq AdS hyp bh D-dim}.

For $D=5$, the function $f(r)$ in the extreme case can be written
as
\begin{eqnarray}
f(r)=-\frac{\Lambda}{3}\frac{1}{r^4}(r-\rho)^2(r+\rho)^2
 \left ( r-\sqrt{-\frac{3}{\Lambda}-2\rho^2}\right )
 \left ( r+\sqrt{-\frac{3}{\Lambda}-2\rho^2}\right )\:.
 \label{Fextreme 5D AdS hyperbolic}
 \end{eqnarray}
The condition $Q^2\geq 0$ requires $\rho^2\geq -3/(2\Lambda)$
which implies that $r=\rho$ is the only real root of the solution.

\subsection{\label{sec:BH D-dim extremal limits AdS}Extremal limits of the higher dimensional AdS black holes}

In this subsection, we will consider the extremal limits of the
near-extreme higher dimensional AdS black holes. This procedure
leads to the generation of the higher dimensional anti-Nariai
solution and higher dimensional AdS Bertotti-Robinson solution.
 To achieve our aim, we first go back to the extreme case of
(\ref{RN:f AdS D-dim}) and rewrite it in the form
$f(r)=A(r)(r-\rho)^2$, where $r=\rho$ is the degenerated horizon
of the black hole, and $A(r)$ is a polynomial function of $r$.
Then, we introduce a new time coordinate $T$, $t=
\frac{1}{\varepsilon A}\,T$, and a new radial coordinate $\chi$,
$r=\rho+\varepsilon \cosh\chi$, where $A\equiv A(\rho)$, and
$\varepsilon<<1$ measures the deviation from degeneracy. Finally,
taking the limit $\varepsilon \rightarrow 0$ in (\ref{AdS bH
D-dim}) yields the gravitational field of the new higher
dimensional solutions \cite{VitorOscarLemos-BHdSDdim},
\begin{eqnarray}
d s^2 = \frac{1}{A} \left (-\sinh^2\chi\, dT^2 +d\chi^2 \right ) +
 \frac{1}{B}\, (d\Omega^{k}_{D-2})^2 \:,
 \label{D-dim:AdS NBR solution}
\end{eqnarray}
where $k=1,0,-1$ in the spherical, cylindrical and hyperbolic
cases, respectively, and $A$ and $B$ are constants related to
$\Lambda$ and $Q$ by
\begin{eqnarray}
 \Lambda =-\frac{3}{(D-2)(D-1)}\left [ A-k\,(D-3)^2 B \right ]\:,
 \quad {\rm and} \qquad
 Q^2=\frac{A+k\,(D-3)B}{(D-3)(D-2)B^{D-2}}\:,
\label{D-dim:range AB NBR AdS}
\end{eqnarray}
In the new coordinate system, the Maxwell field
 (\ref{RN:maxwell D-dim}) of the solutions is
 \begin{eqnarray}
 F=-Q_{\rm ADM}\,\frac{B^{(D-2)/2}}{A}\,\sinh \chi \,dT \wedge d\chi\:.
\label{D-dim:NBR AdS Maxwell}
\end{eqnarray}
Equations
 (\ref{D-dim:AdS NBR solution})-(\ref{D-dim:NBR AdS Maxwell})
describe three exact solutions of the Einstein-Maxwell theory with
$\Lambda<0$ in $D$-dimensions. The $k=+1$ case describes the AdS
Bertotti-Robinson solution with spherical topology. This solution
is the direct topological product of $AdS_2 \times S^{D-2}$, i.e.,
of a (1+1)-dimensional AdS spacetime with a ($D-2$)-sphere of
fixed radius $B^{-1/2}$. The $k=0$ case describes the AdS
Bertotti-Robinson solution with toroidal or cylindrical topology.
This solution is the direct topological product of $AdS_2 \times
\mathbb{E}^{D-2}$, i.e., of a (1+1)-dimensional AdS spacetime with
a ($D-2$) Euclidean space. Finally, the $k=-1$ case describes the
the higher dimensional anti-Nariai solution. This solution is the
direct topological product of $AdS_2 \times H^{D-2}$, i.e., of a
(1+1)-dimensional AdS spacetime with a ($D-2$)-hyperboloid with a
fixed size, $B^{-1/2}$. The $k=-1$ case is the only one that
admits a solution with $Q=0$. This neutral anti-Nariai solution
satisfies $A=-\frac{D-1}{3}\Lambda$ and
$B=-\frac{D-1}{3(D-3)}\Lambda$. The $\Lambda=0$ limit of the
anti-Nariai solution is $D$-dimensional Minkowski spacetime as
occurs with the $D=4$ solution (see subsection \ref{sec:lim L=0
Nariai}).

%% file: Chapter10.tex
\thispagestyle{empty} \setcounter{minitocdepth}{1}
\chapter[Pair creation of black holes in higher dimensional spacetimes]
{\Large{Pair creation of black holes in higher dimensional
spacetimes}}
 \label{chap:Pair creation in higher dimensions}
 \lhead[]{\fancyplain{}{\bfseries Chapter \thechapter. \leftmark}}
 \rhead[\fancyplain{}{\bfseries \rightmark}]{}
  \minitoc \thispagestyle{empty}
\renewcommand{\thepage}{\arabic{page}}

\addtocontents{lof}{\textbf{Chapter Figures \thechapter}\\}


In this chapter, we discuss in detail the creation of a higher
dimensional Tangherlini black hole pair in a dS background
\cite{OscarLemosNuno,VitorOscarLemosPChigherDim}. The instantons
that describe the process are constructed from the dS Tangherlini
solution, which describes a pair of higher dimensional dS black
holes accelerated apart by the cosmological constant expansion. We
compute the pair creation rates of the process. This study is the
only one discussing the black hole pair creation process in a
higher dimensional spacetime.

\section{\label{sec:higher dimensional instantons}The higher
dimensional dS instantons}

In order to evaluate the black hole pair creation rate we need to
find the instantons of the theory, i.e., we must look into the
Euclidean section of the higher dimensional dS solution and choose
only those Euclidean solutions which are regular in a way that
will be explained soon. To obtain the Euclidean section of the dS
solution from the Lorentzian dS solution, (\ref{RN:metric D-dim}),
(\ref{RN:maxwell D-dim}) and (\ref{RN:f cosmolog D-dim}), we
simply introduce an imaginary time coordinate $\tau=-it$.

To have a positive definite Euclidean metric we must require that
$r$ belongs to $r_+ \leq r \leq r_{\rm c}$. In general, when $r_-
\neq r_+$, one then has conical singularities at the horizons
$r=r_+$ and $r=r_{\rm c}$. In order to obtain a regular solution
we have to eliminate the conical singularities at both horizons.
This is achieved by imposing that the period of $\tau$ is the same
for the two horizons, and is equivalent to requiring that the
Hawking temperature of the two horizons be equal. To eliminate the
conical singularity at $r=r_{\rm c}$ the period of $\tau$ must be
$\beta=2 \pi/ k_{\rm c}$ (where $k_{\rm c}$ is the surface gravity
of the cosmological horizon),
\begin{equation}
\beta=\frac{4 \pi}{|f'(r_{\rm c})|}\:.
 \label{PC D-dim:Period tau-yA PCdS}
 \end{equation}
 This choice for the period of $\tau$ also eliminates
simultaneously the conical singularity at the outer black hole
horizon, $r=r_+$, if and only if the parameters of the solution
are such that  the surface gravities of the  black hole and
acceleration horizons are equal ($k_+=k_{\rm c}$), i.e.,
\begin{equation}
 f'(r_+)=-f'(r_{\rm c})\:.
 \label{PC D-dim:k+=kA PCdS}
 \end{equation}
There are two ways to satisfy this condition. One is a regular
Euclidean solution with $r_+ \neq r_{\rm c}$, and will be called
lukewarm instanton. This solution requires the presence of an
electromagnetic charge. The other way is to have $r_+ = r_{\rm
c}$, and will be called Nariai instanton. This last solution
exists with or without charge. When we want to distinguish them,
they will be labelled by charged Nariai and neutral Nariai
instantons, respectively.

We now turn our attention to the case $r_- =r_+$ and $r_{\rm c}
\neq r_+$, which obviously requires the presence of charge. When
this happens the allowed range of $r$ in the Euclidean sector is
simply  $r_+<r\leq r_{\rm c}$. This occurs because when $r_- =r_+$
the proper distance along spatial directions between $r_+$ and
$r_{\rm c}$ goes to infinity. The point $r_+$ disappears from the
$\tau, r$ section which is no longer compact but becomes
topologically $S^1 \times {\mathbb{R}}$. Thus, in this case we
have a conical singularity only at $r_{\rm c}$, and so we obtain a
regular Euclidean solution by simply requiring that the period of
$\tau$ be equal to (\ref{PC D-dim:Period tau-yA PCdS}). We will
label this solution by cold instanton. Finally, we have a special
solution that satisfies
 $r_-=r_+=r_{\rm c}$ and that is regular when condition
(\ref{PC D-dim:Period tau-yA PCdS}) is satisfied. This instanton
will be called ultracold instanton and can be viewed as a limiting
case of both the charged Nariai instanton and cold instanton.

Below, we will describe each one of these four instantons,
following the order:  (A) cold instanton, (B) Nariai  instanton,
(C) ultracold  instanton, and (D) lukewarm  instanton. These
instantons are the higher dimensional counterparts of the
4-dimensional instantons that have been constructed
\cite{MelMos,Rom,MannRoss,BooMann} from the Euclidean section of
the dS$-$Reissner-Nordstr\"{o}m solution, and thus we preserve the
4-dimensional nomenclature. The cold instanton, the Nariai
instanton, and the ultracold instanton are obtained by
euclideanizing solutions found chapter in
 \ref{chap:Black holes in higher dimensions}.
 These four families of instantons will allow
us to calculate the pair creation rate of accelerated
dS$-$Reissner-Nordstr\"{o}m black holes in section
 \ref{sec:Calc-I PC D-dim}.

As is clear from the above discussion, when the charge vanishes
the only regular Euclidean solution that can be constructed is the
neutral Nariai instanton. The same feature is present in the
4-dimensional case where only the neutral Nariai instanton is
available \cite{GinsPerry,MannRoss,BoussoHawk,VolkovWipf}.

\subsection{\label{sec:Cold-inst PC D-dim}The higher dimensional cold instanton}

The gravitational field of the higher dimensional cold instanton
 is given by (\ref{RN:metric D-dim}) and (\ref{RN:f cosmolog
D-dim}), while its Maxwell field is given by (\ref{RN:maxwell
D-dim}), with the replacement $\tau=-it$. Moreover, the
degenerated horizon $\rho$, the mass parameter $M$, and the charge
parameter $Q$ satisfy relations (\ref{mq D-dim}) and
(\ref{ColdDdim:range}). The topology of the higher dimensional
cold  instanton is ${\mathbb{R}}^2 \times S^{D-2}$, since
$r=r_+=\rho$ is at an infinite proper distance ($0 \leq \tau \leq
\beta$, $r_+ < r \leq r_{\rm c}$). The surface $r=r_+=\rho$ is
then an internal infinity boundary that will have to be taken into
account in the calculation of the action of the cold  instanton
(see section \ref{sec:Cold-rate PC D-dim}). The Lorentzian sector
of this cold case describes two extreme ($r_-=r_+$) dS black holes
being accelerated by the cosmological background, and the higher
dimensional cold instanton describes pair creation of these
extreme black holes. To compute the pair creation rate of cold
black holes we need to know the location of the cosmological
horizon, $r_{\rm c}$. This location can be explicitly determined
for $D=4$ and $D=5$. Specifically, for $D=4$ one has $r_{\rm
c}=\sqrt{3/\Lambda-2\rho^2}-\rho$, where, from
(\ref{ColdDdim:range}), one has $0<\rho<1/\sqrt{2\Lambda}$. For
$D=5$ one has $r_{\rm c}=\sqrt{3/\Lambda-2\rho^2}$, where, from
(\ref{ColdDdim:range}), one has $0<\rho<1/\sqrt{\Lambda}$. For
$D\geq 6$ finding explicitly $r_{\rm c}$ requires solving a
polynomial of degree higher than four.

\subsection{\label{sec:Nariai-inst PC D-dim}The higher dimensional Nariai instanton}

The gravitational field of the higher dimensional Nariai instanton
 is given by (\ref{D-dim:Nariai solution}), while the
Maxwell field  is given by (\ref{D-dim:Nariai Maxwell}), with the
replacement $\tau=-iT$. Moreover, the parameter $A$ and
$B=\rho^{-2}$ satisfy relations
 (\ref{D-dim:Nariai solution:range AB}) and (\ref{NariaiDdim:range}).
 The topology of the Nariai
instanton is $S^2 \times S^{D-2}$
 ($0 \leq \tau \leq \beta$, $0\leq \chi \leq \pi$).
 The Lorentzian sector of this solution is  the direct topological product of
 $dS_2 \times S^{D-2}$, i.e., of a
(1+1)-dimensional de Sitter spacetime with a  (D-2)-sphere of
fixed size. To each point in the sphere corresponds a $dS_2$
spacetime.  In the $D=4$ case it has been shown
\cite{GinsPerry,BoussoHawk} that the Nariai solution decays
through the quantum tunnelling process into a slightly non-extreme
dS black hole pair (for a complete review  on this subject see
\cite{Bousso60y}). We then naturally expect that an analogous
quantum instability is present in the higher dimensional Nariai
solution. Therefore, the Nariai instanton describes the creation
of a higher dimensional Nariai universe  that then decays into a
slightly non-extreme ($r_+ \sim r_{\rm c}$) pair of black holes
accelerated by the cosmological constant background.

\subsection{\label{sec:Ultracold-inst PC D-dim}The higher dimensional ultracold instanton}

The gravitational field of the higher dimensional ultracold
instanton is given by (\ref{D-dim:Nariai BertRob solution}) while
the Maxwell field is given by
 (\ref{D-dim:Nariai BertRob Maxwell}), with the replacement $\tau=-iT$.
Moreover the parameters $\rho_{\rm u}$ and $Q$ are defined  in
(\ref{def rho ultra}) and (\ref{UltracoldDdim:range}),
respectively. The topology of the ultracold instanton is
${\mathbb{E}}^{2}\times S^{D-2}$. The Lorentzian sector of this
solution is  the direct topological product of
 ${\mathbb{M}}^{(1,1)}\times
S^{D-2}$, i.e., of a (1+1)-dimensional Minkowski spacetime with a
(D-2)-sphere of fixed size. The ultracold instanton describes the
creation of a higher dimensional Nariai$-$Bertotti-Robinson
universe that then decays into a slightly non-extreme ($r_- \sim
r_+ \sim r_{\rm c}$) pair of black holes accelerated by the
cosmological constant background.

\subsection{\label{sec:Lukewarm-inst PC D-dim}The higher dimensional lukewarm instanton}

The gravitational field of the higher dimensional lukewarm
instanton  is given by (\ref{RN:metric D-dim}) with the
requirement that $f(r)$ satisfies condition
 (\ref{PC D-dim:k+=kA PCdS}) and
$f(r_+)=0=f(r_{\rm c})$. To find the properties of the lukewarm
instanton we note that the function $f(r)$, given by
(\ref{RN:metric D-dim}), can also be written as
\begin{eqnarray}
& & f(r)=-\frac{\Lambda}{3}r^2 \left (  1-\frac{r_+}{r}\right )
\left
(  1-\frac{r_{\rm c}}{r}\right ) \nonumber \\
& & \hspace{1cm}\times {\biggl (}
1+\frac{a_1}{r}+\frac{a_2}{r^2}+\cdots+\frac{a_{2(D-3)}}{r^{2(D-3)}}
 {\biggr )}\:,
 \label{PC D-dim:F-luk}
 \end{eqnarray}
where $a_i$ ($i=1,\cdots,2(D-3)$) are constants that can be found
from the matching between (\ref{RN:metric D-dim}) and
 (\ref{PC D-dim:F-luk}). This matching, together with the extra condition
 (\ref{PC D-dim:k+=kA PCdS}), lead to unique relations between the
parameters $\Lambda$, $M$, $Q$  and the position of the horizons,
$r_+$ and $r_{\rm c}$. Since this procedure involves polynomials
with a high degree, we have not been able to find the general
relations between ($\Lambda$, $M$, $Q$)  and ($r_+$, $r_{\rm c}$)
for any $D$. So, we have to carry this procedure for each $D$. As
examples, we specifically discuss now the $D=4$ and the $D=5$
lukewarm instantons. For $D=4$, the above procedure yields the
relations
\begin{eqnarray}
& & \Lambda =\frac{3}{(r_{\rm c}+r_+)^2} ,
\nonumber \\
& & M=2 \frac{r_{\rm c} r_+}{r_{\rm c}+r_+}\,, \nonumber \\
& &  Q= \frac{r_{\rm c} r_+}{r_{\rm c}+r_+}\:.
 \label{PC D-dim: zeros4D}
 \end{eqnarray}
For $D=5$, the relations are
\begin{eqnarray}
& & \Lambda =3 \left ( 2r_{\rm c}+r_+ -\frac{r_{\rm c}^3(r_{\rm
c}+r_+)}{r_{\rm c}^2+r_{\rm c}r_+ + r_+^2} \right )^{-1},
\nonumber \\
& & M=\frac{r_{\rm c}^2 r_+^2 (2r_{\rm c}^2+r_{\rm c}r_+ +
2r_+^2)}
 {r_{\rm c}^4+r_{\rm c}^3r_+ +3r_{\rm c}^2 r_+^2+r_{\rm c}r_+^3 + r_+^4}\,, \nonumber \\
& &  Q^2= \frac{r_{\rm c}^4 r_+^4}
 {r_{\rm c}^4+r_{\rm c}^3r_+ +3r_{\rm c}^2 r_+^2+r_{\rm c}r_+^3 + r_+^4}\:.
 \label{PC D-dim: zeros5D}
 \end{eqnarray}
These two examples indicate an important difference between the
lukewarm instanton in $D=4$ dimensions and in $D\geq 5$: for $D=4$
the lukewarm instanton has a ADM mass, $M_{\rm ADM}=M/2$, equal to
its ADM charge, $Q_{ADM}=Q$, while for $D\geq 5$ one has $M_{\rm
ADM}\neq Q_{ADM}$.
 Note also that relations (\ref{PC D-dim: zeros4D}) and (\ref{PC D-dim: zeros5D})
and their higher dimensional counterparts define implicitly
$r_{\rm c}$ and $r_+$ as a function of $\Lambda$, $M$, and $Q$.
The location of $r_{\rm c}$ and $r_+$ can be explicitly determined
for $D=4$ and $D=5$.  For $D\geq 6$ finding explicitly $r_{\rm c}$
requires solving a polynomial of degree higher than four.

The topology of the lukewarm instanton is $S^2 \times S^{D-2}$
 ($0\leq \tau \leq \beta$, $r_+ \leq r \leq r_{\rm c}$). The Lorentzian sector
of the lukewarm solution describes two higher dimensional dS black
holes being accelerated apart by the cosmological constant, so
this instanton describes pair creation of nonextreme black holes.

\section{\label{sec:Calc-I PC D-dim}Calculation of the black hole pair creation rates}

The pair creation rate of higher dimensional black holes in a dS
background is given, according to the no-boundary proposal of
\cite{HartleHawk}, by (see subsection \ref{sec:instanton-method})
\begin{eqnarray}
\Gamma \sim \eta \, e^{-2I_{\rm inst}+2I_{\rm dS}} \:,
 \label{PC D-dim:PC-rate}
 \end{eqnarray}
where $\eta$ is the one-loop contribution from the quantum
quadratic fluctuations in the fields that will not be considered
here. $I_{\rm inst}$ is the classical Euclidean action of the
gravitational instanton that mediates the pair creation of black
holes, given by \cite{Brown,HawkRoss}
 \begin{eqnarray}
I_{\rm inst}&=&-\frac{1}{16\pi}\int_{\cal M} d^Dx\sqrt{g} \left (
R-2\lambda-F^{\mu\nu}F_{\mu\nu} \right ) \nonumber \\
& & -\frac{1}{8\pi}\int_{\Sigma=\partial {\cal M}}
d^{D-1}x\sqrt{h}\, K  \nonumber \\
 & & -\frac{1}{4\pi}\int_{\Sigma=\partial {\cal M}}
d^{D-1}x\sqrt{h}\, F^{\mu\nu}n_{\mu}A_{\nu} \:,
 \label{PC D-dim:I-electric}
 \end{eqnarray}
where $\Sigma=\partial {\cal M}$ is the boundary of a compact
manifold ${\cal M}$, $g$ is the determinant of the Euclidean
metric, $h$ is the determinant of the induced metric on the
boundary $\Sigma$, $\lambda$ is proportional to the cosmological
constant as defined in (\ref{D-dim:def lambda}), $R$ is the Ricci
scalar defined in (\ref{D-dim:Ricci Tensor}), $K$ is the trace of
 the extrinsic curvature $K_{ij}$ of the boundary,
 $F_{\mu\nu}=\partial_{\mu}A_{\nu}-\partial_{\nu}A_{\mu}$ is
the Maxwell field strength of the gauge field $A_{\nu}$, and
$n_{\mu}$ is the unit outward normal to $\Sigma$

We note again, since this fact is important for the computation of
the pair creation rates, that in a general $D$-dimensional
background the electromagnetic energy-momentum tensor is not
traceless. Indeed, it is given by (\ref{D-dim:trace T}), which
vanishes only for $D=4$.

$I_{\rm dS}$ is the Euclidean action of the $S^{D}$ gravitational
instanton that mediates the nucleation of a dS space from nothing,
given by
\begin{eqnarray}
 I_{\rm dS}&=& -\frac{1}{16\pi}\int d^D x \sqrt{g} \left (
R-2\lambda \right ) \nonumber \\
&=& -\frac{3^{D/2}}{12\,\Lambda^{(D-2)/2}}\,
\frac{\pi^{(D-1)/2}}{\Gamma[(D-1)/2]} \:.
 \label{PC D-dim:I dS pure}
\end{eqnarray}

\subsection{\label{sec:Cold-rate PC D-dim}The higher dimensional cold pair creation rate}

The Maxwell field of the higher dimensional cold is $F=-i
\frac{Q_{\rm ADM}}{r^{D-2}}\, d\tau \wedge dr$. With this
information we are able to compute all the terms of the Euclidean
action (\ref{PC D-dim:I-electric}). We start with
  \begin{eqnarray}
& & \hspace{-2.0 cm} -\frac{1}{16\pi}\int_{\cal{M}} d^Dx\sqrt{g}
\left ( R-2\lambda \right ) \nonumber \\
&=& \left( -\frac{D-1}{24\pi}\,\Lambda+\frac{(D-4)(D-3)}{16
\pi}\,Q^2 \right ) \int d\Omega_{D-2}
 \int_0^{\beta/2} \!\!\!\!d\tau
\int_{\rho}^{r_{\rm c}} \!\!\!\!dr\: r^{D-2}
 \nonumber \\
&=& \frac{\pi^{(D-3)/2}}{\Gamma[(D-1)/2]}\,\frac{\beta}{8}
 \left [ -\frac{\Lambda}{3}
 \left ( r_{\rm c}^{\,D-1}-\rho^{\,D-1} \right )
+\frac{(D-4)\,Q^2}{2}
 \left (\rho^{\,-(D-3)}-r_{\rm c}^{\,-(D-3)} \right ) \right ]
\:,
 \label{PC D-dim:I1-cold}
   \end{eqnarray}
where  $\int d\Omega_{D-2}=\Omega_{D-2}$ is defined in
(\ref{integratedsolidangle}). The Maxwell term in the action
yields
\begin{eqnarray}
\hspace{-0.4 cm} \frac{1}{16\pi}\int_{\cal{M}} d^D x\sqrt{g} \:F^2
=-\frac{(D-2)\,Q^2\,\beta}{16}
\frac{\pi^{(D-3)/2}}{\Gamma[(D-1)/2]}
 \left [\rho^{\,-(D-3)}-r_{\rm c}^{\,-(D-3)} \right ]\!, \nonumber \\
& &
 \label{PC D-dim:I2-cold}
 \end{eqnarray}
and $\int_{\Sigma} d^{D-1}x\sqrt{h}\, K=0$. In order to compute
the extra Maxwell boundary term in (\ref{PC D-dim:I-electric}) we
have to find a vector potential, $A_{\nu}$, that is regular
everywhere including at the horizons. An appropriate choice in the
cold case is $A_r=- i\,\frac{Q_{\rm ADM}}{r^{D-2}}\,\tau$. The
integral over $\Sigma$ consists of an integration between $\rho$
and $r_{\rm c}$ along the $\tau=0$ surface and back along
$\tau=\beta/2$, and of an integration between $\tau=0$ and
$\tau=\beta/2$ along the $r=r_{\rm c}$ surface and back along the
$r=\rho$ surface. The normal to $\Sigma_{\tau}$ is
 $n_{\mu}=\left (\sqrt{f(r)},0,\cdots,0\right )$, and the normal to $\Sigma_{h}$ is
$n_{\mu}=\left (0,\sqrt{f(r)},0,\cdots,0\right )$. Thus the
non-vanishing contribution comes only from the integration along
the $\tau=\beta/2$ surface. The Maxwell boundary term in
 (\ref{PC D-dim:I-electric}) is then
 \begin{eqnarray}
\hspace{-0.5cm}-\frac{1}{4\pi}\int_{\Sigma_{\tau=\beta/2}}\!\!\!\!\!\!\!\!\!\!\!\!
d^{D-1}x\sqrt{h}\, F^{\tau r}n_{\tau}A_{r} =
-\frac{1}{8\pi}\int_{\cal{M}}\!\!\! d^D x\sqrt{g} \:F^2 .
 \label{PC D-dim:I-electric-cold}
 \end{eqnarray}
Adding all these terms yields the action
 (\ref{PC D-dim:I-electric}) of the higher dimensional cold
instanton (onwards the subscript ``${\rm c}$" means cold)
\begin{eqnarray}
 I_{\rm c}= -r_{\rm c}^{\,D-2}\, \frac{\pi^{(D-1)/2}}{4\,\Gamma[(D-1)/2]}
 \:, \label{PC D-dim:I-total-cold}
 \end{eqnarray}
which, for $D=4$, reduces to the result of \cite{MannRoss}. The
allowed interval of $\rho$ is defined in (\ref{ColdDdim:range}).
As $\rho$ varies from $\rho=\rho_{\rm u}$, defined in
 (\ref{def rho ultra}), to $\rho=0$, the cold action (\ref{PC D-dim:I-total-cold})
varies according to
\begin{eqnarray}
 -\frac{ \rho_{\rm u}^{\, D-2} }{4}
\frac{\pi^{(D-1)/2}}{\Gamma[(D-1)/2]} < I_{\rm c} < I_{\rm dS} \:,
  \label{PC D-dim:rangeI-cold}
 \end{eqnarray}
where the lower limit of this relation is the ultracold action, as
we shall see in (\ref{PC D-dim:I-total-ultracold}), and $I_{\rm
dS}$ is defined in (\ref{PC D-dim:I dS pure}).

The pair creation rate of extreme cold black holes is given by
(\ref{PC D-dim:PC-rate}),
\begin{eqnarray}
\Gamma_{\rm c}=\eta\,e^{-2I_{\rm c}+2I_{\rm dS}}\,,
 \label{PC rate D-dim:cold}
 \end{eqnarray}
where $\eta$ is the one-loop contribution not computed here.

\subsection{\label{sec:Nariai-rate PC D-dim}The higher
dimensional Nariai pair creation rate}

 The first term of the Euclidean action
(\ref{PC D-dim:I-electric}) gives in the Nariai case
 \begin{eqnarray}
& &\hspace{-2.0 cm}  -\frac{1}{16\pi}\int_{\cal{M}} d^Dx\sqrt{g}
\left ( R-2\lambda
\right ) \nonumber \\
&=& \left (-\frac{\Lambda\,(D-1)}{24\pi}+\frac{(D-4)(D-3)}{16\pi}
 \,Q^2 B^{D-2} \right ) \int d\Omega_{D-2}
 \int_0^{2\pi /2} \!\!\!\!d\tau
\int_{0}^{\pi} \!\!\!\!d\chi\:\frac{\sin \chi}{A\,B^{(D-2)/2}} \nonumber \\
&=& \frac{\pi^{(D-1)/2}}{\Gamma[(D-1)/2]}
 \left [ -\frac{\Lambda\,(D-1)}{6}\, \frac{1}{A\,B^{(D-2)/2}}
 + \frac{Q^2 \,(D-4)(D-3)}{4}\, \frac{B^{(D-2)/2}}{A}\right ]\:.
 \label{PC D-dim:I1-Nariai}
 \end{eqnarray}
The Maxwell term in the action yields
\begin{eqnarray}
 \frac{1}{16\pi}\int_{\cal{M}} d^D x\sqrt{g} \:F^2  =-\frac{(D-2)(D-3)\,Q^2}{4}\,
\frac{B^{(D-2)/2}}{A}\, \frac{\pi^{(D-1)/2}}{\Gamma[(D-1)/2]},
 \label{PC D-dim:I2-Nariai}
 \end{eqnarray}
and $\int_{\Sigma} d^{D-1}x\sqrt{h}\, K=0$. In order to compute
the extra Maxwell boundary term in (\ref{PC D-dim:I-electric}) we
have to find a vector potential, $A_{\nu}$, that is regular
everywhere including at the horizons. An appropriate choice in the
lukewarm case is $A_{\chi}= i\,Q_{\rm ADM}\,\frac{B^{(D-2)/2}}{A}
\sin \chi \:\tau$. The integral over $\Sigma$ consists of an
integration between $\chi=0$ and $\chi=\pi$ along the $\tau=0$
surface and back along $\tau=\pi$, and of an integration between
$\tau=0$ and $\tau=\pi$ along the $\chi=0$ surface, and back along
the $\chi=\pi$ surface. The non-vanishing contribution to the
Maxwell boundary term in (\ref{PC D-dim:I-electric}),
$-\frac{1}{4\pi}\int_{\Sigma} d^3x\sqrt{h}\, F^{\mu\nu}n_{\mu}
A_{\nu}$, comes only from the integration along the $\tau=\pi$
surface and is given by
 \begin{eqnarray}
\hspace{-0.5cm}-\frac{1}{4\pi}\int_{\Sigma_{\tau=\pi}}\!\!\!\!\!\!\!\!
d^{D-1}x\sqrt{h}\, F^{\tau \chi}n_{\tau}A_{\chi}=
-\frac{1}{8\pi}\int_{\cal{M}} \!\!\! d^D x\sqrt{g} \:F^2 .
 \label{PC D-dim:I-electric-Nariai}
 \end{eqnarray}
Adding all these terms yields the action
 (\ref{PC D-dim:I-electric}) of the higher dimensional Nariai instanton
(onwards the subscript ``${\rm N}$" means Nariai)
\begin{eqnarray}
 I_{\rm N}= -\frac{1}{2\,B^{(D-2)/2}}
\frac{\pi^{(D-1)/2}}{\Gamma[(D-1)/2]} \:,
  \label{PC D-dim:I-total-Nariai}
 \end{eqnarray}
which, for $D=4$, reduces to the result of
\cite{HawkRoss,MannRoss}. One has $B=\rho^{-2}$, where $\rho$ lies
in the range defined in (\ref{NariaiDdim:range}). Thus, the Nariai
action
 (\ref{PC D-dim:I-total-Nariai}) lies in the range
\begin{eqnarray}
-\frac{\pi^{(D-1)/2}\, \rho_{\rm max}^{D-2}}{2\, \Gamma[(D-1)/2]}
  \leq I_{\rm N}
 < - \frac{\pi^{(D-1)/2}\, \rho_{\rm u}^{D-2}}{2\, \Gamma[(D-1)/2]}
  \:,
  \label{PC D-dim:rangeI-Nariai}
 \end{eqnarray}
where the quantities $\rho_{\rm max}$ and $\rho_{\rm u}$ are
defined, respectively, in (\ref{def rho max}) and (\ref{def rho
ultra}). The equality holds in the neutral Nariai case ($Q=0$),
while the upper limit of (\ref{PC D-dim:rangeI-Nariai}) is the
double of the ultracold action, which will be defined in
 (\ref{PC D-dim:I-total-ultracold}).

The pair creation rate of extreme Nariai black holes is given by
(\ref{PC D-dim:PC-rate}),
\begin{eqnarray}
\Gamma_{\rm N}=\eta\,e^{-2I_{\rm N}+2I_{\rm dS}}\,,
 \label{PC rate D-dim:Nariai}
 \end{eqnarray}
where $I_{\rm dS}$ is given by (\ref{PC D-dim:I dS pure}), and
$\eta$ is the one-loop contribution not computed here. The process
studied in this subsection describes the nucleation of a higher
dimensional Nariai universe that is unstable
\cite{GinsPerry,BoussoHawk,Bousso60y} and decays through the pair
creation of extreme Nariai black holes.

\subsection{\label{sec:Ultracold-rate PC D-dim}The higher
dimensional ultracold pair creation rate}

 The boundary $\Sigma=\partial \cal{M}$ that
appears in (\ref{PC D-dim:I-electric}) consists of an initial
spatial surface at $\tau=0$ plus a final spatial surface at
$\tau=\pi$. We label these two 3-surfaces by $\Sigma_{\tau}$. Each
one of these two spatial $(D-1)$-surfaces is delimitated by a
($D-2$)-surface at the Rindler horizon $\chi=0$ and by a
($D-2$)-surface at the internal infinity $\chi=\infty$. The two
surfaces $\Sigma_{\tau}$ are connected by a timelike
($D-1$)-surface, $\Sigma_{h}$, that intersects $\Sigma_{\tau}$ at
the frontier $\chi=0$ and by a timelike ($D-1$)-surface,
$\Sigma^{\rm int}_{\infty}$, that intersects $\Sigma_{\tau}$ at
the internal infinity boundary $\chi=\infty$. Thus
$\Sigma=\Sigma_{\tau}+\Sigma_{h}+\Sigma^{\rm int}_{\infty}$, and
the region $\cal{M}$ within it is compact. The first term of the
Euclidean action (\ref{PC D-dim:I-electric}) yields
 \begin{eqnarray}
& & \hspace{-2.0 cm} -\frac{1}{16\pi}\int_{\cal{M}} d^Dx\sqrt{g}
\left ( R-2\lambda \right ) \nonumber \\
 & &=  \frac{1}{16\pi}\left [ \Lambda
\,\frac{2(D-1)}{3}
 \, \rho_{\rm u}^{\, D-2} + Q^2\,(D-4)(D-3)\frac{1}{\rho_{\rm u}^{\, D-2} }\right ] \int d\Omega_{D-2}
 \int_0^{2\pi /2} \!\!\!\!d\tau
\int_{0}^{\chi_0\rightarrow \infty} \!\!\!\!d\chi \, \chi \nonumber \\
 & &=  \left [ -\frac{D-1}{24}\, \Lambda\, \rho_{\rm u}^{\, D-2}
 + \frac{(D-4)(D-3)}{16}\,\frac{Q^2}{\rho_{\rm u}^{\, D-2}}\right ]\,
\frac{\pi^{(D-1)/2}}{\Gamma[(D-1)/2]}\, \chi_0^{\,2}
 {\biggl |}_{\chi_0\rightarrow \infty} \:.
 \label{PC D-dim:I1-ultracold}
 \end{eqnarray}
 The Maxwell term in the action
yields
\begin{eqnarray}
 \frac{1}{16\pi}\int_{\cal{M}} d^D x\sqrt{g} \:F^2 =-\frac{(D-3)(D-2)\,Q^2}{16\, \rho_{\rm u}^{\, D-2}}
\frac{\pi^{(D-1)/2}}{\Gamma[(D-1)/2]} \chi_0^{\,2}
 {\biggl |}_{\chi_0\rightarrow \infty} \!\!\!\!\!\!.
 \label{PC D-dim:I2-ultracold}
 \end{eqnarray}

Now, contrary to the other instantons, the ultracold instanton has
a non-vanishing extrinsic curvature boundary term,
$-\frac{1}{16\pi}\int_{\Sigma} d^{D-1}x\sqrt{h}\, K \neq 0$, due
to the internal infinity boundary ($\Sigma^{\rm int}_{\infty}$ at
$\chi=\infty$) contribution. The extrinsic curvature to
$\Sigma^{\rm int}_{\infty}$ is
$K_{\mu\nu}=h_{\mu}^{\:\:\:\alpha}\nabla_{\alpha}n_{\nu}$, where
 $n_{\nu}=(0,1,0,\cdots,0)$ is the unit outward normal to
$\Sigma^{\rm int}_{\infty}$,
$h_{\mu}^{\:\:\:\alpha}=g_{\mu}^{\:\:\:\alpha}-n_{\mu}n^{\alpha}
 =(1,0,1,\cdots,1)$ is the projection tensor onto $\Sigma^{\rm int}_{\infty}$,
 and $\nabla_{\alpha}$ represents the covariant derivative with respect
 to $g_{\mu\nu}$. Thus the trace of the extrinsic
curvature to $\Sigma^{\rm int}_{\infty}$ is
$K=g^{\mu\nu}K_{\mu\nu}=\frac{1}{\chi}$, and
\begin{eqnarray}
 -\frac{1}{8\pi}\int_{\Sigma} d^{D-1}x\sqrt{h}\, K=
 -\frac{ \rho_{\rm u}^{\, D-2} }{4}
\frac{\pi^{(D-1)/2}}{\Gamma[(D-1)/2]} \:.
 \label{PC D-dim:I3-ultracold}
 \end{eqnarray}
In the ultracold case the vector potential $A_{\nu}$, that is
regular everywhere including at the horizon, needed to compute the
extra Maxwell boundary term in (\ref{PC D-dim:I-electric}) is
$A_{\tau}=- i\,\frac{Q_{\rm ADM}}{\rho^{D-2}}\,\chi^2/2$. The
integral over $\Sigma$ consists of an integration between $\chi=0$
and $\chi=\infty$ along the $\tau=0$ surface and back along
$\tau=\pi$, and of an integration between $\tau=0$ and $\tau=\pi$
along the $\chi=0$ surface, and back along the internal infinity
 surface $\chi=\infty$. The non-vanishing contribution to the Maxwell
boundary term in (\ref{PC D-dim:I-electric}) comes only from the
integration along the internal infinity boundary $\Sigma^{\rm
int}_{\infty}$, and is given by
 \begin{eqnarray}
\hspace{-0.5cm} -\frac{1}{4\pi}\int_{\Sigma^{\rm int}_{\infty}}
\!\!\!\!\!\! d^{D-1}x\sqrt{h}\, F^{\chi\tau}n_{\chi}A_{\tau}=
-\frac{1}{8\pi}\int_{\cal{M}} \!\!\! d^D x\sqrt{g} \:F^2 .
 \label{PC D-dim:I-electric-ultracold}
 \end{eqnarray}
Due to the fact that $\chi_0\rightarrow \infty$ it might seem that
the contribution from (\ref{PC D-dim:I1-ultracold}), (\ref{PC
D-dim:I2-ultracold}) and (\ref{PC D-dim:I-electric-ultracold})
diverges. This is not however the case since these three terms
cancel each other. The only contribution to the action
 (\ref{PC D-dim:I-electric}) of the higher dimensional ultracold instanton
(onwards the subscript ``${\rm u}$" means ultracold) comes from
(\ref{PC D-dim:I3-ultracold}) yielding
\begin{eqnarray}
I_{\rm u}=
  -\frac{ \rho_{\rm u}^{\, D-2} }{4}
\frac{\pi^{(D-1)/2}}{\Gamma[(D-1)/2]}\:,
  \label{PC D-dim:I-total-ultracold}
 \end{eqnarray}
which, for $D=4$, reduces to the result of \cite{MannRoss}. The
ultracold action coincides with the minimum value of the cold
action range (\ref{PC D-dim:rangeI-cold}), and is equal to one
half the maximum value of the Nariai action range
 (\ref{PC D-dim:rangeI-Nariai}).

The pair creation rate of extreme ultracold black holes is given
by (\ref{PC D-dim:PC-rate}),
\begin{eqnarray}
\Gamma_{\rm u}=\eta\,e^{-2I_{\rm u}+2I_{\rm dS}}\,,
 \label{PC rate D-dim:ultracold}
 \end{eqnarray}
where $I_{\rm dS}$ is given by (\ref{PC D-dim:I dS pure}), and
$\eta$ is the one-loop contribution not computed here. The process
studied in this subsection describes the nucleation of a higher
dimensional Nariai$-$Bertotti-Robinson universe that is unstable,
and decays through the pair creation of extreme ultracold black
holes.

\subsection{\label{sec:Lukewarm-rate PC D-dim}The higher dimensional lukewarm  pair creation rate}

The evaluation of the Euclidean action of the higher dimensional
lukewarm instanton follows as in the cold case as long as we
replace $\rho$ by $r_+$. Therefore, the action
 (\ref{PC D-dim:I-electric}) of the higher dimensional lukewarm
instanton (onwards the subscript ``$\ell$" means lukewarm) is
given by
\begin{eqnarray}
 I_{\rm \ell}=
 \frac{\pi^{(D-3)/2}}{\Gamma[(D-1)/2]}\,\frac{\beta}{4}
 {\biggl [}
 -\frac{\Lambda \left ( r_{\rm c}^{\,D-1}-r_+^{\,D-1} \right )}
 {(D-2)(D-1)}  +\frac{(D-3)Q^2}{2}
 \left (r_+^{\,-(D-3)}-r_{\rm c}^{\,-(D-3)} \right )
 {\biggr ]}\:. \label{PC D-dim:I-total-luk}
 \end{eqnarray}
and the pair creation rate of nonextreme lukewarm black holes is
given by (\ref{PC D-dim:PC-rate}).

\subsection{\label{sec:Sub-Maximal-rate D-dim}Pair creation rate of higher
dimensional nonextreme sub-maximal black holes}

The cold, Nariai, ultracold and lukewarm instantons are saddle
point solutions free of conical singularities both in the $r_+$
and $r_{\rm c}$ horizons. The corresponding black holes may then
nucleate in the dS background, and we have computed their pair
creation rates in the last four subsections. However, these
particular black holes are not the only ones that can be pair
created. Indeed, it has been shown in
\cite{WuSubMax,BoussoHawkSubMax} that Euclidean solutions with
conical singularities may also be used as saddle points for the
pair creation process. In this way, pair creation of nonextreme
sub-maximal black holes is allowed (by this nomenclature we mean
all the nonextreme black holes other than the lukewarm ones that
are in the region interior to the close line $ONUO$ in Fig.
\ref{range mq dS bh D-dim}), and their pair creation rate may be
computed. In order to calculate this rate, the action is given by
(\ref{PC D-dim:I-electric}) and, in addition, it has now an extra
contribution from the conical singularity (c.s.) that is present
in one of the horizons ($r_+$, say) given by
\cite{ReggeGibbonsPerryAconSing,GinsPerry}
\begin{eqnarray}
\frac{1}{16\pi}\int_{\cal{M}} d^Dx\sqrt{g}
 \:\left (
R-2\lambda \right ){\biggl |}_{{\rm c.s.}\:{\rm at}\:r_+}
 \!\!\!= \frac{ {\cal A}_+\:\delta}{4\,D\,\pi}\:,
 \label{I conical sing D-dim}
 \end{eqnarray}
where ${\cal A}_+=\frac{2\pi^{(D-1)/2}}{\Gamma[(D-1)/2]}
\,r_+^{D-2}$ is the area of the ($D-2$)-sphere spanned by the
conical singularity, and
\begin{eqnarray}
\delta=2\pi \left ( 1-\frac{\beta_{\rm c}}{\beta_+}\right )
 \label{delta concical sing PCdS D-dim}
 \end{eqnarray}
is the deficit angle associated to the conical singularity at the
horizon $r_+$, with $\beta_{\rm c}=4 \pi / |f'(r_{\rm c})|$ and
$\beta_+=4 \pi / |f'(r_+)|$ being the periods of $\tau$ that avoid
a conical singularity in the horizons $r_{\rm c}$ and $r_+$,
respectively. The contribution from (\ref{PC D-dim:I-electric})
follows straightforwardly in a similar way as the one shown in
subsection \ref{sec:Lukewarm-rate PC D-dim} with the period of
$\tau$, $\beta_{\rm c}$,  chosen in order to avoid the conical
singularity at the cosmological horizon, $r=r_{\rm c}$.

\section{\label{sec:Conc:PC-Ddim}Discussion of the results}

We have studied in detail the quantum process in which a pair of
black holes is created in a higher dimensional de Sitter (dS)
background, a process that in $D=4$ was previously discussed in
\cite{MannRoss}. The energy to materialize and accelerate the pair
comes from the positive cosmological constant. The dS space is the
only background in which we can discuss analytically the pair
creation process of higher dimensional black holes, since the
C-metric and the Ernst solutions, that describe respectively a
pair accelerated by a string and by an electromagnetic field, are
not know yet in a higher dimensional spacetime.

In previous works on black hole pair creation in general
background fields it has been well established that the pair
creation rate is proportional to the exponential of the
gravitational entropy $S$ of the system, $\Gamma \propto e^S$,
with the entropy being given by one quarter of the the total area
$\cal{A}$ of all the horizons present in the instanton,
$S={\cal{A}}/4$. It is straightforward to verify that these
relations also hold for the higher dimensional dS instantons.
Indeed, in the cold case, the instanton has a single horizon,  the
cosmological horizon at $r=r_{\rm c}$, in its Euclidean section,
since $r=r_+$ is an internal infinity. So, the total area of the
cold instanton is ${\cal A}_{\rm c}=\Omega_{D-2}\,r_{\rm
c}^{D-2}$. Thus, $S_{\rm c}=-2I_{\rm c}={\cal{A}_{\rm c}}/4$,
where $I_{\rm c}$ is given by (\ref{PC D-dim:I-total-cold}). In
the Nariai case, the instanton has two horizons in its Euclidean
section, namely the cosmological horizon $r=r_{\rm c}$ and the
black hole horizon $r_+$, both at $r=\rho=B^{-1/2}$, and thus they
have the same area. So, the total area of the Nariai instanton is
${\cal A}_{\rm N}=2\Omega_{D-2}\,B^{-(D-2)/2}$. Again, one has
$S_{\rm N}=-2I_{\rm N}={\cal{A}_{\rm N}}/4$, where $I_{\rm N}$ is
given by (\ref{PC D-dim:I-total-Nariai}). In the ultracold  case,
the instanton has a single horizon, the Rindler horizon at
$\chi=0$, in its Euclidean section, since
 $\chi=\infty$ is an internal infinity. The total area of the
ultracold instanton is then ${\cal A}_{\rm
c}=\Omega_{D-2}\,\rho_{\rm u}^{D-2}$. Thus, $S_{\rm u}=-2I_{\rm
u}={\cal{A}_{\rm u}}/4$, where $I_{\rm u}$ is given by (\ref{PC
D-dim:I-total-ultracold}).

The ultracold  instanton is a limiting case of both the charged
Nariai instanton and the cold instanton (see Fig.
 \ref{range mq dS bh D-dim}). Then, as expected, the action of the
cold instanton gives, in this limit, the action of the ultracold
instanton (see \ref{PC D-dim:rangeI-cold}). However, the ultracold
frontier of the Nariai action is given by two times the ultracold
action (see \ref{PC D-dim:rangeI-Nariai}). The reason for this
behavior is clear. Indeed, in the ultracold case and in the cold
case, the respective instantons have a single horizon (the other
possible horizon turns out to be an internal infinity). This
horizon gives the only contribution to the total area,
${\cal{A}}$, and therefore to the pair creation rate. In the
Nariai case, the instanton has two horizons with the same area,
and thus the ultracold limit of the Nariai action is doubled with
respect to the true ultracold action.

A property of the higher dimensional lukewarm instanton is worth
of mention. In $D=4$ it is known that the lukewarm instanton has
an ADM mass equal to its ADM charge. This relation is valid in
several background fields, e.g., in a dS background, in an
electromagnetic background and in a cosmic string background. For
$D\geq 5$ we have shown that the dS lukewarm instanton no longer
has an ADM mass equal to its ADM charge.

In order to clarify the physical interpretation of the results, in
this Appendix we heuristically derive the nucleation rates for the
processes discussed in the main body of the paper. An estimate for
the nucleation probability is given by the Boltzmann factor,
$\Gamma \sim e^{-E_0/W_{\rm ext}}$, where $E_0$ is the energy of
the system that nucleates and $W_{\rm ext}=F \ell$ is the work
done by the external force $F$,  that provides the energy for the
nucleation, through the typical distance $\ell$ separating the
created pair. We can then show that the creation probability for a
black hole pair in a dS background is given by $\Gamma \sim
e^{-M/\sqrt{\Lambda}}$, in agreement with the exact results.
Indeed, one has $E_0 \sim 2M$, where $M$ is the rest energy of the
black hole, and $W_{\rm ext}\sim \sqrt{\Lambda}$ is the work
provided by the cosmological background. To derive $W_{\rm
ext}\sim \sqrt{\Lambda}$ one can argue as follows. In the dS case,
the Newtonian potential is $\Phi=\Lambda r^2/3$ and its derivative
yields the force per unit mass or acceleration, $\Lambda r$, where
$r$ is the characteristic dS radius, $\Lambda^{-1/2}$. The force
can then be written as $F= {\rm mass}\times{\rm acceleration}\sim
\sqrt{\Lambda}\sqrt{\Lambda}$, where the characteristic mass of
the system is $\sqrt{\Lambda}$. Thus, the characteristic work is
$W_{\rm ext}={\rm force}\times{\rm distance}\sim \Lambda
\Lambda^{-1/2}\sim \sqrt{\Lambda}$, where the characteristic
distance that separates the pair at the creation moment is
$\Lambda^{-1/2}$. So, from the Boltzmann factor we indeed expect
that the creation rate of a black hole pair in a dS background is
given by $\Gamma \sim e^{-M/\sqrt{\Lambda}}$. This expression is
in agreement with our results since from (\ref{mq D-dim}), one has
$M \sim \rho^{D-3} \sim \Lambda^{-(D-3)/2}$, and thus $\Gamma \sim
e^{-M/\sqrt{\Lambda}}\sim e^{\Lambda^{-(D-2)/2}}$.

%% file: Chapter11.tex
\thispagestyle{empty} \setcounter{minitocdepth}{1}
\chapter[Gravitational radiation in higher dimensional spacetimes
and energy released during black hole creation]
{\Large{Gravitational radiation in higher dimensional spacetimes
and \\ energy released during black hole creation}}
\label{chap:Grav Radiation}
 \lhead[]{\fancyplain{}{\bfseries Chapter \thechapter. \leftmark}}
 \rhead[\fancyplain{}{\bfseries \rightmark}]{}
  \minitoc \thispagestyle{empty}
\renewcommand{\thepage}{\arabic{page}}

\addtocontents{lof}{\textbf{Chapter Figures \thechapter}\\}

One expects to finally detect gravitational waves in the
forthcoming years. If this happens, and if the observed waveforms
match the predicted templates, General Relativity will have pass a
crucial test. Moreover, if one manages to cleanly separate
gravitational waveforms, we will open a new and exciting window to
the Universe, a window from which one can look directly into the
heart of matter, as gravitational waves are weakly scattered by
matter.  A big effort has been spent in the last years trying to
build gravitational wave detectors, and a new era will begin with
gravitational wave astronomy \cite{schutz1,hughes}.  What makes
gravitational wave astronomy attractive, the weakness with which
gravitational waves are scattered by matter, is also the major
source of technical difficulties when trying to develop an
apparatus which interacts with them. Nevertheless, some of these
highly non-trivial technical difficulties have been surmounted,
and we have detectors already operating \cite{geo,ligo,virgo}.
Another effort is being dedicated by theoreticians trying to
obtain accurate templates for the various physical processes that
may give rise to the waves impinging on the detector. We now have
a well established theory of wave generation and propagation,
which started with Einstein and his quadrupole formula.  The
quadrupole formula expresses the energy lost to gravitational
waves by a system moving at low velocities, in terms of its energy
content.  The quadrupole formalism is the most famous example of
slow motion techniques to compute wave generation. All these
techniques break Einstein's equations non-linearity by imposing a
power series in some small quantity and keeping only the lowest or
the lowest few order terms. The quadrupole formalism  starts from
a flat background and expands the relevant quantities in $R/
\lambda$, where $R$ is the size of source and $\lambda$ the
wavelength of waves. Perturbation formalisms on the other hand,
start from some non-radiative background, whose metric is known
exactly, for example the Schwarzschild metric, and expand in
deviations from that background metric. For a catalog of the
various methods and their description we refer the reader to the
review works by Thorne \cite{thorne} and Damour \cite{damour}. The
necessity to develop all such methods was driven of course by the
lack of exact radiative solutions to Einstein's equations
(although there are some worthy exceptions, like the C-metric
\cite{KW}), and by the fact that even nowadays solving the full
set of Einstein's equations numerically is a monumental task, and
has been done only for the more tractable physical situations. All
the existing methods seem to agree with each other when it comes
down to the computation of waveforms and energies radiated during
physical situations, and also agree with the few available results
from a fully numerical evolution of Einstein's equations.

In this chapter we extend some of these results to higher
dimensional spacetimes. There are several reasons why one should
now try to do it. It seems impossible to formulate in four
dimensions a consistent theory which unifies gravity with the
other forces in nature. Thus, most efforts in this direction have
considered a higher dimensional arena for our universe, one
example being string theories which have recently made some
remarkable achievements. Moreover, recent investigations
\cite{hamed} propose the existence of extra dimensions in our
Universe in order to solve the hierarchy problem, i.e., the huge
difference between the electroweak and the Planck scale, $m_{\rm
EW}/M_{\rm Pl}\sim 10^{-17}$. The fields of standard model would
inhabit a 4-dimensional sub-manifold, the brane, whereas the
gravitational degrees of freedom would propagate throughout all
dimensions. One of the most spectacular consequences of this
scenario would be the production of black holes at the Large
Hadron Collider at CERN
\cite{ArgDimMRussLHC,DimopoulosLandsbergLHC,GiddingsThomasLHC,bhprod}
(for recent relevant work related to this topic we refer the
reader to
\cite{KantiMRusselLHC,IdaOdaParkLHC,FrolovStojkovicLHC,cardosolemos0,cardosolemos}).
Now, one of the experimental signatures of black hole production
will be a missing energy, perhaps a large fraction of the center
of mass energy \cite{cardosolemos0}. This will happen because when
the partons collide to form a black hole, some of the initial
energy will be converted to gravitational waves, and due to the
small amplitudes involved, there is no gravitational wave detector
capable of detecting them, so they will appear as missing. Thus,
the collider could in fact indirectly serve as a gravitational
wave detector.  This calls for the calculation of the energy given
away as gravitational waves when two high energy particles collide
to form a black hole, which lives in all the dimensions. The work
done so far on this subject \cite{eardley,yoshino} in higher
dimensions, is mostly geometric, and generalizes a construction by
Penrose to find trapped surfaces on the union of two shock waves,
describing boosted Schwarzschild black holes. On the other hand,
there are clues \cite{cardosolemos0,cardosolemos} indicating that
a formalism described by Weinberg \cite{weinberg} to compute the
gravitational energy radiated in the collision of two point
particles, gives results correct to a order of magnitude when
applied to the collision of two black holes.  This formalism
assumes a hard collision, i.e., a collision lasting zero seconds.
It would be important to apply this formalism in higher
dimensions, trying to see if there is agreement between both
results. This will be one of the topics discussed in this chapter.
The other topic we study in this chapter is the quadrupole formula
in higher dimensions. Due to the difficulties in handling the wave
tails in odd dimensions we concentrate our study in even
dimensions.

In this chapter we follow \cite{VitOscLem} and it is organized as
follows. In section
 \ref{linear Einst eqs} we linearize Einstein's equations in a flat $D$-dimensional
background and arrive at an inhomogeneous wave equation for the
metric perturbations. The source free equations are analyzed in
terms of plane waves, and then the general solution to the
homogeneous equation is deduced in terms of the $D$-dimensional
retarded Green's function.  In section \ref{quadrupole form} we
compute the $D$-dimensional quadrupole formula (assuming slowly
moving sources), expressing the metric and the radiated energy in
terms of the time-time component of the energy-momentum tensor. We
then apply the quadrupole formula to two cases: a particle in
circular motion in a generic background, and a particle falling
into a $D$-dimensional Schwarzschild black hole. In section
\ref{instantaneous collisions} we consider the hard collision
between two particles, i.e., the collision takes zero seconds, and
introduce a cutoff frequency necessary to have meaningful results.
We then apply to the case where one of the colliding particles is
a black hole. We propose that this cutoff should be related to the
gravitational quasinormal frequency of the black hole, and compute
some values of the scalar quasinormal frequencies for higher
dimensional Schwarzschild black holes, expecting that the
gravitational quasinormal frequencies will behave in the same
manner. Finally, we apply this formalism to compute the generation
of gravitational radiation during black hole pair creation in four
and higher dimensions, a result that has never been worked out,
even for $D=4$. In our presentation we shall mostly follow
Weinberg's \cite{weinberg} exposition.

\section[Linearized $D$-dimensional Einstein's equations]
{Linearized ${\bm D}$-dimensional Einstein's equations}
\label{linear Einst eqs}

Due to the non-linearity of Einstein's equations, the treatment of
the gravitational radiation problem is not an easy one since the
energy-momentum tensor of the gravitational wave contributes to
its own gravitational field. To overcome this difficulty it is a
standard procedure to work only with the weak radiative solution,
in the sense that the energy-momentum content of the gravitational
wave is small enough in order to allow us to neglect its
contribution to its own propagation. This approach is justified in
practice since we expect the detected gravitational radiation to
be of low intensity.
\subsection{The inhomogeneous wave equation}
\label{sec:inhomogeneous wave equation}
We begin this subsection by introducing the general background
formalism (whose details can be found, e.g., in \cite{weinberg})
that will be needed in later sections. Then we obtain the
linearized inhomogeneous wave equation.

Greek indices vary as $0,1,\cdots, D-1$ and latin indices as
$1,\cdots, D-1$ and our units are such that $c \equiv 1$. We work
on a $D$-dimensional spacetime described by a metric $g_{\mu\nu}$
that approaches asymptotically the $D$-dimensional Minkowski
metric $\eta_{\mu\nu}={\rm diag}(-1,+1,\cdots,+1)$, and thus we
can write
\begin{equation}
g_{\mu\nu}=\eta_{\mu\nu}+h_{\mu\nu}\,
\hspace{1cm}\mu,\nu=0,1,\cdots,D-1 \,,
 \label{g}
\end{equation}
where $h_{\mu\nu}$ is small, i.e., $|h_{\mu\nu}|<<1$, so that it
represents small corrections to the flat background. The exact
Einstein field equations,
 $G_{\mu\nu}=8\pi {\cal G} T_{\mu\nu}$ (with ${\cal G}$ being the usual Newton
constant), can then be written as
\begin{equation}
R^{(1)}_{\;\;\;\;\mu\nu}-\frac{1}{2}\eta_{\mu\nu}
 R^{(1)\,\alpha}_{\;\;\;\;\;\;\;\alpha}=8\pi {\cal G}\,\, \tau_{\mu\nu} \,,
 \label{R1}
\end{equation}
with
\begin{equation}
\tau^{\mu\nu} \equiv \eta^{\mu \alpha} \eta^{\nu\beta}
\,(T_{\alpha\beta}+t_{\alpha\beta}).
 \label{tau}
\end{equation}
Here $R^{(1)}_{\;\;\;\;\mu\nu}$ is the part of the Ricci tensor
linear in $h_{\mu\nu}$,
 $R^{(1)\,\alpha}_{\;\;\;\;\;\;\;\alpha}=\eta^{\alpha\beta}
 R^{(1)}_{\;\;\;\;\beta\alpha}$,
and $\tau_{\mu\nu}$ is the effective energy-momentum tensor,
containing contributions from $T_{\mu\nu}$, the energy-momentum
tensor of the matter source, and $t_{\mu\nu}$ which represents the
gravitational contribution. The pseudo-tensor $t_{\mu\nu}$
contains the difference between the exact Ricci terms and the
Ricci terms linear in $h_{\mu\nu}$,
\begin{equation}
t_{\mu\nu}=\frac{1}{8\pi {\cal G}}\left [ R_{\mu\nu}
-\frac{1}{2}g_{\mu\nu} R^{\alpha}_{\;\;\;\alpha}
 -R^{(1)}_{\;\;\;\;\mu\nu}+\frac{1}{2}\eta_{\mu\nu}
 R^{(1)\,\alpha}_{\;\;\;\;\;\;\;\alpha} \right ] \,.
 \label{t}
\end{equation}
The Bianchi identities imply that $\tau_{\mu\nu}$ is locally
conserved,
\begin{equation}
\partial_{\mu} \tau^{\mu\nu}=0 \,.
 \label{constau}
\end{equation}
Introducing the cartesian coordinates $x^{\alpha}=(t,{\bf x})$
with ${\bf x}=x^i$, and considering a $D-1$ volume $V$ with a
boundary spacelike surface $S$ with dimension $D-2$ whose unit
exterior normal is ${\bf n}$, eq. (\ref{constau}) yields
\begin{equation}
\frac{d}{dt}\int_{V} d^{D-1}{\bf x} \;\tau^{0\nu} =-\int_{S}
 d^{D-2}{\bf x} \;n_i \tau^{i\nu} \,.
 \label{constau1}
\end{equation}
This means that one may interpret
\begin{equation}
p^{\nu} \equiv \int_{V} d^{D-1}{\bf x} \;\tau^{0\nu}
 \label{p}
\end{equation}
as the total energy-momentum (pseudo)vector of the system,
including matter and gravitation, and $\tau^{i\nu}$ as the
corresponding flux. Since the matter contribution is contained in
$t^{\mu\nu}$, the flux of gravitational radiation is
\begin{equation}
{\rm Flux}=\int_{S} d^{D-2}{\bf x} \;n_i t^{i \nu} \,.
 \label{fluxgrav}
\end{equation}

In this context of linearized general relativity, we neglect terms
of order higher than the first in $h_{\mu\nu}$ and all the indices
are raised and lowered using $\eta^{\mu\nu}$. We also neglect the
contribution of the gravitational energy-momentum tensor
$t_{\mu\nu}$ (i.e., $|t_{\mu\nu}|<<|T_{\mu\nu}|$) since from
(\ref{t}) we see that $t_{\mu\nu}$ is of higher order in
$h_{\mu\nu}$. Then, the conservation equations (\ref{constau})
yield
\begin{equation}
\partial_{\mu}T^{\mu\nu}=0 \,.
 \label{consT}
\end{equation}
In this setting and choosing the convenient coordinate system that
obeys the harmonic (also called Lorentz) gauge conditions,
\begin{equation}
2\partial_{\mu}h^{\mu}_{\;\;\;\nu}=\partial_{\nu}h^{\alpha}_{\;\;\;\alpha}
 \label{hargauge}
\end{equation}
(where $\partial_{\mu}=\partial/\partial x^{\mu}$), the first
order Einstein field equations (\ref{R1}) yield
\begin{equation}
\square h_{\mu\nu}=-16\pi {\cal G} S_{\mu\nu} \,,
 \label{ineq}
\end{equation}
\begin{equation}
S_{\mu\nu}=T_{\mu\nu}-\frac{1}{D-2}\,\eta_{\mu\nu}\,T^{\alpha}_{\;\;\;\alpha}
\,,
 \label{S}
\end{equation}
where $\square=\eta^{\mu\nu}\partial_{\mu}\partial_{\nu}$ is the
$D$-dimensional Laplacian, and $S_{\mu\nu}$ will be called the
modified energy-momentum tensor of the matter source. Eqs.
(\ref{ineq}) and (\ref{S}) subject to (\ref{hargauge}) allow us to
find the gravitational radiation produced by a matter source
$S_{\mu\nu}$.
\subsection{The plane wave solutions}
\label{sec:plane wave solutions}

In vacuum, the linearized equations for the gravitational field
are $R^{(1)}_{\;\;\;\;\mu\nu}=0$ or, equivalently, the homogeneous
equations $\square h_{\mu\nu}=0$, subjected to the harmonic gauge
conditions (\ref{hargauge}). The solutions of these equations, the
plane wave solutions, are important since the general solutions of
the inhomogeneous equations (\ref{hargauge}) and (\ref{ineq})
approach the plane wave solutions at large distances from the
source.  Setting $k_{\alpha}=(-\omega,{\bf k})$ with $\omega$ and
${\bf k}$ being respectively the frequency and wave vector, the
plane wave solutions can be written as a linear superposition of
solutions of the kind
\begin{equation}
 h_{\mu\nu}(t,{\bf x})=e_{\mu\nu}\,e^{i k_{\alpha} x^{\alpha}}+
 e_{\mu\nu}^*\,e^{-i k_{\alpha} x^{\alpha}} \,,
  \label{homsol}
\end{equation}
where $e_{\mu\nu}=e_{\nu\mu}$ is called the polarization tensor
and $^*$ means the complex conjugate. These solutions satisfy eq.
(\ref{ineq}) with $S_{\mu\nu}=0$ if $k_{\alpha} k^{\alpha}=0$, and
obey the harmonic gauge conditions (\ref{hargauge}) if
$2\,k_{\mu}\,e^{\mu}_{\;\;\;\nu}=k_{\nu}\,e^{\mu}_{\;\;\;\mu}$.

An important issue that must be addressed is the number of
different polarizations that a gravitational wave in $D$
dimensions can have. The polarization tensor $e_{\mu\nu}$, being
symmetric, has in general $D(D+1)/2$ independent components.
However, these components are subjected to the $D$ harmonic gauge
conditions that reduce by $D$ the number of independent
components. In addition, under the infinitesimal change of
coordinates $x'^{\mu}=x^{\mu}+\xi^{\mu}(x)$, the polarization
tensor transforms into
$e'_{\mu\nu}=e_{\mu\nu}-\partial_{\nu}\xi_{\mu}
-\partial_{\mu}\xi_{\nu}$. Now, $e'_{\mu\nu}$ and $e_{\mu\nu}$
describe the same physical system for arbitrary values of the $D$
parameters $\xi^{\mu}(x)$. Therefore, the number of independent
components of $e_{\mu\nu}$, i.e., the number of polarization
states of a gravitational wave in $D$ dimensions is
$D(D+1)/2-D-D=D(D-3)/2$. From this computation we can also see
that gravitational waves are present only when $D>3$. Therefore,
from now on we assume ${\rm D}>3$ whenever we refer to $D$. In
what concerns the helicity of the gravitational waves, for
arbitrary $D$ the gravitons are always spin $2$ particles.

To end this subsection on gravitational plane wave solutions, we
present the average gravitational energy-momentum tensor of a
plane wave, a quantity that will be needed later. Notice that in
vacuum, since the matter contribution is zero ($T_{\mu\nu}=0$), we
cannot neglect the contribution of the gravitational
energy-momentum tensor $t_{\mu\nu}$. From eq. (\ref{t}), and
neglecting terms of order higher than $h^2$, the gravitational
energy-momentum tensor of a plane wave is given by
\begin{equation}
t_{\mu\nu}\simeq \frac{1}{8\pi {\cal G} }\left [
R^{(2)}_{\;\;\;\;\mu\nu}-\frac{1}{2}\eta_{\mu\nu}
 R^{(2)\,\alpha}_{\;\;\;\;\;\;\;\alpha} \right ] \,,
 \label{tplane}
\end{equation}
and through a straightforward calculation (see e.g.
\cite{weinberg} for details) we get the average gravitational
energy-momentum tensor of a plane wave,
\begin{equation}
\langle t_{\mu\nu} \rangle=\frac{k_{\mu}k_{\nu}}{16\pi {\cal
G}}\left [ e^{\alpha\beta}e_{\alpha\beta}^*
-\frac{1}{2}|e^{\alpha}_{\;\;\;\alpha}|^2\right ] \,.
 \label{taverage}
\end{equation}

\subsection[The $D$-dimensional retarded Green's function]
{The $\bm D$-dimensional retarded Green's function}
The general solution to the inhomogeneous differential equation
(\ref{ineq}) may be found in the usual way in terms of a Green's
function as
\begin{equation}
h_{\mu\nu}(t,{\bf x})=-16 \pi {\cal G}\int dt'\int d^{D-1}{\bf x'}
S_{\mu \nu}(t',{\bf x'}) G(t-t',{\bf x - x'})+{\rm homogeneous
\,solutions}\,, \label{insol}
\end{equation}
where the Green's function $G(t-t',{\bf x - x'})$ satisfies
\begin{equation}
\eta^{\mu\nu}\partial_{\mu}\partial_{\nu} G(t-t',{\bf x -
x'})=\delta(t-t')\delta({\bf x - x'})\,, \label{greendef}
\end{equation}
where $\delta(z)$ is the Dirac delta function. In the momentum
representation this reads
\begin{equation}
G(t,{\bf x})=-\frac{1}{(2\pi)^{D}} \int d^{D-1} {\bf k} e^{i {\bf
k}\cdot{\bf x}}\int d\omega \frac{e^{-i\omega t}}{\omega^2-k^2}\,,
\label{greenmomentum}
\end{equation}
where $k^2=k_{1}^2+k_{2}^2+...+k_{D-1}^2$. To evaluate this, it is
convenient to perform the $k$-integral by using spherical
coordinates in the ($D-1$)-dimensional $k$-space.  The required
transformation is given by
\begin{eqnarray}
x_{1} & = & r \prod^{D-2}_{i=1}\sin \theta_i \nonumber\\
&\cdots& \nonumber\\
x_{j} & = & r {\bigg (} \prod^{D-j}_{i=1}\sin \theta_i {\bigg )} \cos \theta_{D+1-j}\:,\:\: {\rm for} \:\: 3\leq j < D \nonumber\\
&\cdots& \nonumber\\
x_{D-1} & = & r\cos\theta_1 \:.
 \label{spher coord D-1}
\end{eqnarray}
The result for the retarded Green's function in these spherical
coordinates is
\begin{equation}
G^{\rm ret}(t,{\bf
x})=-\frac{\Theta(t)}{(2\pi)^{(D-1)/2}}\times\frac{1}{r^{(D-3)/2}}
\int k^{(D-3)/2} J_{(D-3)/2}(kr)\sin(kt)dk \,,
\label{greenretgeral}
\end{equation}
where $r^2=x_{1}^2+x_{2}^2+...+x_{D-1}^2$, and  $\Theta(t)$ is the
Heaviside function defined as
\begin{equation}
\Theta(t)=\left\{ \begin{array}{ll}
             1   & \mbox{if $t>0$}\\
             0    & \mbox{if $t<0$}\,.
\end{array}\right.
\label{Heaviside}
\end{equation}
The function $J_{({\rm D}-3)/2}(kr)$ is a Bessel function
\cite{watson,stegun}.  The structure of the retarded Green's
function will depend on the parity of $D$, as we shall see. This
dependence on the parity, which implies major differences between
even and odd spacetime dimensions, is connected to the structure
of the Bessel function. For even $D$, the index of the Bessel
function is semi-integer and then the Bessel function is
expressible in terms of elementary functions, while for odd $D$
this does not happen. A concise explanation of the difference
between retarded Green's function in even and odd $D$, and the
physical consequences that entails is presented in \cite{courant}
(see also \cite{hadamard,barrow,galtsov}).
 A complete derivation of the Green's function in
higher dimensional spaces may be found in Hassani \cite{hassani}.
The result is
\begin{equation}
G^{\rm ret}(t,{\bf x})=\frac{1}{4\pi}\left[-\frac{\partial}{2\pi r
\partial r} \right]^{(D-4)/2}
\left[\frac{\delta(t-r)}{r}\right]\,,\,\,\,\,D \,\,{\rm even}.
\label{greenfinaleven}
\end{equation}
\begin{equation}
G^{\rm ret}(t,{\bf
x})=\frac{\Theta(t)}{2\pi}\left[-\frac{\partial}{2\pi r \partial
r} \right]^{(D-3)/2}
\left[\frac{1}{\sqrt{t^2-r^2}}\right]\,,\,\,\,\,D \,\,{\rm odd}.
\label{greenfinalodd}
\end{equation}
It is sometimes convenient to work with the Fourier transform (in
the time coordinate) of the Green's function. One finds
\cite{hassani} an analytical result independent of the parity of
$D$
\begin{equation}
G^{\rm ret}(\omega,{\bf x})=\frac{i^D \pi}{2(2\pi)^{(D-1)/2}}
\left(\frac{\omega}{r}\right)^{(D-3)/2} H^1_{(D-3)/2}(\omega r)\,,
\label{greenfinalevenfourier}
\end{equation}
where $H^1_{\nu}(z)$ is a modified Bessel function
\cite{stegun,watson}.  Of course, the different structure of the
Green's function for different $D$ is again embodied in these
Bessel functions.  Equations (\ref{greenfinaleven}) and
(\ref{greenfinalevenfourier}), are one of the most important
results we shall use in this chapter.  For $D=4$
(\ref{greenfinaleven}) obviously reproduce well known results
\cite{hassani}.  Now, one sees from eq. (\ref{greenfinaleven})
that although there are delta function derivatives on the even-$D$
Green's function, the localization of the Green's function on the
light cone is preserved. However, eq. (\ref{greenfinalodd}) tells
us that the retarded Green's function for odd dimensions is
non-zero inside the light cone. The consequence, as has been
emphasized by different authors \cite{courant,galtsov, kazinski},
is that for odd $D$ the Huygens principle does not hold: the fact
that the retarded Green's function support extends to the interior
of the light cone implies the appearance of radiative tails in
(\ref{insol}). In other words, we still have a propagation
phenomenum for the wave equation in odd dimensional spacetimes, in
so far as a localized initial state requires a certain time to
reach a point in space. Huygens principle no longer holds, because
the effect of the initial state is not sharply limited in time:
once the signal has reached a point in space, it persists there
indefinitely as a reverberation.

This fact coupled to the analytic structure of the Green's
function in odd dimensions make it hard to get a grip on radiation
generation in odd dimensional spacetimes. Therefore, from now on
we shall focus on even dimensions, for which the retarded Green's
function is given by eq. (\ref{greenfinaleven}).

\subsection[The even $D$-dimensional retarded solution in the wave zone]
{The even ${\bm D}$-dimensional retarded solution in the wave zone}
The retarded solution for the metric perturbation $h_{\mu\nu}$,
obtained by using the retarded Green's function
(\ref{greenfinaleven}) and discarding the homogeneous solution in
(\ref{insol}) will be given by
\begin{equation}
h_{\mu\nu}(t,{\bf x})= 16 \pi {\cal G}\int dt'\int d^{D-1}{\bf x'}
S_{\mu \nu}(t',{\bf x'}) G^{\rm ret}(t-t', {\bf x - x'})\,,
\label{retsol}
\end{equation}
with $G^{\rm ret}(t-t',{\bf x - x'})$ as in eq.
(\ref{greenfinaleven}). For $D=4$ for example one has
\begin{equation}
G^{\rm ret}(t,{\bf
x})=\frac{1}{4\pi}\frac{\delta(t-r)}{r}\,\,,\,\,\,\,\,D=4\,,
\label{green4}
\end{equation}
which is the well known result. For $D=6$, we have
\begin{equation}
G^{\rm ret}(t,{\bf
x})=\frac{1}{8\pi^2}\left(\frac{\delta'(t-r)}{r^2}+\frac{\delta(t-r)}{r^3}\right)
\,\,,\,\,\,\,\,D=6\,, \label{green6}
\end{equation}
where the $\delta'(t-r)$ means derivative of the Dirac delta
function with respect to its argument. For $D=8$, we have
\begin{equation}
G^{\rm ret}(t,{\bf
x})=\frac{1}{16\pi^3}\left(\frac{\delta''(t-r)}{r^3}+
3\frac{\delta'(t-r)}{r^4}+3\frac{\delta(t-r)}{r^5}\right)
\,\,,\,\,\,\,\,D=8\,. \label{green8}
\end{equation}
We see that in general even-$D$ dimensions the Green's function
consists of inverse integer powers in r, spanning all values
between $\frac{1}{r^{(D-2)/2}}$ and $\frac{1}{r^{D-3}}$, including
these ones. Now, the retarded solution is given by eq.
(\ref{retsol}) as a product of the Green's function times the
modified energy-momentum tensor $S_{\mu \nu}$. The net result of
having derivatives on the delta functions is to transfer these
derivatives to the energy-momentum tensor as time derivatives
(this can be seen by integrating (\ref{retsol}) by parts in the
$t$-integral).

A close inspection then shows that the retarded field possesses a
kind of peeling property in that it consists of terms with
different fall off at infinity. Explicitly, this means that the
retarded field will consist of a sum of terms possessing all
integer inverse powers in $r$ between $\frac{D-2}{2}$ and $D-3$.
The term that dies off more quickly at infinity is the
$\frac{1}{r^{D-3}}$, typically a static term, since it comes from
the Laplacian.  As a matter of fact this term was already observed
in the higher dimensional black hole by Tangherlini
\cite{tangherlini} (see also Myers and Perry \cite{myersperry}).
We will see that the term falling more slowly, the one that goes
like $\frac{1}{r^{(D-2)/2}}$, gives rise to gravitational
radiation. It is well defined, in the sense that the power
crossing sufficiently large hyperspheres with different radius is
the same, because the volume element goes as $r^{D-2}$ and the
energy as $|h|^2 \sim \frac{1}{r^{D-2}}$.

In radiation problems, one is interested in finding out the field
at large distances from the source, $r>>\lambda$, where $\lambda$
is the wavelength of the waves, and also much larger than the
source's dimensions $R$. This is defined as the wave zone.  In the
wave zone, one may neglect all terms in the Green's function that
decay faster than $\frac{1}{r^{(D-2)/2}}$. So, in the wave zone,
we find
\begin{equation}
h_{\mu\nu}(t,{\bf x})= -8 \pi {\cal G} \frac{1}{(2\pi
r)^{(D-2)/2}}
\partial_{t}^{(\frac{D-4}{2})}\left[ \int d^{D-1}{\bf x'}
 S_{\mu \nu}(t-|{\bf x-x'}|,{\bf x'})\right]\,,
\label{retsolwavezone}
\end{equation}
where $\partial_{t}^{(\frac{D-4}{2})}$ stands for the
$\frac{D-4}{2}$th derivative with respect to time. For $D=4$ eq.
(\ref{retsolwavezone}) yields the standard result \cite{weinberg}:
\begin{equation}
h_{\mu\nu}(t,{\bf x})= -\frac{4 {\cal G}}{r} \int d^{D-1}{\bf x'}
 S_{\mu \nu}(t-|{\bf x-x'}|,{\bf x'})\,,\,\,\,\,\,D=4.
\label{retsolwavezoned4}
\end{equation}
To find the Fourier transform of the metric, one uses the
representation (\ref{greenfinalevenfourier}) for the Green's
function. Now, in the wave zone, the Green's function may be
simplified using the asymptotic expansion for the Bessel function
\cite{stegun}
\begin{equation}
H^1_{(D-3)/2}(\omega r) \sim \sqrt{\frac{2}{\pi(\omega r)}}
e^{i\left[\omega r-\frac{\pi}{4}(D-2)\right]}\,\,,\,\,\,\,\,
\omega r \rightarrow \infty. \label{asymbessel}
\end{equation}
This yields
\begin{equation}
h_{\mu\nu}(\omega,{\bf x})= -\frac{8 \pi {\cal G}}{(2\pi
r)^{(D-2)/2}}\omega^{(D-4)/2}e^{i\omega r} \int d^{D-1}{\bf x'}
S_{\mu \nu}(\omega,{\bf x'})\,. \label{retsolwavezonefourier}
\end{equation}
This could also have been arrived at directly from
(\ref{retsolwavezone}), using the rule time derivative
$\rightarrow -i\omega $ for Fourier transforms.  Equations
(\ref{retsolwavezone}) and (\ref{retsolwavezonefourier}) are one
of the most important results derived in this chapter, and will be
the basis for all the subsequent section. Similar equations, but
not as general as the ones presented here, were given by Chen, Li
and Lin \cite{lin} in the context of gravitational radiation by a
rolling tachyon.

To get the energy spectrum, we use (\ref{S}) yielding
\begin{equation}
\frac{d^2E}{d\omega d\Omega}= 2 {\cal G}
\frac{\omega^{D-2}}{(2\pi)^{D-4}} \left( T^{\mu\nu}(\omega,{\bf
k})T_{\mu\nu}^*(\omega,{\bf k})-\frac{1}{D-2}
|T^{\lambda}_{\:\:\:\lambda}(\omega, {\bf k})|^2\right)\,.
\label{powerwavezone}
\end{equation}

\section[The even $D$-dimensional quadrupole formula]
{The even $\bm D$-dimensional quadrupole formula}
\label{quadrupole form}
\subsection[Derivation of the even $D$-dimensional quadrupole formula]
{Derivation of the even $\bm D$-dimensional quadrupole formula}
When the velocities of the sources that generate the gravitational
waves are small, it is sufficient to know the $T^{00}$ component
of the gravitational energy-momentum tensor in order to have a
good estimate of the energy they radiate. In this subsection, we
will deduce the $D$-dimensional quadrupole formula and in the next
subsection we will apply it to (1) a particle in circular orbit
and (2) a particle in free fall into a $D$-dimensional
Schwarzschild black hole.

We start by recalling that the Fourier transform of the
energy-momentum tensor is
\begin{equation}
T_{\mu\nu}(\omega,{\bf k})=\int d^{D-1}{\bf x'} e^{-i {\bf k}\cdot
{\bf x'}} \int dt\, e^{i \omega t}\,T^{\mu\nu}(t,{\bf x})+ {\rm
c.c.} \,, \label{fourierT}
\end{equation}
where ${\rm c.c.}$ means the complex conjugate of the preceding
term. Then, the conservation equations (\ref{consT}) for
$T^{\mu\nu}(t,{\bf x})$ applied to eq. (\ref{fourierT}) yield
$k^{\mu}\,T_{\mu\nu}(\omega,{\bf k})=0$. Using this last result we
obtain $T_{00}(\omega,{\bf k})=\hat{k}^j \,\hat{k}^i
\,T_{ji}(\omega,{\bf k})$ and $T_{0i}(\omega,{\bf k})=-\hat{k}^j\,
T_{ji}(\omega,{\bf k})$, where ${\bf \hat{k}}={\bf k}/\omega$. We
can then write the energy spectrum, eq. (\ref{powerwavezone}), as
a function only of the spacelike components of
$T^{\mu\nu}(\omega,{\bf k})$,
\begin{equation}
\frac{d^2E}{d\omega d\Omega}= 2 {\cal G}
\frac{\omega^{D-2}}{(2\pi)^{D-4}}\, \Lambda_{ij,\,lm}(\hat{k})\,
T^{*\,ij}(\omega,{\bf k})\, T^{\,ij}(\omega,{\bf k}) \,,
\label{powerwavezone2}
\end{equation}
where
\begin{equation}
 \Lambda_{ij,\,lm}(\hat{k})=\delta_{il}\delta_{jm}
 -2\hat{k}_j \hat{k}_m \delta_{il}
 +\frac{1}{D-2}\left (-\delta_{ij}\delta_{lm}
 + \hat{k}_l \hat{k}_m \delta_{ij}
 +\hat{k}_i \hat{k}_j \delta_{lm}\right )
 +\frac{D-3}{D-2}\hat{k}_i \hat{k}_j \hat{k}_l \hat{k}_m \,.
\label{Lambda}
\end{equation}
At this point, we make a new approximation (in addition to the
wave zone approximation) and assume that $\omega R<<1$, where $R$
is the source's radius. In other words, we assume that the
internal velocities of the sources are small and thus the source's
radius is much smaller than the characteristic wavelength $\sim
1/\omega$ of the emitted gravitational waves. Within this
approximation, one can set $e^{-i {\bf k}\cdot {\bf x'}} \sim 1$
in eq. (\ref{fourierT}) (since $R=|{\bf x'}|_{\rm max}$).
Moreover, after a straightforward calculation, one can also set in
eq. (\ref{powerwavezone2}) the approximation $T^{\,ij}(\omega,{\bf
k})\simeq -(\omega^2/2)D_{ij}(\omega)$, where
\begin{equation}
D_{ij}(\omega)=
 \int d^{D-1} {\bf x}\, x^i \,x^j \,T^{00}(\omega,{\bf x})\,.
 \label{Dij}
\end{equation}
Finally, using
\begin{eqnarray}
\int d\Omega_{D-2}\hat{k}_i
\hat{k}_j=\frac{\Omega_{D-2}}{D-1}\delta_{ij}\,, \nonumber \\
\int d\Omega_{D-2}\hat{k}_i \hat{k}_j \hat{k}_l \hat{k}_m=
 \frac{3\Omega_{D-2}}{D^2-1} (\delta_{ij}\delta_{lm}
  +\delta_{il}\delta_{jm}+\delta_{im}\delta_{jl})\,,
\label{intk}
\end{eqnarray}
where $\Omega_{D-2}$ is the $(D-2)$-dimensional solid angle
defined in (\ref{integratedsolidangle}), we obtain the
$D$-dimensional quadrupole formula
\begin{equation}
\frac{dE}{d\omega}=\frac{2^{2-D}\pi^{-(D-5)/2}{\cal G}\,(D-3)D}
{\Gamma[(D-1)/2](D^2-1)(D-2)}\, \omega^{D+2} {\biggl [}
(D-1)D^*_{ij}(\omega)D_{ij}(\omega)-|D_{ii}(\omega)|^2 {\biggr ]}
 \,,
\label{quadw}
\end{equation}
where the Gamma function $\Gamma[z]$ is defined in
 (\ref{Gamma function D-dim}).  As the dimension $D$ grows it is seen that the
rate of gravitational energy radiated increases as $\omega^{D+2}$.
Sometimes it will be more useful to have the time rate of emitted
energy
\begin{equation}
\frac{dE}{dt}=\frac{2^{2-D}\pi^{-(D-5)/2} {\cal G}\,(D-3)D}
{\Gamma[(D-1)/2](D^2-1)(D-2)}\,{\biggl [}
(D-1)\partial_t^{(D+2)/2}D^*_{ij}(t)\partial_t^{(D+2)/2}D_{ij}(t)-
|\partial_t^{(D+2)/2}D_{ii}(t)|^2 {\biggr ]}
 \,.
\label{quad}
\end{equation}
For $D=4$, eq. (\ref{quad}) yields the well known result
\cite{weinberg}
\begin{equation}
\frac{dE}{dt}=\frac{{\cal G}}{5}\, {\biggl [}
\partial_t^{3}D^*_{ij}(t)\partial_t^{3}D_{ij}(t)-
\frac{1}{3}|\partial_t^{3}D_{ii}(t)|^2 {\biggr ]}
 \,.
\label{quad4}
\end{equation}

\subsection{Applications of the quadrupole formula: test particles in a background geometry}
The quadrupole formula has been used successfully in almost all
kind of problems involving gravitational wave generation. By
successful we mean that it agrees with other more accurate
methods. Its simplicity and the fact that it gives results correct
to within a few percent, makes it an invaluable tool in estimating
gravitational radiation emission. We shall in the following
present two important examples of the application of the
quadrupole formula.
\subsubsection{A particle in circular orbit}
The radiation generated by particles in circular motion was
perhaps the first situation to be considered in the analysis of
gravitational wave generation. For orbits with low frequency, the
quadrupole formula yields excellent results. As expected it is
difficult to find in nature a system with perfect circular orbits,
they will in general be elliptic. In this case the agreement is
also remarkable, and one finds that the quadrupole formalism can
account with precision for the increase in period of the pulsar
PSR 1913$+$16, due to gravitational wave emission \cite{pulsar}.
In four dimensions the full treatment of elliptic orbital motion
is discussed by Peters \cite{peters}.  In dimensions higher than
four, it has been shown \cite{tangherlini} that there are no
stable geodesic circular orbits, and so geodesic circular motion
is not as interesting for higher $D$. For this reason, and also
because we only want to put in evidence the differences that arise
in gravitational wave emission as one varies the spacetime
dimension $D$, we will just analyze the simple circular, not
necessarily geodesic motion, to see whether the results are
non-trivially changed as one increases $D$.  Consider then two
bodies of equal mass $m$ in circular orbits a distance $l$ apart.
Suppose they revolve around the center of mass, which is at $l/2$
from both masses, and that they orbit with frequency $\omega$ in
the $x-y$ plane.  A simple calculation \cite{peters,schutzlivro}
yields
\begin{eqnarray}
D_{xx}=\frac{ml^2}{4}\cos({2\omega t}) \,+\,{\rm const}\,\,, \\
D_{yy}=-D_{xx}\,\,, \\
D_{xy}=\frac{ml^2}{4}\sin({2\omega t}) \,+\,{\rm const}\,\,,
\label{moments}
\end{eqnarray}
independently of the dimension in which they are imbedded and with
all other components being zero.  We therefore get from eq.
(\ref{quad})

\begin{equation}
\frac{dE}{dt}=\frac{2{\cal
G}D(D-3)}{\pi^{(D-5)/2}\Gamma[(D-1)/2](D+1)(D-2)}m^2 l^4
 \omega^{D+2}.
\label{totalenecircular}
\end{equation}
For $D=4$ one gets
\begin{equation}
\frac{dE}{dt}=\frac{8{\cal G}}{5}m^2 l^4 \omega^{6}\,,
\label{totalenecircularD4}
\end{equation}
which agrees with known results \cite{peters,schutzlivro}. Eq.
(\ref{totalenecircular}) is telling us that as one climbs up in
dimension number $D$, the frequency effects gets more pronounced.

\subsubsection{A particle falling radially into a higher dimensional
Schwarzschild black hole}
As yet another example of the use of the quadrupole formula eq.
(\ref{quad}) we now calculate the energy given away as
gravitational waves when a point particle, with mass $m$ falls
into a $D$-dimensional Schwarzschild black hole, a metric first
given in \cite{tangherlini}.  Historically, the case of a particle
falling into a $D=4$ Schwarzschild black hole was one of the first
to be studied \cite{zerilli,davis} in connection with
gravitational wave generation, and later served as a model
calculation when one wanted to evolve Einstein's equations fully
numerically \cite{smarr,gleiser}.  This process was first studied
\cite{davis} by solving numerically Zerilli's \cite{zerilli} wave
equation for a particle at rest at infinity and then falling into
a Schwarzschild black hole. Davis et al \cite{davis} found
numerically that the amount of energy radiated to infinity as
gravitational waves was $\Delta E_{\rm num} =0.01 \frac{m^2}{M}$,
where $m$ is the mass of the particle falling in and $M$ is the
mass of the black hole. It is found that the $D=4$ quadrupole
formula yields \cite{quem} $\Delta E_{\rm quad} =0.019
\frac{m^2}{M}$, so it is of the order of magnitude as that given
by fully relativistic numerical results. Despite the fact that the
quadrupole formula fails somewhere near the black hole (the motion
is not slow, and the background is certainly not flat), it looks
like one can get an idea of how much radiation will be released
with the help of this formula.  Based on this good agreement, we
shall now consider this process but for higher dimensional
spacetimes.  The metric for the $D$-dimensional Schwarzschild
black hole in ($t,r,\theta_1,\theta_2,..,\theta_{D-2}$)
coordinates is (see section \ref{sec:BH D-dim flat})
\begin{equation}
ds^2= -\left(1-\frac{16\pi{\cal G}
M}{(D-2)\Omega_{D-2}}\frac{1}{r^{D-3}}\right)dt^2+
\left(1-\frac{16\pi{\cal G}
M}{(D-2)\Omega_{D-2}}\frac{1}{r^{D-3}}\right)^{-1}dr^2
+r^2d\Omega_{D-2}^2. \label{metricmyersperry}
\end{equation}
Consider a particle falling along a radial geodesic, and at rest
at infinity. Then, the geodesic equations give
\begin{equation}
\frac{dr}{dt} \sim \frac{16\pi{\cal G}
M}{(D-2)\Omega_{D-2}}\frac{1}{r^{D-3}}\,, \label{geodesic}
\end{equation}
where we make the flat space approximation $t=\tau$.  We then
have, in these coordinates, $D_{11}=r^2$, and all other components
vanish. From (\ref{quad}) we get the energy radiated per second,
which yields
\begin{equation}
\frac{dE}{dt}= \frac{2^{2-D}\pi^{-(D-5)/2}{\cal G}\,(D-3)}
{\Gamma[(D-1)/2](D^2-1)}D|\partial_{t}^{(\frac{D+2}{2})}
D_{11}|^2\,, \label{enerpersecinfall}
\end{equation}
We can perform the derivatives and integrate to get the total
energy radiated.  There is a slight problem though, where do we
stop the integration?  The expression for the energy diverges at
$r=0$ but this is no problem, as we know that as the particle
approaches the horizon, the radiation will be infinitely
red-shifted. Moreover, the standard picture \cite{quem} is that of
a particle falling in, and in the last stages being frozen near
the horizon.  With this in mind we integrate from $r=\infty$ to
some point near the horizon, say $r=b\times r_+$, where $r_+$ is
the horizon radius and $b$ is some number larger than unit, and we
get
\begin{equation}
\Delta E= A \frac{D(D-2)\pi}{2^{2D-4}} \times b^{(9-D^2)/2} \times
\frac{m^2}{M}\,, \label{totalenergyinfall}
\end{equation}
where
\begin{equation}
A=\frac{(3-D)^2(5-D)^2(7-3D)^2(8-4D)^2(9-5D)^2...
(D/2+4-D^2/2)^2}{\Gamma[(D-1)/2]^2(D-1)(D+1)(D+3)} \label{A}
\end{equation}
To understand the effect of both the dimension number $D$ and the
parameter $b$ on the total energy radiated according to the
quadrupole formula, we list in Table 1 some values $\Delta E$ for
different dimensions, and $b$ between $1$ and $1.3$.

\vskip 1mm
\begin{table}
\caption{\label{tab:zzz} The energy radiated by a particle falling
from rest into a higher dimensional Schwarzschild black hole, as a
function of dimension.  The integration is stopped at $b\times
r_+$ where $r_+$ is the horizon radius.}
  \vskip 4mm \centering
\begin{tabular}{|l|l|l|l|}  \hline
\multicolumn{1}{c}{} & \multicolumn{3}{c}{ $\Delta E \times
\frac{M}{m^2}$}\\ \hline $D$ & $b=1$:  &     $b=1.2$:   &
$b=1.3$:\\ \hline 4   &  0.019  &  0.01 &  0.0076 \\ \hline 6   &
0.576  &  0.05 &  0.0167 \\ \hline 8   &  180    &  1.19 &  0.13
\\ \hline 10  &  24567  &  6.13 &  0.16   \\ \hline 12
&$3.3\times 10^6$ & 14.77 & 0.0665 \\ \hline
\end{tabular}
\end{table}
\vskip 1mm The parameter $b$ is in fact a measure of our ignorance
of what goes on near the black hole horizon, so if the energy
radiated doesn't vary much with $b$ it means that our lack of
knowledge doesn't affect the results very much.  For $D=4$ that
happens indeed. Putting $b=1$ gives only an energy $2.6$ times
larger than with $b=1.3$, and still very close to the fully
relativistic numerical result of $0.01 \frac{m^2}{M}$.  However as
we increase $D$, the effect of $b$ increases dramatically. For
$D=12$ for example, we can see that a change in $b$ from $1$ to
$1.3$ gives a corresponding change in $\Delta E$ of $3\times 10^6$
to $0.0665$. This is $8$ orders of magnitude lower!  Since there
is as yet no Regge-Wheeler-Zerilli \cite{zerilli,regge}
wavefunction for higher dimensional Schwarzschild black holes,
there are no fully relativistic numerical results to compare our
results with. Thus $D=4$ is just the perfect dimension to predict,
through the quadrupole formula, the gravitational energy coming
from collisions between particles and black holes, or between
small and massive black holes.  It is not a problem related to the
quadrupole formalism, but rather one related to $D$. A small
change in parameters translates itself, for high $D$, in a large
variation in the final result. Thus, as the dimension $D$ grows,
the knowledge of the cutoff radius $b\times r_+$ becomes essential
to compute accurately the energy released.

\section[Instantaneous collisions in even $D$-dimensions.
Energy released during black hole pair creation]{Instantaneous
collisions in even $\bm D$-dimensions. Energy released during
black hole pair creation} \label{instantaneous collisions}
In general, whenever two bodies collide or scatter there will be
gravitational energy released due to the changes in momentum
involved in the process. If the collision is hard meaning that the
incoming and outgoing trajectories have constant velocities, there
is a method first envisaged by Weinberg \cite{weinberg,wein1},
later explored in \cite{wein2} by Smarr to compute exactly the
metric perturbation and energy released. The method is valid for
arbitrary velocities (one will still be working in the linear
approximation, so energies have to be low). Basically, it assumes
a collision lasting for zero seconds. It was found that in this
case the resulting spectra were flat, precisely what one would
expect based on one's experience with electromagnetism
\cite{jackson}, and so to give a meaning to the total energy, a
cutoff frequency is needed. This cutoff frequency depends upon
some physical cutoff in the particular problem. We shall now
generalize this construction for arbitrary dimensions.
\subsection{Derivation of the Radiation Formula
in terms of a cutoff for a head-on collision. Energy released
during formation of black hole at LHC} \label{HeadonCollision}
Consider therefore a system of freely moving particles with
$D$-momenta $P_{i}^{\mu}$, energies $E_i$ and ($D-1$)-velocities
${\bf v}$, which due to the collision change abruptly at $t=0$, to
corresponding primed quantities. For such a system, the
energy-momentum tensor is
\begin{equation}
T^{\mu\nu}(t, {\bf v})= \sum \frac{P_{i}^{\mu}P_{i}^{\nu}}{E_i}
\delta^{D-1}({\bf x}-{\bf v}t)\Theta(-t)+
\frac{{P'}_{i}^{\mu}{P'}_{i}^{\nu}}{E'_i} \delta^{D-1}({\bf
x'}-{\bf v'}t)\Theta(t)\,, \label{enmomtenpointpctles}
\end{equation}
from which, using eqs. (\ref{retsolwavezonefourier}) and
(\ref{powerwavezone}) one can get the quantities $h_{\mu\nu}$ and
also the radiation emitted. Let us consider the particular case in
which one has a head-on collision of two particles, particle $1$
with mass $m_1$ and Lorentz factor $\gamma_1$, and particle $2$
with mass $m=m_2$ with Lorentz factor $\gamma_2$, colliding to
form a particle at rest.  Without loss of generality, one may
orient the axis so that the motion is in the $(x_{D-1},x_D)$
plane, and the $x_D$ axis is the radiation direction [see
(\ref{spher coord D-1})]. We then have
\begin{eqnarray}
P_{1}= \gamma_1m_1
(1,0,0,...,v_1\sin\theta_1,v_1\cos\theta_1)\,\,;\,\,\,\,\,
P'_1=(E'_{1},0,0,...,0,0) \label{momenta1}
\\
P_{2}= \gamma_2m_2
(1,0,0,...,-v_2\sin\theta_1,-v_2\cos\theta_1)\,\,;\,\,\,\,\,
P'_2=(E'_{2},0,0,...,0,0). \label{momenta2}
\end{eqnarray}
Momentum conservation leads to the additional relation
$\gamma_1m_1v_1=\gamma_2m_2v_2$.  Replacing (\ref{momenta1}) and
(\ref{momenta2}) in the energy-momentum tensor
(\ref{enmomtenpointpctles}) and using (\ref{powerwavezone}) we
find
\begin{equation}
\frac{d^2E}{d\omega d\Omega}=\frac{2{\cal G}}{(2\pi)^{D-2}}
\frac{D-3}{D-2}\frac{\gamma_{1}^2m_{1}^2v_{1}^2(v_1+v_2)^2
\sin{\theta_1}^4}{(1-v_1\cos\theta_1)^2(1+v_2\cos\theta_1)^2}\times
\omega^{D-4}\,. \label{energypersolidanglefreqinstcol}
\end{equation}
We see that the for arbitrary (even) $D$ the spectrum is not flat.
Flatness happens only for $D=4$. For any $D$ the total energy,
integrated over all frequencies would diverge so one needs a
cutoff frequency which shall depend on the particular problem
under consideration.  Integrating
(\ref{energypersolidanglefreqinstcol}) from $\omega=0$ to the
cutoff frequency $\omega_c$ we have
\begin{equation}
\frac{dE}{d\Omega}=\frac{2{\cal G}}{(2\pi)^{D-2}}
\frac{1}{D-2}\frac{\gamma_{1}^2m_{1}^2v_{1}^2(v_1+v_2)^2
\sin{\theta_1}^4}{(1-v_1\cos\theta_1)^2(1+v_2\cos\theta_1)^2}\times
\omega_{c}^{D-3}\,. \label{energypersolidangleinstcol}
\end{equation}
Two limiting cases are of interest here, namely (i) the collision
between identical particles and (ii) the collision between a light
particle and a very massive one.  In case (i) replacing
$m_1=m_2=m$, $v_1=v_2=v$, eq. (\ref{energypersolidangleinstcol})
gives
\begin{equation}
\frac{dE}{d\Omega}=\frac{8{\cal G}}{(2\pi)^{D-2}}
\frac{1}{D-2}\frac{\gamma^2m^2v^4
\sin{\theta_1}^4}{(1-v^2\cos^2\theta_1)^2}\times
\omega_{c}^{D-3}\,. \label{energypersolidangleinstcolident}
\end{equation}
In case (ii) considering $m_1\gamma_1 \equiv m\gamma
<<m_2\gamma_2$, $v_1\equiv v>>v_2$, eq.
(\ref{energypersolidangleinstcol}) yields
\begin{equation}
\frac{dE}{d\Omega}=\frac{2{\cal G}}{(2\pi)^{D-2}}
\frac{1}{D-2}\frac{\gamma^2m^2v^4
\sin{\theta_1}^4}{(1-v\cos\theta_1)^2}\times \omega_{c}^{D-3}\,.
\label{energypersolidangleinstcoliheavylight}
\end{equation}
Notice that the technique just described is expected to break down
if the velocities involved are very low, since then the collision
would not be instantaneous. In fact a condition for this method to
work would can be stated

 Indeed, one can see from eq.
(\ref{energypersolidangleinstcol}) that if $v\rightarrow 0$,
$\frac{dE}{d\omega}\rightarrow 0$, even though we know (see
Subsection (\ref{HeadonCollision})) that $\Delta E \neq 0$. In any
case, if the velocities are small one can use the quadrupole
formula instead.

\subsection{Applications:
The cutoff frequency when one of the particles is a black hole and
radiation from black hole pair creation}

\subsubsection{The cutoff frequency when one of the head-on colliding
particles is a black hole}

We shall now restrict ourselves to the case (ii) of last
subsection, in which at least one of the particles participating
in the collision is a massive black hole, with mass $M>>m$ (where
we have put $m_1=m$ and $m_2=M$).
 Formulas
(\ref{energypersolidangleinstcol})-
(\ref{energypersolidangleinstcoliheavylight}) are useless unless
one is able to determine the cutoff frequency $\omega_c$ present
in the particular problem under consideration.  In the situation
where one has a small particle colliding at high velocities with a
black hole, it has been suggested by Smarr \cite{wein2} that the
cutoff frequency should be $\omega_c \sim 1/2M$, presumably
because the characteristic collision time is dictated by the large
black hole whose radius is $2M$.  Using this cutoff he finds
\begin{equation}
\Delta E_{\rm Smarr}\sim 0.2 \gamma^2 \frac{m^2}{M}. \label{smarr}
\end{equation}
The exact result, using a relativistic perturbation approach which
reduces to the numerical integration of a second order
differential equation (the Zerilli wavefunction), has been given
by Cardoso and Lemos \cite{cardosolemos}, as
\begin{equation}
\Delta E_{\rm exact} = 0.26 \gamma^2 \frac{m^2}{M}. \label{exact}
\end{equation}
This is equivalent to saying that $\omega_c =\frac{0.613}{M} \sim
\frac{1}{1.63 M}$, and so it looks like the cutoff is indeed the
inverse of the horizon radius.  However, in the numerical work by
Cardoso and Lemos, it was found that it was not the presence of an
horizon that contributed to this cutoff, but the presence of a
potential barrier $V$ outside the horizon. By decomposing the
field in tensorial spherical harmonics with index $l$ standing for
the angular quantum number, we found that for each $l$, the
spectrum is indeed flat (as predicted by eq.
(\ref{energypersolidangleinstcol}) for $D=4$), until a cutoff
frequency $\omega_{c_l}$ which was numerically equal to the lowest
gravitational quasinormal frequency $\omega_{\rm QN}$. For
$\omega> \omega_{c_l}$ the spectrum decays exponentially. This
behavior is illustrated in Fig. 1.
\begin{figure}
\centerline{\includegraphics[width=10 cm,height=6.5 cm]
{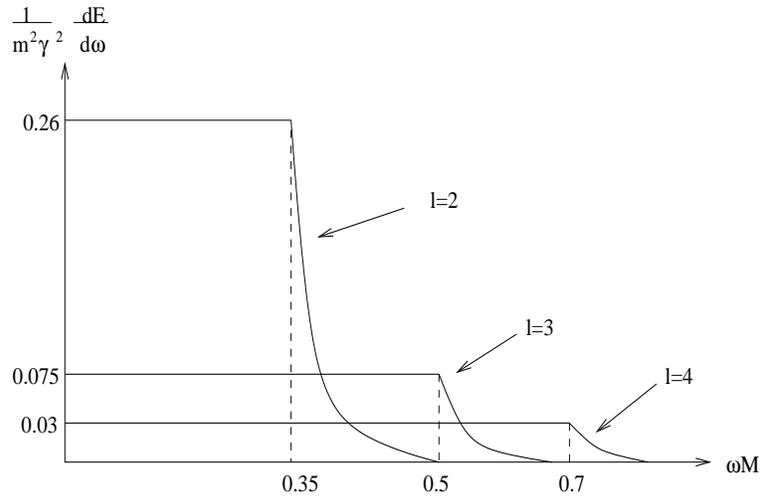}} \caption{The energy spectra as a function of the
angular number $l$, for a highly relativistic particle falling
into a $D=4$ Schwarzschild black hole \cite{cardosolemos}. The
particle begins to fall with a Lorentz factor $\gamma$. Notice
that for each $l$ there is a cutoff frequency $\omega_{c_l}$ which
is equal to the quasinormal frequency $\omega_{\rm QN}$ after
which the spectrum decays exponentially. So it is clearly seen
that $\omega_{\rm QN}$ works as a cutoff frequency. The total
energy radiated is a given by a sum over $l$, which is the same as
saying that the effective cutoff frequency is given by a weighted
average of the various $\omega_{c_l}$.} \label{fig:1}
\end{figure}
The quasinormal frequencies \cite{kokkotas} are those frequencies
that correspond to only outgoing waves at infinity and only
ingoing waves near the horizon.  As such the gravitational
quasinormal frequencies will in general have a real and an
imaginary part, the latter denoting gravitational wave emission
and therefore a decay in the perturbation.  There have been a
wealth of works dwelling on quasinormal modes on asymptotically
flat spacetimes \cite{kokkotas}, due to its close connection with
gravitational wave emission, and also on non-asymptotically flat
spacetimes, like asymptotically anti-de Sitter \cite{horowitz} or
asymptotically de Sitter \cite{abdalla} spacetimes, mainly due to
the AdS/CFT and dS/CFT \cite{maldacena} correspondence conjecture.
We argue here that it is indeed the quasinormal frequency that
dictates the cutoff, and not the horizon radius.  For $D=4$ it so
happens that the weighted average of $\omega_{c_l}$ is
$\frac{0.613}{M}$, which, as we said, is quite similar to
$r_+=\frac{1}{2M}$.  The reason for the cutoff being dictated by
the quasinormal frequency can be understood using some WKB
intuition. The presence of a potential barrier outside the horizon
means that waves with some frequencies get reflected back on the
barrier while others can cross. Frequencies such that $\omega^2$
is lower than the maximum barrier height $V_{\rm max}$ will be
reflected back to infinity where they will be detected. However,
frequencies $\omega^2$ larger than the maximum barrier height
cross the barrier and enter the black hole, thereby being absorbed
and not contributing to the energy detected at infinity.  So only
frequencies $\omega^2$ lower than this maximum barrier height are
detected at infinity.  It has been shown \cite{schutz2} that the
gravitational quasinormal frequencies are to first order equal to
the square root of the maximum barrier height.  In view of this
picture, and considering the physical meaning of the cutoff
frequency, it seems quite natural to say that the cutoff frequency
is equal to the quasinormal frequency. If the frequencies are
higher than the barrier height, they don't get reflected back to
infinity. This discussion is very important to understand how the
total energy varies with the number $D$ of dimensions. In fact, if
we set $\omega_c \sim \frac{1}{r_+}$, we find that the total
energy radiated decreases rapidly with the dimension number,
because $r_+$ increases rapidly with the dimension. This conflicts
with recent results \cite{eardley,yoshino}, which using shock
waves that describe boosted Schwarzschild black holes, and
searching for apparent horizons, indicate an increase with $D$.
So, we need the gravitational quasinormal frequencies for higher
dimensional Schwarzschild black holes. To arrive at an wave
equation for gravitational perturbations of higher dimensional
Schwarzschild black holes, and therefore to compute its
gravitational quasinormal frequencies, one needs to decompose
Einstein's equations in D-dimensional tensorial harmonics, which
would lead to some quite complex expressions.  It is not necessary
to go that far though, because one can get an idea of how the
gravitational quasinormal frequencies vary by searching for the
quasinormal frequencies of scalar perturbations, and scalar
quasinormal frequencies are a lot easier to find.  One hopes that
the scalar frequencies will behave with $D$ in the same manner as
do the gravitational ones.  Scalar perturbations in
$D$-dimensional Schwarzschild spacetimes obey the wave equation
(consult \cite{cardosolemos3} for details)
\begin{equation}
\frac{\partial^{2} \phi(\omega,r)}{\partial r_*^{2}} +
\left\lbrack\omega^2-V(r)\right\rbrack \phi(\omega,r)=0 \,.
\label{scalwavequat}
\end{equation}
The potential $V(r)$ appearing in equation (\ref{scalwavequat}) is
given by
\begin{equation}
V(r)= f(r)\left\lbrack\frac{a}{r^2}+
\frac{(D-2)(D-4)f(r)}{4r^2}+\frac{(D-2)f'(r)}{2r}\right\rbrack \,,
\label{potential}
\end{equation}
where $a=l(l+D-3)$ is the eigenvalue of the Laplacian on the
hypersphere $S^{D-2}$, the tortoise coordinate $r_*$ is defined as
$\frac{\partial r}{\partial r_*}=f(r)=\left(1-\frac{16\pi {\cal G}
M}{(D-2)\Omega_{D-2}}\frac{1}{r^{D-3}}\right)$, and
$f'(r)=\frac{df(r)}{dr}$. We have found the quasinormal
frequencies of spherically symmetric ($l=0$) scalar perturbations,
by using a WKB approach developed by Schutz, Will and
collaborators \cite{schutz2}.  The results are presented in Table
2, where we also show the maximum barrier height of the potential
in eq. (\ref{potential}), as well as the horizon radius. \vskip
1mm
\begin{table}
\caption{\label{tab:zfl} The lowest scalar quasinormal frequencies
for spherically symmetric ($l=0$) scalar perturbations of higher
dimensional Schwarzschild black holes, obtained using a WKB method
\cite{schutz2}. Notice that the real part of the quasinormal
frequency is always the same order of magnitude as the square root
of the maximum barrier height. We show also the maximum barrier
height as well as the horizon radius as a function of dimension
$D$. The mass $M$ of the black hole has been set to 1.}
 \vskip 4mm \centering
\begin{tabular}{|l|l|l|l|l|}  \hline
$D$&${\rm Re}[\omega_{QN}]$:&${\rm Im}[\omega_{QN}]:$&$\sqrt{\rm
V_{\rm max}}$:&$1/r_+$:\\ \hline 4   &  0.10      & -0.12 &  0.16
&  0.5 \\ \hline 6   &  1.033     &  -0.713      &  1.441 &  1.28
\\ \hline 8   &  1.969      &  -1.023       &  2.637 &  1.32   \\
\hline 10  &  2.779     &   -1.158     &  3.64   & 1.25   \\
\hline 12  &  3.49     &    -1.202     &  4.503   & 1.17 \\ \hline
\end{tabular}
\end{table}
\vskip 1mm The first thing worth noticing is that the real part of
the scalar quasinormal frequency is to first order reasonably
close to the square root of the maximum barrier height
$\sqrt{V_{\rm max}}$, supporting the previous discussion.
Furthermore, the scalar quasinormal frequency grows more rapidly
than the inverse of the horizon radius $\frac{1}{r_+}$ as one
increases $D$. In fact, the scalar quasinormal frequency grows
with $D$ while the horizon radius $r_+$ gets smaller. Note that
from pure dimensional arguments, for fixed $D$, $\omega \propto
\frac{1}{r_+}$. The statement here is that the constant of
proportionality depends on the dimension $D$, more explicitly it
grows with $D$,  and can be found from Table 2. Assuming that the
gravitational quasinormal frequencies will have the same behavior
(and some very recent studies \cite{quasispectrum} relating black
hole entropy and damped quasinormal frequencies seem to point that
way), the total energy radiated will during high-energy collisions
does indeed increase with $D$, as some studies
\cite{eardley,yoshino} seem to indicate.

\subsubsection{The gravitational
energy radiated during black hole pair creation}
As a new application of this instantaneous collision formalism, we
will now consider the gravitational energy released during the
quantum creation of pairs of black holes, a process which as far
as we know has not been analyzed in the context of gravitational
wave emission, even for $D=4$.  It is well known that vacuum
quantum fluctuations produce virtual electron-positron pairs.
These pairs can become real \cite{Schwinger} if they are pulled
apart by an external electric field, in which case the energy for
the pair materialization and acceleration comes from the external
electric field energy.  Likewise, a black hole pair can be created
in the presence of an external field whenever the energy pumped
into the system is enough in order to make the pair of virtual
black holes real (see chapter \ref{chap:Pair creation}). If one
tries to predict the spectrum of radiation coming from pair
creation, one expects of course a spectrum characteristic of
accelerated masses but one also expects that this follows some
kind of signal indicating pair creation. In other words, the
process of pair creation itself, which involves the sudden
creation of particles, must imply emission of radiation. It is
this phase we shall focus on, forgetting the subsequent emission
of radiation caused by the acceleration.

Pair creation is a pure quantum-mechanical process in nature, with
no classical explanation. But given that the process does occur,
one may ask about the spectrum and intensity of the radiation
accompanying it. The sudden creation of pairs can be viewed for
our purposes as an instantaneous creation of particles (i.e., the
time reverse process of instantaneous collisions), the violent
acceleration of particles initially at rest to some final velocity
in a very short time, and the technique described at the beginning
of this section applies. This is quite similar to another pure
quantum-mechanical process, the beta decay. The electromagnetic
radiation emitted during beta decay has been computed classically
by Chang and Falkoff \cite{chang} and is also presented in Jackson
\cite{jackson}.  The classical calculation is similar in all
aspects to the one described in this section (the instantaneous
collision formalism) assuming the sudden acceleration to energies
$E$ of a charge initially at rest, and requires also a cutoff in
the frequency, which has been assumed to be given by the
uncertainty principle $\omega_c \sim \frac{E}{\hbar}$. Assuming
this cutoff one finds that the agreement between the classical
calculation and the quantum calculation \cite{chang} is extremely
good (specially in the low frequency regime), and more important,
was verified experimentally. Summarizing, formula
(\ref{energypersolidangleinstcolident}) also describes the
gravitational energy radiated when two black holes, each with mass
$m$ and energy $E$ form through quantum pair creation.  The
typical pair creation time can be estimated by the uncertainty
principle $\tau_{\rm creation}\sim \hbar/E \sim
\frac{\hbar}{m\gamma}$, and thus we find the cutoff frequency as
\begin{equation}
\omega_c\sim \frac{1}{\tau_{\rm creation}}\sim
\frac{m\gamma}{\hbar}\,. \label{cutofffrequency}
\end{equation}
Here we would like to draw the reader's attention to the fact that
the units of Planck's constant $\hbar$ change with dimension
number $D$: according to our convention of setting $c=1$ the units
of $\hbar$ are $[M]^{\frac{D-2}{D-3}}$.  With this cutoff, we find
the spectrum of the gravitational radiation emitted during pair
creation to be given by (\ref{energypersolidanglefreqinstcol})
with $m_1=m_2$ and $v_1=v_2$ (we are considering the pair creation
of two identical black holes):
\begin{equation}
\frac{d^2E}{d\omega d\Omega}=\frac{8{\cal G}}{(2\pi)^{D-2}}
\frac{D-3}{D-2}\frac{\gamma^2m^2v^4
\sin{\theta_1}^4}{(1-v^2\cos^2\theta_1)^2}\times \omega^{D-4}\,,
\label{specpaircreat}
\end{equation}
and the total frequency integrated energy per solid angle is
\begin{equation}
\frac{dE}{d\Omega}=\frac{8{\cal G}}{(2\pi)^{D-2}(D-2)}\frac{v^4
\sin{\theta_1}^4}{(1-v^2\cos^2\theta_1)^2}\times \frac{(m
\gamma)^{D-1}}{\hbar^{D-3}}\,. \label{specpaircreatintfreq}
\end{equation}
For example, in four dimensions and for pairs with $v \sim 1$ one
obtains
\begin{equation}
\frac{dE}{d\omega} =\frac{4{\cal G}}{\pi} \gamma^2m^2 \,,
\label{specpair4d}
\end{equation}
and will have for the total energy radiated during production
itself, using the cutoff frequency (\ref{cutofffrequency})
\begin{equation}
\Delta E =\frac{4{\cal G}}{\pi} \frac{\gamma^3m^3}{\hbar} \,.
\label{totenergpair4d}
\end{equation}
This could lead, under appropriate numbers of $m$ and $\gamma$ to
huge quantities. Although one cannot be sure as to the cutoff
frequency, and therefore the total energy (\ref{totenergpair4d}),
it is extremely likely that, as was verified experimentally in
beta decay, the zero frequency limit (\ref{specpair4d}) is exact.

\section{Summary and discussion}
We have developed the formalism to compute gravitational wave
generation in higher $D$ dimensional spacetimes, with $D$ even.
Several examples have been worked out, and one cannot help the
feeling that our apparently four dimensional world is the best one
to make predictions about the intensity of gravitational waves in
concrete situations, in the sense that a small variation of
parameters leads in high $D$ to a huge variation of the energy
radiated.  A lot more work is still needed if one wants to make
precise predictions about gravitational wave generation in $D$
dimensional spacetimes. For example, it would be important to find
a way to treat gravitational perturbations of higher dimensional
Schwarzschild black holes.  One of the examples worked out, the
gravitational radiation emitted during black hole pair creation,
had not been previously considered in the literature, and it seems
to be a good candidate, even in $D=4$, to radiate intensely
through gravitational waves.